%% file: thesis_astro-ph.tex
\def\fileversion{v1.0n} \def\filedate{1997/09/15}
\documentclass[11pt,oneside]{book}
\pdfoutput=1
\usepackage[fullpage]{ckim_final}

\usepackage{graphicx}
\usepackage{macros}

\begin{document}

\title{Surveys of the Galactic Center and the Nature of the Galactic Center Lobe}
\author{Casey James Law}
\department{Physics and Astronomy}
\phdthesis
\degreeyear{2007}

\maketitlemain

\frontmatter

\setcounter{page}{2}

\include{abstract}

\include{acknowledgements}

\newpage
\leavevmode\vfill
\begin{center}
To my parents, Curtis and Jamie, for showing me how it can be fun to learn \\ and my wife Jessica for pushing me to do so.
\end{center}
\vfill


\tableofcontents
\listoftables
\listoffigures

\mainmatter

\include{introduction_astro-ph}
\include{gcsurvey_gbt_thesis3_astro-ph}
\include{gcl_recomb_thesis2_astro-ph}
\include{gcl_vla_thesis2_astro-ph}

\include{gcl_vlapoln_thesis2_astro-ph}

\include{gcl_all_thesis2_astro-ph}
\include{summary_astro-ph}


\backmatter

\bibliography{thesisbib}

\include{cv}

\end{document}

%% file: abstract.tex
\begin{center}
\leavevmode

ABSTRACT

\vspace{0.5cm}

Surveys of the Galactic Center and the Nature of the Galactic Center Lobe

\vspace{0.5cm}

Casey James Law

\end{center}

The Galactic center (GC) is a dense and chaotic region filled with unusual sources, such as intense star forming regions, dense star clusters, nonthermal radio filaments, and a massive black hole.  The proximity of the GC region makes it an ideal place to study the unusual processes that tend to manifest themselves in Galactic nuclei.  This thesis uses single-dish and interferometric radio continuum, radio recombination line, polarized radio continuum, and mid-IR observations to study the wide variety of physical processes seen in the GC region on physical scales from 0.1 to 100 parsecs.  These observations provide one of the most sensitive and complete studies of the cm-wavelength continuum emission in the central 500 parsecs.  I also study the distribution of radio spectral index and polarized emission from nonthermal radio filaments, which can constrain their origin and the structure of the magnetic field in the GC region.

The presence of massive star clusters and a massive black hole suggest that starburst and AGN phenomena can manifest themselves in our Galaxy.  This thesis explores this possibility by studying a 150-pc-tall, shell-like structure called the Galactic center lobe (GCL).  Our multiwavelength observations of the central few hundred parsecs of the Galaxy examine the spectral index, dust emission, polarized continuum emission, and ionized gas throughout the GCL.  I find strong evidence supporting the idea that the GCL is a true three-dimensional shell located in the GC region with nested layers of ionized, magnetized, and mid-IR--emitting componenets.  I compare the physical conditions of the GCL to proposed models for its origin and find best agreement with starburst outflows seen in other galaxies, yet consistent with the currently observed pressure and star formation rate in the central tens of parsecs of our Galaxy.

\vfill

%% file: acknowledgements.tex
\chapter*{Acknowledgments}

Many people have made significant contributions to this thesis and without them, it could not have been completed.  Although there are more contributors than can be listed here, I would like to single out a few.  Apologies for omissions.

My advisor Farhad has helped throughout this work, from stimulating ideas for observations, advice in writing proposals, contributions to data analysis, and useful comments on writing style.  His consistent support and pressure to work harder has made a positive impact on my thesis and my career as a scientist.

I also acknowledge the staff at the National Radio Astronomy Observatories for their tireless support of observers.  I am especially grateful for contributions from Bill Cotton, Ron Maddalena, Eric Greisen, all of the Analysts, Joan Wrobel, and Miller Goss.

For helping during observations and discussing results of the analysis, I would like to thank Don Backer, Jack Hewitt, Doug Roberts, Mark Morris, and Bryan Gaensler.  Thanks to Jack for making me practice my AAS thesis presentation until it was actually presentable.  Also thanks to my wife, Jessica, for sparing others from my bad grammar and spelling through her diligent editing.

I am grateful to Antonella Fruscione, Aneta Siemiginowska, and everyone at the \chandra\ X-ray Center for supporting my desire to return to graduate school.

Finally, a big ``thank you'' to all my friends at Northwestern for making graduate school a whole lot easier.  I will especially remember the good times shared with my fellow denizens of the round room in Dearborn Observatory.

%% file: introduction_astro-ph.tex
\chapter{Introduction}
\label{introduction}
\begin{figure}
\includegraphics[width=\textwidth]{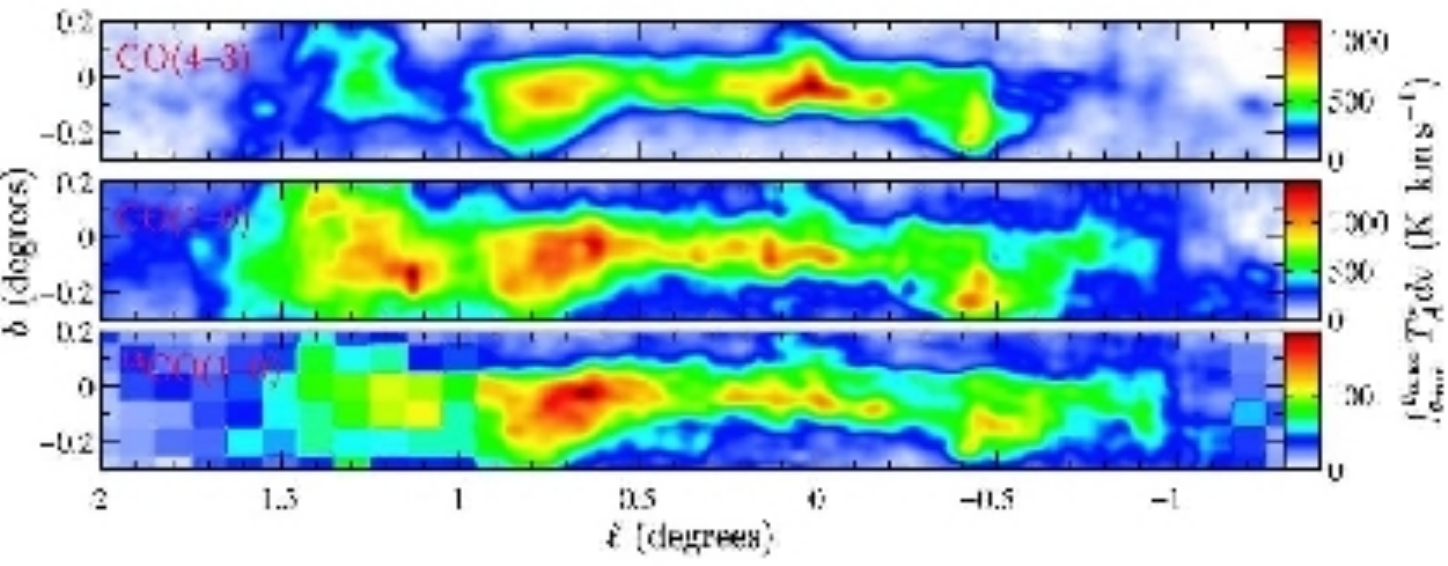}
\caption{Integrated line emission for three molecular transitions in the GC region from Fig. 2 of \citet{ma04}.  The molecular gas traced by these species is relatively dense, so the region occupied by the gas is referred to as the central molecular zone. \label{martin_cmz}}
\end{figure}

The central few hundred parsecs of the Milky Way comprise a region in the Galaxy unique for its high stellar density, intense ionizing radiation field, massive black hole, gas and stellar dynamics, and unusual magnetized structures \citep{f04,r01,g05,b87,y04}.  The extent of the GC region is roughly 400 pc in diameter, defined by a region with relatively high density gas \citep[$n_{\rm{H_2}}\gtrsim10^4$ cm$^{-3}$;][]{b87}.  This region, sometimes called the ``central molecular zone'' (Fig. \ref{martin_cmz}), produces 5\%--10\% of the Galaxy's infrared and Lyman continuum luminosity and contains 10\% of its molecular gas \citep{b87,m96}.  Sometimes the extent of the GC region is defined more narrowly as the region occupied by the most active star-forming regions spanning the central 200 pc \citep[$\sim$1\ddeg4 for a distance of 8 kpc;][]{r93}.

Interest in studying the GC region has been stimulated by the unusual collection of phenomena observed there.  That this region seems unique is a direct consequence of its being the nearest galactic nucleus.  In general, galactic nuclei are known to host intense star formation, massive black holes, and other phenomena not found in galactic disks \citep{j05}.  As discussed below, this is largely due to dynamical processes that cause gas to migrate toward a galaxy's center and collect there \citep{bi91}.  While the GC region is not unique among galactic nuclei in this regard, its distance of 8 kpc is about 100 times closer than the nearest comparable galactic nucleus \citep[M31, at a distance of $\sim800$ kpc;][]{ri05}.  Thus, our galactic nucleus may be observed with unusually high physical resolution.  This unique view explains why the GC region has one of the best examples of a massive black hole and why it is the only place known to have nonthermal radio filaments \citep{g05,y04}.

\section{Observing the GC Region}
Like any topic in astronomy, observations of the GC region have a unique set of constraints and challenges.  Limitations for observing the GC region exist because it is seen through roughly 8 kpc of the Galactic disk, which poses two specific challenges:  extinction and confusion.  The GC extinction has been measured by various techniques to be $A_V\approx 30$ mag \citep{c96}, $A_{K}\approx 3$ mag \citep{f99}, or $N_{\rm{H}}\approx 6\times10^{22}$ cm$^{-2}$ \citep{m04}.  For wavelengths longer than the near-IR and shorter than the UV, the absorption is small enough that current detectors can observe the GC region.  Radio, far-IR, and X-ray observations have made the most significant contributions to our understanding of GC physics.  In particular, the fact that radio observations are essentially unaffected by extinction (though related problems exist) and are sensitive to a wide range of physical processes makes them a popular way to study the GC region.

Confusion limits observations of the central few hundred parsecs of the Galaxy, since that field also includes sources in the 8 kpc of Galactic foreground, plus the Galactic and extragalactic backgrounds.  One way to distinguish between GC and non-GC sources is to measure source velocities.  Galactic surveys of atomic hydrogen and molecular gas have established reliable models of the dynamics of the Galactic disk \citep{st92,d01}.  Most of the inner Galaxy is observed along a line of sight perpendicular to the Galactic rotational motion, so that gas has relatively small apparent velocities \citep[$-60<v_{\rm{LSR}}<20$;][]{ma04} and can be isolated from higher velocity GC gas.  Extinction can also help locate a source in the foreground of the GC region, since there is no known class of source optically luminous enough to be detected in the GC region.  Sources with optical counterparts are more likely to be in the foreground.  Finally, statistical arguments may be used to imply that an object is in the GC region.  The central 400 pc of the Galaxy is home to 10\% of the Galaxy's molecular mass, massive stars, IR flux, and Lyman continuum flux \citep{b87,m96,f04}.  Assuming a uniform mass distribution with no extinction, we can use geometry to estimate the fraction of the Galactic disk seen in projection toward the GC region.  We estimate that roughly 1/30 of the Galactic disk is seen projected against the central 400 pc.  If on the whole, 90\% of the Galaxy's mass is in the disk and 10\% in the GC region, then toward the central 400 pc one would see $90\% \times 1/30 = 3\%$ of the disk and 10\% of the GC mass.  Thus, it is about 3 times more likely that objects seen in projection against the GC region are physically located there than in the foreground.

\section{The GC Environment}
\begin{figure}
\includegraphics[width=\textwidth]{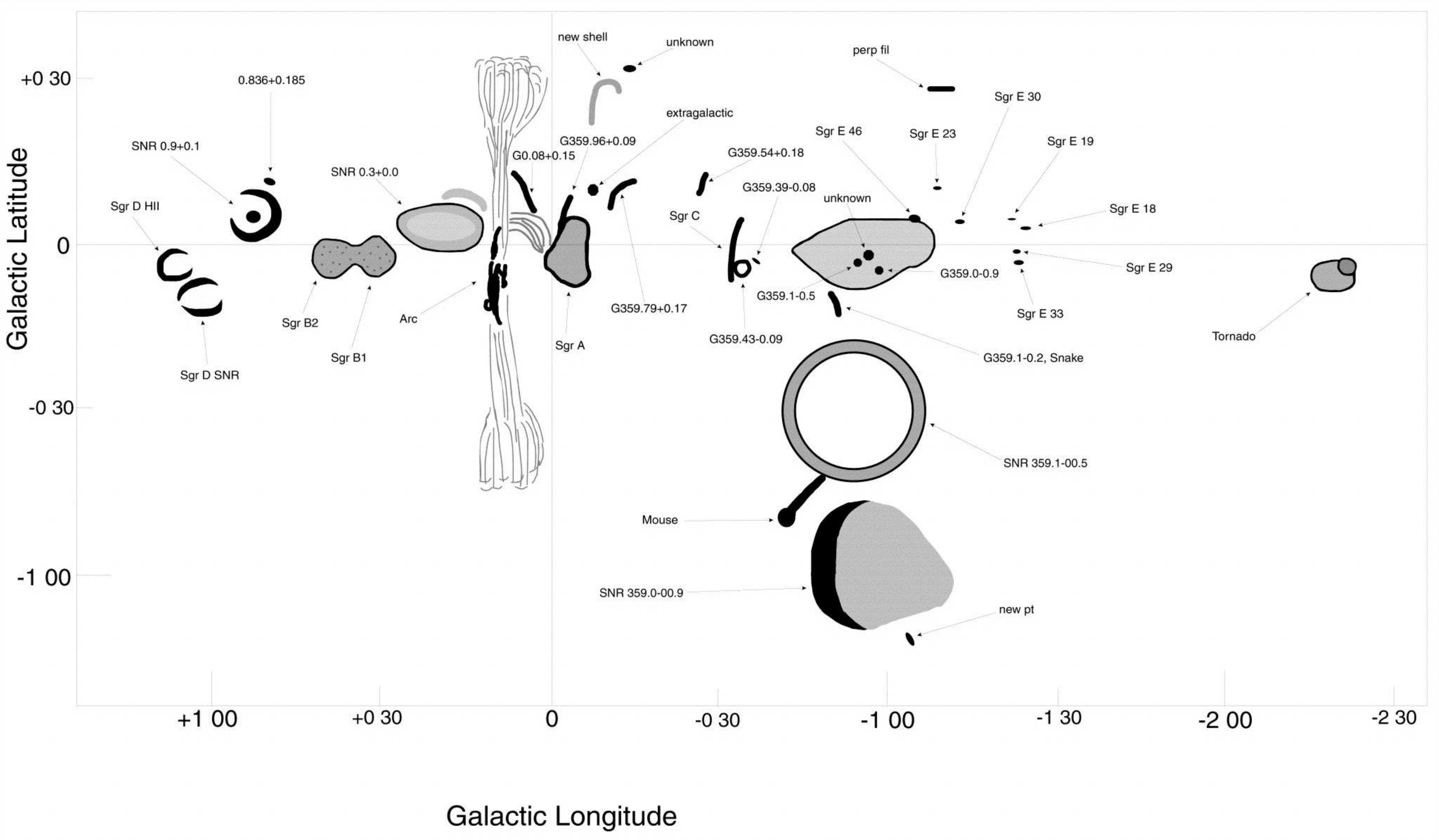}
\caption{Schematic view of the GC region at 90 cm \citep{l00}. \label{larosa_schem}}
\end{figure}

\begin{figure}
\includegraphics[width=\textwidth]{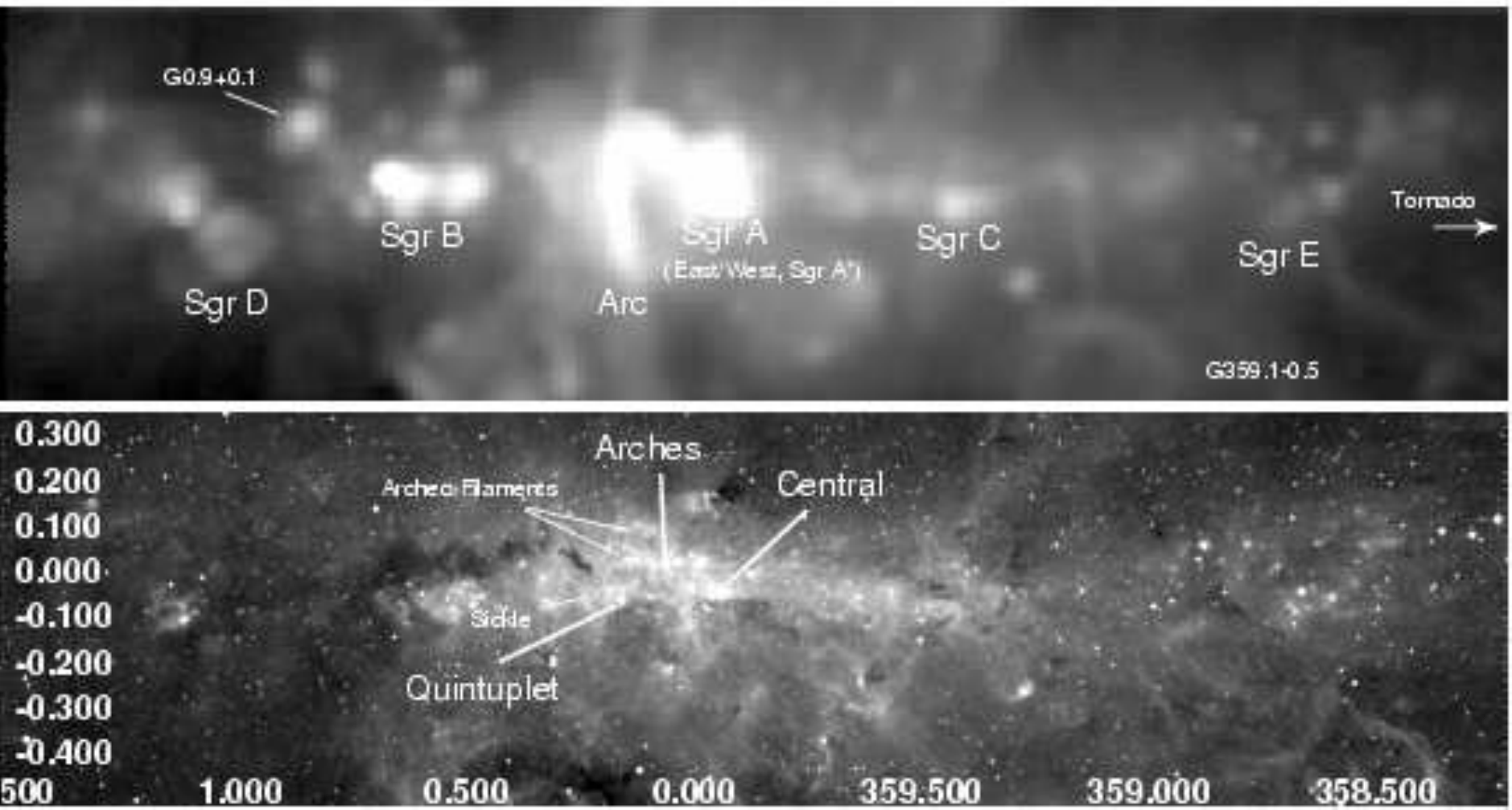}
\caption{\emph{Top}: 6 cm radio continuum emission from the central 3\sdeg\ of the Galaxy ($\sim$420 pc, assuming a distance of 8 kpc), as seen by the GBT.  Major radio continuum structures are labeled.  \emph{Bottom}: \msx 8 $\mu$m emission from an identical region with Galactic coordinates (in degrees) overlaid.  The three young star clusters are labeled.  \label{intro_schem}}
\end{figure}

The fact that the GC region is filled with a wide range of phenomena makes it both interesting and complex to study.  To reduce confusion, this section describes the taxonomy of objects in the GC region, including a discussion of its most interesting objects.  Figures \ref{larosa_schem} and \ref{intro_schem} show schematic and real views of the radio and mid-IR sources in the GC region.

\subsection{The Life Cycle of Stars and Star Clusters}
The GC region has a higher star formation density than elsewhere in the Galaxy.  The star formation rate in the central 100 pc is roughly 0.02 \msol\ yr$^{-1}$, which gives a star formation density more than 2 orders of magnitude larger than the mean Galactic star formation density \citep{f04,ro05}.  A survey of interesting objects in the GC region produces an overview of the stellar life cycle, including ultracompact \hii\ regions, dense star clusters, photoionized molecular clouds, supernova remnants (SNRs), and X-ray transients.  Here we discuss where these phenomena are observed in the GC region, beginning with examples of stellar birth and proceeding through the stellar life cycle.

Early radio continuum observations of the region resolved several bright, extended sources that were later identified as star formation regions \citep{d66,ba74,a79}.  The brightest radio continuum sources were named Sgr A through E, as shown in Figure \ref{intro_schem};  of these, all but Sgr A are extended \hii\ regions with active star formation.  The eastern part of Sgr B, called Sgr B2, is one of the most active star forming regions in the Galaxy with dozens of ultracompact \hii\ regions and star-formation--tracing masers \citep{m93,s00,d05}.  With a molecular mass of $10^7$ \msol, Sgr B2 is dense enough to host many rare molecules and is one of the first places studied in the searches for new molecules \citep{b87,n00}.  Sgr C is composed of a bright central \hii\ region ionized by the equivalent Lyman continuum flux of an O5.5-type star that is surrounded by several smaller \hii\ regions \citep{l95}. Sgr D is dominated by an SNR and bright \hii\ region, indicating ongoing star formation \citep{l92}.  Observations of Sgr E have found a complex of about 20 \hii\ regions and $2\times10^6$ \msol\ scattered through a region a half-degree across \citep{g93}.  The giant molecular clouds and star forming regions in the GC region are a part of a molecular ring that has a radius ranging from 70--175 pc \citep{s04}, with the most intense star formation occurring at GC distances of 70--80 pc (in Sgr B and Sgr C).  The location of the molecular ring is consistent with theoretical models that predict stable, roughly circular orbits inside the inner Linblad resonance, a dynamical feature associated with a barred gravitational potential \citep{bi91,st04,l06}.

In the study of GC star formation, star clusters are of particular interest since they form together and allow statistical studies of their formation environment (mass distribution, metallicity, etc.).  The development of high-resolution infrared imaging led to the discovery of three young ($\sim$5 Myr), massive ($10^4$ \msol) star clusters in the GC region \citep{n90,c96,f99}.  As shown in Figure \ref{intro_schem}, the Arches and Quintuplet clusters are located about 30 pc east of the GC (in projection), while the Central cluster occupies the central parsec of the Galaxy.  These clusters are among the most dense and massive in the Galaxy \citep[e.g.,][]{f99,cl05} but are located within roughly 30 pc of each other.  The star formation rate implied by these three clusters makes up half of the total estimated rate for the central 100 pc \citep{f04}.

Once formed, star clusters are subject to gravitational forces in the densely populated GC region.  A dynamical effect of the motion of the clusters through the ambient stellar field is to migrate closer to the GC \citep{p03,k04,gu05}.  Theoretical modeling and simulations have shown that dynamical friction against the stellar bulge causes the star clusters to lose angular momentum, potentially moving them tens of parsecs over a few Myr \citep[especially for massive clusters or clusters with an intermediate-mass black hole;][]{gu05}.  The gravitational tides across a cluster can also strip its outer members and effectively ``evaporate'' the cluster.  Thus, dynamical processes may create a means for populating the central tens of parsecs with young stars.  As discussed above, observations have found dense, star-forming molecular clouds in a ring at a GC distance of 70--175 pc \citep[e.g.,][]{m93,st04}.  Star clusters that form at these distances could migrate toward the GC while evaporating, spreading young stars through the central 100 parsecs.

The interaction of hot stars with molecular gas creates several unusual \hii\ regions in the GC region.  The Arched Filaments are thermal, filamentary structures near $(l, b)=(0.1, +0.05)$ that seem to ``connect'' Sgr A and the Radio Arc \citep{y84}.  Radio, mm-wavelength, and infrared observations have shown that the Arched Filaments are the near side of a molecular cloud being photoionized by the Arches star cluster \citep{l02,co05}.  The Sickle and Sgr A West are \hii\ regions that are being photoionized by the Quintuplet and Central star clusters, respectively \citep{y84,l97,r93}.  These two \hii\ regions are interesting because both have been found to have the potential to form stars (Christopher et al. 2005; A. Cotera et al. 2006, in preparation).  In the case of the Sickle, finger-like condensations are visible in high-resolution mid-IR images, while molecular line observations of Sgr A West suggest that the gas contains dense clumps that should be able to collapse to form stars.  These objects highlight the possibility that the intense radiation from star clusters can stimulate star formation by shocking molecular clouds \citep[e.g.,][]{w98}.

Radio continuum observations of the GC region also show several examples of the remnants of stellar death.  Figure \ref{intro_schem} shows G359.1--0.5, a shell-like SNR that may be interacting with molecular gas in the GC region \citep{u92,y95}.  G0.9+0.1 is one of the first ``composite'' SNRs discovered, with a shell and a Crab-like pulsar wind nebula inside \citep{h87}.  G357.7--0.1, also called the Tornado is an SNR candidate located in the GC region with an twisted, axisymmetric shape \citep{c76}.  Its unusual morphology has inspired a wide range of hypotheses for its origin, including an SNR, an accreting binary system, or the interaction of an SNR with a biconical stellar outflow \citep{b03a,b85,m87}.  A final example of an interesting GC SNR is Sgr A East, a shell-like SNR that surrounds Sgr A* in projection \citep{e83}.  The coincidence of Sgr A East with Sgr A* has led to speculation that their interaction can explain wide-ranging phenomena, such as Sgr A*'s unusually low luminosity \citep{m02} or the presence of intense gamma-ray and cosmic-ray emission from the central parsecs \citep{m98,a06}.

The GC region has several known X-ray and radio transients, examples of stellar ``afterlife'' \citep{z92,h05,mu05}.  These sources have correlated X-ray and radio emission with orders-of-magnitude intensity variations on time scales of weeks to months.  They are generally believed to be neutron stars or black holes accreting mass from a binary companion \citep[e.g.,][]{f01}.  A remarkable, X-ray/radio transient located roughly 0.1 pc from Sgr A* has been observed to launch jets, which suggests that these objects could interact with Sgr A* or its accretion disk \citep{bo05}.  A recent \emph{Chandra} observing campaign of the central 23 pc of the GC has discovered seven X-ray transients and found evidence for an overdensity (per unit mass) of X-ray transients in the central parsec relative to outside that region \citep{mu05}.  This overdensity of transients is consistent with the idea that dynamical friction brings stellar-mass black holes and neutron stars into the region, such that they contribute roughly 1\% of the stellar number density in the central parsec \citep{mu05}.

Perhaps the most interesting compact object in the GC region is the massive black hole at the center.  Interferometric radio continuum observations with resolutions near an arcsecond discovered Sgr A*, which was the best early evidence for a massive black hole in the GC \citep{b74,go03}.  Initially, the identification of Sgr A* as a massive black hole was based mostly on the fact that the source's brightness temperature was much higher than expected from a thermal source \citep[$T_b>10^7$ K;][]{b74,d76}.  However, the strongest test of the nature of Sgr A* came with high-resolution infrared observations of stars orbiting Sgr A* \citep{sc03,g05}.  Dozens of stars have been tracked for more than 10 years, four of which have orbital periods of order tens of years.  The orbits of these stars require a mass of $3.7\pm0.2\times10^6$ \msol\ contained in a volume of radius 45 AU \citep{g05}.  The orbits of some stars around Sgr A* may be close enough that general relativistic effects, such as precession of the periapse or gravitational redshift, may be observable with future observations \citep{f00}.  An ongoing topic of study is the unusually low luminosity of Sgr A* and the implications for accretion physics in compact objects \citep{m01,li02,bo05,mar05}.  

These objects cover the range of the stellar life cycle, from stellar birth, the lives of stellar systems, and the interaction of stars with molecular clouds, to stellar death and beyond.

\subsection{Nonthermal Radio Filaments}
Early, high-resolution radio continuum imaging of the GC region found a set of remarkable, straight filaments with lengths of tens of pc located about 30 parsecs east of Sgr A \citep{y84}.  This complex (called the Radio Arc) has a nonthermal spectrum and a polarization fraction of tens of percent \citep{y88}.  The individual filaments of the Radio Arc have widths of a fraction of a parsec and are located near the radio-bright Sgr A complex, so high-fidelity, interferometric imaging with the VLA was required to detect them.  Further observations found that the Radio Arc is a part of a larger class of objects called nonthermal radio filaments (NRFs).  NRFs are distinguished by their high aspect ratios (greater than 10:1) and synchrotron emission \citep{y84,n04,y04}.  These filaments have only been seen in the GC region, but have been found there in abundance, with about 15 known and several dozen candidates in the central 2\sdeg\ of the Galaxy.

The unusual nature of the NRFs has inspired a wide range of theoretical models to explain their origin \citep{t86,h88,s94,s99,y03,b06}.  Generally, the models can be separated into two groups based on their assumption for the strength and organization for the GC magnetic field.  One class of models assumes that the GC magnetic field has milligauss strength throughout the GC region and has a poloidal (dipole-like) structure.  In these models, the NRFs are formed where energetic particles are accelerated \citep[e.g., by magnetic field reconnection;][]{s94}.  These models are supported by observations of the interaction of the NRFs with molecular gas \citep{y87,c03} and by the trend for the brightest NRFs to be oriented perpendicular to the Galactic plane \citep{m96}.  A second class of models assumes that the GC magnetic field has a strength of 10--100 $\mu$G on large scales and that the NRFs are regions where the field is locally enhanced.  In these models, the appearance of NRFs are a natural consequence of the interaction of ambient electrons with the enhanced magnetic field \citep[e.g.,][]{s99,b06}.  These models are supported by observations of faint NRFs that have no preferred orientation \citep{y04} and by upper limits on cosmic-ray electron energy density in the GC region \citep{la05}.  The debate between these two models for the GC magnetic field is ongoing.

\subsection{Ambient GC Environment}
The ambient conditions in the GC region also differ from those of the rest of the Galaxy.  Early observations found that the molecular gas was very turbulent and unusually warm \citep[$T\sim70$ K;][and references therein]{m96}.  The high temperature is found throughout the central molecular zone, which suggests that it is a general property of the GC region.  Possible origins for the heating are tidal shearing from rotation and viscous magnetic heating \citep{w82,m96}.  The standard model for the GC region assumes that dense ($n\geq10^4$ cm$^{-3}$) gas has a filling fraction of about 10\%, although more recent observations suggest that the dense, cool gas has a filling factor an order of magnitude lower \citep{m96,o05}.  The new observations require most of the GC molecular gas to have densities $n\sim100$ cm$^{-3}$.

Another significant component of the GC ISM (by volume filling fraction) is hot, X-ray--emitting gas.  Gas with temperatures greater than $10^6$ K emit thermally in the X-ray regime ($0.1\lesssim E\lesssim 100$ keV).  Diffuse X-ray emission has been observed throughout the GC region to have two distinct components with temperatures of roughly 0.8 and 8 keV or $10^7$ and $10^8$ K \citep{k96,m04}.  The hot component is more spatially extended, suggesting that it occupies a larger volume than the cool component by a factor of 10--100 \citep{m04}.  While the properties of the cool component are consisitent with its being heated by supernova shock waves, the hot component seems to require a more exotic (not yet identified) process.  Furthermore, the hot component is too hot to remain bound to the Galaxy [$c_s(8\ \rm{ keV})\approx 1500$ \kms$> v_{\rm{esc}}\approx 900$ \kms; Breitschwerdt et al. 1991], so maintaining it requires at least two orders of magnitude more power than maintaining the cool component \citep{m04,be05}.

The mass distribution as a function of radius from the GC has been measured by several methods.  The distribution of IR light traces the density of K and M giants and increases roughly linearly with radius from one to a few hundred pc \citep[][and references therein]{g87}.  If the enclosed mass varies linearly with $r$, then the mass density varies as $r^{-2}$, which is consistent with a spherically symmetric mass distribution.  Observations of gas dynamics show a roughly flat rotation curve for radii from 10 to 100 parsecs, which also implies a linear increase in the enclosed mass as a function of radius \citep[e.g.,][]{b87}.  Stellar dynamics can be traced by studying the position and velocity distribution of OH/IR stars \citep{li92}.  OH/IR stars are evolved giant stars that are bright in the IR and can be host to OH masers, which makes them convenient tracers of stellar dynamics \citep{b98}.  \citet{li92} used the VLA to study these stars and found results consistent with the gas dynamics and the IR light distribution for $10<r<100$ pc.  This work also shows directly that the stellar number density varies as $r^{-2}$, thus confirming that the stars have a spherically symmetric distribution and dominate the gravitational potential for the region.

\section{The Galactic Center Lobe}
\label{gclintro}
\begin{figure}[htb]
\includegraphics[width=\textwidth]{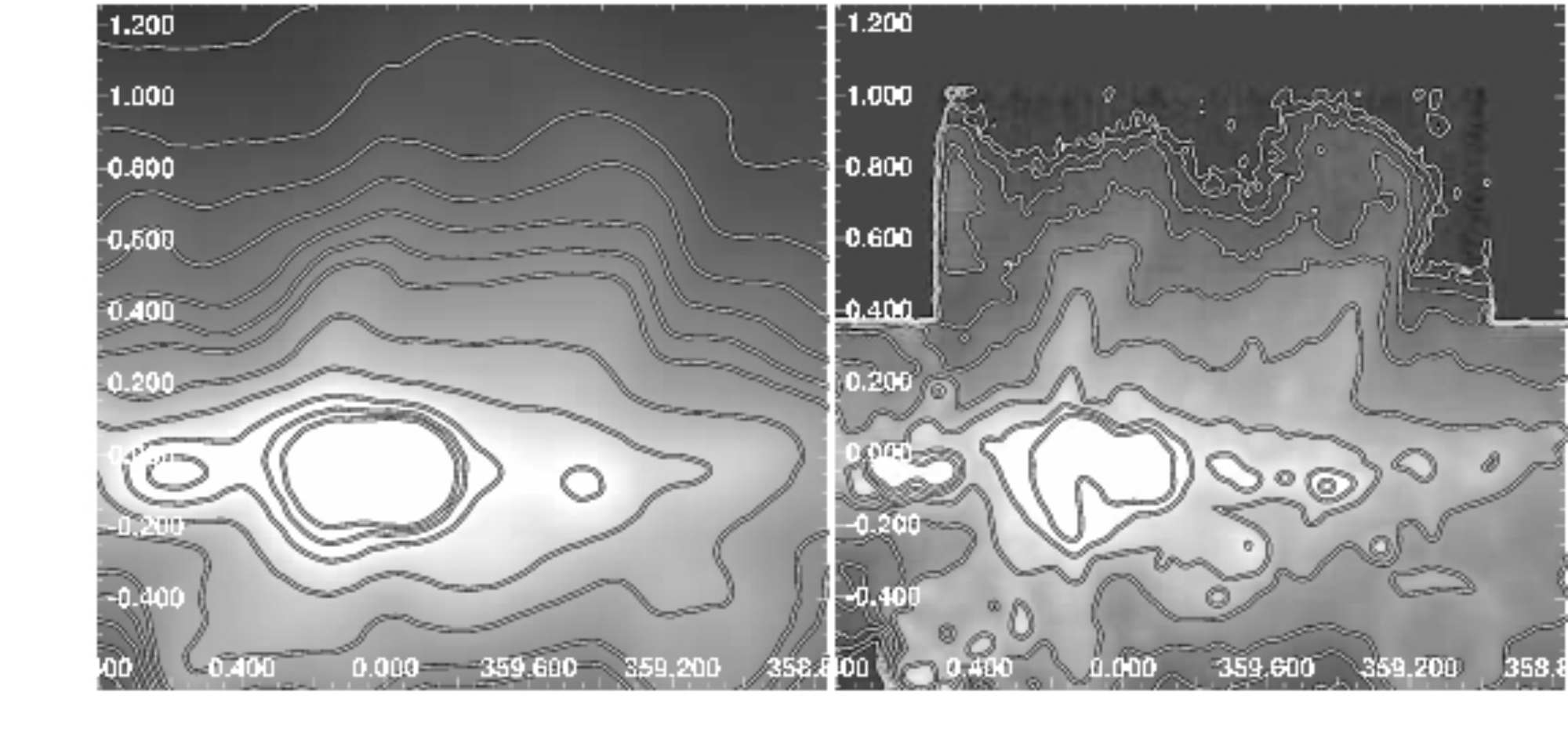}
\caption{\emph{Left}:  20 cm GBT image of the GCL with a resolution of 9\arcmin.  Contours are at levels of 12, 14, \ldots, 24, 30, 40, 60, 80, 100 Jy beam$^{-1}$.  \emph{Right}:  6 cm GBT image of the GCL with a resolution of 2\damin5.  Contours are at levels of $0.02\times2^n$ Jy beam$^{-1}$, with $n=0$--8.  \label{intro_gcl620}}
\end{figure}

Aside from studying the general properties of the GC region, this work has a focus on an object called the Galactic center lobe (GCL).  The GCL is a roughly 1\sdeg\ tall, loop-like structure that spans the central degree of our Galaxy in radio continuum emission (Fig. \ref{intro_gcl620}).  This section briefly describes the history of observations of the GCL, some proposed models for its formation, and current challenges to understanding its nature.

\subsection{Discovery and Initial Work}
The GCL was discovered in a 10 GHz radio continuum survey of the GC region \citep{s84}.  That work noted the distinctive loop- or shell-like morphology that is similar to radio lobes seen in extragalactic outflows and radio galaxies.  The observations used a spatial filtering technique to make high-resolution images of the region and measure spectral indices.  That initial work found a thermal-like spectral index ($\alpha\approx-0.1$, for $S_\nu\propto\nu^\alpha$) for the GCL near 10 GHz.  The shell-like morphology and thermal spectral index suggested that the GCL was thermal gas associated with some kind of outflow from the GC region.  Assuming that the radio continuum emission was thermal, the mass in the gas north of $b=0\ddeg2$ was estimated to be $4\times10^5$ \msol\ \citep{s85}.

The coincidence of the GCL with the massive black hole, Sgr A*, has inspired formation models related to jet activity \citep[e.g.,][]{m01,ga05}.  Accretion of mass onto a black hole generally results in the formation of energetic jets that expel the majority of the accreted mass \citep{m01}.  In principle, if only a small fraction of the mass inflow to the central parsecs of the Galaxy were to accrete onto Sgr A*, it could result in the formation of a significant jet \citep{bi91,m01}.  Although no such jet is observed today, early work suggested that the GCL was formed by a jet at some time in the past, which has since weakened \citep{s85}.  Evidence for a jet on scales as large as the GCL would be interesting, since Sgr A* is generally less luminous and less variable than typical AGNs \citep{m01,m05}.

The GCL also spans the central 100 pc of our Galaxy, the region with the highest stellar density in the Galaxy \citep{m96,f04}.  It has been suggested that the collective energy input from stellar winds and supernovae in the central parsec could create a supersonic flow expanding beyond 100 pc \citep{c92}.  Although the star formation rate in the central 100 pc is much less than observed in starburst nuclei, the observed stellar winds seem to be strong enough to result in some mass outflow, according to the theory of \citet{c85}.  The radial dependence of gas pressure is also consistent with this model \citep{c92}.

Another early model for the formation of the GCL invokes magnetodynamic processes \citep{u85,su87}.  This model predicted that Galactic rotation and infall of matter in the central 100 pc would bend the magnetic field according to the so-called sweeping magnetic twist effect.  This model assumes that the GC magnetic field initially has a polodial orientation, which, once twisted, can lift ionized gas from the Galactic disk by the ${\bf J\times B}$ force.  Numerical simulations show that for typical conditions in the GC region and a flat rotation curve, the magnetic twist effect can propel gas up to $\sim100$ \kms\ and create a shell-like structure like the GCL \citep{su87}.  This model has the advantage of explaining the coincidence of the Radio Arc with the eastern edge of the GCL and its rotation measure, since the model predicts that both structures are a part of the same distorted magnetic field \citep{t86}.

In general, the fact that the GCL is only observed north of the Galactic plane is ascribed to either confusion or intrinsic asymmetry in the outflow.  Confusion certainly contributes to the difficulty of detecting anything that is within a degree of the GC region and south of the plane.  There are two large \hii\ complexes found in that region that are in the foreground of the GC (see Ch. \ref{gcsurvey_gbt}).

\subsection{Contradiction}
After the discovery of the GCL and the development of initial models, there were observations that showed problems with the early interpretations.  The criticisms suggested that either (1) the eastern and western parts of the GCL are unrelated to each other, or (2) the features associated with the GCL are not caused by an outflow.

Several observations of the GC region have found differences in the properties of the GCL-East and GCL-West \citep{t86,h92,u94,b96,s04}.  Polarized radio continuum observations found that the GCL-East had polarization fractions up to 40\% covering Galactic latitudes up to $0\ddeg7$ \citep{t86,h92}.  However, in the GCL-West, the polarization fraction was much smaller, with upper limits around 5\%--10\%.  The relative lack of polarization in was strengthened by observations of absorption in 74 MHz VLA observations coincident with the GCL-West \citep{b03b};  the absorption requires there to be thermal gas in the GCL-West.  Extragalactic radio lobes are predominately nonthermal with a symmetric polarization distribution \citep[e.g.,][]{e98}, so this asymmetry suggested that the GCL-East and GCL-West were unrelated.  

The GCL-West also seemed to be associated with an unusual mid-IR--emitting molecular cloud, called AFGL 5376 \citep{u90,u94}.  Extensive mid-IR and molecular line observations suggested that the molecular gas associated with AFGL 5376 was being shocked and heated, perhaps by collision with another cloud or ambient Galactic gas.  This shock could, in turn, ionize some of the gas and create the radio continuum emission associated with the GCL-West.  This model was consistent with the lack of polarized emission seen in the GCL-West and the lack of a molecular counterpart in the GCL-East \citep{u94}.  The difference between the GCL-East and -West is strengthened by the fact that the GCL-West (with AFGL 5376) is located at the western edge of a kinematic feature in the molecular gas in the central 3\sdeg\ of the GC region \citep[sometimes called the expanding molecular ring;][]{b87}.  The origin of this peculiar kinematic feature is most likely related to the motion of gas in a barred gravitational potential, which helps explain the peculiar kinematics of the molecular gas associated with the GCL-West and AFGL 5376 \citep{bi91,u90}.  Since no such coincidence exists near the GCL-East, it is difficult to explain the radio continuum emission there by any similar process.

Furthermore, it has been suggested that the emission associated with the GCL does not require something as dramatic as an outflow, but may be explained by more benign processes.  The GCL shell is not centered on the GC and has no counterpart in the southern Galactic plane, which suggests that it might not be associated with the GC region.  For example, the GCL could be a foreground SNR that coincides with the GC region.  If it is in the GC region, the mysterious NRFs could create the radio continuum emission of the GCL.  There is no widely accepted model for the origin of the NRFs, but many models for their creation would have nothing to do with a GC outflow \citep[e.g.,][]{b88,d02}.  Finally, early criticism of models for the GCL noted that there is little evidence for unusually energetic behavior in the GC region.  The star formation rate is not as large as observed in starburst galaxies, and Sgr A* is known to have an unusually low luminosity with no direct evidence for a parsec-scale jet \citep{m01,f04}.

\subsection{Resurrection}
The idea that the GCL was a sign of some kind of GC outflow languished until recent, high-resolution, mid-IR observations \citep{b03}.  The \emph{Midcourse Space Experiment} \citep[\emph{MSX};][]{p01} survey of the Galactic center at 8 $\mu$m found filamentary structures that are coincident with the GCL-East and GCL-West.  The mid-IR emission is aligned in filaments several arcminutes long, parallel and adjacent to the radio continuum emission of the GCL.  Interestingly, there is also some indication of a similar mid-IR filament south of the Galactic plane.  The detection of mid-IR emission was the first time that any similarity was found between the GCL-East and GCL-West outside of the radio continuum morphology.  

The mid-IR emission strengthened the idea that the east and west sides of the GCL had a common origin and motivated new thought on its formation and effect on the GC environment.  \citet{b03} modeled the mid-IR emission as shock-heated dust in a starburst outflow.  They estimated a total molecular mass of $5\times10^5$ \msol\ in two hemispheres and, assuming a velocity of 100 \kms, found a lower limit to the kinetic energy of this outflow of about $10^{54}$ ergs.  This mass and energy are comparable to those observed in small- to moderate-sized starburst outflows \citep{v05}.

Despite the strong morphological evidence in radio continuum and mid-IR observations, there are several lingering questions about the nature of the GCL.  For example, are the different polarization properties of the GCL-East and GCL-West consistent with the outflow hypothesis?  What is the kinematics of the gas in the GCL?  Which of the outflow models is most consistent with the observations?  To answer these questions in detail, we undertook new observations of the GCL, including more sensitive radio continuum, recombination line, high-resolution polarized radio continuum, and high-resolution mid-IR observations.

\section{Thesis Organization}
This thesis has two major goals:  (1) to study various aspects of the GC astrophysics through radio continuum surveys and (2) to understand the nature of the GCL.  The thesis begins in chapter \ref{gcsurvey_gbt} with a description of a multiwavelength radio continuum survey of the GC region using the Green Bank Telescope (GBT).  The radio continuum survey revealed new information about the structure and spectrum of the GCL, which inspired the GBT radio recombination line study described in chapter \ref{gcl_recomb}.  Chapter \ref{gcl_vla} describes a Very Large Array (VLA) radio continuum survey of the GCL region with an emphasis on the NRFs found there.  Chapter \ref{gcl_vlapoln} describes the 6 cm polarized radio continuum in the GCL observed by the VLA, including a discussion of the structure of the GC magnetic field.  Finally, Chapter \ref{gcl_all} synthesizes all available observations of the GCL, including a new \emph{Spitzer}/IRAC survey, to elucidate the nature of the GCL.  

%% file: gcsurvey_gbt_thesis3_astro-ph.tex
\chapter{GBT Multiwavelength Survey of the Galactic Center Region}
\label{gcsurvey_gbt}

\section{Introduction}
Single-dish radio continuum observations are a useful tool for studying the GC region.  While the relatively low resolution of single-dish observations makes them more susceptible to confusion, their sensitivity is not limited to small spatial scales, as in interferometric observations.  The lack of a bias simplifies the interpretation of the data, especially for large surveys, where emission is seen on a wide range of spatial scales.  An ideal single-dish telescope can produce images of superior quality, since they have low sidelobes and bright sources do not create image artifacts.  A major challenge to single-dish observations is the calibration of an absolute zero flux level, since radio continuum emission is observed from throughout the sky.  Nontheless, single-dish, multiwavelength surveys can robustly probe the nature of radio-emitting sources, particularly by studying their morphology and spectral index distribution.

Several other single-dish surveys of radio continuum emission from the GC region have been conducted \citep{a79,ha87,r90,h92,d95}.  \citet{a79} observed the Galactic disk from $l=60$\sdeg to the GC region at 4.8 GHz with the 100 m Effelsberg telescope.  With a resolution of 2\damin6, this was one of the first single-dish surveys of the GC region with a resolution of a few arcminutes and discovered many compact and diffuse sources.  Since that time, the Parkes 64 m and Effelsberg 100 m telescopes have been used to create complete surveys of the Galactic disk in the northern and southern celestial skies near 2.5 GHz, revealing dozens of new supernova remnant (SNR) candidates and compact \hii\ regions \citep{r90,d95}.  \citet{ha87} used the 45 m Nobeyama Radio Observatory to survey the Galactic disk (including the GC region) at 10 GHz;  at this higher frequency, Galactic synchrotron emission is weaker and thermal emission is more prominent, making the identification of \hii\ regions easier.  In the GC region, the Parkes 64 m telescope surveyed polarized continuum emission at 3.5 cm and revealed polarized plumes extending up to two degrees long and oriented perpendicular to the Galactic plane \citep{h92}.

This chapter describes the results of a multiwavelength radio continuum survey of the GC region with the GBT.  The GBT's large, unblocked aperature enables the sensitive, unbiased images required for this confused region.  No large-scale survey of the GC region with the GBT has yet been published, so the telescope's unique capabilities can shed new light on the region.  Section \ref{gcsurvey_obs} describes the observations and data reduction.  In \S\ \ref{gcsurvey_res}, the results of the survey are described, including the compilation of compact and extended source catalogs at 3.5 and 6 cm and a detailed discussion of the radio continuum properties for each source in the region.  The high resolution and sensitivity of the GBT observations allow us to identify and integrate the flux from all thermal and nonthermal emitters.  We calculate the percentage of flux from sources in the central degrees of the Galaxy from thermal/nonthermal processes at 3.5 and 6 cm.  Finally, \S\ \ref{gcsurvey_con} summarizes the results of this analysis.  A study of the continuum emission from the GC lobe is not included here, but discussed in detail in chapter \ref{gcl_all}.

\section{Observations and Data Reductions}
\label{gcsurvey_obs} 
In June 2003, we surveyed the central degrees of the Galaxy with the GBT at 3.5, 6, 20, and 90 cm.  The GBT is located in Green Bank, West Virginia at a latitude of 38\sdeg26\arcmin\ N in the National Radio Quiet Zone.  The GBT is unique in that it has a large ($\sim$100 m) unblocked aperature that is fully steerable, with an elevation range of 5--90\sdeg.   At 3.5, 6, and 20 cm, observations were made with the Digital Continuum Receiver in ``on-the-fly'' mapping mode, while at 90 cm the observations were made with the Spectral Processor.  After flagging the 90 cm data for radio frequency interference, the data were averaged in frequency and treated identically to the higher frequency data.  Observations had bandwidths of 320, 320, 20, and 40 MHz at 3.5, 6, 20, and 90 cm.

\begin{figure}[tbp]
\includegraphics[width=\textwidth]{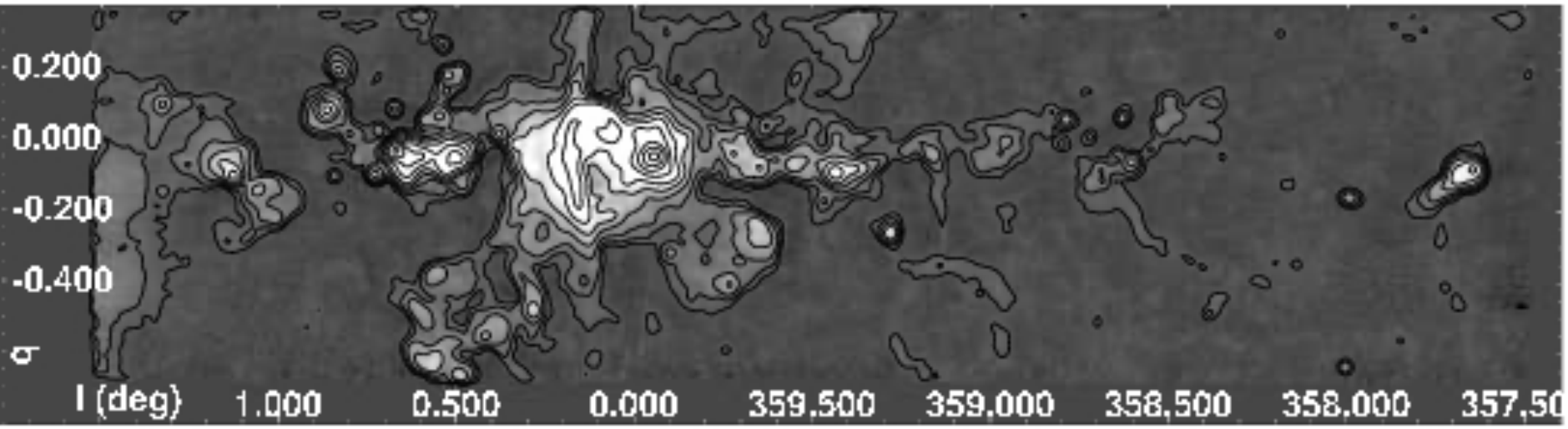}
\includegraphics[width=\textwidth]{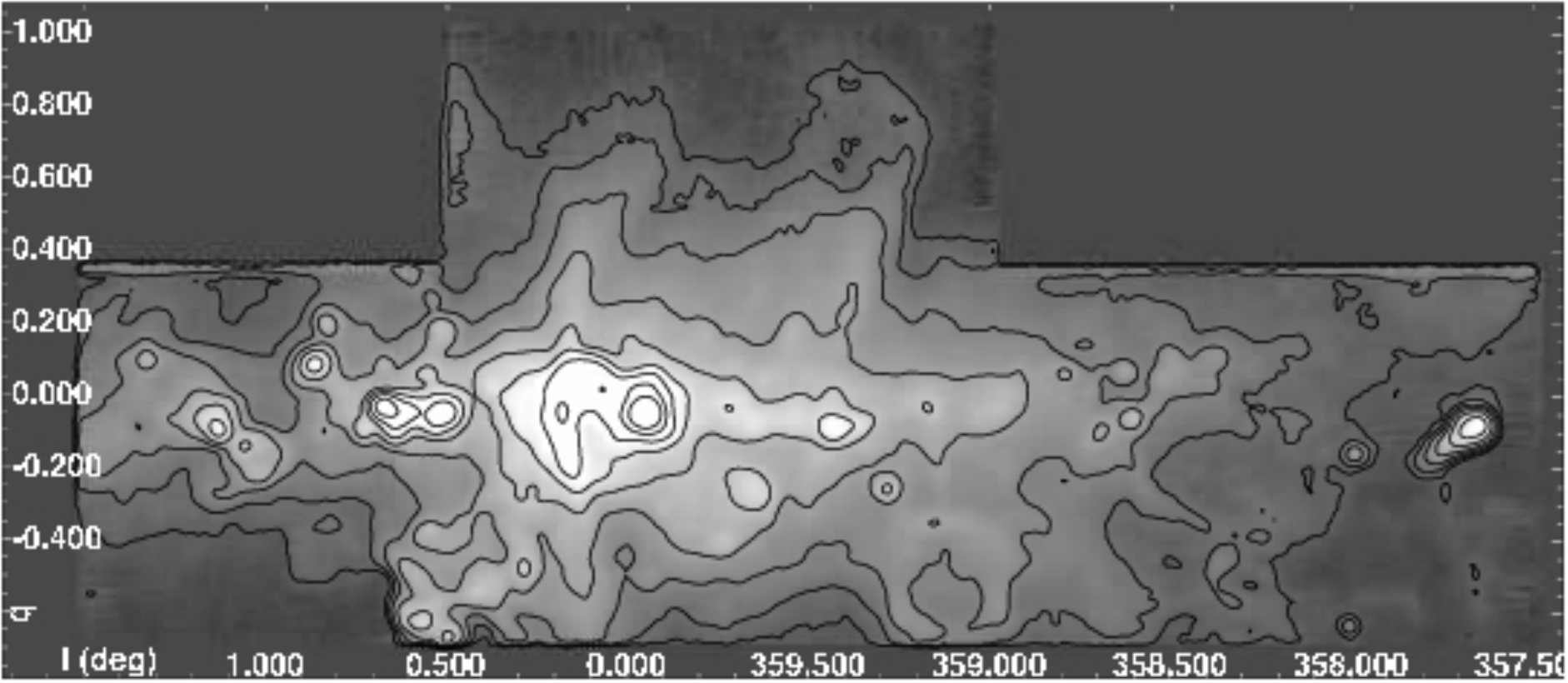}
\caption{\emph{Top and bottom}:  The entire extent of our GBT radio continuum survey of GC region at 3.5 and 6 cm, respectively.  Contours for the 3.5 cm survey are at levels of $0.05*2^n$ Jy beam$^{-1}$, for $n=0-9$ and a beam size of 88\arcsec.  The contours for the 6 cm survey are identical, but for $n=1-9$ and a beam size of 153\arcsec.  Galactic coordinates are shown on each image.  \label{mapscx}}
\end{figure}

\begin{figure}[tbp]
\begin{center}
\includegraphics[width=0.65\textwidth]{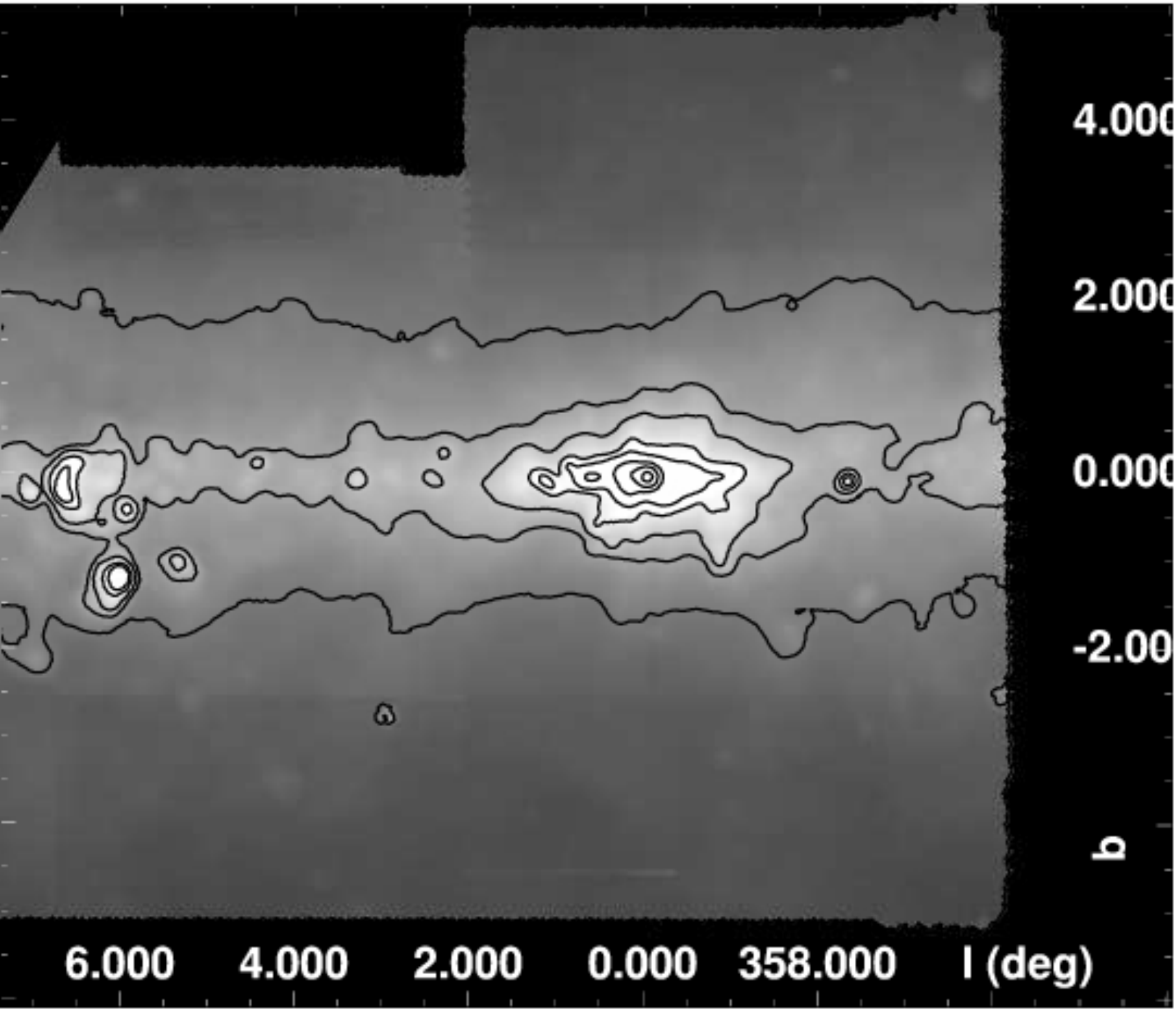}

\includegraphics[width=0.65\textwidth]{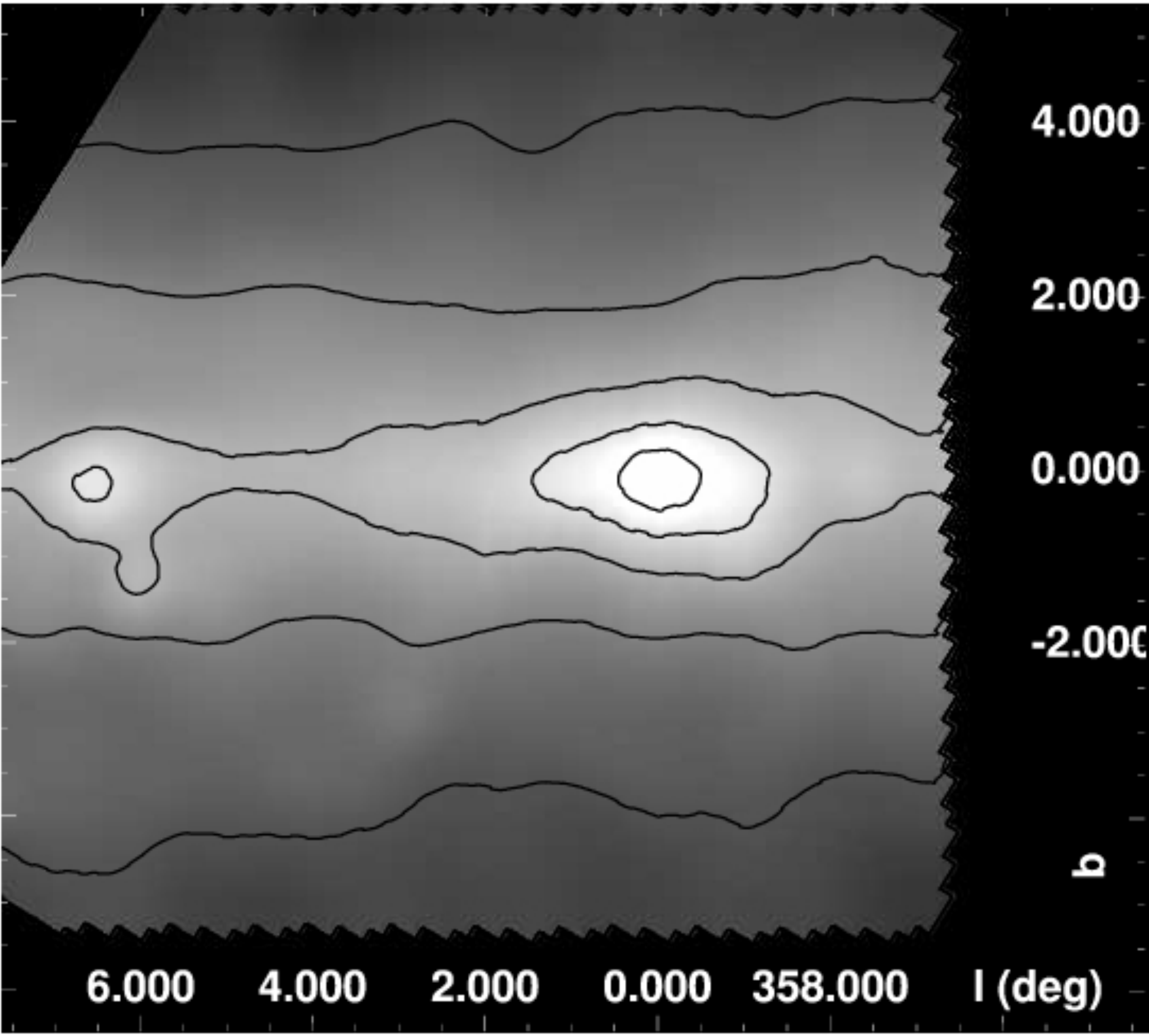}
\end{center}
\caption{\emph{Top and bottom}:  The entire extent of our GBT radio continuum survey of GC region at 20 and 90 cm, respectively.  Contours for the 20 cm survey are at 5, 10, 15, 20, 30, 40, 80, 160, and 320 Jy beam $^{-1}$, with a beam size of 9\arcmin.  Contours for the 90 cm survey are at 80, 160, 320, 640, and 1280 Jy beam $^{-1}$, with a beam size of $38\damin8$.  \label{mapspl}}
\end{figure}

All observations surveyed at least a 4\sdeg$\times$1\sdeg\ area roughly centered on the GC, but the 6, 20, and 90 cm maps also surveyed beyond this region.  The 3.5 and 6 cm surveys, shown in Figure \ref{mapscx}, have similar coverage and are largely used to study sources in the central four degrees of the GC.  The 20 and 90 cm surveys, shown in Figure \ref{mapspl}, cover the central ten degrees
The spatial coverage, resolution, and typical $1\sigma$ sensitivity of these maps are shown in Table \ref{mapstats}.

Flux calibration was done by adding a calibrated noise source to every other integration.  The brightness of the noise source is estimated from observations of 3C286 and gives a conservative absolute flux accuracy of 5\%.  The final amplitude calibration is made from the median brightness of the noise source for each scan.

No simultaneous measurement of the sky brightness was made during these observations, so a few techniques were used to estimate the noncelestial background for each map.  First, an initial esimate of the sky brightness is done by observing a position far from the Galactic plane, near 3C286.  Second, the atmospheric opacity and temperature are estimated from weather monitoring stations and subtracted from the data.  Finally, a low-pass filter is used to estimate sky brightness and ground spillover.  However, this technique cannot distinguish between slow variations in the atmospheric brightness and true celestial changes, which effectively introduces a local ``zero point''.  Thus, measurements of brightnesses and flux densities require the subtraction of a background, especially in the 3.5 and 6 cm maps, where atmospheric effects are stronger.

Each of the radio continuum maps was observed with orthogonal, ``basket-weaving'' patterns in Galactic coordinates.  The redundancy of observing each position in the map in two orthogonal scans is helpful in identifying the flux contribution from noncelestial sources.  The final images were made by convolving the data with a Gaussian kernel and resampling onto the image grid.  Maps were made with ``Obit'', a group of software packages designed to handle single-dish and interferometric radio astronomy data, such as AIPS data disks or FITS files.  ``ObitSD'' is a low-level addition to the software package and is designed for making maps from on-the-fly data.

\begin{deluxetable}{ccccccc}
\tablecaption{Overview of GBT Surveys of GC Region \label{mapstats}}
\tablewidth{0pt}
\tablehead{
\colhead{Band} & \colhead{$\lambda$} & \colhead{$\nu$} & \colhead{Resolution} & \colhead{Long. Range} & \colhead{Lat. Range} & \colhead{Sensitivity\tablenotemark{b}} \\
\colhead{} & \colhead{(cm)} & \colhead{(GHz)} & \colhead{(arcmin)} & \colhead{(deg)} & \colhead{(deg)} & \colhead{(mJy beam$^{-1}$)} \\
}
\startdata
X & 3.5 & 8.50 & 1.5 & 357.5, +1.5 & --0.7, +0.35 & 9 \\
C & 6.2 & 4.85 & 2.5 & 357.5, +1.5\tablenotemark{a} & --0.7, +0.35\tablenotemark{a} & 20 \\
L & 21.3 & 1.42 & 9.0 & 355.5, +7.6 & --5.1, +3.4 & 300 \\
P & 92.3 & 0.325 & 38.8 & 356.4, +7.6 & --5.4, +5.4 & 4000 
\enddata
\tablenotetext{a}{Coverage also extends up to $b=+0\ddeg8$ for $l=359\ddeg0$ to 0\sdeg5}
\tablenotetext{b}{Measured away from obvious sources, but tends to include background synchrotron emission at longer wavelengths.}
\end{deluxetable}
\clearpage 

\section{Results}
\label{gcsurvey_res}
The 3.5 and 6 cm surveys have sensitivities of 8 and 10 mJy beam$^{-1}$, which makes them among the highest-resolution and most sensitive single-dish surveys of the region ever made at these wavelengths.  At wavelengths near 3.5 cm, the GC region has been best surveyed with 45 m Nobeyama Radio Observatory and 64 m Parkes Telescope to sensitivities of 15 and 30 mJy beam$^{-1}$, respectively, with resolutions of about 2\damin5 \citep{ha87,h92};  the present survey surpasses this previous work in resolution and sensitivity.  The 6 cm survey of \citet{a79} used the 100 m Effelsberg telescope, which has a comparable resolution of 2\damin6 and a sensitivity of 5 mJy beam$^{-1}$, or about twice as sensitive as the present work.  An advantage of the present work is that it was planned as a multiwavelength campaign, so the observations at 3.5, 6, 20, and 90 cm were calibrated by similar techniques;  this makes the survey robust and especially suited to studying spectral indices.  With a minimum coverage of 4\sdeg$\times1$\sdeg\ at 3.5 cm, these maps provide a valuable census of a variety of Galactic sources.  In the following sections, catalogs of compact and extended sources are presented, plus a study of the diffuse background emission of the GC region.

\subsection{GBT Compact Source Catalog}
\label{compactsrcsec}
Detection of compact sources was done with JMFIT of AIPS.  JMFIT finds sources by fitting a 2-D Gaussian to an image, assuming and initial size and shape equal to the beam size;  if no source is found, the fit will not converge.  The JMFIT routine also fits for the absolute background flux and its first spatial gradient, which is particularly useful for the 6 cm maps.  The output of the source detection is a set of positions, flux densities, sizes for a best-fit Gaussian, plus a source-free image.  The residuals in the source-free image were inspected as a test of the fit quality and were generally less than about 10\% of the source peak brightness.

Source selection was done by eye, searching for compact sources that were not dominated by extended source flux.  A source is considered to be compact if its best-fit width is less than approximately twice the beam FWHM.  Only one of the 36 sources in the 3.5 cm compact catalog has a Gaussian width approaching roughly twice that of the beam (roughly 3\arcmin).  Any source that is not compact in the highest-resolution image (3.5 cm) is considered ``extended'' and described in \S\ \ref{diffsrcsec} and \S\ \ref{slicediff}.  Other source-fitting algorithms, including the IDL astrolib photometry programs and MIRIAD's sfind, were tested on our images, but neither of these produced as trustworthy results as JMFIT.  In particular, the algorithms did not easily produce source-subtracted residual images to enable visual inspection of the fit quality.  

The catalog of compact sources in the 3.5 and 6 cm images are shown in Tables \ref{srcX} and \ref{srcC}, respectively.  In these tables, columns (2)-(5) give the position of the source, columns (6)-(9) give the source peak brightness and its error, the source flux density and its error, and columns (10)-(12) give the major axis, minor axis, and position angle of the best-fit Gaussian to the source.  For comparison, column (13) shows the flux density given in \citet{h92} for sources that are coincident with sources in the present catalog.  The 20 and 90 cm images do not have any distinctly compact sources in the region covered by the 3.5 and 6 cm surveys, so no compact source catalog is given here for those images.

\begin{deluxetable}{ccccccccccccc}
\tablecaption{GBT 3.5 cm Compact Source Catalog for GC Region \label{srcX}}
\tabletypesize{\scriptsize}
\rotate
\tablewidth{0pt}
\tablehead{
\colhead{\#} & \colhead{l} & \colhead{b} & \colhead{RA} & \colhead{Dec} & \colhead{$S_p$\tablenotemark{a}} & \colhead{$\sigma_{S_p}$\tablenotemark{a}} & \colhead{$S_i$} & \colhead{$\sigma_{S_i}$} & \colhead{bmaj} & \colhead{bmin} & \colhead{bpa} & \colhead{$S_i^{H92}$} \\ 
\colhead{} & \colhead{(deg)} & \colhead{(deg)} & \colhead{(J2000)} & \colhead{(J2000)} & \colhead{(Jy/bm)} & \colhead{(Jy/bm)} & \colhead{(Jy)} & \colhead{(Jy)} & \colhead{(arcsec)} & \colhead{(arcsec)} & \colhead{(deg)} & \colhead{(Jy)} \\ 
}
\startdata
1 & 1.128    &  --0.100 & 17:48:40.29 & --28:01:23.5 & 2.29  & 1.7E-2 & 3.48  & 4.0E-2 & 115.2 & 102.3 &  24 & $8.61\pm$20\% \\
2 & 0.864    &   +0.081 & 17:47:20.97 & --28:09:20.7 & 2.22  & 1.7E-2 & 4.35  & 4.7E-2 & 127.7 & 119.0 & 115 & $7.15\pm$10\% \\
3 & 0.843    &  --0.105 & 17:48:01.39 & --28:16:13.0 & 0.20  & 1.8E-2 & 0.21  & 3.3E-2 &  94.2 &  89.8 & 106 & \\
4 & 0.728    &  --0.099 & 17:47:43.64 & --28:21:55.3 & 1.29  & 1.7E-2 & 1.72  & 3.7E-2 & 104.0 &  98.4 &  67 & \\
5 & 0.719    &   +0.022 & 17:47:14.24 & --28:18:36.1 & 0.15  & 1.8E-2 & 0.18  & 3.5E-2 & 101.2 &  92.6 &  40 & \\
6 & 0.675    &   +0.083 & 17:46:53.71 & --28:18:59.5 & 0.31  & 1.7E-2 & 0.52  & 4.3E-2 & 119.0 & 110.0 &   5 & \\
7 & 0.671    &  --0.034 & 17:47:20.50 & --28:22:51.6 &17.59  & 1.7E-2 &28.10  & 4.1E-2 & 128.5 &  96.2 &  67 & $34.9\pm$10\% \\
8 & 0.605    &  --0.195 & 17:47:48.66 & --28:31:14.0 & 0.09  & 1.7E-2 & 0.17  & 4.6E-2 & 134.4 & 108.9 &  35 & \\
9 & 0.538    &   +0.264 & 17:45:52.33 & --28:20:21.3 & 0.29  & 1.7E-2 & 0.38  & 3.6E-2 & 104.9 &  96.4 &  44 & \\
10& 0.530    &   +0.133 & 17:46:21.55 & --28:24:51.5 & 0.19  & 1.7E-2 & 0.27  & 3.8E-2 & 122.7 &  89.5 & 134 & \\
11& 0.524    &   +0.178 & 17:46:10.24 & --28:23:46.2 & 0.66  & 1.7E-2 & 1.07  & 4.1E-2 & 116.2 & 108.6 &  96 & $1.94\pm$20\% \\
12& 0.475    &   +0.066 & 17:46:29.41 & --28:29:47.6 & 0.13  & 1.7E-2 & 0.16  & 3.6E-2 & 122.9 &  80.2 &  74 & \\
13& 0.380    &   +0.017 & 17:46:27.43 & --28:36:09.1 & 1.20  & 1.8E-2 & 1.27  & 3.2E-2 &  92.8 &  88.7 &  15 & \\
14& 0.326    &  --0.015 & 17:46:27.26 & --28:39:57.1 & 0.90  & 1.8E-2 & 0.95  & 3.2E-2 &  93.3 &  87.8 & 121 & \\
15& 359.896  &  --0.319 & 17:46:37.11 & --29:11:27.2 & 0.22  & 1.7E-2 & 0.42  & 4.5E-2 & 125.0 & 115.6 &  30 & \\
16& 359.782  &   +0.035 & 17:44:57.84 & --29:06:12.4 & 0.14  & 1.7E-2 & 0.39  & 5.9E-2 & 216.8 &  97.3 &  53 & \\
17& 359.742  &  --0.594 & 17:47:19.94 & --29:27:55.3 & 0.07  & 1.7E-2 & 0.12  & 4.3E-2 & 139.9 &  94.9 &  80 & \\
18& 359.722  &  --0.040 & 17:45:06.72 & --29:11:40.0 & 0.61  & 1.7E-2 & 1.06  & 4.3E-2 & 137.5 &  97.9 &  49 & \\
19& 359.694  &   +0.003 & 17:44:52.63 & --29:11:42.9 & 0.38  & 1.7E-2 & 0.77  & 4.8E-2 & 138.4 & 113.7 &  32 & \\
20& 359.628  &   +0.039 & 17:44:34.60 & --29:13:59.3 & 0.22  & 1.6E-2 & 0.77  & 7.2E-2 & 193.3 & 142.1 & 134 & \\
21& 359.467  &  --0.173 & 17:45:01.28 & --29:28:51.6 & 0.22  & 1.8E-2 & 0.23  & 3.1E-2 &  97.5 &  81.8 & 172 & \\
22& 359.433  &   +0.001 & 17:44:15.64 & --29:25:10.0 & 0.16  & 1.6E-2 & 0.62  & 7.6E-2 & 218.6 & 134.6 & 179 & \\
23& 359.281  &  --0.261 & 17:44:55.23 & --29:41:09.7 & 0.89  & 1.7E-2 & 1.71  & 4.6E-2 & 125.4 & 119.1 & 167 & $1.84\pm$10\% \\
24& 358.801  &  --0.009 & 17:42:46.24 & --29:57:45.2 & 0.14  & 1.8E-2 & 0.17  & 3.5E-2 & 106.3 &  90.8 &  71 & \\
25& 358.789  &   +0.059 & 17:42:28.52 & --29:56:13.3 & 0.40  & 1.7E-2 & 0.52  & 3.6E-2 & 108.4 &  91.7 &  83 & $0.6\pm$30\% \\
26& 358.723  &   +0.008 & 17:42:30.98 & --30:01:10.2 & 0.15  & 1.8E-2 & 0.17  & 3.4E-2 & 107.5 &  82.6 & 148 & \\
27& 358.690  &  --0.087 & 17:42:48.30 & --30:05:52.3 & 0.19  & 1.7E-2 & 0.52  & 5.9E-2 & 154.7 & 137.1 &  10 & \\
28& 358.689  &  --0.125 & 17:42:57.26 & --30:07:08.1 & 0.19  & 1.7E-2 & 0.32  & 4.3E-2 & 127.2 & 103.4 & 176 & \\
29& 358.646  &  --0.036 & 17:42:29.95 & --30:06:31.3 & 0.15  & 1.8E-2 & 0.17  & 3.3E-2 & 106.5 &  80.8 &  53 & \\
30& 358.631  &   +0.062 & 17:42:04.58 & --30:04:10.5 & 0.41  & 1.8E-2 & 0.49  & 3.4E-2 & 104.6 &  89.0 & 157 & $0.45\pm$20\% \\
31& 358.605  &  --0.062 & 17:42:30.11 & --30:09:25.7 & 0.57  & 1.7E-2 & 0.98  & 4.3E-2 & 126.4 & 104.4 & 142 & $1.18\pm$10\% \\
32& 358.548  &  --0.022 & 17:42:12.44 & --30:11:04.3 & 0.10  & 1.8E-2 & 0.13  & 3.5E-2 & 130.5 &  73.7 & 140 & \\
33& 358.383  &  --0.481 & 17:43:36.98 & --30:33:57.3 & 0.12  & 1.7E-2 & 0.25  & 4.9E-2 & 154.3 & 105.9 & 129 & \\
34& 358.187  &   +0.302 & 17:40:02.83 & --30:19:09.3 & 0.05  & 1.7E-2 & 0.13  & 5.5E-2 & 159.0 & 121.2 &  91 & \\
35& 358.003  &  --0.635 & 17:43:17.72 & --30:58:12.3 & 0.24  & 1.8E-2 & 0.29  & 3.4E-2 &  97.6 &  93.5 &   0 & \\
36& 357.992  &  --0.160 & 17:41:23.32 & --30:43:44.0 & 0.40  & 1.7E-2 & 0.70  & 4.3E-2 & 134.8 &  99.7 &  81 & 
\enddata
\tablenotetext{a}{GBT beam FWHM at 3.5 cm is 88\arcsec.}
\end{deluxetable}

\begin{deluxetable}{cccccccccccccc}
\tablecaption{GBT 6 cm Compact Source Catalog for GC Region \label{srcC}}
\tabletypesize{\scriptsize}
\tablewidth{0pt}
\tablehead{
\colhead{\#} & \colhead{l} & \colhead{b} & \colhead{RA} & \colhead{Dec} & \colhead{$S_p$\tablenotemark{a}} & \colhead{$\sigma_{S_p}$\tablenotemark{a}} & \colhead{$S_i$} & \colhead{$\sigma_{S_i}$} & \colhead{bmaj} & \colhead{bmin} & \colhead{bpa} \\ 
\colhead{} & \colhead{(deg)} & \colhead{(deg)} & \colhead{(J2000)} & \colhead{(J2000)} & \colhead{(Jy/bm)} & \colhead{(Jy/bm)} & \colhead{(Jy)} & \colhead{(Jy)} & \colhead{(arcsec)} & \colhead{(arcsec)} & \colhead{(deg)} \\ 
}
\startdata
1 & 0.862 &  +0.081   & 17:47:20.61 & --28:09:25.4 & 3.81 & 1.5E-1 & 7.35 & 4.2E-1 & 224.0 & 200.8 & 123 \\
2 & 0.724 & --0.090   & 17:47:41.07 & --28:21:51.9 & 1.79 & 1.5E-1 & 3.81 & 4.5E-1 & 223.4 & 212.8 & 138 \\
3 & 0.670 & --0.034   & 17:47:20.33 & --28:22:51.2 & 21.73& 1.6E-1 & 32.8 & 3.6E-1 & 199.5 & 176.5 & 68  \\
4 & 0.673 &  +0.084   & 17:46:53.31 & --28:19:05.4 & 0.54 & 1.5E-1 & 1.49 & 5.5E-1 & 266.4 & 243.0 & 113  \\
5 & 359.282 & --0.258 & 17:44:54.68 & --29:40:59.6 & 1.34 & 1.6E-1 & 1.79 & 3.3E-1 & 177.9 & 175.9 & 92  \\
6 & 358.791 & +0.063  & 17:42:27.76 & --29:55:59.2 & 0.45 & 1.6E-1 & 0.52 & 3.1E-1 & 192.2 & 139.0 & 78  \\
7 & 358.630 & +0.066  & 17:42:03.57 & --30:04:05.8 & 0.46 & 1.6E-1 & 0.57 & 3.2E-1 & 182.0 & 157.5 & 84  \\
8 & 358.606 & --0.061 & 17:42:30.04 & --30:09:20.1 & 0.86 & 1.5E-1 & 1.60 & 4.1E-1 & 239.6 & 180.9 & 116 \\
9 & 358.004 & --0.634 & 17:43:17.56 & --30:58:07.7 & 0.29 & 1.6E-1 & 0.29 & 2.8E-1 & 160.5 & 148.1 & 51  \\
10& 357.990 & --0.158 & 17:41:22.48 & --30:43:47.2 & 0.54 & 1.6E-1 & 0.77 & 3.5E-1 & 197.7 & 168.3 & 94  
\enddata
\tablenotetext{a}{GBT beam FWHM at 6 cm is 153\arcsec.}
\end{deluxetable}
\clearpage

All compact sources detected at 6 cm are also detected at 3.5 cm, which allows the calculation of the spectral index for each source.  Table \ref{psspix} shows the spectral index for all compact sources detected at 3.5 and 6 cm.  The spectral index is calculated from the integrated flux densities according to the standard equations.

\begin{equation}
\label{alpha}
S_\nu \propto \nu^{\alpha}
\end{equation}

\begin{equation}
\label{alpha2}
\alpha_{12} = \rm{log}(S_1/S_2)/\rm{log}(\nu_1/\nu_2)
\end{equation}

\begin{equation}
\label{alpha3}
\sigma_{\alpha_{12}} = (1/\rm{log}(\nu_1/\nu_2))*\sqrt{(\sigma_1/S_1)^2 + (\sigma_2/S_2)^2}
\end{equation}

\noindent where ``1'' and ``2'' refer to the observing frequencies, which are abbreviated as X for 3.5 cm, C for 6 cm, L for 20 cm, and P for 90 cm.  The images were not convolved to the same resolution for measuring the compact source flux densities, since tests with JMFIT show that the integrated flux density of the 3.5 cm image is the same at its natural resolution and after being convolved to match the 6 cm GBT resolution.

\begin{deluxetable}{ccccc}
\tablecaption{Spectral Index for GBT Compact Sources Detected at 3.5 and 6 cm \label{psspix}}
\tablewidth{0pt}
\tablehead{
\colhead{\#} & \colhead{l} & \colhead{b} & \colhead{$\alpha_{CX}$} & \colhead{$\sigma_\alpha$\tablenotemark{a}} \\ 
\colhead{} & \colhead{(deg)} & \colhead{(deg)} & \colhead{} & \colhead{} \\ 
}
\startdata
1 & 0.862 &  +0.081   & --0.93 & 0.10 \\
2 & 0.724 & --0.090   & --1.42 & 0.21 \\
3 & 0.670 & --0.034   & --0.28 & 0.02 \\
4 & 0.673 &  +0.084   & --1.88 & 0.67 \\
5 & 359.282 & --0.258 & --0.08 & 0.33 \\
6 & 358.791 & +0.063  &   0.00 & 1.07 \\
7 & 358.630 & +0.066  & --0.27 & 1.01 \\
8 & 358.606 & --0.061 & --0.87 & 0.46 \\
9 & 358.004 & --0.634 &   0.00 & 1.73 \\
10& 357.990 & --0.158 & --0.17 & 0.82 
\enddata
\tablenotetext{a}{Error in spectral index based on statistical errors and do not account for absolute flux calibration errors.  For the expected 5\% absolute flux errors in the 3.5 and 6 cm maps, a spectral index uncertainty of $\sim0.13$ should be added in quadrature to these errors.}
\end{deluxetable}
\clearpage

\citet{h92} used the Parkes 64 m telescope to survey the GC region at 3.5 cm and produce a point source catalog.  Those data have a resolution and sensitivity roughly two to three times larger than the present observations (2\damin8 beam and 30 mJy beam$^{-1}$, respectively).  Sources in that catalog that are coinicident with the present 3.5 cm survey are shown in Table \ref{srcX} for comparison.  The integrated flux densities observed here are systematically 24\% less than in \citet{h92}, which is more than expected from the calibration uncertainty of 6\% for their work and 5\% expected for the present work.  The sources with the biggest differences in flux density are in the Sgr B and Sgr D regions, which are filled with extended \hii\ regions.  It is likely that the larger beam used by \citet{h92} includes more of this extended emission in their compact source catalog.

The 6 cm catalog in Table \ref{srcC} was compared to catalogs in the literature to confirm those results, but no single-dish survey at this wavelength could be found.  In the VLA study of Sgr E by \citet{g93}, two sources (J174203--300405 and J174227--295559) are associated with relatively unconfused sources in the present 6 cm catalog.  Also, in the 6 cm VLA survey of \citet{b94}, three sources can be compared to our survey.  In all these cases, the VLA flux densities are about range from 2--70\% of that measured by our 6 cm survey.  It is likely that there is flux missing from the interferometric observations.

\subsection{GBT Extended Sources}
\label{diffsrcsec}
A survey of radio continuum emission in the central degrees of the Galaxy provides a useful catalog of the \hii\ regions, supernova remnants, and extragalactic radio sources.  The central four degrees of the Galaxy covered by this survey include the central molecular zone, a 200-pc region where molecular mass collects due to dynamical processes in a barred gravitational potential \citep{m96,s04}.  Radio continuum observations are an effective way to study this region because it can separate the thermal and nonthermal emission by measuring the spectral index.  The following sections describe the cataloging and analysis of extended sources observed in the survey region.  The spectral index is studied in two ways:  through integrated and slice analysis.  First, \S\ \ref{srcex} describes how sources are identified and how the integrated flux density at 3.5 and 6 cm is used to find the spectral index.  Second, \S\ \ref{slicediff} describes how a 1-D strip or ``slice'' of the flux in an image can be compared between two frequencies to measure the spectral index of objects.

\paragraph{Extended Source Catalog}
\label{srcex}
As a first step in studying the distribution of the radio continuum emission at 3.5 and 6 cm surveys, we have constructed a catalog of the extended sources.  All sources not considered ``compact'' were included in the extended source catalog shown in Table \ref{diffsrc} and discussed in Section \ref{compactsrcsec}.  

\begin{figure}[tbp]
\includegraphics[width=\textwidth]{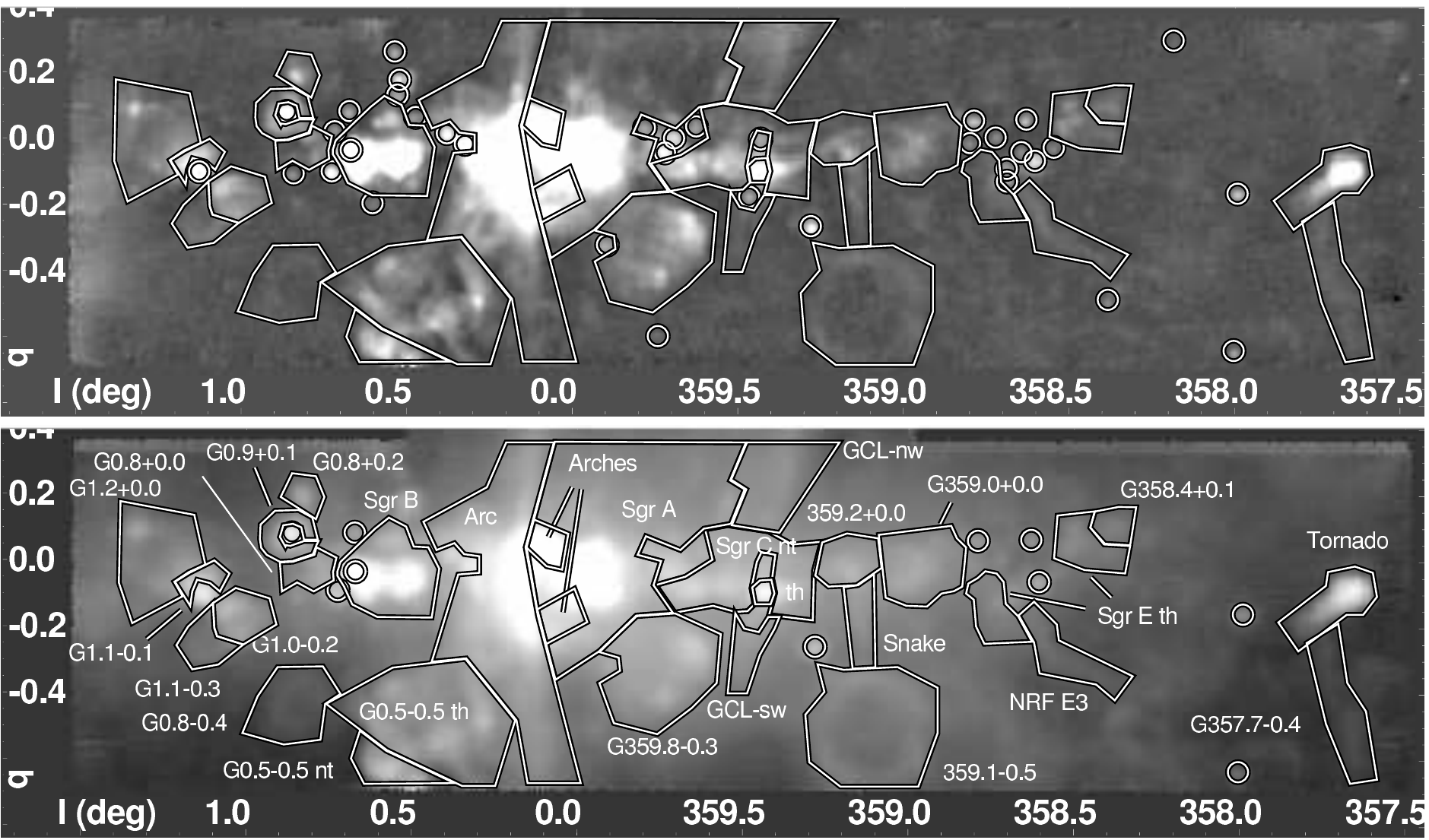}
\caption{\emph{Top}: GBT map of 3.5 cm continuum emission showing the extended sources listed in Table \ref{diffsrc} and the compact sources given in Table \ref{srcX}.  The brightness of the image is shown in a logarithmic scale from 0 to 5 Jy beam$^{-1}$.  \emph{Bottom}:  GBT map of 6 cm continuum emission with the same regions overlaid, except the circles show 6 cm compact sources given in Table \ref{srcC}.  The brightness of the image is shown in a logarithmic scale from 0 to 10 Jy beam$^{-1}$.  The polygon regions define extent of extended sources and are identified with their common names.  The flux density, location, and other properties of the extended sources are given in Tables \ref{diffsrc} and \ref{diffsrcspix}.  \label{diffreg}}
\end{figure}

Figure \ref{diffreg} shows the 3.5 and 6 cm images with extended and compact regions overlaid.  Regions were defined to enclose all flux from an object (in both the 3.5 and 6 cm images) that is believed to have a similar origin.  Often this is simple, such as for the supernova remnant G359.1--0.5, which has a distinctive ring-like shape and has been extensively studied.  Otherwise, the regions were simply defined by whether it had a thermal or nonthermal spectral index between 6 and 3.5 cm (see \S\ \ref{slicediff}).  In a few cases, the extended sources in Table \ref{diffsrc} have names appended with ``th'' or ``nt'', according to whether the region contains the thermally- or nonthermally-emitting parts of the complex.  The source regions for Sgr A and Sgr B include other sources, so the background-subtracted source flux also subtracts these other contributions, as noted in the table.

Properties of extended sources in the 3.5 and 6 cm images are given in Table \ref{diffsrc}.  Columns (1)-(4) give the commonly-used source name, position, and effective radius, calculated from the area ($R_{eff}=\sqrt{A/\pi}$).   Columns (5)-(8) give the 3.5 cm raw source flux density, the rms uncertainty measured in a background region, and the background-subtracted flux density and its error.  Columns (9)-(12) give the same quantities for the sources at 6 cm.  The source brightness is integrated over all pixel values and then scaled by ratio of the pixel area to the beam area to get a flux density in Jy.  The scale factor is $(30\arcsec^2)/(1.1331*FWHM^2)$ (as used in AIPS), with $FWHM=88$\arcsec\ and 153\arcsec\ at 3.5 and 6 cm, respectively.  The background flux density is measured over a nearby region.  The rms in the background is scaled to the source area to estimate the uncertainty in the integrated source flux density.  Often this method overestimates the error in the flux density, since the rms in the background is dominated by other sources or Galactic emission (e.g., the error for the Arched filaments is large for this reason).  With the flux density measured at 3.5 and 6 cm, the spectral index and its error is calculated and shown in Table \ref{diffsrcspix}.  The sixth column shows our conclusion on the nature of the radio continuum emission (thermal vs. nonthermal), based on the integrated and slice spectral index analysis (see \S\ \ref{slicediff}).

\begin{deluxetable}{cccccc|cccc|cccc}
\rotate
\tablecaption{Extended Source Catalog for 3.5 and 6 cm GBT Observataions of GC Region \label{diffsrc}}
\tabletypesize{\scriptsize}
\tablewidth{0pt}
\tablehead{
 & & & & & \multicolumn{4}{c}{3.5 cm} & \multicolumn{4}{c}{6 cm} \\
\hline
\colhead{Name} & \colhead{l} & \colhead{b} & \colhead{RA} & \colhead{Dec} & \colhead{$R_{eff}$} & \colhead{raw $S_{i,X}$} & \colhead{$\sigma_{bg,X}$} & \colhead{$S_{i,X}$} & \colhead{$\sigma_{S_{i,X}}$} & \colhead{raw $S_{i,C}$} & \colhead{$\sigma_{bg,C}$} & \colhead{$S_{i,C}$} & \colhead{$\sigma_{S_{i,C}}$} \\ 
 & \colhead{(deg)} & \colhead{(deg)} & \colhead{(J2000)} & \colhead{(J2000)} & \colhead{(arcmin)} & \colhead{(Jy)} & \colhead{(Jy/bm)} & \colhead{(Jy)} & \colhead{(Jy)} & \colhead{(Jy)} & \colhead{(Jy/bm)} & \colhead{(Jy)} & \colhead{(Jy)} \\ 
}
\startdata
Tornado & 357.66 & --0.09 & 17:40:17.0 & --30:58:12 &      6.5 & 14.16 & 9.7E-3 & 13.81 & 0.51 & 18.45 & 1.6E-2     & 18.27 & 0.41 \\ 
G357.7--0.4 & 357.70 & --0.44 &  17:41:46.6 & -31:07:35 & 7.0 & 2.09 & 1.0E-2 & 1.55 & 0.71   & 1.30 & 1.6E-2      & 1.05 & 0.55 \\ 
G358.4+0.1 & 358.38 & +0.12 & 17:41:15.9 & --30:15:01 &    4.0 & 1.52 & 7.8E-3 & 1.24 & 0.14   & 3.41    & 1.9E-2   & 1.85 & 0.16 \\
RF E3 & 358.54 & --0.29 & 17:43:14.1 & --30:19:56 &        6.5 & 1.99 & 9.0E-3 & 1.51 & 0.45   & 4.28 & 2.3E-2      & 1.99 & 0.56 \\ 
Sgr E th\tablenotemark{a} & 358.50 & +0.05 & 17:41:49.0 & --30:11:09 &      8.0 & 4.54 & 8.0E-3 & 3.67 & 0.61   & 12.12 & 7.2E-2     & 3.79 & 2.64 \\ 
G359.0+0.0 & 358.91 & --0.03 & 17:43:07.9 & --29:52:54 &   8.0 & 8.13 & 9.3E-3 & 7.15 & 0.71   & 31.06   & 5.1E-2   & 13.52 & 1.86 \\
G359.1--0.5 & 359.11 & --0.51 & 17:45:29.3 & --29:57:35 & 12.0 & 6.24 & 1.0E-2 & 3.29 & 1.84   & 28.84 & 4.0E-2     & 6.81 & 2.46 \\ 
Snake & 359.14 & --0.19 & 17:44:17.9 & --29:46:00 &        5.0 & 1.21 & 1.1E-2 & 1.05 & 0.35   & 7.97 & 3.6E-2      & 2.99 & 0.55 \\ 
G359.2+0.0 & 359.18 & +0.01 & 17:43:37.5 & --29:38:04 &    5.5 & 3.24 & 1.0E-2 & 2.81 & 0.37   & 13.47 & 3.6E-2     & 7.71 & 0.64 \\ 
GCL-NW & 359.38 & +0.27 & 17:43:04.1 & --29:19:26 &        8.0 & 6.10 & 1.0E-2 & 5.07 & 0.85   & 19.38 & 5.2E-2     & 7.59 & 2.13 \\ 
Sgr C th\tablenotemark{a} & 359.44 & --0.10 & 17:44:39.6 & --29:27:53.6 &   2.5 & 7.56 & 8.2E-3 & 7.51 & 0.05   & 8.77 & 1.2E-1      & 6.82 & 0.34 \\ 
GCL-SW & 359.49 & --0.28 & 17:45:31.3 & --29:31:05 &       4.5 & 0.89 & 1.0E-2 & 0.51 & 0.27   & 7.40 & 1.8E-1      & 2.00 & 2.30 \\ 
Sgr C nt\tablenotemark{b} & 359.56 & --0.07 & 17:44:50.9 & --29:21:16 &     9.5 & 22.78 & 8.2E-3 & 21.79 & 0.99 & 67.69 & 1.2E-1     & 28.19 & 6.96 \\ 
G359.8--0.3 & 359.77 & --0.32 & 17:46:19.3 & --29:18:04 & 10.5 & 20.31 & 1.5E-2 & 18.721 & 2.12 & 56.08 & 1.8E-1     & 27.57 & 12.20 \\ 
Sgr A & 359.95 & --0.06 & 17:45:42.9 & --29:00:39 & 17.5 & 332.18\tablenotemark{c} & 1.3E-2 & 289.14 & 5.54 & 529.07\tablenotemark{c} & 9.1E-2 & 461.17 & 18.57 \\ 
Arched filaments & 0.09 & +0.05 & 17:45:39.6 & --28:49:53    &       6.0 & 66.80 & 6.2E-1 & 37.53 & 24.55 & 84.23 & 1.3E+0     & 8.81 & 24.66 \\ 
Arc & 0.18 & --0.06 & 17:46:16.9 & --28:48:52       &     16.5 & 175.34 & 1.4E-2 & 171.31 & 4.59& 277.25 & 9.1E-2    & 231.87 & 14.28 \\
G0.5--0.5 th\tablenotemark{a} & 0.42 & --0.49 & 17:48:31.7 & --28:49:33 &  12.5 & 22.08 & 1.2E-2 & 20.39 & 2.39 & 51.73 & 1.2E-1     & 22.82 & 11.47 \\ 
G0.5--0.5 nt\tablenotemark{b} & 0.57 & --0.63 & 17:49:24.0 & --28:46:32 &   5.5 & 8.01 & 1.2E-2 & 7.72 & 0.40   & 14.78 & 4.7E-2     & 14.60 & 0.76 \\ 
Sgr B & 0.62 & --0.06 & 17:47:18.9 & --28:25:58 & 8.5 & 103.60\tablenotemark{d} & 1.1E-2 & 72.23 & 1.07 & 123.34\tablenotemark{d} & 1.2E-1 & 64.67 & 5.59 \\ 
G0.8+0.0 & 0.81 & --0.03 & 17:47:38.9 & --28:15:42 &       4.0 & 2.58 & 9.1E-3 & 2.42 & 0.20   &  9.51 & 2.8E-2     & 6.42 & 0.29 \\
G0.8+0.2 & 0.82 & +0.20 & 17:46:46.1 & --28:07:44 &        3.5 & 2.90 & 1.5E-2 & 2.44 & 0.25   &  4.91 & 1.1E-1     & 2.99 & 1.48 \\
G0.8--0.4 & 0.83 & --0.42 & 17:49:14.1 & --28:26:24 &      7.5 & 1.49 & 1.2E-2 & 0.95 & 0.84   &  3.00 & 2.7E-2     &  1.98 & 0.90 \\ 
G0.9+0.1 & 0.87 & +0.08 & 17:47:22.5 & --28:09:19 &        4.5 & 5.91 & 1.3E-2 & 5.46 & 0.35   & 12.46 & 4.1E-2     & 9.08 & 1.61 \\
G1.0--0.2 & 1.02 & --0.17 & 17:48:42.5 & --28:09:11 &      5.5 & 9.25 & 1.5E-2 & 8.52 & 0.53   & 15.95 & 7.0E-2     &  9.90 & 1.18 \\
G1.1--0.3 & 1.13 & --0.28 & 17:49:22.5 & --28:06:55 &      5.0 & 2.50 & 1.5E-2 & 1.92 & 0.41   &  4.99 & 2.9E-2     &  2.37 & 0.38 \\
G1.1--0.1 & 1.13 & --0.07 & 17:48:33.4 & --28:00:06 &      3.5 & 5.04 & 1.1E-2 &  4.93 & 0.15  &  6.84 & 9.2E-2     &  5.59 & 0.58 \\
G1.2+0.0 & 1.23 & +0.01 & 17:48:28.9 & -27:52:37 &        8.5 & 10.99 & 1.1E-2 & 10.17 & 1.09 & 18.71 & 9.2E-2     & 9.33 & 4.37 
\enddata
\tablenotetext{a}{The ``th'' appended to the source name emphasizes that the source region is defined for the thermal-emitting part of the complex.}
\tablenotetext{b}{The ``nt'' appended to the source name emphasizes that the source region is defined for the nonthermal-emitting part of the complex.}
\tablenotetext{c}{Flux includes Arched filaments extended source, which is subtracted with the background.}
\tablenotetext{d}{Flux includes the Sgr B2 point source, which is subtracted with the background.}
\end{deluxetable}
\clearpage

\begin{deluxetable}{cccccc}
\tablecaption{Spectral Indices for Extended Sources \label{diffsrcspix}}
\tabletypesize{\scriptsize}
\tablewidth{0pt}
\tablehead{
\colhead{Name} & \colhead{l} & \colhead{b} & \colhead{$\alpha_{CX}$} & \colhead{$\sigma_{\alpha}$} & \colhead{Thermal/Nonthermal} \\ 
 & \colhead{(deg)} & \colhead{(deg)} & & & \\ 
}
\startdata
Tornado     & 357.66 & --0.09 & --0.50 & 0.07 & NT \\ 
G357.7--0.4 & 357.70 & --0.44 &  +0.69 & 1.04 & T \\ 
G358.4+0.1  & 358.38 &  +0.12 & --0.71 & 0.23 & NT \\
Sgr E th\tablenotemark{a}    & 358.50 &  +0.05 & --0.06 & 0.91 & T \\ 
RF E3       & 358.54 & --0.29 & --0.49 & 0.63 & T \\ 
G359.0+0.0  & 358.91 & --0.03 & --1.23 & 0.25 & NT \\
G359.1--0.5 & 359.11 & --0.51 & --1.30 & 1.19 & NT \\ 
Snake       & 359.14 & --0.19 & --1.86 & 0.64 & NT \\ 
G359.2+0.0  & 359.18 &  +0.01 & --1.80 & 0.26 & NT \\ 
GCL-NW      & 359.38 &  +0.27 & --0.72 & 0.46 & NT \\ 
Sgr C th\tablenotemark{a}    & 359.44 & --0.10 &  +0.17 & 0.06 & T \\ 
GCL-SW      & 359.49 & --0.28 & --2.44 & 1.70 & NT \\ 
Sgr C nt\tablenotemark{b}    & 359.56 & --0.07 & --0.46 & 0.32 & NT \\ 
G359.8--0.3 & 359.77 & --0.32 & --0.69 & 0.58 & T \\ 
Sgr A       & 359.95 & --0.06 & --0.83 & 0.06 & NT \\ 
Arched filaments &   0.09 &  +0.05 &  +2.58 & 3.65 & T \\ 
Arc         &   0.18 & --0.06 & --0.54 & 0.09 & NT \\
G0.5--0.5 th\tablenotemark{a} &  0.42 & --0.49 & --0.20 & 0.66 & T \\ 
G0.5--0.5 nt\tablenotemark{b} &  0.57 & --0.63 & --1.14 & 0.11 & NT \\ 
Sgr B       &   0.62 & --0.06 &  +0.20 & 0.11 & T \\ 
G0.8+0.0    &   0.81 & --0.03 & --1.74 & 0.16 & NT \\
G0.8+0.2    &   0.82 &  +0.20 & --0.36 & 0.64 & NT \\
G0.8--0.4   &   0.83 & --0.42 & --1.31 & 1.67 & T \\ 
G0.9+0.1    &   0.87 &  +0.08 & --0.91 & 0.25 & NT \\
G1.0--0.2   &   1.02 & --0.17 & --0.27 & 0.18 & NT \\
G1.1--0.3   &   1.13 & --0.28 & --0.38 & 0.43 & NT \\
G1.1--0.1   &   1.13 & --0.07 & --0.22 & 0.14 & T \\
G1.2+0.0    &   1.23 &  +0.01 &  +0.15 & 0.60 & T
\enddata
\tablenotetext{a}{The ``th'' appended to the source name emphasizes that the source region is defined for the thermal-emitting part of the complex.}
\tablenotetext{b}{The ``nt'' appended to the source name emphasizes that the source region is defined for the nonthermal-emitting part of the complex.}
\end{deluxetable}

The spectral index values shown in Table \ref{diffsrcspix} are generally consistent with the spectral indices measured from slices of the data presented in \S\ \ref{slicediff}.  There is a trend for the integrated spectral index values to be lower than the more precise slice analysis.  The integrated spectral index values for G359.2+0.0, the Snake, GCL-NW, G359.8--0.3, and G0.9+0.1 are all at the low end of 1$\sigma$ errors from the slice analysis.  This may be caused by imperfect subtraction of Galactic synchrotron emission, which is stronger at 6 cm than at 3.5 cm and near the Galactic plane.

\paragraph{Known and Candidate Supernova Remnants}
The extended source catalog includes eight sources that have previously been identified as known or candidate SNRs.  The types of SNRs represented in the GC region vary from traditional shell-types to those having pulsar-wind neblae to mixed-mophology types.  Table \ref{snrsrc} summarizes the observed characteristics of all supernova remnants in the region \citep[e.g.,][]{gr04}, although some are unresolved or confused with other sources.  The positions, extent, flux densities, and measured spectral indices are given.

The results presented here are generally consistent with the high-resolution, 843 GHz observations presented in \citet{g94} and associated papers.  The spectral index values measured by the slice analysis is generally more accurate than the integrated spectral index values.  Between 3.5 and 6 cm, the spectral indices range from --2 to +0.08.  A detailed discussion of the sources and comparison to previous work is given in \S\ \ref{slicediff}.

\begin{deluxetable}{ccccccccccc}
\tablecaption{Catalog of Known and Candidate Supernova Remnants Observed at 3.5, 6, and 20 cm \label{snrsrc}}
\tabletypesize{\scriptsize}
\rotate
\tablewidth{0pt}
\tablehead{
\colhead{Name} & \colhead{RA} & \colhead{Dec} & \colhead{Size} & \colhead{$S_{X}$} & \colhead{$S_{C}$} & \colhead{$\alpha_{CX}^{\rm{int}}$} & \colhead{$\alpha_{CX}^{\rm{slice}}$} & \colhead{$\alpha_{LC}^{\rm{slice}}$} & \colhead{Figures\tablenotemark{a}} & \colhead{References\tablenotemark{b}} \\ 
               & \colhead{(J2000)} & \colhead{(J2000)} & \colhead{(arcmin)} & \colhead{(Jy)} & \colhead{(Jy)} & & & & & \\ 
}
\startdata
G357.7--0.1 (Tornado)\tablenotemark{c} & 17:41:46.6 & -31:07:35 & 4$\times$2.5 & $13.81\pm0.51$ & $18.27\pm0.28$ & $-0.50\pm0.07$ & $\sim-0.45$              & $\sim-0.63$ & \ref{sptornado},\ref{sptornadocl} & 1,2 \\ 
G357.7+0.3\tablenotemark{d}            & 17:38:37.6 & --30:40:32 & 22           &                &                &                & $\sim0.04$ to $\sim-1.5$ &             & \ref{sptornado} & 1,3 \\
G359.1--0.5                            & 17:45:29.3 & --29:57:35 & 19          & $3.29\pm1.84$  & $6.81\pm2.46$ & $-1.30\pm1.19$ & $-2$ to $-0.5$           & $\sim-0.8$ & \ref{spg359.1-0.5},\ref{spg359.1-0.5cl}& 1,3,4 \\ 
G0.0+0.0 (Sgr A)\tablenotemark{d}      & 17:45:42.9 & --29:00:39 & $3\times4$   &                &                &                & $-0.44\pm0.02$           & $-0.42\pm0.02$ & \ref{spsgra},\ref{spsgracl} & 1,7,8 \\ 
G0.33+0.04\tablenotemark{d}            & 17:46:15   & --28:38:00 & $14\times8$  &                &                &                & $-2.0$ to $-1.5$         &             & \ref{speastarc}                & 9 \\
G0.9+0.1\tablenotemark{e}              & 17:47:22.5 & --28:09:19 & 3            & $5.46\pm0.35$  & $9.08\pm1.12$  & $-0.91\pm0.25$ & $-0.35$ to 0.08          & $-0.35$ & \ref{spg0.9+0.1},\ref{spg359.1-0.5cl} & 1,5 \\
G1.0--0.2\tablenotemark{e}             & 17:48:42.5 & --28:09:11 & 3$\times$4   & $8.52\pm0.53$  & $9.90\pm0.82$  & $-0.27\pm0.18$ & $-0.92$ to $-0.55$       & $-0.47$ & \ref{spg0.9+0.1},\ref{spg359.1-0.5cl} & 1,6 
\enddata
\tablenotetext{a}{The figure number featuring the SNR is listed here.}
\tablenotetext{b}{References that discuss object --- 1: \citet{g94}, 2: \citet{b85}, 3: \citet{r84}, 4: \citet{u92}, 5: \citet{h87}, 6: \citet{l92}, 7: \citet{p89}, 8: \citet{y87}, 9: \citet{ka96}}
\tablenotetext{c}{The Tornado is a candidate supernova remnant.}
\tablenotetext{d}{Source is only partially surveyed or confused with other sources, so no integrated characteristics are given.}
\tablenotetext{e}{Flux densities are given for the shell component only.}
\end{deluxetable}

\paragraph{Thermal/Nonthermal Flux Fractions}
\label{fluxfracsec}
Measuring the spectral indices of objects allows us to separate the thermally- and nonthermally-emitting processes.  One application of such a census of objects toward the GC region, is in understanding how such objects contribute to the appearance of unresolved extragalactic nuclei \citep{s78,k83}.  Separating the thermal and nonthermal processes also allows an estimate of their basic properties \citep[$n_e$, $B$;][]{h96}.

Each extended source in Table \ref{diffsrcspix} has been classified as either thermal or nonthermal according to the spectral index analysis (see \S\ \ref{slicediff}), which allows the fraction of total flux contributed by these processes to be calculated.  Table \ref{fluxfrac} shows the sum of the flux densities for each extended source, with thermal and nonthermal flux fractions of 24\%/76\% at 3.5 cm and 15\%/85\% at 6 cm, with uncertainties of about 4\% in both bands.  The flux density of sources with compact sources with measured spectral indices also contribute significantly, with 39$\pm$0.4 Jy detected at 3.5 cm and 51$\pm$4 Jy at 6 cm (see Tables \ref{srcX} and \ref{srcC}).  Furthermore, there is about 46$\pm$1 Jy of flux density in sources that are compact at 3.5 cm, but are confused with extended emission at 6 cm (and are thus not in the 6 cm compact source catalog).  Assuming that these other compact sources have a similar distribution of spectral indices, then we can estimate their contribution to the total 3.5 and 6 cm emission toward the GC region.  Table \ref{fluxfrac} shows the best estimate of the thermal and nonthermal flux fractions toward the GC region is 28\%/72\% at 3.5 cm and 19\%/81\% at 6 cm.  The total flux densities for all compact and extended sources are $783\pm52$ Jy and $1063\pm93$ Jy at 3.5 and 6 cm, respectively.

\begin{deluxetable}{l|ccc|ccc}
\tablecaption{Flux Contributions by Thermal and Nonthermal Sources at 3.5 and 6 cm \label{fluxfrac}}
\tabletypesize{\scriptsize}
\tablewidth{0pt}
\tablehead{
               & \multicolumn{3}{|c|}{Extended Catalog} & \multicolumn{3}{|c|}{Extended and Compact Catalogs\tablenotemark{a}} \\
\hline
\colhead{Type} & \colhead{6 cm} & \colhead{3.5 cm} & \colhead{$\alpha_{CX}$} & \colhead{6 cm} & \colhead{3.5 cm} & \colhead{$\alpha_{CX}$} \\ 
\colhead{} & \colhead{(Jy)/(\%)} & \colhead{(Jy)/(\%)} & \colhead{} & \colhead{(Jy)/(\%)} & \colhead{(Jy)/(\%)} & \colhead{} \\ 
}
\startdata
thermal    &$147\pm43$ ($15\pm4$)  &$173\pm33$ ($24\pm5$) &$0.29\pm0.63$  &$202\pm48$ ($19\pm4$) &$220\pm33$ ($28\pm4$) &$0.15\pm0.50$ \\
nonthermal &$841\pm44$ ($85\pm4$)  &$552\pm19$ ($76\pm3$) &$-0.75\pm0.11$ &$862\pm46$ ($81\pm4$) &$563\pm19$ ($72\pm2$) &$-0.76\pm0.11$ \\
total      &$987\pm87$ (100)       &$725\pm52$ (100)      &$-0.55\pm0.20$ &$1063\pm93$ (100)     &$783\pm52$ (100)      &$-0.54\pm0.20$ 
\enddata
\tablenotetext{a}{Includes all sources listed in Tables \ref{srcX}, \ref{srcC}, and \ref{diffsrc}.  See text in \S\ \ref{fluxfracsec} for details.}
\end{deluxetable}

The first caveat in the interpretation of these results is that these flux fractions are measured for discrete sources and does not consider the diffuse Galactic emission.  Overall, the Galactic emission is dominated by extended synchrotron emission that fills the field observed here, particularly at low frequencies.  Second, the distinction of ``thermal'' and ``nonthermal'' sources by the measured spectral index may not be accurate for sources with a mixture of thermal and nonthermal processes.  For example, it would not be accurate to describe a supernova remnant embedded in a star forming complex as one of these two categories.  While the source regions used to make Table \ref{diffsrc} were defined to separate sources with different structure and emission mechanisms as best as possible, there may be weak emission that is miscategorized by this analysis.

A previous study of the thermal/nonthermal flux distribution in the GC region was done with the Effelsberg 100 m telescope \citep{s78}.  That work found that the central 2\sdeg5 had roughly equal contributions from thermal and nonthermal emission at 6 cm, with a total flux density of about 2000 Jy \citep{k83,l94}.  The total flux density is twice that observed here for a similar region, so that work probably included the extended background emission that this work excluded (the work is only available in a thesis, so it has not been directly studied).  The present work finds that a much higher proportion of the emission from discrete sources is nonthermal.  This low fraction of thermal flux density from the GC region confirms other studies that found that the mm-wavelength continuum emission has a negligible contribution from free-free emission \citep[i.e., the extrapolated 6 cm flux density is much less than the observed 0.8 mm flux density of 13.6 kJy;][]{l94}.

\subsection{Slices of Extended Sources in GBT Images}
\label{slicediff}
Extended sources in the GC region, such as \hii\ regions, SNRs, and nonthermal radio filaments (NRFs), dominate the radio brightness of the GC region.  For a more detailed study of these sources, flux density slices were taken at different positions and orientations through extended sources in order to measure brightnesses and the corresponding spectral index.  Slices were taken from two images convolved to the same resolution;  the convolution size was 2\damin5 for the 3.5/6 cm slices and 9\arcmin\ for the 6/20 cm slices.  To estimate the spectral index of a source, a background must be subtracted prior to taking the ratio of the source brightnesses.  This is done by a reduced chi squared fit of a line to a portion of the slice data.  All structure in the slice that does not look like a background was ignored in fitting the background.  The background was fit to each slice independently, although the same background region was used for both slices.  The best-fit line was then subtracted from the data prior to calculation of the spectral index.  This was all accomplished with an IDL program called ``slicealpha''.

\begin{figure}[tbp]
\begin{center}
\includegraphics[width=0.8\textwidth]{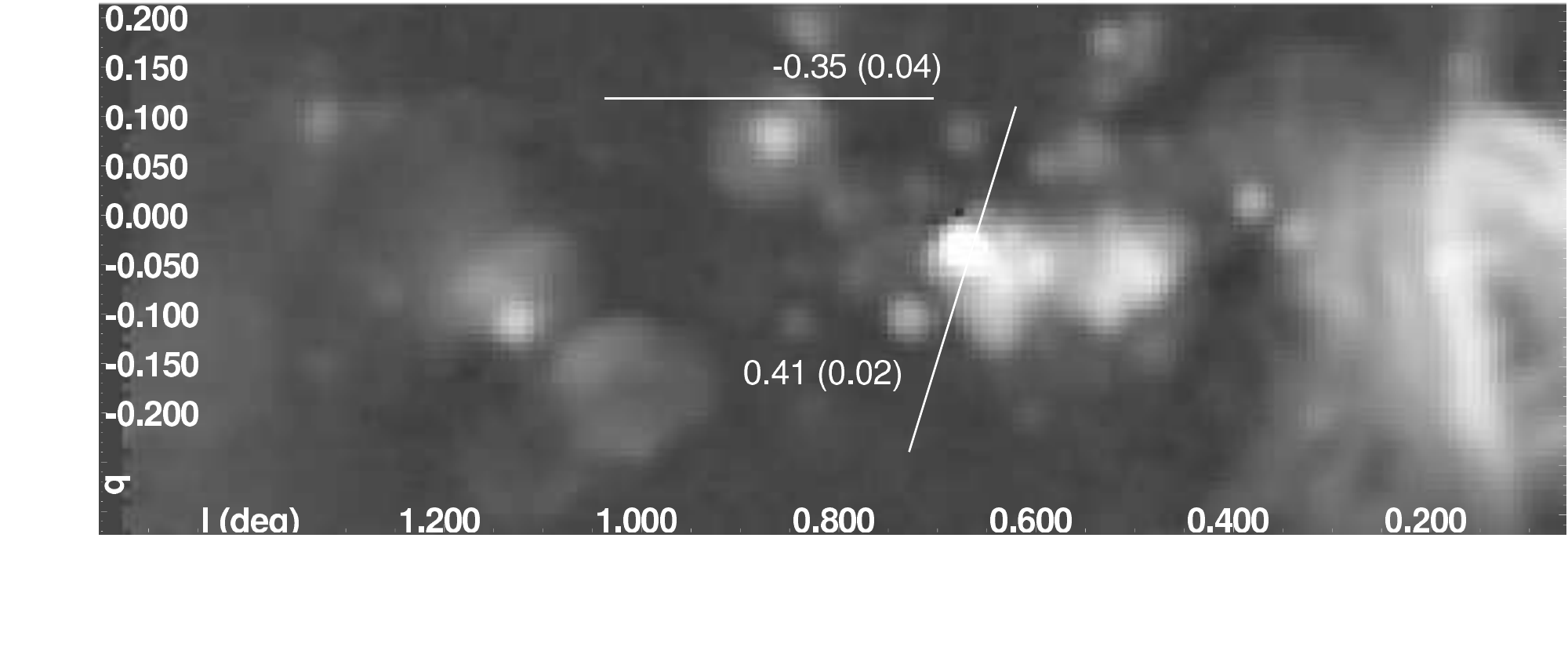}

\includegraphics[width=0.8\textwidth]{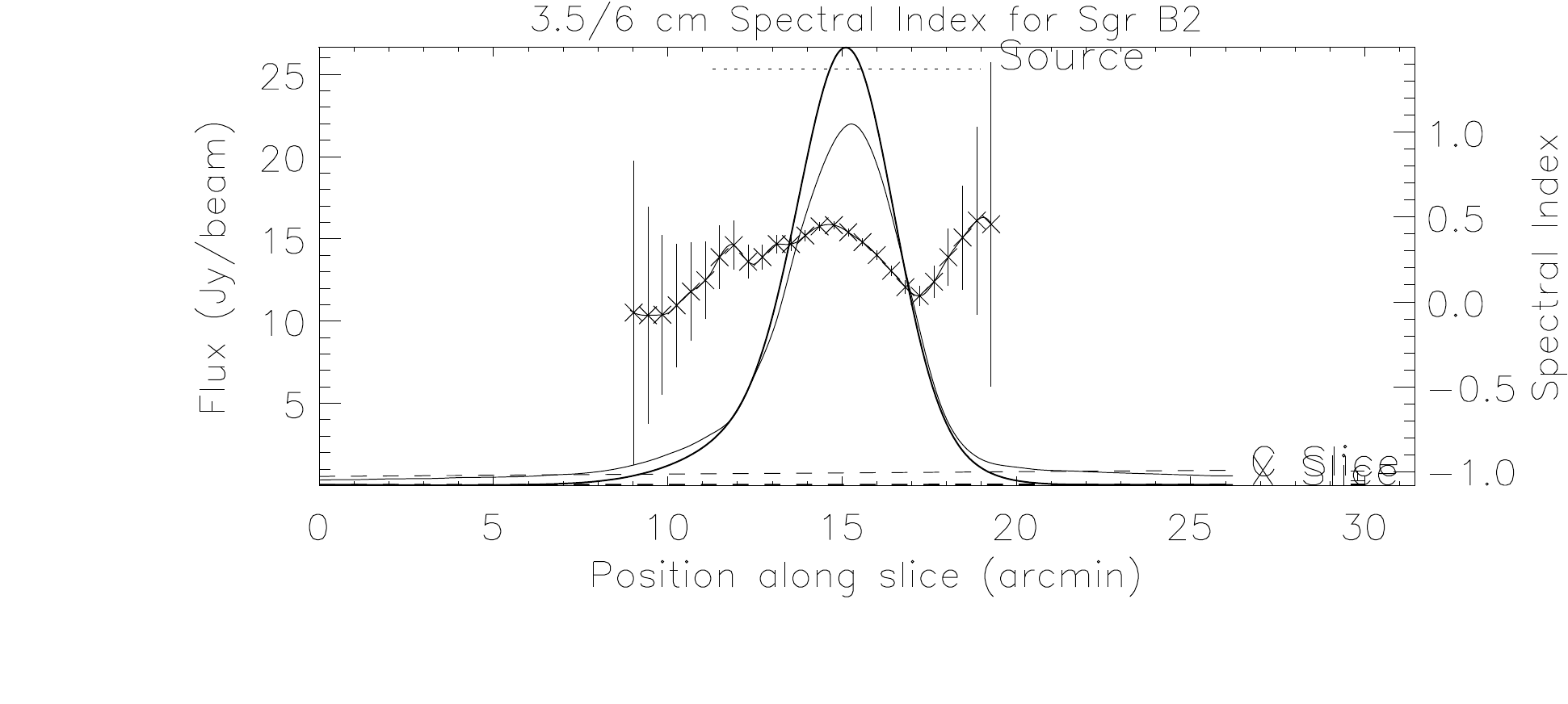}

\includegraphics[width=0.8\textwidth]{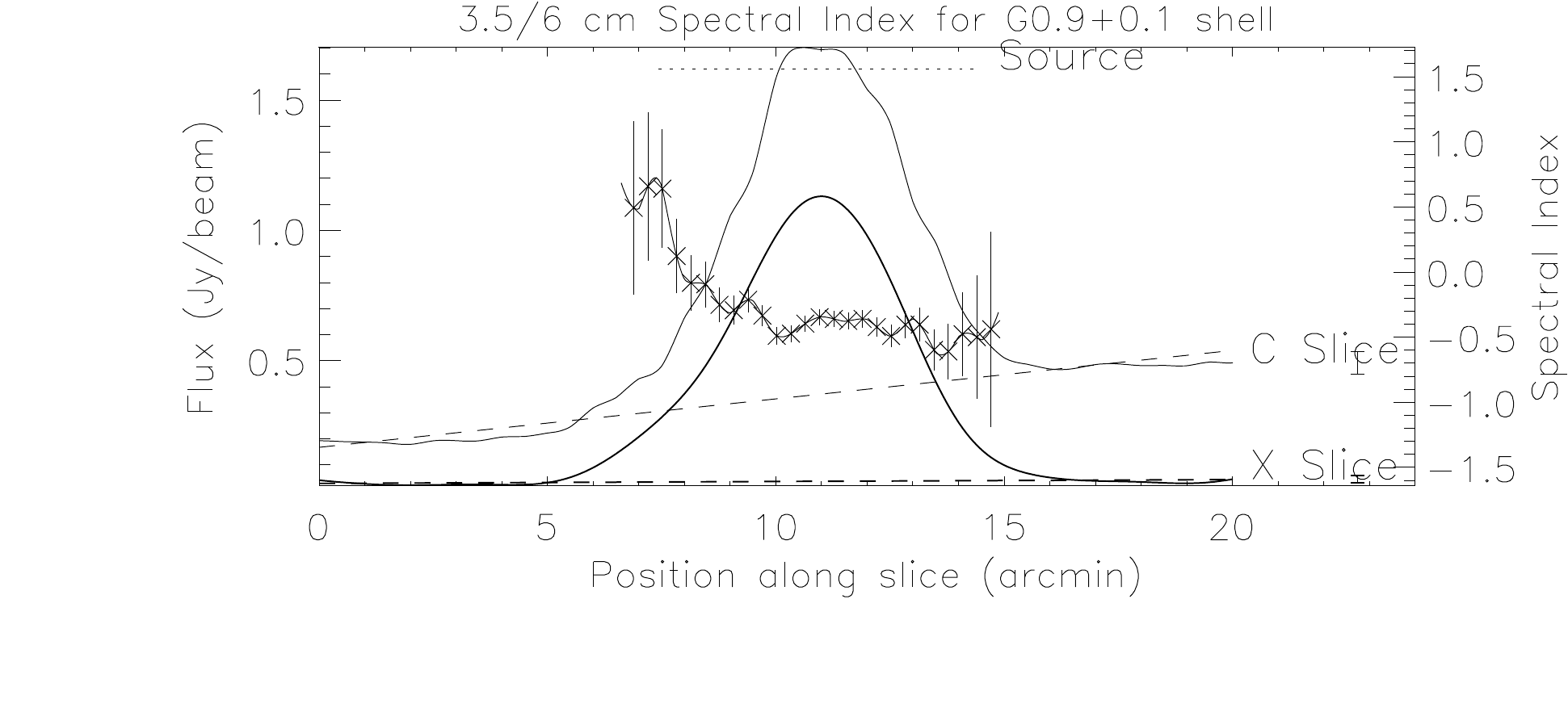}
\end{center}
\caption{\emph{Top}: 3.5 cm image with the locations of two slices used to demonstrate the spectral index slice analysis method.  Both slices labeled with the 6/3.5 cm spectral index at the brightest part of the slice.  \emph{Middle and bottom}:  The flux and spectral index for the slice across Sgr B2 and G0.9+0.1 SNR, respectively.  Each plot shows slices for two frequencies with the 3.5 cm (``X'') slice below the 6 cm (``C'') slice.  The x axis shows the entire length of the slice in units of arcmin.  The slice brightnesses are indicated by the left axis and the spectral index value is shown on the right axis.  The best-fit background to each slice is shown as a dashed line.  The dotted line shows the ``source'' region, which is ignored in the determination of the background.  The rms deviation of the background data about the best-fit line is shown as an error bar to the right of each slice.  Parts of the slice with spectral index error, $\sigma_\alpha<1$, are plotted.  \label{slicetest}}
\end{figure}

Figure \ref{slicetest} demonstrates the slice analysis technique on two sources with relatively simple morphologies.  For each slice shown in the image, the flux and spectral index are shown in a plot.  The value of the spectral index shown in the image is measured at the peak brightness of the shorter wavelength image (3.5 cm for 6/3.5 cm comparison and 6 cm for 20/6 cm comparison).  The dashed lines show the best-fit background for the slices, made by ignoring the source emission, which is shown with a dotted line;  sometimes the ``source'' includes parts of the slice that are confused with other sources.  The spectral index measurements were considered trustworthy only if they were found not to vary much with $\sim10$\% variations in the background region extent.  The rms deviation of the background data relative to the best-fit line is used as an estimate of the uncertainty in the background-subtracted flux density.  The flux density for each position on the slice and its error are used to calculate the spectral index according the standard relations given in Equations \ref{alpha} through \ref{alpha3}.

One slice in Figure \ref{slicetest} covers the brightest part of Sgr B2, which is optically thick and is expected to have a positive (``inverted'') spectral index \citep{ga95}.  The plot shows how the background is fit to the parts of the slice with no emission and, after being subtracted from the slice, finds $\alpha_{CX}=0.41\pm0.02$, consistent with expectations.  The second slice shown in the figure measures the spectral index of the G0.9+0.1 SNR shell.  The slice analysis shows that the source has a nonthermal spectral index of $\alpha_{CX}=-0.35\pm0.04$, as had been previously observed \citep{h87}.  Note that the spectral index at the flux peak is used to represent the source's spectral index at the location, but it is calculated for every point along the slice.  In some extended sources, the spectral index away from the peak flux is different than at the peak;  in these cases, the slices are discussed in more detail in the text.

\clearpage

\begin{figure}[tbp]
\begin{center}
\includegraphics[width=6.5in]{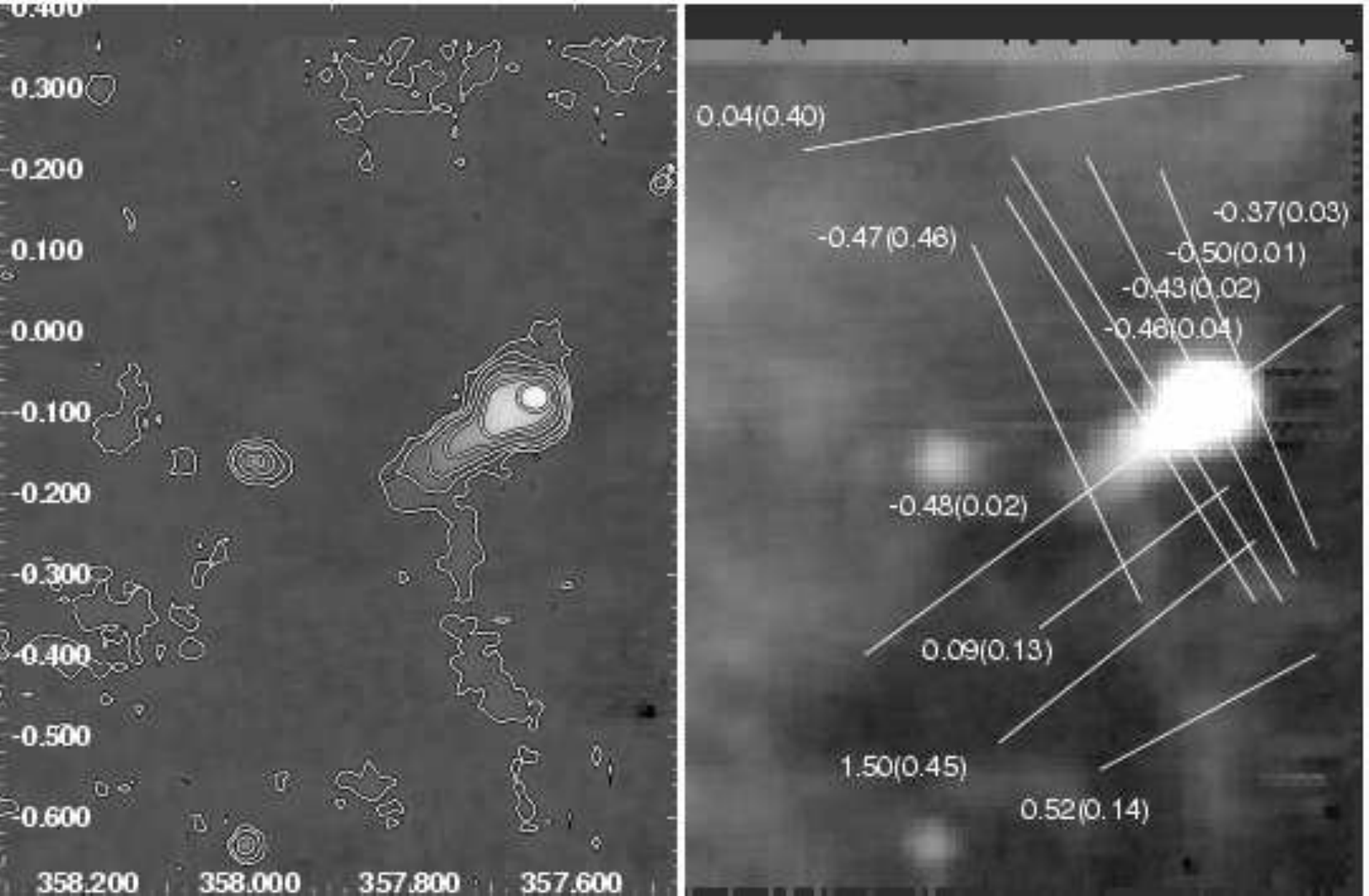}
\includegraphics[width=6.5in]{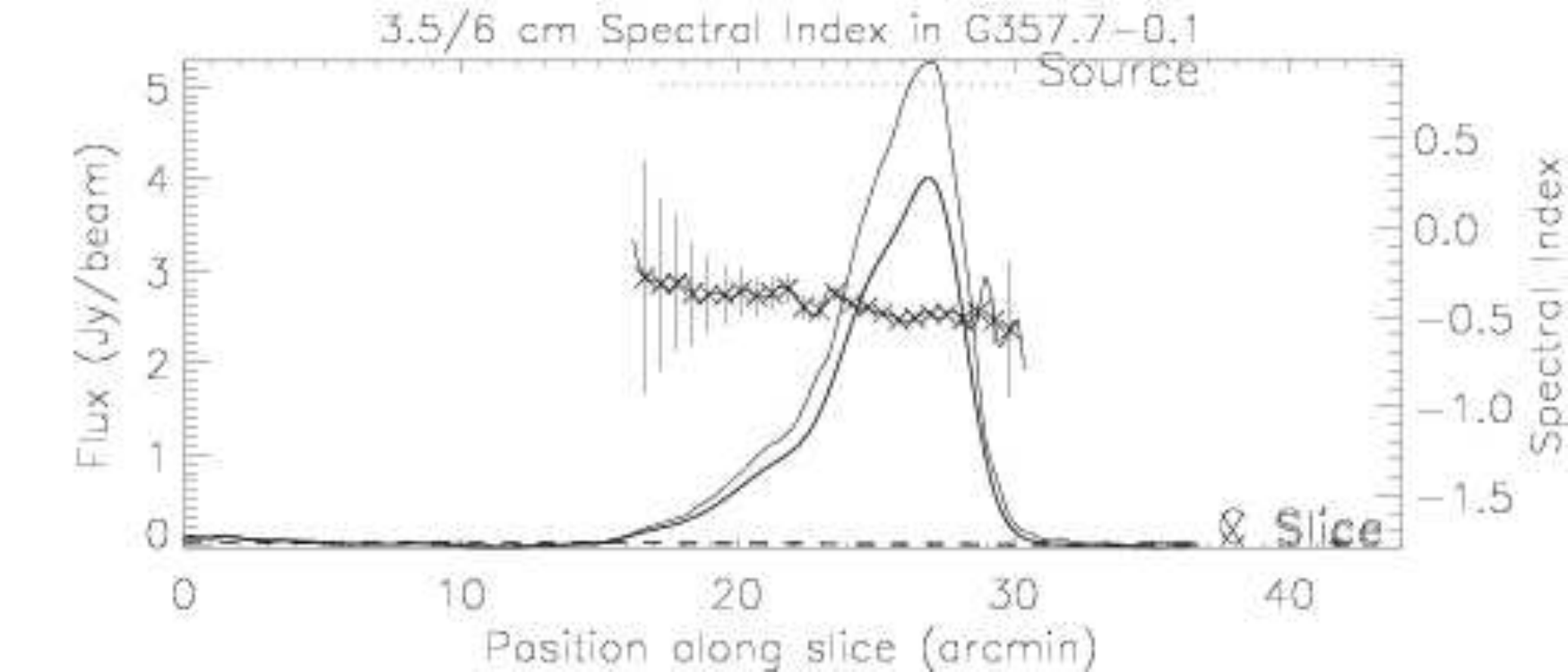}
\end{center}
\caption{\emph{Top, left}:  3.5 cm image of the region around G357.7--0.1 (Tornado) with contours at levels of $0.04*2^n$ Jy beam$^{-1}$, for $n=0-9$.  \emph{Top, right}:  Grayscale shows the 6 cm image of the same region with slices and corresponding spectral index at the peak brightness of each slice.  \emph{Bottom}:  Brightnesses and the corresponding spectral index are shown for the slice with $\alpha_{CX}=-0.48\pm0.02$ in the top right figure.   \label{sptornado}}
\end{figure}

\begin{figure}[tbp]
\includegraphics[width=6.5in]{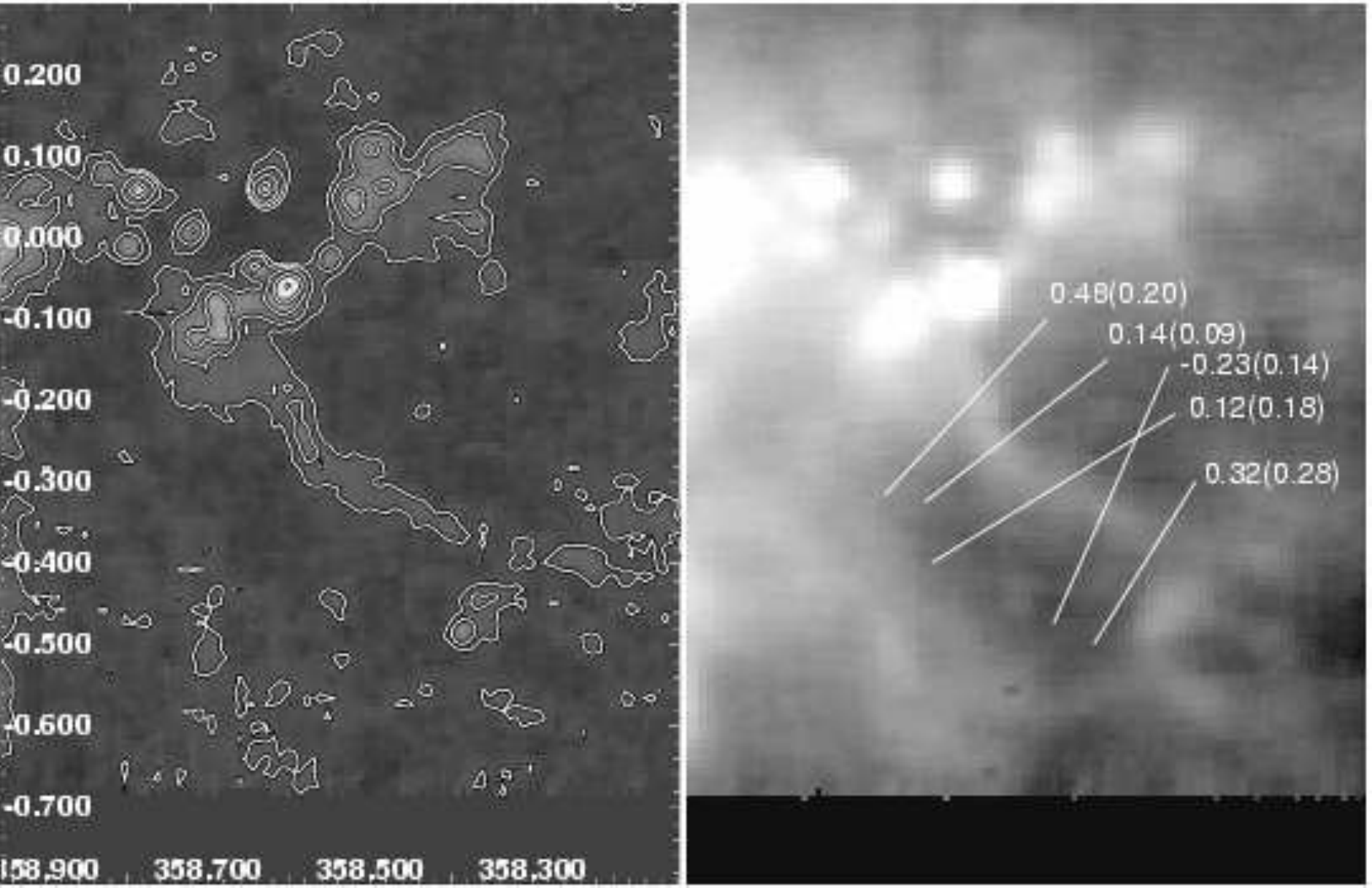}
\includegraphics[width=6.5in]{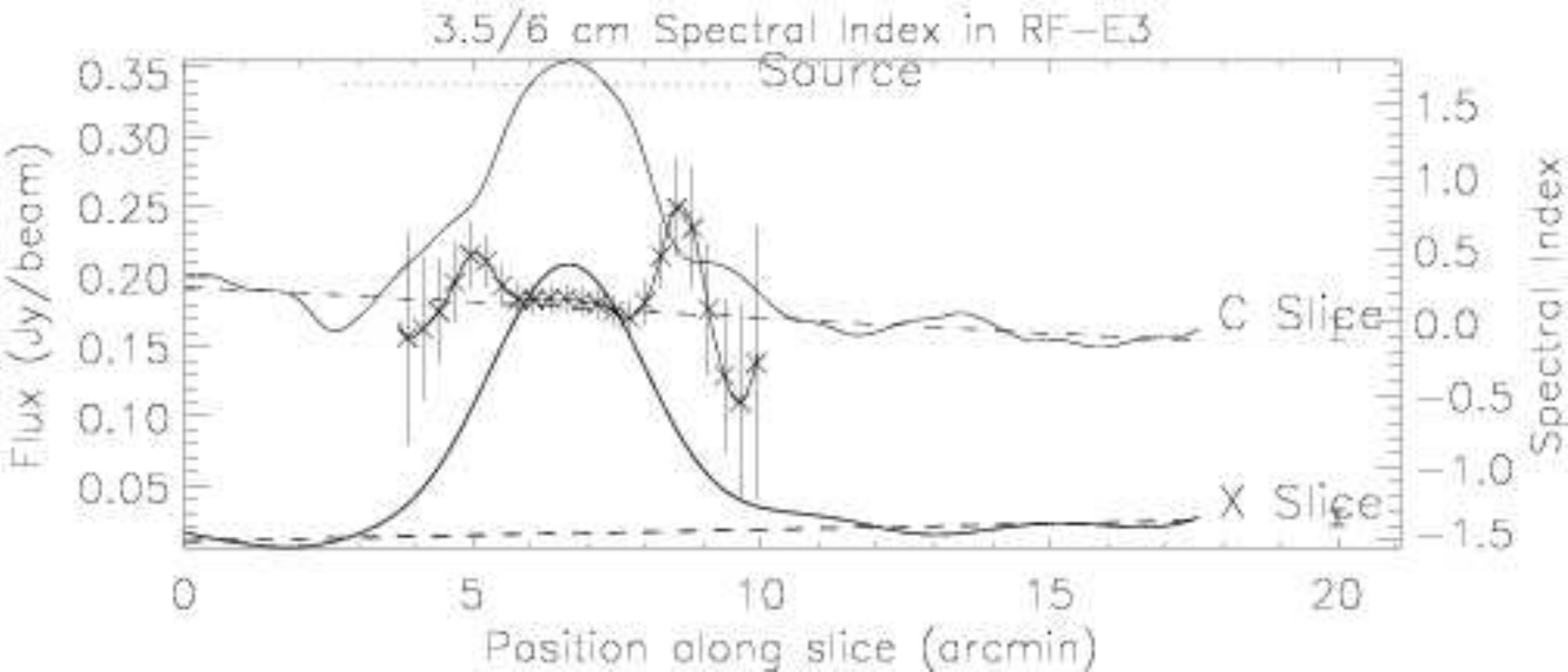}
\caption{Same as for Fig. \ref{sptornado}, but for the E3 radio filament (G358.60-0.27). The plotted slice values correspond to the slice with $\alpha_{CX}=0.12\pm0.18$. \label{spe3}}
\end{figure}

\begin{figure}[tbp]
\includegraphics[width=6.5in]{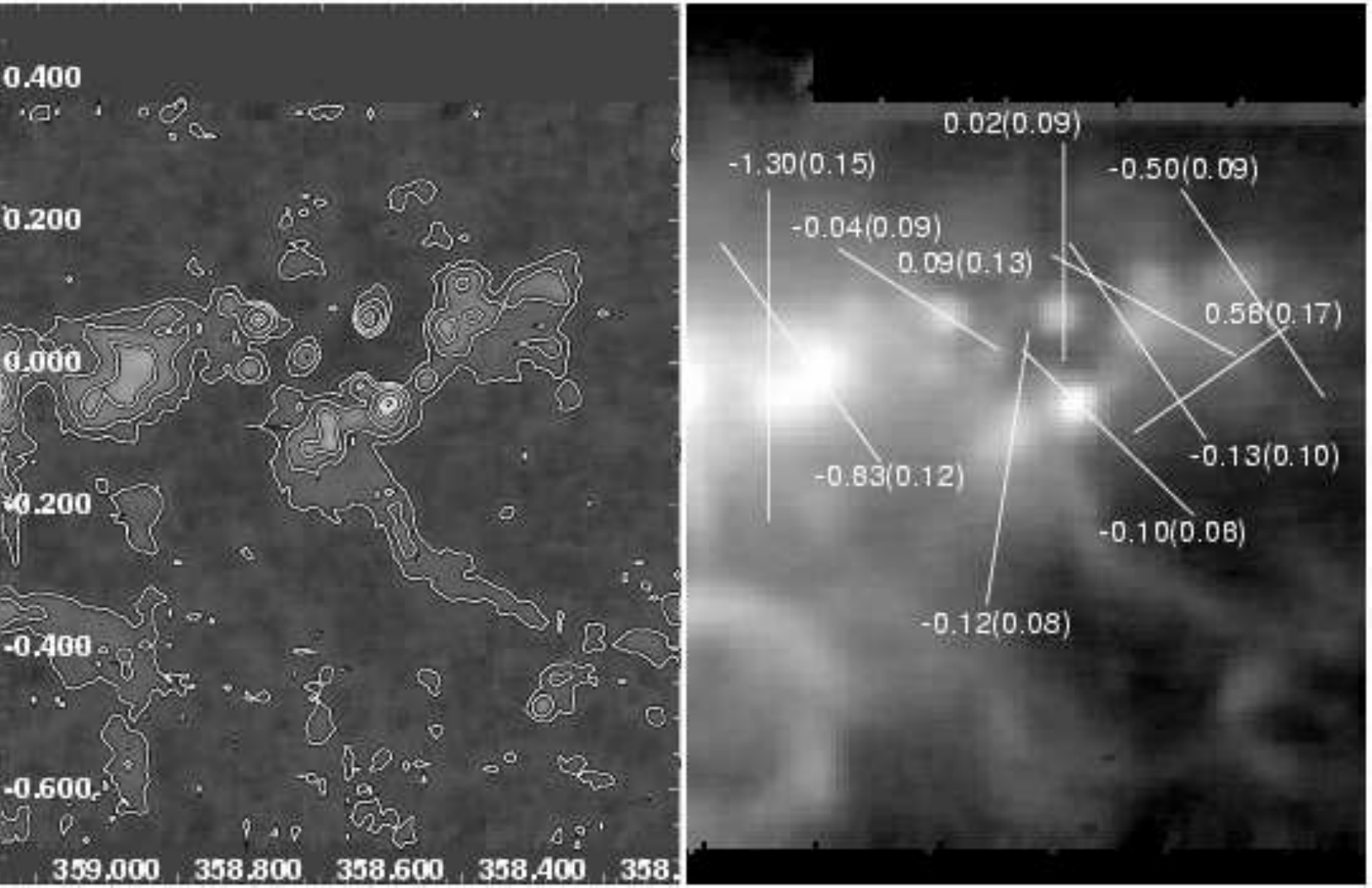}
\includegraphics[width=6.5in]{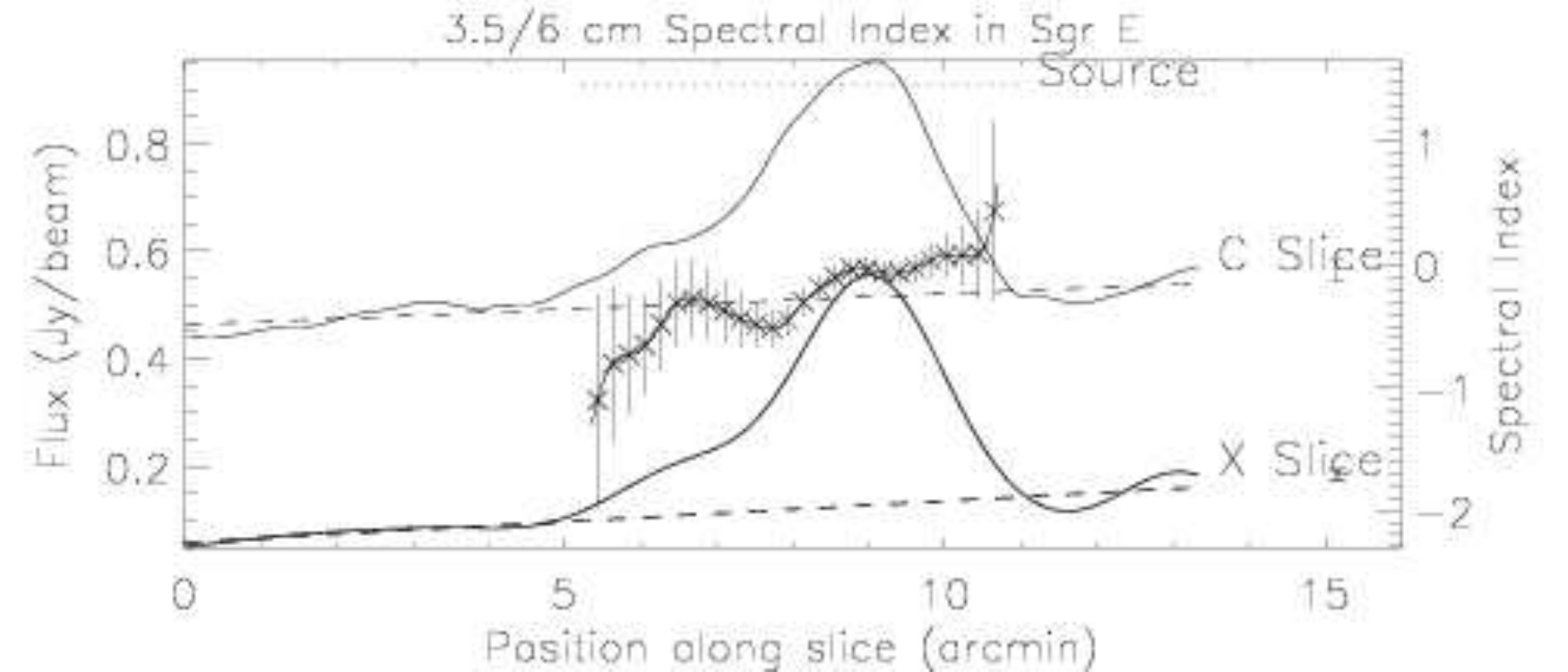}
\caption{Same as for Fig. \ref{sptornado}, but for the Sgr E complex (G358.7-0.0).  The plotted slice values correspond to the slice with $\alpha_{CX}=-0.04\pm0.09$. \label{spsgre}}
\end{figure}

\begin{figure}[tbp]
\includegraphics[width=6.5in]{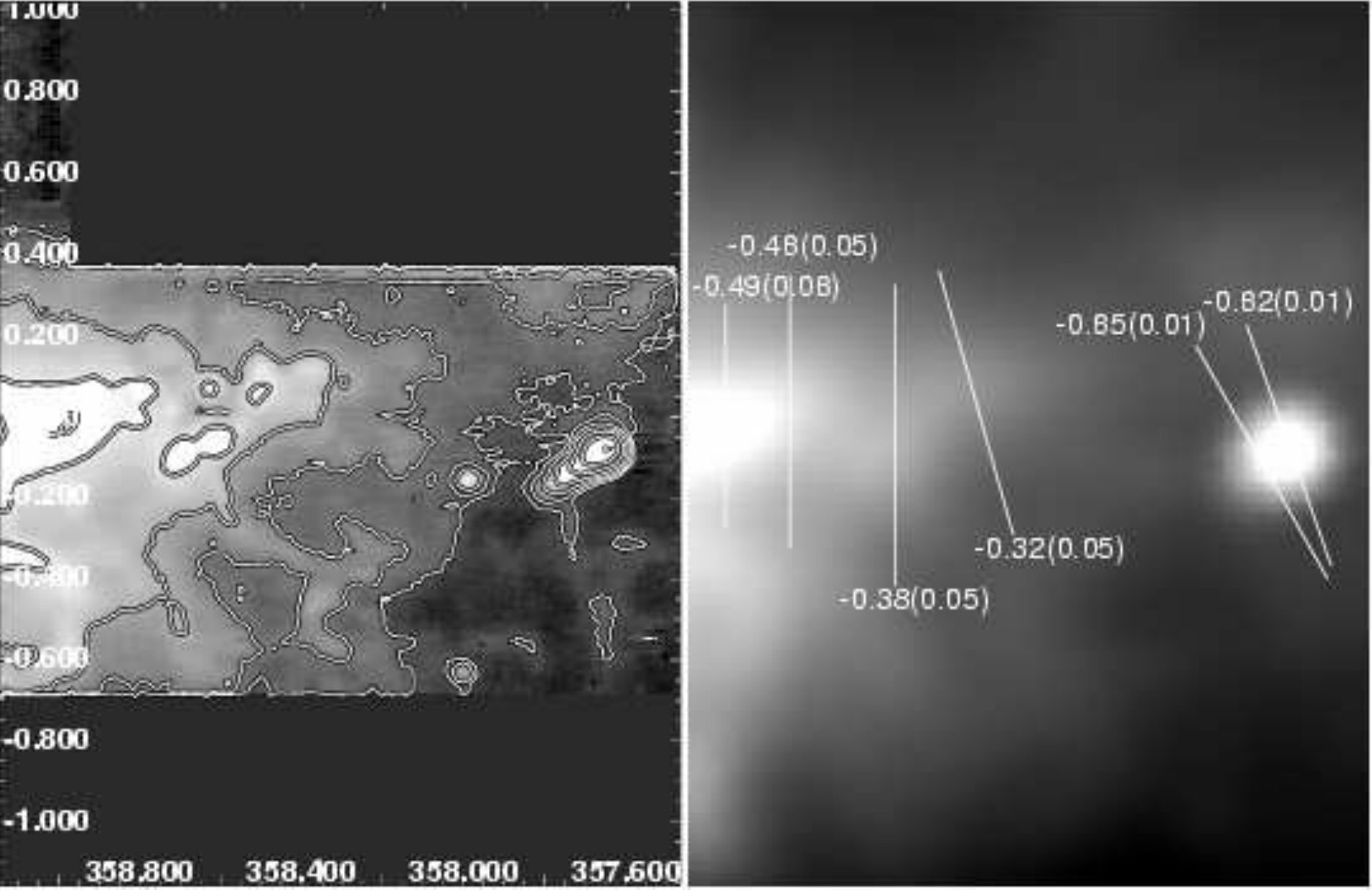}
\includegraphics[width=6.5in]{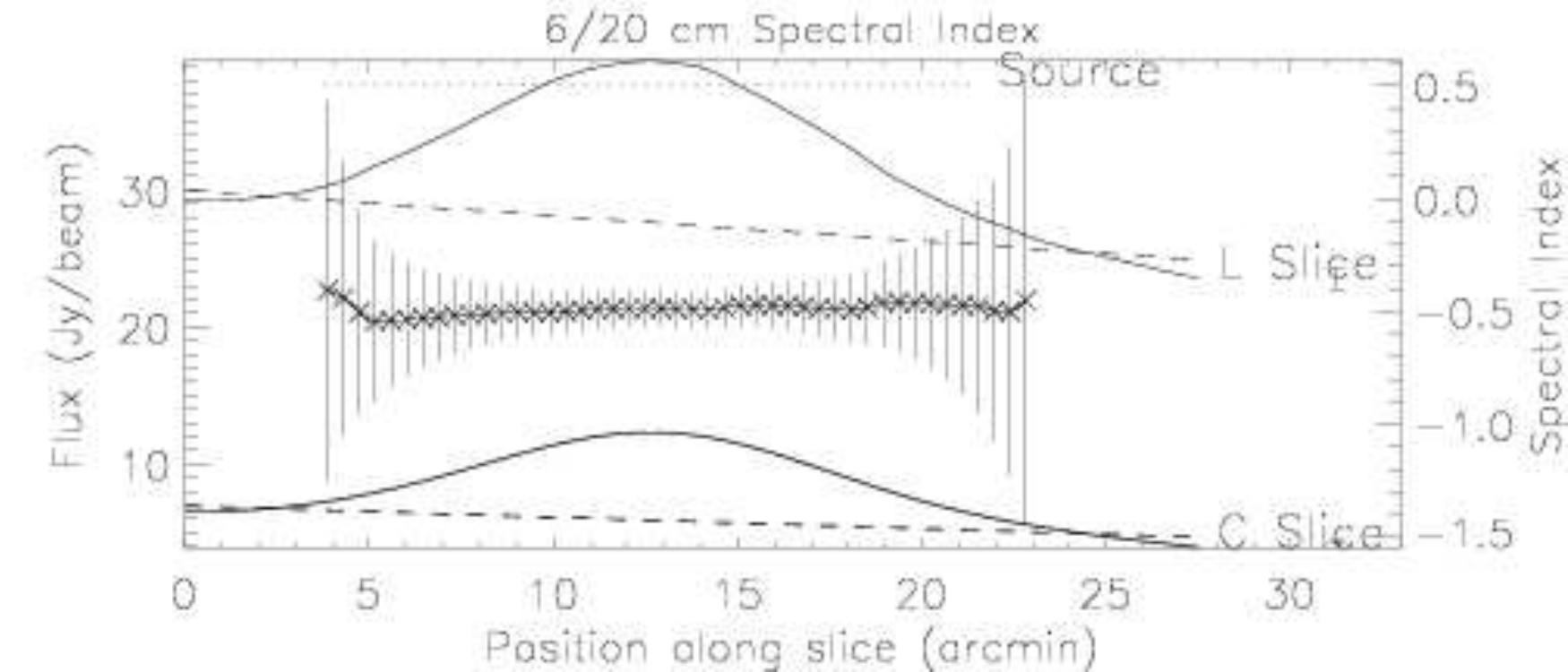}
\caption{Same as for Fig. \ref{sptornado}, but the left and right images show 6 and 20 cm emission in the west of the survey, respectively. Contours on the 6 cm survey are at $0.08*2^n$ Jy beam$^{-1}$, for $n=0-10$.  The plotted slice values correspond to the slice with $\alpha_{LC}=-0.49\pm0.06$.  \label{sptornadocl}}
\end{figure}

\begin{figure}[tbp]
\includegraphics[width=6.5in]{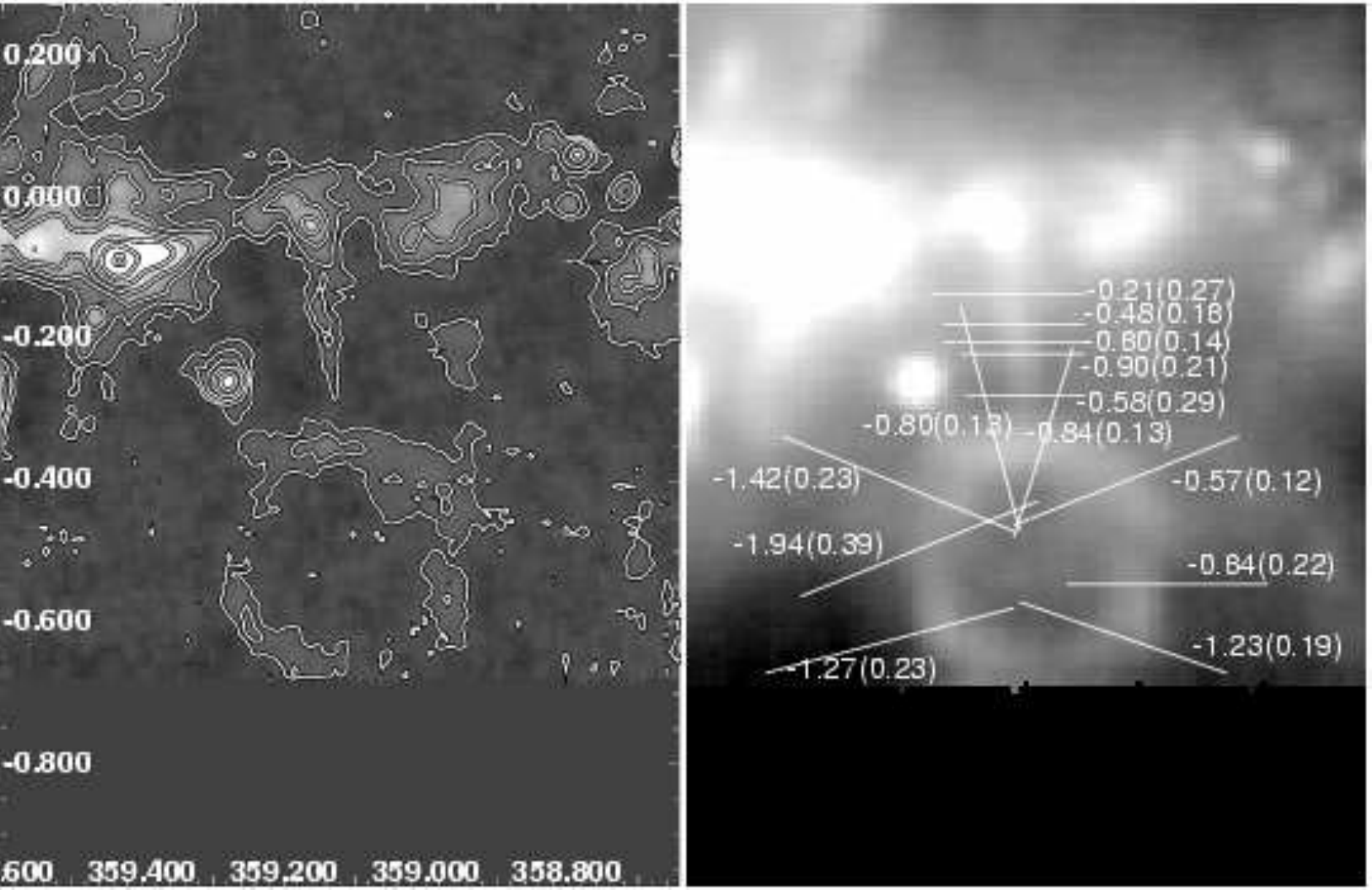}
\includegraphics[width=6.5in]{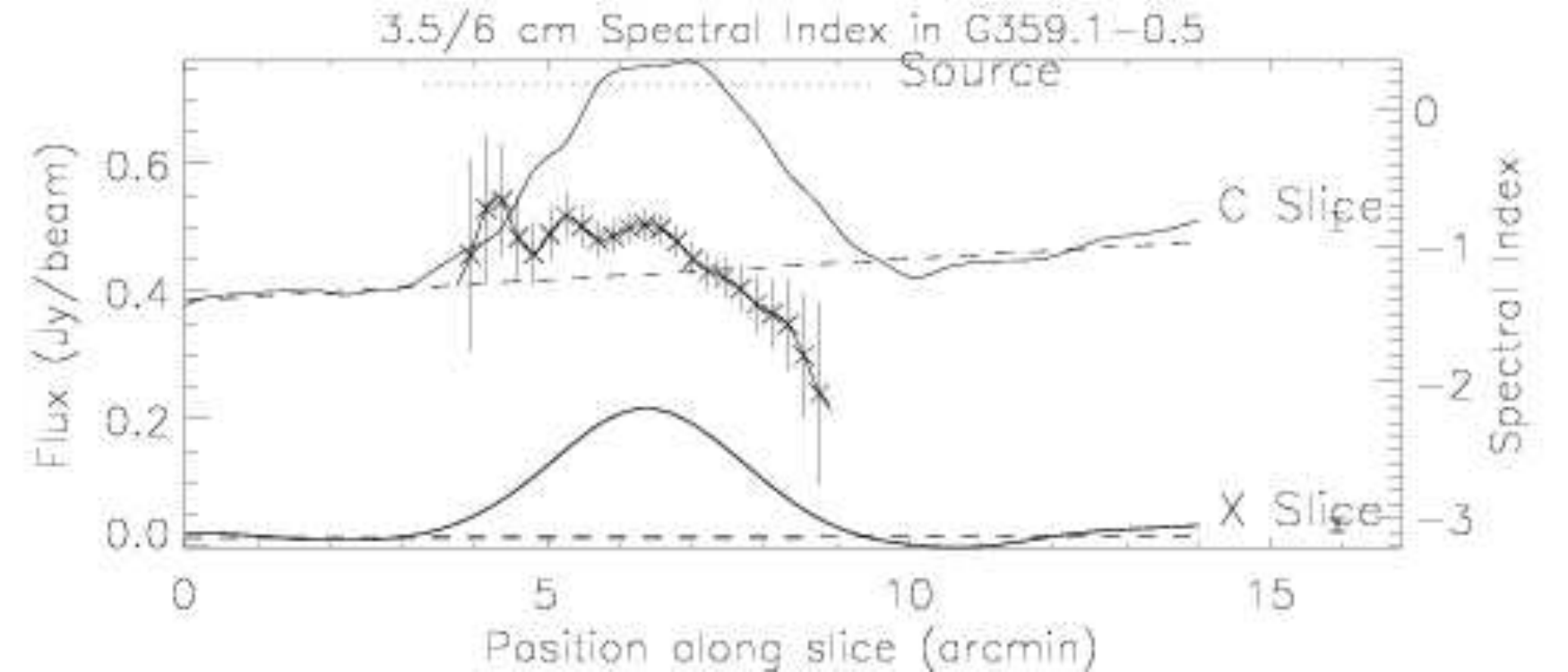}
\caption{Same as for Fig. \ref{sptornado}, but for G359.1--0.5 and the Snake (G359.1-0.2).  The plotted slice values correspond to the slice with $\alpha_{CX}=-0.84\pm0.13$. \label{spg359.1-0.5}}
\end{figure}

\begin{figure}[tbp]
\includegraphics[width=6.5in]{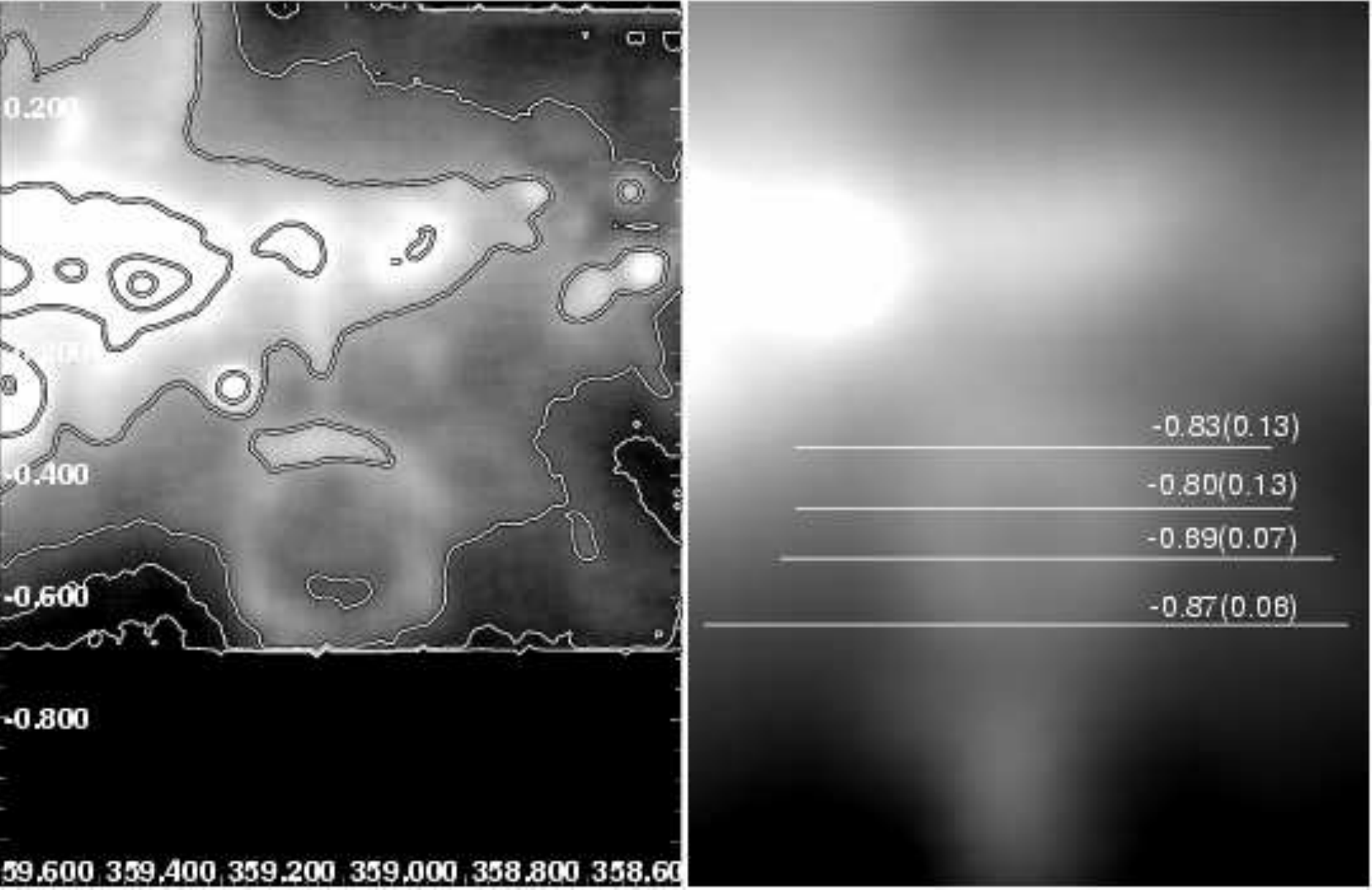}
\includegraphics[width=6.5in]{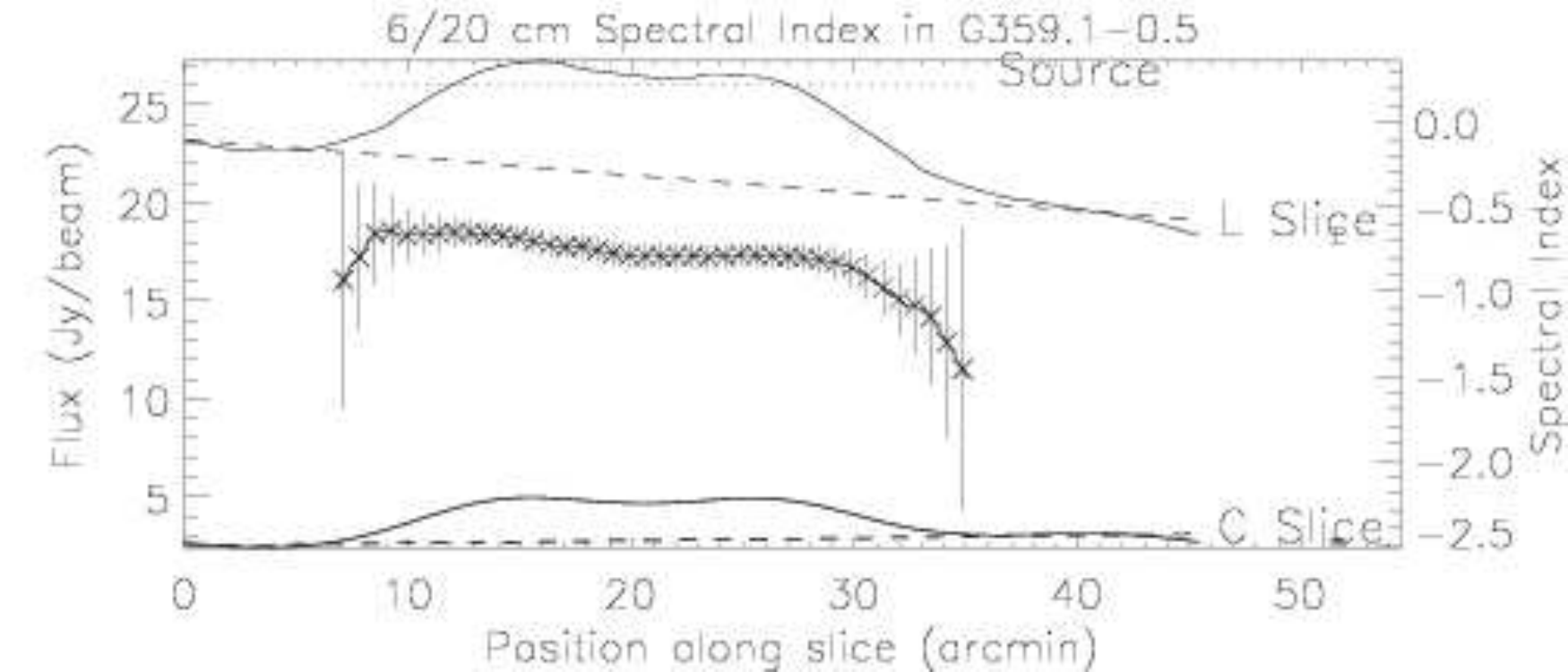}
\caption{Same as for Fig. \ref{sptornadocl}, showing images at 6 and 20 cm for the G359.1--0.5 SNR.  The plotted slice values correspond to the slice with $\alpha_{LC}=-0.89\pm0.07$.  \label{spg359.1-0.5cl}}
\end{figure}

\begin{figure}[tbp]
\includegraphics[width=6.5in]{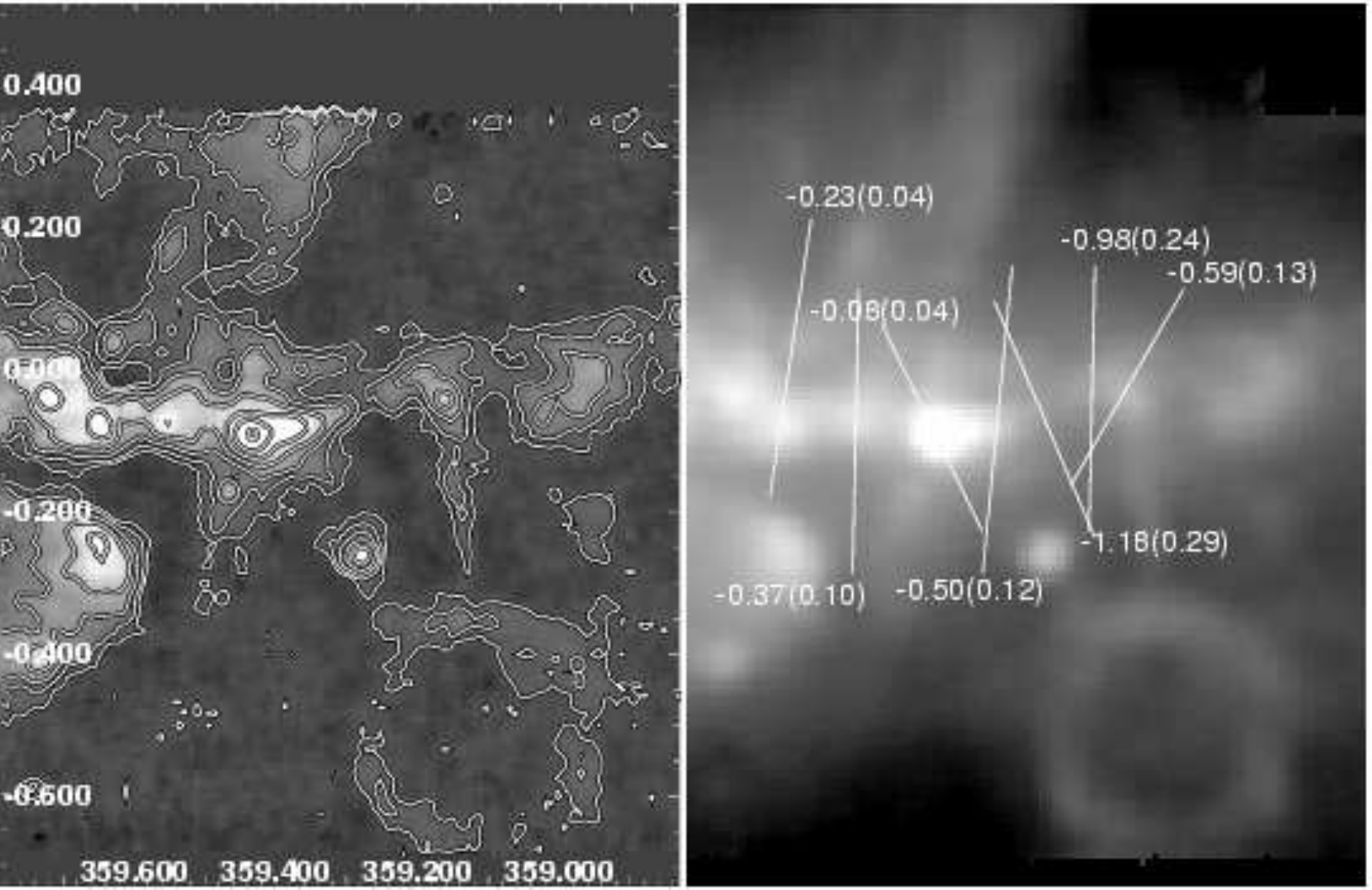}
\includegraphics[width=6.5in]{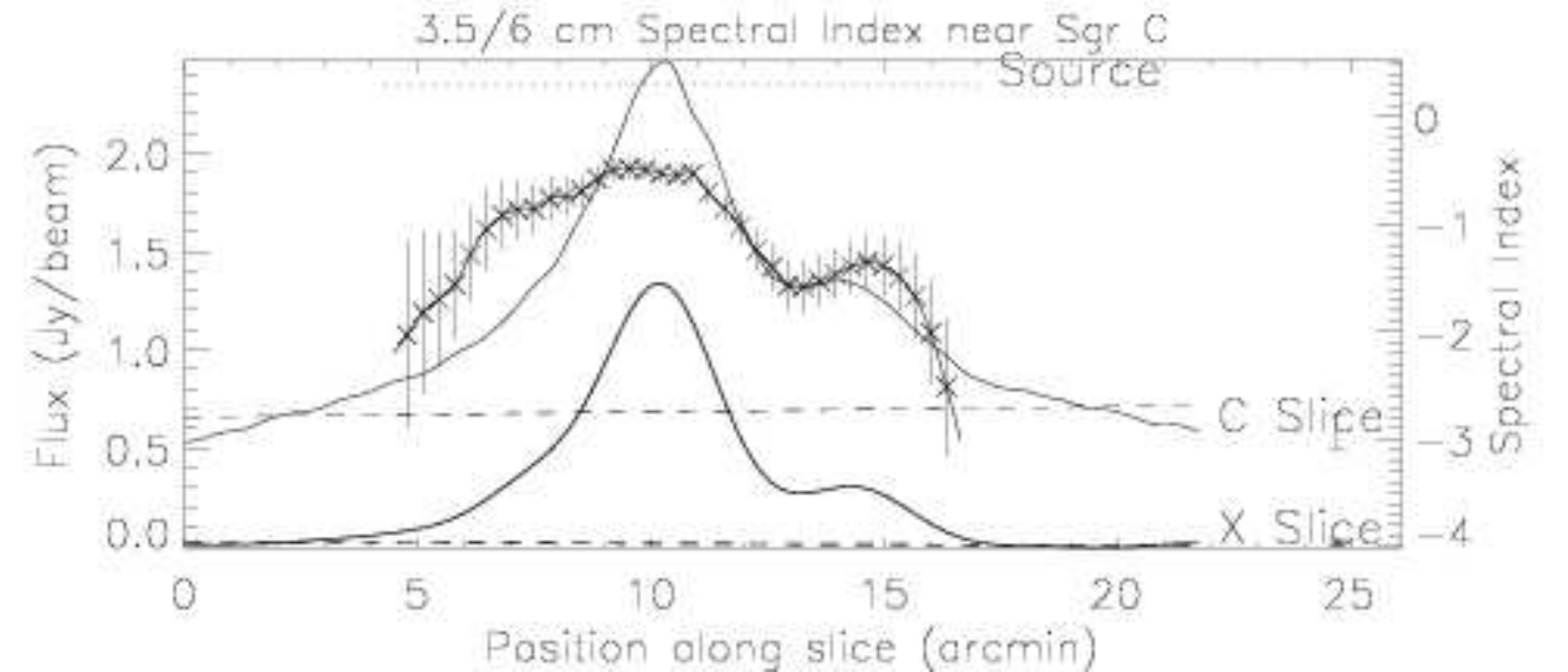}
\caption{Same as for Fig. \ref{sptornado}, but for the Sgr C complex (G359.5-0.0).  The plotted slice values correspond to the slice with $\alpha_{CX}=-0.59\pm0.13$. \label{spsgrc}}
\end{figure}

\begin{figure}[tbp]
\includegraphics[width=6.5in]{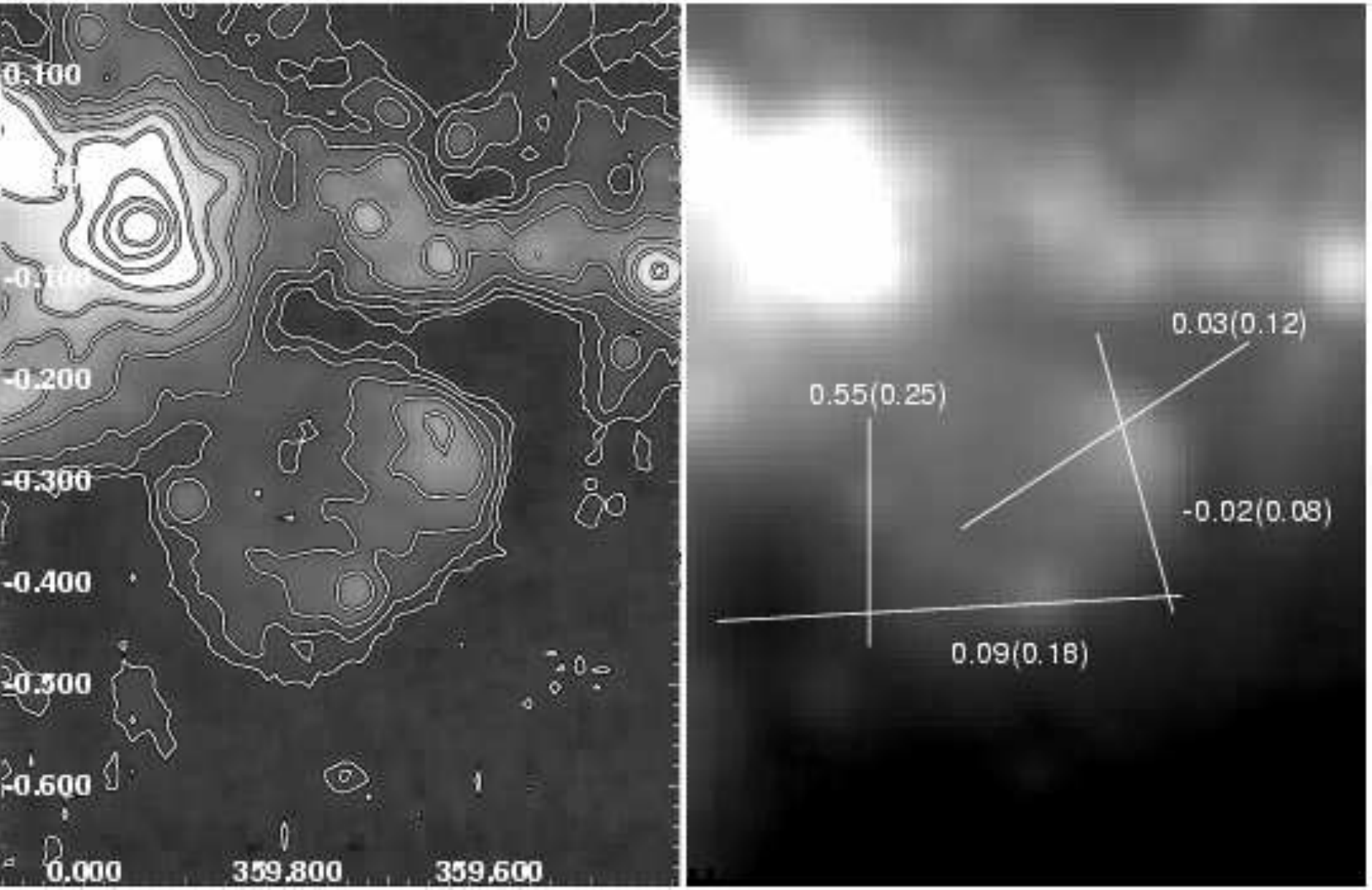}
\includegraphics[width=6.5in]{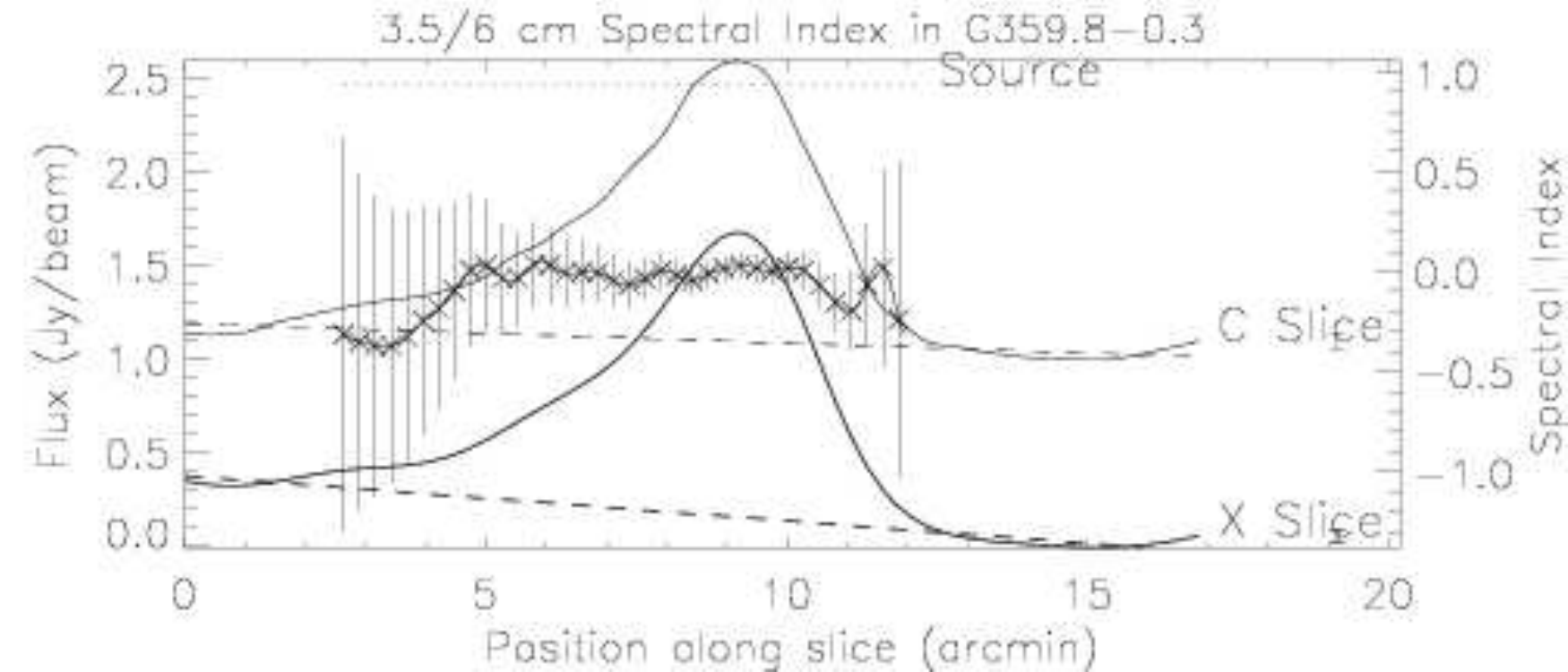}
\caption{Same as for Fig. \ref{sptornado}, but for the G359.8--0.3 complex.  The plotted slice values correspond to the slice with $\alpha_{CX}=0.03\pm0.12$.  \label{spg359.8-0.3}}
\end{figure}

\begin{figure}[tbp]
\includegraphics[width=6.5in]{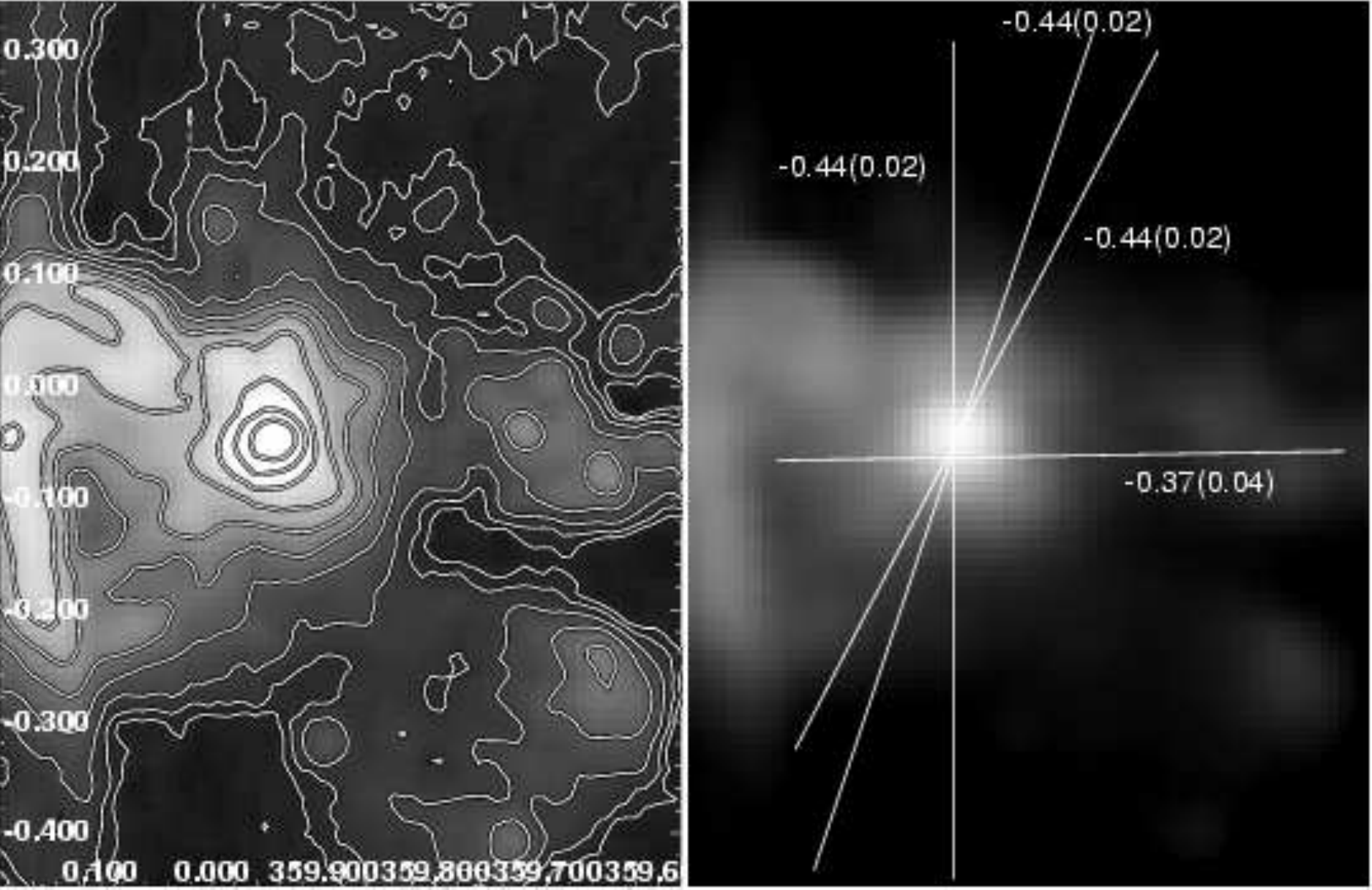}
\includegraphics[width=6.5in]{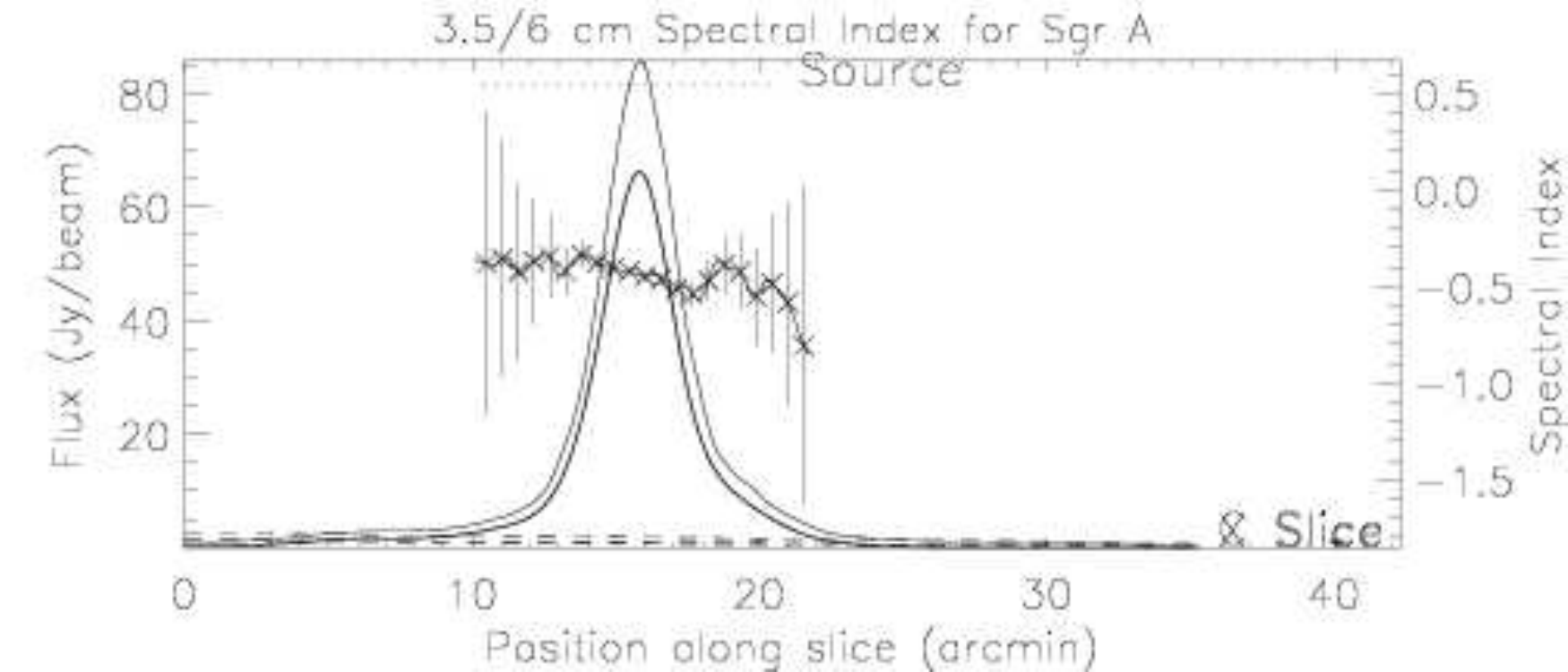}
\caption{Same as for Fig. \ref{sptornado}, but for the Sgr A complex (G0.07+0.04).  The plotted slice values correspond to the slice with $\alpha_{CX}=-0.44\pm0.02$ and the rightmost label.  \label{spsgra}}
\end{figure}

\begin{figure}[tbp]
\includegraphics[width=6.5in]{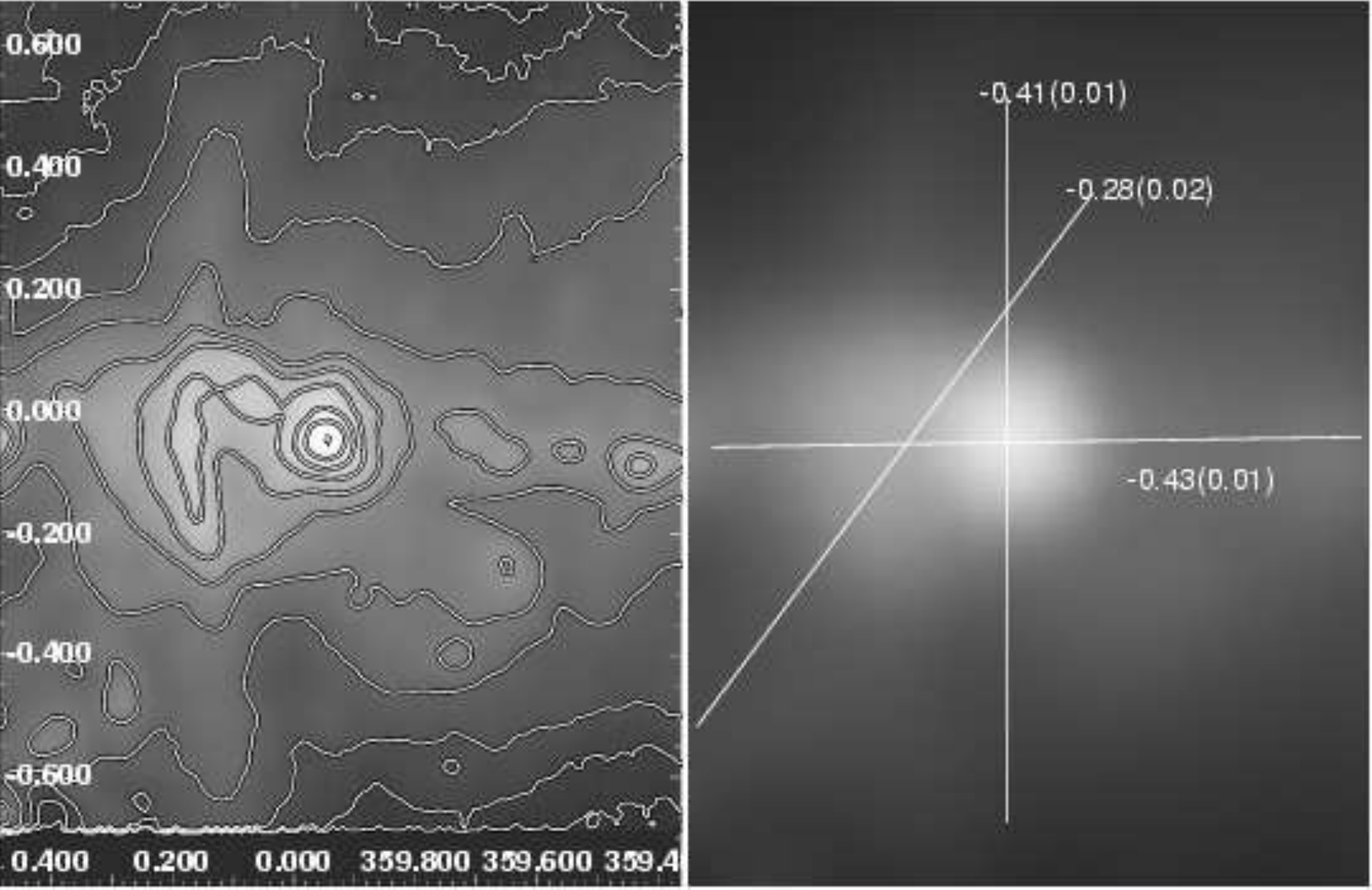}
\includegraphics[width=6.5in]{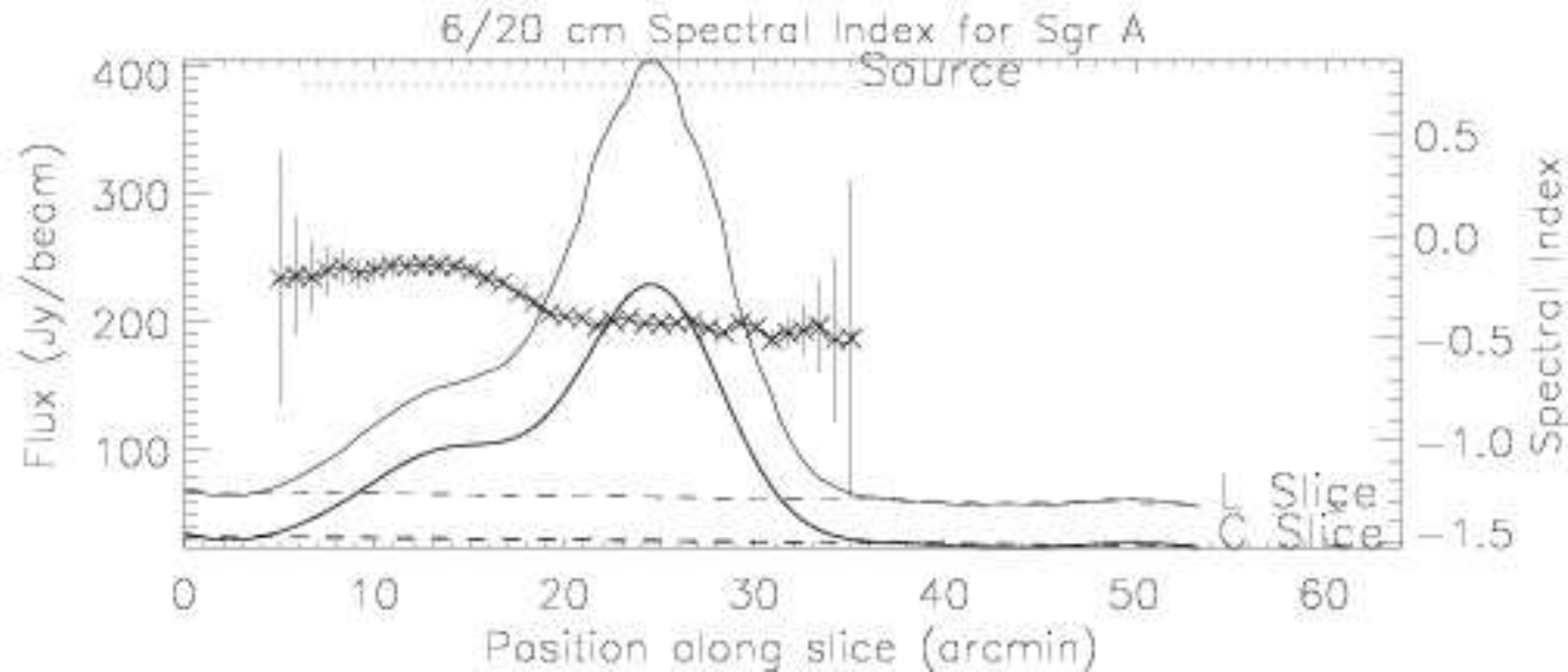}
\caption{Same as for Fig. \ref{sptornadocl}, showing images at 6 and 20 cm for the Sgr A complex (G0.0+0.0).  The plotted slice values correspond to the slice with $\alpha_{LC}=-0.43\pm0.01$.  \label{spsgracl}}
\end{figure}

\begin{figure}[tbp]
\includegraphics[width=6.5in]{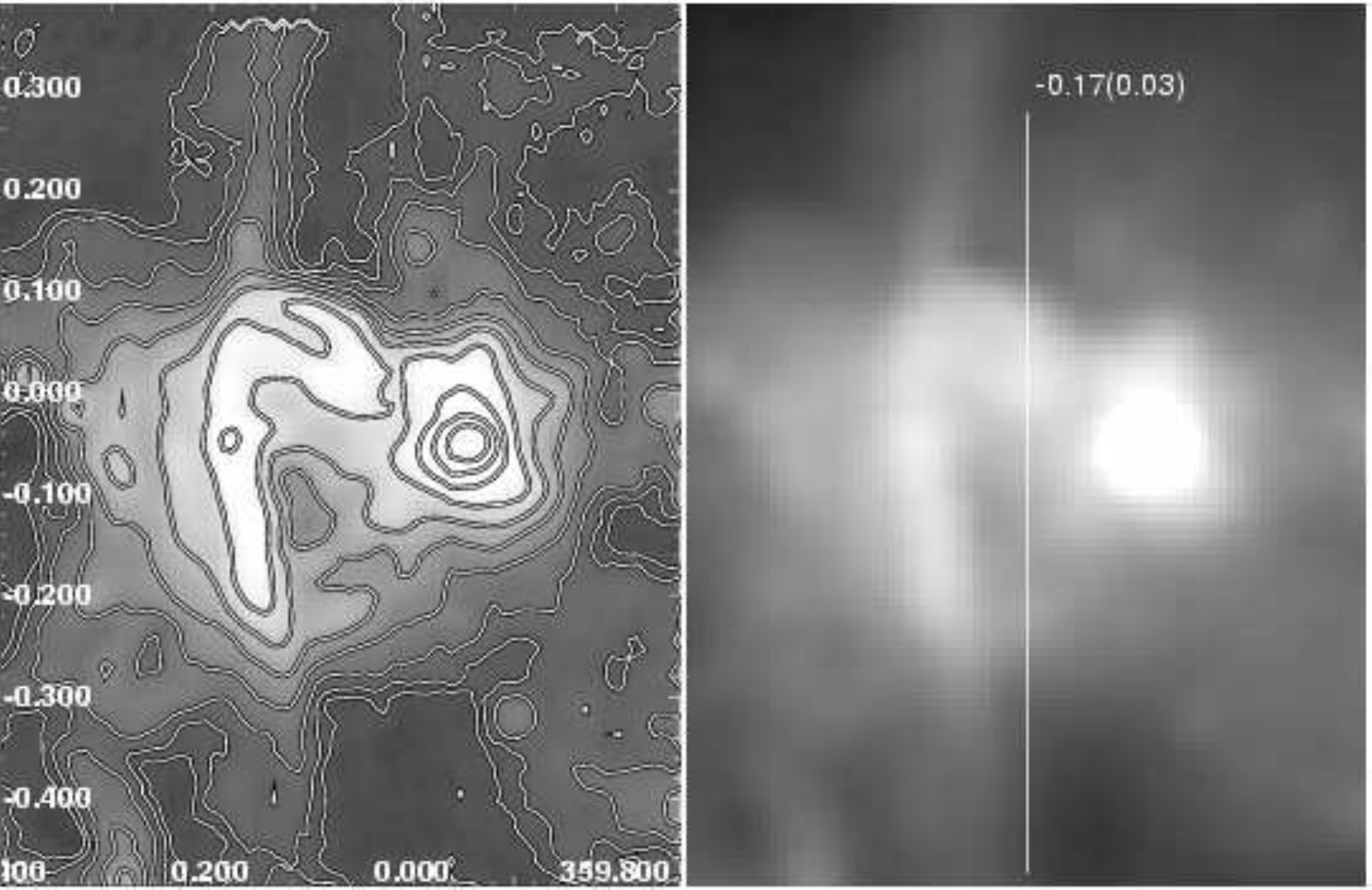}
\includegraphics[width=6.5in]{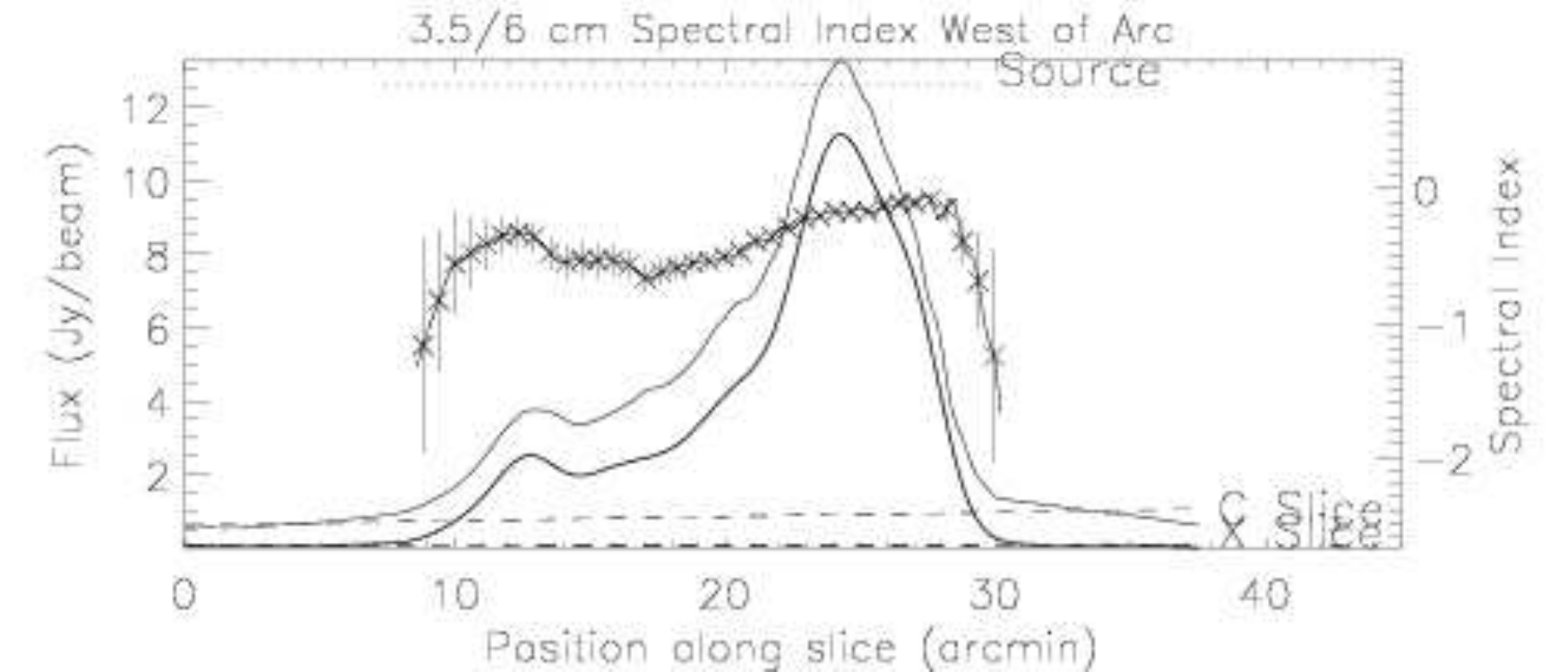}
\caption{Same as for Fig. \ref{sptornado}, but for the Arched filaments complex (G0.07+0.04).  The plotted slice values correspond to the slice shown at the top right. \label{spwestarc}}
\end{figure}

\begin{figure}[tbp]
\includegraphics[width=6.5in]{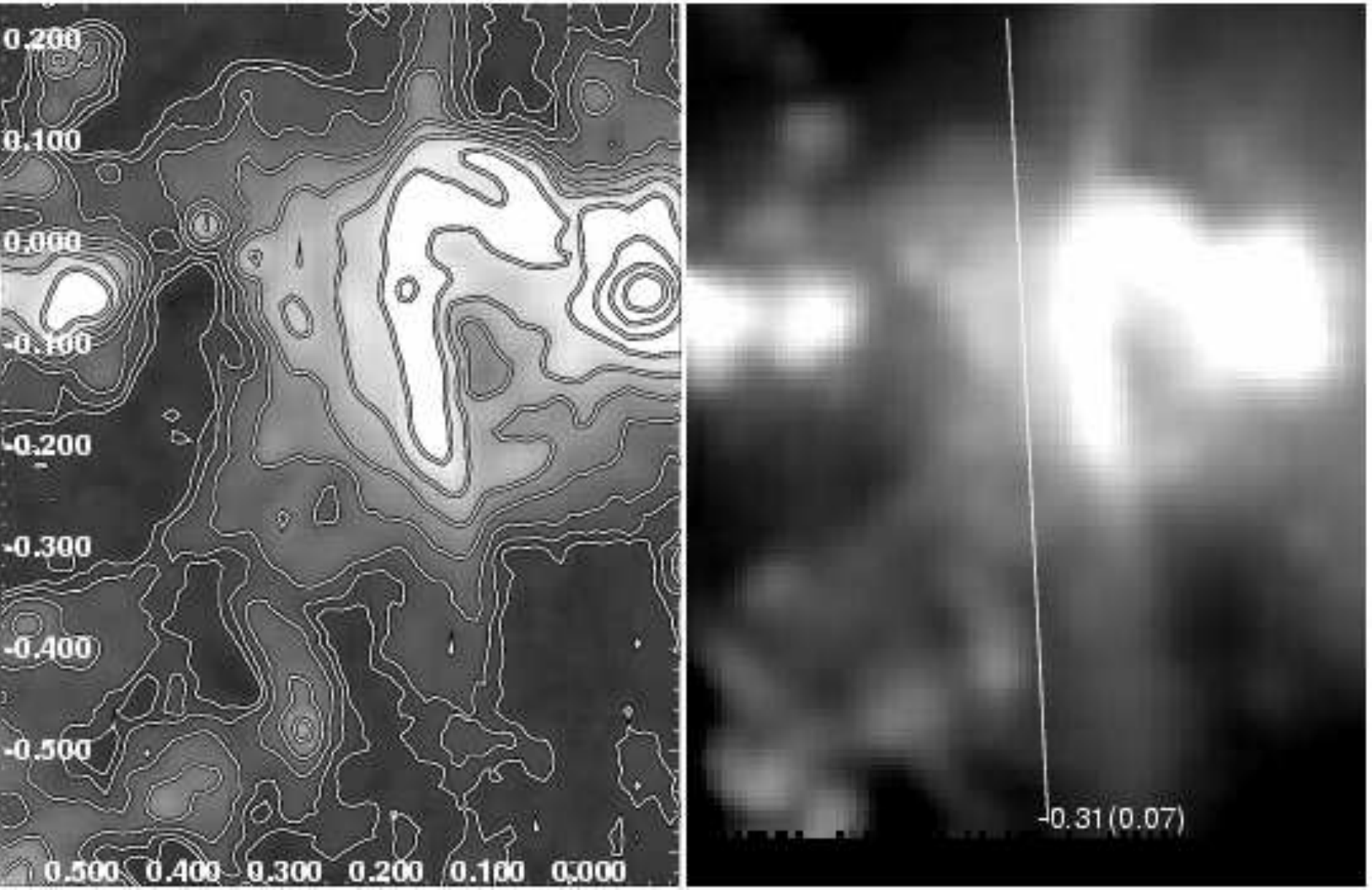}
\includegraphics[width=6.5in]{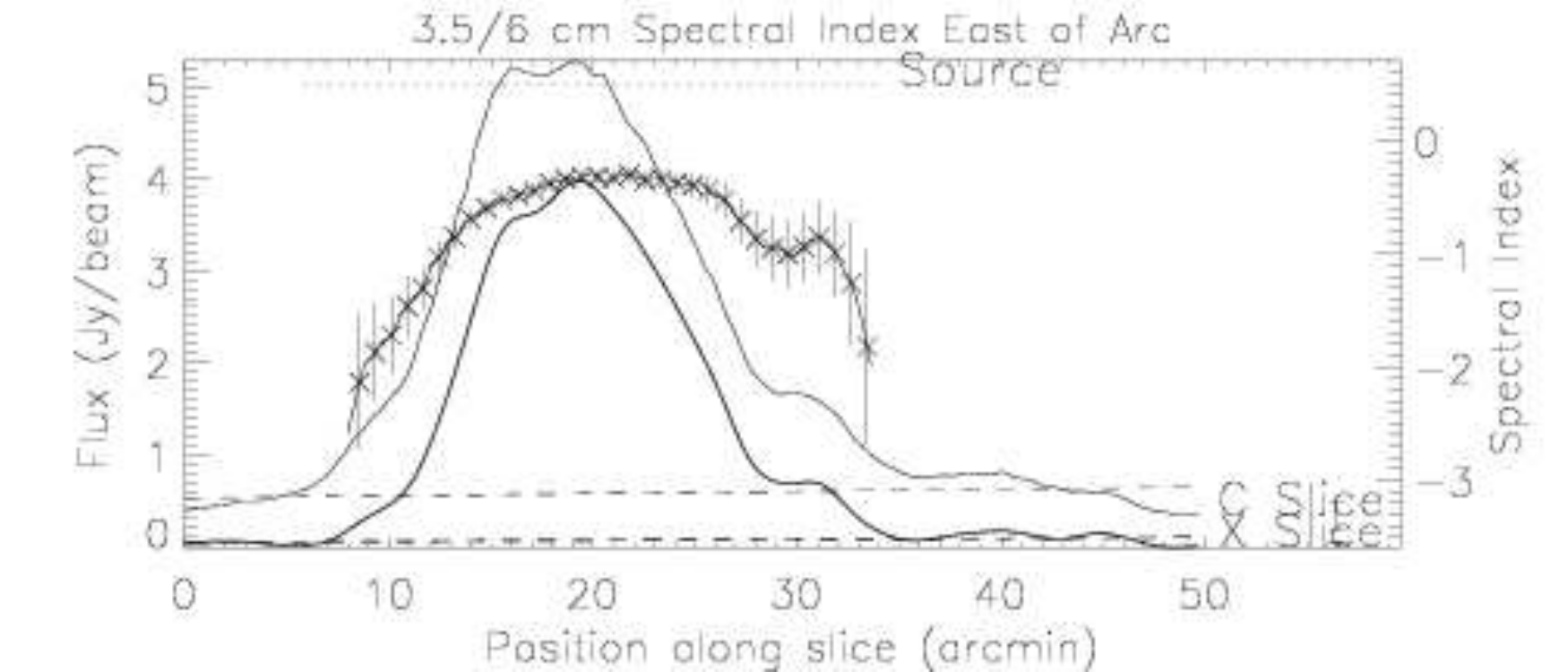}
\caption{Same as for Fig. \ref{sptornado}, but for the emission east of the Radio Arc (G0.2-0.0).  The plotted slice values correspond to the slice shown at the top right.  At position near 10\arcmin, the slice crosses the SNR G0.33+0.04 and has a spectral index $\alpha_{CX}$ of roughly --2.0 to --1.5.  \label{speastarc}}
\end{figure}

\begin{figure}[tbp]
\includegraphics[width=6.5in]{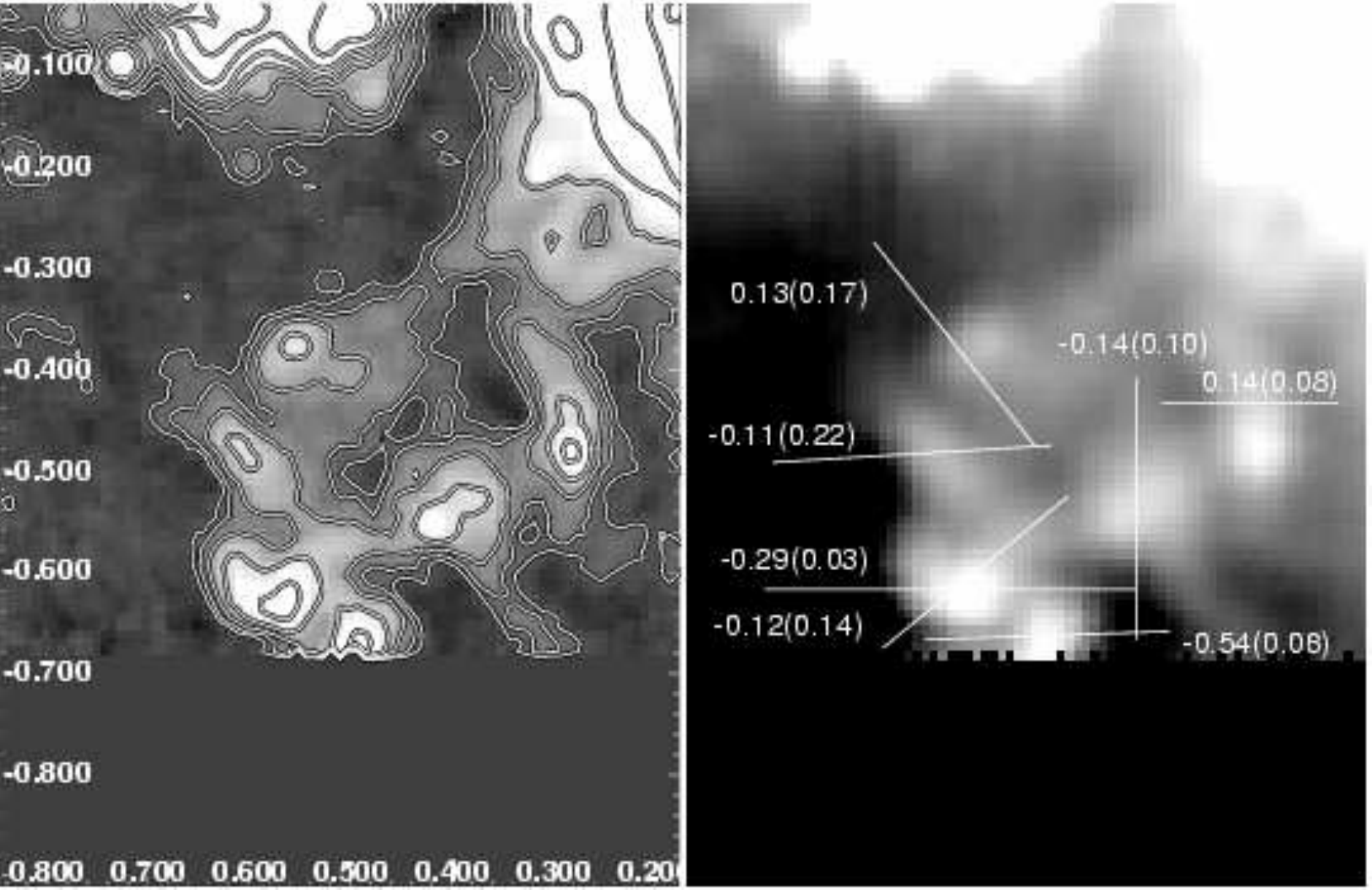}
\includegraphics[width=6.5in]{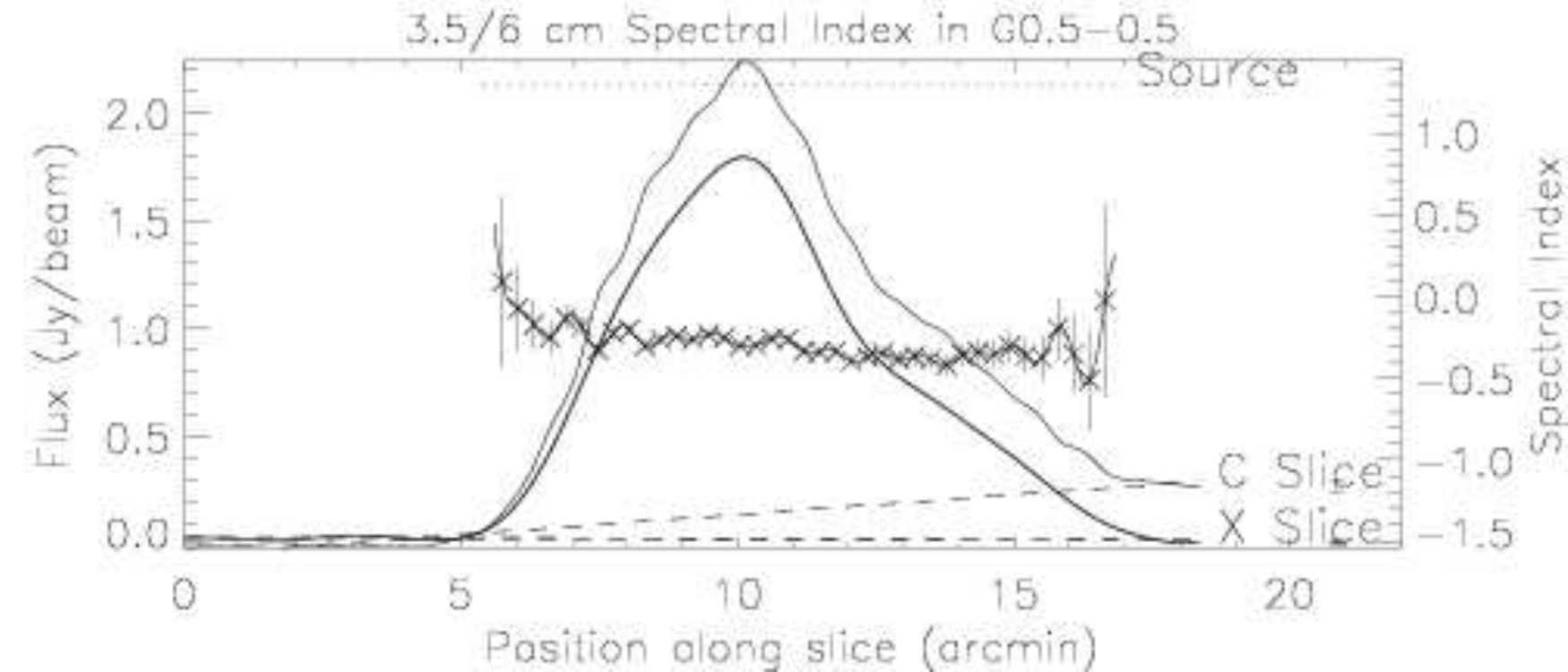}
\caption{Same as for Fig. \ref{sptornado}, but for the G0.5--0.5 complex.  The plotted slice values correspond to the slice with $\alpha_{CX}=-0.29\pm0.03$.  \label{spg0.5-0.5}}
\end{figure}

\begin{figure}[tbp]
\includegraphics[width=6.5in]{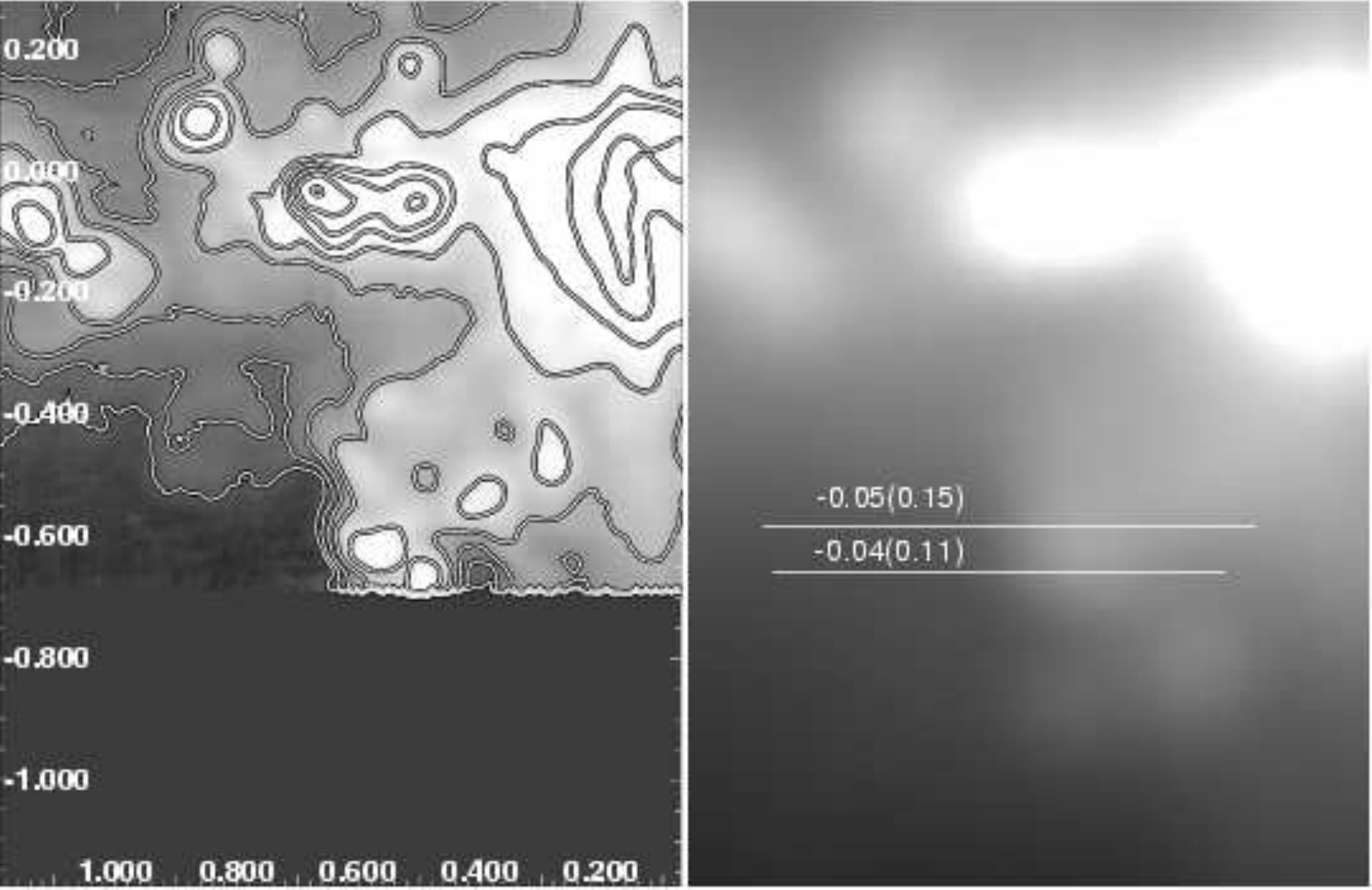}
\includegraphics[width=6.5in]{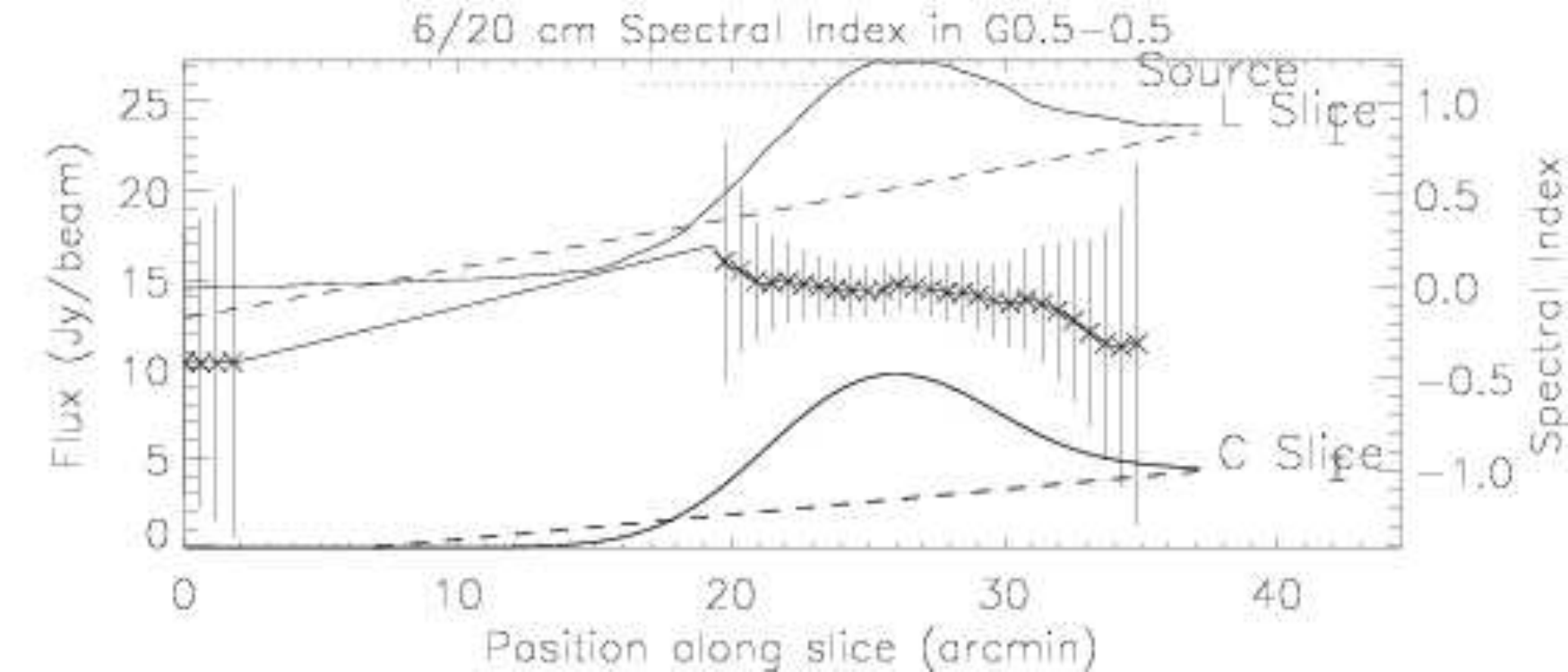}
\caption{Same as for Fig. \ref{sptornadocl}, showing images at 6 and 20 cm for the G0.5--0.5 complex. The plotted slice values correspond to the slice with $\alpha_{LC}=-0.04\pm0.11$.  \label{spg0.5-0.5cl}}
\end{figure}

\begin{figure}[tbp]
\includegraphics[width=6.5in]{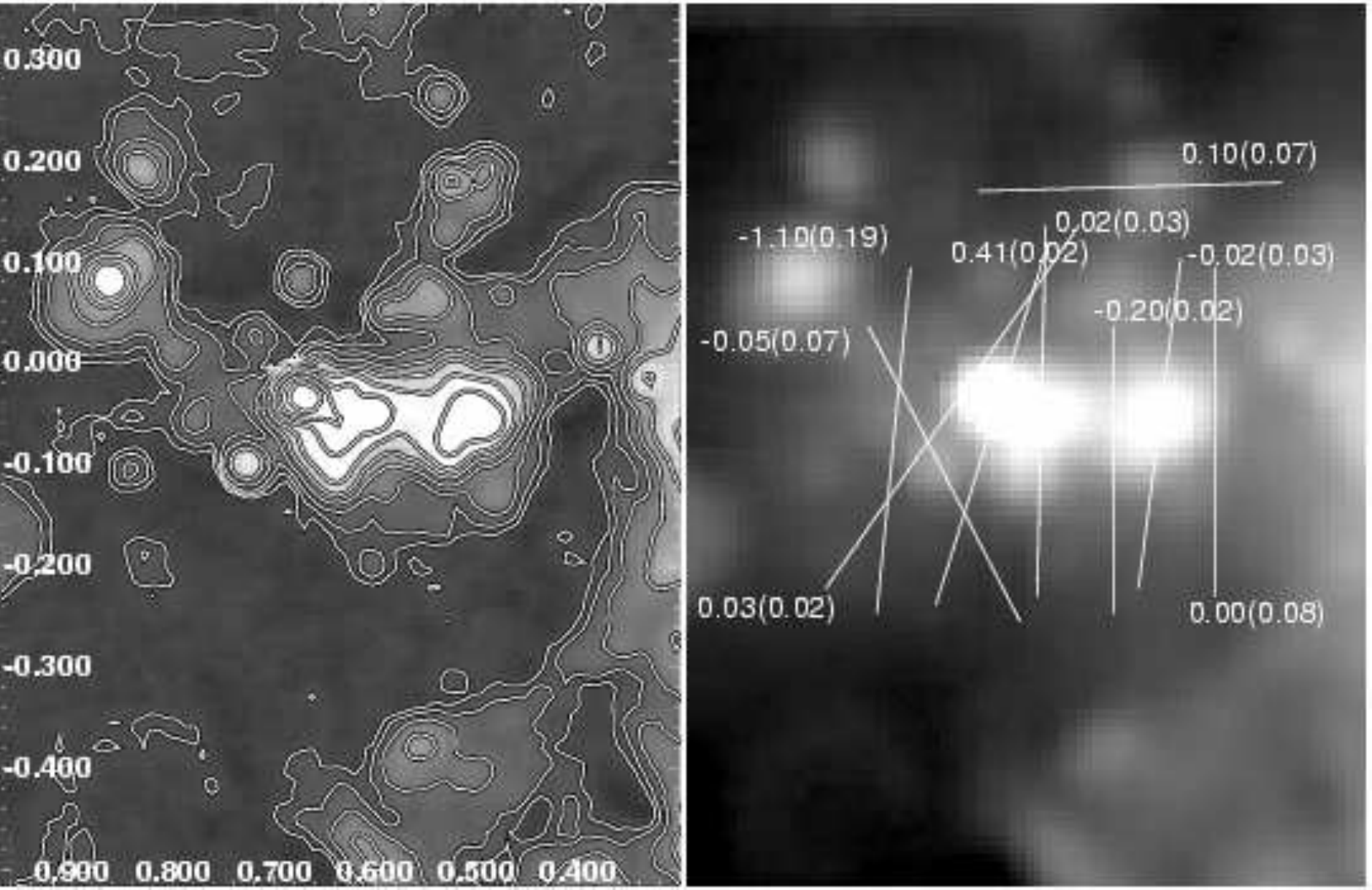}
\includegraphics[width=6.5in]{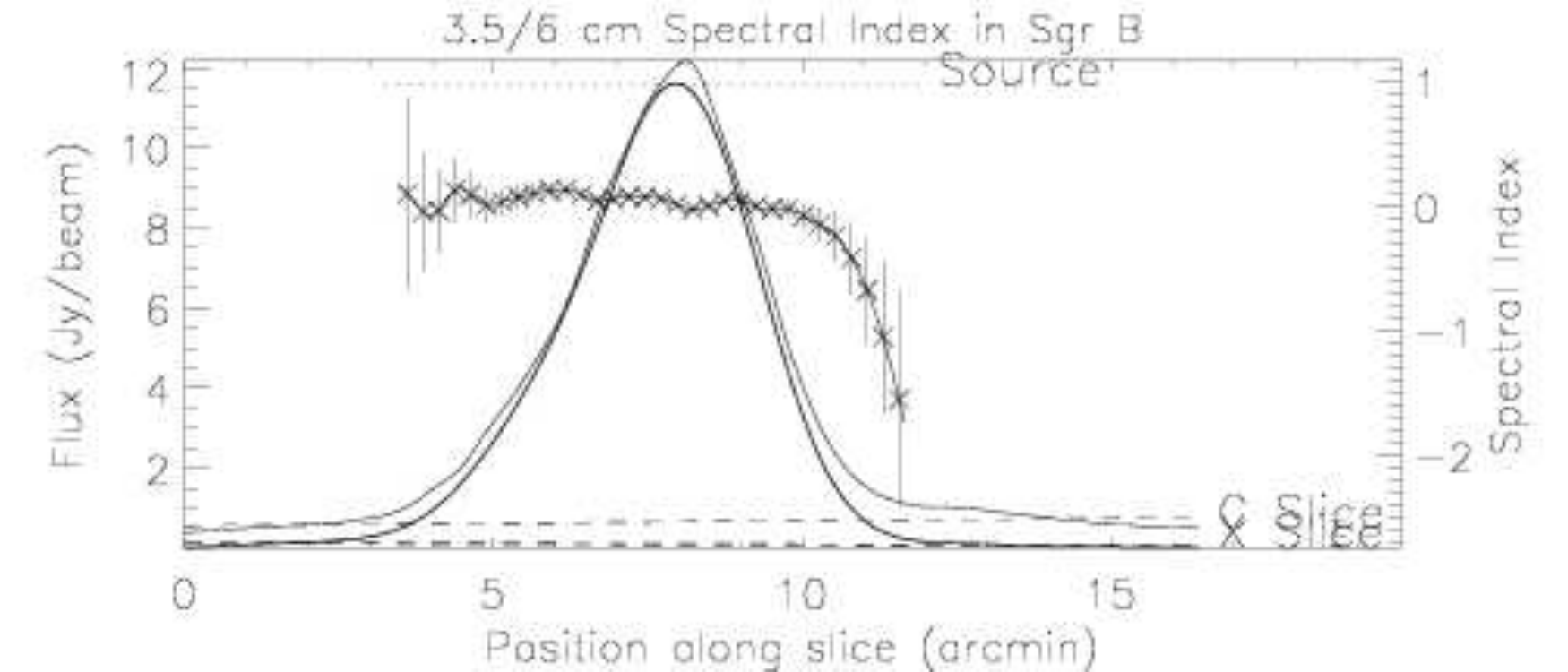}
\caption{Same as for Fig. \ref{sptornado}, but for the Sgr B complex (G0.5--0.0 and G0.7--0.0).  The plotted slice values correspond to the slice with $\alpha_{CX}=0.02\pm0.03$.  \label{spsgrb}}
\end{figure}

\begin{figure}[tbp]
\includegraphics[width=6.5in]{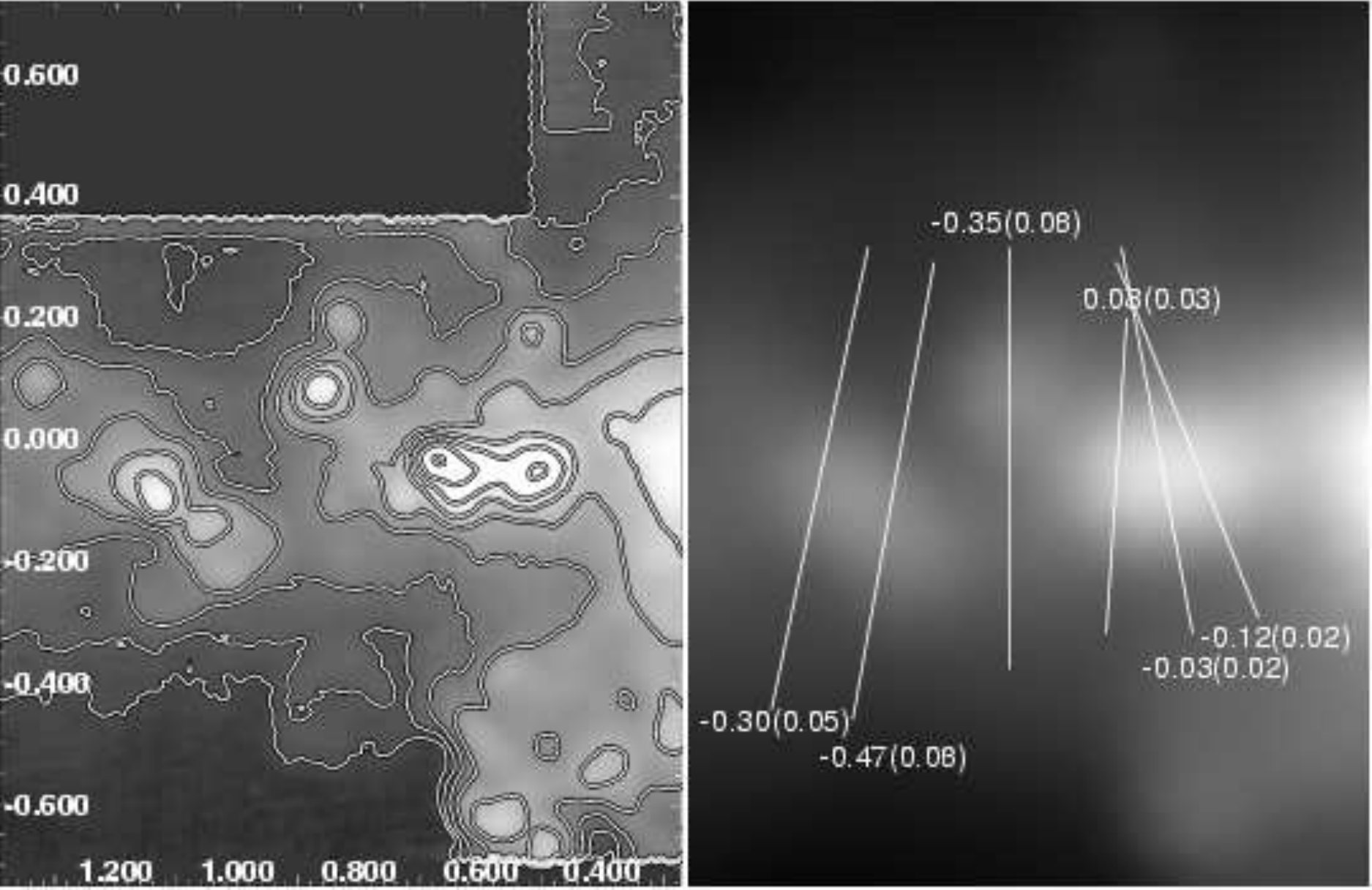}
\includegraphics[width=6.5in]{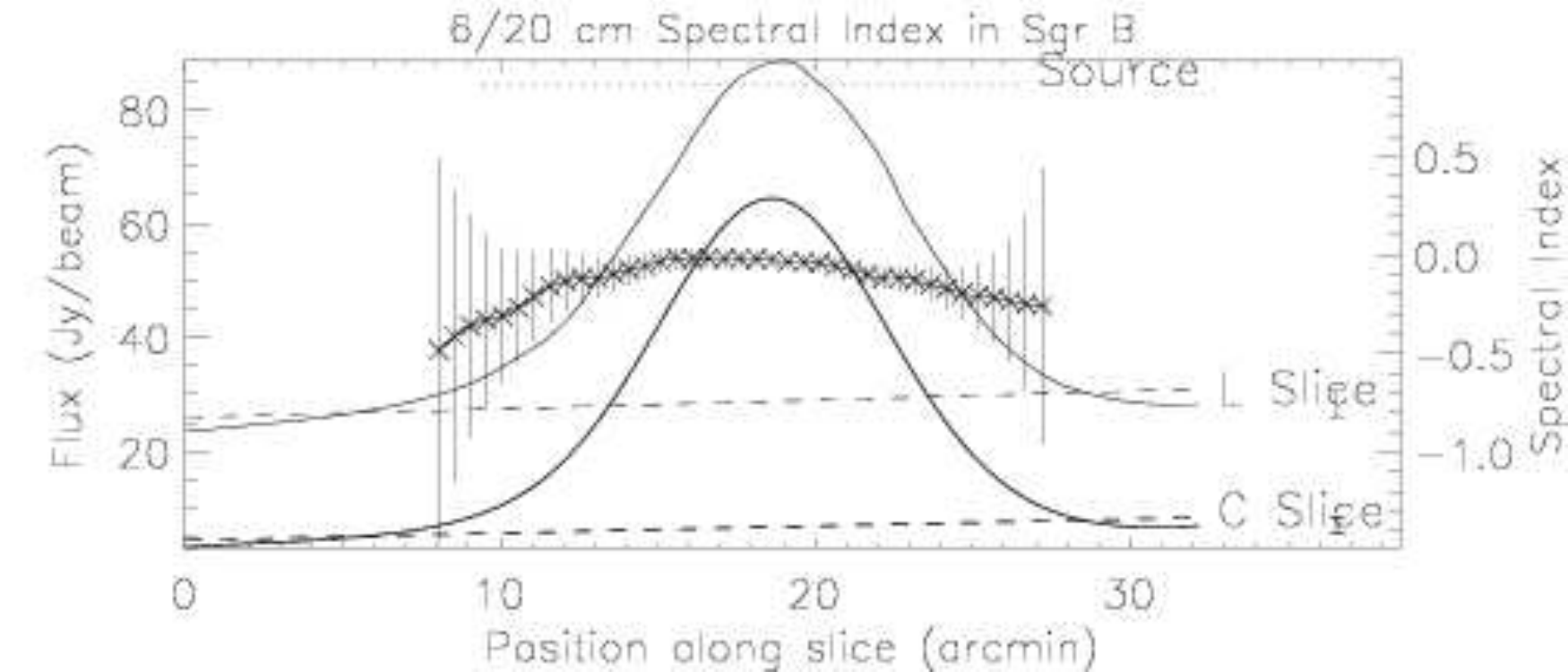}
\caption{Same as for Fig. \ref{sptornadocl}, showing images at 6 and 20 cm for the Sgr B (G0.5--0.0 and G0.7--0.0) complex and eastern sources. The plotted slice values correspond to the slice with $\alpha_{LC}=-0.03\pm0.02$.  \label{spsgrbcl}}
\end{figure}

Figures \ref{sptornado} through \ref{spsgrbcl} show details of the 3.5, 6, and 20 cm surveys and the slices used to study the spectral indices of objects between these frequencies.  The figures are ordered with increasing Galactic longitude, starting from the western edge of the 3.5 and 6 cm surveys.  Each figure shows two images convolved to the same resolution with slice positions overlaid.  For each figure, one representative slice is plotted below the images.

\begin{figure}[tbp]
\includegraphics[width=6.5in]{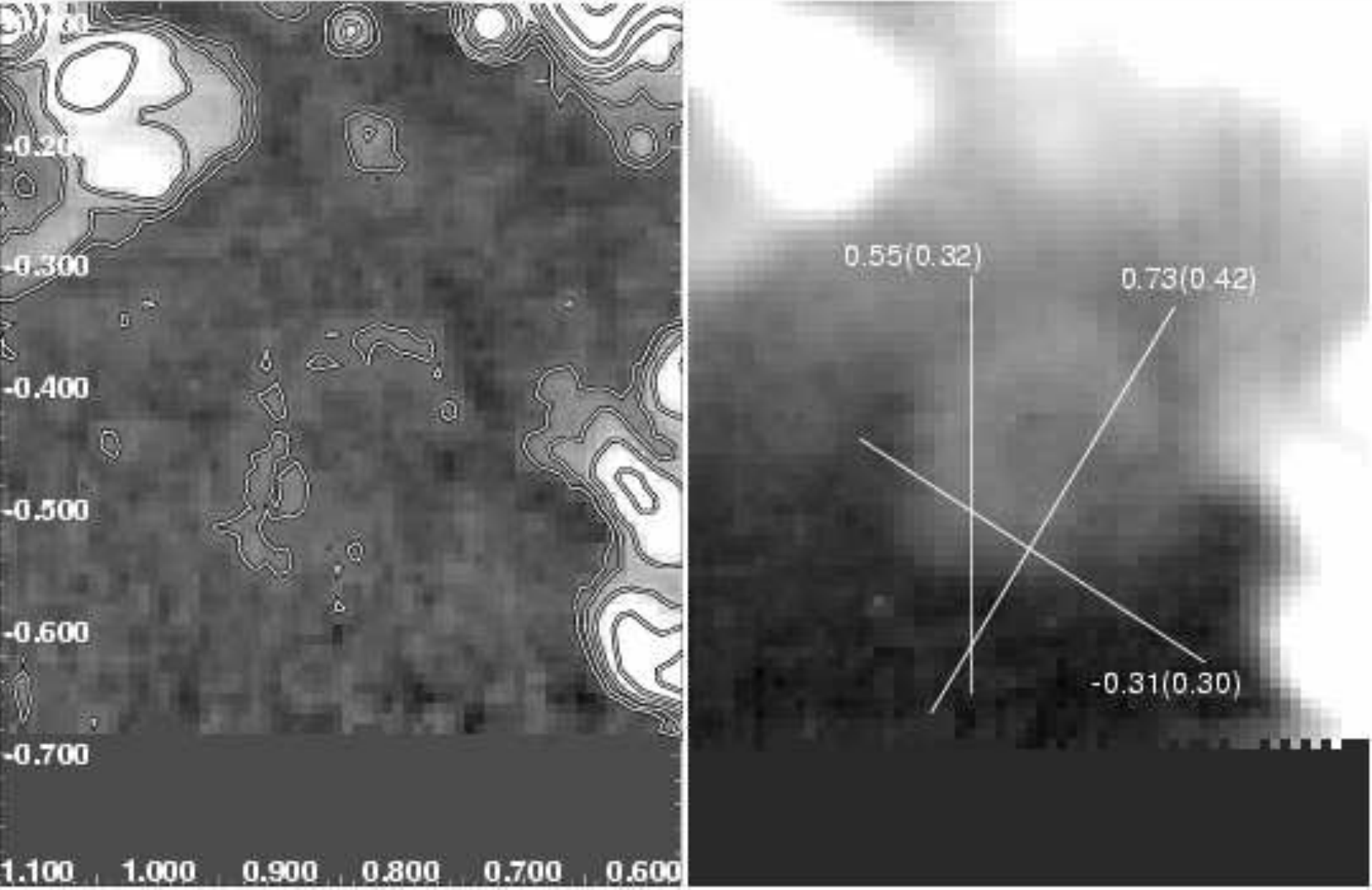}
\includegraphics[width=6.5in]{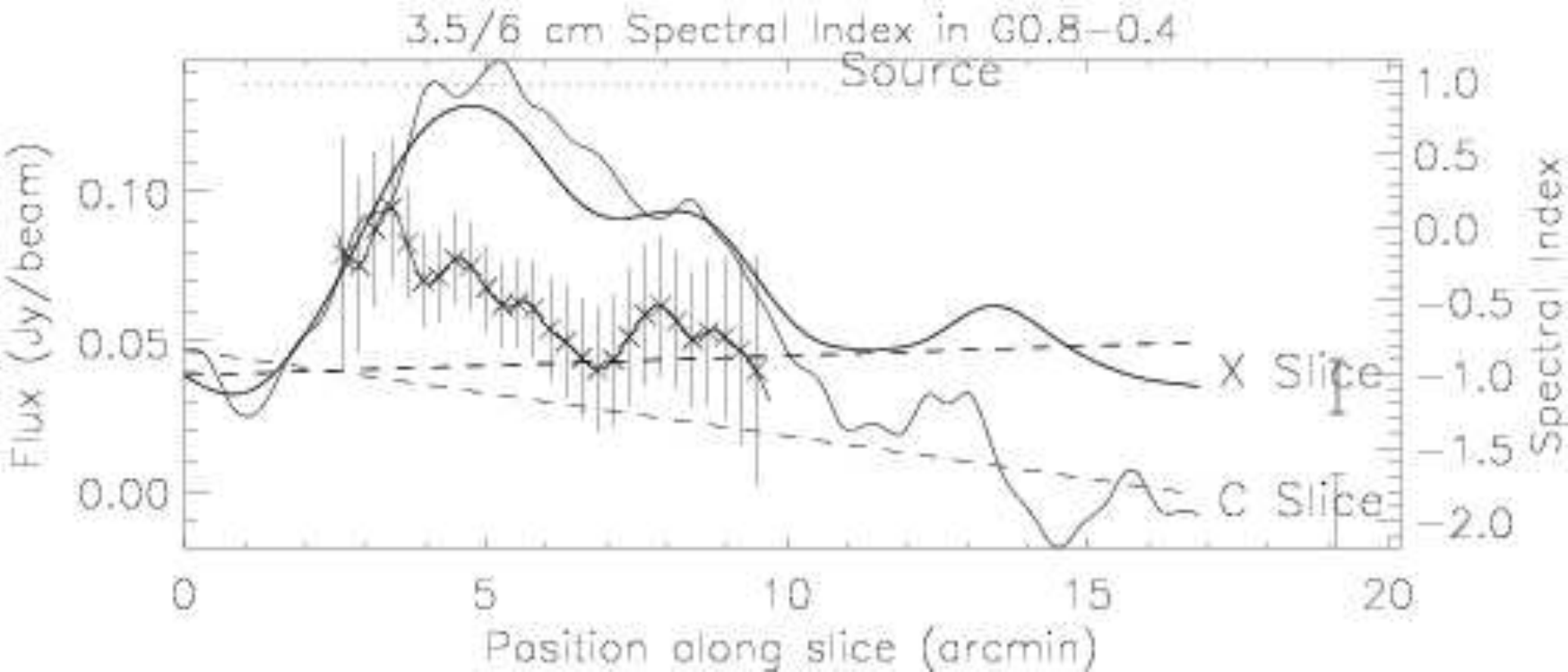}
\caption{Same as for Fig. \ref{sptornado}, but for the G0.8--0.4 complex.   The plotted slice values correspond to the slice with $\alpha_{CX}=-0.31\pm0.30$. \label{spg0.8-0.4}}
\end{figure}

\begin{figure}[tbp]
\includegraphics[width=6.5in]{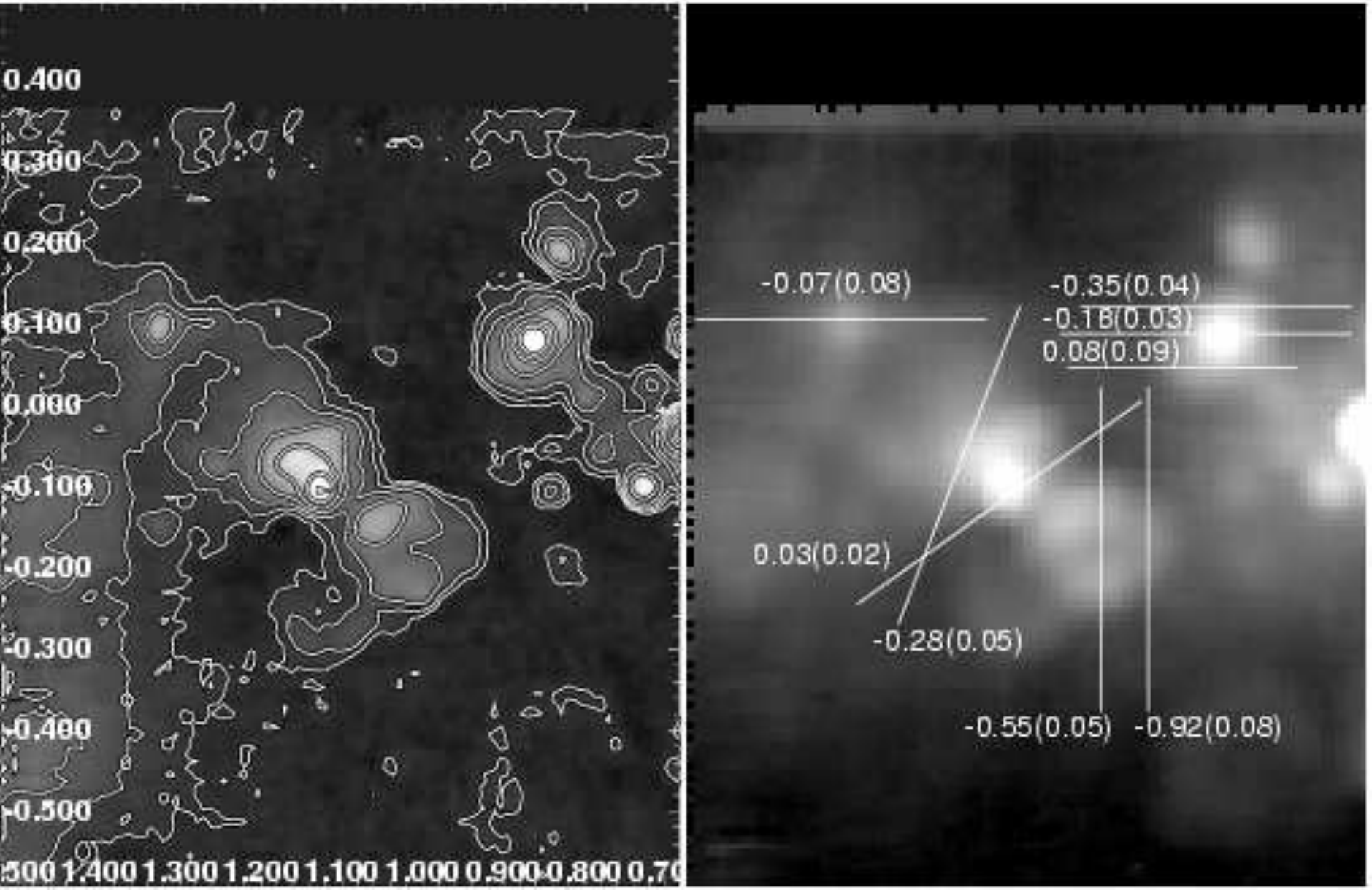}
\includegraphics[width=6.5in]{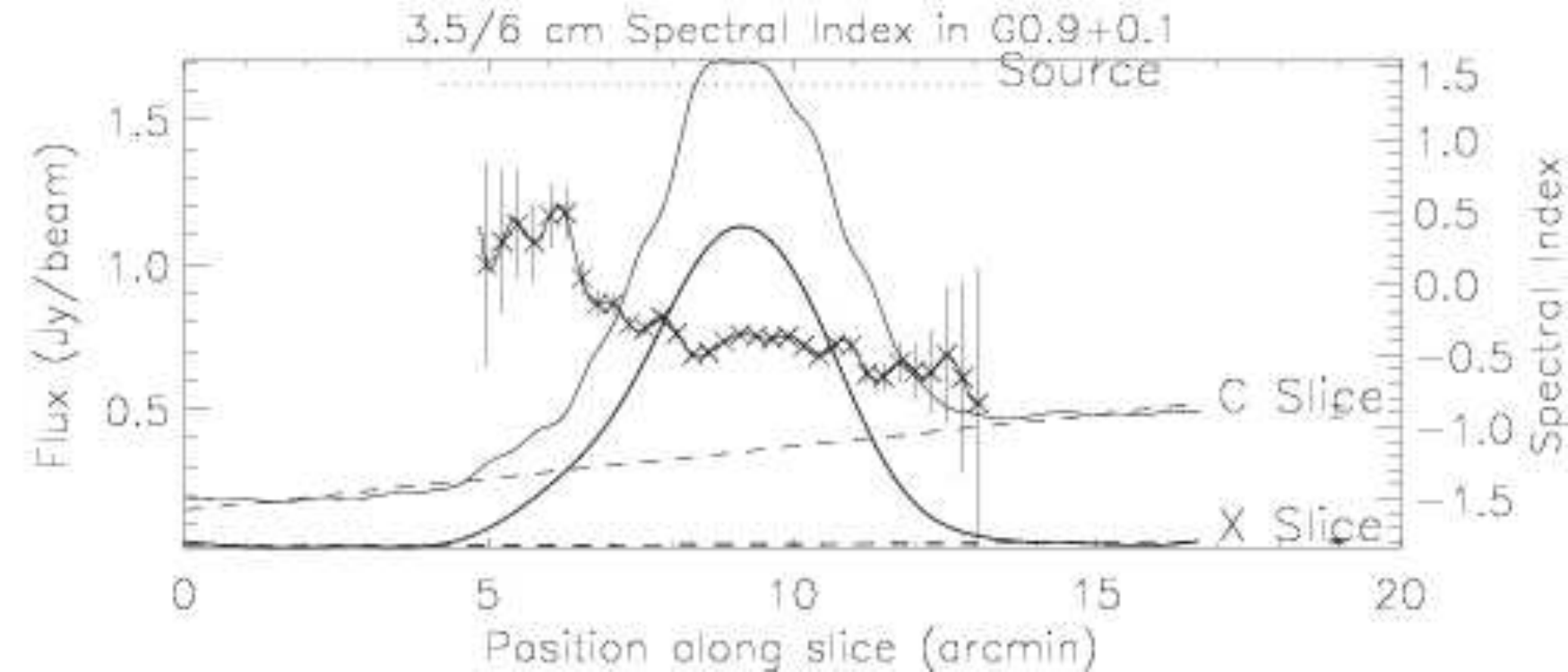}
\caption{Same as for Fig. \ref{sptornado}, but for G0.9+0.1 and eastern complex.  The plotted slice values correspond to the slice with $\alpha_{CX}=-0.35\pm0.04$.  \label{spg0.9+0.1}}
\end{figure}

\clearpage

\paragraph{G357.7--0.1 (The Tornado)}
G357.7--0.1, also known as ``The Tornado'' for its unusual, twisted morphology, is found near the western edge of the 3.5 and 6 cm maps.  In Figure \ref{sptornado}, the 3.5 and 6 cm morphology looks like a head-tail source with integrated flux densities of about 14, 18, and 37 Jy at 3.5, 6 cm, and 20 cm, respectively.  The Tornado is believed to be a mixed-mophology SNR at a distance of about 12 kpc \citep{g03,b03a}.   Mixed-morphology SNRs are characterized by a radio continuum shell filled with thermal, x-ray--emitting gas \citep[e.g.,][]{yu03}.  The elongated morphology of the Tornado is unsual for a SNR, but can be explained by a high proper motion and the fact that the Tornado is interacting with a molecular cloud \citep{f96}.

The spectral index between 6 and 3.5 cm is shown in Figure \ref{sptornado} and between 20 and 6 cm in Figure \ref{sptornadocl}.  The 6/3.5 cm spectral index for most slices perpendicular to the long axis are equal within their $3\sigma$ errors except for one slice on the western side (the ``head'') of the Tornado.  The slice along the length of the Tornado shows a gradual flattening of the spectral index from $\alpha_{CX}=-0.48\pm0.02$ near the head to $-0.33\pm0.07$ in the tail.  The typical spectral index values are $\alpha_{CX}\sim-0.45$ and $\alpha_{LC}\sim-0.63$, with uncertainties of about 0.03 and 0.01, respectively.  The 6/3.5 cm spectral index from the slice analysis is consistent with the integrated spectral index of $-0.50\pm0.07$, given in Table \ref{diffsrcspix}.  The 20/6 cm spectral index is steeper than the 6/3.5 cm index, which suggests that the spectral index steepening at lower frequencies.  From \citet{gr04} and \citet{b85}, the 1 GHz flux of the Tornado is 37 Jy and spectral index is --0.4, while \citet{g94} finds $S_{843 MHz}=49$ Jy.  Extrapolating from our measured 20 cm flux density of 37 Jy and $\alpha_{LC}\sim-0.63$, we predict an integrated flux density of 46 and 51 Jy at 1 GHz and 843 MHz.  Thus, the present observations overpredict the flux given in \citet{gr04}, but are consistent with the observations of \citet{g94}.  The present single-dish--derived flux and spectral index are less likely to be biased than the previous interferometric values, which are more likely to underestimate the flux density.

The spectral index and flux density can be used to calculate an equipartition magnetic field.  \citet{b05} give a new derivation for the equipartition magnetic field strength as
\begin{equation}
B_{eq}={4\pi (2\alpha+1) (K_0+1) I_\nu E_p^{1-2\alpha} (\nu/2c_1)^\alpha / [(2\alpha-1) c_2(\alpha) l c_4(i)]}^{1/(\alpha+3)}
\end{equation}
\noindent where $K_0$ is the proton to electron number density ratio, $c_i$ are constants that depend on the spectral index and magnetic field inclination angle, which hereafter is assumed to be equal to 0 (in the plane of the sky).  Alternatively, the classical formulation of the equipartition magnetic field is
\begin{equation}
B_{class}=(8\pi G (K+1) L_\nu/V)^{2/7}
\end{equation}
where $G$ is a function of the energy range considered and spectral index, $K$ is the energy density ratio between protons and electrons, and $V$ is the volume of the emitting region \citep{p70,b05}.

Assuming a proton to electron energy density ratio of 40-100 \citep{b05} and a path length through the Tornado equivalent to its 2\arcmin\ width (10.4 pc assuming $D=12$ kpc), the 6 cm peak brightness of the Tornado of 5 Jy beam$^{-1}$ and spectral index --0.48 gives $B_{class}=100-130\mu$G.  This calculation integrates over frequencies from 10 MHz to 10 GHz, but changes by less than 10\% for an upper limit of 100 GHz.

The spatial dependence of the spectral index is apparent in Figure \ref{sptornado}, which shows the flux density and spectral index for a slice through the elongated portion of the Tornado.  This slice shows a regular change in $\alpha_{CX}$ from $-0.48\pm0.02$ near the brightest emission to $-0.33\pm0.07$ in the tail.  The spectral index is a direct measure of the energy distribution of the electrons and suggests that that the electrons in the tail region are more energetic than in the head region.  

A new source, G357.7--0.4 is found near the Tornado in projection as shown in Figure \ref{sptornado}.  G357.7--0.4 is an elongated, wavy structure with a thermal spectral index.  The morphology and difference in spectral index suggests that it is unrelated to the Tornado.  The thermal emission from G357.7--0.4 should not affect the measurements of spectral index from the tail of G357.7--0.1, since they are significantly separated and oriented perpendicular to each other.

At the top of Figure \ref{sptornado} and north of the Tornado is G357.7+0.3, which has long been known as a supernova remnant from its ring-like morphology, linearly polarized radio emission, and soft X-ray emission \citep{r84,le89,g94}.  The slice across the southeast portion of G357.7+0.3 shown in Figure \ref{sptornado} has a flat, but uncertain spectral index.  The spectral index is significantly steeper toward the southwest, with $\alpha_{CX}\lesssim-1.5$;  the 3.5 cm emission is absent, but the 6 cm brightness is similar to the southeast of the SNR.  This work is consistent with previous work within their large uncertainties.  G357.7+0.3 is only half covered by the 3.5 and 6 cm surveys, and so the integrated flux density meausrements are no available.

\paragraph{Sgr E Region (G358.7-0.0)}
Figures \ref{spe3} and \ref{spsgre} show the radio continuum emission from the E3 filament (G358.60-0.27) and the Sgr E complex (G358.7-0.0), respectively \citep{y04}.  The E3 filament is about 25\arcmin\ long and takes a twisting path from the Sgr E star-forming complex toward the southwest.  If the E3 filament is near the GC, it has a length of about 58 pc.  The Sgr E complex is made of a collection of compact sources within a 20\arcmin\ region near (358.7,0.0), surrounded by extended sources near (359.0,+0.0) and (358.4,+0.1).

The spectral index between 6 and 3.5 cm are shown in Figures \ref{spe3} and \ref{spsgre} and between 20 and 6 cm in Figure \ref{sptornadocl}.  High-resolution 20 cm continuum observations with the VLA show the region filled with compact \hii\ regions \citep{y04}, so much of the extended emission here could be unresolved compact sources.  Consistent with this expectation, the 6/3.5 cm spectral index for most of this emission is consistent with a thermal origin.  The E3 filament also has a thermal index throughout.

The easternmost and westernmost slices in Figure \ref{spsgre} show nonthermal 6/3.5 cm indices.  The easternmost slice passes through G359.0+0.0, which is seen in high-resolution, 20 cm images as extended, filamentary structures \citep{l92,y04}.  The morphology is remeniscent of an evolved \hii\ region, but the slice and integrated spectral indices are consistent with nonthermal emission.  The peak 6 cm brightness of 1 Jy beam$^{-1}$ and 6/3.5 cm spectral index of --0.63 implies a revised equipartition magnetic field strength of $65-85\mu$G, assuming it is in the GC and has an number density ratio of 40--100 \citep{b05}.

\paragraph{G359.1--0.5 SNR and the Snake NRF (G359.1-0.2)}
\label{g359.1-0.5sec}
Figure \ref{spg359.1-0.5} shows the 6 and 3.5 cm emission and slices across the G359.1--0.5 SNR and the G359.1--0.2 (also known as ``The Snake'').  G359.1--0.5 appears here as a ring of radius 9\damin5 (22 pc).  The ring of emission is noticably more irregular at 3.5 cm than at 6 cm, although it seems to be continuous at both wavelengths.  The flux density of G359.1--0.5 is $6.8\pm2.5$ Jy at 6 cm (see Table \ref{diffsrc}), which is consistent with the published value of $8.1\pm0.5$ \citep{r84}.  The SNR catalog of \citet{gr04} gives a 1 GHz flux density of 14 Jy and spectral index of approximately --0.4, which is also consistent with this flux density.  As shown in Table \ref{snrsrc}, the 6/3.5 cm spectral index in both our slice and integrated analysis methods range from --2.0 to --0.5, which is much steeper than that observed between 6 and 11 cm \citep{r84}.  The peak 6 cm brightness (0.3 Jy beam$^{-1}$; see Figure \ref{spg359.1-0.5}) and associated 6/3.5 cm spectral index (--0.84) are consistent with a revised equipartition magnetic field of $66-83\mu$G, assuming it is located near the GC and has a number density ratio of 40--100 \citep{b05}.

\begin{figure}[tbp]
\includegraphics{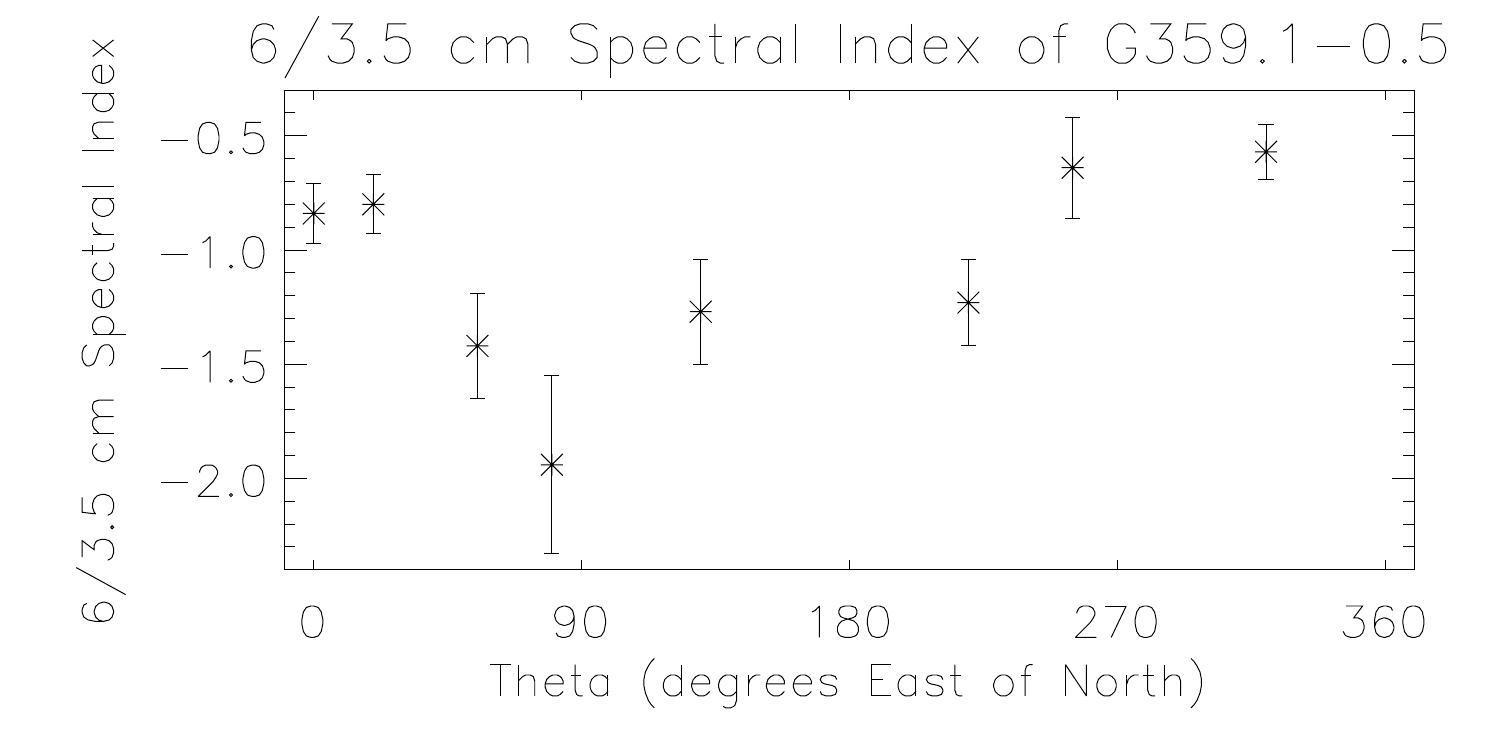}
\caption{Plot of the 6/3.5 cm spectral index values as a function of position on the circular supernova remnant, G359.1--0.5.  Position is represented by an angle measured in degrees, increasing counterclockwise starting at galactic north. \label{plsp}}
\end{figure}

Studies of molecular gas and tracers of shocked gas have found that G359.1--0.5 is interacting with a molecular cloud \citep{u92}.  Interestingly, G359.1--0.5 has a statistically significant spatial variation in its spectral index.  Figure \ref{plsp} plots its 6/3.5 cm spectral index as a function of theta, the position on the ring of emission relative to galactic north.  The 20 cm image does not have the resolution to allow a similar study of the 20/6 cm spectral index.  The 6/3.5 cm spectral index for G359.1--0.5 is significantly flatter for $\theta=250-10$\sdeg\ (the galactic west through north side).  The region with flatter spectral index is nearly identical to the region with most intense emission from the surrounding HI, $^{12}$CO, and OH maser emission \citep{u92,y95}.  This strengthens the idea that G359.1--0.5 is interacting with the surrounding molecular cloud, as proposed by \citet{u92}.

The Snake is a long ($\sim20$\arcmin\ or $\sim46$ pc at 8 kpc), nonthermal filament that runs from the Galactic plane to the G359.1--0.5 SNR \citep{u92,g95,y04}.  The Snake is unusual among NRFs because it has two sharp kinks along its length \citep[not visible with this data, but see][]{g95}.  The integrated flux densities given in Table \ref{diffsrc} imply a 6/3.5 cm spectral index of $-1.86\pm0.64$, which is steeper than the slice spectral index values.  Figure \ref{spg359.1-0.5} shows five slices across the Snake with spectral indices ranging from --0.2 to --0.9, tending to become more negative toward the south.  The slice analysis is more trustworthy than the integrated spectral index analysis, since this region is highly confused.  The slice spectral index values are similar to that measured in interferometric observations between 20 and 6 cm \citep{g95}, although the previous work found some locations with a higher, even inverted, spectral index.  For a 6/3.5 cm spectral index of --0.6, a 6 cm brightness of 0.2 Jy beam$^{-1}$, and a depth equal to its width of 9\dasec4 \citep{g95}, the revised equipartition field strength is 137-176 $\mu$G, for $K_0=40-100$.  This field strength is somewhat larger than that of \citet{g95}, since we assume a larger value of $K_0$, which is more consistent with current estimates \citep{b05}.

\paragraph{Sgr C Region (G359.5-0.0)}
Sgr C is one of the brightest radio continuum complexes in the GC region, being host to a bright \hii\ region and many NRFs \citep{l95,y04}.  Figure \ref{spsgrc} shows the 6 and 3.5 cm flux densities, slices, and spectral indices in this complex region.  Toward the west of this image is G359.2+0.0, located near where the Snake meets the Galactic plane.  This extended structure has flux densities of about 3 and 8 Jy at 3.5 and 6 cm, respectively.  The Sgr C \hii\ region around (359.5,--0.1) has flux densities of about 8 and 7 Jy at 3.5 and 6 cm, respectively, making it the brightest radio continuum source in the western half of the survey.  To the north of Sgr C near (359.4,+0.3) and possibly just south of Sgr C is the radio continuum counterpart of the GC lobe;  this is discussed in more detail in chapter \ref{gcl_all}.

The slice and integrated flux densities show that nonthermal emission dominates the extended emission in the region, outside of the Sgr C \hii\ region.  G359.2+0.0 has a very steep spectral index between 6 and 3.5 cm.  The slice plotted in the bottom of Figure \ref{spsgrc} corresponds to the vertical slice just west of Sgr C;  it shows how the emission surrounding Sgr C is predominately nonthermal.

It is interesting that much of the extended structure seen in the present survey is resolved into NRFs in high resolution observations \citep{y04,n04}.  This is consistent with the observed distribution of NRF flux density, which shows an increasing number of NRFs down to flux densities at present detection limits \citep{n04}.  This suggests that there could be many more faint NRFs than are currently observed.  Although interferometric observations may not yet be sensitive enough to see them, the present single-dish observations should detect the integrated flux density of all as-yet-undetected NRFs.  An estimate of this undetected NRF flux can be obtained by considering the observed 20 cm flux of NRFs to the total flux observed by the GBT.  Near Sgr C, the total NRF flux density is equal to a few Jy at 20 cm \citep{y04}, but the nonthermal, 20 cm flux density observed by GBT is about 10--20 Jy (after background subtraction).  It is possible that some of this nonthermal emission could be resolved into NRFs with higher-sensitivity radio continuum observations.  Regions with radio continuum morphology like Sgr C may be good places to search for NRFs with future observations;  other regions with similar radio continuum morphologies and spectral indices include G359.0+0.0, G359.2+0.0, and G0.8+0.0.

\paragraph{G359.8--0.3}
Figure \ref{spg359.8-0.3} shows the continuum emission and slices in G359.8--0.3.  This source has a lumpy, shell-like morphology in this data, with flux densities of about 19 and 28 Jy at 3.5 and 6 cm, respectively.  

The slices and integrated spectral indices of G359.8--0.3 are consistent with a thermal origin, suggesting that it is an \hii\ region.  G359.8--0.3 has an X-ray counterpart in the \emph{Chandra} survey of the GC region \citep{w02}.  The absorption of the X-ray gas is smaller than expected from the GC region, implying that it is in the foreground of the GC.  Similarly, low-frequency radio continuum absorption and H$\alpha$ emission are observed from G359.8--0.3, as would be expected from a thermal source in the foreground to the GC region \citep{b03b,ga01}.

\paragraph{Sgr A (G0.0+0.0)}
As shown in Figures \ref{spsgra} and \ref{spsgracl}, the Sgr A region is a very complex region with emission from the compact radio source at the dynamical center of the Galaxy, Sgr A*, the Sgr A West \hii\ region, and the Sgr A East SNR \citep{y87,p89,g05}.  Sgr A* is the radiative manifestation of a $3-4\times10^6$ \msol\ black hole that is accreting mass from stellar winds in the central parsec of the Galaxy \citep{m01}.  Sgr A West is the edge of a molecular cloud that is falling toward Sgr A* and being ionized by the hot stars in the central parsec \citep{r96}.  Sgr A East is a mixed-morphology SNR that appears to encompass Sgr A* in projection and whose western edge is coincident with Sgr A West \citep{g94,m01,m02}.

The peak radio brightness on arcminute scales in this region includes emission from Sgr A East and Sgr A West, the two brightest radio continuum sources in the central several arcminutes.  Sgr A East is roughly 4\arcmin$\times$3\arcmin, so no spatial structure is seen in the present survey and only the flux from the whole Sgr A complex is measured.  The peak brightness in this region is 39 Jy beam$^{-1}$ at 3.5 cm and 85 Jy beam$^{-1}$ at 6 cm.  Convolving the 3.5 cm map to the resolution of the 6 cm map gives a brightness of 66 Jy per 2\damin5 beam at 3.5 cm.  At 6 cm, VLA observations of the radio continuum find $70\pm10$ Jy comes from Sgr A East and $21\pm2$ Jy from Sgr A West, with a total flux density of about $88\pm10$ Jy \citep{b83,m89,p89}.  The results of the present survey are consistent with the interferometric observations, which suggests that they do not resolve much flux out \citep{p89}.

The 6/3.5 cm spectral index for the three slices across Sgr A*, and for the integrated brightnesses given above are consistent with $\alpha_{CX}=-0.44\pm0.02$ for the central $\sim$2\damin5.   The spectral index measured in the 6 and 20 cm slices give $\alpha_{LC}=-0.42\pm0.02$ for the central $\sim$9\arcmin.  High resolution observations of Sgr A East measure a spectral index of about --1.1 between 20 and 6 cm, which is believed to extend through the cm-wavelength regime \citep{p89}.  Sgr A West has a marginally optically thick ($\alpha\sim0$) spectrum at wavelengths between 3.6 and 20 cm \citep{m89,p89}.  The combination of these two effects can explain the value of $\alpha_{CX}=-0.44$ observed in the present survey, although at longer wavelengths, the spectral index should become significantly steeper, since Sgr A East begins to dominate the apparent flux.  It is likely that the 6/20 cm spectral index measures a significant amount of the ``halo'' flux around the Sgr A complex, since the 20 cm map has a beam size of 9\arcmin.  The halo emission has a flat spectral index between 20 and 90 cm that may extend to shorter wavelengths \citep{p89}.

\paragraph{Radio Arc (G0.2-0.0) and Arched Filaments (G0.07+0.04)}
The Arched filaments/Radio Arc complex has been the subject of intense speculation since it was first observed at high resolution with the VLA \citep{y84,t86,l02,y02}.  The slices shown in Figures \ref{spsgracl}, \ref{spwestarc}, and \ref{speastarc} confirm the results of early observations that found that the Arched filaments are thermal and the vertical component of the Radio Arc is nonthermal.  The slice in Figure \ref{spwestarc} shows a slightly negative index, but this is likely due to a mixture of the thermal Arched filaments and the ambient nonthermal emission near the Radio Arc;  using a closer part of the slice to estimate the background of the Arched filaments gives $\alpha_{CX}\sim-0.1$.  The integrated arched filament fluxes and spectral index are difficult to evaluate due to confusion with surrounding emission, but are consistent with a thermal origin (see Table \ref{diffsrc}).  The Radio Arc spectral index is generally quite flat, as is shown in Figure \ref{spsgracl}, where $\alpha_{LC}=-0.14\pm0.04$.

\begin{figure}[tbp]
\includegraphics[width=\textwidth]{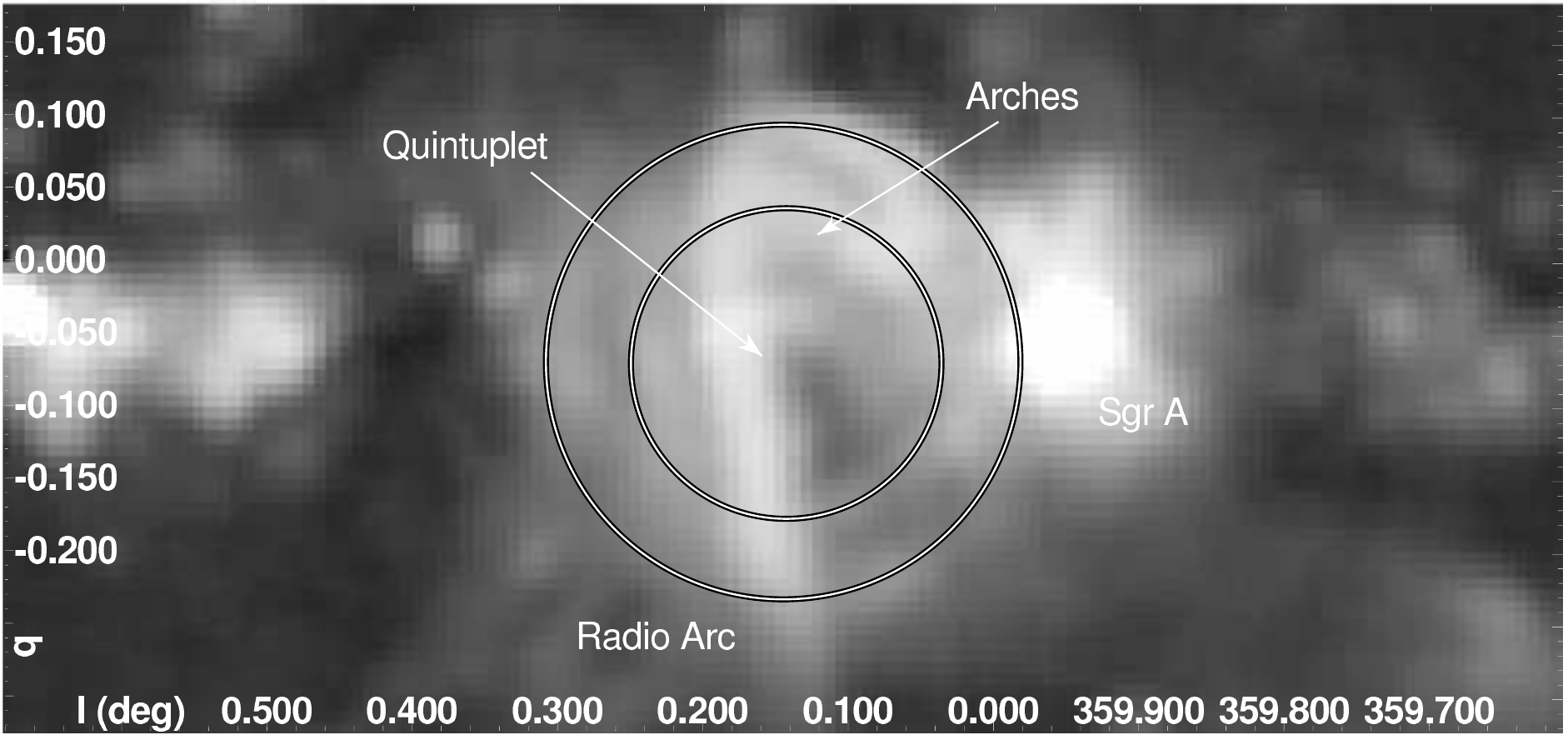}
\caption{GBT 3.5 cm image of the Radio Arc/arched filament region.  The positions of the Radio Arc (G0.2-0.0), Sgr A (G0.0+0.0), and Arches and Quintuplet clusters are labeled.  The circles schematically show how the Arched filaments (G0.07+0.04) connect to the souther Arched filaments (G0.07--0.2) and other radio continuum structures to make a nearly continuous ring around the brightest portion of the Radio Arc. \label{radioring}}
\end{figure}

The morphology of the radio continuum emission at 3.5 cm shows that the Arched filaments seem to be a part of a larger, ring-like structure.  Figure \ref{radioring} shows the GBT 3.5 cm image of the Arched filaments/Radio Arc region with the rings drawn schematically.  The Arched filaments can be connected to other radio continuum structures by two circles with similar center locations, near the brightest portion of the Radio Arc.  This ring-like radio continuum structure has been noted in filtered, 20 cm VLA images and attributed to multiple ``starbursts'' within the past million years \citep{s03}.  That work suggests that the energetic events had a total thermal energy of $10^{51}$ ergs and originated near the Quintuplet cluster, although they find more shells with more irregular shapes.

The morphology of the radio ring in the present survey is similar to that of \citet{s03}.  The lack of any spatial filtering of the present survey gives some confidence in the identification of the ring structure.  The slice in Figure \ref{spwestarc} crosses the southwest portion of this ring and shows that the 6/3.5 cm spectral index is flatter there, suggesting there is a thermal contribution to the emission;  defining the background region closer shows that these ``southern Arched filaments'' (G0.07--0.2) have $\alpha_{CX}=0.11\pm0.13$, similar to the Arched filaments.  The center of the ring is near the Quintuplet cluster, which is known to be irradiating the Sickle nebula \citep{r01};  meanwhile, the Arches cluster has been shown to be the cause of the ionization of the Arched filaments \citep{l02}.  No study of the ionization properties of the southern Arched filaments has been done.

There is also a coincidence between the brightness of the Radio Arc and the extent of the radio ring.  As seen in Figure \ref{radioring}, the 3.5 cm emission from the Radio Arc is brightest inside the ring made by the Arched filaments and southern Arched filaments, but fades rapidly outside of that ring.  A more complete study of this region, including \emph{Spitzer/IRAC} observations will be given elsewhere \citep{c07}.

The SNR G0.33+0.04 is located adjacent to (and partially confused with) the Radio Arc in Figure \ref{speastarc}.  Because it is so confused with the Radio Arc emission, an integrated spectral analysis is not done.  A multiwavelength study of G0.33+0.04 finds a spectral index of --0.56 for frequencies higher than a GHz \citep{ka96}.  The slice shown in Figure \ref{speastarc} crosses a portion of this SNR and finds a spectral index $\alpha_{CX}=-2$ to $-1.5$, significantly steeper than previous claims.  The highest frequency detections of this source are at 5 and 15.5 GHz (6 and 2 cm) \citep{a79,g74}.  At these frequencies the emission is dominated by emission from the Radio Arc and a few \hii\ regions and the morphology does not resemble the clear SNR-like morphology seen at lower frequencies \citep{ka96}.  It is possible that the previous detections at 5 and 15.5 GHz confused this other emission with G0.33+0.04 and overestimated its flux density, thus predicting a flatter spectral index than seen in the present observations.

\paragraph{G0.5--0.5}
Figure \ref{spg0.5-0.5} shows G0.5--0.5, an irregular complex of clumpy extended emission with peak brightnesses around 2 Jy beam$^{-1}$ at 6 cm.  The 6/3.5 cm spectral index measurements shown in Figure \ref{spg0.5-0.5} are near zero, showing that G0.5--0.5 is predominately thermal.  Like G359.8--0.3, this region is seen in absorption in 74 MHz continuum and emission in optical H$\alpha$, indicating that it is in the foreground of the GC region \citep{b03b,ga01}.  One slice in Figure \ref{spg0.5-0.5} shows a significantly nonthermal spectrum, although it appears morphologically similar to the rest of the G0.5--0.5 complex.  The 6 and 3.5 cm integrated fluxes and spectral index (Table \ref{diffsrc}) are consistent with nonthermal emission in the southern portion of G0.5--0.5.  All other 20/6 cm spectral index measurements across G0.5--0.5 shown in Figure \ref{spg0.5-0.5cl} are consistent with a thermal origin.

\paragraph{Sgr B (G0.5--0.0 and G0.7--0.0)}
Sgr B is the most active region of star formation in the GC region, being host to dozens of compact and ultracompact \hii\ regions \citep{m93,ga95}, many star-formation--tracing masers \citep[e.g.,][]{s00}, and a massive molecular cloud \citep[$\sim10^7$\msol;][]{l89}.  As shown in Figures \ref{spsgrb} and \ref{spsgrbcl}, Sgr B is the brightest radio continuum source in the eastern half of the survey.  The emission is brightest in two regions, called Sgr B1  (G0.5--0.0) and Sgr B2 (G0.7--0.0), the western and eastern halves, respectively.  The entire Sgr B region has flux densities of 90 Jy and 85 Jy at 3.5 and 6 cm, respectively, of which roughly 60\% comes from Sgr B2.  Sgr B is also surrounded by a few isolated \hii\ regions that are detected as compact sources in Tables \ref{srcX} and \ref{srcC}.

The 6/3.5 cm and 20/6 cm spectral index measurements for most of the Sgr B complex are near zero, which is consistent with it being thermal.  As shown in Table \ref{diffsrc}, the 6 and 3.5 cm spectral index values are near zero with three exceptions:  one significantly larger than zero and two much less than zero.  The positive (a.k.a. ``inverted'') spectral index is measured at the brightest portion of Sgr B2, a region known to be filled with ultracompact \hii\ regions \citep{ga95}.  These \hii\ regions are dense enough to be optically thick, thus causing the inverted spectral index.  The nonthermal spectral index measured east of Sgr B2 is in a region with extended emission that is not clearly associated with the Sgr B complex.  The other nonthermal spectral index is measured between Sgr B1 and B2, two regions with active star formation, but apparently having different ages.  Sgr B1 has more extended radio continuum morphology and has fewer compact \hii\ regions that Sgr B2, which suggests that the region is relatively older than Sgr B2 \citep{m92}.  The nonthermal index measured between the two may indicate that a supernova has occurred there, although no clear morphological signatures of a supernova in the Sgr B complex have ever been reported.  Another possibility is that faint (as yet undetected) NRFs are in that region.  G0.43+0.01 and G0.39+0.05 are two candidate NRFs located adjacent to Sgr B that have spectral indices of about --0.5 \citep{y04,n04}.  The nonthermal index measured between Sgr B1 and Sgr B2 is --0.2, flatter than that of the candidate NRFs; this spectral index may be an average of that expected from NRFs and the \hii\ regions in the region.

\paragraph{Sgr D (G1.1--0.1), G1.0--0.2, and G0.9+0.1}
The eastern edge of the survey has a mixture of thermal and nonthermal sources, as seen in the images and slices in Figures \ref{spsgrbcl} and \ref{spg0.9+0.1}.  Sgr D (called ``G1.1--0.1'' in Table \ref{snrsrc}) appears here as a compact source (at 3.5 cm; see Table \ref{srcX}) surrounded by a shell.  Previous work has suggested that the Sgr D shell is ionized by the compact source at its center and that the extended radio continuum emission toward the east (called ``G1.2+0.0'' here) may be gas escaping from Sgr D \citep{l92}.  G0.9+0.1 and G1.0--0.2 \citep[a.k.a. ``G1.05--0.1''][]{g94} are supernova remnants with clear shell-like structures and nonthermal 6/3.5 and 20/6 cm spectral indices.  G0.9+0.1 also has a Crab-like source inside the SNR (with a similar flux density as the shell at 6 cm), making it one of the first SNRs categorized as a ``composite'' \citep{h87}.

The integrated spectral index of the Sgr D extended emission of $\alpha_{CX}=-0.22\pm0.14$ is consistent with its identification as an ionized shell of gas.  Although the slice through G1.1--0.1 (the extended Sgr D emission) is nominally nonthermal ($\alpha_{CX}=-0.28\pm0.05$), parts of that slice have $\alpha_{CX}=-0.18\pm0.10$, which suggests that there is a mixture of thermal and nonthermal emission in the region.  The slice through the compact source at (1.1,--0.1) clearly has a thermal-like index of $0.03\pm0.02$.  The compact source spectral index is consistent with previous work that identified that source as an \hii\ region with radio recombination line emission with $v_{LSR}\approx25$ \kms \citep[source G1.127-1.04 of][]{l92}.  The source is detected as a comapct source at 3.5 cm at (17:48:40.29,--28:01:23.5) with a peak brightness of 2.3 Jy and flux density of 3.5 Jy, as compared to the flux density at 1.6 GHz is 1.7 Jy \citep{l92}.

G0.9+0.1 is an excellent example of a composite supernova remnant, which has a flat spectrum pulsar wind nebula in its center and a steep-spectrum shell \citep{h87,g94}.  The slices shown in Figure \ref{spg0.9+0.1} are generally consistent with this picture, showing flat 6/3.5 cm indices in the core ($\sim-0.16$) and a steeper spectral index in the shell ($\sim-0.35$).  Although the compact source inside the SNR has a flat spectral index, it also has linear polarization, which is consistent with suggestions that it is a pulsar wind nebula \citep{h87}.  The peak brightness for the unresolved core emission is about 4.6 and 4.0 Jy per 2\damin5-beam at 6 and 3.5 cm, respectively, which is similar to the 6 cm flux density of 4.16 Jy observed by \citet{h87}.  The flux density of the shell is difficult to estimate, since it is confused with the core in the 2\damin5-resolution images;  the shell brightness is about 1.3 and 1.1 Jy per 2\damin5-beam, at 6 and 3.5 cm, respectively.  The flux density of the shell is given in Table \ref{diffsrc} as 9.08 Jy and 5.46 Jy, at 6 and 3.5 cm, respectively.  The 6/3.5 cm spectral indices for the core and shell from the slice analysis are $-0.16\pm0.03$ and $-0.35\pm0.04$, respectively, which is consistent with the values of $-0.1$ and $-0.45$ found by \citet{h87}.

G1.0--0.2 is a supernova remnant with a shell morphology and nonthermal radio spectral index \citep[called G1.05--0.1 in][]{g94}.  Unlike G0.9+0.1, there is no compact source inside G1.0--0.2.  Figure \ref{spg0.9+0.1} shows two slices through G1.0--0.2 with $\alpha_{CX}=-0.55\pm0.05$ and $-0.92\pm0.08$, consistent with previous observations that found an index that ranged from $-0.6$ to $-0.7$ between 57.5 MHz and 1616 MHz.  Table \ref{diffsrc} shows the measured flux densities of G1.0--0.2 are 8.52 and 9.90 Jy at 3.5 and 6 cm, respectively, and an integrated spectral index $\alpha_{CX}=-0.27\pm0.18$, which is roughly similar to the slice results.  \citet{l92} report that G1.0--0.2 has a flux density of 12.3 Jy from interferometric observations at 1616 MHz; this gives an upper limit to the spectral index of $\alpha_{LC}<-0.2$.

\section{Discussion and Conclusions}
\label{gcsurvey_con}
This paper has shown results from a new survey of the radio continuum emission from the central degrees of the Galaxy at 90, 20, 6, and 3.5 cm with the GBT.  The 6 and 3.5 cm surveys are the most sensitive, highest resolution, single-dish radio surveys of the central degrees of the GC ever made at these wavelengths.  The 6 and 3.5 cm surveys are also valuable in resolving and cataloging many of the compact sources and complexes within two degrees of the GC.  Key results presented here include:  catalogs of compact and extended sources with a spectral index analysis of all sources in these catalogs.

In \S\ \ref{compactsrcsec}, compact source catalogs from the 6 and 3.5 cm surveys are presented.  At these frequencies and resolutions, most of the sources are likely to be Galactic \hii\ regions.  Indeed, most of the sources are found near well-known \hii\ complexes, such as Sgr B and Sgr E.  About one quarter of the sources detected at 3.5 cm are also detected at 6 cm;  most of these sources have thermal spectral indices.

Section \ref{slicediff} shows the results of analysis of the flux and spectral index distributions of extended sources in the 20, 6, and 3.5 cm surveys.  These observations find evidence of extended nonthermal emission on the scale of a tens of arcminutes associated with star-forming regions in the GC region, particularly between the Sgr E and Sgr C complexes.  These extended nonthermal sources have similar morphology and spectral indices as regions known to be filled with nonthermal radio filaments.  This similarity suggests that much of this diffuse flux may be related to the filaments.  The brightness-distribution function of these filaments estimated by high resolution observations indicates that they are much more numerous at low flux densities, so the extended nonthermal emission may yet be resolved as new, nonthermal radio filaments.  We also find a structure south of the Arched filaments that is thermal and seems to be morphologically connected to the well-known Arched filaments.  The Arched filaments and the new southern Arched filaments (G0.07--0.2) form a thermal, radio-continuum ring that surrounds the brightest part of the nonthermal emission from the Radio Arc and the dense star clusters, the Quintuplet and Arches clusters;  it is not clear how these objects are related to each other.  Another interesting result from our spectral index study is that the 6/3.5 cm spectral index distribution around the G359.1--0.5 SNR suggests that it is interacting with a neighboring molecular cloud;  this cloud is believed to be in the GC region, which also implies the SNR is in the GC region.  

A catalog of the the 6 and 3.5 cm properties of extended sources are compiled in \S\ \ref{diffsrcsec}.  Through a combination of slice and integrated flux analysis, all objects are catagorized as having a thermal or nonthermal origin.  Thus, the distribution of flux between thermal and nonthermal sources can be quantified.  Within the central 4\sdeg$\times$1\sdeg\ of the GC, the thermal to nonthermal flux fractions for all discrete emission are 28\%/72\% at 3.5 cm and 19\%/81\% at 6 cm.  This does not include the contribution from synchrotron emission from the Galaxy or the GC region, which begins to dominate the Galaxy's flux density for frequencies below 5 GHz.  Also, some of these sources in the field are likely to be in the foreground of the GC region, although the density of gas and stars is generally much higher in the GC region, so most sources are likely to truly be in the central kpc.


%% file: gcl_recomb_thesis2_astro-ph.tex
\chapter{Radio Recombination Line Observations of the GCL}
\label{gcl_recomb}

\section{Introduction}
Since their prediction and discovery more than 40 years ago \citep{k59,d64,s64}, radio recombination lines have been useful tools for probing the physical conditions of the ISM.  The properties of the line and continuum emission can constrain fundamental parameters such as gas temperature, density, and volume filling factor \citep[e.g.,][]{r92}.  Furthermore, the spectral line can provide valuable information about the kinematics of the emitting gas.  In the case of the GCL, information about the gas conditions and kinematics would provide a valuable test of the outflow hypothesis for its origin.  Many star-forming regions and extragalactic outflows have been observed in radio recombination lines, so the GCL can be compared directly to them \citep[e.g.,][]{m93,v05}.

A common use of radio recombination lines in the GC region is to separate thermal \hii\ regions from the widespread nonthermal gas and study its kinematics.  \citet{l97} used the VLA to study recombination lines in the Pistol and Sickle \hii\ regions and identify the source of their ionizing photons as local massive stars.  On the larger scale, radio recombination line observations have found an abundance of low-velocity emission throughout the GC region with characteristics consistent with stimulated emission \citep{l73,p75,a97}.  The coincidence of stimulated emission at low velocities suggests that much of the emission may not trace the properties of gas in the GC region, but that it originates in the cold gas of the ambient ISM.  However, these observations are typically performed at frequencies below 2 GHz, which is considerably lower than the present observations and much more likely to find stimulated emission \citep[e.g.,][]{r92}.

This chapter describes observations of the GCL with the GBT and Hat Creek Radio Observatory (HCRO).  Section \ref{recomb_obs} shows how the observations were made and reduced.  Section \ref{recomb_res} describes the results of the observations, including the morphology, line characteristics, and kinematics of the recombination line emission in the GCL.  The GBT observations find unusually narrow recombination line emission throughout the GCL, which places useful constraints on the gas temperature and other properties.  In \S\ \ref{recomb_dis}, we discuss the implications of the radio recombination line emission on the nature of the GCL.  The mass, ionization, and energetics are found to be consistent with values observed toward dwarf starburst outflows.  We also discuss the results of a search for high-velocity gas in the GCL, as might be expected from an energetic nuclear outflow. 

\section{Observations and Data Reductions}
\label{recomb_obs}
\subsection{HCRO Survey}
In 1985, the H106$\alpha$ recombination line emission from the GCL was mapped by D. Backer with the HCRO 26 m telescope.  The spectrometer at HCRO was tuned to a central frequency of 5008.923 MHz and had a 512 lag, 20 MHz bandwidth.  This bandwidth covered the H109$\alpha$ and H137$\beta$ transitions, although only the H109$\alpha$ transition was studied.  Spectra were Hanning smoothed and a seventh-order baseline fit before Gaussian line fitting.  The velocity uncertainty of the fits are about 1 \kms.  The system temperature was assumed to be 60 K, which is likely to be a 20-50\% overestimate.


At 5 GHz, the HCRO beam size is 10\arcmin\ and its Nyquist sample size is 3\damin3.  The HCRO gain was calibrated by comparing the brightness temperature to that of the GBT for a few identical pointings, which gives a gain of $\sim9$ Jy K$^{-1}$.  Pointings were made over a roughly 1\sdeg square region of the GCL on a 6\arcmin\ grid;  neighboring points in the grid were skipped to form a checkerboard pattern.  Spectra were calibrated by position switching to a distant, fixed position, with a typical integration time of 2 hours for each spectrum.  Several scattered points outside the radio continuum emission associated with the GCL were also observed to constrain the background emission.

Figure \ref{hcroimg} shows images of the H109$\alpha$ line antenna temperature and velocity from the HCRO data.  The line brightness is not corrected for atmospheric absorption.  Since the checkerboard observing pattern makes visualization awkward, the data were interpolated onto a regular grid using a custom IDL program called ``interp\_recomb.pro''.  The program linearly interpolates between all pixels with four neighbors that have measured values.  The program then interpolates between points with progressively fewer neighbors.  The order in which this interpolation is done creates some bias in the interpolated pixels, but the general properties of the emission on spatial scales larger than the beam (10\arcmin) are not affected.

\subsection{GBT Observations}
Inspired by the detection of recombination line emission in the GCL by HCRO, we used the GBT Spectrometer to study the line emission in more detail.  The observations were conducted over a 7 hour period in 2005 August.  Four windows of width 200 MHz were set up to observe eight H$\alpha$ and He$\alpha$ transitions from $n=$106 to 113.   Each pointing produced dual-circular polarizations of four 8196 channel spectra.  The spectral windows, centered near 5.37, 5.08, 4.81, and 4.56 GHz, also covered other transitions with more than 20 MHz any window edge, including H$\beta$ (7; $n=$134--142), H$\gamma$ (8; $n=$152--162), H$\delta$ (8; $n=$167--178), and H$\epsilon$ (8; $n=$180--191).  Three-level sampling was used and the spectra had 24.4 kHz channels, equivalent to 1.5 \kms\ at 4.9 GHz.  All velocities in this work are given in the local standard of rest (LSR).  Calibration for each scan was done by position switching to a position two degrees north in Galactic latitude.  

\begin{figure}[tbp]
\includegraphics[width=\textwidth]{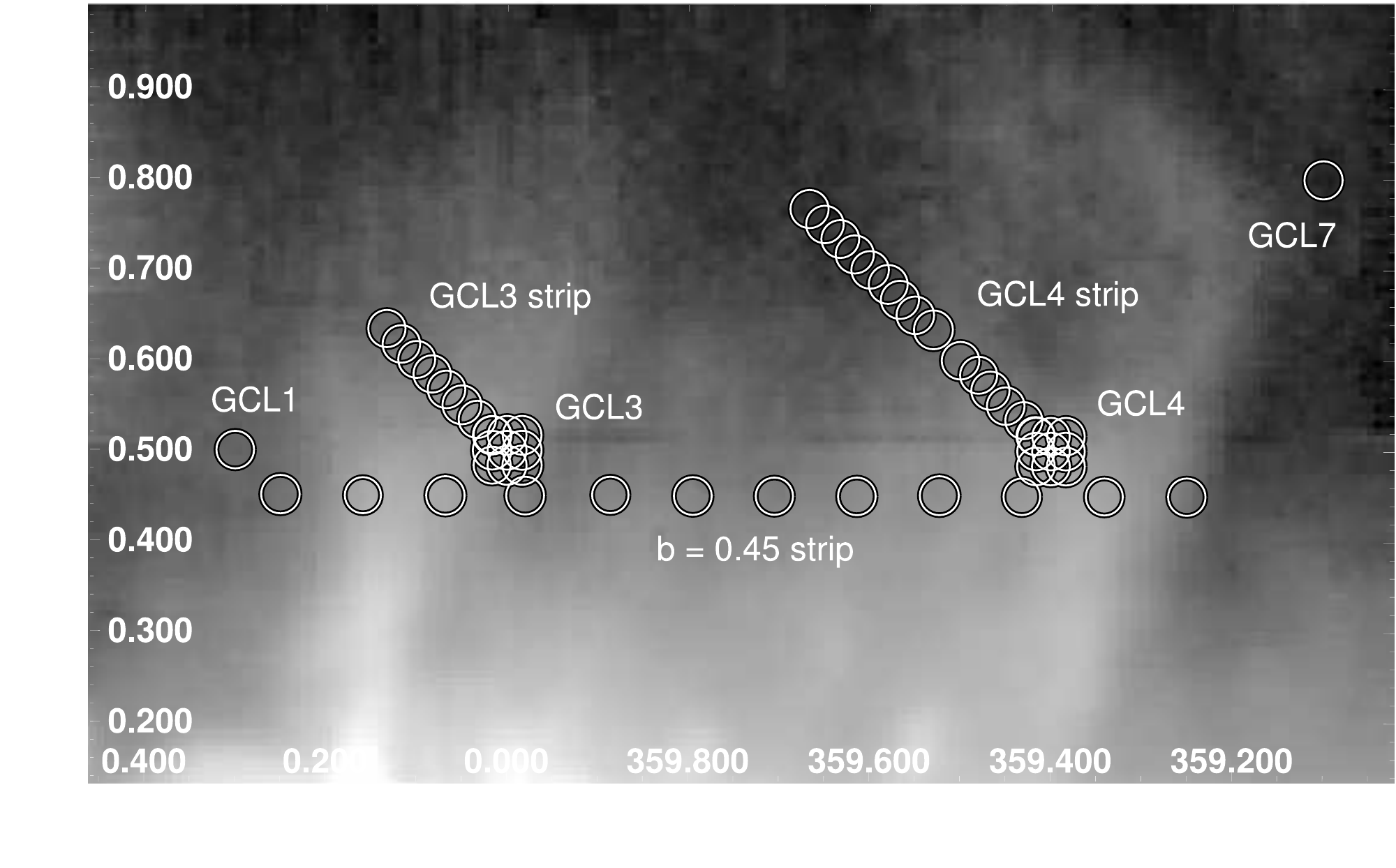}
\caption{Positions of the GBT observations toward the GCL are shown as circles on a GBT 6 cm radio continuum survey of the region (Ch. \ref{gcsurvey_gbt}).  The circles have a diameter of 2\damin5, which is the FWHM of the GBT beam at 6 cm. \label{positions}}
\end{figure}

The locations of the pointings are shown in Figure \ref{positions} and described in Table \ref{gbtpointings}.  The locations named ``GCL3'' and ``GCL4'' are at the peak recombination line flux of the HCRO data, near the GCL-E and GCL-W, respectively.  These regions were mapped in a 3$\times$3, Nyquist-sampled (1\arcmin\ near 6 cm) grid, with on-source integration times of 90 s per pointing.  Diagonal strips were observed starting at these positions and going to the Galactic northeast to sample interesting gradients in the velocity field. \footnote{The GCL3 strip was intended to observe toward the northwest, but the author screwed up.}  A sparse horizontal strip was observed at $b=0\ddeg45$, to sample the changes in the east-west direction.  Finally, two positions outside the GCL, ``GCL1'' and ``GCL7'', were observed to measure the background of the GCL.  The strips and individual pointings had integration times of 60 s per pointing.

\begin{deluxetable}{lcclcc}
\tablecaption{GBT Radio Recombination Line Pointings \label{gbtpointings}}
\tablewidth{0pt}
\tablehead{
\colhead{Name} & \colhead{$l$} & \colhead{$b$} & \colhead{Map type} & \colhead{Exposure} & \colhead{Spacing}\\
\colhead{} & \colhead{(deg)} & \colhead{(deg)} & \colhead{} & \colhead{(s)} & \colhead{(arcmin)} \\
}
\startdata
GCL3 & 0.0 & 0.5 & 3x3 grid & 90 & 1 \\
GCL4 & 359.4 & 0.5 & 3x3 grid & 90 & 1 \\
GCL3 strip & +0.016--0.133 & 0.516--0.633 & diagonal strip & 60 & 1.4 \\
GCL4 strip & 359.416--359.666\tablenotemark{a} & 0.516--0.766 & diagonal strip & 60 & 1.4 \\
$b=0.45$ strip & 0.25--359.25 & 0.45 & horizontal strip & 60 & 5.4 \\
GCL1 & 0.3 & 0.5 & single point & 60 & -- \\
GCL7 & 359.1 & 0.8 & single point & 60 & -- \\
\enddata
\tablenotetext{a}{One pointing, near $b=-0\ddeg48$, was accidentally skipped.}
\end{deluxetable}

Calibration and analysis was done using tools released with GBTIDL v1.2.1 and custom IDL programs written by the author.  Spectra were calibrated using the standard (on-off)/off method.  An IDL program called ``offoncal'' was written to do basic calibration, averaging of lines, and fitting of line profiles.  All line emission studied here is averaged over both polarizations and all $n$ values of a given transition (e.g., eight transitions with $n=$134--142 for H$\beta$) to improve the signal to noise ratio.  Rest frequencies for all transitions were taken from \citet{l68}.  Unless otherwise noted, a third-order baseline was fit to each spectrum prior to fitting a Gaussian to the line profile.

All spectra presented here are corrected for the beam efficiency and atmospheric opacity, and thus are in units of main beam brightness temperature.  The beam efficiency is assumed to be equal to 1.37 times the aperture efficiency at 6 cm (65\%), or 89\%. \footnote{See \url{http://wwwlocal.gb.nrao.edu/gbtprops/man/GBTpg/GBTpg\_tf.html}.}  The zenith atmospheric opacity around 6 cm ranged from 8.5 to 9$\times10^{-3}$ during the observing period.  This work corrects for atmospheric absorption by assuming a mean opacity of 8.6$\times10^{-3}$, which increases brightnesses by 2--5\% for the elevations of these observations (25--8\sdeg).

The GC region is seen at low elevations from Green Bank, which makes the calibration of the continuum levels difficult.  One important aspect of the calibration of the continuum levels is to account for the frequency dependence of the calibrator (a.k.a. ``$T_{\rm{cal}}$'').  The antenna temperature and system temperature are defined as:
\begin{equation}
T_{\rm{a}}(i) = \langle T_{\rm{sys}}(i)\rangle ({\rm{Sig}}(i) - {\rm{Ref}}(i)) / {\rm{Ref}}(i),
\end{equation}
\noindent where $\rm{Sig} = (\rm{Sig}_{\rm{calon}} + \rm{Sig}_{\rm{caloff}})/2$ (and same for Ref), and 
\begin{equation}
\langle Tsys(i)\rangle = \langle T_{\rm{cal}}(i)\rangle * \langle \rm{Ref}(i)/(\rm{Ref}_{\rm{calOn}(i)}-\rm{Ref}_{\rm{calOff}(i)})\rangle,
\end{equation}
\noindent where $T_{\rm{cal}}(i)$ has been measured previously and angle brackets denote a boxcar smoothing of width 500 channels, which was found to reduce statistical noise without smoothing bandpass shape.  This calibration scheme dramatically reduces the ``ripples'' in the radio continuum flux that is caused by errors in $T_{\rm{cal}}$.  

A second correction factor that is applied is to the data is to subtract the difference in sky brightness between the signal and reference pointings.  The sky brightness is assumed to depend on the elevation as $T_{\rm{sky}} = T_{\rm{atm}} * \tau_z/\sin ({\rm{elevation}})$, with $T_{\rm{atm}}$ assumed to be 260 K.  This correction factor typically reduces the continuum flux by 0.2--1 K, which is roughly 20--50\% of the continuum flux before the correction.  

Unfortunately, the calibration method and sky brightness correction do not account for changes in the sky brightness on the timescale of the switching between the signal and reference positions (around 1 minute).  These kinds of changes in the sky brightness seem to affect the slope and intensity of the continuum brightness for many of the scans, especially at low elevation and with large changes in elevation between signal and reference positions.  This effect is most evident by comparing the observed continuum flux with the continuum flux and spectral index measured in chapter \ref{gcsurvey_gbt}.  After the calibration steps outlined above, the continuum level observed by the present observations were significantly higher than observed in chapter \ref{gcsurvey_gbt}.  Since those observations were made with a calibration method optimized for continuum levels, we choose to use the continuum observed in chapter \ref{gcsurvey_gbt} when continuum fluxes are needed.

\section{Results}
\label{recomb_res}

\subsection{HCRO Survey}
\begin{figure}[tbp]
\includegraphics[width=\textwidth]{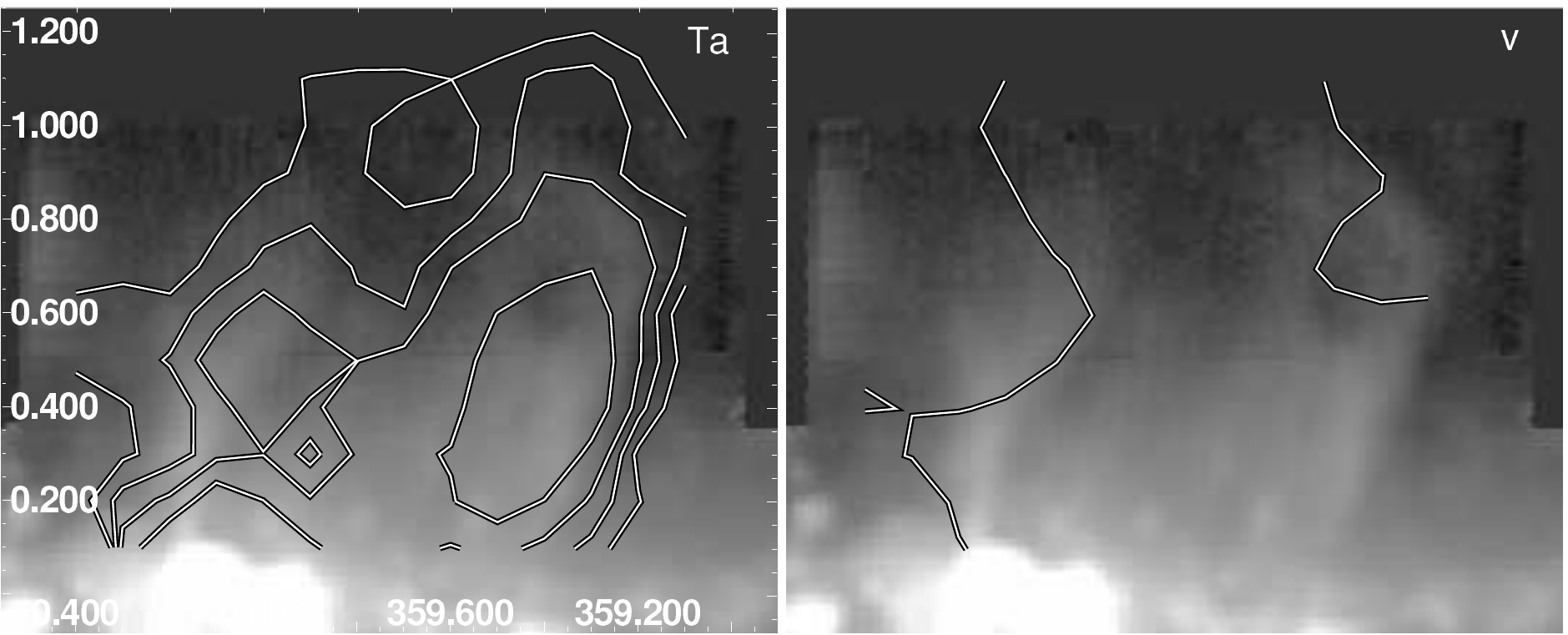}
\caption{\emph{Left}: The gray scale shows GBT 6 cm radio continuum emission toward the GCL with contours of H109$\alpha$ brightness from the HCRO at $T_{\rm{a}}=$20, 30, 40, and 60 mK. \emph{Right}: Same as the left panel, but with a contour of H109$\alpha$ line velocity at 0 \kms.  The areas to the far left and right have negative velocities and and between the lines are generally positive velocities. \label{hcroimg}}
\end{figure}

Figure \ref{hcroimg} shows the HCRO line intensity and velocity compared to 6 cm continuum emission.  The radio recombination line is brightest along two vertical ridges near ($l, b$)=(0\ddeg0, 0\ddeg5) and (359\ddeg4, 0\ddeg5).  There is also a slight increase in brightness at the top of the line-emitting region, near ($l, b$)=(359\ddeg6, 1\ddeg0).  The longitude of the line emission north of the plane is centered near $l=359\ddeg6$ and the central longitude shifts slightly to the west at higher latitudes.  

The distribution of emission is consistent with a limb-brightened shell, where the changes in brightness correspond to changes in column density.  \citet{b03} model the \emph{MSX} mid-IR emission from this region as a ``telescope dome'' shell that extends in latitude more than in longitude.  If the line emission seen by HCRO originates in this kind of shell, the radius of this shell is about 40 pc and it seems to have an apparent thickness comparable to the HCRO beam size of 10\arcmin\ (23 pc, assuming a distance of 8 kpc).  The discussion of the multiwavelength observations and morphological modeling of the GCL are described in more detail in chapter \ref{gcl_all}.

The line emission is significantly brighter inside the GCL than outside.  Although the locations of the brightest HCRO line emission is only known to within 10\arcmin, the recombination line emission tends to lie closer to the GC than the 6 cm continuum emission (``inside'', if you think of the entire structure as a single shell).  Several pointings were made outside the radio continuum shell of the GCL and constrain the line brightness to be significantly less than observed inside the GCL.  The observations in the eastern half of the GCL show a significant offset of about 0\ddeg1 between the peak line emission and continuum emission.  In the western half of the GCL, the two components appear coincident.  The distance between the east and west regions of peak brightness is about $0\ddeg6\approx80$ pc, which is significantly smaller than the extent of the radio continuum emission, or $0\ddeg8\approx110$ pc.  There is a similar trend at the top of the GCL, where the recombination line emission tends to lie inside the radio continuum shell of the GCL seen by the GBT at 20 cm (Ch. \ref{gcsurvey_gbt}).

The HCRO observations analyzed the H109$\alpha$ line.  With some assumptions this line can be used to estimate gas properties of the GCL.  As described in \S\ \ref{gbt:lines}, the GBT observations provide a much more accurate picture of the ionized gas conditions.

The peak brightness of a radio recombination line depends on physical properties of the gas.  The basic equation that relates the local thermodynamic equilibrium (LTE) line brightness to the electron density, temperature, and other properties is
\begin{equation}
T_L (\rm{K}) = 10^4 Z^2 \eta_A \frac{m-n}{\Delta\nu_D} \frac{f_{nm}}{n} T_e^{-3/2} e^{-X_n} \rm{EM}
\end{equation}
\noindent where $\eta_A$ is the aperture efficiency, $m$ and $n$ are the initial and final electronic levels, $f_{nm}$ is the oscillator strength, $\Delta\nu_D$ is the line width, $e^{-X_n}$ is a correction factor near 1, and EM is the emission measure \citep{d70,r92}.  Although no complete analysis of line widths was produced from the HCRO observations, a H109$\alpha$ line profile from a pointing has a line width of about 22 \kms.  The line brightness in the GCL ranges from 10--50 mK inside the GCL to 108 mK at the peak in the western half.

Assuming the gas is in LTE with a typical electron temperature of $5000$ K, the EM ranges from 1400--7000 pc cm$^{-6}$ inside and 15,000 pc cm$^{-6}$ at the peak in the GCL-West.  If the emission toward the center of the GCL comes from the front and back side of a shell of thickness $<23$ pc, then the rms electron density is constrained from $\sqrt{\langle n_e^2\rangle}>6-12$ cm$^{-3}$ (angle brackets denote an average).  If the shell has a thickness of $0\ddeg1\approx15$ pc (see \S\ \ref{gbt:morph}), then $\sqrt{\langle n_e^2\rangle}=7-15$ cm$^{-3}$.  Toward the brightest emission in the GCL-West, assuming a path length of 50 pc gives $\sqrt{\langle n_e^2\rangle}=17$ cm$^{-3}$.  As shown in \S\ \ref{gbt:lines}, the assumption that the gas is in LTE seems to be accurate, although the electron temperature is probably significantly lower than 5000 K.  The rms electron density is a lower limit to the true electron density with a dependence on the filling fraction, $f$, as $n_e = \sqrt{\langle n_e^2\rangle}/\sqrt{f}$.

The right side of Figure \ref{hcroimg} shows the velocity of the H109$\alpha$ line across the GCL.  Two striking characteristics of the line velocities in this region are their small values and the absence of any simple pattern or gradient.  The largest velocity is +27 \kms\ at (0.1, 0.1), near the Arches thermal filaments \citep{l02}, but north of $b=0\ddeg3$ the line velocity does not exceed the range of -5 to +5 \kms.  This is consistent with previous observations of radio recombination lines in the region \citep{p75,l73,a97}.

The spatial structure in the line velocities observed by the HCRO is complex.  There is no signature of an east-west gradient with a magnitude greater than about 3 \kms across the roughly 100-pc width of the GCL.  The velocity structure in the north-south direction has a wavy pattern, with predominately positive velocities around $b=0\ddeg3$, then negative around $b=0\ddeg7$, and back to positive velocities at higher latitudes.  This wavy velocity structure seems shifted slightly to lower latitudes in the eastern half of the GCL.  The velocities at the center of the GCL (around $l=359\ddeg7$) have less of a wavy pattern and the line velocity tends to decrease with increasing latitude;  the magnitude of the velocity goes from $\sim2.5$ to 0 \kms\ for $b$ from $0\ddeg3$ to $1\ddeg0$.

\subsection{GBT Observations}
\label{recomb_gbt}
The GBT observations were more limited in spatial coverage than the HCRO survey, but much more sensitive and with multiple transitions.  Thus, the goal of the GBT observation was not only to confirm the results of the HCRO observations, but to study the gas conditions in more detail.  Generally, the GBT and HCRO observations are in agreement.  However, as described below, the GBT observations have found unusual line widths, which provides a measure of the electron temperature and density.

\begin{figure}[tbp]
\includegraphics[angle=270,scale=0.6]{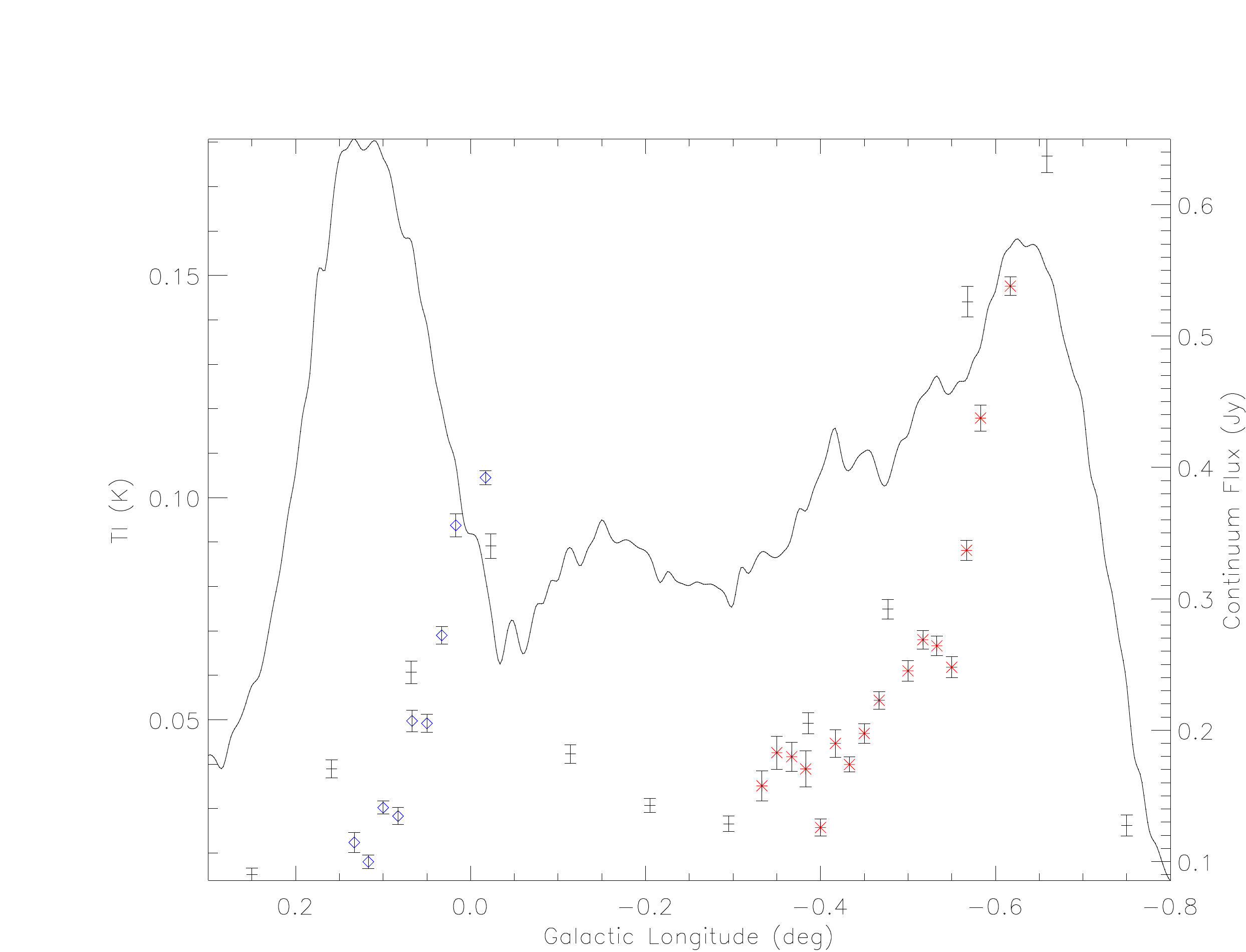}
\caption{Peak brightness temperature of H$\alpha$ as a function of Galactic longitude for three strips of GBT observations through the GCL.  The $b=0.45$ strip is shown with black crosses, diagonal ``GCL 4 strip'' shown with red stars, and diagonal ``GCL 3 strip'' shown with blue diamonds (see Table \ref{gbtpointings} for strip specifications).  Error bars show 1 $\sigma$ uncertainties.  The line shows the GBT 6 cm continuum flux described in chapter \ref{gcsurvey_gbt}, which defines the extent of the radio continuum GCL. \label{tastrips}}
\end{figure}

\label{gbt:morph}
Figure \ref{tastrips} shows plots of the average H$\alpha$ line brightness for $n=106-113$ for the three strips of observations through the GCL.  The horizontal strip at $b=0\ddeg45$ shows a similar longitude structure in the GCL as seen in the HCRO data, with two recombination line peaks near the radio continuum peaks.  The eastern peak in the recombination line emission is offset from the peak of the radio continuum emission by roughly $0\ddeg15$ or 21 pc.  The western peak of the line and continuum emission seem to fall at roughly the same position ($l\sim359.35-359.4$), within the uncertainties of the recombination line spatial sampling.  This is consistent with the morphology observed by the HCRO.

The line flux toward the center of the GCL is significantly brighter than outside of it and has different line characteristics.  The GCL1 and GCL7 pointings sampled the emission toward lines of sight outside the GCL.  The best-fit Gaussian to the H$\alpha$ line in the GCL1 position had $(T_l, v, \Delta v)=(0.0103\pm0.0016\ \rm{K}, -9.5\pm2.2$\ \kms$, 29.2\pm5.2$\ \kms$)$.  No H$\alpha$ line was detected toward GCL7 with a 1 $\sigma$ upper limit of 0.0059 K.  The line brightness for the weakest lines inside the GCL have $T_l\sim0.025$\ K and, as discussed below, have much narrower H$\alpha$ line widths.  The line emission outside the GCL has line widths consistent with that expected from foreground emission, but far fainter than the emission found within the radio continuum shell of the GCL.  

The two diagonal strips of observations were designed to study the velocity structure seen in the HCRO observations, so they do not probe the spatial structure well.  However, like the $b=0\ddeg45$ strip, they show a decrease in peak line brightness as they move away from the brightest regions near the GCL-E and GCL-W.  This suggests that the recombination line emission has a similar $l$ dependence for latitudes of $0\ddeg45$ and near $0\ddeg6$, consistent with the HCRO images.  Assuming the recombination line emission from the GCL has a shell-like morphology intrinsically, the width of the GCL ``wall'' can be estimated from the $b=0\ddeg45$ strip of scans.  The longitude width of the east and west peaks of the recombination line emission are roughly $0\ddeg2\approx28$ pc at brightnesses roughly half the peak brightness.

\label{gbt:lines}
To better constrain the gas conditions, a long integration was made by averaging spectra observed throughout the GCL.  The lack of velocity structure (discussed in \S\ \ref{gbt:vel}), makes this possible by allowing the averaging of multiple integrations together.  The two 3$\times$3, Nyquist-sampled patterns near the brightest regions of the GCL were found to have not only similar line velocities and line widths, but also similar line ratios (e.g., H$\beta$/H$\alpha$), suggesting that these regions have similar physical conditions.  The total on-source integration time for this deep spectrum was 27 minutes.  

\begin{deluxetable}{ccccccc}
\tablecaption{Detected Lines and their Properties in Deep Integration Toward GCL Recombination Line Peaks \label{lines}}
\tabletypesize{\small}
\tablewidth{0pt}
\tablehead{
\colhead{Transition} & \colhead{$n$ Range} & \colhead{$T_l$} & \colhead{$v_{\rm{LSR}}$} & \colhead{$\Delta v$} & \colhead{$I^{\rm{obs}}$\tablenotemark{a}} & \colhead{$R^{\rm{obs}}_\alpha$\tablenotemark{b}} \\
\colhead{} & \colhead{} & \colhead{(K)} & \colhead{(\kms)} & \colhead{(\kms)} & \colhead{(K \kms)} & \colhead{} \\
}
\startdata
H$\alpha$ & 106--113 & $0.1257\pm0.0024$ & $0.71\pm0.13$ & $13.45\pm0.30$ & $1.802\pm0.052$ & 1 \\
He$\alpha$ & 106--113 & $0.0110\pm0.0006$ & $0.97\pm0.32$ & $11.50\pm0.74$ & $0.135\pm0.012$ & $0.075\pm0.007$ \\
H$\beta$ & 134--142 & $0.0339\pm0.0008$ & $0.75\pm0.15$ & $13.72\pm0.36$ & $0.495\pm0.017$ & $0.275\pm0.012$ \\
H$\gamma$ & 152--162 & $0.0144\pm0.0007$ & $0.36\pm0.35$ & $15.26\pm0.83$ & $0.233\pm0.017$ & $0.130\pm0.010$ \\
H$\delta$ & 167--178 & $0.0091\pm0.0006$ & $1.98\pm0.40$ & $12.47\pm1.05$ & $0.120\pm0.013$ & $0.067\pm0.007$ \\
H$\epsilon$ & 180--191 & $0.0030\pm0.0003$ & $1.67\pm1.34$ & $24.47\pm3.15$ & $0.073\pm0.012$ & $0.041\pm0.007$ \\
\enddata
\tablenotetext{a}{$I^{\rm{obs}} = \int_{\rm{line}} T_{\rm{mb}} dv \approx 1.065 T_l \Delta v$}
\tablenotetext{b}{$R^{\rm{obs}}_\alpha = I^{\rm{obs}}_{x}/I^{\rm{obs}}_\alpha$}
\end{deluxetable}

\begin{figure}[tbp]
\begin{center}
\includegraphics[scale=0.15]{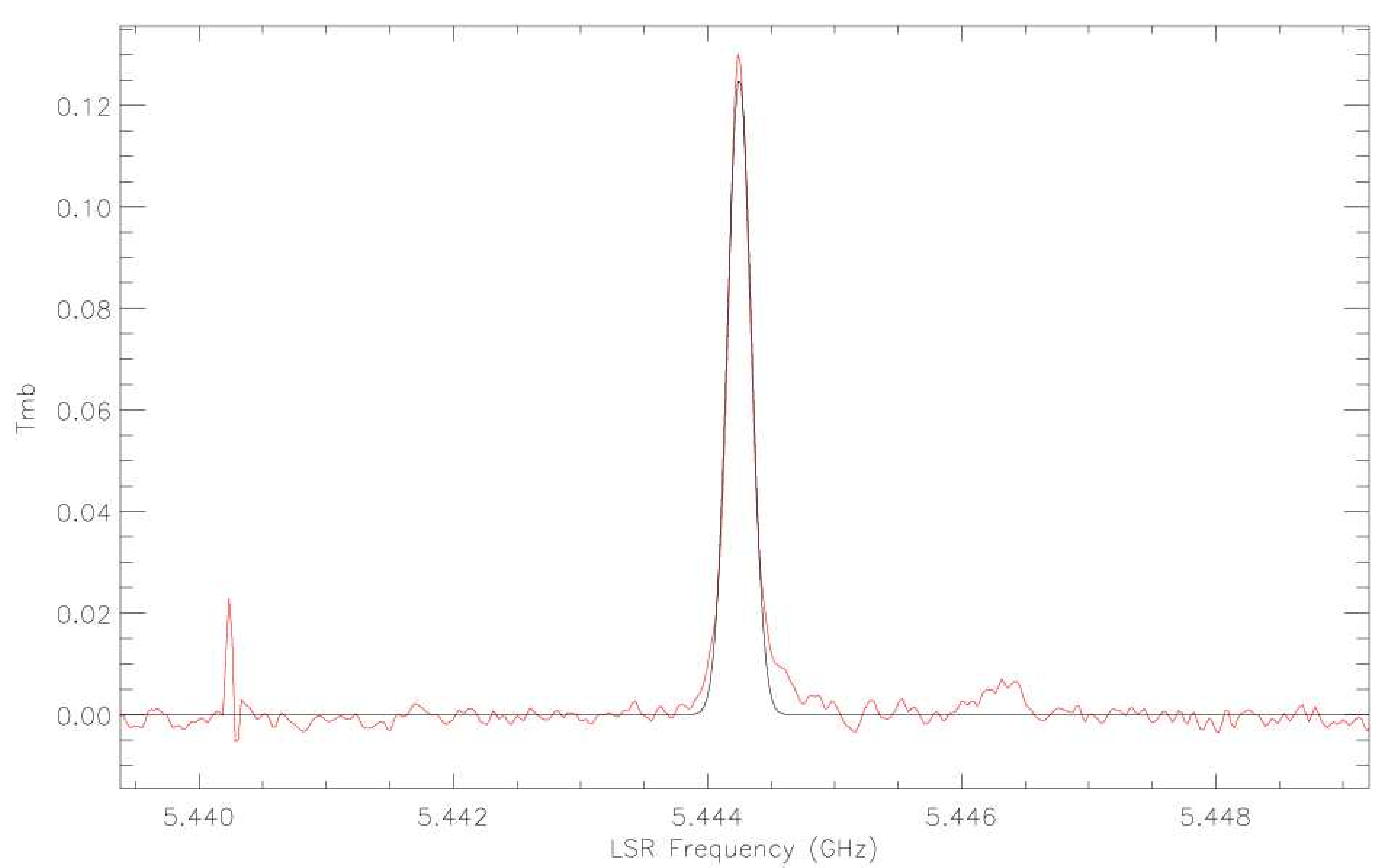}
\hfil
\includegraphics[scale=0.15]{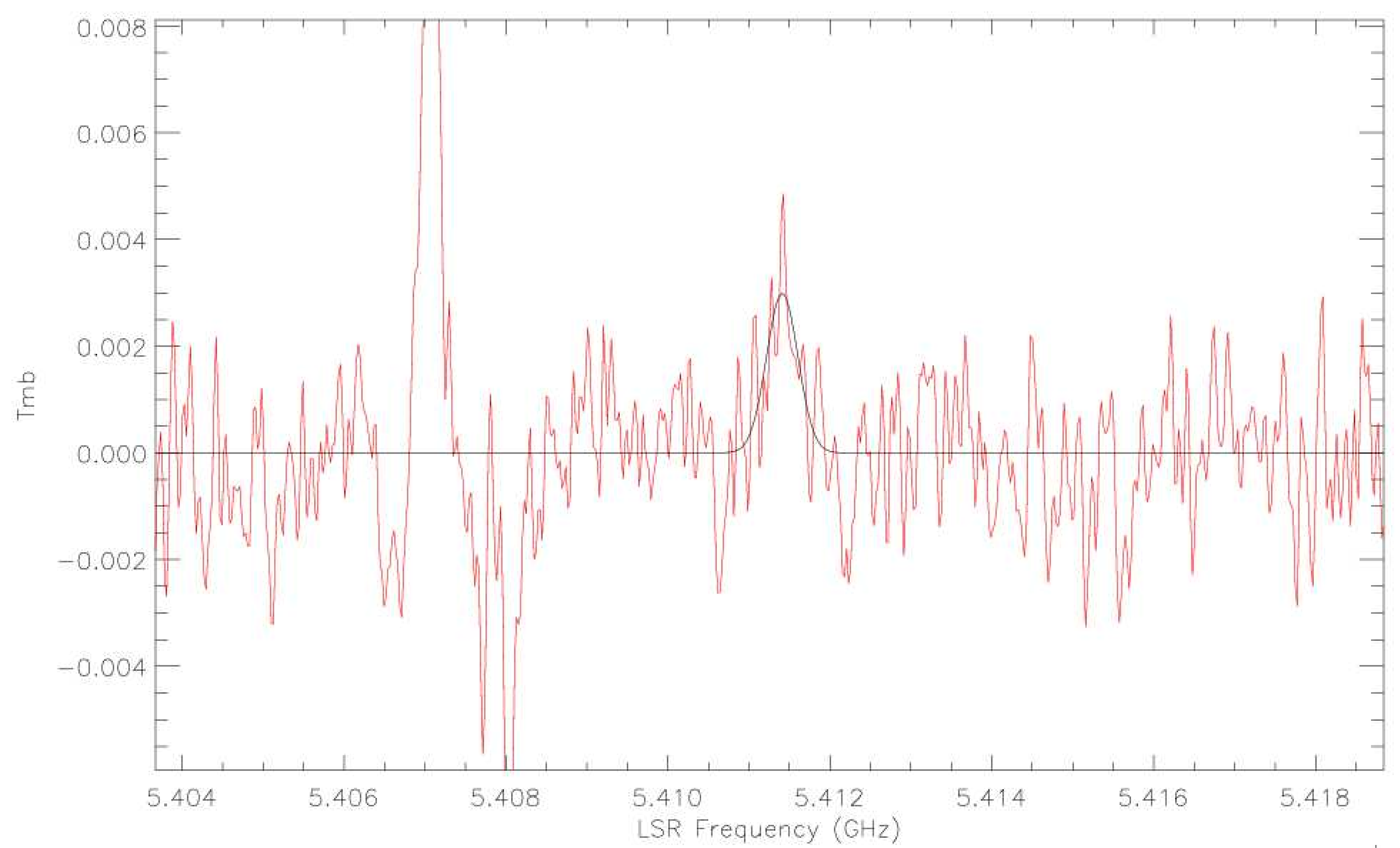}
\end{center}
\caption{\emph{Top}: Average H$\alpha$ spectrum for eight transitions from $n=$106 to 113 in the deep average spectrum of the GCL.  The best-fit Gaussian line described in Table \ref{lines} is overlaid.  The lump near $\nu=5.446$ GHz is the smeared average of the nearest He$\alpha$ transitions.   \emph{Bottom}:  Same as above, but for eight H$\epsilon$ transitions with $n=$180 to 191.  The brightest line, at $\nu=5.407$ GHz, is H108$\alpha$, which is about 4 MHz from H183$\epsilon$.  Note that the frequencies of the spectrum shown here correspond to that of the highest-frequency transition observed, but that all velocities and line properties are calculated by assuming an idealized ``mean'' of all transitions with a given $\Delta n$. \label{lineplots}}
\end{figure}

Table \ref{lines} lists the line measurements and Figure \ref{lineplots} shows the highest and lowest significance fits in the deep spectrum toward the GCL.  A Gaussian line is fit to the average line profile for all transitions of a given $\Delta n$ (e.g., $\Delta n=1$ for ``$\alpha$''), which includes 7 or 8 transitions in the range given in Table \ref{lines}.  Generally, these averages over several $n$ values are treated like a single transition with $n$ and $\nu$ equal to the mean of the range of transitions averaged ($\overline{\nu}\approx4.95$ GHz for all transitions).  A custom third-order baseline is subtracted prior to Gaussian fitting for all spectra.  The best-fit line parameters have errors derived from standard error analysis routines.  However, it is worth noting that these errors do not account for the uncertainty in fitting a baseline to the spectra, and are likely to underestimate the true errors slightly. The baseline fitting uncertainty depends on the line brightness in a nontrivial way.  Empirically we find that the baseline-induced uncertainty is at most equal to the statistical errors in the lowest significance fit considered here (H$\epsilon$ line).

One of the most surprising characteristics of the lines shown in Table \ref{lines} is their unusually small widths.  The H$\alpha$, H$\beta$, H$\gamma$, and H$\delta$ transitions have widths that are similar to each other and have an error-weighted mean of $\Delta v=13.5\pm0.2$ \kms.  The H$\epsilon$ line is significantly wider and is discussed in more detail below.  Typical \hii\ regions have much larger line widths, with $\Delta v\approx20$ \kms\ \citep{s83,a96}.  A hydrogen recombination line width of $13.5\pm0.2$ \kms\ is equivalent to a Doppler temperature of $3960\pm120$ K \citep{r92}, which is also significantly less than values seen in typical \hii\ regions ($T_D\approx5000-10000$K).  The Doppler temperature represents the combined effects of bulk (turbulent) gas motion and thermal broadening, so it is a firm upper limit on the electron temperature.  The line width is roughly constant for transitions from H$\alpha$ to H$\delta$, which represent transitions with mean $n$ ranging from 109.5 to 172.5.

The narrow line widths with central velocities near 0 \kms\ are highly suggestive of stimulated emission by ionized gas along the line of sight to the GC region \citep{c90,a97}.  The present observations can test for the possibility of stimulated line emission by comparing the line ratios of transitions with different $n$.  All transitions are observed with a similar beam size, which means that they are probing the same volume of space and can be meaningfully compared \citep{d70}.  In this case, the integrated line ratio between transitions to states $n$ and $m$ takes on the form
\begin{equation}
R \approx R_{\rm{LTE}} \frac{b_n}{b_m} \frac{T_e - T_c \beta_{n}}{T_e - T_c \beta_{m}},
\end{equation}
\noindent where $R_{\rm{LTE}}= n^2 f_{n,n+\Delta n}/m^2 f_{m,m+\Delta m}$, $b$ and $\beta$ are the departure coefficients for non-LTE effects, and $T_e$ and $T_c$ are the electron and background continuum brightness temperatures \citep{d70,b72,r92}.  The departure coefficients vary with $n_e$, $T_e$, and $n$, and have been calculated for many transitions under a wide range of gas conditions \citep{d72,s79}.  To estimate non-LTE effects, we calculated departure coefficients for $T_e=20-10000$ K and $n_e=1-1000$ cm$^{-3}$.  The lowest temperature and density used in the calculations correspond to the conditions that may cause stimulated emission observed toward the GC region at 20 cm \citep{l73}.  The LTE line ratios depend on the oscillator strengths for the transitions, which have relative values known to high precision \citep{d69}.

Table \ref{lineratios} shows the integrated line ratios and the predictions for LTE and non-LTE conditions.  The ratios show excellent agreement with LTE for all transitions observed, representing $n$ values ranging from $\sim$109.5 to 185.5.  The predicted line ratios for several other models generally do not all agree with the observed line ratios.  One reason that there are few non-LTE models that have similar line ratios as the LTE model is that the $\beta$ deviation coefficient has a nonmonotonic dependence on $n$, such that some $n$ values are stimulated more than others.  This is why observing a wide range of $n$ values helps constrain the amount of non-LTE emission.

One model, with $T_e=1000$ K and $n_e=1000$ cm$^{-3}$, predicts line ratios similar to the observed and LTE values.  The amount of stimulated emission can be calculated as
\begin{equation}
T_l/T_l^{\rm{LTE}} \approx b_n [1 - \beta_n (T_c/T_e)],
\end{equation}
\noindent where the background continuum brightness, $T_c\approx1-2$ K (Ch. \ref{gcsurvey_gbt}).  For $T_e=1000$ K and $n_e=1000$ cm$^{-3}$, the amplification of the line is only a few percent.  A much stronger background continuum is required to cause significant stimulated emission, with 10\% amplification for $T_c\sim25$ K;  this is more than 10 times brighter than the observed continuum in the GCL.  Thus, the line ratios severely constrain the contribution of stimulated emission to the recombination line emission for the brightest emission observed toward the GCL.

There are two other points that suggest that the recombination line emission in the GCL is not stimulated.  First, if the radio continuum from the GCL was stimulating the recombination line emission, the morphology of the line emission would closely follow the background continuum emission \citep{l73,c90}.  The HCRO and GBT observations show that the peak line emission, particularly in the eastern half of the GCL, is significantly offset from the peak continuum emission.  The similar line width, line velocity, and line ratios toward the eastern and western recombination line peaks suggest that the entire structure emits by a similar mechanism.  Second, the width of the average H$\epsilon$ line is significantly larger than that of the other lines, which is inconsistent with the idea that it is stimulated.  However, the H$\epsilon$ line represents the highest $n$ states, where stimulated emission is expected to be strongest.  These points strengthen the case for an LTE origin for the recombination line emission in the brightest portions of the GCL, which implies that the narrow line widths with velocities nearly at rest are related to intrinsic properties of the ionized gas.

\begin{deluxetable}{lcc|c|cccccccccc}
\tablecaption{Comparing Line Ratios to LTE and Stimulated Emission Models \label{lineratios}}
\tabletypesize{\scriptsize}
\rotate
\tablewidth{0pt}
\tablehead{
\colhead{Transition} & \colhead{$n$ Range} & \colhead{$R^{\rm{obs}}_\alpha$} & \colhead{$R^{\rm{LTE}}_\alpha$} & \colhead{$R^{\rm{20,1}}_\alpha$} & \colhead{$R^{\rm{100,10}}_\alpha$} & \colhead{$R^{\rm{100,100}}_\alpha$} & \colhead{$R^{\rm{100,1000}}_\alpha$} & \colhead{$R^{\rm{1000,10}}_\alpha$} & \colhead{$R^{\rm{1000,100}}_\alpha$} & \colhead{$R^{\rm{1000,1000}}_\alpha$} & \colhead{$R^{\rm{10000,10}}_\alpha$} & \colhead{$R^{\rm{10000,100}}_\alpha$} & \colhead{$R^{\rm{10000,1000}}_\alpha$} \\
}
\startdata
H$\alpha$ & 106--113  & 1               & 1      & 1     & 1     & 1     & 1     & 1     & 1     & 1     &  1    & 1     & 1 \\
H$\beta$ & 134--142 & 0.275$\pm$0.012   & 0.279  & 0.322 & 0.381 & 0.300 & 0.295 & 0.373 & 0.281 & 0.275 & 0.417 & 0.242 & 0.245 \\
H$\gamma$ & 152--162 & 0.130$\pm$0.010  & 0.127  & 0.224 & 0.198 & 0.137 & 0.133 & 0.173 & 0.123 & 0.124 & 0.193 & 0.098 & 0.103 \\
H$\delta$ & 167--178 & 0.067$\pm$0.007  & 0.073  & 0.146 & 0.118 & 0.083 & 0.077 & 0.099 & 0.071 & 0.072 & 0.107 & 0.053 & 0.058 \\
H$\epsilon$ & 180--191 & 0.041$\pm$0.007 & 0.047 & 0.112 & 0.077 & 0.055 & 0.049 & 0.068 & 0.053 & 0.048 & 0.055 & 0.051 & 0.047 \\
\enddata
\tablecomments{The line ratios for LTE conditions correspond to the ``mean'' transition of the range of $n$ values shown.  Oscillator strengths were taken at the mean $n$ or the average of the nearest two $n$ values, for fractional $n$.  Stimulated emission models labeled with ($T_e, n_e$) in K and cm$^{-3}$.  \citet{d72} and \citet{s79} were used for $T\leq100$ K and $>100$ K, respectively.}
\end{deluxetable}

Using the LTE assumption, the line properties give useful constraints on the intrinsic gas properties.  As mentioned before, the width of the H$\epsilon$ line is significantly larger than the other transitions.  Widths of lines that are not stimulated are generally dominated by thermal and doppler broadening, although these effects are not expected to change between the different transitions observed here.  The most common $n$-dependent broadening effect is collisional broadening (a.k.a. ``impact'' broadening), in which inelastic electron collisions preferentially broaden lines with large $n$ \citep{l78,r92}.  The ratio of collisional and doppler broadening widths of hydrogen lines is:
\begin{equation}
\frac{\Delta v_c}{\Delta v_d} = \frac{0.142}{\Delta n} \left(\frac{n}{100}\right)^{7.4} \left(\frac{T_e}{10^4\ \rm{K}}\right)^{-0.1} \left(\frac{n_e}{10^4\ \rm{cm}^{-3}}\right) \left(\frac{T_d}{2\times10^4\ \rm{K}}\right)^{-0.5}
\end{equation}
\noindent or
\begin{equation}
\frac{\Delta v_{\rm{tot}}}{\rm{km\ s}^{-1}} = \sqrt{\left[\frac{4.31}{\Delta n} \left(\frac{n}{100}\right)^{7.4} \frac{n_e}{10^4\ \rm{cm}^{-3}} \left(\frac{T_e}{10^4\ \rm{K}}\right)^{-0.1}\right]^2 + \Delta v_d^2},
\label{widtheqn}
\end{equation}
\noindent where the total line width is the quadrature sum of the Doppler and impact line widths in \kms \citep{r92}.

\begin{figure}[tbp]
\begin{center}
\includegraphics[angle=270,scale=0.7]{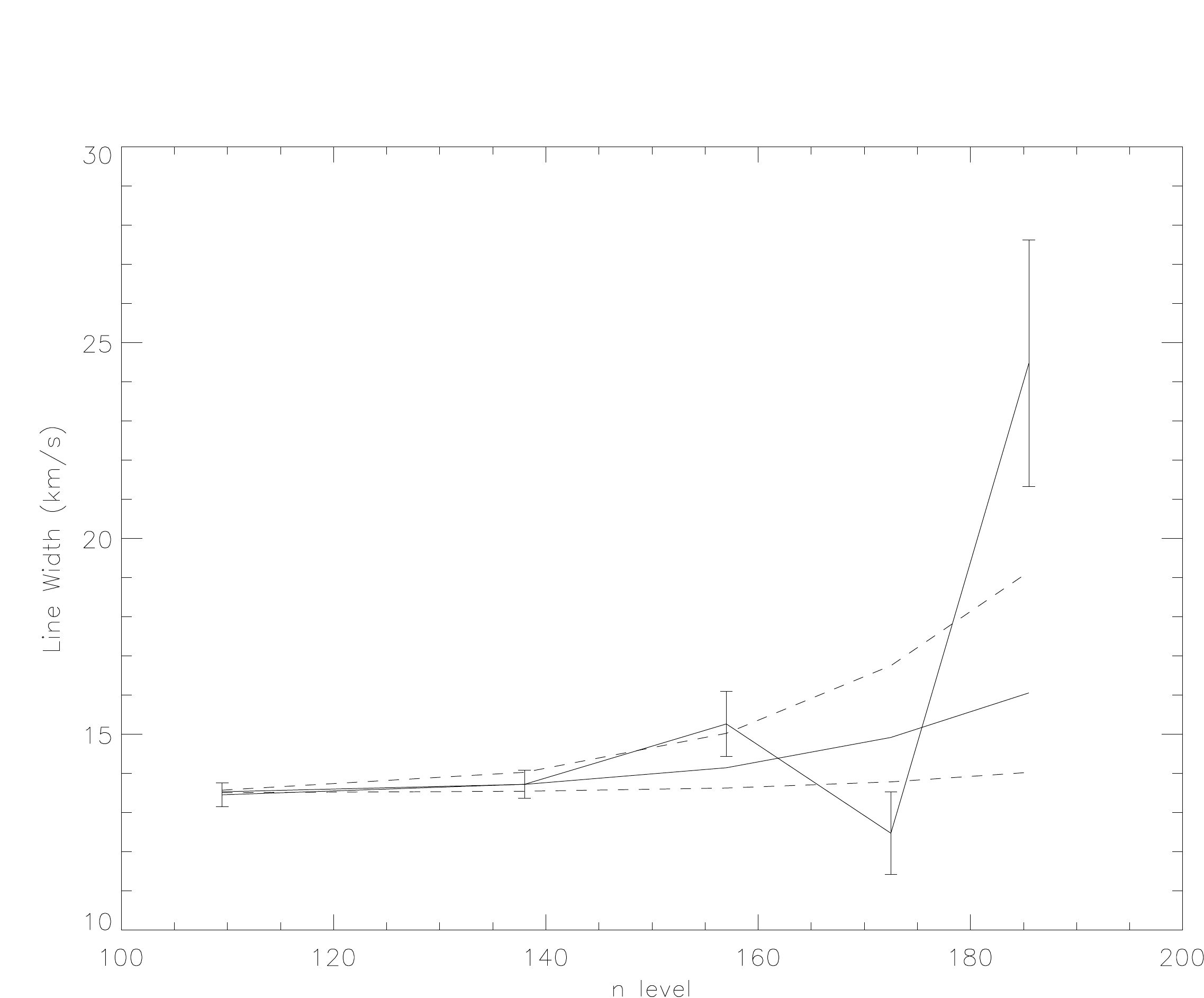}
\end{center}
\caption{Plot of the distribution of hydrogen line widths observed by the GBT as a function of electronic state $n$.  The solid line shows the best-fit model to the widths, considering the effects of collisional broadening; the dotted lines show the $\pm1 \sigma$ fits.  The normalization of the curves corresponds to an electron density $n_e = 950\pm270$ cm$^{-3}$, assuming $T_e=3960$ K. \label{widths}}
\end{figure}

The strong dependence of collisional broadening on $n$ can be used to constrain $n_e$, since $\Delta v_d$ is known and the dependence on $T_e$ is weak.  Equation \ref{widtheqn} was fit to the distribution of Hydrogen line widths assuming $\Delta v_d=13.5$ \kms\ (the error-weighted mean of the H$\alpha$ to H$\delta$ line widths\footnote{Allowing $\Delta v_d$ to vary does not significantly change the fit values or its quality.}) and $T_e = T_d = 3960$ K.  Figure \ref{widths} shows the best-fit line width model with $n_e=950\pm270$ cm$^{-3}$.  The fit quality is relatively poor, with $\chi^2/\nu = 14.4/4$, which suggests that the errors may be underestimated somewhat.  If baseline-subtraction uncertainties scale inversely as the peak line significance and were to double the line width uncertainties for the H$\epsilon$ line, the best fit would have $n_e=1210\pm420$ cm$^{-3}$ with $\chi^2/\nu=5.0/4$.  These fits are equal within their uncertainties, so the results would not change dramatically if the line width error included baseline-fitting uncertainty.  Note that technically, collisional broadening takes the shape of a Lorentz distribution, which, when convolved with a Gaussian, produces a Voigt profile \citep{r92}.  However, at the low signal to noise ratio observed for the H$\epsilon$ transition, no line wings are apparent and a Gaussian is a close approximation to the expected line shape.

The He$\alpha$ line was detected in the deep spectrum toward the GCL peaks with 18$\sigma$ significance.  The ratio of the He$\alpha$ to H$\alpha$ lines is a measure of the ionized helium abundance, $Y^+=I_{\rm{He}\alpha}/I_{\rm{H}\alpha}$.  As shown in Table \ref{lines}, the integrated line ratio toward the GCL is $0.075\pm0.007$, which is somewhat less than the solar abundance ratio \citep[$\sim$0.1;][]{l74,r92}.  In subsequent calculations involving $Y^+$, a value of 7.5\% is assumed for the GCL.

Typically, the line-to-continuum ratio for a radio recombination line is a useful way to measure the electron temperature of the emitting region.  The LTE electron temperature is:
\begin{equation}
T_e^{\rm{LTE}}=\left[6943 \frac{\nu}{\rm{GHz}}^{1.1} \frac{T_c}{T_l \Delta V_G} \frac{1}{1+Y^+}\right]^{0.87} \rm{K}
\end{equation}
\noindent which is related to the actual electron temperature by:
\begin{equation}
T_e^{\rm{LTE}} = T_e \left[b_n \left(1 - \frac{\beta_n \tau_c}{2}\right)\right]^{-0.87}
\end{equation}
\noindent However, much of the continuum emission in the GCL is nonthermal (Ch. \ref{gcsurvey_gbt}), so the line-to-continuum ratio merely provides an upper limit on the electron temperature.  For the deep spectrum, the continuum flux was estimated from the 5 GHz map described in chapter \ref{gcsurvey_gbt} to be $T_c\approx0.8$ K, which gives $T_l \Delta v/T_c=2.0$ \kms, for the H$\alpha$ line.
This line-to-continuum ratio leads to an upper limit on the LTE electron temperature of $T_e^{\rm{LTE}}\lesssim5220$ K, which is consistent with the upper limit on $T_e$ from the narrow line widths.  For the largest non-LTE effects that are consistent with the observed line ratios ($n_e=100$\ cm$^{-3}$ and $T_e=1000$\ K), $T_e = (0.855)^{0.87} T_e^{\rm{LTE}}\lesssim4550$ K.  However, for higher densities and temperatures, $T_e$ is within a few percent of $T_e^{\rm{LTE}}$.  The upper limit on the electron temperature from the observed line width gives a lower limit on the line-to-continuum ratio of $T_l \Delta v/T_c>2.75$ \kms\ and an upper limit on the thermal continuum in the observed 5 GHz continuum brightness of 0.6 K.  Thus, at most about 70\% of the radio continuum toward the recombination line peaks of the GCL is thermal.

The upper limit to the electron temperature constrains the emission measure.  Assuming $\tau_L\ll1$ and LTE for the H109$\alpha$ transition, the continuum emission measure is:
\begin{equation}
\rm{EM}_c = 8.5\times10^{-3} (1+Y^+) \Delta v T_l T_e^{3/2} e^{-X_n} \rm{pc}\ \rm{cm}^{-6},
\end{equation}
\noindent where $\Delta v$ is in units of \kms and $e^{-X_n}\approx1$, for $T_e\gtrsim1000$ K \citep{d70}.  Using the average H$\alpha$ line parameters for the deep GCL spectrum gives EM$_c\approx3850 (T_e/3960\ \rm{K})^{3/2}$ pc cm$^{-6}$.  The eastern and western peak line brightnesses are somewhat different, giving EM$_c\approx3080 (T_e/3960\ \rm{K})^{3/2}$ and $4570 (T_e/3960\ \rm{K})^{3/2}$ pc cm$^{-6}$ for the east and west, respectively.  Assuming a shell geometry for the GCL recombination line emission (see \S\ \ref{disc:morph}), the path length through the shell edge is about 50 pc, which gives an rms electron density, $\sqrt{\langle n_e^2\rangle}\approx8.8 (T_e/3960\ \rm{K})^{3/4}$ cm$^{-3}$.  Mild non-LTE effects from the $b_n$ departure coefficient effectively scale $T_l$ up by $1/b_n$, so for the largest non-LTE deviations allowed ($n_e=100$ cm$^{-3}$ and $T_e=1000$ K), EM$\approx4570 (T_e/3960\ \rm{K})^{3/2}$ pc cm$^{-6}$ and $\sqrt{n_e^2}\approx9.6 (T_e/3960\ \rm{K})^{3/4}$ cm$^{-3}$.

\label{gbt:vel}
The GBT observations generally confirm the velocity structure observed by HCRO, finding no emission with velocities $|v|>20$ \kms.  The rms noise in the baseline level of the deep GBT spectrum gives an upper limit to the peak line brightness of about 1 mK for $v=\pm500$ \kms\ and 2 mK for $v=\pm1500$ \kms;  these limits are 1\% and 2\% of the peak H$\alpha$ emission with $v\sim0$ \kms, respectively.

\begin{figure}[tbp]
\begin{center}
\includegraphics[angle=270,scale=0.7]{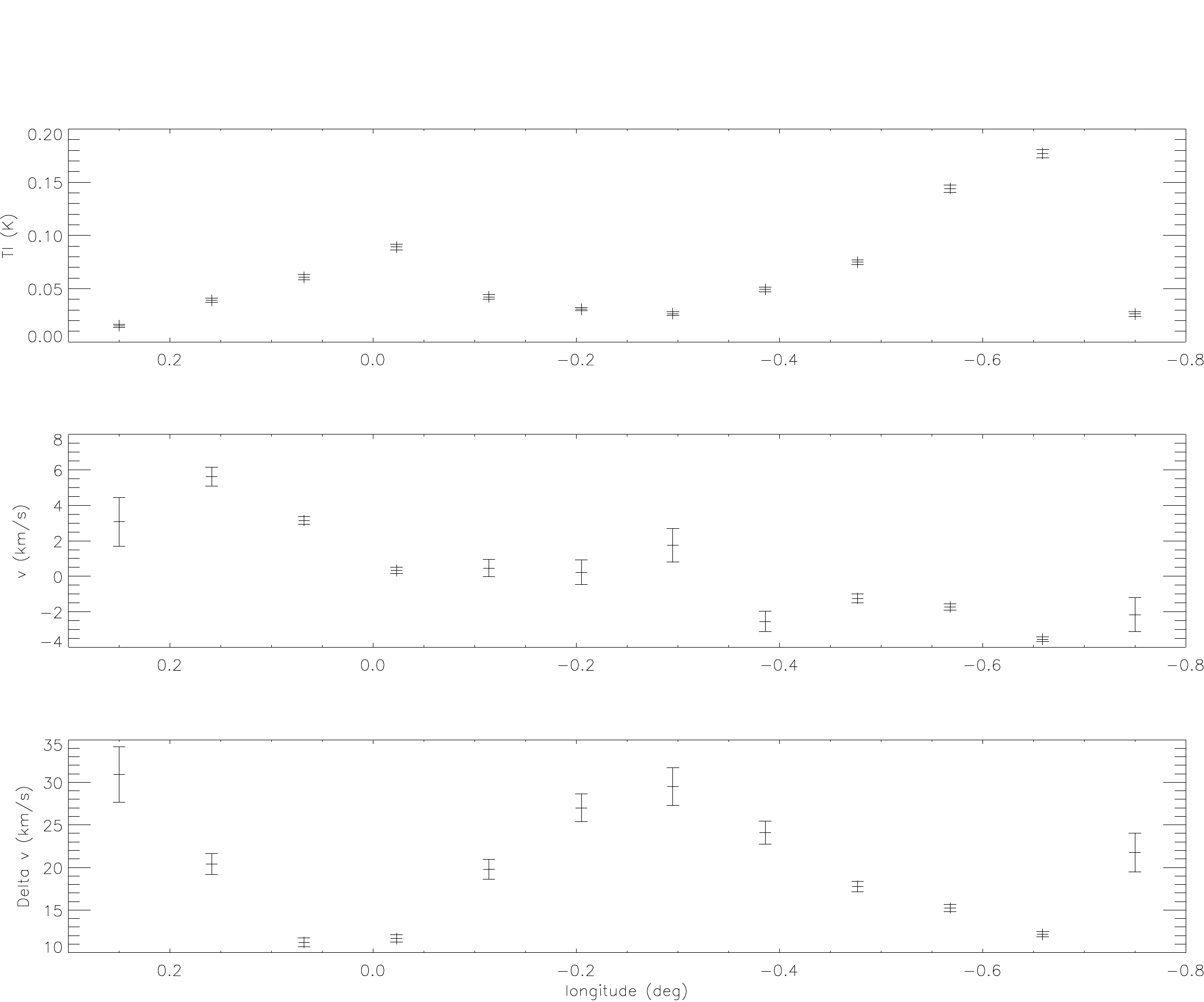}
\end{center}
\caption{Plots of the properties of the average H$\alpha$ recombination line for the series of pointings across the GCL at $b=0\ddeg45$.  Line properties found by fitting a single Gaussian to the profile.  The top plot shows the peak line brightness temperature, the middle plot shows the line velocity, and the bottom plot shows the line width.  The brightest recombination line emission (near $l\sim0$\sdeg\ and $-0\ddeg6$) has the narrowest line width. \label{polstart}}
\end{figure}

At the brightest parts of the recombination line emission, the line velocities observed by the GBT are consistent with those of the HCRO.  Figure \ref{polstart} shows how the line properties change as a function of Galactic longitude for the scans at $b=0\ddeg45$.  Comparing the GBT and HCRO line velocities in Figures \ref{polstart} and \ref{hcroimg} shows some significant differences.  Some of these differences may be due to confusion between multiple components of the line (see below).  Unfortunately, the HCRO data are not available for detailed comparison, so the present work focuses on the GBT results, which have a more reliable velocity calibration.

One interesting trend in Figure \ref{polstart} is the east-west asymmetry in the line velocities.  There is a tendency for positive velocities on the east side and negative on the west.  Averaging over the four easternmost and westernmost scans in the $b=0\ddeg45$ strip (with average longitudes of $0\ddeg1$ and $-0\ddeg6$, respectively) finds mean velocities of $2.3\pm0.2$ and $-2.4\pm0.2$ \kms, respectively.  These averages show relatively little line structure, so a single Gaussian is a good fit to the profile.  The pattern of positive and negative velocities in the east and west is the same as that observed in the Galactic disk, although the molecular gas in the disk reaches velocities up to $\pm$100 \kms\ and rotates about at $l=0$\sdeg \citep{b87}.  The velocity gradient with Galactic longitude observed by GBT is consistent with Galactic rotation and the upper limit on the gradient from HCRO observations.

\begin{figure}[tbp]
\includegraphics[scale=0.15]{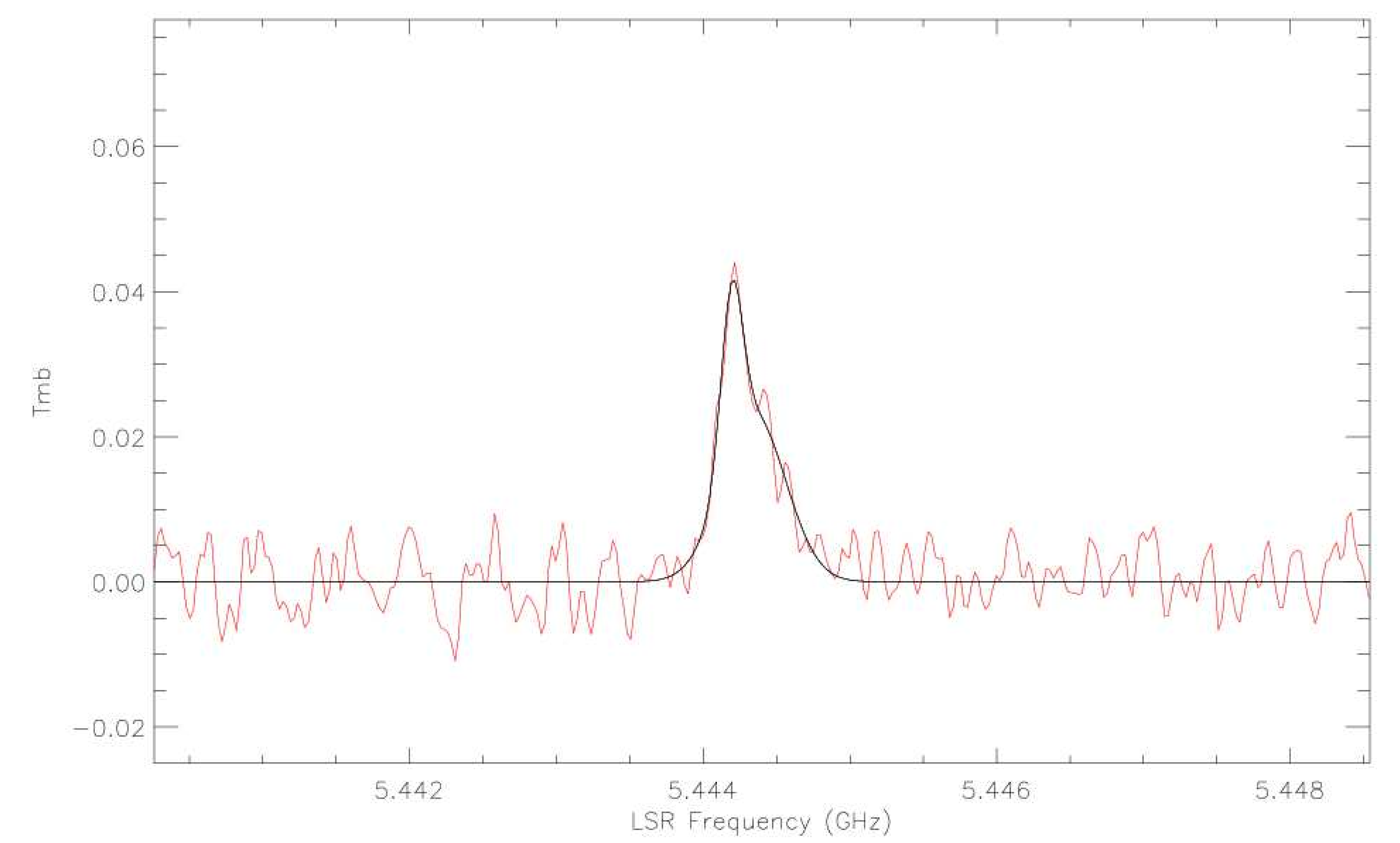}
\caption{Average H$\alpha$ spectrum for four scans near the center of the GCL with a line showing the best-fit model described in the text.  Fitting one Gaussian gives a moderately wide fit, but fitting two Gaussians gives a narrow and wide component of similar amplitude.  This shows how the assumption of a single Gaussian component can lead to a inverse relationship between line width and amplitude and that the unusually narrow line is observed throughout the GCL. \label{polcen}}
\end{figure}

One striking correlation in Figure \ref{polstart} is the inverse relationship between peak line brightness and best-fit line width:  the brightest line emission has the narrowest widths and vice versa.  Two potential explanations for this trend are that (1) the line width is intrinsically tied to its brightness, or (2) the line has two components (a wide and narrow line) that can affect the best-fit line width according to their relative strengths.  
In an attempt to understand the inverse relationship between line brightness and width, an average of the central four scans of the $b=0\ddeg45$ strip was made.  Figure \ref{polcen} shows the average H$\alpha$ profile for the four scans located in the center of the GCL that, separately, have best-fit line widths of 20--30 \kms.  The average spectrum shows a slightly more complex line profile than seen near the brightest recombination line emission.  If the profile is fit with two Gaussians, their best-fit components are $(T_l, v, \Delta v)=(22\pm3\ \rm{mK}, 5.2\pm0.5\ $\kms$, 9.3\pm1.5\ $\kms$)$ and $(25\pm2\ \rm{mK}, -5.0\pm1.3$\ \kms$, 30.6\pm1.9$\ \kms$)$.  The narrow component is similar to that observed near the brightest parts of the GCL, but its amplitude is comparable to the wide component, such that a single-Gaussian fit is moderately wide.  In averages of other scans, there is a tendency for a similar, wide (20--30 \kms), low-level (10--20 mK) line to appear.  The width and central velocity of the wide component are consistent with an origin in the diffuse ionized gas in the Galactic disk.  However, the amplitude of this line is larger than the $\sim10$ mK seen in the background outside the GCL, so the wide line seems to be associated with the GCL.  Regardless of its origin, it seems that the changes in best-fit, single-Gaussian line width shown in Figure \ref{polstart} are likely due to the relative strength of the narrow GCL line and this wider component.  Thus, there seems to be narrow ($9-14$ \kms) recombination line emission for all lines of sight through the GCL.

\begin{figure}[tbp]
\includegraphics[scale=0.6,bb=0 600 375 593]{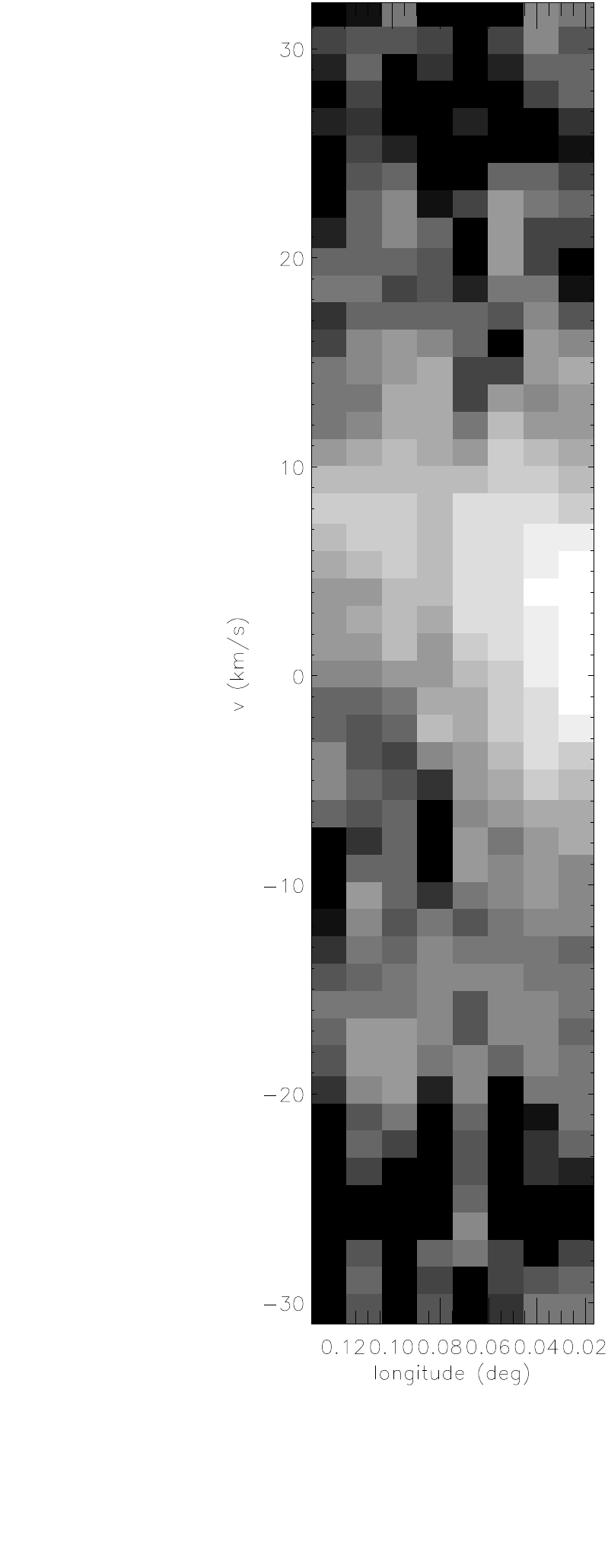}
\hfil
\includegraphics[scale=0.6,bb=0 600 375 593]{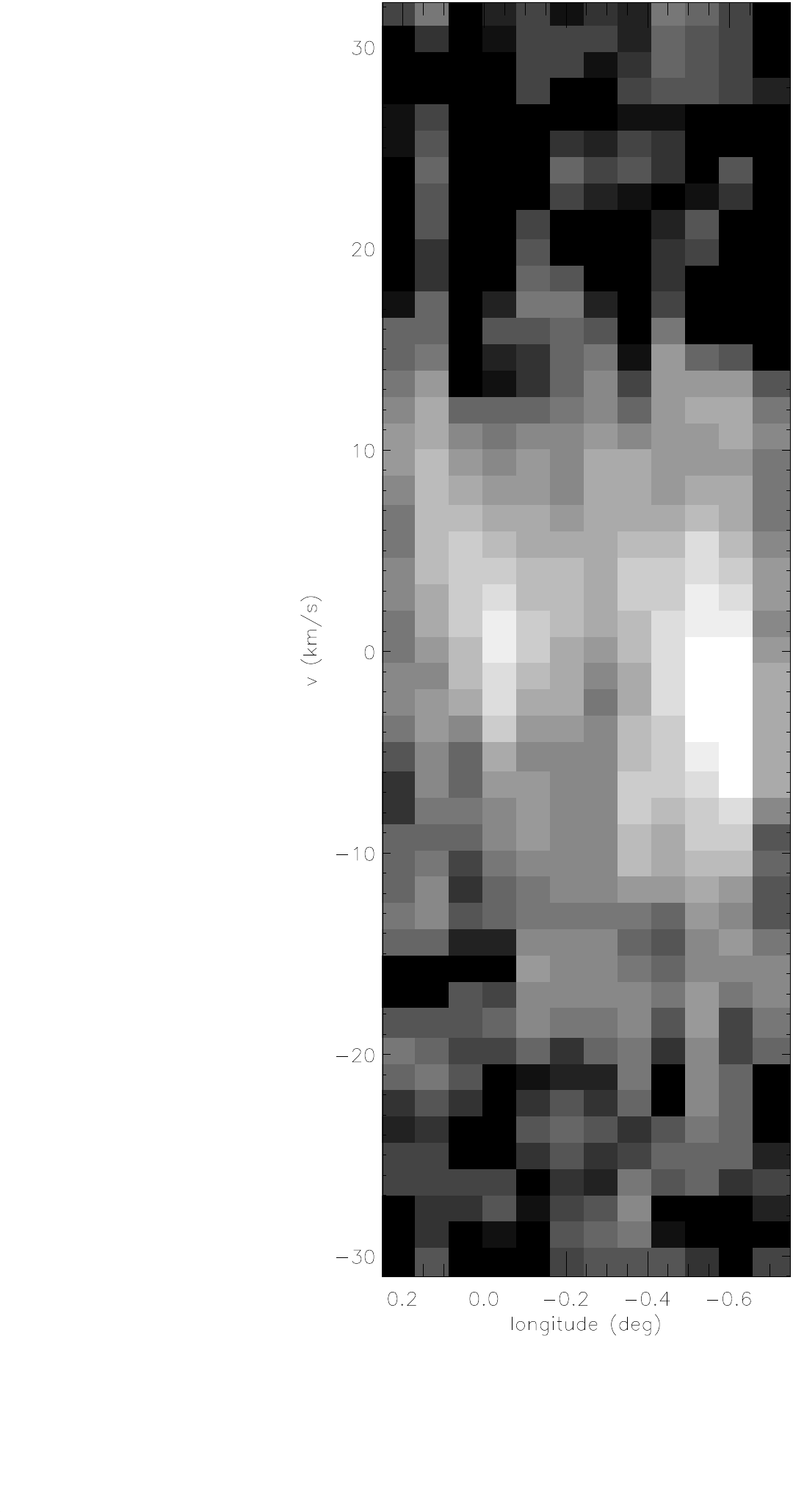}
\hfil
\includegraphics[scale=0.6,bb=-100 0 375 593]{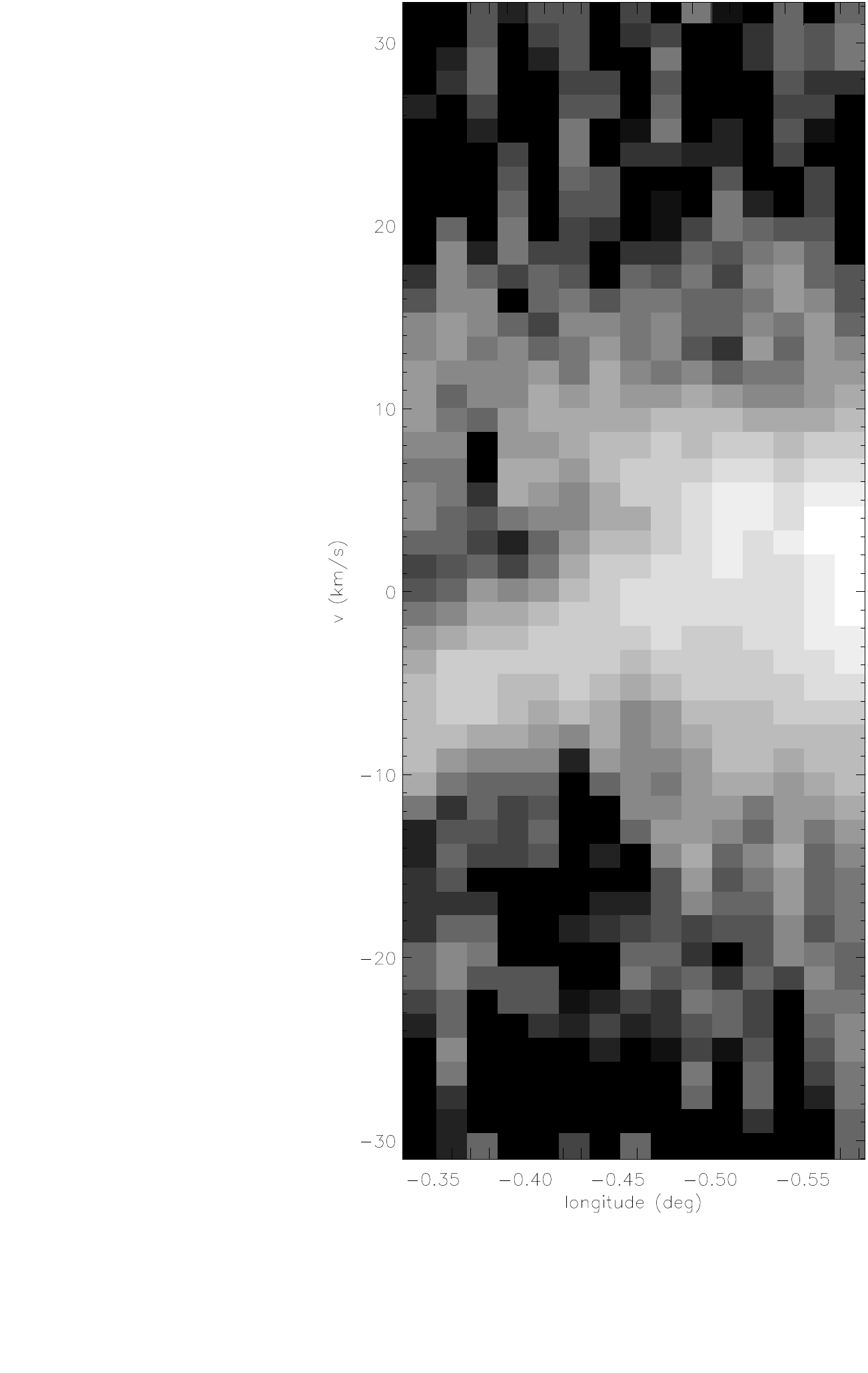}
\caption{$l$--$v$ diagrams of the average H$\alpha$ radio line brightness for the GCL3 strip (\emph{left}), $b=0\ddeg45$ strip (\emph{center}), and GCL4 strip (\emph{right}).  The GCL3 strip starts near the eastern peak of the recombination line emission of the GCL and goes toward the northeast.  The $b=0\ddeg45$ strip spans the entire east-west extent of the GCL.  The GCL4 strip starts near the western peak of the line emission in the GCL and goes toward the northeast.  Detailed descriptions of these strips of observations are given in Table \ref{gbtpointings}.  The right $l$--$v$ diagram does not account for the missing scan at $l=-0\ddeg48$, which gives errors in $l$ of order $1\damin4$ near that longitude. \label{lvimg}}
\end{figure}

The above work shows that the line kinematics are more complex in the center of the GCL.  Figure \ref{lvimg} shows an alternative way of visualizing the recombination line velocity structure using $l$--$v$ diagrams.  Three strips of observations show the change in the line velocity as a function of Galactic longitude.  These plots show clearly how the brightest emission is always within 20 \kms\ of being at rest, and that the lines are typically about 10 \kms\ wide.  The plots also help show how the line profile has multiple components inside the GCL, for $l=-0\ddeg2$ to $-0\ddeg45$.  In particular, the GCL4 $l$--$v$ diagram shows two diagonal structures near $l=0\ddeg4$, which indicate that there are two narrow components to the line profile with velocities approaching $\pm10$ \kms\ toward the eastern end of the strip.

\begin{figure}[tbp]
\includegraphics[scale=0.15]{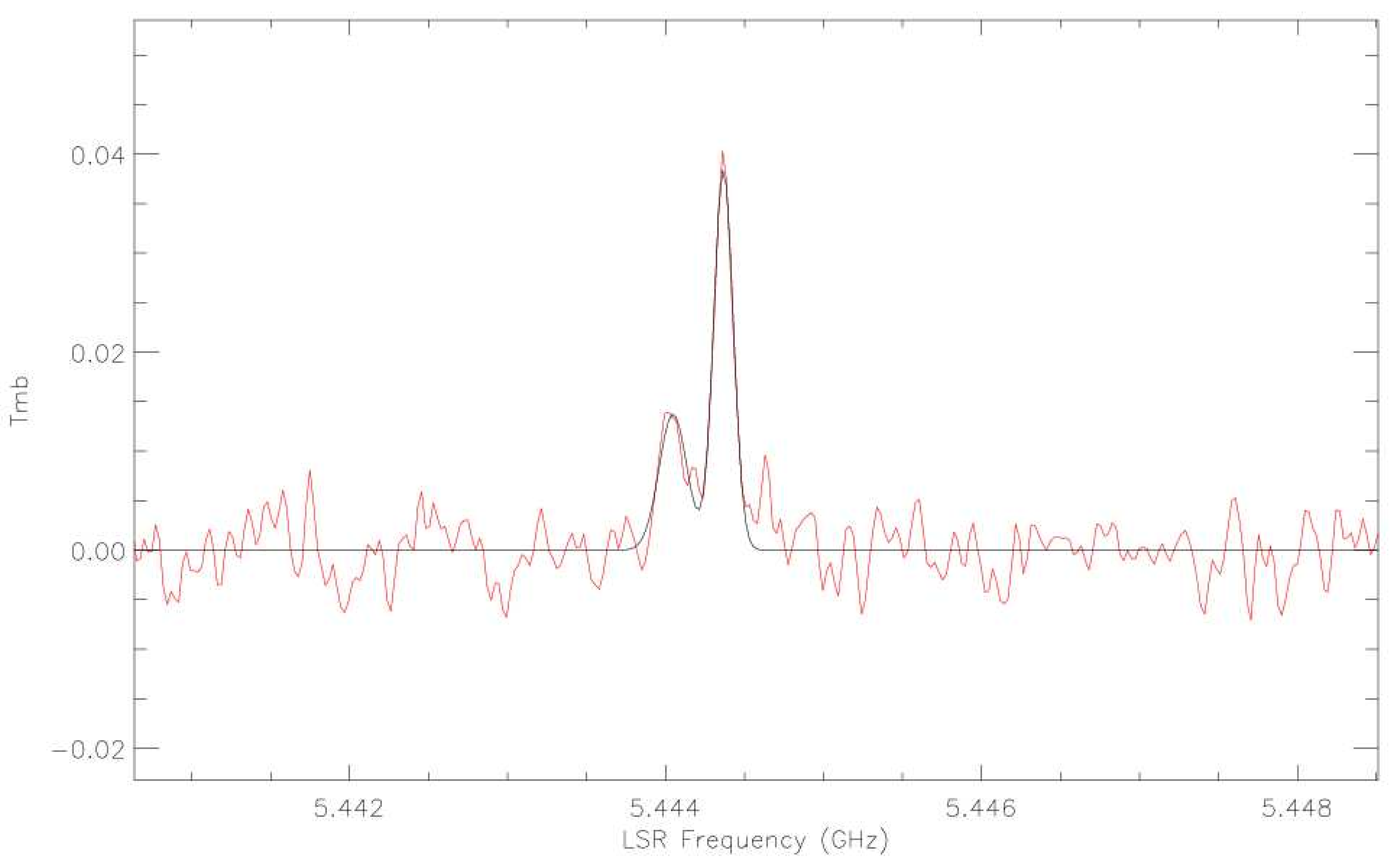}
\caption{Plot of the average H$\alpha$ spectrum for three scans inside the GCL, on the eastern edge of the GCL4 strip.  The profile requires two narrow Gaussians for an adequate fit, with velocities of roughly +14 and --6 \kms. \label{polcen2}}
\end{figure}

To study the line structure more carefully, an average of the H$\alpha$ spectra for the three scans at the eastern edge of the GCL4 strip was made.  Figure \ref{polcen2} shows how the average spectrum confirms the visual impression from the $l$--$v$ diagram:  the best-fit spectrum has two distinct components with $(T_l, v, \Delta v)=(13.8\pm1.4\ \rm{mK}, 13.0\pm1.6$ \kms$, 12.5\pm1.5$ \kms$)$ and $(38.5\pm1.6\ \rm{mK}, -6.4\pm0.2$ \kms$, 8.8\pm0.4$ \kms$)$.  Evidence for this split line is more difficult to see in individual scans, but is also apparent in the $b=0\ddeg45$ strip, near $l=-0\ddeg35$ (see Fig. \ref{lvimg}).  This splitting of the line is most apparent in the GCL4 strip, since it is more densely sampled and neighboring scans can be averaged together.  Both components of the split line are narrow like that observed in the deep spectrum, which suggests that they originate in regions with similar density and temperature as toward the GCL peaks and is not unrelated foreground emission.  Thus, it is clear that the recombination line emission in the center of the GCL is not dominated by a foreground, but is similar to the emission at the edges of the GCL.  Furthermore, a split line with velocities up to around $\pm10$ \kms\ tends to be found inside the GCL.

The $l$--$v$ plot for the GCL3 strip shown in Figure \ref{lvimg} shows two line components toward its eastern end.  Averaging over the four easternmost scans of the GCL3 strip (with $l=0\ddeg083-0\ddeg133$ shows two Gaussian components to the average H$\alpha$ profile.  The two components have $(T_l, v, \Delta v)=(26.9\pm1.0\ \rm{mK}, 7.6\pm0.3$ \kms$, 17.2\pm0.8$ \kms$)$ and $(11.1\pm 1.4\ \rm{mK}, -18.3\pm0.5$ \kms$, 8.6\pm1.3$ \kms$)$.  Both components are narrow, especially when considering the broadening caused by averaging over multiple scans, so the emission is likely to be associated with the GCL.

\section{Discussion}
\label{recomb_dis}

\subsection{The Shell Model for the GCL}
\label{disc:morph}
The GBT and HCRO have found radio recombination line emission toward the GCL with the strongest emission near its eastern and western edges.  The unusually narrow line width and LTE line ratios for the line emission throughout the GCL suggests that all the line emission observed is a part of the same structure.  The detection of emission well above the background from everywhere inside the GCL suggests that the line emission is associated with the radio continuum GCL.  As described in chapter \ref{gcl_all}, the apparent edge-to-center contrast is consistent with an analytical expression and simulations of the appearance of a shell.  The analysis presented here assumes that the recombination line emission in the GCL comes from a hollow shell with radius 40 pc and thickness 15 pc.

\subsection{Narrow Line Widths and Ionized Gas Conditions}
\label{disc:gas}
If the gas emits in LTE the line widths can be used to derive intrinsic gas conditions.  Previous observations of radio recombination lines in the GC region have found narrow lines, particularly at lower frequencies, and found that the emission was most likely stimulated \citep{l73,p75,a97}.  However, there are several characteristics of the line emission observed here that show that the gas is in LTE, including  well-constrained line ratios consistent with LTE, the lack of correlation between background continuum brightness and line brightness, and the relative broadening of the transition with the highest $n$ values (H$\epsilon$).

Assuming the emission is not stimulated, the Doppler widths of these lines provide a strict upper limit to the electron temperature of $3960\pm120$ K.  This limit to the electron temperature is among the lowest values observed in the Galaxy, although radio recombination line observations of other \hii\ regions have found electron and Doppler temperatures as low as 4000 K \citep[][M. Goss 2006, private communication]{s83,a96}.  Some of the best-fit line widths discussed in this work are as low as 9 \kms, which corresponds to a Doppler temperature of $\sim$1800 K, much smaller than observed elsewhere in the Galaxy.

The simplest explanation for low electron temperature in the GCL is that the gas has a large metal abundance that allows it to cool efficiently \citep{m85,a96}.  Studies of recombination line emission from \hii\ regions have found a gradient in $T_e$ with galactocentric radius, showing significantly lower temperatures nearer to the GC \citep{s83,a96}.  This gradient has been directly traced to an abundance gradient in some of the elements responsible for cooling \hii\ regions, such as oxygen \citep{m00}.  The study of the $T_e$ gradient by \citet{a96} predicts $T_e\approx5500$ K in the GC region, while \citet{s83} predicts $T_e\approx3100$ K.  These studies predict different values because they study different types of sources and the efficiency of cooling depends on $n_e$ \citep{a96}.  The gas densities of the sources studied by \citet{s83} have $n_e=2$ cm$^{-3}$ to $2.5\times10^3$ cm$^{-3}$, which includes the range of gas densities found for the GCL ($\sim1000$ cm$^{-3}$);  the sources studied by \citet{a96} have $n_e>8\times10^3$ cm$^{-3}$, which is less appropriate for the GCL.  Thus, the constraint on $T_e$ is consistent with the Galactic $T_e$ gradient.  The abundance gradient is generally believed to be caused by enhanced star formation in the high-pressure, high-density gas nearer to the center of the Galaxy.  However, predicting the magnitude of the gradient from models of star formation and Galactic dynamics is an area of active research \citep{g92,c01,c05}.

The small line width also constrains the amount of turbulent motion in the ionized gas.  If the observed line width is dominated by thermal broadening, the turbulent gas motion is significantly less than 13.5 \kms.  This line width is much less than observed in \hii\ regions associated with active star formation, where stellar winds and other processes stir the surrounding ISM.  For example, the hydrogen recombination line observed toward 30 Dor has line widths dominated by Doppler broadening \citep[$\Delta v\gtrsim30$ \kms;][]{c94,p97}.  In the GC region, VLA observations have found broad ($\sim50$ \kms) and narrow ($<20$ \kms) hydrogen recombination line widths in the Pistol and Arches \hii\ regions, respectively \citep{l97,l02}.  These \hii\ regions are ionized by the Quintuplet and Arches star clusters.


The broadening of the line width with increasing electronic state, $n$, is best explained by collisional broadening.  The best-fit model to the line widths gives $n_e=950\pm270$ cm$^{-3}$.  The best-fit $n_e$ assumes $T_e=T_d$, but since the dependence varies as $T_e^{0.1}$, an electron temperature of half the Doppler temperature would decrease $n_e$ by only 7\%.  The best-fit density is consistent with the lack of stimulated emission in the GCL.  As discussed in \S\ \ref{gbt:lines}, simulated emission would significantly change the line ratios from LTE values for $n_e\lesssim100$ cm$^{-3}$.

The volume filling factor for the gas in the deep spectrum of the GCL can be found from $f = \langle n_e^2\rangle/n_e^2$.  Assuming LTE, the average H$\alpha$ line brightness at the peak of the GCL emission gives $\sqrt{\langle n_e^2\rangle}\approx8.8 (T_e/3960\ \rm{K})^{3/4} (50\ {\rm{pc}}/L)^{0.5}$ cm$^{-3}$, where $L$ is the integrated path length that is assumed to be through the edge of the shell.  The collisional broadening observed in the deep spectrum of the GCL requires the true electron density to be $n_e=950\pm270 (T_e/3960\ \rm{K})^{0.1}$ cm$^{-3}$.   Thus, the volume filling factor for the deep spectrum at the peaks of the GCL is $f=(9\pm2)\times10^{-5} (T_e/3960\ \rm{K})^{1.3} (50\ {\rm{pc}}/L)$ cm$^{-3}$.  The deep spectrum toward the peaks of the GCL cover an area of two square beams ($2\damin5$) in the east and west, equivalent to a physical volume of $\sim5500 (L/50\ \rm{pc})$ pc$^{-3}$.  In this volume, the ionized gas occupies about $0.5 (T_e/3960\ \rm{K})^{1.3}$ pc$^{-3}$.  The small filling factor also suggests that high-resolution radio recombination line observations should find structure on smaller spatial scales.

The thermal pressure implied by the constraint on $T_e$ and measurement of $n_e$ is $P/k=3.8\times10^6 (T_e/3960\ \rm{K})$ K cm$^{-3}$.  Although this pressure is about 100 times larger than the total gas pressure near the Sun \citep{bl87}, it is not unusual for the GC region \citep{s92}.  Observations of molecular gas in the central degrees of the Galaxy show velocity dispersions that imply virial pressures around $P_{\rm{vir}}/k=nT_{\rm{vir}}=6\times10^6$ K cm$^{-3}$ \citep{ma04}.  The two components of the X-ray--emitting gas characterized by temperatures of $\approx10^7$ and $\approx10^8$ K have pressures in the range of $(1-5)\times10^6$ K cm$^{-3}$ \citep{k96,m04}.  This pressure is also similar to that of the equipartition magnetic field in the GCL of $\approx0.1$ mG \citep[$P_B/k\approx3\times10^6$ K cm$^{-3}$, chapter \ref{gcsurvey_gbt};][]{m96}.  Thus, the pressures in typical GC molecular clouds, X-ray gas, equipartition magnetic field, and the ionized gas in the GCL are roughly in equilibrium.  The observed pressure and its equilibrium between different components of the GC region may have interesting implications for the origin of the GCL \citep{h90,v05}.  These ideas are discussed in more detail in chapter \ref{gcl_all}.

\subsection{Mass and Ionization of the GCL}
\label{derived}
Knowing that the recombination line is in LTE allows an estimate of the thermal continuum flux, and hence, the mass of the ionized gas in the GCL.  The mass estimate is parameterized like this:
\begin{equation}
M = 0.419 (T_e/10^4\ {\rm{K}})^{0.175} (S_{5 {\rm{GHz}}}/{\rm{Jy}})^{0.5} (D/{\rm{kpc}})^{2.5} (\theta_G/{\rm{arcmin}})^{1.5}
\end{equation}
\noindent where $D$ is the distance to the GCL, $\theta_{\rm{G}}$ is the angular size of the emission, and the geometry is assumed to have a cylindrical shape \citep{p75,p78}.  Since the gas is in LTE, the line-to-continuum ratio is $T_l \Delta v/T_c=2.75 (T_e/3960\ \rm{K})^{-0.87}$ \kms, as shown in \S\ \ref{gbt:lines}.  The integrated line intensity over the GCL from HCRO observations is about $1.8\ \rm{K} * 13.5 $ \kms$ = 24 $ K \kms, for $b\ge0\ddeg2$ and subtracting 10\% for the background.  This gives a 5 GHz continuum flux of 80 Jy and a ionized mass $M=3\times10^5 (T_e/3960\ \rm{K})^{0.61}$ \msol\ in the GCL.  Note that the mass estimate includes the implicit dependence of the flux on the line to continuum ratio and, hence, $T_e$.  Uncertainties in the mass limit are probably dominated by incomplete sampling of the HCRO survey and are estimated to be about 20\%.

The mean electron density constrains the number of Lyman continuum photons required to ionize the gas.  The Lyman continuum photon flux is $N_{\rm{Ly}} = V n_e n_{\rm{H}} \alpha^{(2)}$, with $\alpha^{(2)}\approx6.6\times10^{-13}$ cm$^3$ s$^{-1}$, for $T_e=3\times10^3$\ K \citep{s88}, and the volume, $V\approx4\pi/3 ((r+\delta/2)^3-(r-\delta/2)^3) = 3.0\times10^5$ pc$^{-3}$, assuming $r=40$ pc and $\delta=15$ pc.  The ionizing flux is:
\begin{equation}
N_{Ly} = 1.6\times10^{49} (T_e/10^4\ {\rm{K}})^{0.35} (S_{5 {\rm{GHz}}}/{\rm{Jy}}) (D/{\rm{kpc}})^{-1} (\theta/{\rm{arcmin}})^{-3} (1/1+Y^+) (V/{\rm{pc}}^{-3}) \rm{s}^{-1}
\end{equation}
\noindent This equation is based on the relations of \citet{m67}, assuming $\theta = 1.47 \theta_G$.  For the observed flux and parameters of the GCL, the ionizing flux is $N_{Ly} = 1.0\times10^{50} (T_e/3960\ \rm{K})^{1.22})$ s$^{-1}$.  This method also results in a electron density of about 4.4 cm$^{-3}$ averaged over the entire 1\sdeg-region, which is consistent with the slightly higher mean densities of 7--10 cm$^{-3}$ observed at the shell edge.

The ionizing flux is 10 times the Lyman continuum flux of an O7 V star \citep{sm02}.  If this structure is ionized by stars inside the GCL in the plane, then at most half of the stellar flux would reach the GCL, so the required ionizing flux is equivalent to at least 20 07V-type stars (ignoring extinction), or $2\times10^{50}$ s$^{-1}$.  The Arches, Quintuplet, and Central star clusters, all located inside the GCL, have measured Lyman continuum fluxes of $10^{51.0}$, $10^{50.9}$, and $10^{50.5}$ s$^{-1}$ \citep{f99}.  Even if extinction is large, these clusters should easily be able to ionize the gas in the GCL.  Also, four regions of isolated star formation have been observed inside the GCL with ionizing fluxes of $\sim10^{50}$ s$^{-1}$ \citep{ca99}, which should contribute significantly to the ionization of the GCL;  other such sources are likely to be discovered in the region.  Thus, the ionized gas associated with the GCL is most likely photoionized.

\subsection{Kinematics}
\label{kinematics}
The radio recombination line is a useful tracer of the kinematics of ionized gas.  Our GBT observations include three strips of pointings across different parts of the ionized gas in the GCL, called the GCL3, GCL4, and $b=0\ddeg45$ strips.  Here we discuss the trends in the line velocities and whether they are consistent with an outflow.  A more general comparison of the observed gas kinematics to extragalactic outflows is given in chapter \ref{gcl_all}.

The HCRO and GBT recombination line observations confirm previous observations that found line velocities near zero toward the GCL \citep{p75}.  Small line velocities toward the GC region are typically interpreted as emission from the 8 kpc of Galactic foreground \citep{ma04}.  However, the narrow line widths and strong morphological connection to the radio continuum shell of the GCL \citep[constrained to be in the GC by HI absorption and RM measurements, chapter \ref{gcl_vlapoln};][]{l89,t86} are two compelling indications that the emission originates in the GC region.

The GBT recombination line has more complex structure than seen previously.  The $l$--$v$ diagrams show that the line velocities tend to be much more complex inside the GCL (from $l=0\ddeg0$ to $-0\ddeg5$).  Although the spectra at the brightest parts of the GCL are adequately-fit with a single Gaussian with $v\sim0$ \kms, the average spectra near $l=-0\ddeg35$ clearly show two components with $v=14$ and --6 \kms.  The two components of the recombination line are brighter than the background and very narrow, showing that they are associated with the low-$T_e$ gas in the GCL.  Even the broad line ($\sim30$ \kms) found inside the GCL is much brighter than the background and and may simply be a kinematically-blended narrow line.  One explanation for the splitting of the line is that the two components trace the front and back sides of the expanding shell of an outflow.  This idea is explored below.

If the velocity implied by the split recombination line is related to the expansion of the GCL, it can be used to estimate its kinetic energy.  In \S\ \ref{derived}, the mass of the ionized gas in the GCL is estimated as $3\times10^5$ \msol, which gives a kinetic energy of KE$=1/2 m v^2 = 3\times10^{50} (m/3\times10^5$ \msol$) (v/10$ \kms$)$ ergs.  This estimate is a lower limit since the expansion of the shell may be much greater perpendicular to the line of sight \citep{b03}.  Also, the molecular gas in the GCL will contribute significantly to the mass of the outflow \citep[$>3\times10^5$ \msol;][]{s96,b03}, although there is no molecular line survey sensitive enough to find gas north of $l\sim0\ddeg3$ associated with the GCL.  As described in chapter \ref{gcl_all}, the minimum kinetic energy of the GCL is similar to that observed in extragalactic outflows from dwarf galaxies, with energies from $10^{50}$ to $10^{54}$ ergs \citep{v05}.  However, the observed velocity is far smaller than the escape velocity for the GC region of about 900 \kms\ for galactocentric radii less than 500 pc \citep{b91}.  


Alternatively, the line split may not be related to expansion and the GCL may be static.  A split line of about 25 \kms\ is observed at the eastern edge of the GCL, coincident with the radio continuum emission.  The outflow model does not predict any line structure at the edge of the expanding shell, so this line split casts doubt on the outflow model.  More observations are needed to show that the line split is found throughout the GCL and is morphologically consistent with the outflow model.

Finally, we note that the strip of GBT pointings across the GCL at $b=0\ddeg45$ shows a velocity gradient with a similar sense as the Galactic rotation (positive on the east, negative on the west).  If there is a trend for the ionized gas in the GCL to rotate like the disk gas, it would be consistent with the idea that the gas comes from the disk and was entrained in the outflow, as is observed in other outflows \citep{s98,wa02}.  From east to west across the GCL, the mean line line velocity changes from +2.5 to --2.5 \kms; this is about a factor of 40 less than the velocity range of gas in the central 100 pc of the Galaxy \citep{b87,s04}.  Under the outflow model for the GCL, this gas likely originated in the plane and was carried away by the expansion of hot gas.  In extragalactic outflows, conservation of angular momentum reduces the velocity gradient from galactic rotation as the outflow expands \citep{s98,s01}.  If angular momentum is conserved in the GCL, then the current angular momentum and rotation curve implies that the gas originated at a radius, $r_o \approx (2.5\ \rm{km\ s}^{-1}) 40\ \rm{pc}/100\ \rm{km\ s}^{-1} = 1\ \rm{pc}$.  However, if the gas came from gas on the inner ``$x2$'' orbits \citep{bi91} or if angular velocity was reduced in any way (e.g., turbulence, resistance by magnetic fields), then the $r_o$ would increase.  More sensitive observations at a range of latitudes are needed to see whether this trend extends over the whole GCL.

%% file: gcl_vla_thesis2_astro-ph.tex
\chapter{VLA Radio Continuum Observations of the Galactic Center Lobe at 6 and 20 cm Wavelengths}
\label{gcl_vla}

\section{Introduction}
Interferometric radio observations are useful in studying the GC region, since they are able to spatially separate its densely-packed sources from each other.  The earliest VLA observations of the GC region demonstrated this by discovering a new type of source, the nonthermal radio filament \citep[NRF;][]{y84}.  Sensitive radio continuum observations also have the potential to serendipitously find pulsars and other unusual objects \citep[e.g.,][]{n04,la05}.  In particular, multiwavelength observations can study the spectral index distribution of sources, which can help determine their nature.

NRFs are long (tens of parsecs), polarized filamentary structures seen in radio continuum emission \citep{y84}.  These filaments have only been seen in the GC region, but have been found there in abundance, with about 15 known and several dozen candidates in the central two degrees of the Galaxy \citep{n04,y04}.  The nonthermal spectra and linearly-polarized radio emission from the NRFs make it clear that they emit synchrotron radiation.  Aside from the need to understand how they are formed, NRFs are interesting to study because they are one of the few direct tracers of the GC magnetic field.  However, translating the observed properties of NRFs into an unambiguous measurement of the GC magnetic field has proven controversial, with several competing models for NRF formation and the implied GC magnetic structure \citep[e.g.,][]{m96,l04,y03}.

There have been several previous studies of the spatial and spectral changes in the spectral index of NRFs \citep{l00,l01,so92,l95,l99}.  Of the few NRFs that have been studied in detail, most are found to have a radio spectral index that does not change with position \citep{l00,l95,l99}.  An exception to this trend is in the Radio Arc, which seems to be connected to the GCL-East and its gradually-changing spectral index \citep[see Ch. \ref{gcl_all};][]{s84}.  However, the Radio Arc is the most luminous NRF and is unusual in other ways, so its physical characteristics may not be expected elsewhere.  A better example of a spatial gradient in an NRF spectral index would be G359.85+0.39, between wavelengths of 20 and 90 cm \citep{l01}.  If this gradient can be observed in other NRFs, it may help understand their origin and tell us something about the intrinsic structure of the GC magnetic field.

Studying point-like radio continuum sources in the GC region can probe a variety of sources.  The high electron density in the GC region disperses pulsed emission from the region \citep{c04}.  This is the main reason that no pulsars have been detected from the central few hundred parsecs of the Galaxy, despite the overabundance of X-ray binaries and relatively high star formation rate there \citep{mu05,f04}.  Sensitive surveys of the spectral index of compact sources can be used to search for candidate pulsars that may guide future searches for pulsations \citep[e.g.,][]{n04,l98b}.  Compact thermal sources toward the GC region can also be useful to constraining the Galactic distribution of star formation.  The scale height of compact \hii\ regions can be measured by studying the distribution of sources with thermal spectral indices \citep{g05b}.

This chapter describes the results of a high-resolution radio continuum observations of the GCL that use data from the VLA and the GBT to create images with uniform sensitivity to 6 and 20 cm emission on all spatial scales.  One goal of the present work is to expand on the catalog of known NRFs and to measure their spectral indices in a systematic fashion.  We also want to study thermal and nonthermal compact sources to search for pulsar candidates and compact \hii\ regions.  The observations and data analysis techniques are described in \S\ \ref{vla_obs}.  In \S\ \ref{vla_res}, the results from the total intensity images at 6 and 20 cm, plus the polarized intensity at 20 cm, are presented, including a detailed discussion of the spectral index distribution of nonthermal radio filaments. (Note:  The 6 cm polarized intensity results are presented in Ch. \ref{gcl_vlapoln}.)  Finally, \S\ \ref{vla_dis} discusses the nature of the compact and extended sources in the survey, including an examination of the spectral indices of nonthermal radio filaments.

\section{Observations and Data Reductions}
\label{vla_obs}
\subsection{VLA Data}
The observations at 6 and 20 cm were designed to have similar sensitivities, resolutions, and \emph{uv} coverage, so that accurate spectral index measurements could be made for the surveyed region.  Table \ref{vla_params} summarizes the parameters of the 6 and 20 cm surveys.  The 6 cm observations were conducted in the compact, hybrid, DnC-array configuration.  The 20 cm observations were conducted in three configurations:  the CnB configuration was observed for two days, the DnC configuration and D configurations for one day each.  The ratios of the sizes of neighboring configurations (e.g., CnB and DnD) are similar to the ratios of the observing frequencies, such that observations have similar spatial sensitivities to extended emission.  The hybrid configuration of the VLA has the northern arm extended further, such that the configuration (and hence, the beam) is more circular at low declinations, such as toward the GC region.  For the 6 and 20 cm observations, the default continuum observing set up was used, with two, 50-MHz bandpasses centered at (4.885, 4.835 GHz) and (1.465, 1.385 GHz), respectively.  Two of the 27 antennas were out of commission during all of the 6 and 20 cm observations.

\begin{deluxetable}{lcc}
\tablecaption{Parameters for VLA Continuum Observations \label{vla_params}}
\tablewidth{0pt}
\tablehead{
\colhead{Parameter} & \colhead{6 cm} & \colhead{20 cm} \\
}
\startdata
Dates & June 2004 & January--August 2004 \\
Frequency & 4.85 GHz & 1.4 GHz \\
Configuration & DnC & CnB, DnC, D \\
Integration time per pointing & $\sim$30 min & $\sim$110 min \\
Typical beam size & 14\arcsec$\times$9\arcsec & 11\arcsec$\times$8\arcsec \\
Survey area & 0.8 sq. degrees & 2 sq. degrees \\
Bandwidth & 2$\times$50 MHz & 2$\times$50 MHz \\
\emph{uv} sensitivity range & 0.1 to 30 k$\lambda$ & 0.4 to 30 k$\lambda$ \\
\enddata
\end{deluxetable}

\begin{figure}[tbp]
\includegraphics[width=\textwidth]{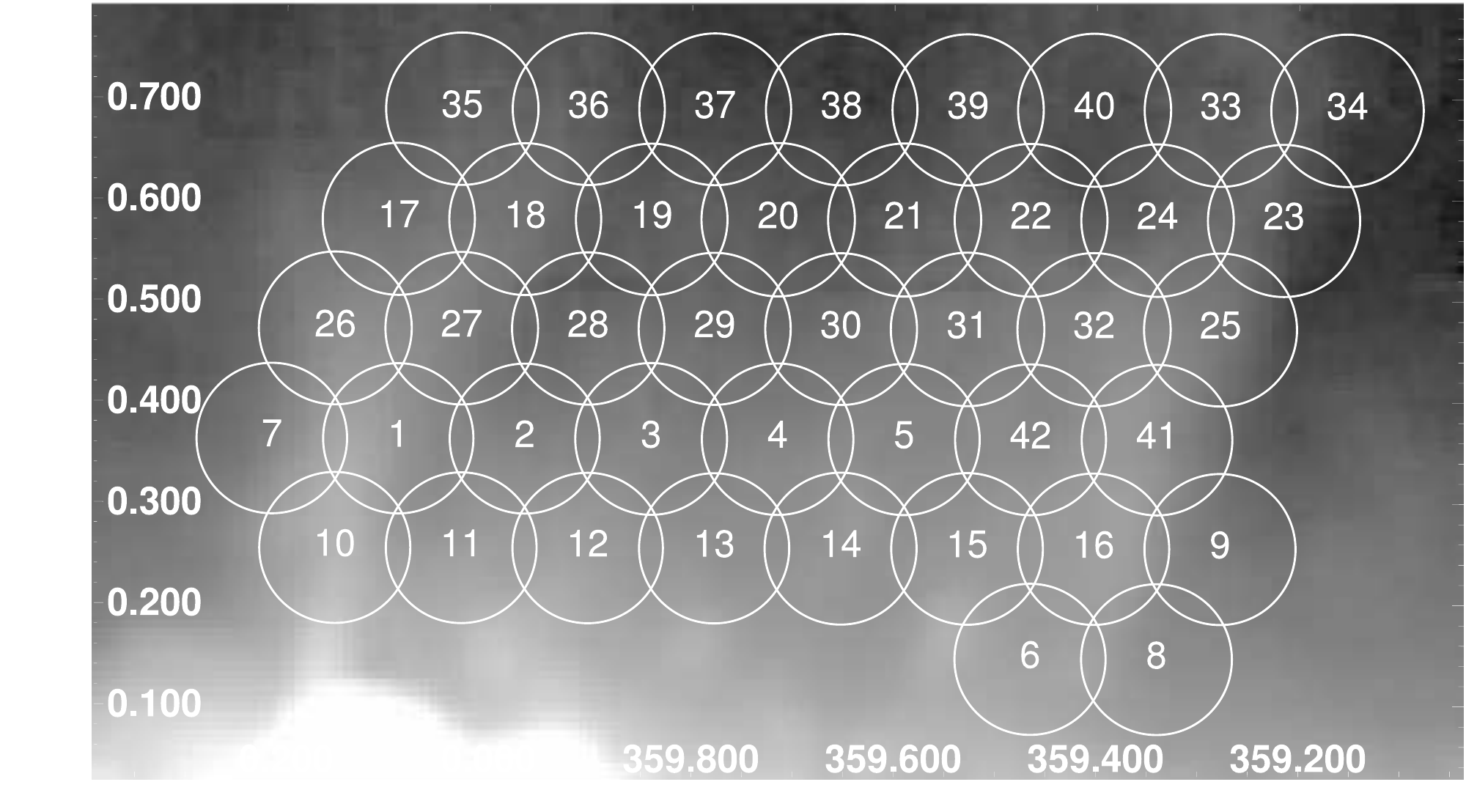}
\caption{The coverage of the 6 cm, DnC-config VLA observations of the GCL are shown in white circles.  The circles have diameters of 9\arcmin, the size of the VLA primary beam at 6 cm.  The gray scale shows a 6 cm, GBT continuum survey of the region, described in detail in Ch. \ref{gcsurvey_gbt}.  Field numbers show the order of observation. \label{coverage6}}
\end{figure}

\begin{figure}[tbp]
\includegraphics[width=\textwidth]{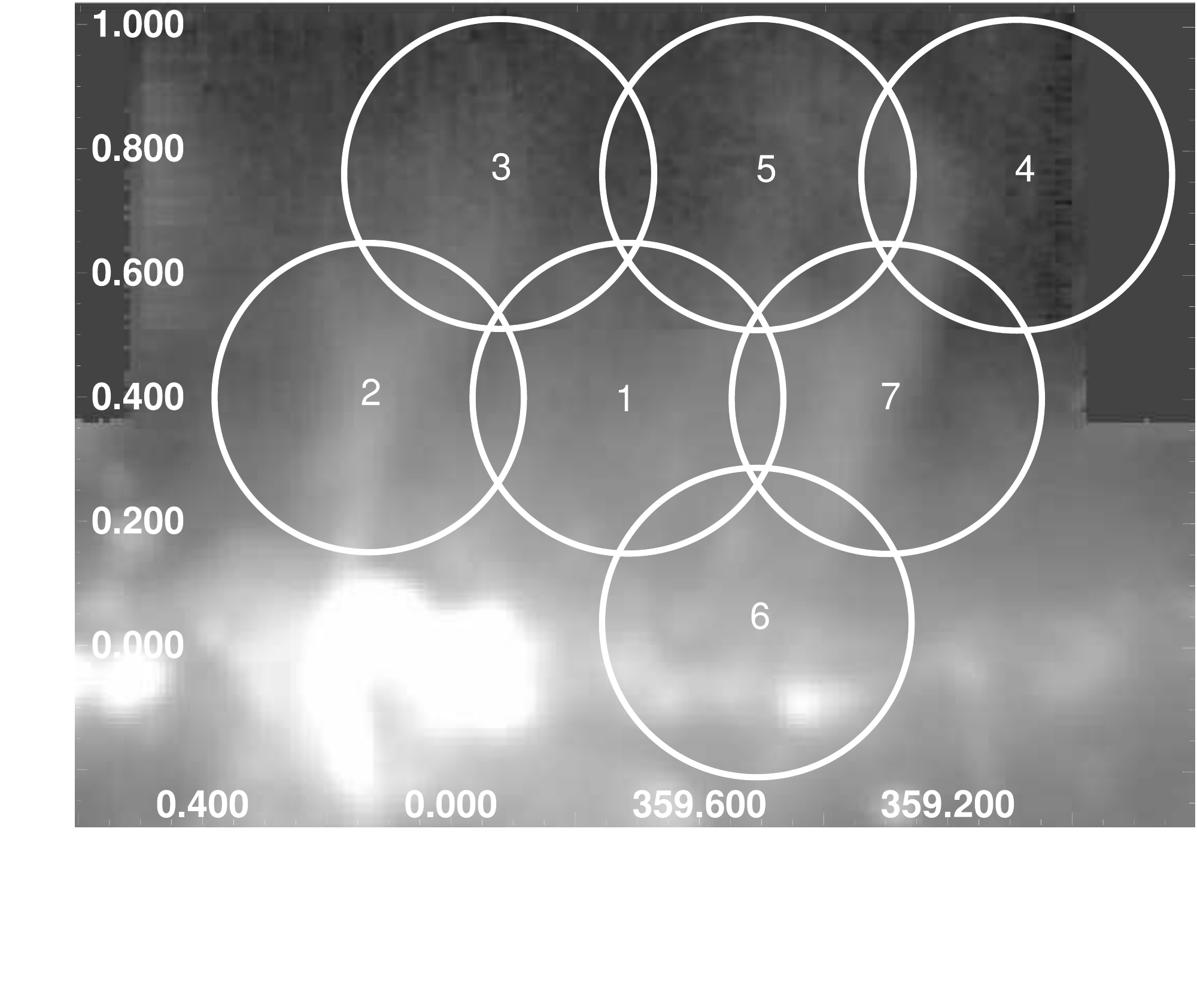}
\caption{The coverage of the 20 cm, CnB-configuration survey of the GCL is shown with circles.  The circle diameter of 30\arcmin\ is equal to the size of the VLA primary beam at 20 cm.  Gray scale shows a 6 cm GBT image of the GCL.  Fields numbers show the order of observation. \label{coverage20}}
\end{figure}

Figure \ref{coverage6} shows the 6 cm survey region, which was chosen to cover most of the GCL.  The survey consisted of 42 pointings arranged in a hexagonal pattern with neighboring pointings separated by 7.5\arcmin.  This pattern is an efficient way to cover an area with a circular beam and makes the sensitivity relatively uniform.  The sensitivity varied by about a factor of two across the inner part of the survey, with values ranging from 0.1--0.2 mJy beam$^{-1}$.  At 20 cm, the survey consisted of seven pointings arranged in a hexagonal pattern spaced by 24\arcmin, as shown in Figure \ref{coverage20}.  The sensitivity at 20 cm is described in more detail below.

Standard flux and phase calibration was done by observing 1331+305 (3C286) and 1751--253, respectively.  Phase calibration observations were done roughly every 24 minutes at 6 cm and every 40 minutes at 20 cm.  The phase calibrator was found to have a flux of 0.46 Jy and 1.13 Jy at 6 and 20 cm, respectively.  To improve \emph{uv} coverage, each pointing was observed in several intervals throughout the 7-hour observing period for a total integration time of roughly 30 and 110 minutes at 6 and 20cm, respectively.  The \emph{uv} data were rotated into galactic coordinates prior to imaging.

At 6 cm, imaging was done using IMAGR in AIPS.  The point source catalog was made from VLA data with \emph{uv} spacings $>0.8$ k$\lambda$ and a resolution of roughly 12\arcsec $\times$ 8\dasec5;  removing short \emph{uv} spacings reduces sensitivity to extended emission and simplifies the detection of compact sources.  VLA images for studying extended structures and for feathering with GBT images have no uv cut.  The data corresponding to Field 32 was self-calibrated using a $\sim$40-mJy source before the final images were made.  The self-calibration process uses a bright source to reduce antenna phase calibration errors.

To study emission from the entire region surveyed, mosaicked images (using FLATN in AIPS) were created from images convolved to have identical beamsizes.  The 6 cm fields were convolved to the minimum beam size shared by all 42 fields of 14\arcsec\ $\times$ 9\arcsec\ with $\theta_{PA}=70$\sdeg\ (East of celestial North; roughly 128\sdeg\ East of Galactic North).  Prior to convolution a typical resolution was 12\arcsec\ $\times$ 8\dasec5.  One $\sigma$ noise values in the mosaic image are typically about 0.1 mJy/beam, but range from 0.08 mJy beam$^{-1}$ (at field center far from Sgr A*) to 0.5 mJy beam$^{-1}$ (at field edge near Sgr A*).

At 20 cm, imaging was done using the VLALB task\footnote{VLALB is a part of the 4MASS software add-on to AIPS written by W. Cotton; \url{http://www.cv.nrao.edu/$\sim$bcotton/}}.  VLALB iteratively images and self-calibrates the data using the imaging results for sources brighter than 40 mJy in successive refinements of the phase calibration.  As was done at 6 cm, two sets of images were created:  one set using all data to be used in the study of extended sources and one optimized for point sources, using \emph{uv} spacings $>0.8$ k$\lambda$.  The imaging for the point source detection used only the 20 cm data in BnC configuration;  this configuration has similar \emph{uv} coverage as the 6 cm, DnC configuration data and is thus sensitive to similar spatial scales.  For most fields, Sgr A was imaged simultaneously with the rest of the field to reduce the affect of its sidelobes.  The stokes Q and U parameters were also imaged in the seven fields with the same \emph{uv} range and robustness and made into a polarized intensity image for studying polarized sources.

The seven fields were convolved to the beam size of the 6 cm survey (14\arcsec\ $\times$9\arcsec) with $\theta_{PA}=60$\sdeg\ (East of celestial North, or $\sim118$\sdeg\ East of Galactic North).  Prior to convolution, a typical resolution was 11\arcsec\ by 8\arcsec.  The images were then merged together (using FLATN) prior to point source detection.  The sensitivity of the images change from field to field, mostly due to confusion introduced by Sgr A* side lobes.  Noise values for fields 1--7 are 0.8, 1.2, 0.6, 0.2, 0.5, 1.7, and 0.3 mJy/beam, respectively.

\subsection{Feathering VLA and GBT data}
The mosaicked VLA images were also combined with GBT surveys of the region using the ``feathering'' technique to study extended emission.  Feathering involves the combination of multiple data sets with sensitivity to different spatial scales to produce a single image with sensitivity to a large range of spatial scales.  Feathering interferometric observations with single-dish observations can correct the ``missing flux'' problem of interferometric images and improve the appearance of the image \citep{b79}.  Ideally, the result is an image that has uniform sensitivity to flux at all spatial scales.  Using all baselines, the VLA data is sensitive to emission on angular scales up to 5\arcmin\ and 15\arcmin\ at 6 and 20 cm, while the GBT has resolutions of 2\damin5 and 9\arcmin\ at 6 and 20 cm, respectively.  For best results, the diameter of the single dish must be greater than 2 times the shortest baseline of the interferometer, so the relative amplitudes of the two datasets can be measured from the data \citep{s02}.  The shortest baseline separation for the VLA C and D configurations is 35 m, comfortably less than half the GBT diameter of 100 m, so feathering these data sets should work well.

The 6 and 20 cm VLA images are feathered with images from the GBT (see Ch. \ref{gcsurvey_gbt}).  The technique combines the images in the \emph{uv}-plane, using a mask to select the spatial frequencies that each image samples best.  In total, the images are Fourier-transformed, multiplied by the mask, combined, and then inverse-Fourier-transformed to create the final, feathered image.  The entire process is performed within AIPS by a series of tasks written by W. Cotton.

\begin{figure}[tbp]
\begin{center}
\includegraphics[width=0.95\textwidth]{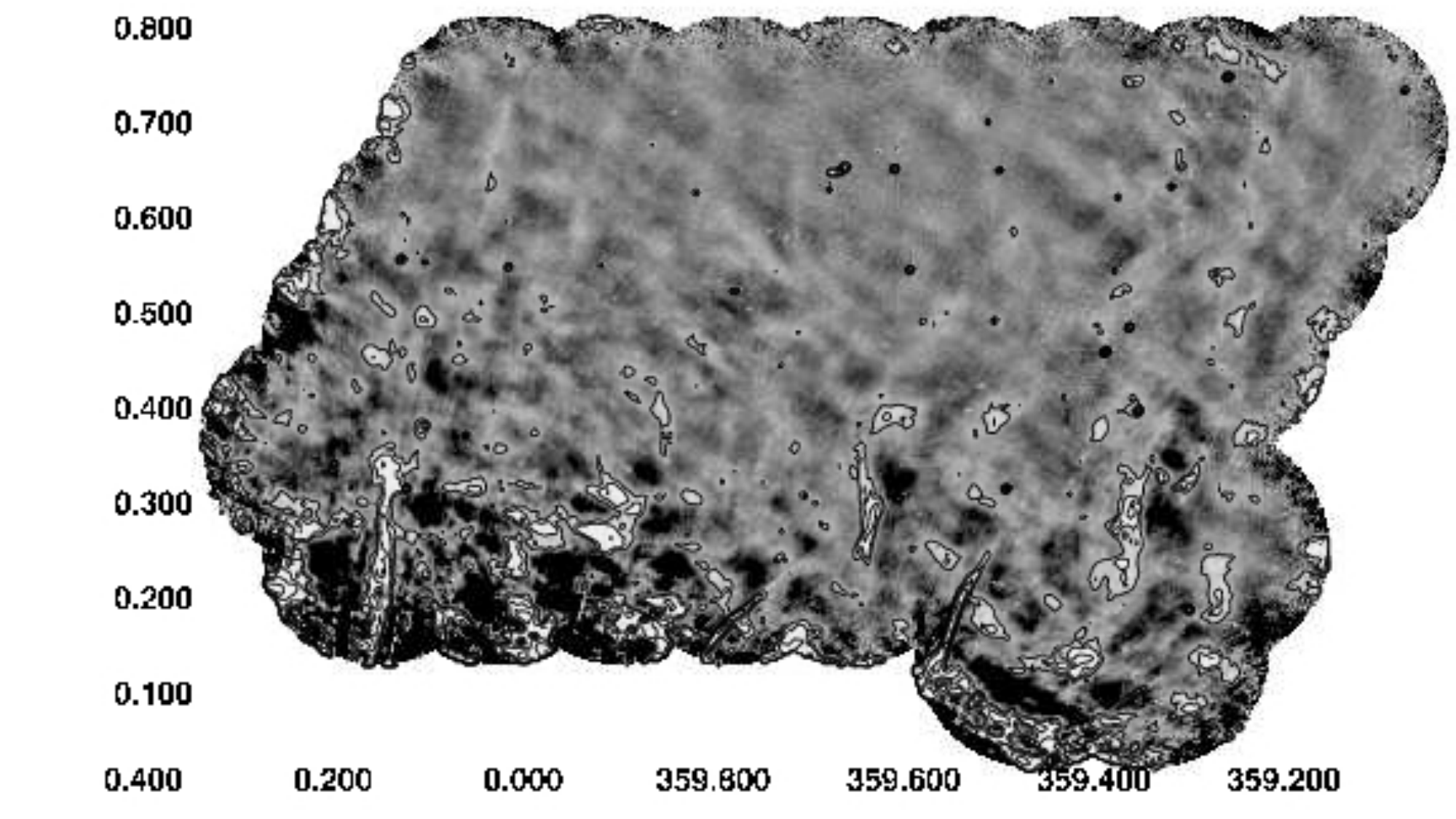}

\includegraphics[width=0.95\textwidth]{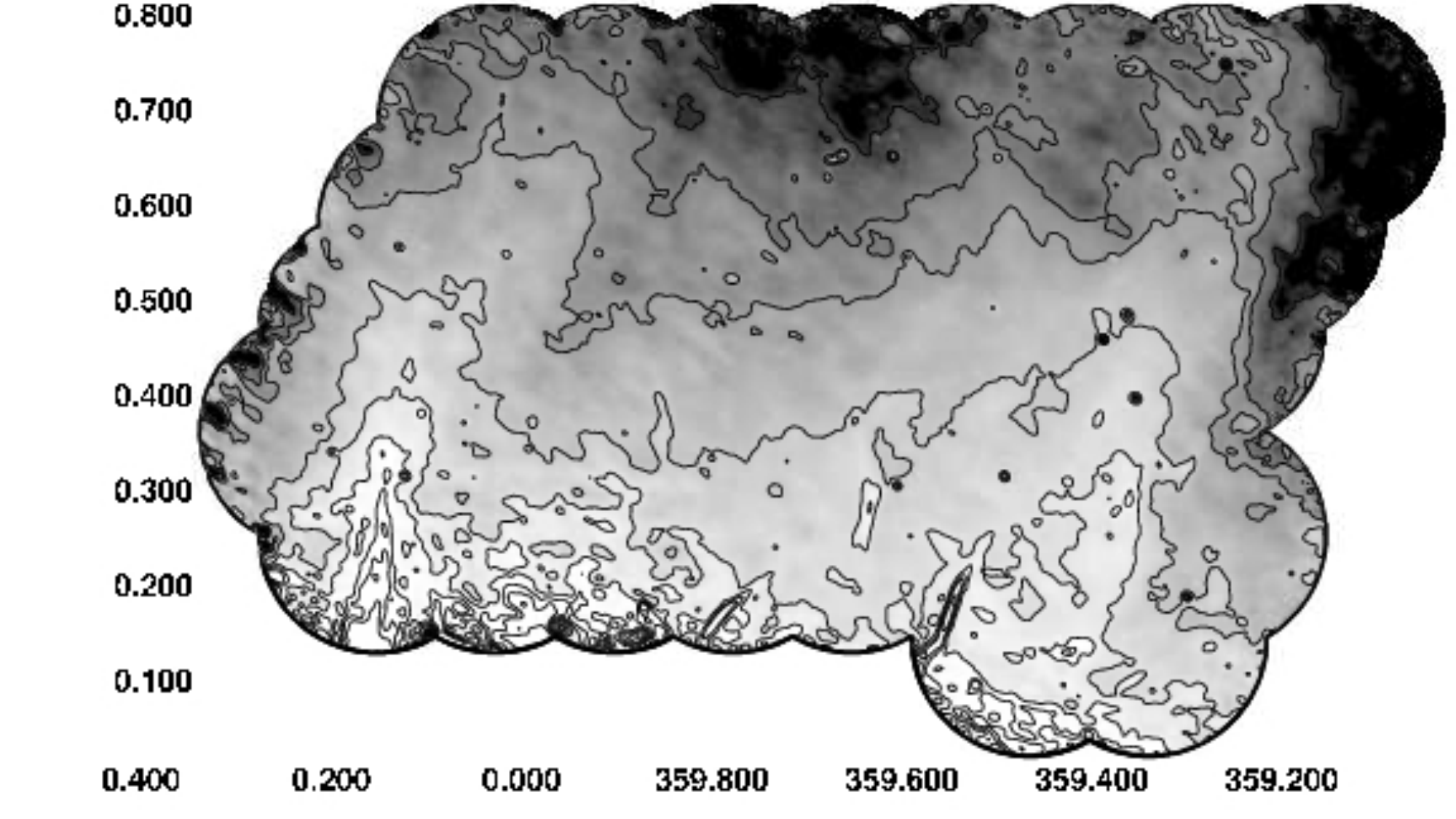}
\end{center}
\caption{\emph{Top}: Mosaic of 42, 6 cm VLA fields in the GCL with Galactic coordinates.  Contours show flux density levels at $0.5 * 2^n$ mJy beam$^{-1}$, with $n=0-7$.  \emph{Bottom}: Same data as above, but feathered with 6 cm GBT observations of the region and convolved to a beam size of 26\arcsec$\times$18\arcsec.  Contours show flux density levels of 1, 3, 5, 10, 15, 20, 25, 30, and 40 mJy beam$^{-1}$.  This image is used in the comparison of 20 and 6 cm fluxes of extended sources with slice analysis. \label{6large}}
\end{figure}

\begin{figure}[tbp]
\begin{center}
\includegraphics[width=0.75\textwidth]{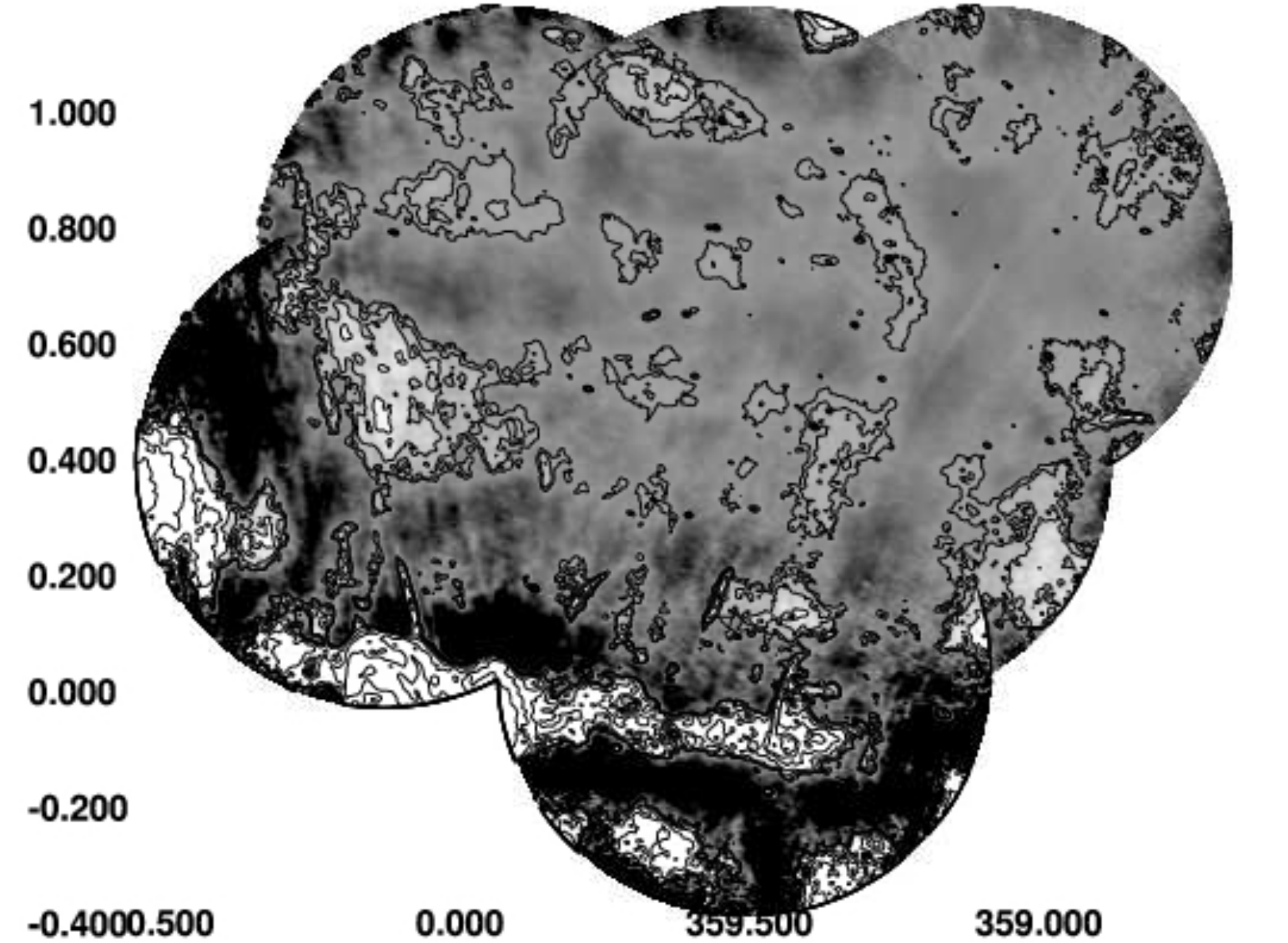}

\includegraphics[width=0.75\textwidth]{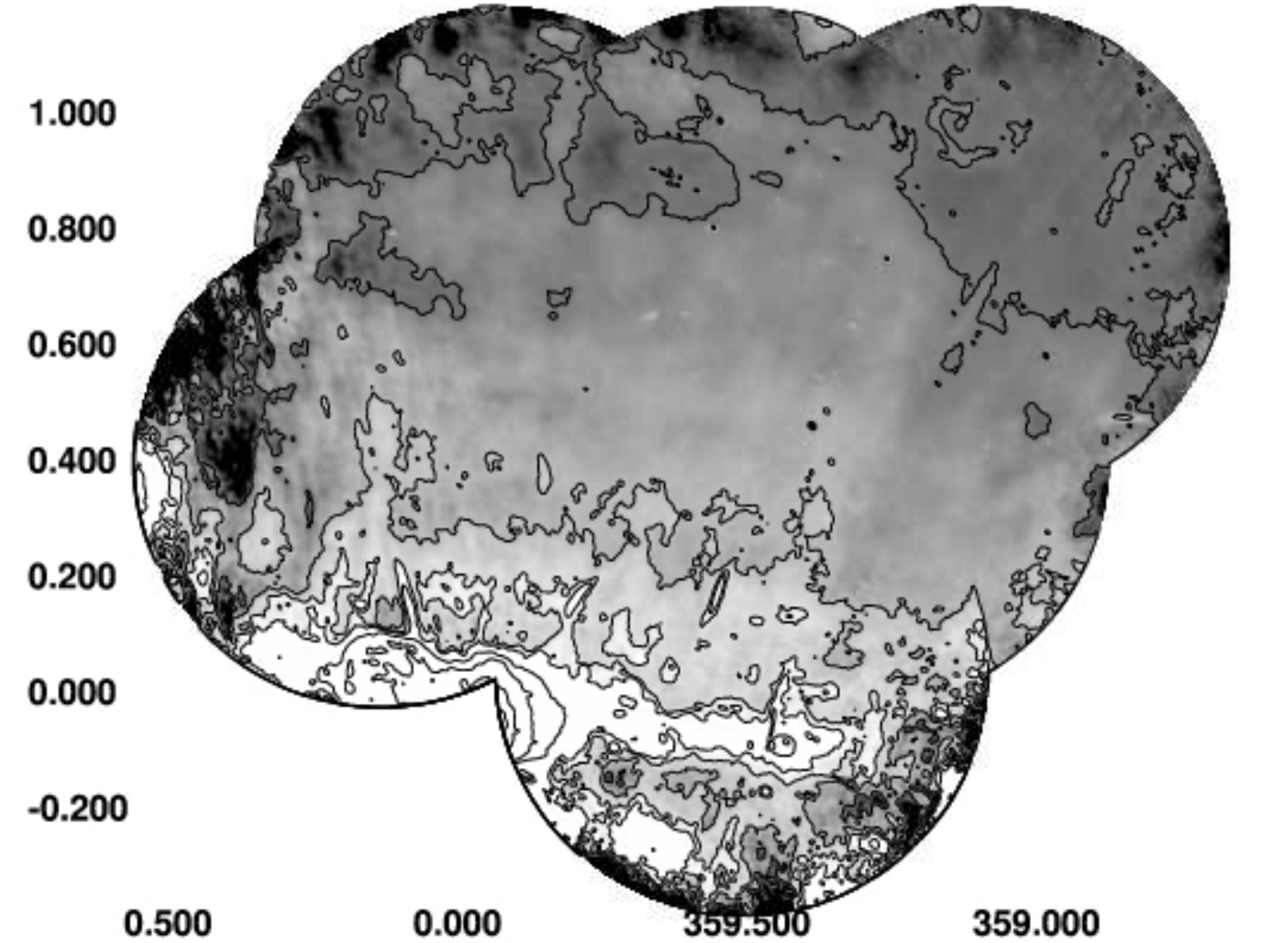}
\end{center}
\caption{\emph{Top}  Mosaic of the seven, 20 cm VLA fields in the GCL with Galactic coordinates.  Contours show flux density levels at $2 * 2^n$ mJy beam$^{-1}$, with $n=0-7$.  \emph{Bottom}: Same data as above, but feathered with 20 cm, GBT observations of the region and convolved to a beam size of 26\arcsec$\times$18\arcsec.  Contours show flux density levels of 25, 50, 75, 100, 200, 400, 800, 1600, and 3200 mJy beam$^{-1}$.  This image is used in the comparison of 20 and 6 cm fluxes of extended sources with slice analysis. \label{20large}}
\end{figure}

Figures \ref{6large} and \ref{20large} show the distribution of 6 and 20 cm continuum emission on large scales with both VLA and VLA/GBT feathered maps.  The biggest difference between VLA and VLA/GBT feathered images is the strength of the continuum emission from the GCL, especially at 6 cm.  The feathered images also fill in the negative ``bowls'' in the interferometric images to produce a more uniform flux distribution near bright sources.  

\subsection{Spectral Index Measurements}
The 6/20 cm spectral index distribution was measured by taking slices across the 6 and 20 cm feathered images convolved to a beam size of 26\arcsec$\times$18\arcsec.  The spectral index is measured by fitting a first-order polynomial to a source-free part of the slice to estimate the background and noise in the slice.  The spectral index and its error are calculated according to the following equations:
\begin{equation}
\alpha = \rm{log}(S_1/S_2)/\rm{log}(\nu_1/\nu_2)
\end{equation}
\begin{equation}
\sigma_\alpha = \sqrt{(\sigma_1/S_1)^2 + (\sigma_2/S_2)^2}/\rm{log}(\nu_1/\nu_2),
\end{equation}

\noindent where $S_i$ are the peak flux densities and $\sigma_i$ are the standard deviations of the background regions about the best-fit first-order polynomial.  These relations assume $S_\nu\propto\nu^{\alpha}$.  Note that the uncertainty in the spectral index values are a slight underestimate, since they assume that the background has only noise-like emission.  It is possible that the best-fit background (and, hence, the spectral index) is biased by emission in the field.  However, the spectral index measurements were considered trustworthy only if they were found not to vary significantly with $\sim10$\% variations in the definition of the background region.

\begin{figure}[tbp]
\includegraphics[width=6.5in]{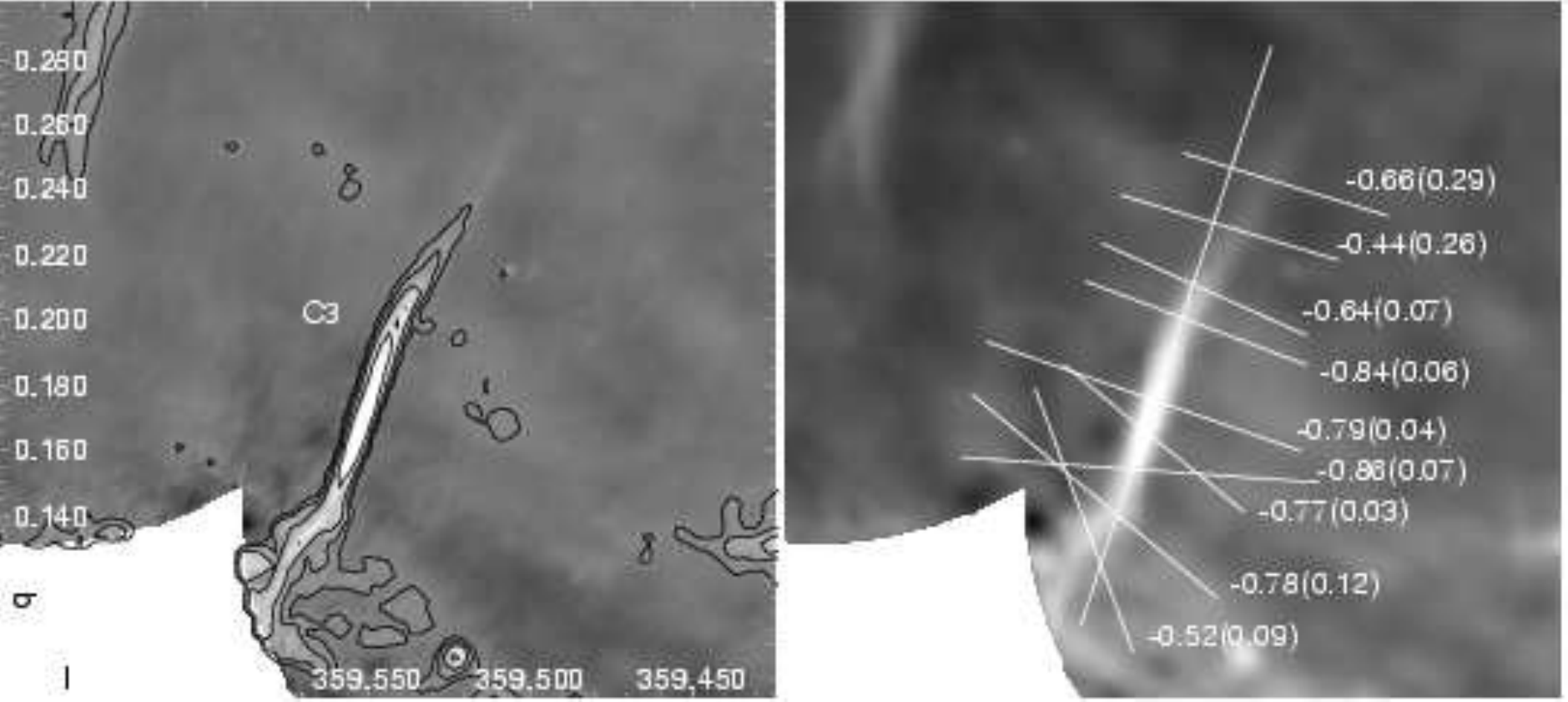}
\includegraphics[width=6.5in]{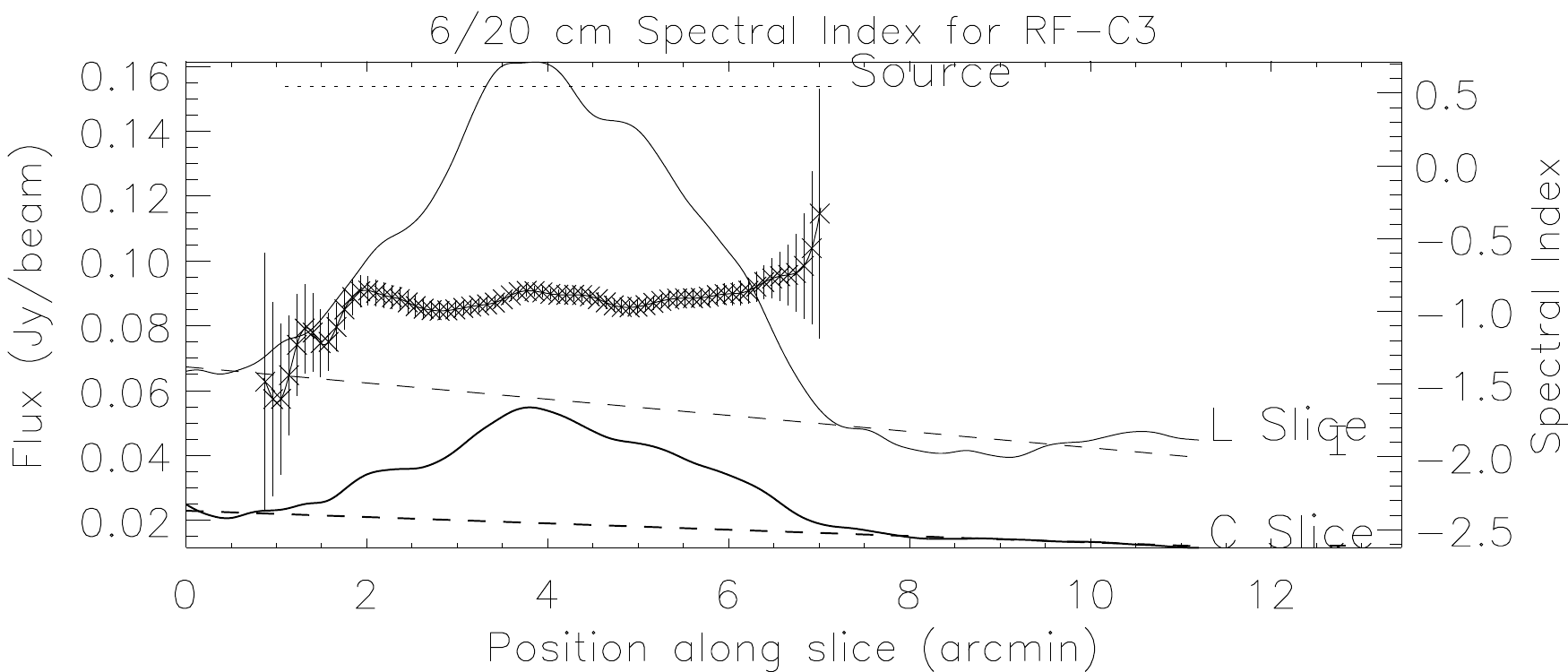}
\caption{\emph{Left}: High-resolution, VLA-only image of the C3 radio filament at 6 cm.  Contours levels are at 1, 2, 4, and 8 mJy beam$^{-1}$.  The high-resolution image has a beamsize of 15\arcsec$\times$10\arcsec.  \emph{Right}:  Feathered, 6 cm image of the region, convolved to the resolution of the 20 cm feathered image of 26\arcsec$\times$18\arcsec.  The lines show the positions of slices across the 6 and 20 cm images for the spectral index study.  The measured value and error of $\alpha_{LC}$ is shown near each slice.  \emph{Bottom}: 6 and 20 cm flux densities with their spectral index as a function of position from the southern edge of the slice along the length of the filament.  The best-fit, first-order polynomial background is shown with a dashed line for both flux slices.  The background is fit ignoring the portion of the slice labeled ``source''.  \label{c3fig}}
\end{figure}

As a demonstration of the spectral index measuring technique, Figure \ref{c3fig} shows the spectral index measured from slices across the G359.54+0.18 (RF-C3).  For each slice shown in the figure, another slice was made with a different orientation;  all paired slices had equal spectral index measurements within their errors, which helps validate the method.  As another test of the accuracy of the slice analysis, slices of the radio filaments discussed in \S\ \ref{nrfsec} were taken from feathered and VLA-only images.  In all but a few cases, the spectral index values measured on the two sets of images were equal within their errors;  the exceptions are discussed in detail in \S\ \ref{nrfsec}.  

Finally, bandwidth smearing causes a radial smearing that reduces peak (per beam) brightness away from the field center, which can affect slice analysis.  Although bandwidth smearing is not commonly thought to affect extended sources, the widths of some NRFs are smaller than the beam in our observations, which means that smearing can affect their peak brightness in some cases.  In theory, bandwidth smearing increases with $\beta=\frac{\Delta\nu}{\nu_0}\frac{\theta_0}{\theta_{\rm{FWHM}}}$, the product of the fractional bandwidth and offset from the phase center in beamwidths \citep{b99}.  This effect is only noticeable in our 20 cm, where the primary beam size and bandwidth are relatively large.  The reduction in peak brightness at 20 cm can be estimated from Figures 18-1 and 18-2 of \citet{b99} and accounting for the relative orientation of the filament to the radial distance from the nearest phase center.  The 6/20 cm spectral index of four point sources near the edge of the nearest 20 cm primary beam were measured with crossed slices, to compare with the results from integrated spectral index measured from point source detection (see \S\ \ref{vla_pssec}).  These sources are closer to the edge of the 20 cm field than the radio filaments C3, C6, C7, C11, C12, and N11 (see \S\ \ref{nrfsec}), so the bandwidth-smearing effect from the point sources is an upper limit to that expected from these filaments.  Three of the point sources have slice spectral indices consistent with the integrated indices.  One point source, G359.872,+0.178, has the most bandwidth smearing of any point source in the survey, with a source smeared in size by a factor of about 1.6.  This source gives an upper limit to the bandwidth-smearing bias in the 6/20 cm spectral index measured from slices of $\Delta\alpha_{LC}\sim+0.4$.  However, this bias is far higher than seen in the other three point sources studied, which have no apparent bandwidth-smearing.  These three compact sources are more representative of the typical source studied by slice analysis;  exceptions to this rule are discussed individually.

\subsection{Point Source Detection}
As described above, the 6 and 20 cm point source detection was performed on the VLA data optimized for sensitivity to compact emission.  This is done by removing short baselines before imaging the data.  The final beam size of the 6 and 20 cm images were 14\arcsec $\times$9\arcsec.

At 6 cm, point sources were detected with the sfind routine of Miriad\footnote{see \url{http://www.atnf.csiro.au/computing/software/miriad/}}.  The ``old'' algorithm was used to find sources with significance greater than $5\sigma$.  A few sources were removed from the catalog by eye if they were heavily confused with extended emission, or if they appeared to be noise amplified by the primary beam correction at the survey edge.  The SAD and JMFIT algorithms of AIPS were used on sources that sfind had difficulty with, such as obvious double sources or confused sources.  Studying the 20 cm point source catalog found by AIPS, we find that the integrated flux density uncertainty follows this relation to within 5\%: $\sigma_{S_i} \approx \sigma_{S_p} * (A_{src}/A_{beam} + 0.75)$.  We use these relations to calculate the 6 cm integrated flux density and its error to ensure accurate comparison between the 20 and 6 cm catalogs.  

At 20 cm, all potential sources were selected by eye and detected with SAD or JMFIT.  The final source list was made by cataloging all apparent sources and keeping those with $S_p>5\sigma$.  Fluxes are not corrected for bandwidth smearing, which reduces the peak brightness and slightly elongates the sources at field edges (mostly at 20 cm).  In the case of two sources, the bandwidth smearing caused artifacts in the mosaicked image, so source detection was done on the individual, primary-beam corrected images.  Three sources could only be detected by fixing the source size at the beam size of 14.0\arcsec\ by 9.0\arcsec, due to confusion or bandwidth smearing.  

\section{Results}
\label{vla_res}
\subsection{Extended Sources}
This section discusses the extended sources visible in the mosaicked images of the GCL at 20 and 6 cm.  Individual objects are described, including detailed imaging and spectral index analysis.  Point sources are described in detail in \S\ \ref{vla_pssec}.

\paragraph{Nonthermal Radio Filaments}
\label{nrfsec}
The 6/20 cm spectral index for the radio filaments presented here is believed to be robust to observational effects.  First of all, the indices were measured on both feathered and VLA-only images and found to be equal within their errors, except in 2 of the 74 slices compared.  Thus, the slice values are 97.3\% reliable to changes caused by the feathering process.  Another possible problem is that sensitivity of the 6 cm mosaic image may change across the field, since each field is $\sim$9\arcmin\ across, comparable to the lengths of some filaments.  Comparing the measurements of $\alpha_{CX}$ to an image of the theoretical noise level for the 6 cm mosaic, no correlation is found between the value of $\alpha_{CX}$ and noise levels, which suggests that the noise and primary-beam correction do not bias the measurements.  Furthermore, there is no correlation between the spectral index uncertainty and the theoretical noise, suggesting that the noise is dominated by confusion.  

\begin{figure}[tbp]
\includegraphics[width=6.5in]{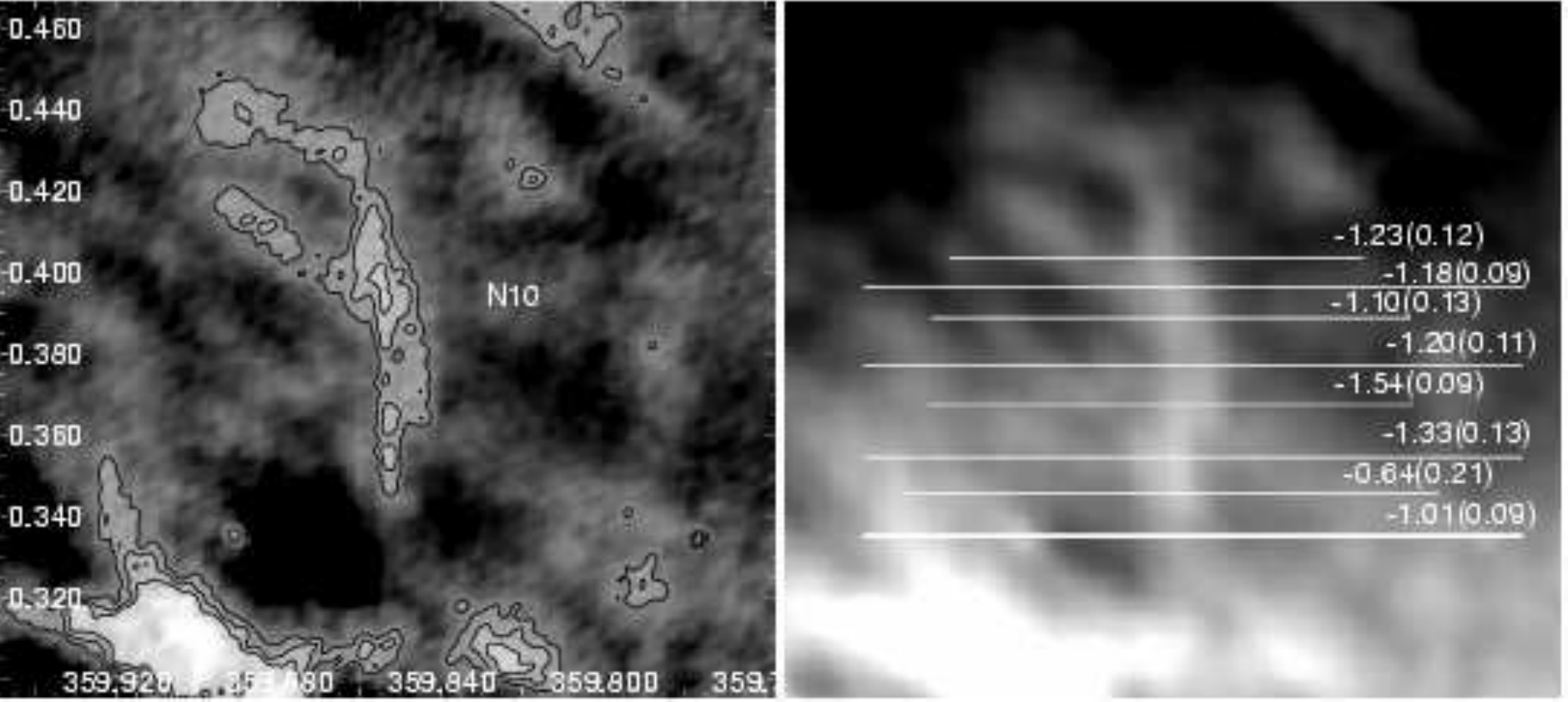}
\includegraphics[width=6.5in]{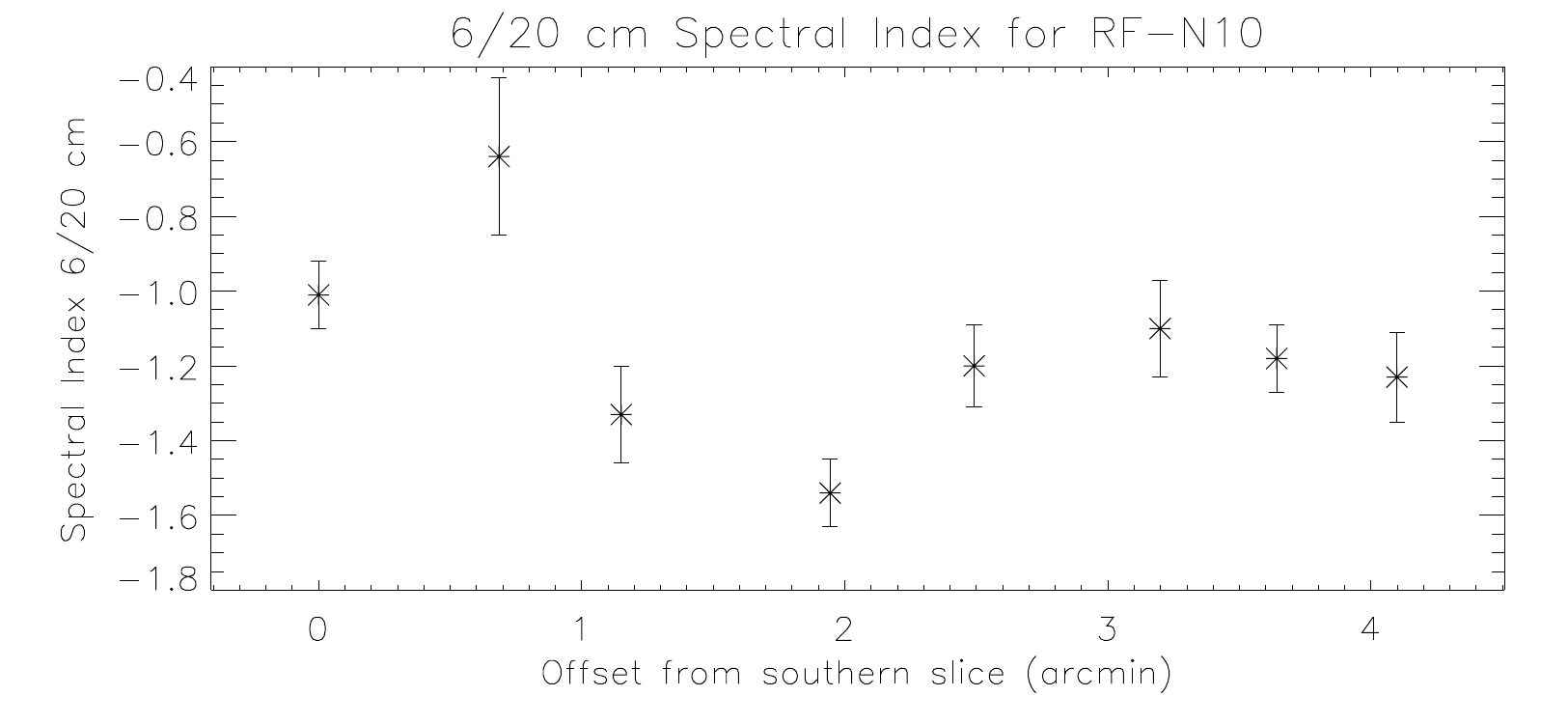}
\caption{This figure shows the high-resolution 6 cm emission and 6/20 cm spectral indices toward the N10 radio filament.  The left image has a resolution of 15\arcsec$\times$8\arcsec\ and contour levels at 0.4, 0.6, and 0.8 mJy beam$^{-1}$.  The right image shows the feathered, 6 cm image of the region, convolved to the resolution of the 20 cm feathered image of 26\arcsec$\times$18\arcsec, as in Fig. \ref{c3fig}.  The plot at bottom shows the 6/20 cm spectral index measured for each slice shown in the right image.  The thick slice in the right image shows the southernmost slice, which is also the origin x-axis in the plot. \label{n10fig}}
\end{figure}

To demonstrate robustness of the spectral index analysis of the filaments, we present the analysis of G359.85+0.39 (RF-N10).  We compare our results to a previous study with the VLA by \citet{l01}.  Figure \ref{n10fig} shows a 6 cm, high-resolution image and the spectral index distribution across the N10 radio filament, which is located relatively far north of the Galactic plane and oriented nearly perpendicular to it.  The N10 filament has subfilamentation at 6 cm, with at least two, parallel, narrow components.  This filament has an integrated 6 cm flux density of about 40 mJy and is polarized at 6 cm.

\begin{figure}[tbp]
\includegraphics[width=6.5in]{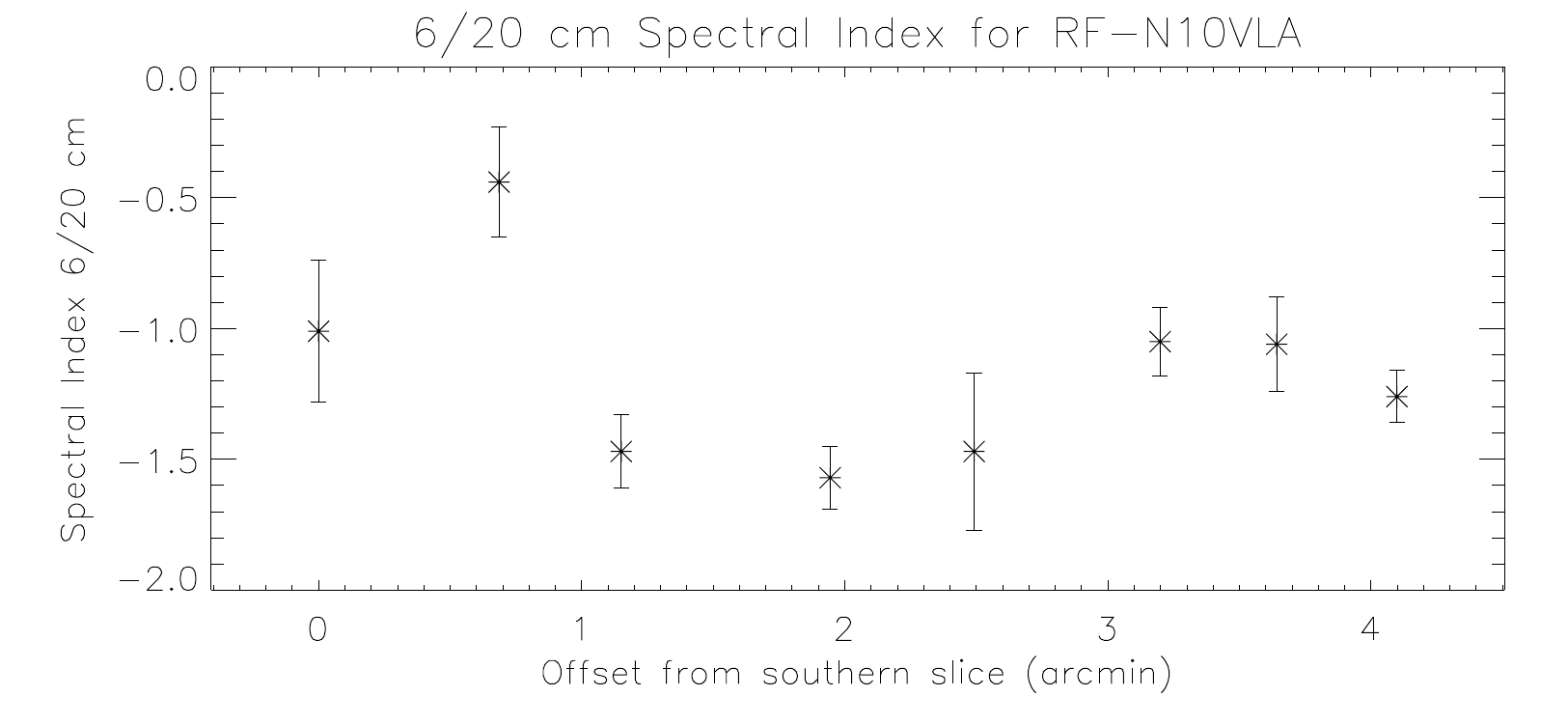}
\caption{Plot of the 6/20 cm spectral index measured for slices across the N10 radio filament.  The plot is similar to that shown in Fig. \ref{n10fig}, except that only the VLA data was used to make the images.  Thus, no zero spacing information from the GBT is included the image used for the slice analysis.  \label{n10tstfig}}
\end{figure}

The 6/20 cm spectral index changes observed in the feathered images of N10 are shown in Figure \ref{n10fig}.  The index shows no significant trend with position along the filament, but does become significantly steeper at its mid-point in both feathered and VLA-only images.  This result is surprising considering that the 20/90 cm spectral index has been observed to change from --0.15 to -1.1 from the south to the north of N10 \citep{l01}.  The 6/20 cm spectral index presented here ranges from --0.6 to --1.5 with no systematic trend.  This is similar to the result of \citet{l01}, which find an index that ranges from --0.9 to --1.3 with no systematic trend.  Since their 6 and 20 cm data did not have matching \emph{uv} coverage and were missing short spacings, those authors suggested that their 6/20 cm spectral index measurements were biased.  However, the 6/20 cm spectral index measurements shown in Figure \ref{n10fig} were taken from data with matching configurations that include the zero spacing information from GBT and should be sensitive to emission on all size scales.  As shown in Figure \ref{n10tstfig}, the distribution of spectral index values measured in the VLA-only data are similar to that measured in the feathered data, which gives confidence in the lack of a trend.  The 6/20 cm spectral index from the integrated fluxes is $\alpha_{LC}=-1.40\pm-0.17$, similar to that measured from the slices.  The internal consistence and the consistency between the present work and \citet{l01} give some confidence in the present spectral index analysis.

\begin{figure}[tbp]
\includegraphics[width=\textwidth]{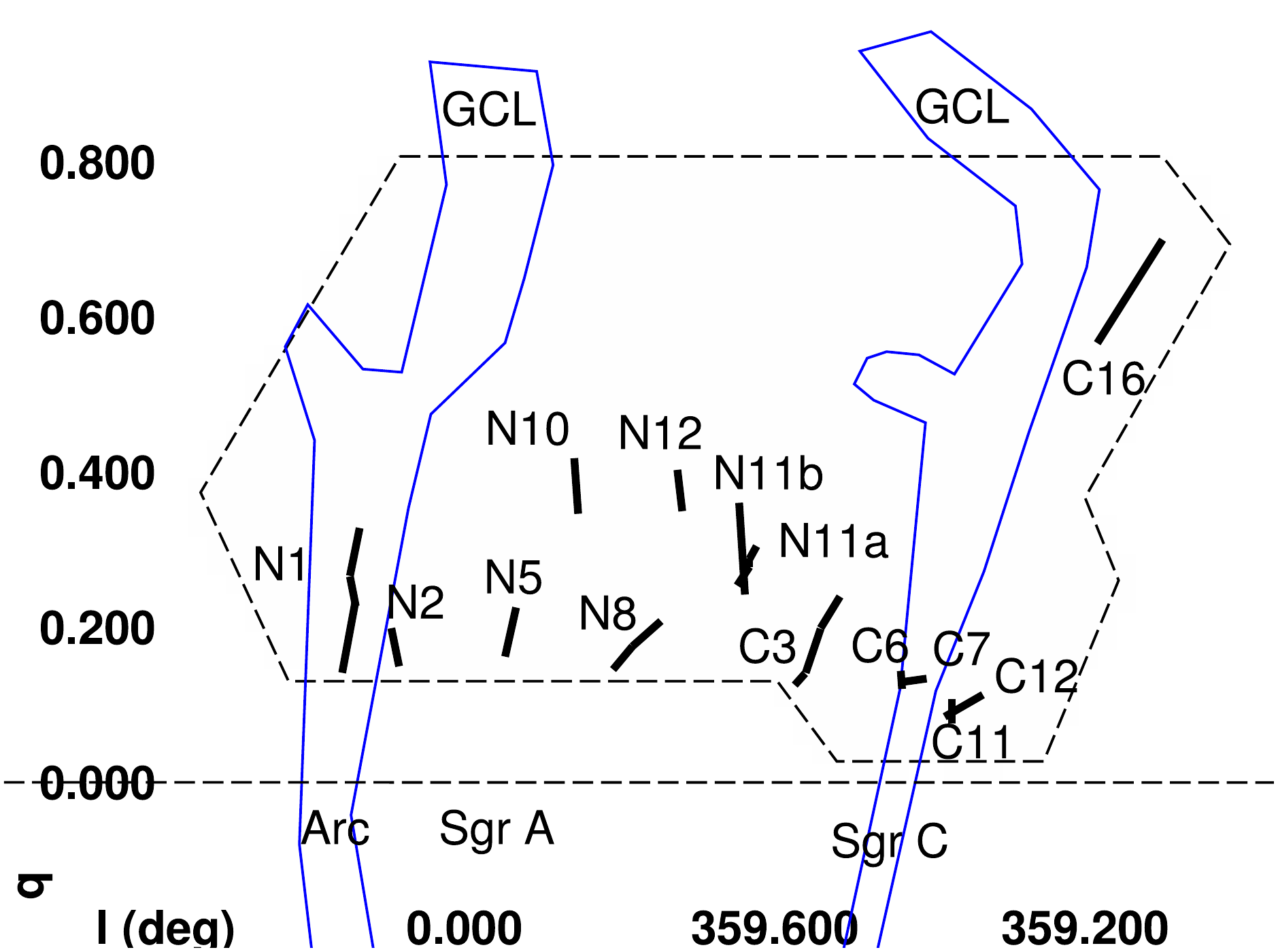}
\caption{Schematic of the NRFs and other major features of the GC region.  Each NRF detected at 6 cm is labeled in the figure.  The outline of the radio continuum emission from the GCL is shown in blue.  The extent of the 6 cm survey region is shown with the enclosed dashed region, while the Galactic plane is shown with a dashed line. \label{vla_nrfschem}}
\end{figure}

Aside from the N10 filament, several other filamentary structures are apparent in our 6 and 20 cm images of the GCL.  Figure \ref{vla_nrfschem} shows a schematic view of the survey region with major features and NRFs labeled.  Table \ref{nrftab} summarizes the properties of the known and candidate NRFs from the 6 cm data, including the 6/20 cm spectral index from the slice analysis presented below and calculated by comparing integrated flux densities with more sensitive 20 cm observations \citep{y04}.  The 6 cm observations have detected polarized emission from several of the filaments, include three for the first time, which confirms their identification as synchrotron-emitting sources.  Below, the analysis of each filament is discussed, including high-resolution imaging with the VLA and spectral index analysis with VLA/GBT feathered images.  The C1, N5, and N12 filaments are not discussed because they are only marginally detected, or are too near the survey edge to measure reliable properties.

\begin{deluxetable}{ccccccccc}
\tablecaption{Radio Filaments Detected at 6 cm \label{nrftab}}
\tablewidth{0pt}
\tabletypesize{\scriptsize}
\tablehead{
\colhead{Identifier} & \colhead{} & \colhead{Length} & \colhead{$S_\nu^p$\tablenotemark{b}} & \colhead{$S_\nu^I$\tablenotemark{b}} & \colhead{} & \colhead{} & \colhead{}\\
\colhead{($l$,$b$)} & \colhead{Name\tablenotemark{a}} & \colhead{(arcmin)} & \colhead{mJy beam$^{-1}$} & \colhead{mJy} & \colhead{$\alpha_{LC}^{int}$\tablenotemark{a}} & \colhead{$\alpha_{LC}^{slice}$} & \colhead{Pol'd?}\\
}
\startdata
G359.45--0.06 & C1 &    &                &                &                 & & N \\
G359.54+0.18  & C3 & 11 &  18.0$\pm$0.5  & 486.4$\pm$61.7 & --0.48$\pm$0.10 & --0.5 to --0.8 & Y \\
G359.44+0.14  & C6 & 2  &   1.5$\pm$0.1  &   6.9$\pm$1.4  & --1.49$\pm$0.17 & & N \\
G359.42+0.13  & C7 & 3  &   7.0$\pm$0.5  &  28.2$\pm$8.3  & --1.13$\pm$0.24 & --0.7 to --1.1 & N \\
G359.37+0.11  & C11& 2  &   1.0$\pm$0.02 &   6.3$\pm$0.2  & --0.17$\pm$0.02 & & N \\
G359.36+0.10  & C12& 4  &   4.0$\pm$0.2  &  24.8$\pm$4.5  &  +0.14$\pm$0.15  & --0.5 to --1.8 & Y\tablenotemark{c} \\
G359.21+0.54  &C16&$\sim10$&0.2$\pm$0.09 &   0.6$\pm$6.5  &                 & --1.2 to --1.8 & N \\
G0.15+0.23    & N1 &    &                &                &                 & +0.2 to --0.5 & Y\tablenotemark{c} \\
G0.08+0.15    & N2 &    &                &                &                 & --0.8 to --1.2 & Y \\
G359.96+0.09  & N5 &    &                &                &                 & & Y \\
G359.79+0.17  & N8 & 9  &   11.4$\pm$0.5 & 226.7$\pm$24.8 & --1.91$\pm$0.09 & --0.9 to --1.3 & Y \\
G359.85+0.39  & N10& 5  &    0.9$\pm$0.1 &  42.9$\pm$9.0  & --1.40$\pm$0.17 & --0.6 to --1.5 & Y \\
G359.62+0.28  & N11a& 4 &    4.0$\pm$0.1 & 104.8$\pm$6.2  & --0.64$\pm$0.05 & --0.1 to --0.3 & Y\tablenotemark{c} \\
G359.64+0.30  & N11b& 8 &    2.5$\pm$0.2 &  62.0$\pm$17.8 &                 & $>-0.1$ & N \\
G359.71+0.40  & N12&$\sim3$& 0.4$\pm$0.06&   2.7$\pm$1.7  & --1.60$\pm$0.51 & & N \\
\enddata
\tablenotetext{a}{Name and 20 cm flux density from \citet{y04}.  Integrated flux density not given for C16 filament.}
\tablenotetext{b}{Peak and integrated flux density measured only for filaments that are completely covered by the 6 cm survey.}
\tablenotetext{c}{First confirmation that this filament is an NRF.}
\end{deluxetable}

$G359.54+0.18 (RF-C3)$ --- The 6 cm VLA image and the 6/20 cm spectral index distribution of the C3 filament are shown in Figure \ref{c3fig}.  The filament is brightest near its midpoint and has a clear bend at its southern end, where it may be interacting with a molecular cloud \citep{b89,r03}.  The 6 cm integrated flux density is about $490\pm60$ mJy, consistent with previous observations \citep{b89}.  The C3 filament is observed to have significant polarized emission at 6 cm (see Ch. \ref{gcl_vlapoln}), as has been noted before \citep{b89}, making it a confirmed NRF.  

The 6/20 cm spectral index is relatively constant across C3, with a typical value of $\alpha_{LC}=-0.8\pm0.05$.  There is some suggestion in the slice analysis of feathered and VLA-only images that the spectral index is flatter in the fainter portions near the northern and southern ends of C3.  The 6/20 cm spectral index is $\alpha_{LC}=-0.48\pm0.10$, based on the integrated fluxes at 6 cm (see Table \ref{nrftab}) and 20 cm \citep{y04}; this is slightly flatter than the spectral index measured from the slices.  The slice spectral index is more trustworthy, since it compares 6 and 20 cm data sets that were designed for comparison.  The integrated spectral index could be in error because the 20 cm data are missing zero-spacing flux or because the flux was integrated over a different region than done here.

\begin{figure}[tbp]
\includegraphics[width=6.5in]{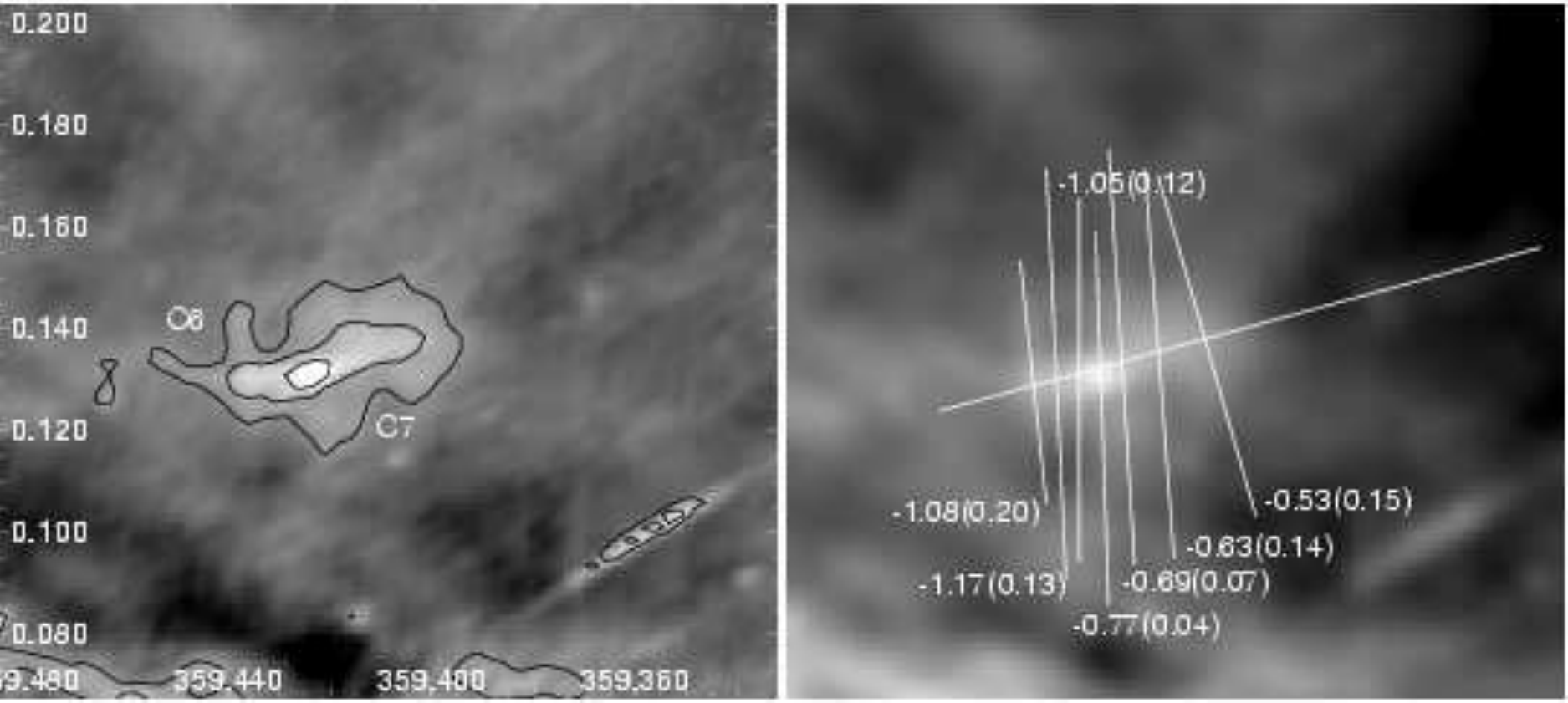}
\includegraphics[width=6.5in]{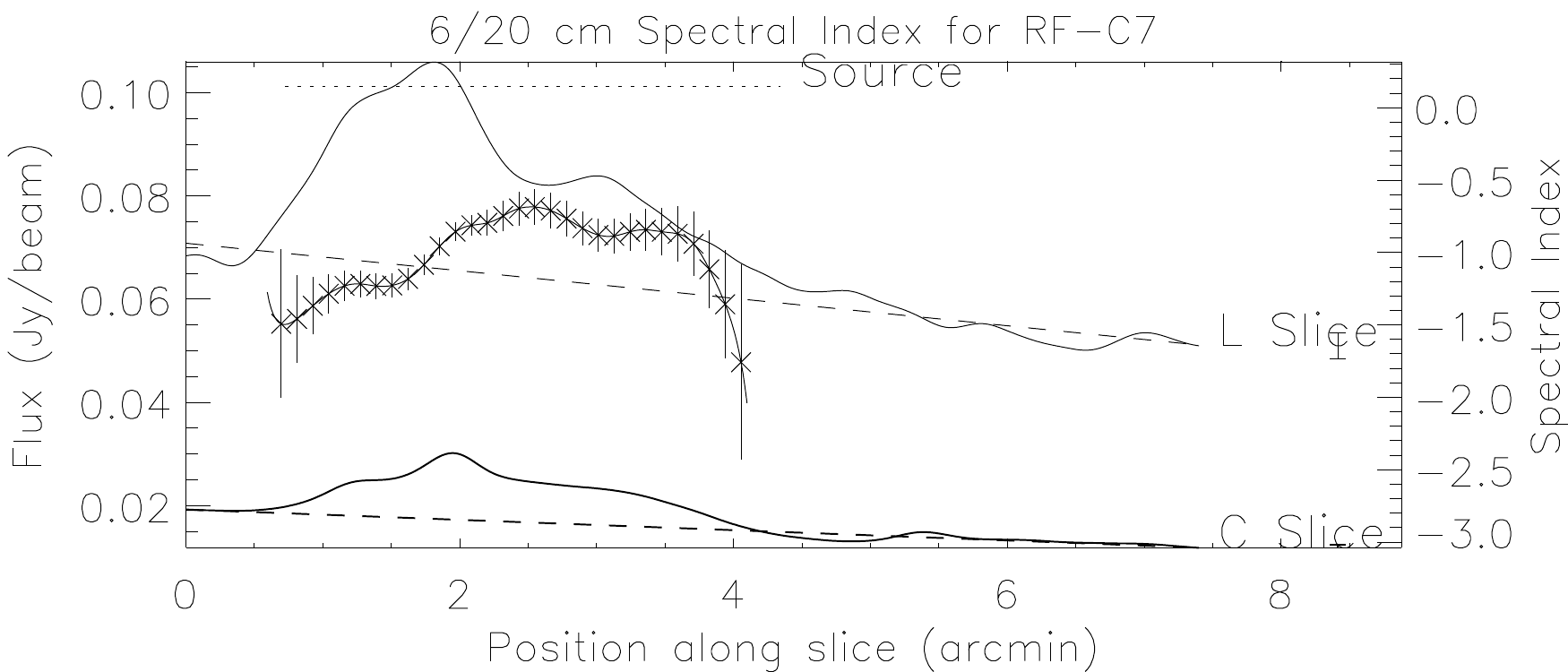}
\caption{This figure shows the 6 cm emission and 6/20 cm spectral indices toward the C6 and C7 radio filaments, as described in Figure \ref{c3fig}. Contour levels are at 1, 2, and 4 mJy beam$^{-1}$.  The plot shows the 6 and 20 cm flux densities with their spectral index as a function of position from the southern edge of the slice along the length of the filament.  \label{c7fig}}
\end{figure}

$G359.42+0.13 (RF-C7)$ --- Figure \ref{c7fig} shows the 6 cm continuum and 6/20 cm spectral index distribution of the C7 radio filament.  The filament is relatively straight and short, oriented parallel to the Galactic plane with a significant brightening at its midpoint.  The integrated 6 cm flux is around 30 mJy and no polarized emission is found toward this filament brighter than a confusion limit of 0.4 mJy beam$^{-1}$.  The spectral index seems to steepen significantly toward the easternmost slices of both the feathered and VLA-only images, with values changing from $\sim-0.7$ to $\sim-1.1$.  The most abrupt change in the spectral index occurs near the brightest portion of the filament.  The integrated 6/20 cm spectral index for C7 is --1.13$\pm$0.24.

$G359.44+0.14 (RF-C6)$ --- This source is detected in 6 cm images (see Fig. \ref{c7fig}) at the eastern edge of the C7 filament, but is not seen in the 20 cm images presented here.  The C6 filament is similar to the C11 filament in that it is oriented perpendicular to the Galactic plane and crosses a horizontal filament that is brighter than its integrated flux density of about 7 mJy.  The integrated 6/20 cm spectral index is $\alpha_{LC}=-1.49\pm0.17$, steeper than that of the C7 filament.

\begin{figure}[tbp]
\includegraphics[width=6.5in]{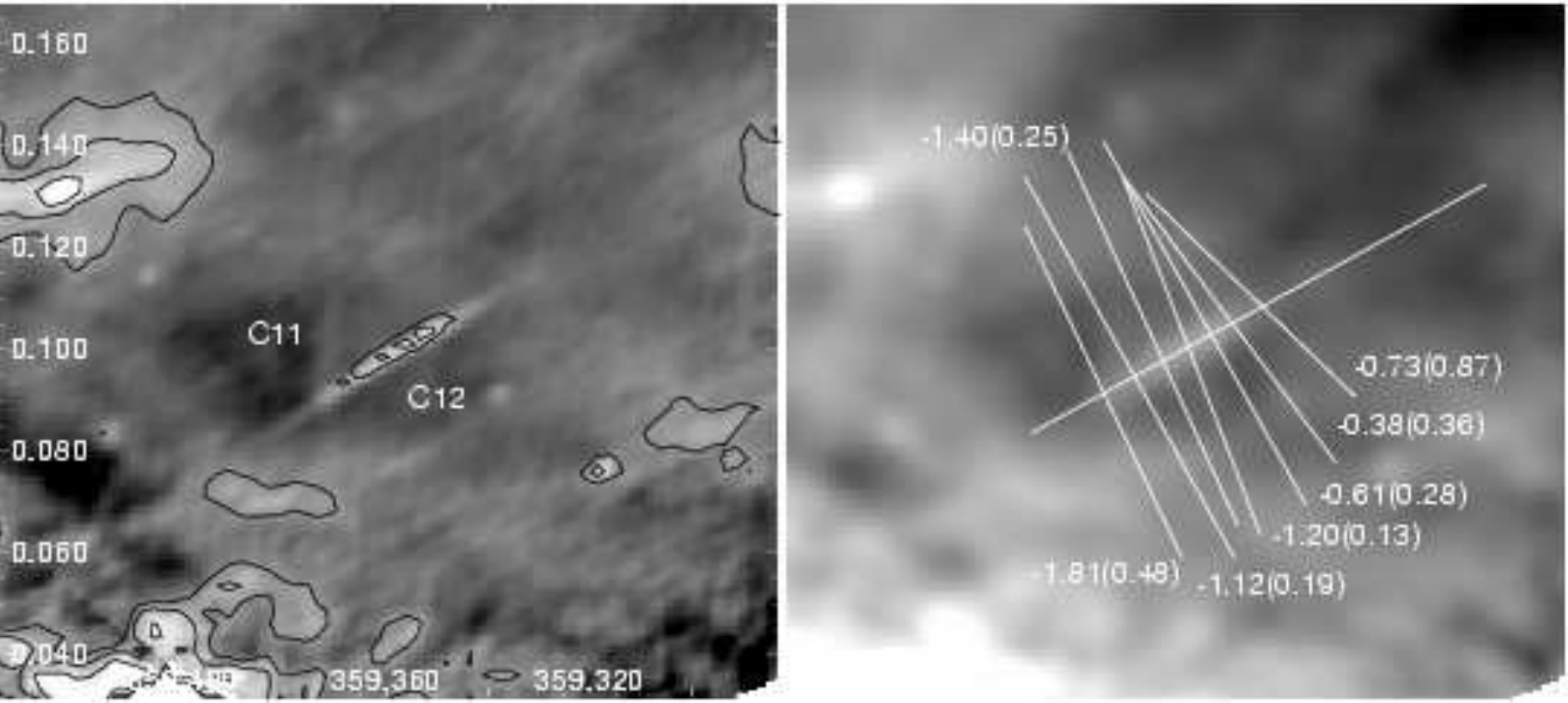}
\includegraphics[width=6.5in]{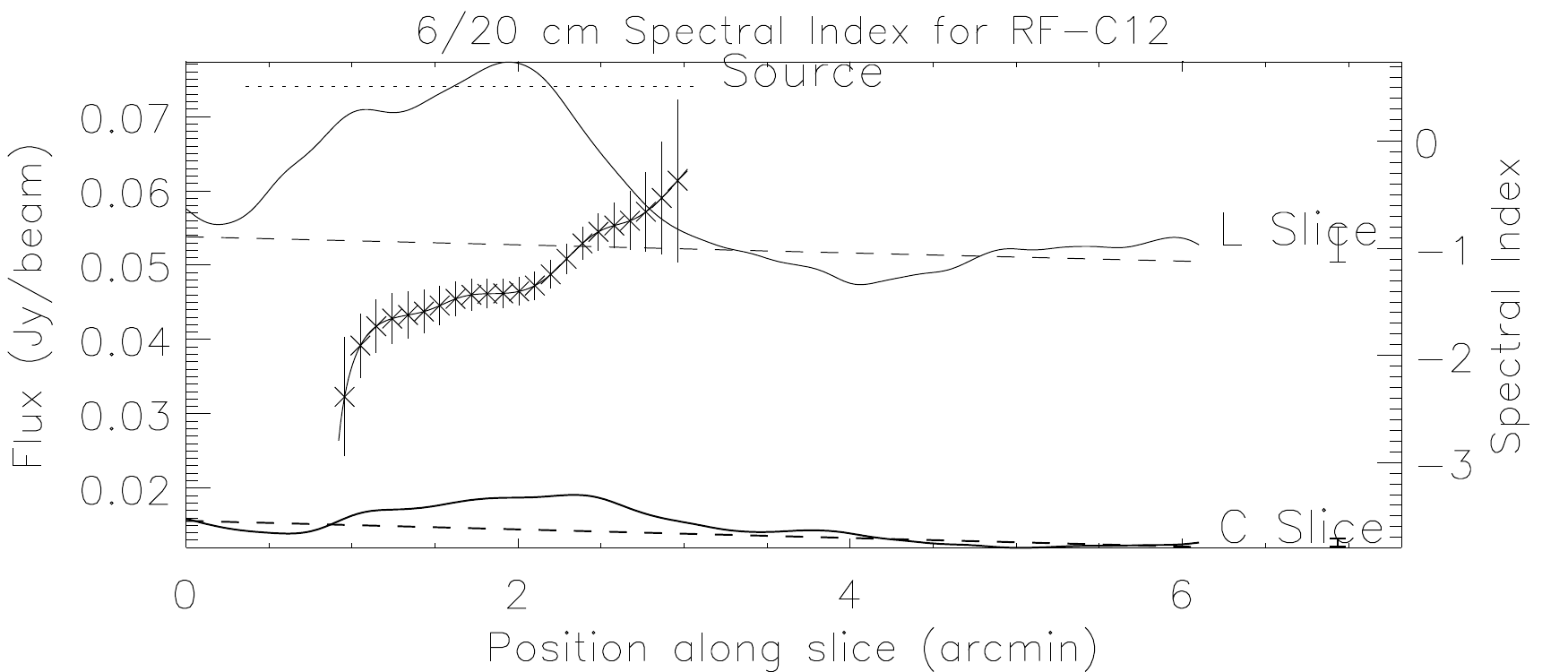}
\caption{This figure shows the 6 cm emission and 6/20 cm spectral indices toward the C11 and C12 radio filaments, as described in Fig. \ref{c3fig}. Contour levels are at 0.5, 1, and 2 mJy beam$^{-1}$.  The plot shows the 6 and 20 cm flux densities with their spectral index as a function of position from the southern edge of the slice along the length of the filament.  \label{c12fig}}
\end{figure}

$G359.36+0.10 (RF-C12)$ --- Figure \ref{c12fig} shows an overview of the radio continuum emission of the C12 filament.  The 6 cm data find polarized emission from this filament for the first time, confirming that it is an NRF;  the polarized emission is discussed further elsewhere (see Ch. \ref{gcl_vlapoln}).  The C12 NRF is almost 4\arcmin\ long with a relatively constant brightness and no significant curvature.  The filament is oriented at an angle of about 30\sdeg\ relative to the plane.  \citet{n04} refer to this filament as G359.36+0.09 and find a very faint bridge of 90 cm emission between the Sgr C filament and G359.36+0.09;  the 20 cm maps presented here do not show this very well, but there is a marginal indication of this bridge emission.  The spectral index for G359.36+0.10 does not change steadily across its length, but does show a significant steepening in $\alpha_{LC}$ toward the east.  Figure \ref{c12fig} also shows that there is significant 20 cm emission near the eastern edge of the filament, although no 6 cm emission is seen.  Although uncertainties in the background make it difficult to estimate the spectral index at the eastern edge of the filament, the physical offset of the 6 and 20 cm-emitting regions is similar to that observed by \citet{l01} for NRF-N10.  The morphology, flux, and spectral index changes across this filament are similar to that seen in the C7 filament.

$G359.37+0.11 (RF-C11)$ --- This source is detected in 6 cm images at the eastern edge of C12, but is not seen in the 20 cm images presented here.  As noted above, this filament and its relation to the C12 filament is similar to the C6/C7 filament complex in their orientation and relative fluxes.

\begin{figure}[tbp]
\includegraphics[width=6.5in]{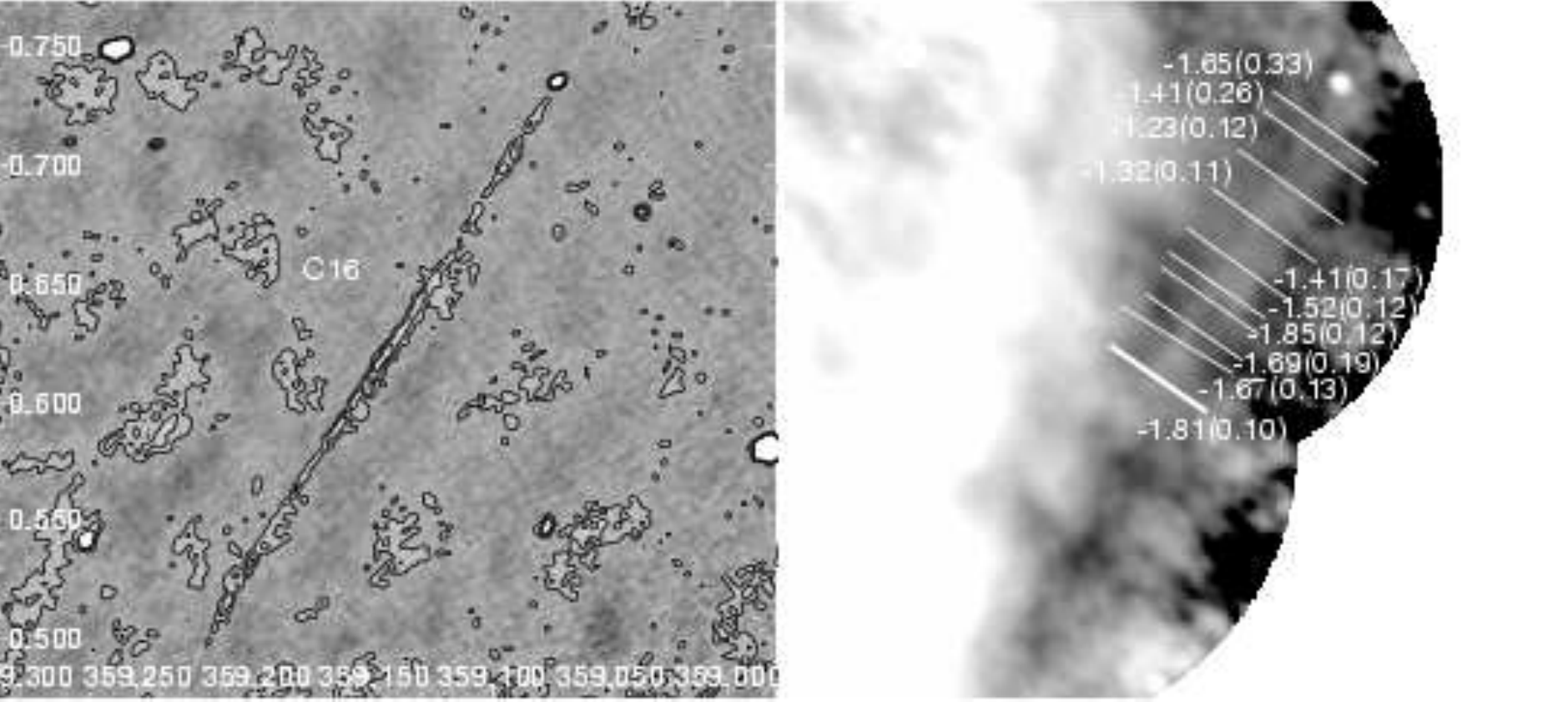}
\includegraphics[width=6.5in]{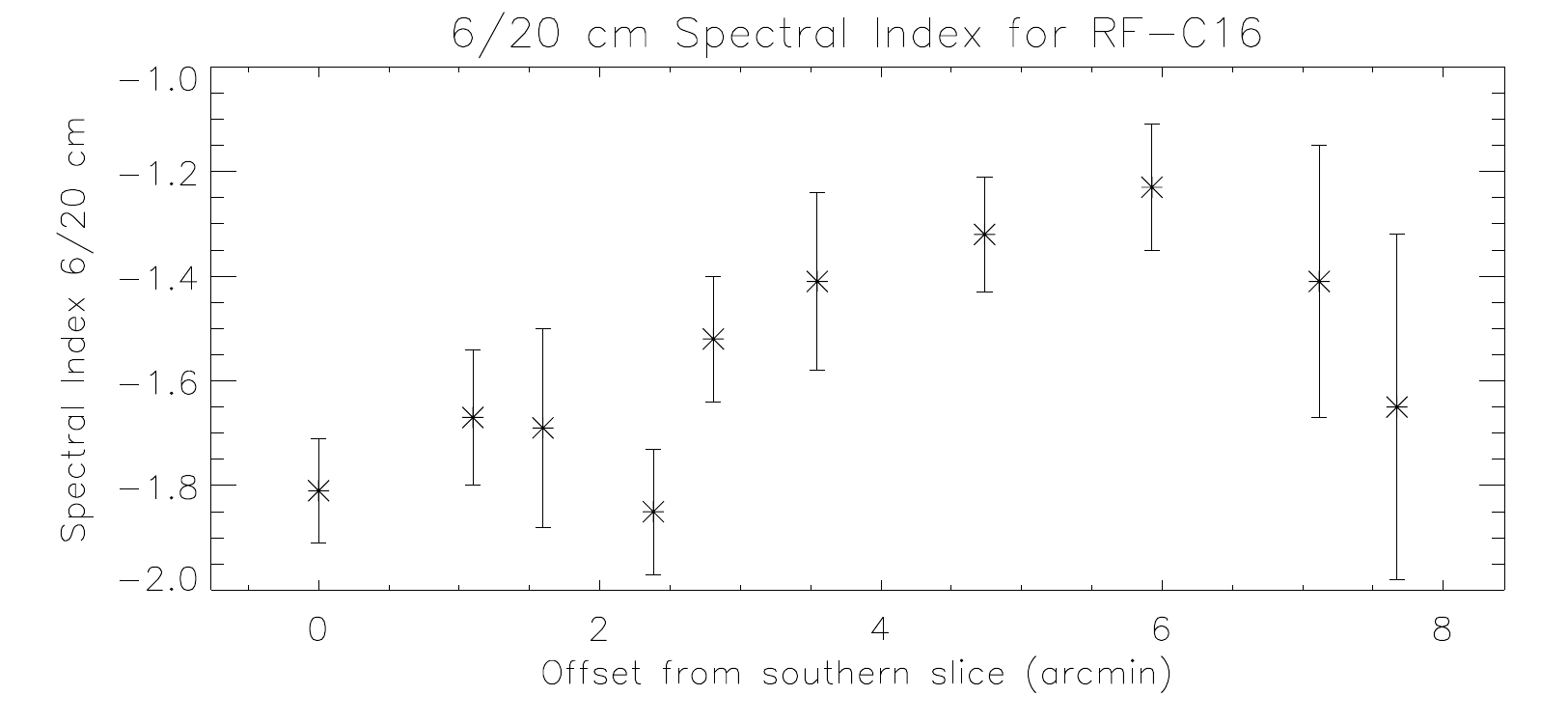}
\caption{The left image shows a high-resolution, 20 cm image and the right image shows the 6 cm, feathered image with 6/20 cm spectral indices overlaid, near the C16 radio filament, as described in Fig. \ref{c3fig}.  The 20 cm contour levels are at 0.3, 0.6, and 0.9 mJy per 14\arcsec$\times$9\arcsec-beam.  The plot shows the 6/20 cm spectral index for the slices shown in the right image.  The thick slice in the right image shows the southernmost slice, which is also the origin x-axis in the plot. \label{c16fig}}
\end{figure}

$G359.21+0.54 (RF-C16)$ --- Figure \ref{c16fig} shows the 20 cm continuum and 6/20 cm spectral index distribution for the C16 filament \citep[detected at 90 cm and called G359.12+0.66 by][]{n04}.  This filament is unusual in that it is very long ($\sim$10\arcmin) and located far from the Galactic plane ($0\ddeg6\approx80$ pc, in projection).  If confirmed as an NRF, this would increase the volume of space occupied by NRFs dramatically, out to (l,b)=(357\ddeg1,0\ddeg75).  The C16 filament is also interesting because it is oriented nearly parallel to the GCL in this region, as seen on large scales in Figure \ref{20large}.  There is some indication of a second filament, or an extended region of emission, parallel and just west of the brightest part of C16.  No polarized emission is evident at 6 or 20 cm to a level of 0.4 and 0.3 mJy beam$^{-1}$, respectively.

The spectral index shown in Figure \ref{c16fig} changes steadily across the entire length of C16, ranging from $\alpha_{LC}\approx-1.2$ to $-1.8$.  The large negative values of the spectral index suffer from uncertainty not shown in the figure, since it is difficult to separate such a faint source from background noise in the slice analysis.  The spectral index is steepest next to the southern end and flattens toward the north, with a suggestion of steepening again at the northern end.  Although the trend is seen in both feathered and VLA-only data, it is more irregular in the analysis of VLA-only images.

There is a compact source detected at 6 and 20 cm that happens to lie along the C16 filament, with RA,Dec = 17:40:31.410,--29:20:19.11 (J2000).  The compact source has $\alpha_{LC}=-0.07\pm0.18$, consistent with thermal emission.  The density of compact sources near the C16 filament is about 250 per square degree (ten sources in a 0\ddeg2$\times$0\ddeg2 region); the area covered by the filament is about 9.8$\times10^{-4}$ square degrees.  The ratio of these two areas is a rough estimate of the chance of alignment between the filament and a random compact source, which is 3.9$\times10^{-6}$ or 0.0004\%.  Note that the spectral index changes seem to have no correlation with the position of the compact source, which is just north of the northernmost slice.

\begin{figure}[tbp]
\includegraphics[width=6.5in]{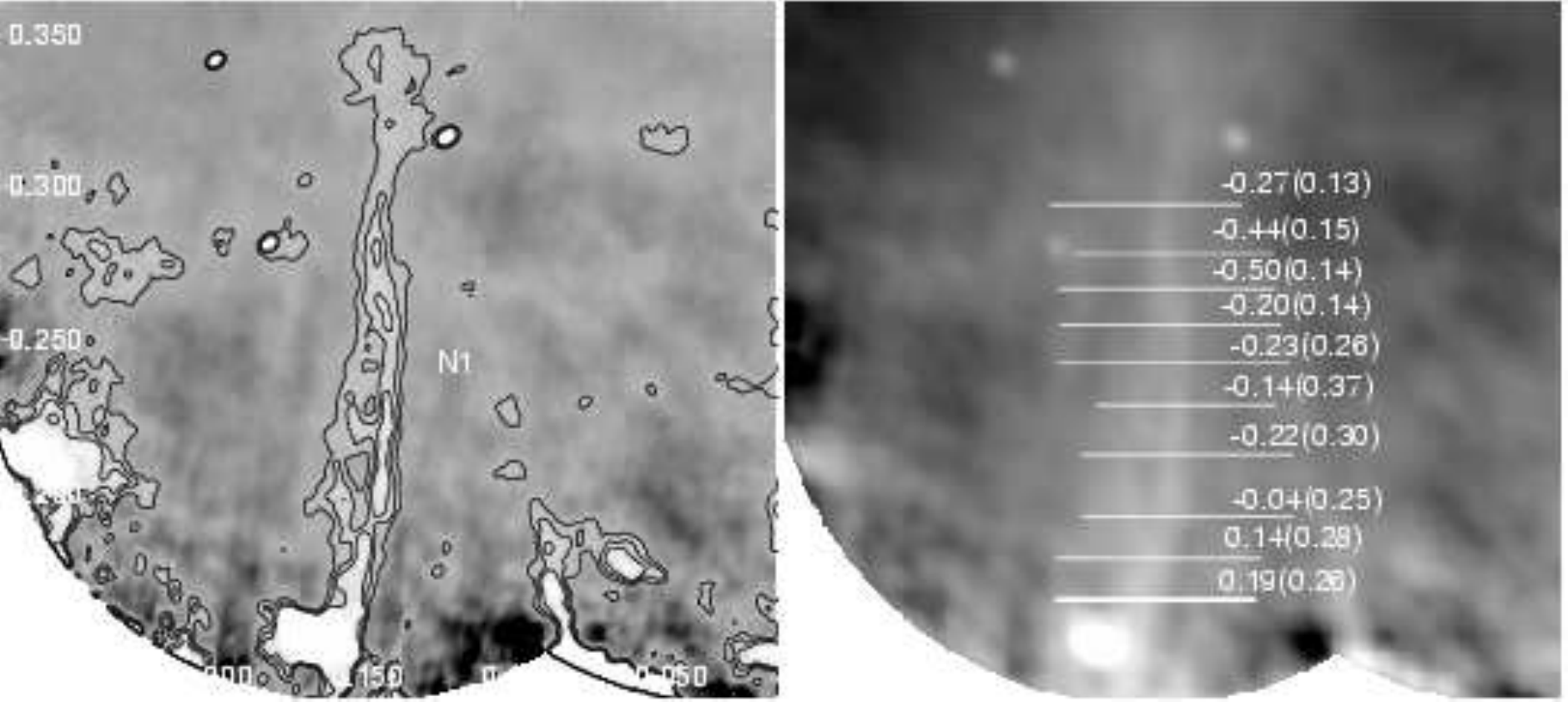}
\includegraphics[width=6.5in]{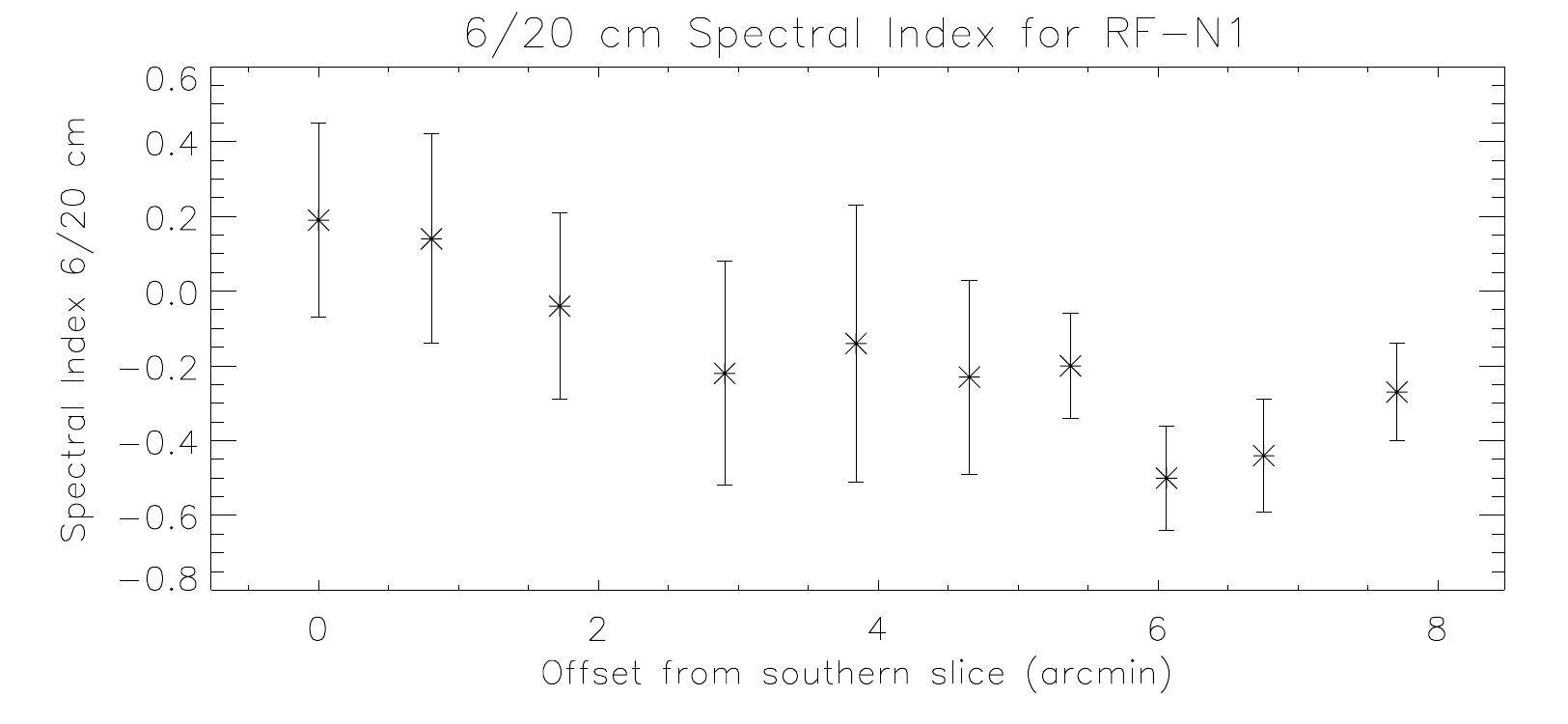}
\caption{This figure shows the high-resolution 6 cm emission and 6/20 cm spectral indices toward the N1 radio filament.  The left image has a resolution of 15\arcsec$\times$8\arcsec\ and contour levels at 1, 2, and 3 mJy beam$^{-1}$.  The right image shows the feathered, 6 cm image of the region, convolved to the resolution of the 20 cm feathered image of 26\arcsec$\times$18\arcsec, as in Fig. \ref{c3fig}.  The plot at bottom shows the 6/20 cm spectral index measured for each slice shown in the right image.  The thick slice in the right image shows the southernmost slice, which is also the origin x-axis in the plot.  \label{n1fig}}
\end{figure}

$G0.15+0.23 (RF-N1)$ --- The N1 filament is shown in Figure \ref{n1fig}.  It is embedded in the emission from the Radio Arc, the brightest and most complex collection of NRFs \citep{y84,m96}.  The N1 filament is north of the brightest part of the Radio Arc, but is contiguously connected to that region and the GCL in images of high-resolution polarized emission \citep{y88}.  The shape of the N1 filament is interesting because it seems to bend around the \hii\ region, G0.17+0.15 (located at the bottom of the images in Figure \ref{n1fig}).  This is another suggestion that these nonthermal structures interact with star-forming regions or their associated molecular and ionized gas \citep{s94,y04}.  The 6 cm data show for the first time that there is significant polarized emission along the N1 filament, confirming that it is an NRF (polarized emission fills this region, but has never been definitively associated with N1).

The N1 NRF spectral index between 6 and 20 cm is studied by taking slices across the feathered images, as described before.  Note that N1 is embedded in diffuse nonthermal emission, which the slice analysis subtracts in order to calculate $\alpha_{LC}$.  Figure \ref{n1fig} shows the spectral index steepening steadily toward the northernmost slices, with values ranging from +0.2 to --0.5.  Studying the VLA-only images shows that the northernmost slice is somewhat flatter than in the feathered images, but the general trend for steepening is the same in both sets of images.  The relatively flat spectral index is unusual for NRFs in general, but not so for the brightest NRFs, like in the Radio Arc, which has a spectral index $\sim0$ for GHz frequencies \citep{y86,p92}.  Interestingly, the spectral index does show moderate steepening toward the north;  this trend is seen in the large-scale emission of the GCL, as observed by the GBT (see Ch. \ref{gcl_all}).

\begin{figure}[tbp]
\includegraphics[width=6.5in]{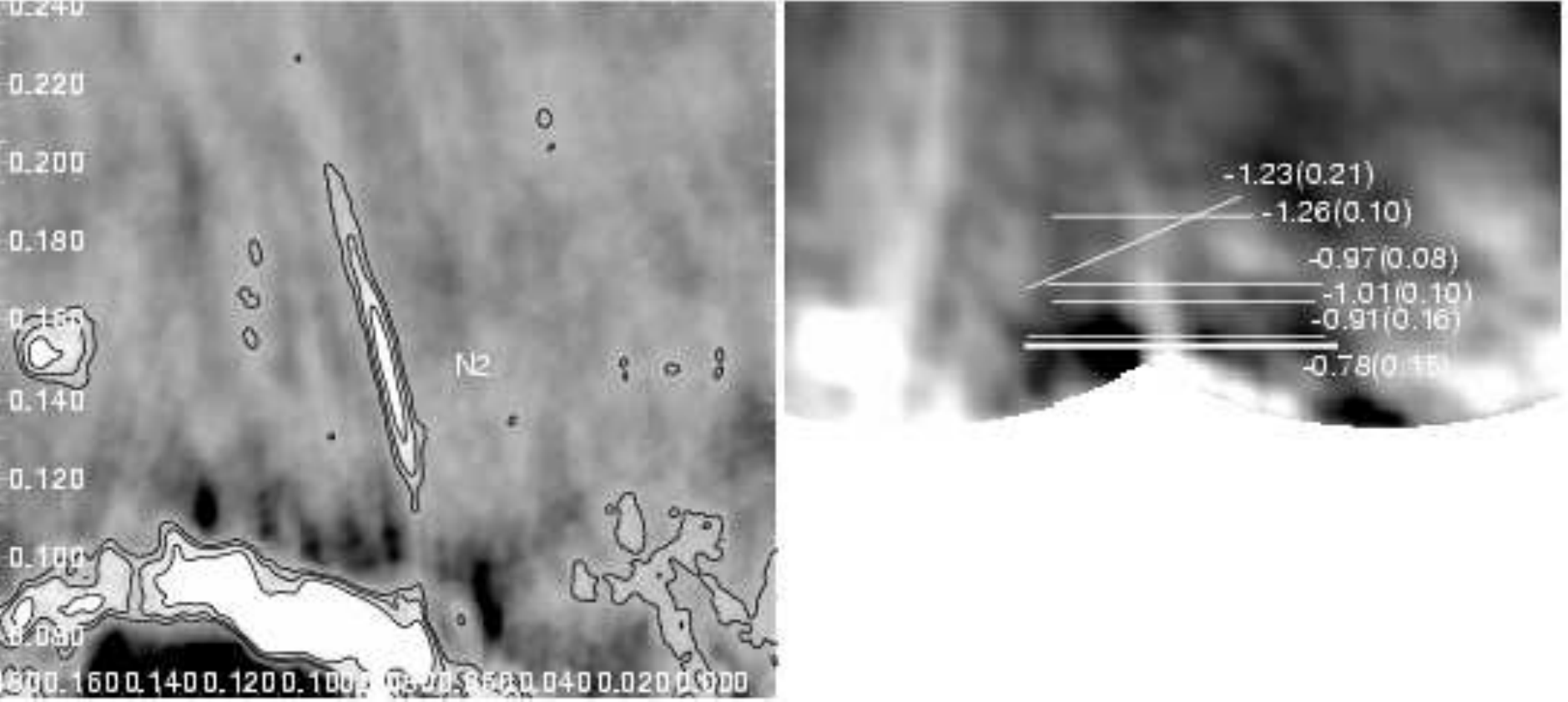}
\includegraphics[width=6.5in]{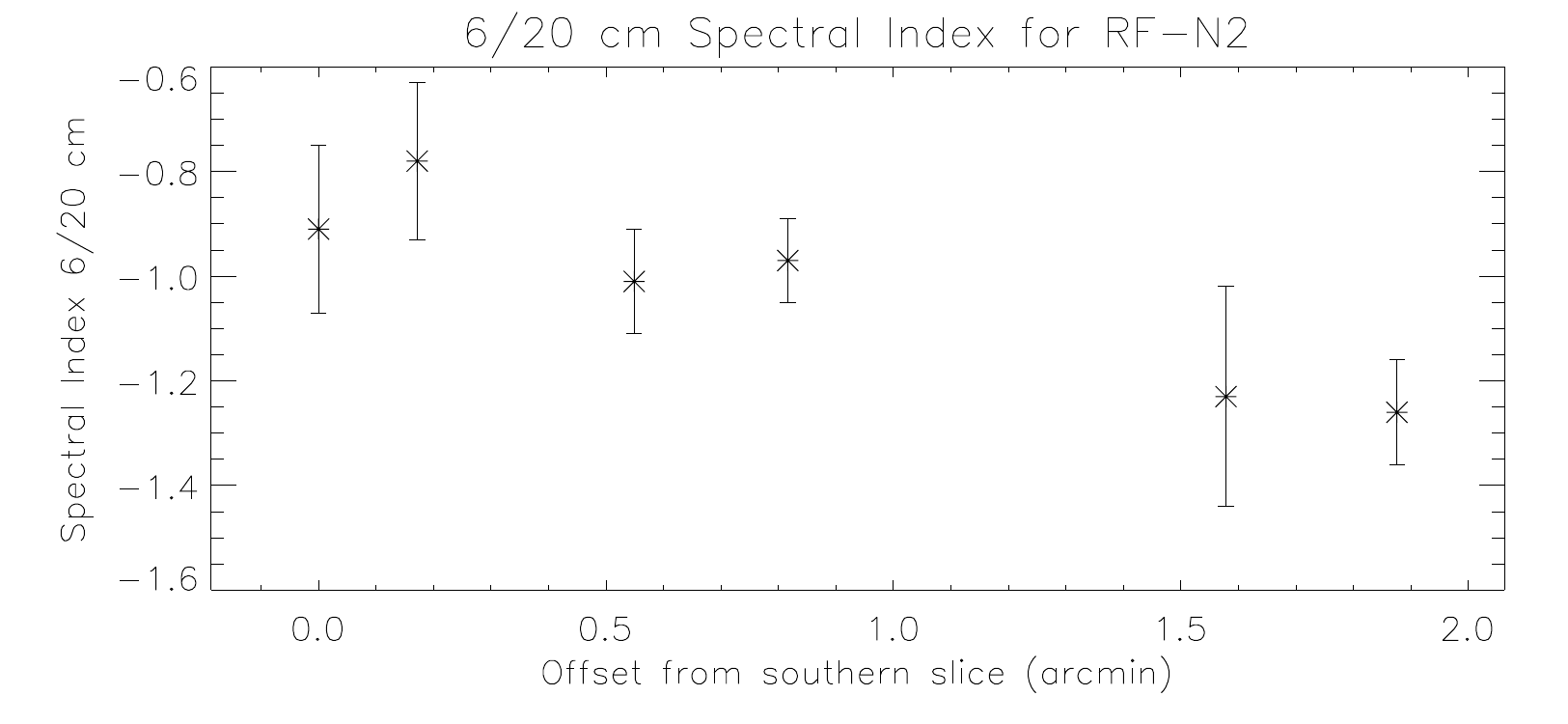}
\caption{This figure shows a high-resolution 20 cm image and 6/20 cm spectral indices toward the N2 radio filament.  The left image shows 20 cm brightness with contours levels at 5, 10, and 20 mJy per 14\arcsec$\times$9\arcsec-beam.  The right image shows the feathered, 6 cm image of the region, convolved to the resolution of the 20 cm feathered image of 26\arcsec$\times$18\arcsec, as in Figure \ref{c3fig}.  The plot at bottom shows the 6/20 cm spectral index measured for each slice shown in the right image.  The thick slice in the right image shows the southernmost slice, which is also the origin x-axis in the plot.  \label{n2fig}}
\end{figure}

$G0.08+0.15 (RF-N2)$ --- Figure \ref{n2fig} shows the 20 cm emission and 6/20 cm spectral index distribution along the N2 radio filament \citep[also called the ``Northern Thread'';][]{l99}.  This long (15\arcmin), vertical filament is fairly typical of NRFs, and the detection of linearly polarized emission at 3.6, 6, and 20 cm is consistent with this classification \citep{l99,y04}.  This filament is located at the edge of the 6 cm survey region, so only a fraction of the filament is seen and no integrated 6 cm flux densities are given in Table \ref{nrftab}.  

The 6/20 cm spectral index measured in slices across feathered and VLA-only images of N2 show a clearly nonthermal spectral index that steepens towards higher latitudes.  The southernmost slices are near of the edge of the primary beam of the 6 cm VLA observations; primary-beam corrections at 6 cm vary by a factor of 2--3 across N2, so it is possible that the spectral index at the edge is affected by uncertainties in this correction.  The 6/20 cm spectral index of N2 has been measured south of the region covered by the present survey to have a flatter spectral index with little spatial variation, ranging from 0.0 to --0.5 \citep{l99}.  Note that the value of $\alpha_{LC}$ at the origin in our Figure \ref{n2fig} is off the plot, at an offset of $\sim230$\arcsec\ in Figure 11 of \citet{l99}.  \citet{l00} measured the 20/90 cm spectral index for N2 and find that it has no significant change across a 4\arcmin\ length, with a mean value of --0.6.  It isn't clear where these measurements are in relation to the 6/20 cm index measured in the present work, but the 20/90 cm measurements are likely south of the present ones, since the N2 filament is more than 10\arcmin\ long and the present work only detects the northern 2--3\arcmin.

\begin{figure}[tbp]
\includegraphics[width=6.5in]{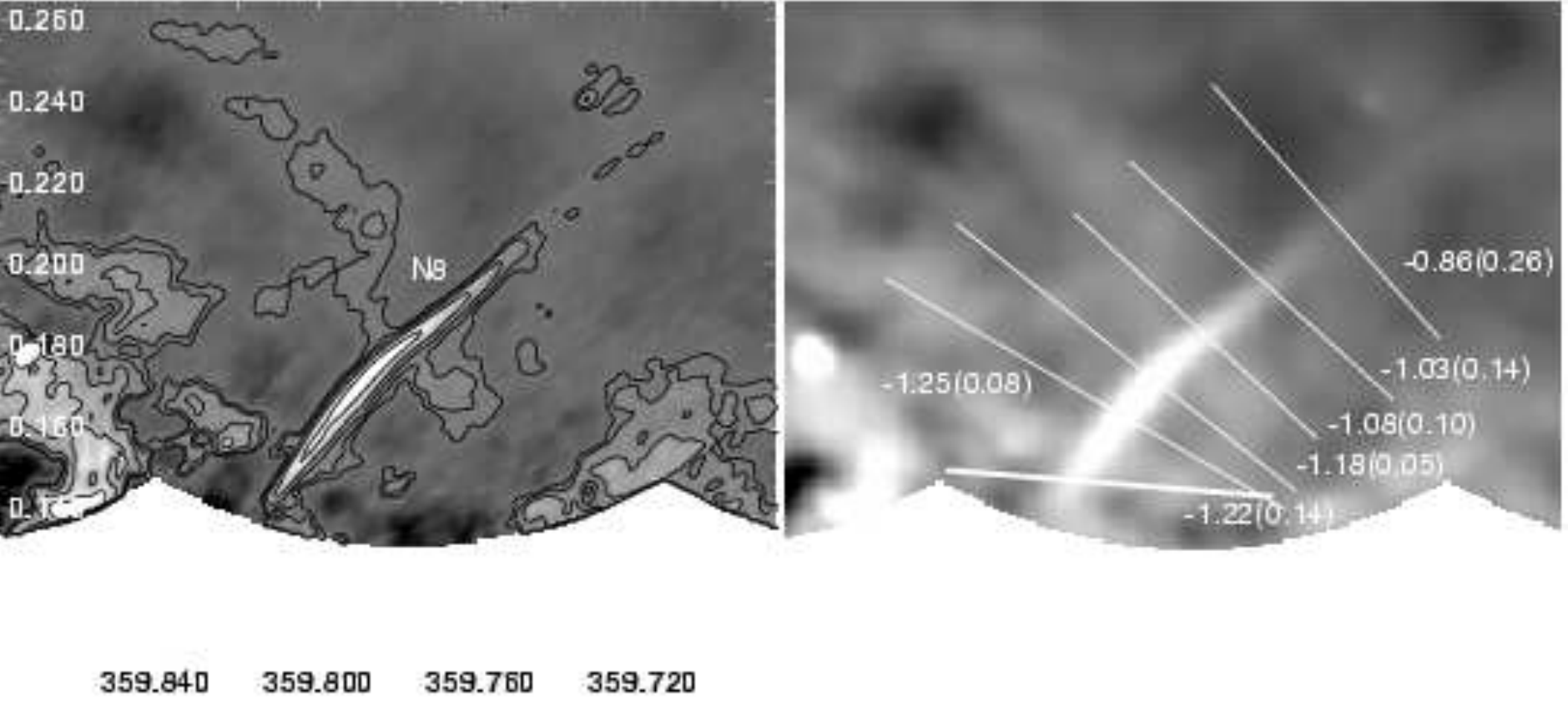}
\includegraphics[width=6.5in]{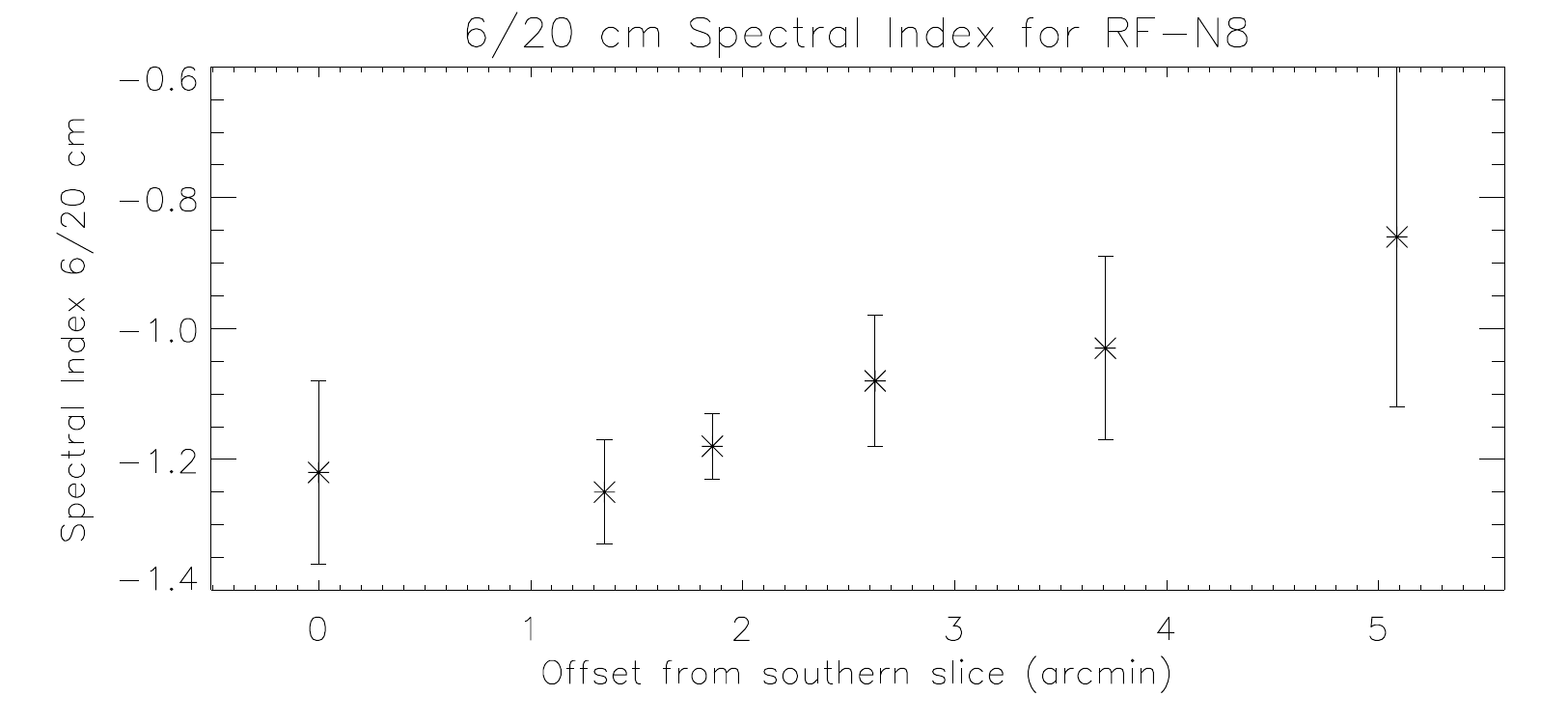}
\caption{This figure shows the high-resolution 6 cm emission and 6/20 cm spectral indices toward the N8 radio filament.  The left image has a resolution of 15\arcsec$\times$8\arcsec\ and contour levels at 0.5, 1, 2, 4, and 8 mJy beam$^{-1}$.  The right image shows the feathered, 6 cm image of the region, convolved to the resolution of the 20 cm feathered image of 26\arcsec$\times$18\arcsec, as in Figure \ref{c3fig}.  The plot at bottom shows the 6/20 cm spectral index measured for each slice shown in the right image.  The thick slice in the right image shows the southernmost slice, which is also the origin x-axis in the plot.  \label{n8fig}}
\end{figure}

$G359.79+0.17 (RF-N8)$ --- The N8 filament, as shown in Figure \ref{n8fig}, has a similar morphology to the C3 NRF in that it is relatively long ($\sim8$\arcmin) and has a slightly wavy shape.  The high-resolution 6 cm image shows that the brightest part of N8 forms an curve that opens downward.  North of the brightest emission the curve opens upward and then downward again in the faintest northern parts of the filament.  Subfilamentation is apparent in the faint northern and southern portions of N8.  The 20 cm image shows a broader, fan-like shape in the southern part of the filament \citep[see also Fig. 19b of][]{y04}.  This filament has an integrated 6 cm flux density of $226.7\pm24.8$ Jy and was previously found to have polarized continuum emission at 20 cm, which is also found in the present 6 cm observations.

The spectral index found by slice analysis of feathered and VLA-only images shows a slight flattening in $\alpha_{LC}$ toward the west, ranging from --1.3 to --0.9.  The N8 index has the opposite latitude dependence as seen in N1 and N2, but similar to that seen in C3 and C12.  The N8 NRF is also similar to C3 and C12 (and different from N1 and N2) in the angle that the filament makes with the Galactic plane.  N8, C3, and C12 are tilted west of Galactic north, while N1 and N2 are tilted east of north.  These two groups of NRFs are also located on west and east sides of Sgr A, respectively.  

Bandwidth smearing seems to bias the 6/20 cm spectral index measured for the N8 and N10 filaments.  The bandwidth-smearing bias for slice analysis was estimated by studying the nearby point source, G359.872,+0.178, which finds a biased of $\Delta\alpha\sim+0.4$. This is an upper limit to the bias in N8 and N10, since the filaments are partially resolved and the bias is likely to be significantly less than that seen in the point source.  Comparing the integrated 6 cm brightness to that of \citep{y04} gives $\alpha_{LC}=-1.91\pm0.09$.  This is more consistent with the slice spectral index values for N8 of $\alpha_{LC}=-0.9$ to $-1.3$, if there is a bandwidth-smearing bias like that seen in the nearby compact source.

\begin{figure}[tbp]
\includegraphics[width=6.5in]{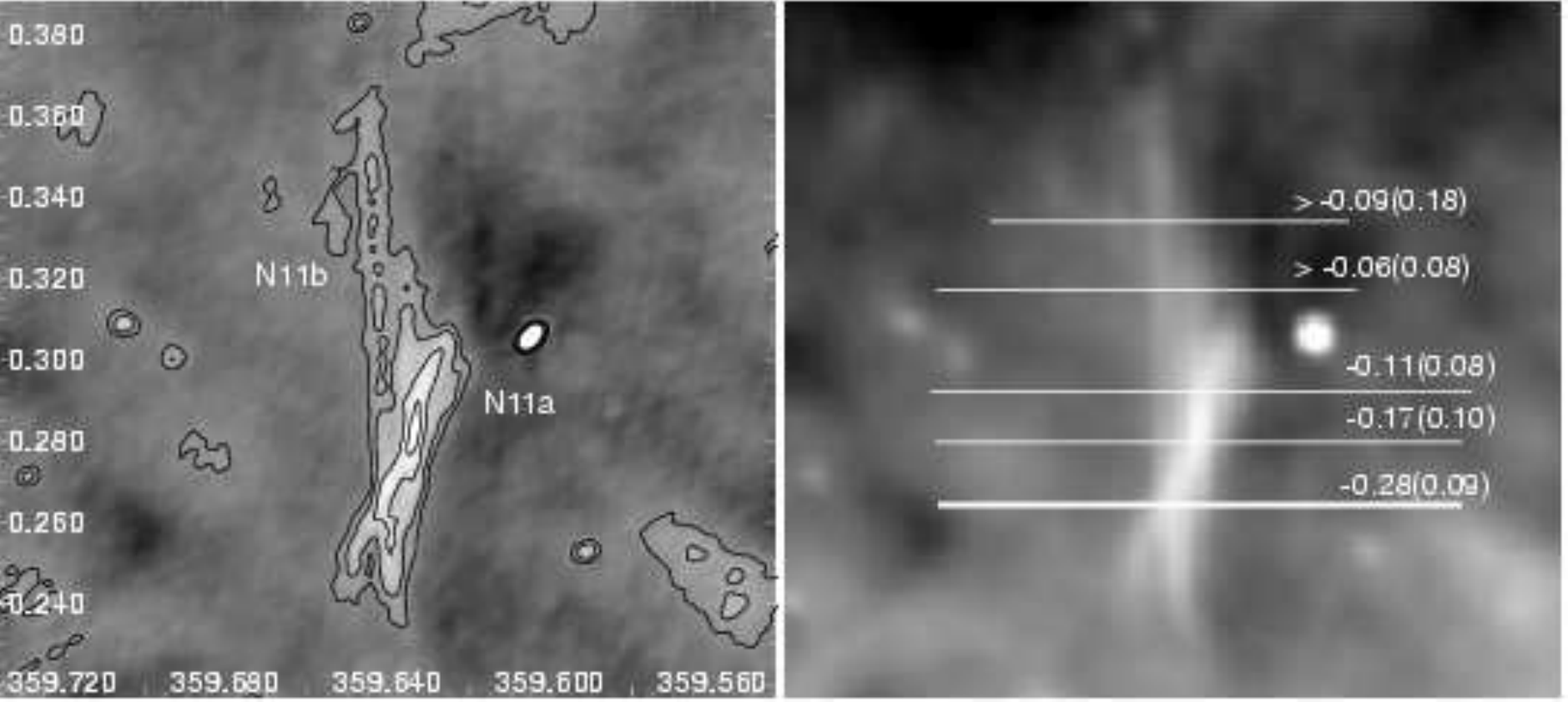}
\includegraphics[width=6.5in]{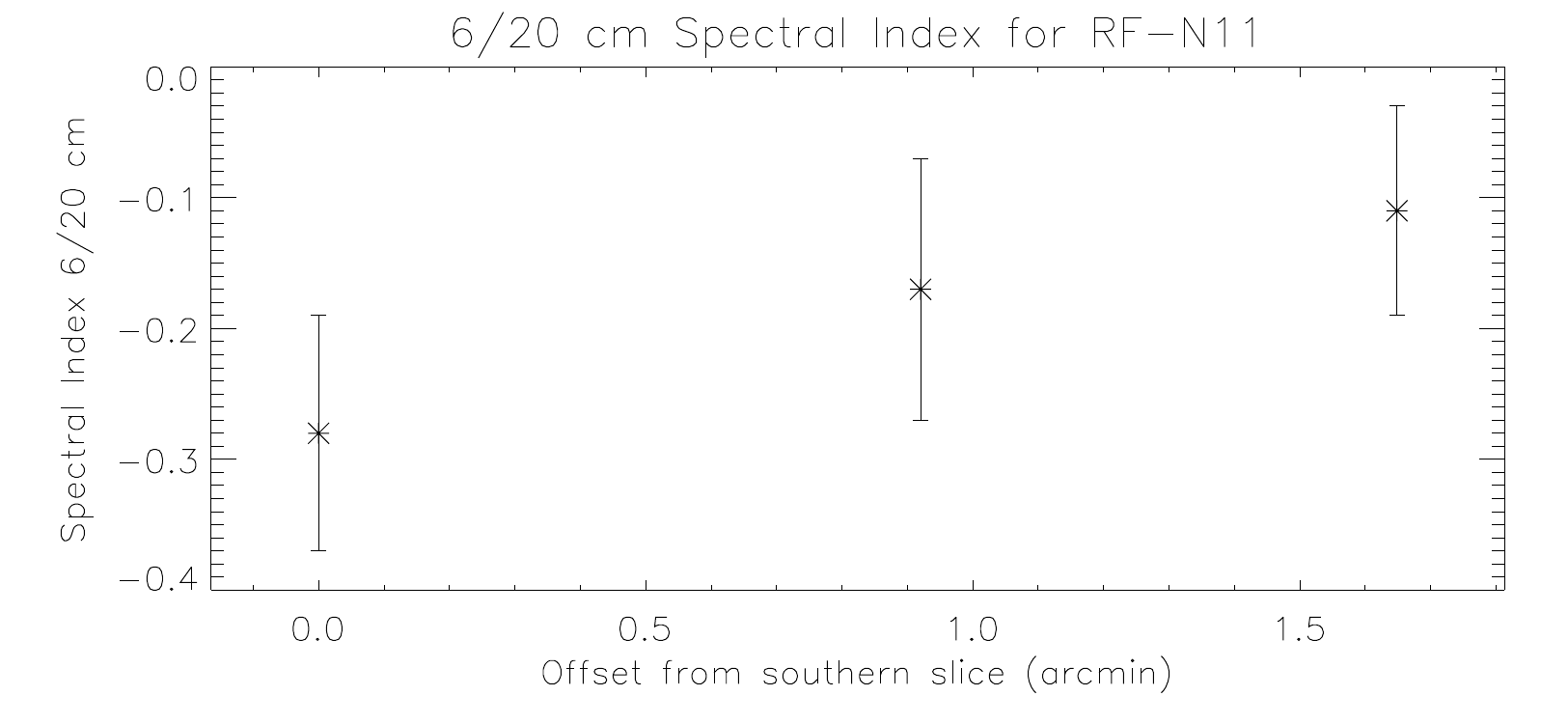}
\caption{This figure shows the high-resolution 6 cm emission and 6/20 cm spectral indices toward the N10 radio filament.  The left image has a resolution of 15\arcsec$\times$8\arcsec\ and contour levels at 0.5, 1, 2, and 3 mJy beam$^{-1}$.  The right image shows the feathered, 6 cm image of the region, convolved to the resolution of the 20 cm feathered image of 26\arcsec$\times$18\arcsec, as in Fig. \ref{c3fig}.  The plot at bottom shows the 6/20 cm spectral index measured for each slice shown in the right image.  The thick slice in the right image shows the southernmost slice, which is also the origin of the x-axis in the plot. \label{n11fig}}
\end{figure}

$G359.62+0.28 (RF-N11a)$ --- Figure \ref{n11fig} shows a complex of filaments near ($l$,$b$) = (359.6,+0.3).  The emission seems to be organized into two sets of crossed filaments:  one, 100-mJy, kinked filament running southeast to northwest, and then one or two parallel filaments running north-south with a flux density of about 60 mJy.  The kinked filament was previously identified as ``RF-N11'' by \citet{y04};  here it is referred to as N11a and the vertical filaments are called N11b.  The separation of the two groups of filaments and their large-scale structure is probably most obvious in the feathered image shown at right in Figure \ref{n11fig}.  The N11a filament is clearly polarized in the present 6 cm continuum observations, showing for the first time that this is an NRF.

The N11a filament is seen in the present 20 cm survey and that of \citet{y04}.  The 6/20 cm spectral index is found to be relatively flat across N11a, with values ranging from --0.28 to -0.11.  There is a trend for the index values to flatten with increasing latitude, but the trend has low significance.  The integrated 6/20 cm spectral index is steeper than measured by slice analysis, with $\alpha_{LC}=-0.64\pm0.05$.

$G359.64+0.30 (RF-N11b)$ --- The vertical filaments next to N11a in Figure \ref{n11fig} are here referred to as N11b.  N11b seems to be composed of two parallel filaments that cross the N11a filament near its midpoint.  N11b is seen in multiple VLA pointings, and thus is unlikely to be an imaging artifact.  As discussed in chapter \ref{gcl_vlapoln}, there is polarized emission throughout this region, but there is no clear morphological association with the N11b filament.  

The N11b filaments have no counterpart at 20 cm to a level of 0.2 mJy beam$^{-1}$ in the present observations nor in \citet{y04}, which constrains the spectral index to be greater than about --0.09.  This 6/20 cm spectral index is unusually flat for NRFs, but is more common in NRFs found in groups \citep{m96}.  The limit on the 6/20 cm spectral index is not plotted at the bottom of Figure \ref{n11fig}, but the values are consistent with a further flattening of the spectral index in N11a with increasing Galactic latitude.

\paragraph{Galactic Center Lobe}
Figures \ref{6large} and \ref{20large} show that the emission from the GCL dominates the images made from the feathered data.  The flux from the GCL appears strongest on spatial scales around 0\ddeg2, which is near the largest angular scale that the 6 cm observations can resolve ($\theta_{LAS}^{6 \rm{cm}}\approx0\ddeg1$).  The largest angular scale resolvable by the 20 cm observations is about three times larger ($\theta_{LAS}^{6 \rm{cm}}=900$\arcsec), so some of the GCL emission is visible in the 20 cm VLA-only image.  Since the GCL emission is not fully sampled by both data sets, no spectral index study can be done with the VLA data.  The radio spectral index distribution of the GCL from the GBT data are discussed in detail elsewhere (see Ch. \ref{gcl_all}).  Nonetheless, it is noteworthy that these first high spatial resolution observations of the GCL have constrained the spatial scale on which it emits.  Prior to these observations, single-dish observations could not determine if the GCL was a single, shell-like structure on 0\ddeg2-scales, or composed of smaller-scale filamentary structures.

\begin{figure}[tbp]
\includegraphics[width=6.5in]{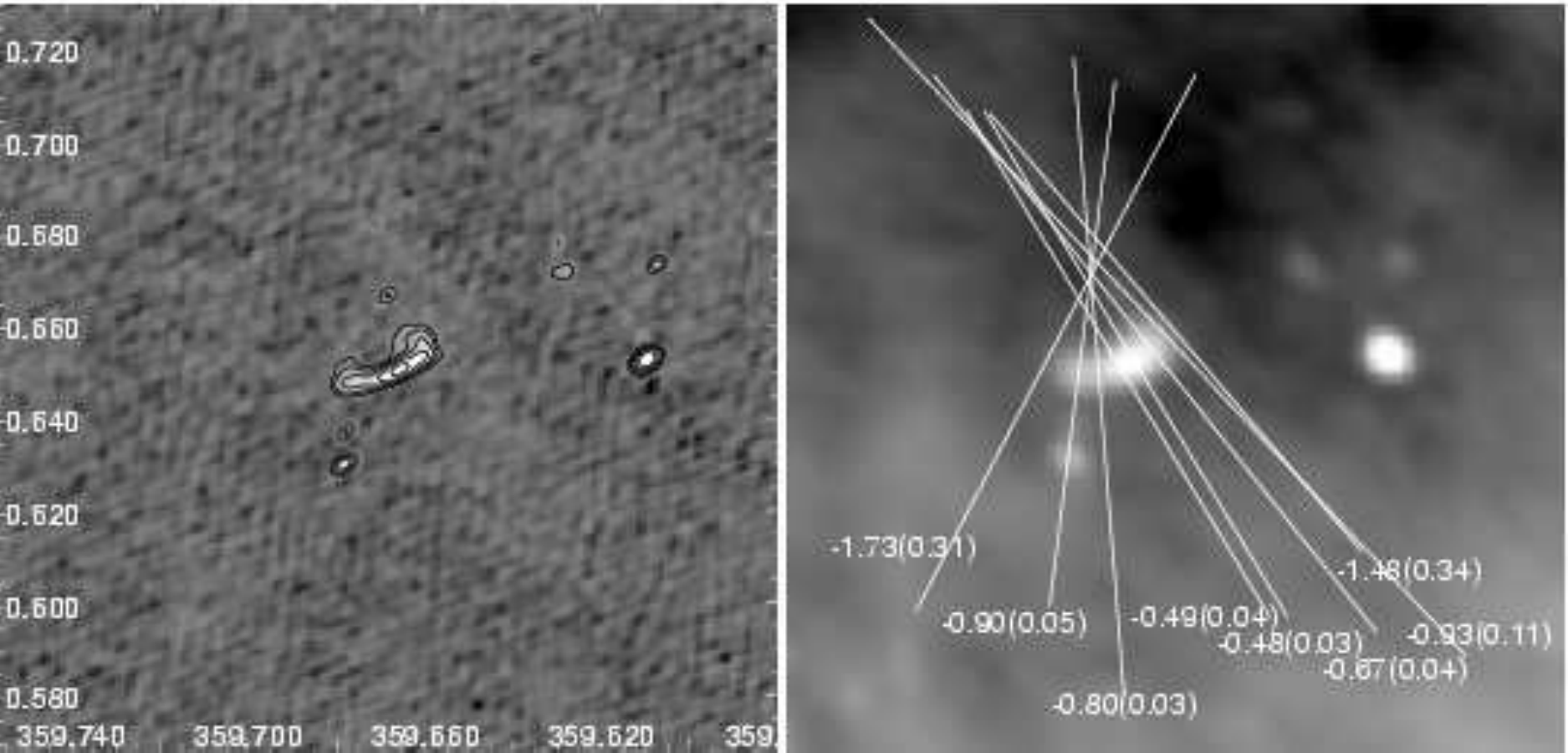}
\includegraphics[width=6.5in]{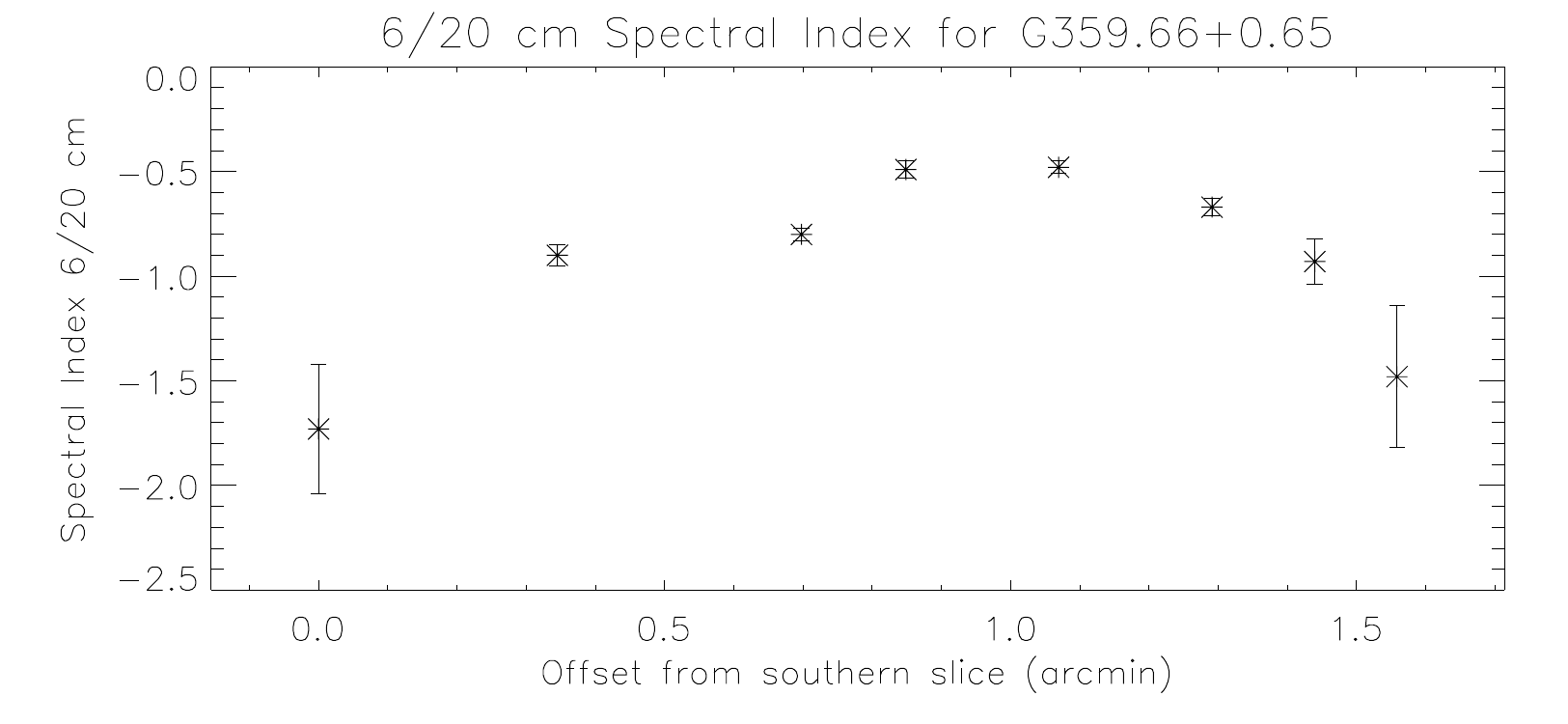}
\caption{This figure shows a high-resolution 6 cm image and 6/20 cm spectral indices toward G359.66+0.65, as described in Fig. \ref{c3fig}. Contour levels are at 0.25, 0.5, 1, 2, and 4 mJy beam$^{-1}$. \label{359.66+0.65fig}}
\end{figure}

\paragraph{Other Extended Objects}
$G359.66+0.65$ --- Figure \ref{359.66+0.65fig} shows the 6 cm emission and 6/20 cm spectral index across G359.66+0.65, an elongated, curving source roughly 1\damin5 long.  G359.66+0.65 has an integrated total intensity flux density of $29.8\pm2.0$ mJy in the 6 cm feathered image.  The 6/20 cm spectral index distribution is shown for a series of slices across G359.66+0.65 in Figure \ref{359.66+0.65fig}.  The index shows a steady change with position along the source ranging from --1.73 in the faintest part of the lobe-like region to --0.49 in the bright core region.  The integrated 6/20 cm spectral index from fluxes measured from the feathered images is $-0.92\pm0.12$.

G359.66+0.65 has a polarized counterpart with a flux density of $23.0\pm3.3$ mJy.  Comparing the polarized and total integrated fluxes for the VLA data give a polarized fraction of about 72\% for G359.66+0.65, which is near the theoretical maximum of $\sim$74\% for synchrotron emission polarization fraction with $\alpha=-0.9$ \citep{r79}.  No polarized source is seen at 20 cm down to a level of 0.2 mJy beam$^{-1}$; depolarization effects are about 11 times higher at 20 cm than at 6 cm.

The symmetry of the object about the central bright core, in addition to its high Galactic latitude argue against its identification as a radio filament.  Instead, it most closely resembles the morphology of an FR I type galaxy, which often have radio jets that have a swept-back shape like that seen in G359.66+0.65 \citep{f74,ge05}.  

\begin{figure}[tbp]
\includegraphics[width=6.5in]{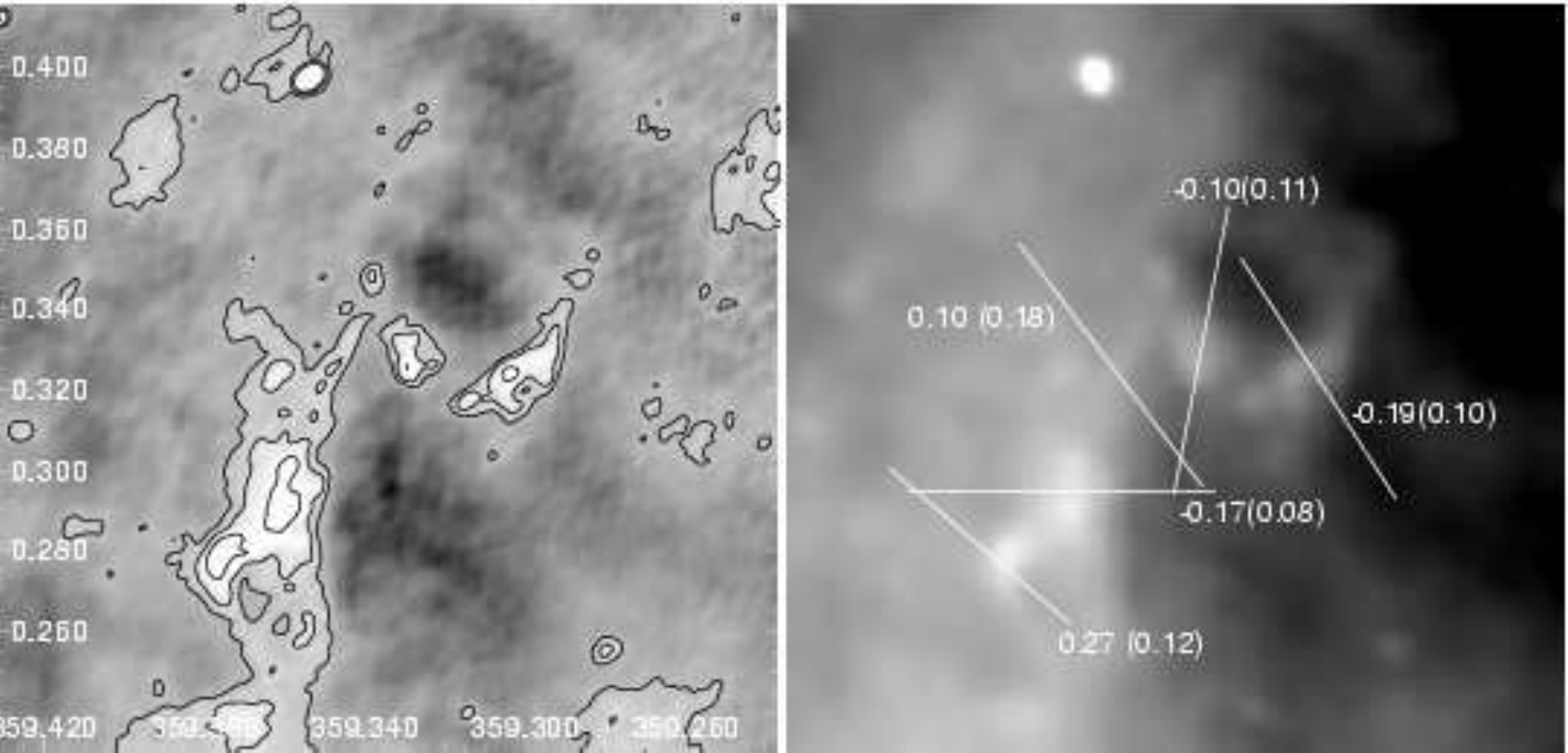}
\caption{\emph{Left}: 6 cm, VLA-only mosaic toward G359.34+0.31 with a beam size of 15\arcsec$\times$10\arcsec\ with $\theta_{PA}=70$\sdeg.  Contour levels are at 0.5, 1, and 2 mJy beam$^{-1}$.  \emph{Right}: 6 cm feathered image of the same region with 6/20 cm spectral indices overlaid. \label{359.34+0.31fig}}
\end{figure}

$G359.34+0.31$ --- This extended source is located within the western half of the radio continuum emission associated with the GCL.  Figure \ref{359.34+0.31fig} shows the 6 and 20 cm emission, which has a wispy morphology with the brightest emission along the GCL radio continuum ridge, but also extending outside that structure to lower longitudes.  G359.34+0.31 is a few arcminutes south of AFGL 5376 and a peak in radio recombination line brightness near $b=0\ddeg4-0\ddeg5$ \citep[][; Ch. \ref{gcl_recomb}]{u94}.  The 6/20 cm spectral index for several slices across the feathered images of G359.34+0.31 are consistent with thermal emission.

It is not clear that G359.34+0.31 is a part of the GCL or not.  One possibility is that it is portion of the GCL that happens to have emission on scales small enough to be seen in the VLA 6 and 20 cm observations.  However, most of the GCL radio continuum emission is on larger scales and has nonthermal radio spectral indices.  Also, G359.34+0.31 has a counterpart at mid-IR wavelengths, whereas much of the mid-IR emission associated with the GCL falls outside of the GCL.  The mid-IR and radio continuum observations of the GCL are discussed more completely in chapter \ref{gcl_all}.

\begin{figure}[tbp]
\includegraphics[width=6.5in]{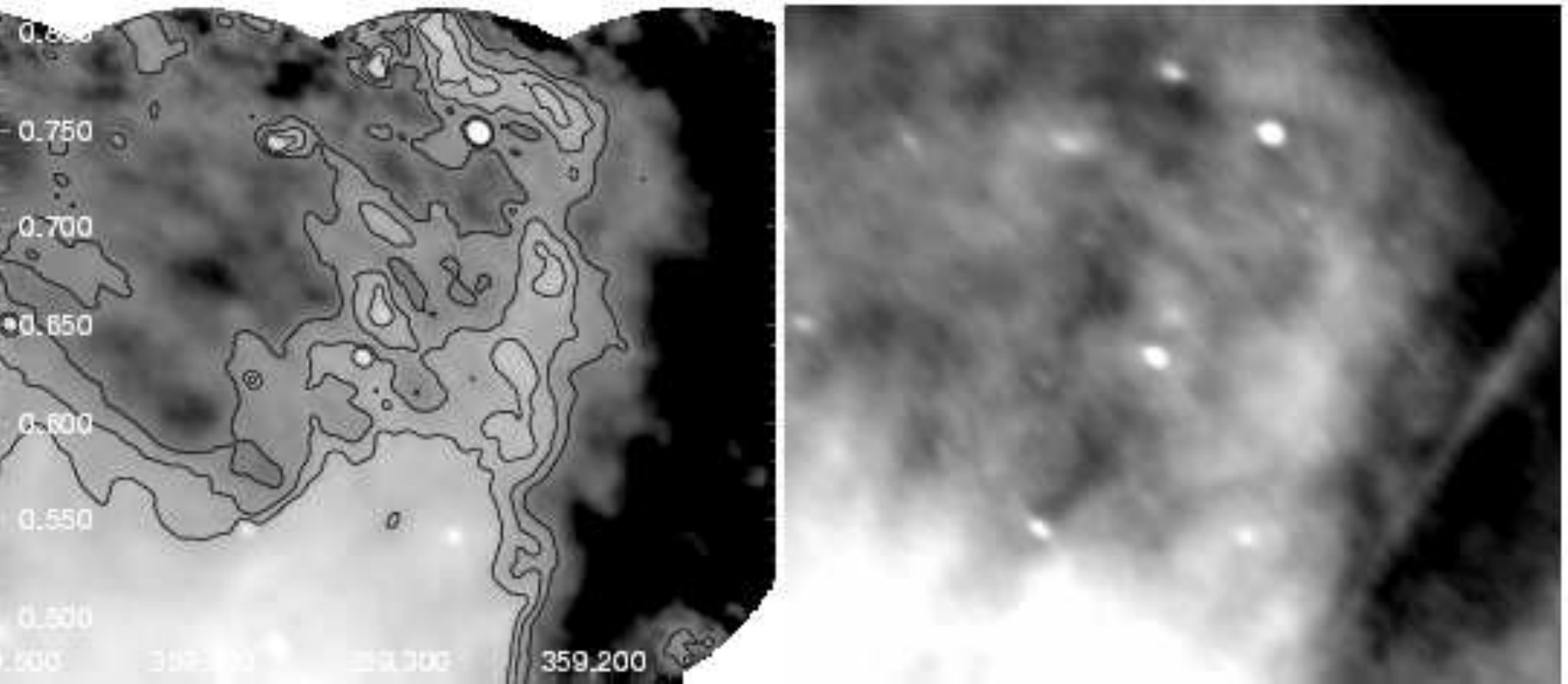}
\caption{The left and right panel show the 6 and 20 cm feathered radio continuum images of the region around G359.2+0.75, or the bend in the northwest corner of the GCL.  The contours in the left panel are at 3, 4, and 5 mJy per 18\arcsec$\times$26\arcsec-beam.  \label{359.2+0.75fig}}
\end{figure}

$G359.2+0.75$ --- At the northwest of the GCL, the radio continuum morphology bends toward higher longitudes, making a hooked shape similar to that seen in the 6 cm GBT image.  Figure \ref{359.2+0.75fig} shows the 6 and 20 cm feathered images of this structure, here referred to as G359.2+0.75.  The emission is visible in the VLA-only data, but the spatial scales of G359.2+0.75 are comparable to the largest angular scales visible to the 6 and 20 cm observations, so some negative bowls are present in that data.  

The 6/20 cm spectral index varies from --0.6 to +0.4 with the steepest index measured south of the bend the inverted index measured north of the bend.  The slice values are not shown in Figure \ref{359.2+0.75fig}, since the variation in $\alpha_{LC}$, including strongly inverted index values, are inconsistent with the GBT only slice analysis described in chapter \ref{gcl_all}.  It may be that the feathered image fluxes are not well calibrated at the VLA's largest resolvable angular scales of 5\arcmin\ and 15\arcmin\ at 6 and 20 cm, respectively.  Any problems in the calibration of the feathered image would likely be worst near G359.2+0.75 and G359.1+0.9 (described below), since they have complex distributions of emission on scales around 5\arcmin.  The slice analysis of all other sources are more trustworthy, since they have simpler emission distributions on size scales smaller than 5\arcmin.

\begin{figure}[tbp]
\includegraphics[width=6.5in]{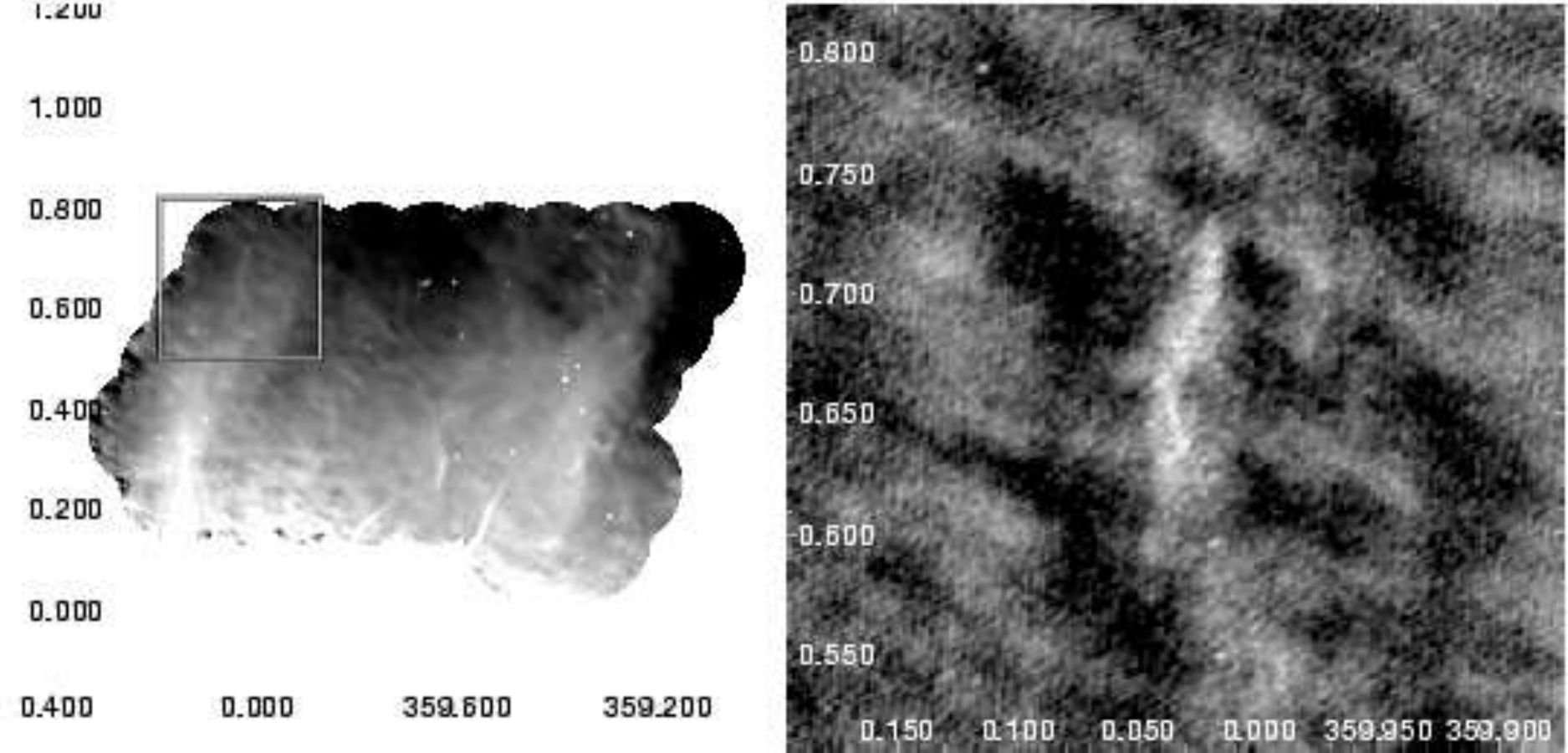}
\caption{\emph{Left}: Large-scale, 6 cm feathered image of the entire the GCL with Galactic coordinates.  The box shows the region covered by the image at right.  \emph{Right}: Image of the VLA-only data centered at G0.03+0.66 with a beam size of 9\dasec4$\times$12\dasec8 with $\theta_{PA}=66$\sdeg.  \label{0.03+0.66fig}}
\end{figure}

$G0.03+0.66$ --- Figure \ref{0.03+0.66fig} shows large- and small-scale views of G0.03+0.66, a 6\arcmin-long, wavy structure seen in 6 cm VLA data.  There is no counterpart to G0.03+0.66 in the 20 cm map, which suggests that it has a relatively flat spectral index, most consistent with thermal emission.  G0.03+0.66 is located within the GCL-East radio continuum emission ridge, similar to G359.34+0.31 in GCL-West.  In this case, however, the structure seen in the VLA data are oriented parallel to the structure seen by the GBT, so they are more likely to be associated with each other.  However, the radio continuum follows the wavy shape of the 8 $\mu$m emission seen by \emph{Spitzer} that has been identified as the ``Double Helix Nebula'' \citep{m06}.  The nature of this source and its possible association with the GCL is discussed in chapter \ref{gcl_all}.

\begin{figure}[tbp]
\includegraphics[width=\textwidth]{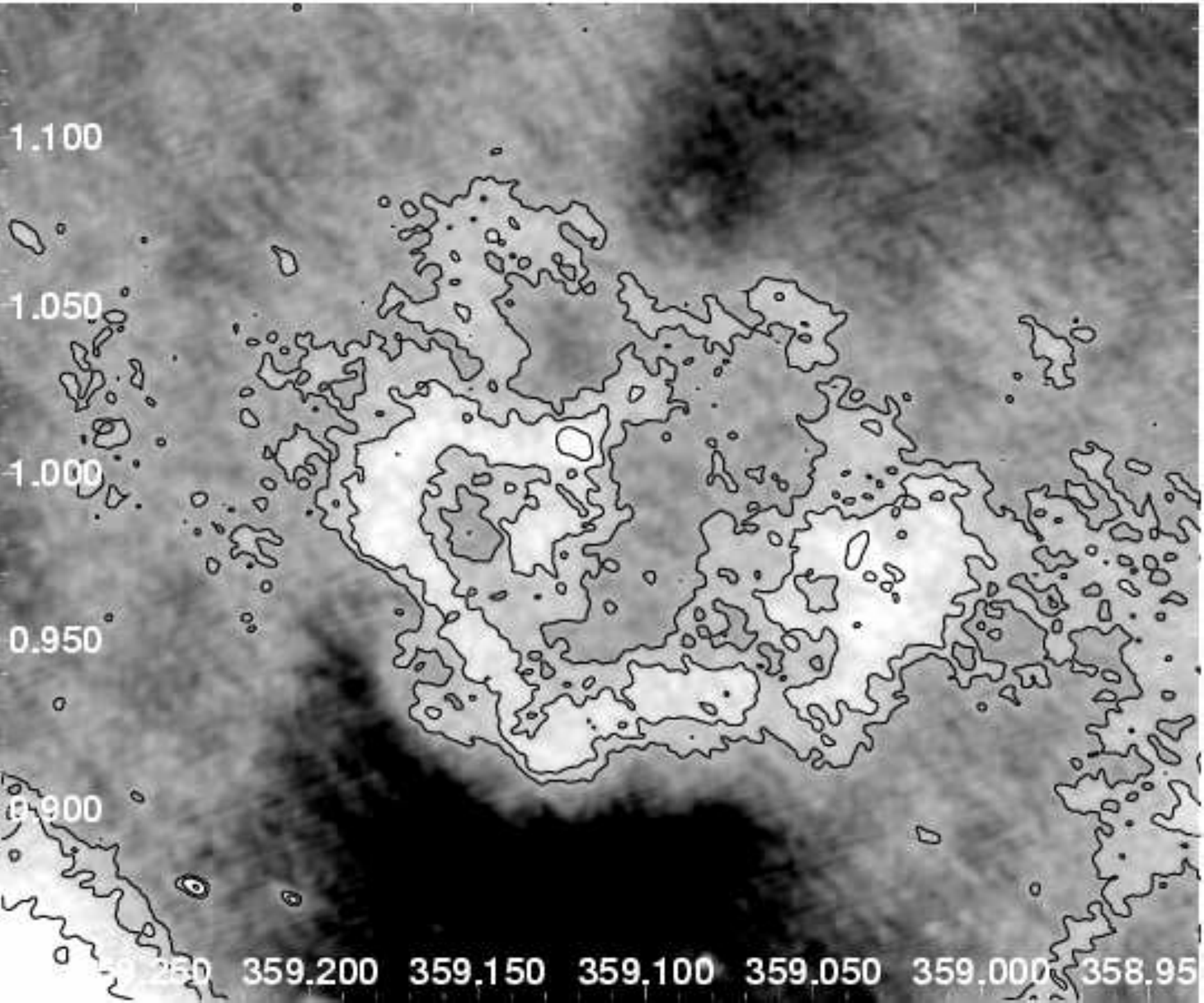}
\caption{Image of the 20 cm continuum emission of G359.1+0.9 seen by the VLA in Galactic coordinates. Contours are at brightness levels of 0.3, 0.6, and 1.2 mJy per 10\dasec9$\times$15\dasec9-beam \label{359.1+0.9fig}}
\end{figure}

$G359.1+0.9$ --- The 20 cm continuum emission from G359.1+0.9 is shown in Figure \ref{359.1+0.9fig}.  G359.1+0.9 is a shell-like SNR that extends 10\arcmin$\times$9\arcmin\ with an irregular ring.  This SNR is only covered by the present 20 cm survey, so no 6 cm map or 6/20 cm spectral index analysis is done.  The 20 cm integrated flux density is 1.3$\pm$0.5 Jy, which, when compared to the known 843 MHz (36 cm) flux density, gives $\alpha_{1400/843 \rm{MHz}}=-2.4\pm1.1$ \citep[assuming $\sigma_{843 \rm{MHz}}=2$ Jy;][]{g94}.

\subsection{Point Source Catalogs}
\label{vla_pssec}
\paragraph{Source Detection}
Tables \ref{ps6cm5} and \ref{ps20cm5} show the 6 and 20 cm catalogs of point sources detected on the mosaicked, high-resolution image, respectively.  The parameters found with the sfind, SAD, and JMFIT routines were compared on several sources.  In all cases tested, the integrated and peak brightnesses for all methods were equal within their quoted errors.  There is a slight trend for SAD/JMFIT to find higher peak flux uncertainties than sfind.  However, none of the several sources detected with both techniques had less than 3$\sigma$ significance in the peak brightness using either technique.  Thus, the 6 cm sources listed in Table \ref{ps6cm5} have at least 3$\sigma$ significance, while the 20 cm sources in Table \ref{ps20cm5} have at least $5\sigma$ significance, and both lists have flux densities that are robust to changes in the detection algorithm.

\begin{deluxetable}{ccccccccc}
\rotate
\tablecaption{Point Sources in 6 cm Survey \label{ps6cm5}}
\tablewidth{0pt}
\tablehead{
\colhead{RA} & \colhead{Dec} & \colhead{$S_p$} & \colhead{$\sigma_{S_p}$} & \colhead{$S_i$} & \colhead{$\sigma_{S_i}$} & \colhead{$\theta_M$} & \colhead{$\theta_m$} & \colhead{$\theta_{PA}$ \tablenotemark{a}} \\  
\colhead{(J2000)} & \colhead{(J2000)} & \colhead{mJy beam$^{-1}$} & \colhead{mJy beam$^{-1}$} & \colhead{mJy} & \colhead{mJy} & \colhead{arcsec} & \colhead{arcsec} & \colhead{degrees} \\
}
\startdata
17:42:07.917 & -29:18:51.92 & 0.707 & 0.142 & 1.278 & 0.372 & 16.4 & 13.9 & 9.9 \\
17:44:48.775 & -28:28:13.20 & 9.237 & 0.356 & 10.911 & 0.702 & 14.8 & 10.4 & 54.9 \\
17:44:45.917 & -28:35:15.24 & 10.657 & 0.172 & 12.697 & 0.337 & 15.1 & 10.1 & 69.1 \\
17:44:57.379 & -28:38:02.30 & 8.092 & 0.288 & 12.749 & 0.685 & 17.1 & 12.0 & 71.6 \\
17:43:45.577 & -28:32:05.88 & 10.302 & 0.139 & 11.393 & 0.258 & 14.7 & 9.5 & 70.7 \\
17:43:34.257 & -28:30:42.55 & 0.876 & 0.109 & 1.470 & 0.240 & 19.4 & 9.4 & 62.1 \\
17:44:40.866 & -28:39:56.24 & 16.338 & 0.274 & 19.177 & 0.537 & 14.8 & 10.3 & 67.1 \\
17:43:41.627 & -28:32:47.90 & 0.847 & 0.106 & 1.257 & 0.236 & 17.1 & 10.9 & 61.1 \\
17:44:22.882 & -28:38:51.19 & 6.230 & 0.210 & 12.977 & 0.606 & 19.8 & 13.6 & 68.4 \\
17:43:42.470 & -28:33:28.05 & 3.615 & 0.089 & 4.061 & 0.169 & 14.6 & 9.9 & 67.7 \\
17:44:54.553 & -28:44:59.10 & 2.281 & 0.393 & 3.187 & 0.836 & 16.7 & 10.4 & 67.9 \\
17:44:18.971 & -28:41:24.26 & 6.945 & 0.139 & 6.830 & 0.241 & 13.9 & 8.9 & 67.8 \\
17:43:20.500 & -28:36:33.38 & 1.354 & 0.084 & 1.349 & 0.147 & 13.9 & 9.1 & 72.0 \\
17:43:31.215 & -28:38:07.69 & 5.674 & 0.085 & 6.078 & 0.155 & 14.5 & 9.3 & 65.1 \\
17:44:31.059 & -28:47:16.36 & 2.531 & 0.408 & 5.968 & 1.347 & 19.6 & 16.4 & 75.9 \\
17:43:47.704 & -28:41:49.36 & 0.838 & 0.188 & 2.566 & 0.690 & 25.2 & 14.6 & 56.6 \\
17:43:39.770 & -28:41:51.71 & 0.738 & 0.105 & 0.729 & 0.187 & 13.4 & 9.7 & 79.0 \\
17:43:17.041 & -28:43:01.67 & 2.075 & 0.091 & 2.854 & 0.200 & 15.7 & 11.6 & 70.4 \\
17:43:22.961 & -28:44:52.30 & 0.840 & 0.077 & 0.626 & 0.116 & 12.0 & 7.9 & 71.2 \\
17:43:12.349 & -28:44:33.67 & 0.479 & 0.088 & 0.427 & 0.149 & 12.6 & 9.4 & 72.7 \\
17:43:10.118 & -28:45:17.28 & 0.433 & 0.059 & 0.260 & 0.079 & 10.9 & 6.9 & 64.1 \\
17:44:37.053 & -28:57:09.45 & 48.66 & 1.285 & 54.08 & 2.356 & 15.0 & 9.1 & 68.8 \\
17:42:39.650 & -28:41:44.50 & 0.894 & 0.074 & 1.948 & 0.204 & 21.6 & 11.7 & 57.9 \\
17:43:23.117 & -28:47:53.97 & 0.558 & 0.104 & 0.857 & 0.245 & 16.7 & 12.1 & 73.3 \\
17:42:28.738 & -28:41:48.90 & 0.688 & 0.080 & 0.962 & 0.177 & 15.9 & 11.6 & 70.2 \\
17:42:44.729 & -28:45:48.66 & 0.834 & 0.062 & 1.200 & 0.141 & 15.8 & 12.2 & 91.0 \\
17:42:44.519 & -28:45:37.68 & 0.614 & 0.063 & 0.955 & 0.130 & 18.8 & 8.8 & 92.0 \\
17:42:04.905 & -28:42:10.41 & 1.489 & 0.177 & 1.880 & 0.369 & 14.9 & 11.3 & 79.7 \\
17:43:02.749 & -28:51:09.16 & 2.885 & 0.129 & 4.455 & 0.291 & 17.7 & 10.7 & 48.0 \\
17:43:03.459 & -28:50:55.42 & 2.103 & 0.133 & 2.184 & 0.234 & 14.5 & 8.8 & 78.0 \\
17:43:46.895 & -28:57:05.44 & 0.973 & 0.099 & 0.897 & 0.166 & 13.4 & 8.7 & 72.6 \\
17:42:24.478 & -28:45:57.63 & 0.436 & 0.058 & 0.447 & 0.107 & 13.1 & 10.6 & -78.4 \\
17:43:05.476 & -28:52:21.88 & 0.649 & 0.092 & 0.958 & 0.200 & 17.4 & 10.3 & 55.2 \\
17:43:45.604 & -28:58:56.53 & 0.766 & 0.104 & 1.490 & 0.300 & 17.1 & 15.7 & -13.0 \\
17:42:42.902 & -28:50:55.25 & 0.626 & 0.105 & 0.478 & 0.162 & 11.8 & 8.5 & 71.2 \\
17:44:02.597 & -29:02:14.04 & 1.989 & 0.169 & 3.107 & 0.393 & 17.4 & 11.4 & 63.8 \\
17:43:14.130 & -28:56:01.58 & 1.706 & 0.129 & 2.058 & 0.256 & 15.1 & 10.3 & 73.1 \\
17:43:54.715 & -29:01:32.42 & 1.111 & 0.100 & 1.503 & 0.218 & 15.5 & 11.6 & 62.3 \\
17:43:42.562 & -29:01:33.72 & 1.473 & 0.136 & 4.710 & 0.540 & 24.9 & 16.3 & 16.5 \\
17:43:42.788 & -29:02:26.58 & 0.955 & 0.108 & 3.987 & 0.573 & 25.5 & 22.5 & 71.0 \\
17:42:23.776 & -28:52:48.41 & 2.980 & 0.103 & 2.901 & 0.178 & 13.8 & 8.9 & 65.0 \\
17:42:13.815 & -28:52:10.78 & 0.636 & 0.077 & 0.861 & 0.161 & 16.4 & 10.3 & 64.7 \\
17:42:42.652 & -28:52:04.75 & 0.634 & 0.136 & 0.755 & 0.245 & 16.2 & 8.2 & 71.2 \\
17:43:17.039 & -29:02:12.27 & 1.130 & 0.130 & 1.286 & 0.245 & 15.0 & 9.5 & 76.0 \\
17:42:06.927 & -28:54:03.26 & 0.655 & 0.074 & 1.023 & 0.179 & 16.4 & 12.8 & 56.9 \\
17:42:53.589 & -29:00:53.94 & 0.824 & 0.083 & 0.997 & 0.166 & 15.0 & 10.5 & 63.4 \\
17:42:08.606 & -28:55:36.43 & 15.15 & 0.101 & 16.59 & 0.187 & 14.6 & 9.5 & 69.9 \\
17:42:03.496 & -28:55:05.04 & 1.024 & 0.074 & 1.158 & 0.140 & 14.8 & 9.7 & 62.5 \\
17:43:28.881 & -29:06:47.19 & 14.85 & 0.142 & 21.44 & 0.320 & 19.0 & 9.6 & 83.0 \\
17:42:31.068 & -28:59:45.85 & 7.804 & 0.111 & 10.44 & 0.236 & 15.8 & 11.0 & 64.7 \\
17:43:39.267 & -29:09:07.81 & 1.575 & 0.143 & 3.982 & 0.478 & 21.8 & 15.0 & 46.7 \\
17:42:41.764 & -29:02:10.19 & 2.087 & 0.125 & 2.101 & 0.220 & 14.0 & 9.1 & 69.1 \\
17:41:26.162 & -28:53:29.12 & 16.06 & 0.148 & 16.77 & 0.266 & 14.2 & 9.3 & 79.0 \\
17:42:57.261 & -29:07:30.73 & 0.666 & 0.124 & 0.777 & 0.239 & 15.0 & 9.9 & 56.7 \\
17:43:43.443 & -29:13:57.65 & 1.538 & 0.147 & 1.840 & 0.285 & 15.8 & 9.5 & 59.0 \\
17:43:42.827 & -29:14:33.81 & 1.105 & 0.231 & 1.884 & 0.549 & 18.8 & 10.9 & 61.9 \\
17:42:49.196 & -29:07:25.20 & 0.699 & 0.121 & 1.985 & 0.443 & 23.1 & 15.9 & 57.5 \\
17:41:42.851 & -28:59:02.78 & 2.901 & 0.085 & 3.020 & 0.151 & 14.4 & 9.0 & 71.9 \\
17:43:36.037 & -29:14:39.50 & 2.637 & 0.174 & 1.965 & 0.261 & 12.0 & 7.9 & 68.9 \\
17:42:30.636 & -29:05:59.66 & 1.669 & 0.099 & 4.466 & 0.350 & 22.1 & 15.9 & 60.6 \\
17:41:53.045 & -29:01:17.97 & 4.966 & 0.074 & 6.027 & 0.148 & 15.1 & 10.4 & 70.8 \\
17:43:34.857 & -29:15:04.12 & 0.783 & 0.144 & 0.752 & 0.250 & 13.4 & 9.3 & 77.3 \\
17:43:10.184 & -29:12:14.49 & 19.99 & 0.115 & 20.79 & 0.207 & 14.2 & 9.3 & 69.3 \\
17:44:03.763 & -29:19:53.25 & 2.899 & 0.341 & 4.710 & 0.769 & 18.6 & 10.2 & 70.7 \\
17:43:37.410 & -29:16:36.86 & 0.773 & 0.198 & 3.718 & 1.050 & 31.7 & 18.1 & 52.8 \\
17:44:09.038 & -29:21:33.34 & 4.234 & 0.607 & 3.583 & 0.973 & 12.8 & 8.4 & 73.1 \\
17:41:23.090 & -29:01:04.19 & 1.426 & 0.116 & 1.722 & 0.237 & 14.3 & 11.4 & 74.1 \\
17:42:38.843 & -29:12:17.97 & 2.418 & 0.105 & 2.975 & 0.216 & 14.6 & 11.3 & 80.8 \\
17:43:02.015 & -29:15:49.07 & 0.858 & 0.124 & 3.035 & 0.555 & 25.1 & 18.7 & 44.3 \\
17:43:53.398 & -29:23:30.23 & 1.794 & 0.319 & 3.245 & 0.853 & 17.7 & 13.7 & -87.7 \\
17:42:21.455 & -29:13:00.29 & 39.431 & 0.206 & 41.158 & 0.370 & 14.3 & 9.2 & 71.1 \\
17:41:59.993 & -29:10:47.53 & 4.282 & 0.148 & 4.168 & 0.255 & 13.8 & 8.9 & 67.7 \\
17:41:41.632 & -29:08:31.65 & 1.937 & 0.067 & 2.425 & 0.138 & 15.1 & 10.9 & 66.7 \\
17:42:11.828 & -29:13:29.73 & 18.926 & 0.125 & 19.050 & 0.220 & 14.0 & 9.1 & 69.3 \\
17:41:21.951 & -29:07:17.07 & 0.520 & 0.079 & 0.493 & 0.130 & 14.1 & 8.0 & 64.0 \\
17:42:31.096 & -29:16:42.26 & 19.799 & 0.138 & 23.023 & 0.264 & 15.1 & 9.7 & 69.3 \\
17:43:40.559 & -29:27:09.47 & 1.163 & 0.206 & 1.233 & 0.385 & 13.7 & 10.3 & 60.6 \\
17:41:30.698 & -29:11:02.97 & 5.720 & 0.100 & 6.047 & 0.182 & 14.3 & 9.4 & 66.1 \\
17:43:41.001 & -29:28:41.05 & 3.037 & 0.323 & 5.713 & 0.830 & 19.6 & 11.7 & 67.4 \\
17:43:02.312 & -29:23:42.34 & 1.127 & 0.084 & 1.615 & 0.188 & 16.2 & 11.6 & 65.3 \\
17:40:55.269 & -29:06:53.64 & 2.872 & 0.469 & 4.312 & 1.047 & 17.3 & 10.8 & 57.3 \\
17:41:24.268 & -29:10:52.57 & 1.424 & 0.168 & 6.593 & 0.964 & 27.7 & 22.7 & 76.1 \\
17:43:12.190 & -29:26:03.67 & 22.286 & 0.230 & 23.561 & 0.418 & 14.3 & 9.4 & 66.1 \\
17:42:53.789 & -29:24:57.01 & 0.900 & 0.117 & 2.508 & 0.428 & 22.5 & 16.3 & 55.8 \\
17:41:07.573 & -29:10:35.18 & 1.160 & 0.121 & 1.638 & 0.261 & 16.7 & 10.6 & 54.1 \\
17:41:45.418 & -29:16:20.83 & 4.051 & 0.076 & 5.403 & 0.159 & 16.1 & 10.5 & 63.1 \\
17:40:55.280 & -29:10:25.13 & 29.683 & 0.245 & 32.406 & 0.454 & 14.5 & 9.6 & 69.6 \\
17:42:10.672 & -29:20:53.95 & 1.385 & 0.125 & 1.483 & 0.230 & 14.3 & 9.6 & 58.7 \\
17:43:09.332 & -29:28:58.20 & 2.557 & 0.190 & 2.357 & 0.305 & 14.0 & 7.7 & 59.5 \\
17:43:10.195 & -29:29:52.04 & 3.293 & 0.384 & 3.935 & 0.751 & 15.2 & 10.0 & 62.7 \\
17:41:02.201 & -29:12:33.24 & 1.231 & 0.112 & 1.659 & 0.244 & 15.3 & 11.8 & 60.4 \\
17:41:28.400 & -29:16:34.27 & 0.687 & 0.084 & 1.318 & 0.227 & 19.1 & 12.9 & 79.4 \\
17:40:31.410 & -29:20:19.11 & 4.993 & 0.213 & 5.476 & 0.410 & 13.6 & 10.9 & 54.9 \\
\enddata
\tablecomments{Sources have greater than 5$\sigma$ peak brightness significance according to sfind and better than $\sim3\sigma$ confidence according to the AIPS SAD/JMFIT routines.  Columns 1--2 give the position of the source in J2000 coordinates; columns 3--4 give the peak flux and its uncertainty; columns 5--6 give the integrated flux and its uncertainty; columns 7--9 give the major and minor beam FWHM and its position angle.}
\tablenotetext{a}{Position angle increases to the east from celestial north.}
\end{deluxetable}

\begin{deluxetable}{ccccccccc}
\rotate
\tablecaption{Point Sources in 20 cm Survey \label{ps20cm5}}
\tablewidth{0pt}
\tablehead{
\colhead{RA} & \colhead{Dec} & \colhead{$S_p$} & \colhead{$\sigma_{S_p}$} & \colhead{$S_i$} & \colhead{$\sigma_{S_i}$} & \colhead{$\theta_M$} & \colhead{$\theta_m$} & \colhead{$\theta_{PA}$ \tablenotemark{b}} \\  
\colhead{(J2000)} & \colhead{(J2000)} & \colhead{mJy beam$^{-1}$} & \colhead{mJy beam$^{-1}$} & \colhead{mJy} & \colhead{mJy} & \colhead{arcsec} & \colhead{arcsec} & \colhead{degrees} \\
}
\startdata
17:44:48.755 & -28:28:14.85 & 15.34 & 0.614 & 30.95 & 1.750 & 21.8 & 11.7 & 33.3 \\ 
17:44:45.929 & -28:35:14.33 & 8.190 & 0.648 & 6.920 & 1.000 & 13.3 & 8.0 & 56.5 \\ 
17:44:40.820 & -28:39:55.55 & 12.58 & 0.648 & 13.11 & 1.150 & 13.3 & 9.9 & 51.2 \\ 
17:44:23.011 & -28:38:51.75 & 5.555 & 0.604 & 14.56 & 2.100 & 21.8 & 15.2 & 28.2 \\ 
17:44:19.104 & -28:41:21.82 & 5.668 & 0.6480 & 4.319 & 0.9400 & 11.2 & 8.6 & 55.9 \\ 
17:43:45.718 & -28:32:06.04 & 5.423 & 0.6460 & 6.190 & 1.220 & 13.8 & 10.4 & 93.3 \\ 
17:42:48.347 & -28:25:28.35 & 14.03 & 0.6190 & 25.24 & 1.620 & 20.5 & 11.1 & 28.6 \\ 
17:42:39.714 & -28:41:40.81 & 4.523\tablenotemark{a} & 0.6480 & 4.523 & 0.6480 & 14.0 & 9.0 & 60 \\ 
17:41:28.597 & -28:36:55.54 & 5.158\tablenotemark{a} & 0.6480 & 5.158 & 0.6480 & 14.0 & 9.0 & 60 \\ 
17:41:19.063 & -28:24:19.68 & 9.224\tablenotemark{a} & 0.6480 & 9.224 & 0.6480 & 14.0 & 9.0 & 60 \\ 
17:41:26.196 & -28:53:30.04 & 32.86 & 0.6300 & 47.69 & 1.410 & 16.3 & 11.2 & 53.4 \\ 
17:40:41.808 & -28:48:13.61 & 44.57 & 0.6100 & 100.1 & 1.880 & 21.2 & 13.4 & 121.3 \\ 
17:40:35.331 & -28:41:42.40 & 5.342 & 0.6310 & 7.664 & 1.400 & 18.9 & 9.6 & 131.8 \\ 
17:40:33.419 & -28:44:05.08 & 4.652 & 0.6450 & 5.359 & 1.230 & 16.6 & 8.7 & 131.3 \\ 
17:40:07.992 & -28:42:04.50 & 55.81 & 0.5910 & 246.3 & 3.140 & 39.5 & 14.1 & 126.9 \\ 
17:40:55.599 & -28:56:23.58 & 3.303 & 0.6340 & 4.502 & 1.360 & 16.5 & 10.4 & 78.8 \\ 
17:40:58.392 & -28:57:31.87 & 7.884 & 0.6350 & 10.55 & 1.340 & 17.0 & 9.9 & 70.5 \\ 
17:40:55.212 & -29:10:25.42 & 20.04 & 0.6120 & 42.43 & 1.810 & 22.9 & 11.7 & 42.7 \\ 
17:40:20.487 & -29:13:44.37 & 8.125 & 0.6330 & 11.28 & 1.370 & 14.6 & 12.0 & 56.4 \\ 
17:40:02.069 & -29:23:36.46 & 5.122 & 0.6410 & 6.249 & 1.270 & 17.1 & 9.0 & 53.1 \\ 
17:39:39.583 & -29:31:22.58 & 4.132 & 0.6020 & 11.59 & 2.210 & 23.9 & 14.8 & 47.9 \\ 
17:39:02.718 & -29:26:06.50 & 2.764 & 0.4350 & 49.08 & 8.140 & 54.7 & 40.9 & 69.4 \\ 
17:40:31.357 & -29:20:19.10 & 5.180 & 0.6450 & 5.992 & 1.230 & 15.3 & 9.5 & 61.3 \\ 
17:40:23.233 & -29:32:15.25 & 7.719 & 0.6120 & 16.51 & 1.820 & 22.0 & 12.3 & 14.7 \\ 
17:40:54.433 & -29:29:46.78 & 41.14 & 0.6060 & 102.1 & 2.020 & 19.7 & 15.9 & 157.1 \\ 
17:41:30.683 & -29:11:02.82 & 7.827 & 0.6060 & 19.35 & 2.020 & 18.5 & 16.9 & 44.8 \\ 
17:43:14.193 & -28:56:01.29 & 3.251 & 0.6480 & 3.294 & 1.130 & 13.4 & 9.5 & 70.7 \\ 
17:42:08.486 & -28:55:36.00 & 8.290 & 0.6280 & 12.64 & 1.450 & 19.2 & 10.0 & 76.6 \\ 
17:42:31.205 & -28:59:45.20 & 4.278 & 0.6230 & 7.106 & 1.540 & 19.0 & 11.0 & 77.1 \\ 
17:43:28.869 & -29:06:46.42 & 19.26 & 0.6080 & 45.78 & 1.960 & 18.2 & 16.4 & 47.9 \\ 
17:43:10.112 & -29:12:13.24 & 6.100 & 0.6260 & 9.607 & 1.490 & 18.7 & 10.6 & 41.7 \\
17:43:12.193 & -29:26:01.47 & 7.928 & 0.6190 & 14.21 & 1.620 & 20.4 & 11.1 & 89.4 \\ 
17:44:05.855 & -29:17:42.77 & 8.387 & 0.5850 & 50.12 & 4.040 & 29.2 & 25.8 & 0.0 \\ 
17:44:05.761 & -29:28:38.51 & 11.31 & 0.5910 & 48.38 & 3.060 & 25.7 & 21.0 & 30.1 \\ 
17:44:57.168 & -28:38:02.72 & 6.354 & 0.6020 & 17.99 & 2.230 & 20.8 & 17.1 & 94.4 \\ 
17:44:36.688 & -28:57:10.19 & 79.92 & 1.7600 & 171.8 & 5.260 & 26.3 & 10.3 & 82.1 \\ 
17:41:42.870 & -28:59:02.10 & 4.145 & 0.6480 & 2.859 & 0.8820 & 11.5 & 7.5 & 67.0 \\ 
17:40:55.084 & -29:06:57.22 & 4.970 & 0.3260 & 18.60 & 1.510 & 34.9 & 13.5 & 155.7 \\ 
17:41:45.520 & -29:16:20.15 & 3.362 & 0.6080 & 8.018 & 1.960 & 21.1 & 14.2 & 92.4 \\ 
17:43:31.081 & -28:38:07.83 & 5.047 & 0.5950 & 18.00 & 2.650 & 26.2 & 17.1 & 93.0 \\ 
17:41:27.917 & -28:52:51.74 & 8.169 & 0.6120 & 17.10 & 1.790 & 18.3 & 14.4 & 52.0 \\ 
17:42:21.477 & -29:13:01.49 & 64.72 & 0.6240 & 104.0 & 1.500 & 16.4 & 12.4 & 40.0 \\ 
17:42:11.868 & -29:13:31.04 & 19.13 & 0.6280 & 28.50 & 1.430 & 14.5 & 12.9 & 35.0 \\ 
17:42:31.138 & -29:16:41.55 & 16.64 & 0.6320 & 23.40 & 1.380 & 16.9 & 10.5 & 61.0 \\ 
17:42:00.060 & -29:10:49.40 & 5.118 & 0.6180 & 9.320 & 1.630 & 17.3 & 13.3 & 159.0 \\ 
17:42:10.778 & -29:20:54.23 & 3.803 & 0.6280 & 5.740 & 1.450 & 17.0 & 11.2 & 61.0 \\ 
17:43:03.318 & -28:50:57.45 & 6.631 & 0.6130 & 13.70 & 1.770 & 17.2 & 15.1 & 144.0 \\ 
17:43:02.797 & -28:51:12.56 & 5.660 & 0.6270 & 8.590 & 1.450 & 14.7 & 13.0 & 2.0 \\ 
\enddata
\tablecomments{Sources have greater than 5$\sigma$ peak brightness significance according to the AIPS SAD/JMFIT routines.  Columns 1--2 give the position of the source in J2000 coordinates; columns 3--4 give the peak flux and its uncertainty; columns 5--6 give the integrated flux and its uncertainty; columns 7--9 give the major and minor beam FWHM and its position angle.}
\tablenotetext{a}{Source parameters found by fixing the source size to the beam size of 14\arcsec\ by 9\arcsec.  This assumption may result in the peak brightness being overestimated by about 50\%.}
\tablenotetext{b}{Position angle increases to the east from celestial north.}
\end{deluxetable}

Polarized point sources were detected by running SAD and JMFIT tasks on the polarized-continuum mosaic images.  All sources with peak polarized fluxes greater than $3\sigma$ are shown in Tables \ref{6poln} and \ref{20poln} for 6 and 20 cm, respectively.  Imperfections in the feed system can also induce false polarization by putting up to 1\% of the total intensity flux into the polarized flux \citep{c99}.  Another important systematic effect is the antenna polarization that induces radially-oriented, linear polarization with increasing offset from the phase center \citep{c99}.  Since this bias affects each field near its edge, the mosaicked images sometimes average biases with perpendicular orientations, thus cancelling the bias out.  Furthermore, the parallactic angle for each source changes during the observation, which reduces antenna-based errors.  During the observations, the parallactic angle changed by about 80\sdeg, which reduces antenna-based polarization by about $1-\sin(\Delta\theta)/\Delta\theta\approx30$\%.  

\begin{deluxetable}{cccccccc}
\tablecaption{Point Sources with 6 cm Polarization \label{6poln}}
\tablewidth{0pt}
\tablehead{
\colhead{RA$_6$} & \colhead{Dec$_6$} & \colhead{$S_i$} & \colhead{$\sigma_{S_i}$} & \colhead{$pS_i$} & \colhead{$\sigma_{pS_i}$} & \colhead{$f_{pol}$} & \colhead{$\sigma_{f_{pol}}$} \\  
\colhead{(J2000)} & \colhead{(J2000)} & \colhead{mJy} & \colhead{mJy} & \colhead{mJy} & \colhead{mJy} & \colhead{} & \colhead{} \\  
}
\startdata
17:40:55.280  & -29:10:25.13  & 32.410 & 0.454 & 3.37 & 0.31 &  0.104 &  0.010 \\  
17:42:21.455  & -29:13:00.29  & 41.160 & 0.370 & 1.47 & 0.32 &  0.036 &  0.008 \\  
\enddata
\tablecomments{Columns 1--2 give the position of the 6 cm source in J2000 coordinates; columns 3--4 give the integrated flux and its uncertainty; columns 5--6 give the integrated polarized flux and its uncertainty; columns 7--8 give the polarization fraction and its error.}
\end{deluxetable}

\begin{deluxetable}{cccccccc}
\tablecaption{Point Sources with 20 cm Polarization \label{20poln}}
\tablewidth{0pt}
\tablehead{
\colhead{RA$_{20}$} & \colhead{Dec$_{20}$} & \colhead{$S_i$} & \colhead{$\sigma_{S_i}$} & \colhead{$pS_i$} & \colhead{$\sigma_{pS_i}$} & \colhead{$f_{pol}$} & \colhead{$\sigma_{f_{pol}}$} \\  
\colhead{(J2000)} & \colhead{(J2000)} & \colhead{mJy beam$^{-1}$} & \colhead{mJy beam$^{-1}$} & \colhead{mJy} & \colhead{mJy} & \colhead{} & \colhead{} \\  
}
\startdata
17:41:26.196  &  -28:53:30.04  &  47.69  & 1.41 &  9.07 & 2.15 & 0.190 & 0.045 \\ 
17:40:41.808  &  -28:48:13.61  &  100.1  & 1.88 &  3.79\tablenotemark{a} & 0.28 & 0.038 & 0.003 \\ 
17:40:07.992   &  -28:42:4.50   &  246.3  & 3.14 &  6.08\tablenotemark{a} & 0.28 & 0.025 & 0.001 \\ 
17:40:54.433  &  -29:29:46.78  &  102.1  & 2.02 & 24.84 & 2.82 & 0.243 & 0.028 \\ 
17:42:21.477  &  -29:13:1.49   &  104.0  & 1.50 & 22.65 & 2.84 & 0.218 & 0.027 \\ 
17:42:11.868  &  -29:13:31.04  &  28.50  & 1.43 &  1.60\tablenotemark{a} & 0.28 & 0.056 & 0.010 \\ 
17:42:31.138  &  -29:16:41.55  &  23.40  & 1.38 &  1.18\tablenotemark{a} & 0.28 & 0.051 & 0.012 \\ 
\enddata 
\tablenotetext{a}{Source parameters found by fixing the source size to the beam size of that field.}
\tablecomments{Columns 1--2 give the position of the 20 cm source in J2000 coordinates; columns 3--4 give the integrated 20 cm flux and its uncertainty; columns 5--6 give the integrated 20 cm polarized flux and its uncertainty; columns 7--8 give the polarization fraction and its error.}
\end{deluxetable}

\paragraph{Positional Accuracy}
The absolute coordinate alignment of the point source catalogs was tested by measuring the mean offset between compact radio sources unambiguously matched with existing catalogs.  Since the source density is low, most sources with counterparts in existing surveys were unambiguous.  Sources from the present 6 and 20 cm catalogs were compared to each other and to those of \citet{n04}, \citet{y04}, \citet{l98}, and \citet{b94}.  These catalogs were chosen for comparison because they have the highest resolutions and best sky coverage at these wavelengths.  No significant offset was found between the present 6 and 20 cm catalogs and that of \citet{l98}, which gave the smallest rms in the offset angles of 0\dasec36 for the 6 cm catalog (5 sources), and 2\dasec2 for the 20 cm catalog (7 sources).  These values are the best estimates of the absolute pointing accuracy of our catalogs.  There was no significant offset between fifteen 20 cm sources matched to the catalog of \citet{y04} (only relatively unconfused sources, north of $b=$0\ddeg1, were considered).  Between the 6 and 20 cm catalogs presented here, no significant offset was found, with ($\Delta$RA,$\Delta$Dec) = (0\dasec09$\pm$1\dasec5,0\dasec01$\pm$1\dasec5) for 32 sources.  There was a $\sim1\sigma$ offset between the eleven 6 cm sources matched to the catalog of \citet{b94}, with offsets of (--0\dasec31$\pm$0\dasec52,0\dasec99$\pm$0\dasec87);  this offset is only marginally significant and is ignored.  There was a significant RA offset between both the 6 and 20 cm catalogs with the 90 cm catalog of \citet{n04}.  The mean RA difference between the 19 sources common between the present surveys catalog and the 90 cm catalog is $RA_{20,6 \rm{cm}}-RA_{90 \rm{cm}}=-4$\dasec74$\pm$2\dasec82.  This error is about twice their quoted astrometric accuracy \citep[2\dasec1;][]{n04} and has since been attributed to an error \citep{n04}.

\paragraph{Spectral Indices}
Table \ref{ps6-20} shows the positions and spectral indices (assuming $S\propto\nu^{\alpha}$) of sources found in both the 6 and 20 cm surveys.  Sources found only in the 6 cm survey are shown in Table \ref{ps6no20}; lower limits on $\alpha$ are based on the $3\sigma$ upper limit on the 20 cm nondetection.  There are 33 sources that are found in both the 6 and 20 cm catalogs, 60 sources found only in the 6 cm catalog, and no sources found only in the 20 cm catalog.  The significance limit for matching sources in the two catalogs is lowered to $3\sigma$.

\begin{deluxetable}{cccccc}
\tablecaption{Point Sources in both 6 and 20 cm Catalogs \label{ps6-20}}
\tablewidth{0pt}
\tablehead{
\colhead{RA$_6$} & \colhead{Dec$_6$} & \colhead{RA$_{20}$} & \colhead{Dec$_{20}$} & \colhead{$\alpha$ \tablenotemark{a}} & \colhead{$\sigma_\alpha$} \\
\colhead{(J2000)} & \colhead{(J2000)} & \colhead{(J2000)} & \colhead{(J2000)} & \colhead{} & \colhead{} \\
}
\startdata
17:42:07.917 & -29:18:51.92 & 17:42:07.825 & -29:18:47.95 & -1.08 & 0.38 \\
17:40:31.410 & -29:20:19.11 & 17:40:31.357 & -29:20:19.10 & -0.07 & 0.18 \\
17:40:55.269 & -29:06:53.64 & 17:40:55.084 & -29:06:57.22 & -1.19 & 0.21 \\
17:40:55.280 & -29:10:25.13 & 17:40:55.212 & -29:10:25.42 & -0.22 & 0.04 \\
17:41:30.698 & -29:11:02.97 & 17:41:30.683 & -29:11:02.82 & -0.95 & 0.09 \\
17:41:45.418 & -29:16:20.83 & 17:41:45.520 & -29:16:20.15 & -0.32 & 0.20 \\
17:41:59.993 & -29:10:47.53 & 17:42:00.060 & -29:10:49.40 & -0.66 & 0.15 \\
17:42:10.672 & -29:20:53.95 & 17:42:10.778 & -29:20:54.23 & -1.10 & 0.24 \\
17:42:31.096 & -29:16:42.26 & 17:42:31.138 & -29:16:41.55 & -0.01 & 0.05 \\
17:42:11.828 & -29:13:29.73 & 17:42:11.868 & -29:13:31.04 & -0.33 & 0.04 \\
17:42:21.455 & -29:13:00.29 & 17:42:21.477 & -29:13:01.49 & -0.76 & 0.01 \\
17:43:12.190 & -29:26:03.67 & 17:43:12.193 & -29:26:01.47 & 0.41 & 0.09 \\
17:43:10.184 & -29:12:14.49 & 17:43:10.112 & -29:12:13.24 & 0.622 & 0.126 \\
17:43:28.881 & -29:06:47.19 & 17:43:28.869 & -29:06:46.42 & -0.61 & 0.04 \\
17:44:37.053 & -28:57:09.45 & 17:44:36.688 & -28:57:10.19 & -0.94 & 0.04 \\
17:43:14.130 & -28:56:01.58 & 17:43:14.193 & -28:56:01.29 & -0.38 & 0.30 \\
17:43:02.749 & -28:51:09.16 & 17:43:02.797 & -28:51:12.56 & -0.54 & 0.15 \\
17:43:03.459 & -28:50:55.42 & 17:43:03.318 & -28:50:57.45 & -1.50 & 0.14 \\
17:42:39.650 & -28:41:44.50 & 17:42:39.714 & -28:41:40.81 & -0.69 & 0.14 \\
17:41:53.045 & -29:01:17.97 & 17:41:52.987 & -29:01:16.62 & -0.04 & 0.23 \\
17:41:42.851 & -28:59:02.78 & 17:41:42.870 & -28:59:02.10 & 0.04 & 0.26 \\
17:42:08.606 & -28:55:36.43 & 17:42:08.486 & -28:55:36.00 & 0.22 & 0.09 \\
17:42:31.068 & -28:59:45.85 & 17:42:31.205 & -28:59:45.20 & 0.31 & 0.18 \\
17:41:26.162 & -28:53:29.12 & 17:41:26.196 & -28:53:30.04 & -0.85 & 0.03 \\
17:43:31.215 & -28:38:07.69 & 17:43:31.081 & -28:38:07.83 & -0.89 & 0.12 \\
17:43:20.500 & -28:36:33.38 & 17:43:20.635 & -28:36:32.21 & -1.25 & 0.26 \\
17:43:45.577 & -28:32:05.88 & 17:43:45.718 & -28:32:06.04 & 0.50 & 0.16 \\
17:44:18.971 & -28:41:24.26 & 17:44:19.104 & -28:41:21.82 & 0.37 & 0.18 \\
17:44:22.882 & -28:38:51.19 & 17:44:23.011 & -28:38:51.75 & -0.09 & 0.12 \\
17:44:40.866 & -28:39:56.24 & 17:44:40.820 & -28:39:55.55 & 0.31 & 0.07 \\
17:44:45.917 & -28:35:15.24 & 17:44:45.929 & -28:35:14.33 & 0.49 & 0.12 \\
17:44:57.379 & -28:38:02.30 & 17:44:57.168 & -28:38:02.72 & -0.28 & 0.11 \\
17:44:48.775 & -28:28:13.20 & 17:44:48.755 & -28:28:14.85 & -0.85 & 0.07 \\
\enddata
\tablecomments{Comparing all sources with greater than $3\sigma$ confidence according to the AIPS SAD/JMFIT routines.  Columns 1--2 give the position of the 6 cm source in J2000 coordinates; columns 3--4 give the position of the 20 cm source; columns 5--6 give the 6/20 cm spectral index from the integrated fluxes of the source and its uncertainty.}
\tablenotetext{a}{Assumes  $S_\nu\propto\nu^{\alpha}$.}
\end{deluxetable}

\begin{deluxetable}{ccc}
\tablecaption{Point Sources in 6 cm Catalog Only \label{ps6no20}}
\tablewidth{0pt}
\tablehead{
\colhead{RA$_6$} & \colhead{Dec$_6$} & \colhead{min $\alpha$ \tablenotemark{a}} \\
\colhead{(J2000)} & \colhead{(J2000)} & \colhead{} \\
}
\startdata
17:43:34.257 & -28:30:42.55 & -0.55 \\
17:43:41.627 & -28:32:47.90 & -0.70 \\
17:43:42.470 & -28:33:28.05 & -0.15 \\
17:44:54.553 & -28:44:59.10 & -0.33 \\
17:44:31.059 & -28:47:16.36 & 0.07 \\
17:43:47.704 & -28:41:49.36 & 0.75 \\
17:43:39.770 & -28:41:51.71 & -0.71 \\
17:43:17.041 & -28:43:01.67 & 0.24 \\
17:43:22.961 & -28:44:52.30 & -0.85 \\
17:43:12.349 & -28:44:33.67 & -0.86 \\
17:43:10.118 & -28:45:17.28 & -1.41 \\
17:43:23.117 & -28:47:53.97 & -0.53 \\
17:42:28.738 & -28:41:48.90 & -0.20 \\
17:42:44.729 & -28:45:48.66 & 0.18 \\
17:42:44.519 & -28:45:37.68 & -0.03 \\
17:42:04.905 & -28:42:10.41 & 0.43 \\
17:43:46.895 & -28:57:05.44 & -0.62 \\
17:42:24.478 & -28:45:57.63 & -1.03 \\
17:43:05.476 & -28:52:21.88 & -0.33 \\
17:43:45.604 & -28:58:56.53 & -0.18 \\
17:42:42.902 & -28:50:55.25 & -0.86 \\
17:44:02.597 & -29:02:14.04 & 0.27 \\
17:43:54.715 & -29:01:32.42 & -0.55 \\
17:43:42.562 & -29:01:33.72 & 0.65 \\
17:43:42.788 & -29:02:26.58 & 0.27 \\
17:42:23.776 & -28:52:48.41 & 0.61 \\
17:42:13.815 & -28:52:10.78 & -0.11 \\
17:42:42.652 & -28:52:04.75 & -0.20 \\
17:43:17.039 & -29:02:12.27 & -0.23 \\
17:42:06.927 & -28:54:03.26 & 0.29 \\
17:42:53.589 & -29:00:53.94 & -0.09 \\
17:42:03.496 & -28:55:05.04 & 0.13 \\
17:43:39.267 & -29:09:07.81 & 0.26 \\
17:42:41.764 & -29:02:10.19 & 0.44 \\
17:42:57.261 & -29:07:30.73 & -0.45 \\
17:43:43.443 & -29:13:57.65 & -0.48 \\
17:43:42.827 & -29:14:33.81 & -0.46 \\
17:42:49.196 & -29:07:25.20 & 0.28 \\
17:43:36.037 & -29:14:39.50 & -0.43 \\
17:42:30.636 & -29:05:59.66 & 0.96 \\
17:43:34.857 & -29:15:04.12 & -1.14 \\
17:44:03.763 & -29:19:53.25 & -0.08 \\
17:43:37.410 & -29:16:36.86 & 0.16 \\
17:44:09.038 & -29:21:33.34 & -0.20 \\
17:41:23.090 & -29:01:04.19 & 0.36 \\
17:42:38.843 & -29:12:17.97 & 0.92 \\
17:43:02.015 & -29:15:49.07 & 0.43 \\
17:43:53.398 & -29:23:30.23 & 1.34 \\
17:41:41.632 & -29:08:31.65 & 0.81 \\
17:41:21.951 & -29:07:17.07 & -0.66 \\
17:43:40.559 & -29:27:09.47 & -0.81 \\
17:43:41.001 & -29:28:41.05 & 0.07 \\
17:43:02.312 & -29:23:42.34 & -0.17 \\
17:41:24.268 & -29:10:52.57 & 1.60 \\
17:42:53.789 & -29:24:57.01 & 0.47 \\
17:41:07.573 & -29:10:35.18 & 0.57 \\
17:43:09.332 & -29:28:58.20 & -0.38 \\
17:43:10.195 & -29:29:52.04 & -0.15 \\
17:41:02.201 & -29:12:33.24 & 0.65 \\
17:41:28.400 & -29:16:34.27 & 0.46 \\
\enddata
\tablenotetext{a}{Assumes  $S_\nu\propto\nu^{\alpha}$.}
\tablecomments{Columns 1--2 give the position of the 6 cm source in J2000 coordinates; column 3 gives the limit on the 6/20 cm spectral index.}
\end{deluxetable}

\paragraph{Consistency Between 20 cm catalog and Literature}
As a test of the accuracy of the fluxes in the point source catalog, fluxes of sources common with the catalog of \citet{y04} were compared.  The fifteen sources (only those north of $b=$0\ddeg1 were considered for simplicity)  common to the two catalogs are shown in Table \ref{yhc}.  The weighted mean offset from exactly equal fluxes for these two catalogs is $-0.18\pm0.94$ mJy.  No trend is evident in the distribution of flux density difference for these sources;  all sources are within 3 sigma of the linear trend.

\begin{deluxetable}{ccccccc}
\tablecaption{20 cm Sources Matching the Catalog of \citet{y04} with $b>$0\ddeg1 \label{yhc}}
\tablewidth{0pt}
\tablehead{
\colhead{RA$_{20}$} & \colhead{Dec$_{20}$} & \colhead{$S_i$} & \colhead{$\sigma_{S_i}$} & \colhead{Previous Name} & \colhead{$S$} & \colhead{$\sigma_{S}$} \\  
\colhead{(J2000)} & \colhead{(J2000)} & \colhead{mJy} & \colhead{mJy} & \colhead{} & \colhead{mJy} & \colhead{mJy} \\
}
\startdata
17:44:40.820 & -28:39:55.55 & 13.11 & 1.150 & G0.12+0.32   & 17.004 & 2.79 \\
17:44:19.104 & -28:41:21.82 & 4.319 & 0.940 & G0.06+0.37   & 4.7929 & 1.46 \\
17:44:36.688 & -28:57:10.19 & 171.8 & 5.260 & G359.87+0.18 & 153.7  & 1.92 \\
17:43:12.193 & -29:26:01.47 & 14.21 & 1.620 & G359.29+0.19 & 17.809 & 1.816 \\
17:43:28.869 & -29:06:46.42 & 45.78 & 1.960 & G359.60+0.31 & 55.205 & 2.898 \\
17:43:10.112 & -29:12:13.24 & 17.05 & 2.370 & G359.49+0.32 & 9.588  & 1.659 \\
17:42:31.138 & -29:16:41.55 & 23.40 & 1.380 & G359.35+0.40 & 22.46  & 1.954 \\
17:42:10.778 & -29:20:54.23 & 5.740 & 1.450 & G359.26+0.42 & 4.9947 & 1.95 \\
17:42:21.477 & -29:13:01.49 & 104.0 & 1.500 & G359.39+0.46 & 102.00 & 1.826 \\
17:42:11.868 & -29:13:31.04 & 28.50 & 1.430 & G359.36+0.49 & 32.231 & 1.917 \\
17:41:45.520 & -29:16:20.15 & 8.018 & 1.960 & G359.27+0.54 & 8.503  & 2.120 \\
17:40:54.433 & -29:29:46.78 & 102.1 & 2.020 & G358.98+0.58 & 103.34 & 2.352 \\
17:41:30.683 & -29:11:02.82 & 19.35 & 2.020 & G359.32+0.63 & 20.059 & 2.006 \\
17:40:55.212 & -29:10:25.42 & 42.43 & 1.810 & G359.26+0.75 & 33.785 & 2.205 \\
17:42:00.060 & -29:10:49.40 & 9.320 & 1.630 & G359.38+0.55 & 14.124 & 1.967 \\
\enddata
\tablecomments{Columns 1--2 give the position of the 20 cm source in J2000 coordinates; columns 3--4 give the integrated 20 cm flux and its uncertainty; column 5 gives the name of the source in the survey of \citet{y04};  columns 6--7 give the integrated 20 cm flux measured by \citet{b94} and its uncertainty.}
\end{deluxetable}

As another consistency check, the present survey was correlated with that of \citet{l98}, as shown in Table \ref{1lc}.  That work observed in the BnA configuration with a similar sensitivity as the present 20 cm survey (0.5--2 mJy beam$^{-1}$).  The present survey had a mean frequency at 1425 MHz, compared to the two bands at 1658 and 1281 MHz for \citet{l98}.  The expectation is that our flux density measurement should fall near to the mean of their two fluxes.  This is true for the following sources, all of which have matching fluxes within 3$\sigma$: 359.076+0.547, 359.259+0.74, 359.073+0.735, 359.388+0.460, 359.088+0.426.  However, excluding one source, 359.388+0.460, for having a source detection problem\footnote{The source detection of \citet{l98} shows this source to be unresolved at 1658 MHz and highly resolved at 1281 MHz.  Considering that there is unlikely to be any difference in source structure over this narrow range of frequencies, this is likely an error in source detection, which makes the flux density unreliable.}, the fluxes presented here tend to be higher than that of \citet{l98}.  The average flux difference between this survey and that of \citet{l98} for the four sources with similar fluxes is $2.7\pm1.7$ mJy ($1\sigma$ error).  Two sources, 0.305+0.394 and 359.872+0.178, are significantly brighter in the present survey than in \citet{l98}.  The observations described in \citet{l98} may resolve out some of the flux that is scattered into an arcmin-scale halo \citep{l98b}, as described in detail in \S\ \ref{vla_compact}.  In this case, the flux density differences cannot be due to sensitivity differences between the two surveys, since the survey as identical sensitivities.

\begin{deluxetable}{ccccccc}
\tablecaption{20 cm Sources Matching the Catalog of \citet{l98} \label{1lc}}
\tablewidth{0pt}
\tablehead{
\colhead{RA$_{20}$} & \colhead{Dec$_{20}$} & \colhead{$S_i$} & \colhead{$\sigma_{S_i}$} & \colhead{Previous Name} & \colhead{$S_{i1658}$} & \colhead{$S_{i1281}$} \\  
\colhead{(J2000)} & \colhead{(J2000)} & \colhead{mJy} & \colhead{mJy} & \colhead{} & \colhead{mJy} & \colhead{mJy} \\
}
\startdata
17:41:16.049 & -29:26:06.66 & 3.090 & 0.848 & 359.076+0.547 & 2.7 & 2.9 \\
17:44:48.755 & -28:28:14.85 & 30.950 & 1.750 & 0.305+0.394 & 9.4 & 12.8 \\
17:40:55.212 & -29:10:25.42 & 42.430 & 1.810 & 359.259+0.749 & 36.5 & 39.4 \\
17:40:31.357 & -29:20:19.10 & 5.992 & 1.230 & 359.073+0.735 & 5.3 & 4.3 \\
17:44:36.688 & -28:57:10.19 & 171.800 & 5.260 & 359.872+0.178 & 83.1 & 81.2 \\
17:42:21.477 & -29:13:01.49 & 104.000 & 1.500 & 359.388+0.460 & 60.6 & 155.8 \\
17:41:46.037 & -29:29:16.26 & 8.960 & 2.340 & 359.088+0.426 & 5.0 & 3.6 \\
\enddata
\tablecomments{Columns 1--2 give the position of the 20 cm source in J2000 coordinates; columns 3--4 give the integrated 20 cm flux and its uncertainty; column 5 gives the name of the source in the survey of \citet{l98};  columns 6--7 give the integrated fluxes measured in two bands near 20 cm by \citet{l98}.}
\end{deluxetable}

\paragraph{Consistency Between 6 cm catalog and Literature}
As a consistency check, the flux densities of sources detected in both the present 6 cm catalog and that of \citet{b94} were compared.  \citet{b94} observed with the CnB array with a resolution of 4\arcsec\ for a typical sensitivity of 2.5mJy.  That survey observed in a hexagonal pattern with neighboring pointings spaced by 10\arcmin\ with integration times of about 90 s.

Table \ref{gpsr5} shows that, of the 13 sources detected in both surveys, the flux densities measured here are uniformly higher than that of the \citet{b94}, often by more than a factor of two.  We have considered several possible origins for this difference.  The systematic offset between the two surveys and the large area over which sources are detected ($\sim$1\sdeg) argues against intrinsic variability or changes in any sort of Galactic foreground.  The best-fit peak brightness is similar to the integrated flux density in our survey, so there doesn't seem to be problems with confusion.  In the present survey, the peak brightness is similar to the max pixel value near the point source and the fit residuals are small compared to the source peak brightness, as expected for a good fit.  To test for possible calibration problems, we reimaged one of the four days of 6 cm data in one field;  the sources in that field had a similar best-fit flux density as found in the image of all the data.  Both surveys used the same flux calibrator, 3C286, and list integrated flux brightnesses corrected for primary beam attenuation.  The most likely explanation for the difference in fluxes is the difference in integration times for the two surveys.  The shorter integration times used by \citet{b94} reduces the sensitivity of the observations and may bias the measured fluxes.  Another possibility for a difference in flux density between the two surveys is that they were conducted in different configurations of the VLA and have sensitivities to emission on different angular scales.  \citet{b94} imaged their data while excluding the shortest ($<90$ m) baselines, which makes it insensitive to emission on angular scales larger than $\sim$120\arcsec.  The present survey was observed in a more compact configuration and excluded fewer of the short baselines, so it is sensitive to emission on angular scales roughly twice that of \citet{b94}.

\begin{deluxetable}{cccccc}
\tablecaption{6 cm Sources Matching the Catalog of \citet{b94} \label{gpsr5}}
\tablewidth{0pt}
\tablehead{
\colhead{RA$_6$} & \colhead{Dec$_6$} & \colhead{$S_{i6}$} & \colhead{$\sigma_{S_{i6}}$} & \colhead{Previous Name} & \colhead{$S_{i}$} \\  
\colhead{(J2000)} & \colhead{(J2000)} & \colhead{mJy} & \colhead{mJy} & \colhead{} & \colhead{mJy} \\  
}
\startdata
17:44:48.775 & -28:28:13.20 & 10.911 & 0.702 & 0.306+0.394 & 6.8 \\
17:44:45.917 & -28:35:15.24 & 12.697 & 0.337 & 0.201+0.342 & 5.2 \\
17:44:57.379 & -28:38:02.30 & 12.749 & 0.685 & 0.185+0.282 & 4.5 \\
17:44:40.866 & -28:39:56.24 & 19.177 & 0.537 & 0.127+0.317 & 13.6 \\
17:44:18.971 & -28:41:24.26 & 6.830 & 0.241 & 0.059+0.372 & 5.9 \\
17:44:37.053 & -28:57:09.45 & 54.077 & 2.356 & 359.873+0.178 & 37.2 \\
17:43:28.883\tablenotemark{a} & -29:06:47.50 & 21.44 & 0.320 & 359.606+0.305 & 7.6 \\
17:43:28.883\tablenotemark{a} & -29:06:47.50 & 21.44 & 0.320 & 359.604+0.307 & 3.3 \\
17:43:10.184 & -29:12:14.49 & 20.791 & 0.207 & 359.491+0.316 & 10.1 \\
17:42:31.096 & -29:16:42.26 & 23.023 & 0.264 & 359.354+0.398 & 17.4 \\
17:43:12.190 & -29:26:03.67 & 23.561 & 0.418 & 359.299+0.189 & 17.9 \\
17:43:09.332 & -29:28:58.20 & 2.357 & 0.305 & 359.253+0.172 & 1.4 \\
17:43:10.195 & -29:29:52.04 & 3.935 & 0.751 & 359.242+0.162 & 1.5 \\
\enddata
\tablenotetext{a}{Two sources in \citet{b94} are unresolved here.}
\tablecomments{Columns 1--2 give the position of the 6 cm source in J2000 coordinates; columns 3--4 give the integrated 6 cm flux and its uncertainty; column 5 gives the name of the source in the survey of \citet{b94};  column 6 gives the integrated 6 cm flux measured by \citet{b94}.}
\end{deluxetable}

\begin{figure}[tbp]
\includegraphics[width=\textwidth]{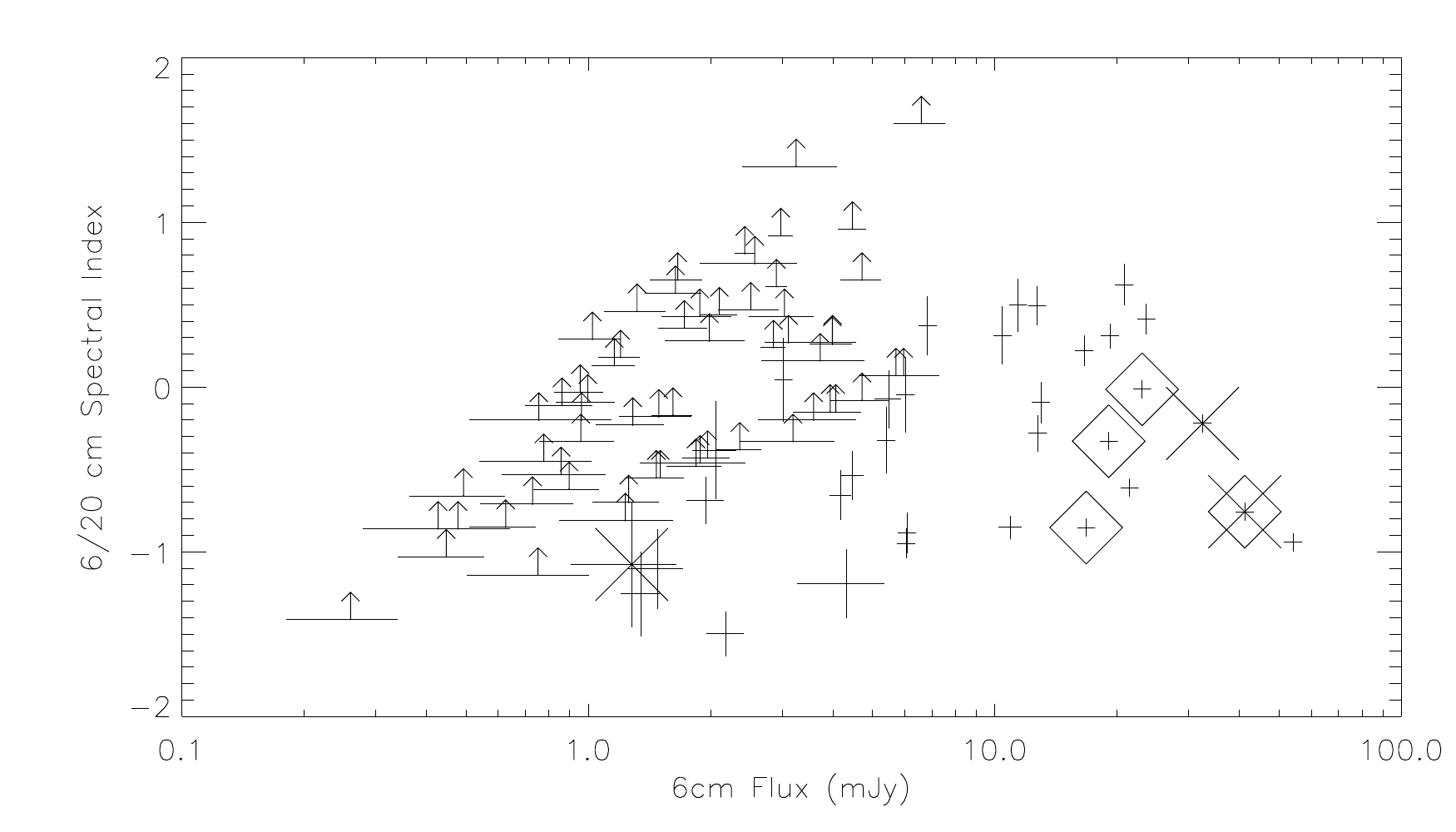}
\caption{A flux-spectral index diagram for point sources detected at 6 and 20 cm.  Lower limits to the spectral index are shown with arrows.  The two sources polarized at 6 cm are shown with crosses and the four 6/20 cm sources with 20 cm polarization are shown with diamonds. \label{cmd206}}
\end{figure}

\paragraph{Spectral Index Distribution Between 6, 20, and 90 cm}
The distribution of 6/20 cm spectral index of point sources is shown in Figure \ref{cmd206}.  Sources with polarization are indicated with larger symbols.  All sources are detected at 6 cm, so they at least have a lower limit to $\alpha_{LC}$.


Tables \ref{6poln} and \ref{20poln} list the point sources detected with 3$\sigma$-confidence in linearly polarized flux at 6 cm and 20 cm, respectively.  All sources with polarized emission at 6 or 20 cm are among the brightest total intensity sources, as seen in Figures \ref{cmd206} and \ref{cmd9020}.  Two sources are linearly polarized at 6 cm, and seven sources at 20 cm.  Both 6 cm polarized sources have 20 cm total intensity counterparts, three sources are only polarized at 20 cm, and one source is polarized at both 20 cm and 6 cm.  The source polarized at 6 and 20 cm is G359.388+0.460.  \citet{r05} find the source has 3\% polarization fraction at 8.5 GHz, which is similar to that observed here at 5 GHz;  the polarization fractions increase at longer wavelengths with $3.6\pm0.8$\% at 6 cm (5 GHz) and $21.8\pm2.7$\% at 20 cm.  From that study, linearly polarized sources G359.604+0.306 and G359.871+0.179 are also in the field of the current survey, but no significant linearly polarized emission is detected; this is consistent with an extrapolation of their observed polarization fractions and flux densities at 8.5 GHz.

\begin{figure}[tbp]
\includegraphics[width=\textwidth]{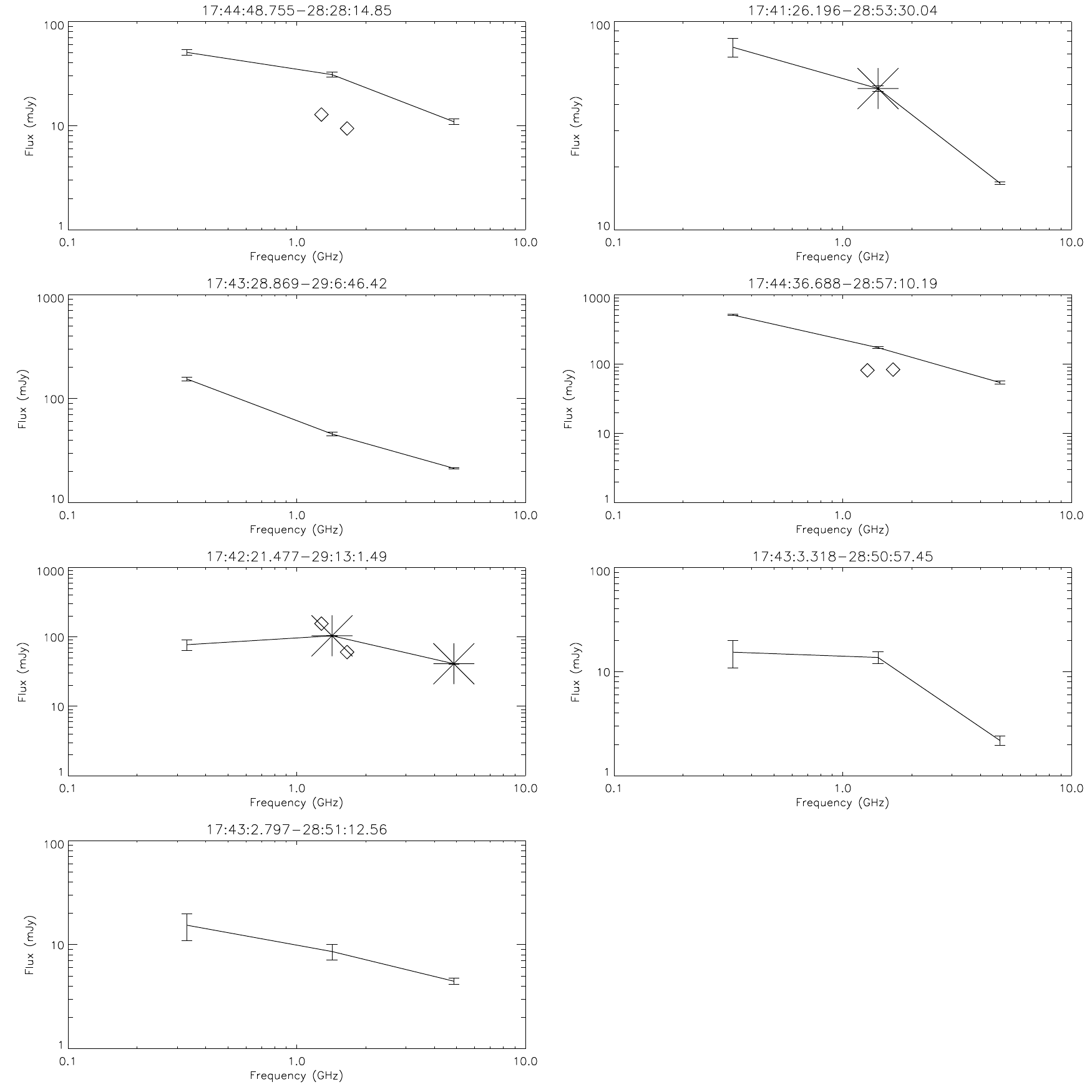}
\caption{Plots of flux density distribution for seven sources that have 90, 20, and 6 cm flux densities measured. Small diamonds show fluxes from \citet{l98}, where available, and large stars show sources that have linear polarization here. 
\label{spectra90}}
\end{figure}

\begin{figure}[tbp]
\includegraphics[width=\textwidth]{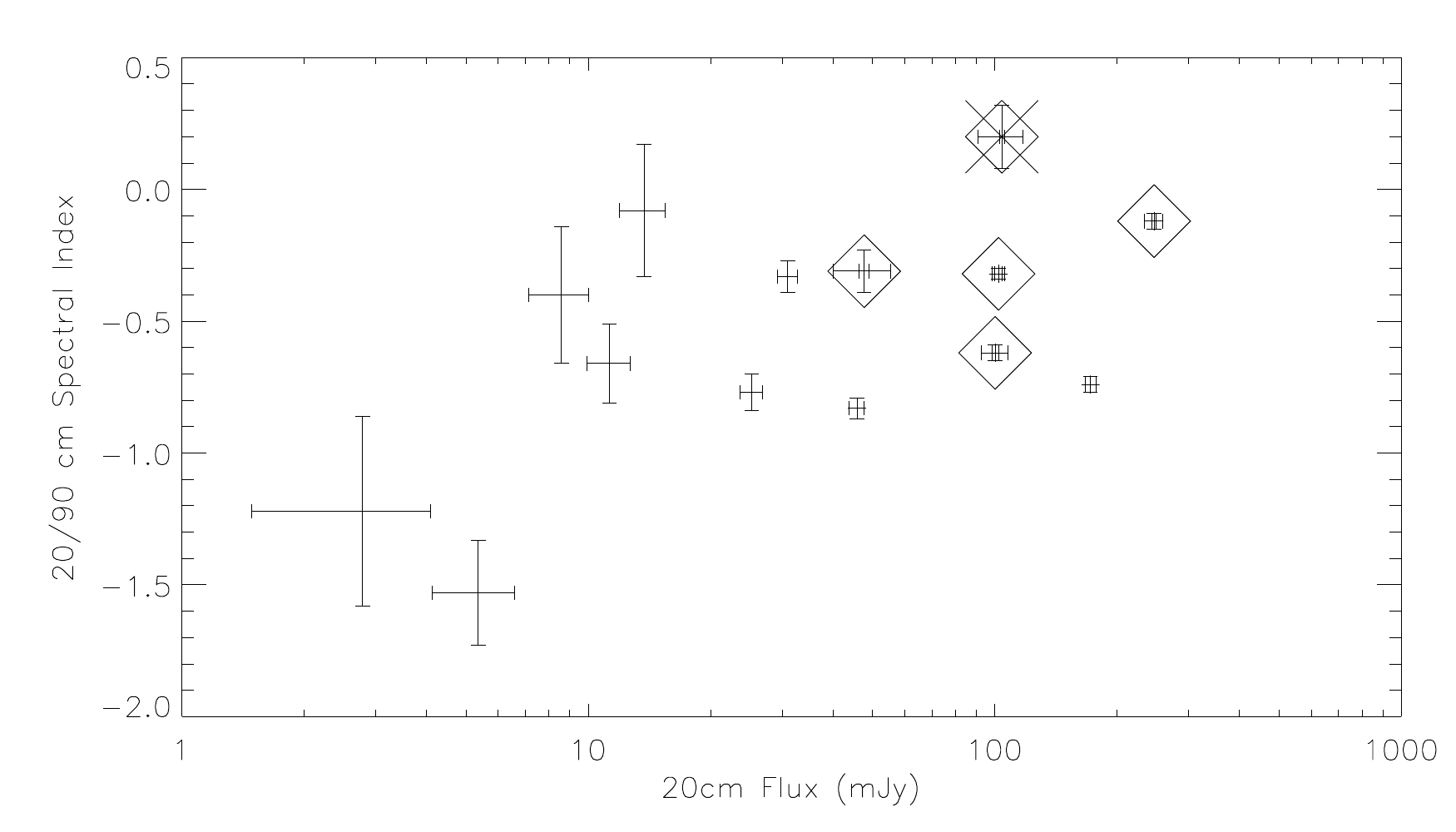}
\caption{A flux-spectral index diagram for the fourteen 20 cm sources with 90 cm counterparts in \citet{n04}. The five, 20/90 cm sources with 20 cm polarization are shown with diamonds and the one source polarized at 6 cm is shown with a cross. \label{cmd9020}}
\end{figure}

Table \ref{gcps} shows the spectral indices for all sources detected in either of the present two surveys, which are also detected in the survey of \citet{n04}.  All 6 cm sources with 90 cm counterparts also have 20 cm counterparts here.  Note that the resolution of the 90 cm survey (7\arcsec$\times$12\arcsec) is almost identical to that of the surveys presented here, so the spectral indices should be fairly accurate.  Figure \ref{spectra90} shows a more detailed look at the fluxes for the seven sources that have counterparts at 90, 20, and 6 cm.  Most of them (5/7) have spectra steepening with increasing frequency, while one has a nearly constant spectral index, and one has a flattening spectrum.

\begin{deluxetable}{cccccc}
\tablecaption{Spectral Indices for Sources in the Catalog of \citet{n04} \label{gcps}}
\tablewidth{0pt}
\tablehead{
\colhead{RA$_{20}$} & \colhead{Dec$_{20}$} & \colhead{$\alpha_{20/90}$} & \colhead{$\sigma_{\alpha_{20/90}}$} & \colhead{$\alpha_{6/20}$} & \colhead{$\sigma_{\alpha_{6/20}}$} \\  
\colhead{(J2000)} & \colhead{(J2000)} & \colhead{} & \colhead{} & \colhead{} & \colhead{} \\  
}
\startdata
17:44:48.755 & -28:28:14.85 & 0.33 & 0.06 & 0.85 & 0.07 \\
17:42:48.347 & -28:25:28.35 & 0.77 & 0.07 & -- & -- \\
17:41:49.359 & -28:43:20.45 & 1.22 & 0.36 & -- & -- \\
17:41:26.196 & -28:53:30.04 & 0.31 & 0.08 & 0.85 & 0.03 \\
17:40:41.808 & -28:48:13.61 & 0.62 & 0.03 & -- & -- \\
17:40:33.419 & -28:44:05.08 & 1.53 & 0.20 & -- & -- \\
17:40:07.992 & -28:42:04.50 & 0.12 & 0.03 & -- & -- \\
17:40:20.487 & -29:13:44.37 & 0.66 & 0.15 & -- & -- \\
17:40:54.433 & -29:29:46.78 & 0.32 & 0.02 & -- & -- \\
17:43:28.869 & -29:06:46.42 & 0.83 & 0.04 & 0.61 & 0.04 \\
17:44:36.688 & -28:57:10.19 & 0.74 & 0.03 & 0.94 & 0.04 \\
17:42:21.477 & -29:13:01.49 & -0.20 & 0.12 & 0.76 & 0.01 \\
17:43:03.318 & -28:50:57.45 & 0.08 & 0.25 & 1.50 & 0.14 \\
17:43:02.797 & -28:51:12.56 & 0.40 & 0.26 & 0.54 & 0.15 \\
\enddata
\tablecomments{Columns 1--2 give the position of the 20 cm source in J2000 coordinates; columns 3--4 give the 20/90 cm spectral index and its uncertainty; columns 5--6 give the 6/20 cm spectral index and its uncertainty.}
\end{deluxetable}

Figure \ref{cmd9020} shows the distribution of $\alpha_{20/90}$ as a function of 20 cm flux density for the fourteen 20 cm sources detected at 90 cm by \citet{n04}.  Most of the sources have 20/90 cm spectral indices between --0.3 to --0.8 for flux densities greater than 10 mJy.  This range is flatter than the range of --0.6 to --1.2 observed in other 90 cm-limited samples of spectral index \citep[for a similar extrapolated flux range][]{g04,n04}.  This range of spectral indices is typically dominated by extragalactic sources like radio galaxies, as has been argued in the field of M31 \citep{g04}.  One difference between the previous surveys and the present work is that the present survey is limited to the central degree of the GC, which is believed to be host to a region of hyperstrong scattering \citep{l98b}.  In \S\ \ref{hyper}, we discuss the effects of the putative hyperstrong scattering screen believed to exist near the GC region on source detection.  This screen may limit our ability to detect low-frequency emission from extragalactic sources through the central degree, which may bias our 90 cm flux-limited sample.

Of the pulsars and pulsar candidates listed in \citet{n04}, only one is detected in the present surveys, mostly due to our smaller spatial coverage.  This source is called 174020--291344 \citep[a.k.a. ``359.145+0.826'';][]{n04}.  The pulsar candidate has an estimated spectral index of $\alpha_{20/90}=-1.4$ by comparing 20 and 90 cm fluxes from \citet{c98} and \citet{n04}, respectively.  However, the 20 cm flux density of 174020--291344 measured by the present survey implies a spectral index $\alpha_{20/90}=-0.6\pm0.15$.  This value is significantly flatter than measured previously and is flatter than that expected from typical radio pulsars \citep[$\alpha_{20/90}\lesssim-1.0$;][]{ma05}.

\section{Discussion and Conclusions}
\label{vla_dis}
\subsection{Radio Filaments}
In total 15 radio filaments were detected in the present survey, with eight found to have polarized flux;  three of these NRFs were found to have polarized flux for the first time, thus confirming that they are NRFs.  One NRF in particular, called N11 (G359.64+0.30), is found to have a radically different morphology at 6 cm than at 20 cm.  N11 is found to split into two filaments, one of which is only detected at 6 cm, and crosses the other filament with a twisting or wavy pattern.  

Six radio filaments have spatial gradients in their 6/20 cm spectral indices on arcminute scales, while three show no clear gradients.  Radio filaments C7, C12, C16, N1, N2, and N8 show steady spectral index changes over their length in both feathered and VLA-only images.  C7 and C12 are two of the brighter filaments that are oriented more parallel than perpendicular to the Galactic plane and they both have steeper indices toward their southeast ends.  N1 and N2 are more vertical and tend to steepen toward the north.  N8 is oriented at a roughly 45\sdeg\ angle to the plane and has a steepening of the spectral index toward its southeast side, similar to C7 and C12.  The C3, N10, and N11 filaments have no significant change in their 6/20 cm spectral index values.  Oddly, one of these (N10) is one of the few NRFs to have been previously shown to have a gradient in its spectral index between 20/90 cm.  However, the lack of a 6/20 cm index gradient is consistent with a previous study \citep{l01}.

There are a few observational effects that could create a spatial gradient in a spectral index, although they cannot explain gradients seen in all of the filaments.  Most observational effects are strongest at the edge of each field, where bandwidth smearing is strongest (see below) and primary beam corrections are largest.  However, as discussed before, the survey pointings were organized in a hexagonal pattern to reduce sensitivity variations to at most a factor of two between pointings.  For C7 and C12, the spectral index change does not occur across a field edge, where primary beam corrections might raise concerns.  More importantly, the noise measured in the slices is larger than the noise measured for each individual field.  This suggests that confusion dominates the spectral index uncertainty, and since confusion has no spatial structure (or at least is not correlated with the gradients in the filament spectral indices), it cannot explain the observed spectral index gradients.

Bandwidth smearing does not affect the slice analysis of the radio filaments significantly.  Filaments N1, N2, C12, and C16 are oriented radially to their nearest 20 cm field phase center, so smearing does not noticeably reduce their peak fluxes.  As described in \S\ \ref{nrfsec}, Filaments N8 and N10 are oriented perpendicular to the radial line from the nearest 20 cm field, so the spectral index values could be biased by $\Delta\alpha_{LC}<+0.4$.  However, there should be no bias to the gradient (or lack thereof) in the spectral index of these filaments, since the entire filament is offset from the 20 cm phase center by the same amount and thus should have a constant smearing throughout.  Comparing slice and photometric analysis of point sources near the edge of the 20 cm primary beam covering Filaments C3, C6, C7, C11, C12, and N11 shows that they are not significantly affected by bandwidth smearing.

Most previous studies of NRFs have found that the spectral index does not change across their lengths, but have not covered the same region as the present work \citep{l99,l00,l95}.  Specifically, previous work has studied the brightest NRFs near the Galactic plane, while the present work has studied filaments at higher latitudes ($b>$0\ddeg1).  Thus, it seems that spatial gradients in the 6/20 cm spectral index are more common away from the Galactic plane.  This is consistent with the fact that the Radio Arc has a relatively constant spectral index distribution near the plane, but steepens at high latitudes \citep[e.g., Fig. 9; see also Ch. \ref{gcsurvey_gbt};][]{p92}.  The trend for steepening of spectral indices away from the plane is consistent with existing observations of the high-latitude ($b\sim0$\ddeg4) filament N10 \citep{l01}.


Spatial gradients in a source's synchrotron spectral index has traditionally been explained by the relative ``age'' of parts of the plasma \citep[e.g.,][]{j73}.  This is caused by the fact that electrons with higher energies emit more power and cool more efficiently through a process called synchrotron cooling.  If a synchrotron spectrum initially extends up to infinite energy, synchrotron cooling will preferentially reduce the flux at higher frequencies, creating a break in the spectrum at a frequency $\nu_b\propto B^{-3}t^{-2}$ \citep{b00}.  Under this model, the spectral index spatial gradient in the filaments is an effect of electrons propagating from a single acceleration region and the regions with steeper spectral indices are interpreted as being further from the acceleration region.  For a 1 mG magnetic field and a spectral break between 6 and 20 cm, synchrotron aging implies an age of $2\times10^4$ yr \citep{l01}.  This age for an NRF, combined with estimates of diffusive electron transport velocity, are consistent with the length scales over which spectral index gradients have been observed, although a detailed analysis of electron diffusion in an NRF has not yet been done \citep{l01}.

Alternatively, \citet{l01} note that the gradient in the filament spectral index may be a consequence of how a curved electron energy distribution emits in a spatially changing magnetic field.  Since the observed frequency of synchrotron emission depends on the electron energy and magnetic field as $\nu_{\rm{syn}}=c B^2 E_{el}$, electrons observed at a given frequency will have different implied energies if the magnetic field strength changes.  A spatial gradient in the spectral index could indicate a weaker magnetic field, as follows:  $\delta\alpha=a_c log(B_1/B_2)$, where $a_c$ is the curvature in the electron energy distribution, and $B_1$ and $B_2$ are the magnetic field strengths across the region where the spectral index is observed to change \citep{b00}.  Spectral index values for filaments are often observed to steepen from $\alpha\sim-0.5$ to $-1.5$ for frequencies from 0.3 to 5 GHz \citep{l99,l01}.  Using the longest-wavelength spectral index as a baseline, a change of $\Delta\alpha=-1$ from $\nu=0.3$ to 5 GHz corresponds to a spectral curvature $a_c\approx1.5$.  Assuming this curvature for the electron energy distribution and a spectral index spatial gradient across the filaments of $\delta\alpha\sim0.5$ (similar to that seen here), the magnetic field strength across a filament falls by about a factor of two.

Unfortunately, very few filaments have clear detections of spectral curvature.  The 90 cm survey of filaments in the region do not include short-spacing data, so their fluxes are underestimates \citep{n04}; this provides an upper limit on the long-wavelength spectral index and no constraint on spectral curvature when compared to the present 6/20 cm spectral indices.  A reliable survey of 20/90 cm spectral index values in filaments could be compared to the measurements presented here for an estimate of the spectral curvature and, under the model of \citet{l01}, an estimate of the change in the magnetic field across several radio filaments.  Alternatively, the synchrotron aging hypothesis can be used to interpret the spectral index gradients as ages for the filaments.

\subsection{Nature of Compact Sources}
\label{vla_compact}
\paragraph{Spectral Index and Polarization Properties}
One way to discern the nature of the point sources is from the distribution of flux density versus spectral index.  Figure \ref{cmd206} shows that the 6/20 cm spectral index distribution follows the roughly bimodal distribution seen in radio point source surveys \citep{b94}.  Flat or positive (``inverted'') spectrum sources, with $\alpha_{LC}\gtrsim0.0$ are likely to be \hii\ regions or planetary nebulae \citep{g05b,i81}.  Positive spectral indices are more likely to be optically-thick, thermal emission.  A flat radio spectral index is also observed toward pulsar wind nebulae \citep{h87,ge05}.  Sources with $\alpha\approx-0.6$ are more likely to be FR I and FR II radio galaxies or supernova remnants \citep{f74,g04,ge05}.  

Since the 6 cm survey was more sensitive than the 20 cm survey ($\sigma_6\approx0.1$ mJy beam$^{-1}$, $\sigma_{20}\approx0.5$ mJy beam$^{-1}$), there are a significant number of lower limits to the 6/20 cm spectral indices that are consistent with thermal emission.  An optically-thin \hii\ region in our Galaxy (within 20 kpc) excited by a B0-type star or earlier would have $S_{\rm{6 cm}}>26$ mJy and $\alpha_{LC}=-0.1$ \citep{g05a}.  These values of flux density and spectral index correspond to a region of the plot in Figure \ref{cmd206} that is basically empty, so we are not detecting any such objects in this survey.  However, there are about two dozen sources with fluxes between 1 and 10 mJy and positive spectral indices, which would be expected for \hii\ regions that are optically thick at 6 cm.  Alternatively, planetary nebulae are expected to have a similar distribution of fluxes and spectral indices as \hii\ regions \citep{i81}, although they are much less numerous in the Galaxy \citep{g05a}.

Only two sources presented here have spectral indices approaching values commonly associated with pulsars, with $\alpha\lesssim-1.5$ \citep{lo95}: 174033-284405 (between 90 and 20 cm) and 174303-285057 (between 20 and 6 cm).  Taking a higher index threshold of $\alpha<-1.0$, includes five more candidates: 174207-291851, 174055-290653, 174210-292053, 174320-283633 (at 20 and 6 cm) and 174149-284320 (at 90 and 20 cm).  Steep radio continuum spectra can also be found in high-$z$ radio galaxies and relic radio galaxies, where synchrotron aging steepens the electron energy (and hence radio emission) spectrum \citep{k00,ge05}.  H{\small I} absorption observations may help localize the source location and constrain its nature.


There are significantly more sources polarized at 20 cm, (with 7; 3$\sigma$ detection) as compared to 6 cm (with 2).  Both surveys have similar sensitivity to polarized emission, with typical errors in the integrated, polarized, point source flux density of 0.3 mJy beam$^{-1}$.  Since depolarization effects increase as $\lambda^2$, the polarized flux from these sources must increase more rapidly than $\lambda^2$.  This suggests that the sources have fairly steep, nonthermal spectra and/or an intrinsically increasing polarization fraction with increasing wavelength.  As mentioned above, sources with radio spectral indices steeper than --1.3 are often identified with ``relic'' radio galaxies, which have radio lobes with steep spectra due to synchrotron cooling of high radio frequency emission \citep{ge05}.  Pulsars are also known to have steep radio spectra and emit polarized radio continuum, so they may contribute to the compact polarized sources observed in this survey.

\paragraph{Spatial Distribution}
There is no evidence for variation in the number density of sources (thermal- or nonthermal-like) over the range of Galactic latitudes covered in this survey, from $l=0$\ddeg1--0\ddeg8.  Considering that the scale height of Galactic \hii\ regions is 24\arcmin--28\arcmin\ \citep{g05b}, one might expect variation in the number density of sources with Galactic latitude, particularly flat- or inverted-spectrum (likely Galactic) sources.  One possible explanation for the lack of variation is that the sources are extragalactic in nature and have no Galactic dependence (although hyperstrong scattering may reduce the number of extragalactic sources visible in the region;  see \S\ \ref{hyper}).  It is important to note that the number density of thermal-like sources in this survey is much higher than that seen away from the Galactic plane.  Matching the flux limit to that of \citet{b94}, the number density of thermal-like, compact sources is about 37 per square degree, which is about twice the source density of all sources away from the Galactic plane \citep[see Fig. 2 of][]{b94}.  Therefore, many of the sources observed in the present survey are Galactic.  The uniform number density of these Galactic sources is more likely caused by changes in the survey's sensitivity with Galactic latitude.  The 6 cm noise level near the Galactic plane is about four times higher than the typical 6 cm noise level because of confusion with sidelobes from Sgr A*.

\paragraph{Hyperstrong Scattering Affecting Flux Measurements?}
\label{hyper}
For low radio frequencies (typically below a few GHz), the propagation of radio waves can be diffracted and scattered by variations in electron density in the ISM \citep{co98}.  Brightness variability of compact radio sources is sometimes caused by refractive interstellar scintillation, which is a change in the refractive index of the ISM from the turbulent motion of the inhomogeneities \citep{co98,ge05}.  The scintillation is visible when observed with bandwidths less than the correlation frequency, $f_c\propto \nu^4/(\int n_e dl)^2$ \citep{l71}.  Of the 20 cm sources observed previously (shown in Tables \ref{yhc} and \ref{1lc}), two sources observed by \citet{l98} have dramatically different fluxes:  174448-282814 and 174436-285710 have fluxes higher by factors of three and two in the present observations, respectively.  While inhomogeneities in the ISM can give year-scale variability at GHz frequencies, this is typically only observed over bandwidths up to several MHz \citep{ge05}.  The bandwidth of 50 MHz used for the present observations suggest that scintillation do not affect the observed intensities significantly.

The scattering of radiation by the ISM can also affect the apparent size of sources observed through the GC region.  Multiwavelength observations of Sgr A* have seen the $\nu^{-2}$-dependence expected from strong scattering by variations in the electron density along the propagation path \citep[e.g.,][]{d76}.  Extragalactic sources seen through this scattering medium undergo a much stronger angular broadening, depending on the relative location of the screen and the GC, with
\begin{equation}
\theta_{xgal} = \frac{D_{GC}}{\Delta_{GC}} \theta_{gal}
\end{equation}
\noindent where $D_{GC}$ is the distance to the GC region and $\Delta_{GC}$ is the distance between the GC and scattering region \citep{v92}.  Combining scattering observations and other constraints on the probes of the ISM along the line of sight to the GC region, \citet{l98b} find that the hyperstrong scattering is best described by a screen with $\Delta_{GC}=133^{+200}_{-80}$ pc that covers the central 1--2\sdeg\ (or 140--280 pc) of the GC.  The coincidence of the physical extent of this screen and its distance from the GC region (combined with constraints on its physical conditions) suggests that the scattering occurs at the ionized surfaces of GC molecular clouds.  Regardless of the physical model, extragalactic radio sources seen through the central degree of the Galaxy are expected to be broadened by as much as 500\arcsec\ at 20 cm \citep{v92,l98,l98b}.  

Hyperstrong scattering broadens sources at low frequencies to a degree that interferometric observations can resolve the flux out.  This effect has been shown to bias against the detection of 20 and 90 cm point sources in the central degree of the GC \citep{l98b,n04}.  This ``scattering bias'' can also explain why the typical spectral index of sources detected at 90 and 20 cm in the central degree of the GC region is significantly flatter than those found toward M31 and over a larger portion of the GC region (see \S\ \ref{vla_pssec}).  No strong scattering screen has been observed in M31 and any scattering is expected to be small, since it is much further away than the GC.  However, 90 cm extragalactic sources seen though the scattering screen in the central degree are expected to have sizes greater than 850\arcsec\ \citep[or $90*(1 \rm{GHz}/0.325 \rm{GHz})^2$\arcsec;][]{l98b}.  The 90 cm observations of the GC region are not sensitive to emission larger than $\sim$540\arcsec\ \citep{n04}, so these sources would be resolved out.  Thus, most sources detected at 90 cm toward the scattering screen are located inside the galaxy.

It may even be possible that hyperstrong scattering alters the fluxes of our compact source catalog.  In \S\ \ref{vla_pssec}, it was discussed how the 6 and 20 cm flux densities are generally higher than observed previously, in observations less sensitive to large-scale emission.  The data used for the point-source detection is sensitive to emission at the largest angular scale of about 240\arcsec, while \citet{b94} and \citet{l98} used less compact configurations and cut inner \emph{uv} spacings more severely, giving sensitivities to angular scales up to only 120\arcsec.  The interferometric observations of \citet{y04} have multiple configurations and are sensitive to emission up to $\sim500$\arcsec, which could explain the consistency between that and the present work.  If these flux densities are discrepant because of scattering, the wavelength dependence of the scattering suggests that the spectral index measurements may be positively biased, despite the fact that the 6 and 20 cm observations have similar \emph{uv} coverages and resolutions.  However, the idea that scattering can bias the flux densities is unproven.

%% file: gcl_vlapoln_thesis2_astro-ph.tex
\chapter{VLA Observations of Polarized 6 cm Continuum Toward the GCL}
\label{gcl_vlapoln}

\section{Introduction}
Polarization properties of radio continuum emission can reveal a great deal of information about the source of the emission and the intervening medium.  Synchrotron emission has an intrinsic, maximum linear polarization fraction of about 75\%.  Since thermal radiation at radio wavelengths is not polarized, the detection of linearly polarized emission from a source can be a useful confirmation of its nonthermal nature \citep[e.g.,][]{g01,y04}.  Another well-studied characteristic of polarized radiation is the rotation of linearly-polarized radiation as it propagates through the magnetized, ionized, interstellar medium (ISM).  This effect is called Faraday rotation and it varies with wavelength as $\Theta=\Theta_0+RM\lambda^2$ \citep{b66,g01}, with the proportionality constant referred to as the rotation measure, or $RM$.  The rotation measure is $RM = 0.81 \int B_{los} n_e dl$ pc rad m$^{-2}$, where $B_{los}$ is the line-of-sight magnetic field strength in $\mu$G and $n_e$ is the electron density in cm$^{-3}$.  Thus, measuring the change in polarization angle with wavelength allows an estimate of the magnetic field and electron density of the ISM \citep{f44,s80,r05}.

Nonthermal radio filaments (NRF) have been observed throughout the central degrees of the galaxy.  They are characterized by an aspect ratio of greater than 10:1, with lengths up to several parsecs \citep{y84,l01,y04}.  They emit synchrotron radiation and, as such, are linearly polarized;  this is a key characteristic in identifying an NRF.  NRFs are useful because they are highly polarized, such that the $RM$ and its variation in the region can constrain the physical conditions along the line of sight to the NRF \citep{y97}.  NRFs also provide one of the few indirect constraints on the strength and geometry of the magnetic field in the GC region, although the inferred structure depends heavily on the model used to explain the NRFs \citep{y84,m96,l01,b05}.

Studies of the $RM$ of extragalactic sources seen through the central degree of the Galaxy show that values are typically less than 1500 rad m$^{-2}$ \citep{r05} and the largest $RM$ values for objects known to be in the GC region are around 2500--3000 rad m$^{-2}$, at the position where the Radio Arc crosses the Galactic plane \citep{y84,t86,t95}.  However, the $RM$ toward the GC region observed in the present survey isn't expected to exceed roughly 2000 rad m$^{-2}$ \citep{t86,r05}.  

This paper describes results from a high-resolution survey of polarized radio continuum emission toward the GCL.  In \S\ \ref{poln_obs}, the 6 cm radio continuum observations are described; \S\ \ref{poln_analysis} discusses some of the analysis techniques used to estimate the polarization properties, in particular an estimate of the Faraday rotation.  Section \ref{poln_results} describes the detection of extended, polarized emission throughout the region and the unusual, large-scale gradient in the $RM$ toward this emission.  Section \ref{poln_discussion} discusses the implications of the polarized emission and its Faraday rotation, including a model for the structure of the GC magnetosphere.

\section{Observations and Data Reductions}
\label{poln_obs}
\begin{figure}[tbp]
\includegraphics[width=\textwidth]{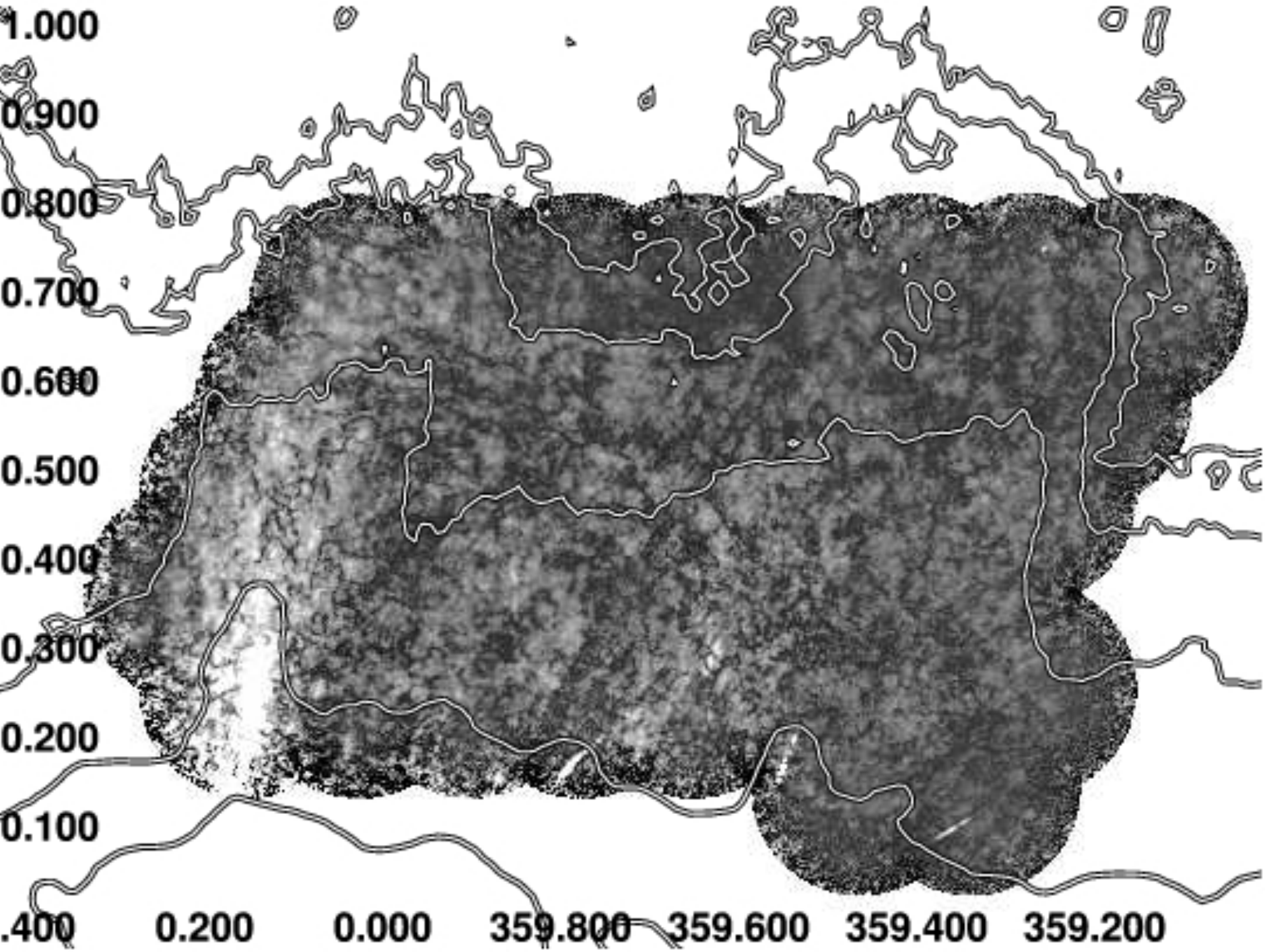}
\caption{Image of 6 cm polarized intensity observed with the VLA with contours of 6 cm total intensity from the GBT with Galactic coordinates (see Ch. \ref{gcsurvey_gbt}).  Polarized intensity image made from 42 separate pointings made with the multi-resolution imaging technique and corrected for noise bias, as described in the text.  The beam size is 15\arcsec$\times$10\arcsec\ with a position angle of 70\sdeg.  Image brightness ranges from 0 to 2 mJy; contours show brightness levels of $33*3^n$ mJy per 150\arcsec\ beam, for $n=0-4$. \label{poln_polc}}
\end{figure}

This work is based on observations conducted with the VLA at 6 cm, in the DnC array configuration.  As shown in Figure \ref{poln_polc}, the observations were conducted such that a mosaic image of the GCL, from $l=$359\ddeg2-0\ddeg2,$b=$0\ddeg2-0\ddeg7, could be constructed.  The default continuum observing parameters were used, with two, adjacent 50 MHz bandpasses centered at 4.8851 and 4.8351 GHz.  Flux calibration was done by observing 1331+305 (3C286) and phase calibration was done by observing 1751--253.  The polarization angle was calibrated by assuming a position angle of 66\sdeg\ for 1331+305.  A more detailed discussion of the observation strategy and calibration is given in chapter \ref{gcl_vla}.

The Stokes Q and U images were cleaned using the IMAGR task in the multi-resolution mode, which independently cleans images with resolutions of roughly 1, 3, and 9 times the beam size and then combines them into a single image.  Similar to the first set of images, the central 16\damin6 was cleaned until the peak flux inside the clean box was less than the noise level outside the clean box.  For consistency and simplicity, the same number of iterations was used to clean both IFs and the Q and U Stokes images.  Images were restored with a beam of 15\arcsec$\times$10\arcsec\ with $PA=70$\sdeg, which is typical of the beam for these observations (see Ch. \ref{gcl_vla}).  The peak polarized flux in these images ranged from $\sim$0.25 to $\sim$4 mJy.  A second set of images was made using the standard clean algorithm and cleaning much more deeply.  In general, the resulting images and their derived quantities are the same for both cleaning techniques, which suggests that the results are not very sensitive to the cleaning method.

After cleaning, the Stokes Q and U images (for both IFs) were primary-beam corrected and added together (using FLATN) to form a single image of the entire region.  Figure \ref{poln_polc} shows the polarized intensity averaged over both IFs, accounting for noise bias.  The noise level for the images was evaluated by measuring the noise level outside the primary beam of individual pointings.  
For the Q and U images, the noise level was measured for each field to have values ranging from 50 to 120 $\mu$Jy, mostly depending on the field's proximity to Sgr A*.  A noise image was then constructed by flattening 42 images together with each field set to a constant value equal to the noise level of that field.  Polarized intensity and position angle images and their errors were made for each IF using the AIPS task COMB.  In general, the polarized intensity images have a lower noise level than the total intensity data, since sources (and their sidelobes) are fainter and there is less confusion.  
The noise values are measured from cleaned images in an area outside the primary beam, since the center of images are filled with emission.  The theoretical prediction of the noise level for the integration time used here ($\sim$30 min; Chap. \ref{gcl_vla}) is 30 $\mu$Jy.  Thus, sidelobe noise and confusion increase the noise in polarized intensity by factors of two to four in these images.


There are several sources of systematic error intrinsic in the VLA that could affect the mosaicked images of the polarized intensity and polarization angle.  One important systematic effect is the antenna polarization that induces radially-oriented, linear polarization with increasing offset from the phase center \citep{c99}.  The 42 fields were observed in a pattern with fields overlapping beyond 3\damin75 from their centers, which helps average these biases in some of the areas where they are strongest.  The change in parallactic angle during the observation also reduces the antenna-based polarization bias.  Each of the 42 pointings in the present survey was observed at several times throughout a 7-hour observing period that the Galactic center is visible to the VLA.  During the observations, the parallactic angle changed by about 80\sdeg, which reduces antenna-based polarization by about $1-\sin(\Delta\theta)/\Delta\theta\approx30$\%.  

Imperfections in the feed system can also induce false polarization by confusing up to 1\% of the total intensity with the polarized intensity \citep{c99}.   The brightest source in total intensity is about 50 mJy beam$^{-1}$, so the brightest sources may have false-polarized signals of about 500 $\mu$Jy, about 5--10 times the polarized intensity noise level.  However, there is no total intensity counterpart to the polarized flux seen filling the survey region in Figure \ref{poln_polc}, so this instrumental effect will do little to change the characteristics of the extended emission.  

Finally, it is important to consider the fact that interferometric observations are not sensitive to emission on the largest angular scales.  \citet{h04} present a detailed study of the effect of missing a spatially-constant offset in the Q and U fluxes;  a constant offset to Q and U can create spurious small-scale polarization and a nonquadratic dependence in wavelength for the Faraday rotation.  There are two ways to show that these problems do not affect the present data.  First, the largest angular scale resolvable at 6 cm in the DnC configuration is 5\arcmin.  Figure \ref{poln_polc} shows that the typical size scale for polarized flux is less than 5\arcmin, and in fact, the Q and U images vary on scales roughly half the size scale seen in the polarized flux.  Second, the appendix of \citet{h04} quantifies the amount of flux that may be missing, depending on the variation in the $RM$ across the field, assuming that the $RM$ originates in a screen and applies a random Gaussian offset to the polarization angle.  In the worst case\footnote{This is a worst case scenario for the analysis presented in this paper, considering size scales greater than 100\arcsec.}, the effective $RM$ distribution (described in detail in \S\ \ref{padianalysis})  has a width of about 500 rad m$^{-2}$, which at 6 cm can lead to fractional offsets in Q and U up to 0.2\%, much less than the 10\% offset that is considered acceptable \citep{h04}.  Thus, the strong and variable Faraday rotation in the present survey assures that the measured polarization properties are not affected by the problems associated with missing large-scale emission.

The accuracy of the polarization properties of the present survey is also tested by comparing it to other surveys in \S\ \ref{poln_comparison}.

\section{Data Analysis}
\label{poln_analysis}

\subsection{Polarization Angle Difference Across IFs}
\label{padianalysis}
To look for any possible signature of Faraday rotation in the data, and image of the change in polarization angle between the two IFs was made.  The difference of these images is hereafter referred to as the polarization angle difference or \dt image.  The difference of the two polarization angle images with angles ranging from --90\sdeg\ to 90\sdeg\ will give an image with a range from -180\sdeg\ to 180\sdeg.  This image was renormalized by adding 180\sdeg\ to all pixels with $-180$\sdeg$<\theta<-90$\sdeg and subtracting 180\sdeg\ from all pixels with angles 90\sdeg$<\theta<$180\sdeg.  This renormalization correctly maps each value of \dt to the range --90\sdeg to 90\sdeg.  

Figure \ref{poln_padi} shows the \dt image and its error.  This image is made by subtracting the polarization angle image in IF 1 at 4.885 GHz from that of IF 2 at 4.835 GHz.  

\begin{rotate}
\begin{figure}[tbp]
\hspace{-2cm}
\includegraphics[width=\textheight]{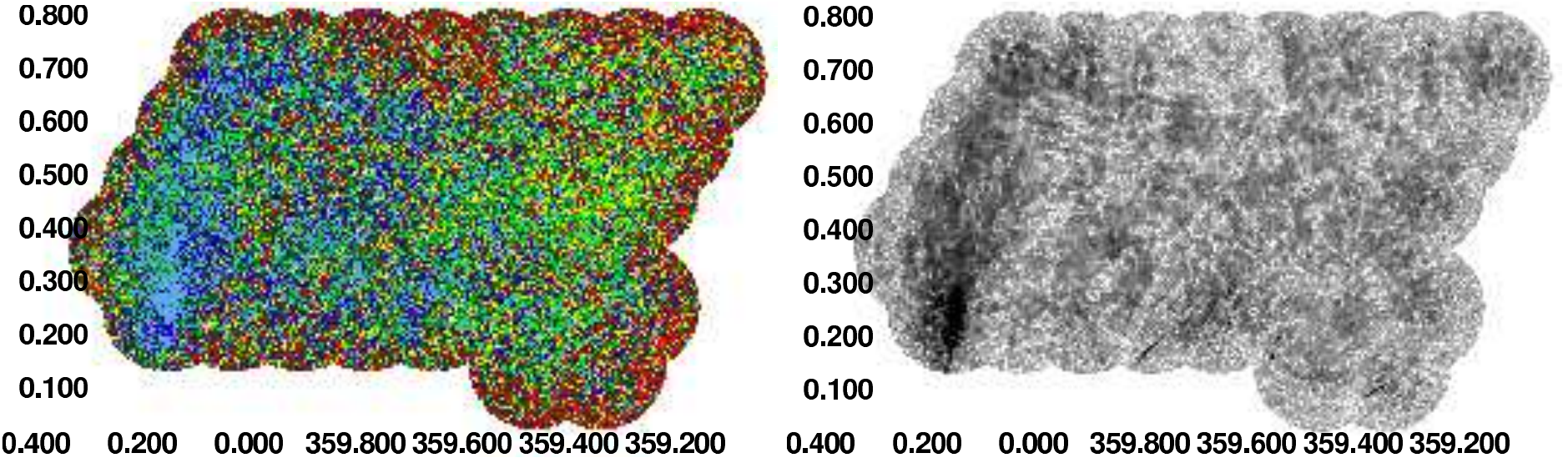}
\caption{\emph{Left}: Image of the polarization angle difference, or \dt, for the 6 cm survey of the GCL.  Pixel values range from $\Delta\theta<-45$\sdeg\ (shown in black) to --45\sdeg$<\Delta\theta<$0\sdeg\ (purple through light blue) to $0<\Delta\theta<+45$\sdeg\ (green through orange) to red $\Delta\theta>+45$\sdeg.  The edges of the mosaicked image appears dominated by red ($\Delta\theta>45$\sdeg), but are actually a mixture of mostly black and red pixels, as is expected in the noisy areas at the survey edge.  The effective $RM$ has the opposite sign of \dt shown here.  \emph{Right}:  Image of the error in \dt, with values ranging from 1\ddeg2 to 180\sdeg.)\label{poln_padi}}
\end{figure}
\end{rotate}

In the simplest case, the change in the polarization angle with wavelength due to Faraday rotation is:
\begin{equation}
\Delta\theta = RM (\lambda_1^2-\lambda_2^2)
\end{equation}
\noindent $RM$ studies often use three or more wavelengths and fit the position angle as a function of wavelength \citep[e.g.,][]{t86,h00}.  In the present observations, we use two IFs with central frequencies separated by 50 MHz.  In this case, a position angle difference of one degree corresponds to a rotation measure of --220 rad m$^{-2}$.  The calculation of an $RM$ from \dt assumes that the Faraday rotation has a $\lambda^2$ dependence, which is not always true \citep{b66}.  Thus, this work assumes that the Faraday rotation follows a $\lambda^2$ dependence and considers the \dt values to be \emph{effective} $RM$ values, or \rmeff.

Studying Faraday rotation across two, adjacent 50-MHz IFs is not common \citep[for one example see][]{y88}, but there are a few reasons why it can give useful results in this application.  First, the $RM$ expected in the GC region covered by this survey is typically less than 2000 rad m$^{-2}$, with a maximum of about 3000 rad m$^{-2}$ in the Galactic plane \citep{y84,t86,t95,r05}.  For $RM=2000$ rad m$^{-2}$, the change in polarization angle between the two IFs of the present survey is about 9\sdeg.  The polarization angle change would be degenerate for $\Delta\theta>180$\sdeg, which is unlikely in the present data, since this would require twenty times higher $RM$ than is expected ($RM>4\times10^4$ rad m$^{-2}$).  Secondly, the bandwidth depolarization expected in this data is roughly $1-\sin(\Delta\theta)/\Delta\theta=0.4$\% for $RM$ of 2000 rad m$^{-2}$, which shouldn't significanly affect the results.  However, Figure \ref{poln_polc} shows that some of the emission is completely depolarized;  if these regions have $RM$ greater than $4\times10^4$ rad m$^{-2}$, these observations would not be sensitive to it.  The accuracy of the \rmeff\ measured in this work is compared to other, more complete observations in \S\ \ref{poln_comparison}.

\begin{figure}[tbp]
\begin{center}
\includegraphics[scale=0.5]{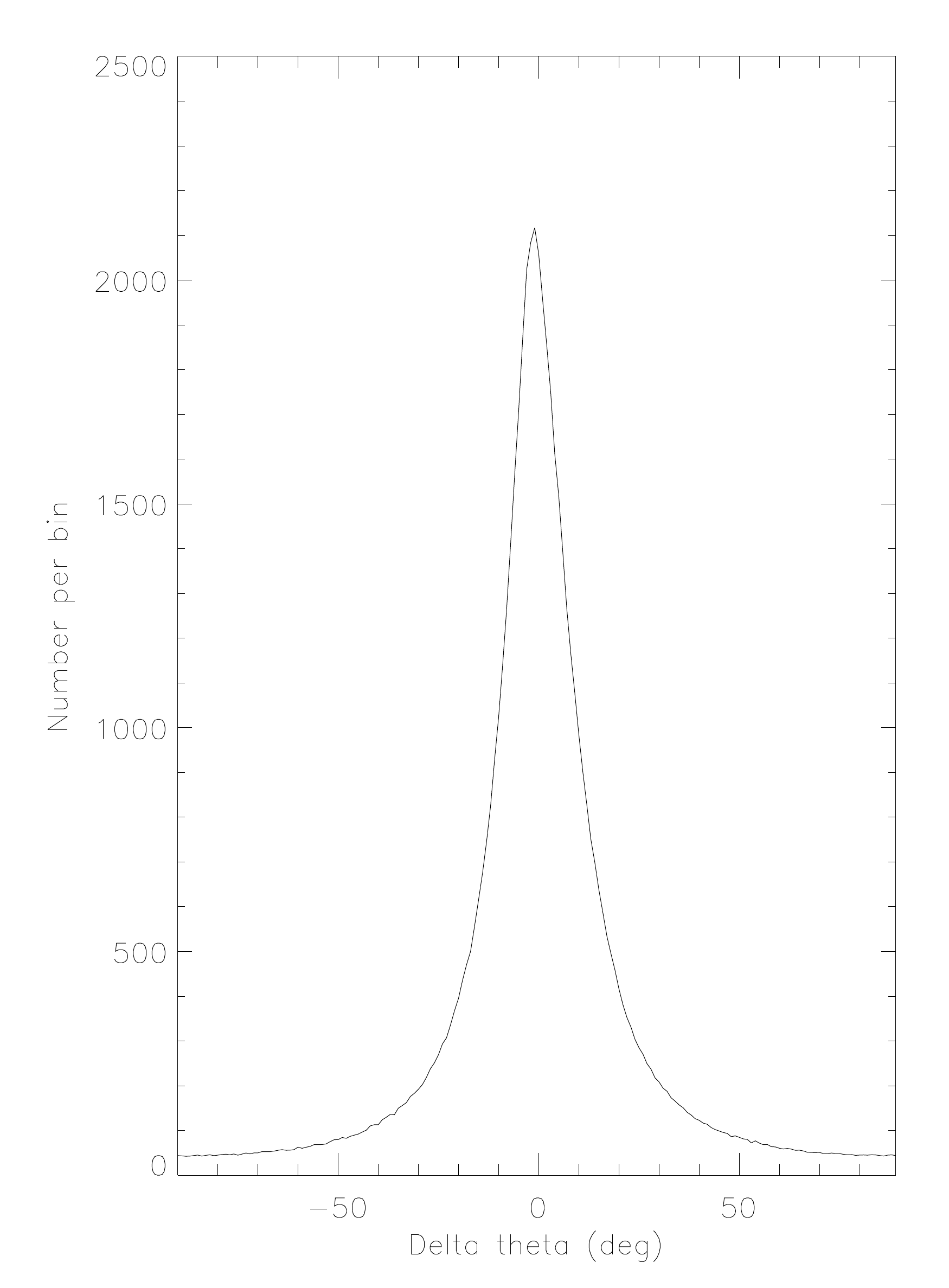}
\end{center}
\caption{The histogram shows the distribution of values of the polarization angle difference (\dt) between the two, 50 MHz IFs of the 6 cm mosaicked image of the GCL.  The histogram shows the number of beams that have a given \dt with a bin size of 1 degree. \label{hist}}
\end{figure}

Figure \ref{hist} shows the distribution of \dt in the mosaicked 6 cm image of the GCL.  The \dt distribution has a Lorentzian shape with a peak of 2045 beams per degree, a half-width of about 10 degrees, plus a constant background of about 13 beams per degree.  Most values of \dt measured have small offsets from 0\sdeg, with $\sim$50\% within $\pm$10\sdeg, $\sim$75\% $\pm$20\sdeg, and $\sim$90\% $\pm$45\sdeg.

\subsubsection{Averaging of \dt Images}
The images of \dt have a sensitivity that varies across the field of view and can make it difficult to visualize the most significant data.  To improve the visualization of the data, two techniques were developed for finding the mean \dt across the survey.  The first method of finding the mean \dt is the most basic:  an error weighted mean value of \dt for a given region, hereafter referred to as a ``simple average''.  We developed a script to calculate the error-weighted average and its uncertainty for a grid of square regions or latitudinal strips.

There is one major caveat to the simple-average method described above:  averaging of angles is not proper, since the angle represents a vector, not scalar, quantity.  For example, the average of a vector with a polarization angle of $\theta=0$\sdeg\ with another with $\theta=180$\sdeg\ would average to $\theta=90$\sdeg, when in fact these vectors are essentially identical.  Generally, the proper way of averaging polarization properties would be to average the Stokes parameters (fluxes), which are well-behaved, linear quantities.  However, the averaging of angles is approximately correct for small angles, which is usually true for the \dt images presented here (see Fig. \ref{hist}).  Furthermore, the large values of theta in the \dt map are generally in the noisiest parts of the image and are down-weighted by large errors.  

\subsubsection{Fitting Histograms of \dt Images}
A second, more robust method of finding averages of the \dt images is to fit the distribution of \dt values with a model.  We wrote a script to fit a model to histograms of \dt values extracted from either a grid of box-like regions or latitudinal strips.  The uncertainty in the bin value was assumed to be Poissonian, which in the low-count approximation is $\sigma = 1 + \sqrt{N + 0.75}$, where N is the number of counts in a bin \citep{g86}.  

\begin{figure}[tbp]
\begin{center}
\includegraphics[width=0.4\textwidth]{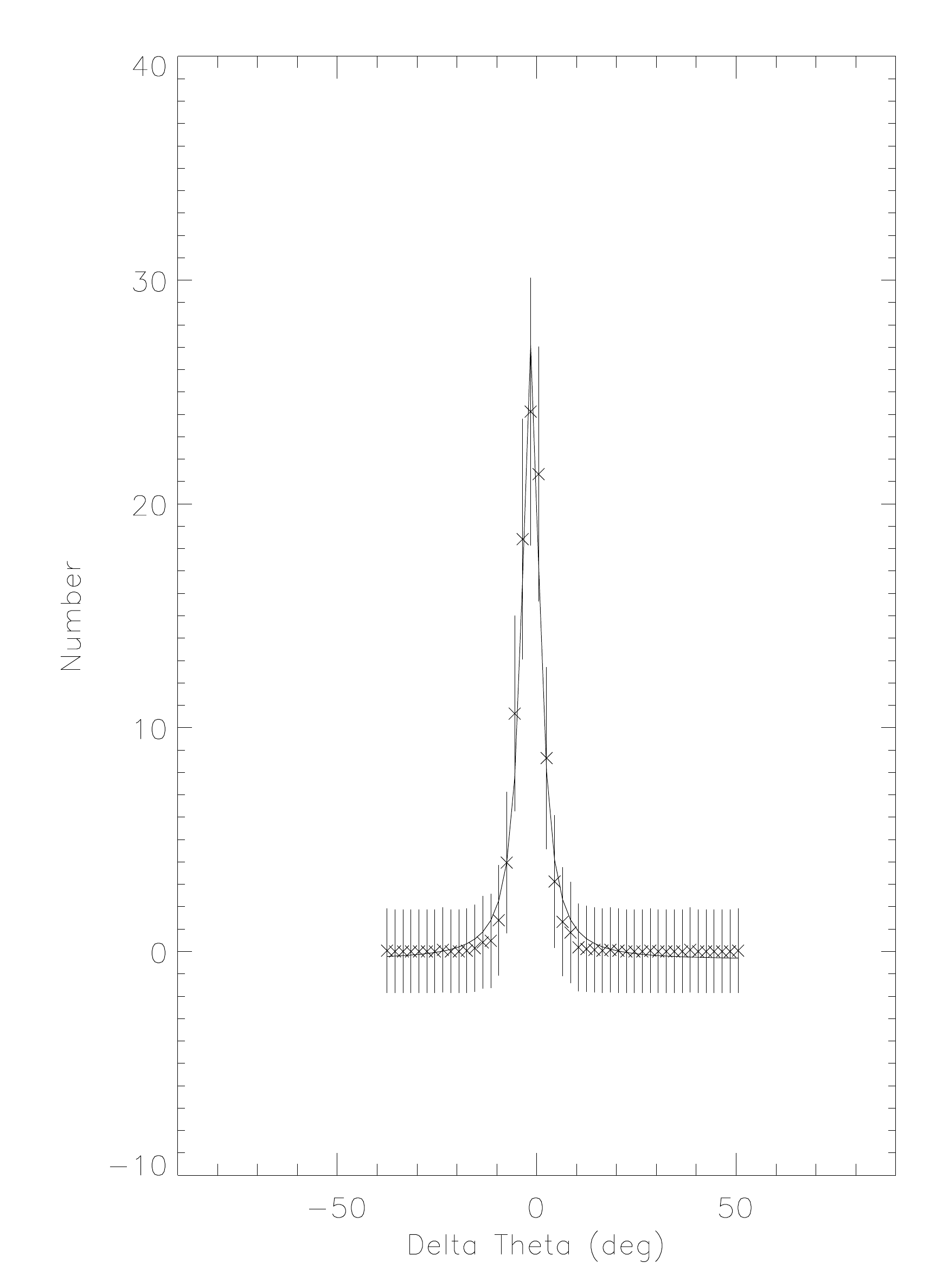}
\includegraphics[width=0.4\textwidth]{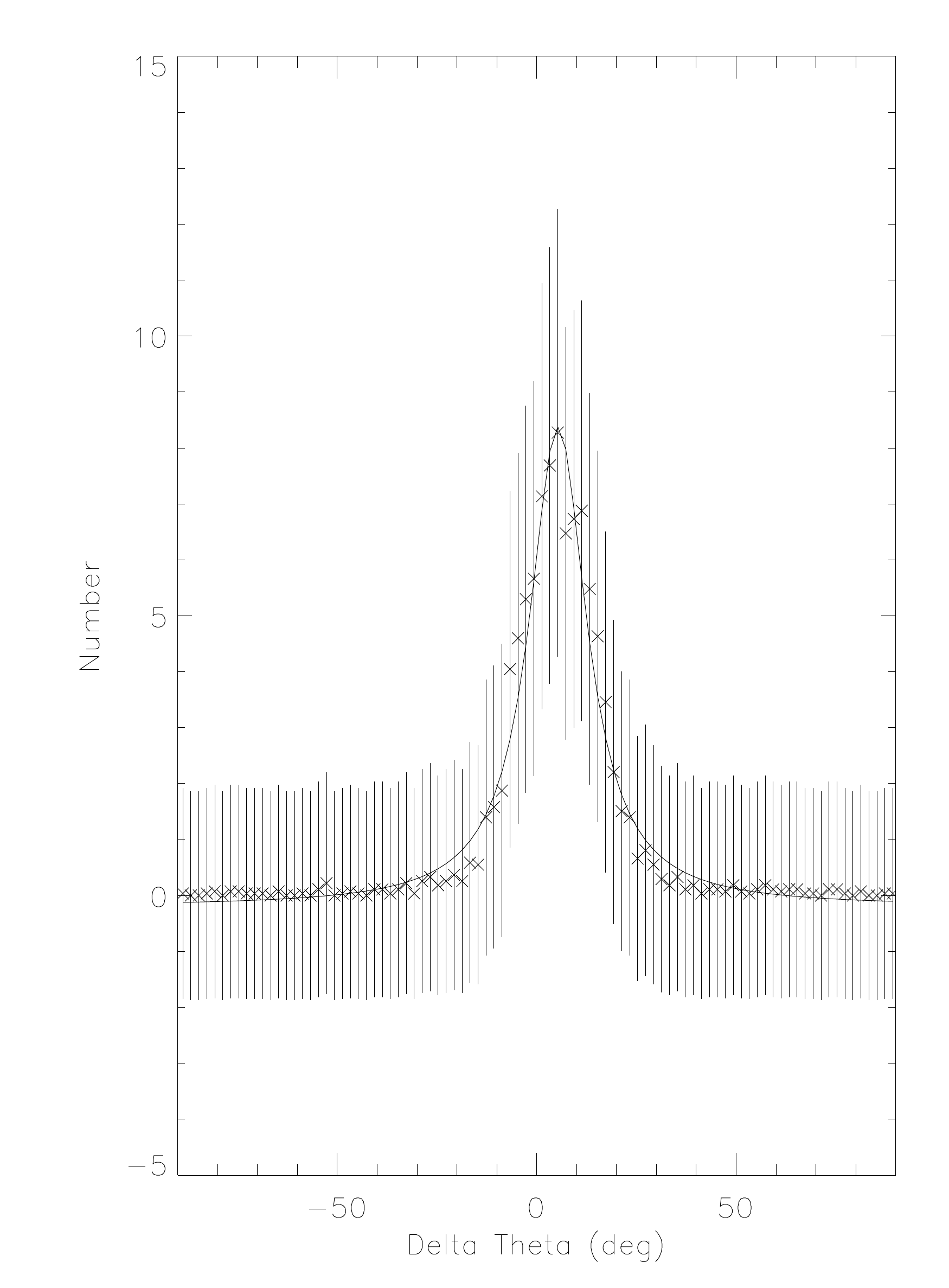}
\end{center}
\caption{\emph{Left}: A plot of the histogram of \dt values for a 125\arcsec$\times$125\arcsec\ box in the eastern half of the survey.  The line shows the best-fit Lorentzian model for the histogram, which has a peak at $\Delta\theta=-1\ddeg5\pm0\ddeg5$.  \emph{Right}: A similar plot as shown on the left, but for a region on the western half of the survey.  The best-fit model has a peak at $\Delta\theta=+5\ddeg3\pm1\ddeg9$. \label{histcomp}}
\end{figure}

The best model to fit the histograms of \dt values is a Lorentzian plus a constant background.  Figure \ref{histcomp} shows two examples of histograms extracted from regions on the east and west sides of the survey region with best-fit models.  The regions are 125\arcsec$\times$125\arcsec\ boxes, which are small enough to give some morphological information, but also have accurate fit results.  For each fit, the best-fit peak and its error are recorded.  One advantage of the histogram-fitting method is that it calculates the mean and error in \dt from the image of \dt alone; no noise image is required, as it is for the simple-average method.  Another advantage is that the quality of the fits can be checked interactively.

One disadvantage of the histogram-fitting method is that it assumes that all pixels in the region sampled are a part of the same distribution.  In fact, there are times when multiple sources, with different distributions of \dt (i.e., different $RM$s) are sampled by the same histogram.  In this case, the source that occupies the most pixels will dominate the histogram distribution and the best-fit value of \dt.  This is different from what happens when calculating an error-weighted average of \dt, in which the average is dominated by the pixels with the lowest-noise (i.e., the brightest sources).

\subsubsection{Comparing Simple-average and Histogram-Fitting Methods}
\begin{figure}[tbp]
\includegraphics[width=\textwidth]{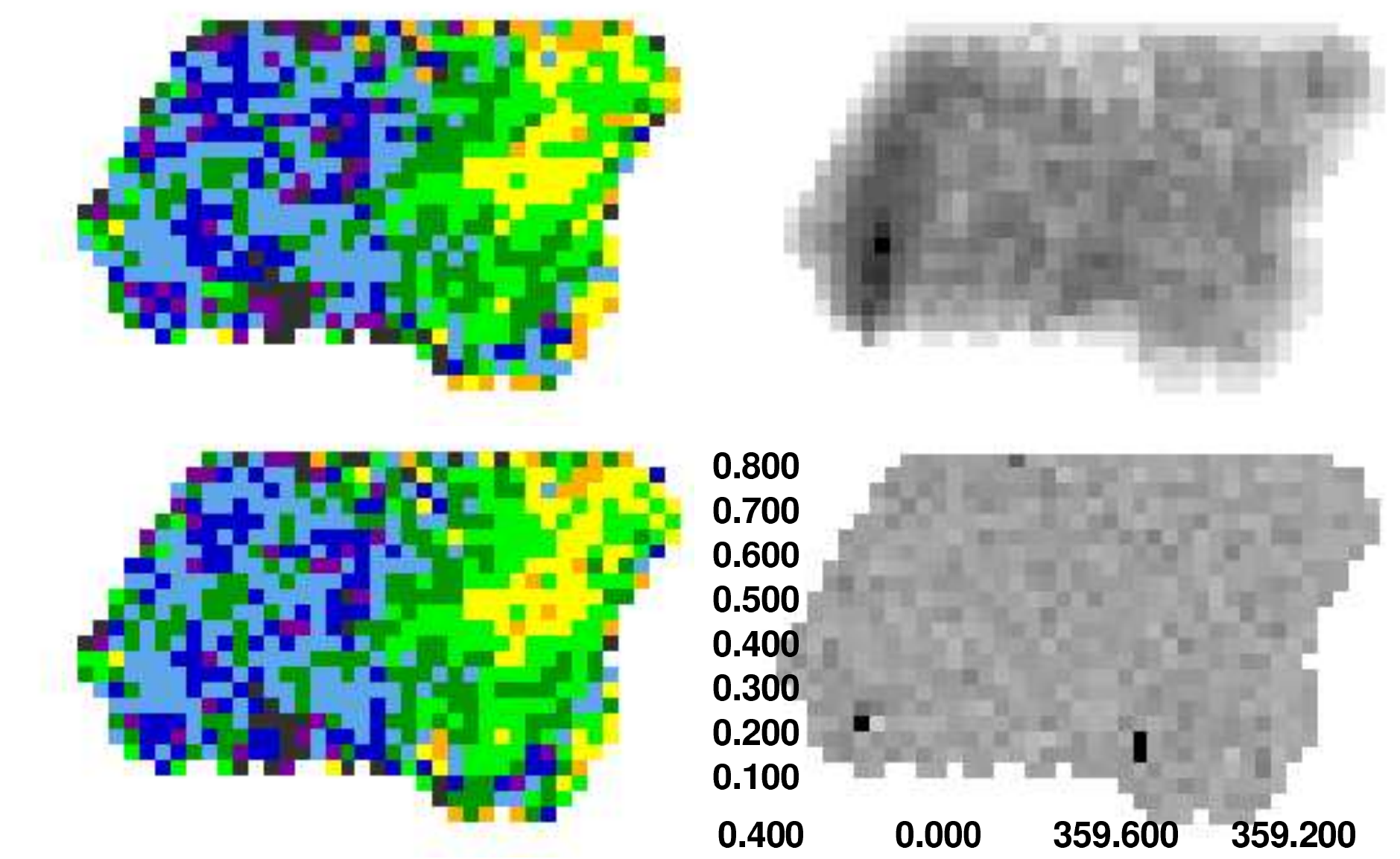}
\caption{\emph{Top left}: An image of the mean \dt in 125\arcsec$\times$125\arcsec\ tiles from the histogram-fitting method.  The color-scale used and in the bottom left panel is similar that used in Fig. \ref{poln_padi}, with colors with wavelengths equivalent to light blue and shorter showing negative values and those with wavelengths equivalent to yellow and longer showing positive values.  \emph{Top right}: An image of the error in \dt for the image shown in the top left.  The values range from 0\ddeg08 to 20\sdeg.  \emph{Bottom left}:  An image of the mean \dt like in the top left panel, but for the mean value is found by the simple-average method. \emph{Bottom right}: An image of the significance of difference between the histogram-fitting and simple-average methods with values ranging from -1 to +1.  The simple-average and histogram-fitting methods are generally in good agreement. \label{padicomp}}
\end{figure}

Figure \ref{padicomp} compares images of \dt using the simple-average and histogram-fitting methods for a grid of 125-arcsec boxes.  An image showing the significance of differences between the two methods is also shown, with each pixel showing the ratio of the difference between the images, divided by its uncertainty.  The average difference between the two images is 0\ddeg002 and the standard deviation is about 0\ddeg2.  So there is no systematic difference between the two methods and, in fact, about 99.5\% of the differences are within the 1$\sigma$ errors.  The similarity of the average \dt images from two independent methods gives confidence to the average values and the methods used to calculate them.

The difference-significance image reaches significance greater than 1$\sigma$ at two positions:  (359\ddeg55,0\ddeg15) and (0\ddeg15,0\ddeg2).  This offset could be explained by considering the biases of the simple-average and histogram-fitting methods, which favor brightness and areal coverage, respectively.  In both of these locations, there is a polarized, localized source (the northern extension of the Radio Arc and the Ripple nonthermal filament) that could have different \dt values from the background polarized emission.  If there is a difference between the \dt values for a compact, bright source and the surrounding, low-brightness polarized emission, the simple-average and histogram-fitting methods will find a difference such as is seen in these two regions.  While this is an important caveat to keep in mind, Figure \ref{padicomp} shows that, in general, the two methods are in good agreement.

The errors for the average \dt in the 125-arcsec tiles differ according to the method used.  The simple-average and histogram-fitting methods produce errors within 50\% of each other for errors less than 3\sdeg.  About 55\% of the 125-arcsec tiles shown in Figure \ref{padicomp} have errors smaller than 3\sdeg.  For errors greater than 3\sdeg, the errors found with the simple-average method are progressively less than those found with histogram fitting.  The maximal error found by the simple-average is about 20\sdeg, while the histogram-fitting method has a maximal error of about 200\sdeg.  The averaging of angles is known have biases when $\theta\not\ll 1$ rad $\approx57$\sdeg;  the mean \dt measured is generally much smaller than 1 rad, but the errors are not, so they are more likely to show problems.  In summary, the histogram fitting errors are a more accurate (and more conservative) estimate of the true errors in the mean value of \dt.


When calculating the mean \dt over size-scales around 100\arcsec, the accuracy of the simple-average and histogram-fitting methods are typically around 1\sdeg.  A change of 1\sdeg\ between the two IFs corresponds to a rotation measure of about --220 rad m$^{-2}$, which is equivalent to a total Faraday rotation of about 49\sdeg.  Thus, the typical uncertainty in \dt makes the calculation of the intrinsic polarization angle highly uncertain.  Unfortunately, binning over a larger region tends to include regions with different polarization angles, for which a single \dt is not appropriate.  Thus, no results on the intrinsic polarization angle of the linearly polarized intensity are presented here.

\subsection{Comparison to Earlier Work}
\label{poln_comparison}
The \dt maps presented here cover regions that have been studied before in polarized emission.  As a test of the accuracy of the images and their derived properties, this section compares the polarized intensity images and $RM$ measurements of the present work to that presented in the literature \citep{l01,t86,h92,r05,y97}.

\begin{figure}[tbp]
\begin{center}
\includegraphics[scale=0.35]{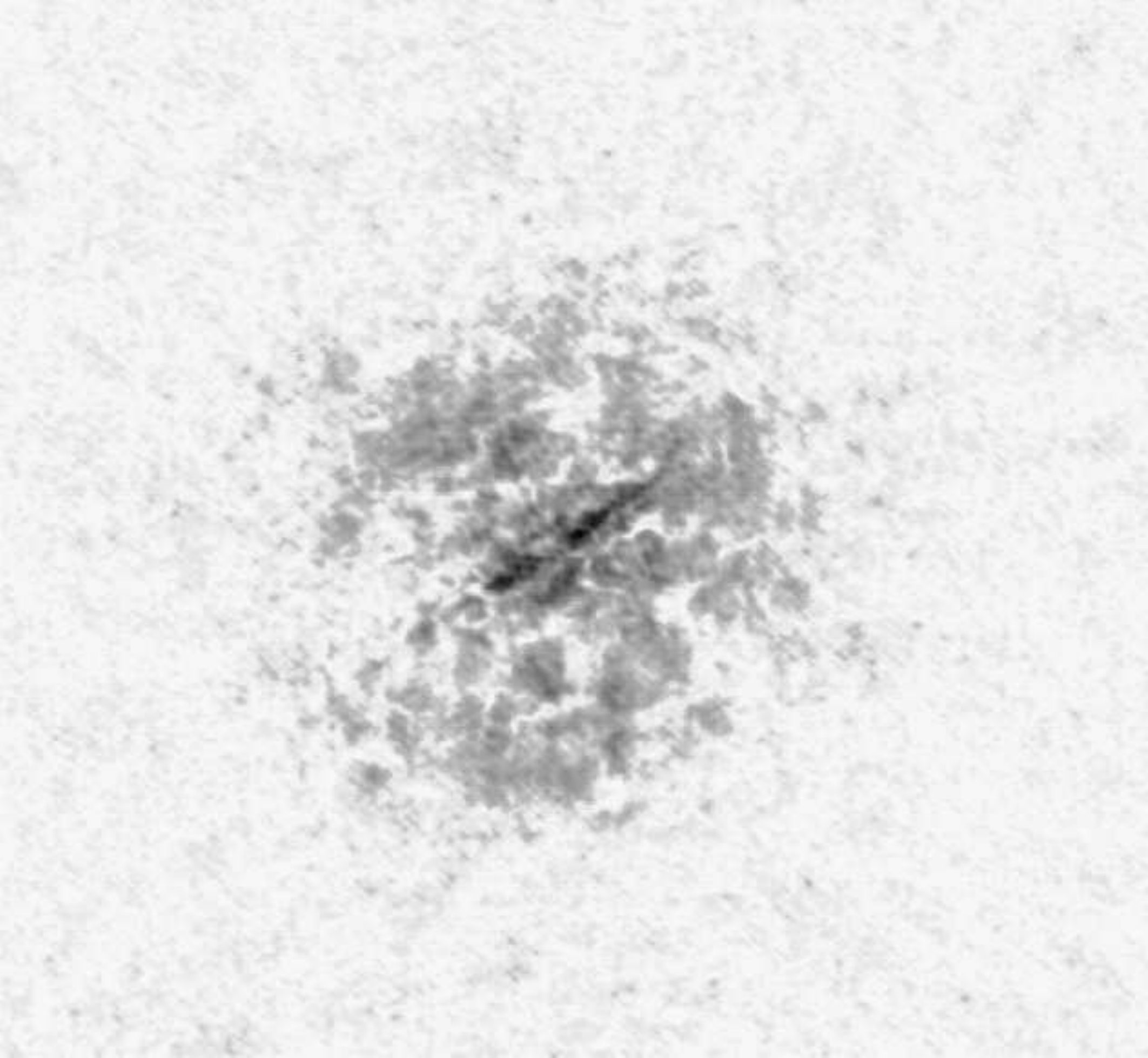}
\hfil
\includegraphics[scale=0.4]{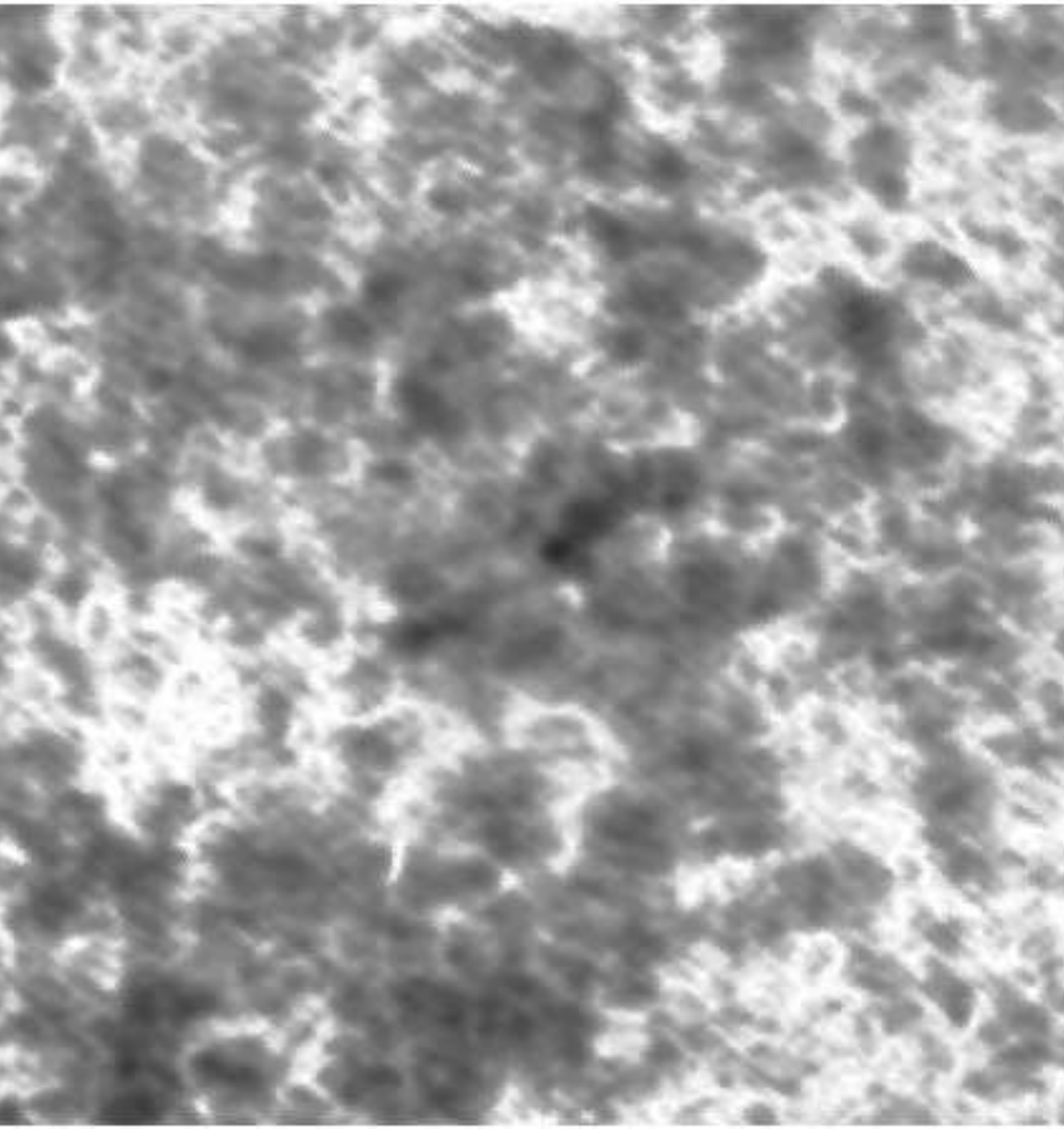}
\end{center}
\caption{\emph{Left and right}: 6 cm, polarized emission from the region near G359.85+0.39 from Figure 5 of \citep{l01} and from the present survey, respectively.  In the middle of the field is the ``Cane'' nonthermal radio filament, which appears in these images as an enhancement of polarized emission surrounded by extended polarized emission.  The images are oriented with celestial north up and east to the left and shown on an inverse gray scale.  The range of fluxes and the morphology of the polarized and depolarized emission are similar between the two (similar) observations, giving confidence to the calibration and imaging methods used. \label{comppolc}}
\end{figure}

\citet{l01} present images of 6 cm polarized intensity near the nonthermal radio filament G359.85+0.39 from VLA data with similar sensitivity and resolution as the present survey.  Figure \ref{comppolc} shows an image of 6 cm, polarized intensity near G359.85+0.39 as shown in Figure 5 of \citet{l01} and in the present survey.  The two images are similar in the structure of the polarized emission, particularly the depolarized regions on the southeast and northeast sides of G359.85+0.39.  The polarized filament is less distinct in the present survey, but the typical brightness varies from 0.8 to 1.2 mJy beam$^{-1}$, similar to the brightness of the filament in Figure 5 of \citet{l01}.  The similarity of the two images suggests that the calibration and imaging process is consistent between the present survey and that of \citet{l01}.

\citet{h92} and \citet{t86} conducted independent, single-dish surveys near 3 cm, covering a few square degrees of the GC region.  Although depolarization effects are lower near 3 cm and the beam is larger for these single-dish observations, there is general agreement between their maps of polarized intensity and those presented here (see Fig. \ref{poln_polc}).  Specifically, \citet{h92} shows a similar distribution of polarized emission from the Radio Arc, which extends up to the top of the present survey at $b=0\ddeg8$.  There is also agreement in the detection of extended polarized emission in the following regions: for $l=359\ddeg5$ to 359\ddeg6 and $b=0\ddeg1$ to 0\ddeg5 and near (359\ddeg3,0\ddeg45).

\begin{figure}[tbp]
\begin{center}
\includegraphics[scale=0.7]{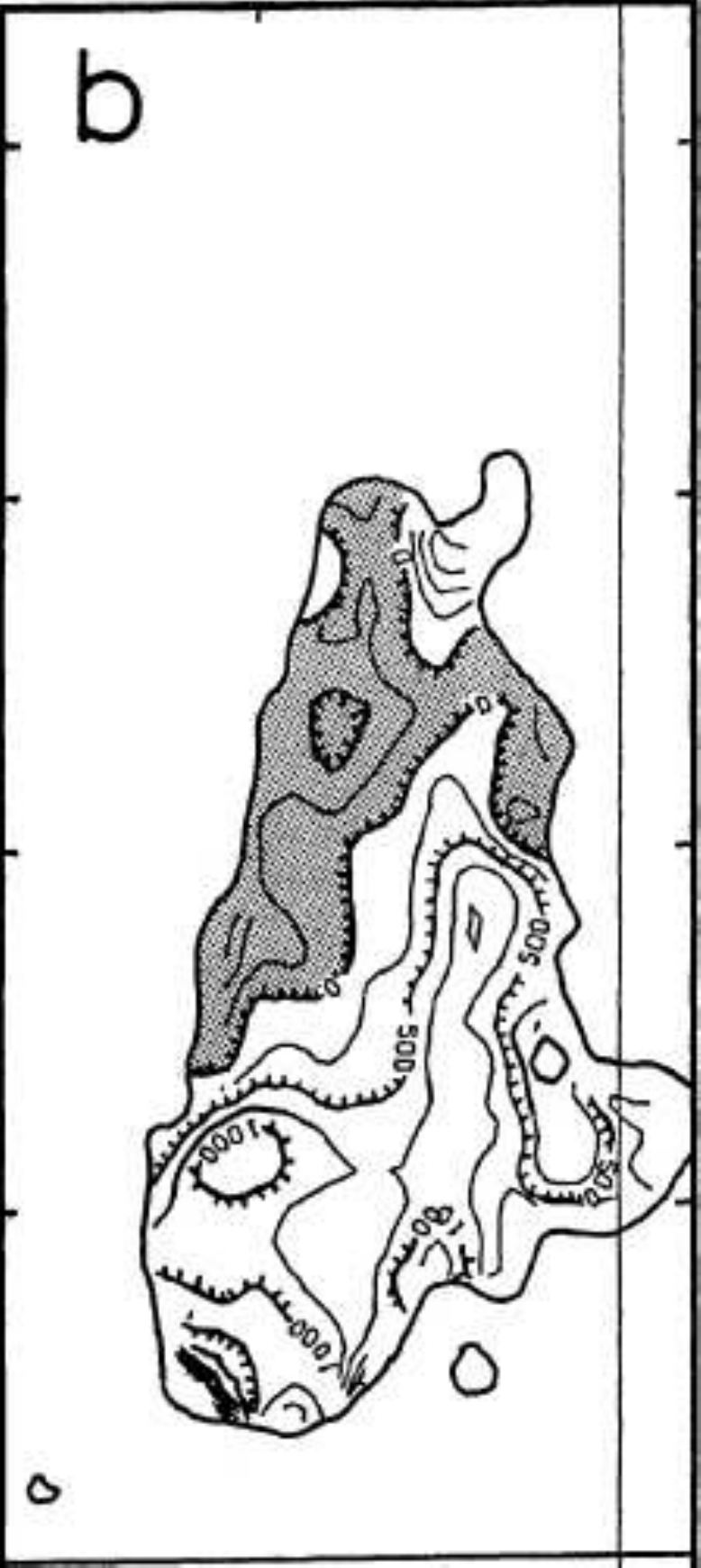}
\hfil
\includegraphics[scale=0.5]{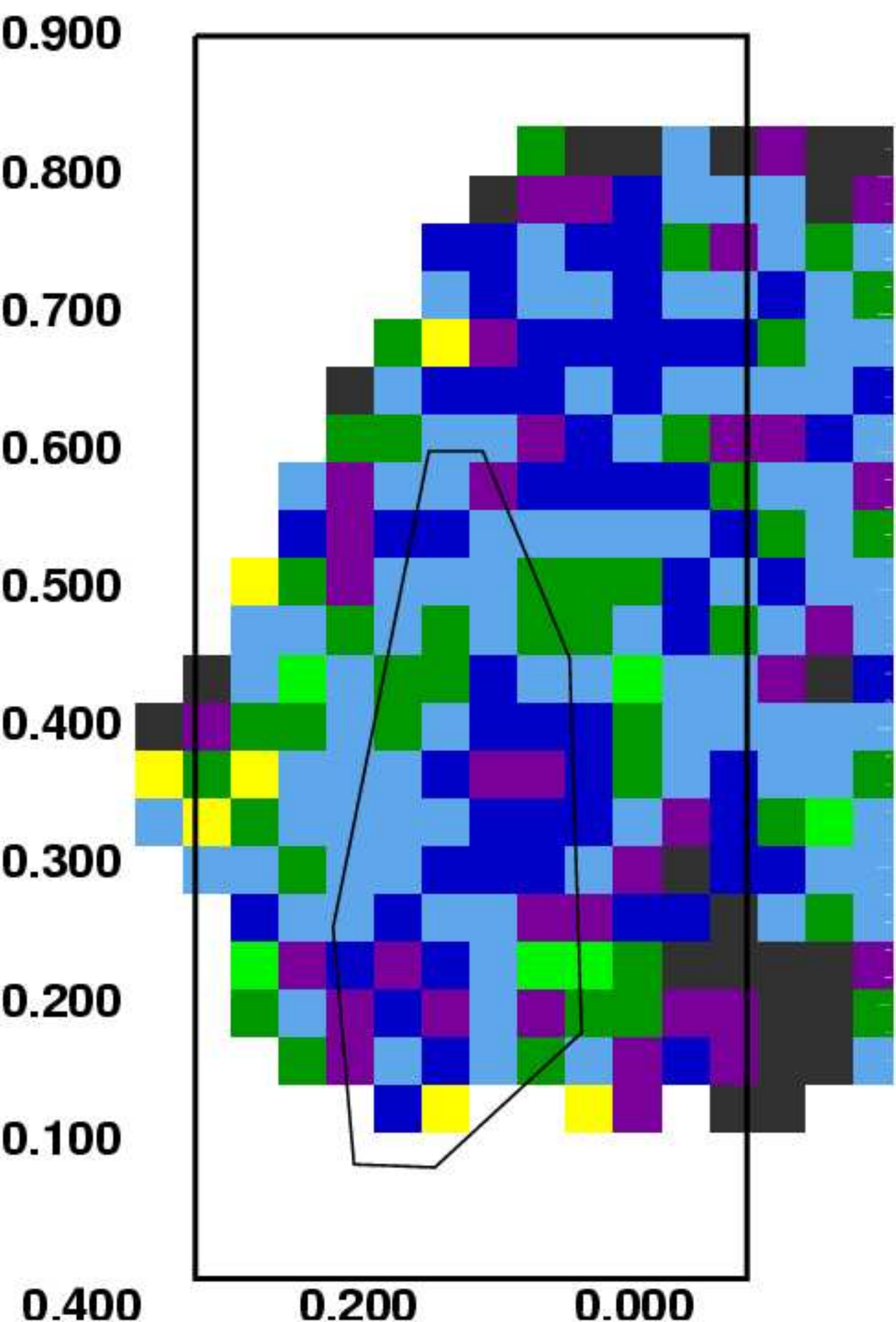}
\end{center}
\caption{\emph{Left}: Copy of Fig. 4 from \citet{t86} showing the $RM$ for the northern extension of the Radio Arc, as observed by the Nobeyama Radio Observatory near 3 cm with a beam size of 2\damin7.  The contours show steps of 250 rad m$^{-2}$, with the shaded region showing negative $RM$ and unshaded region showing positive $RM$. \emph{Right}: Image of \dt from the present 6 cm, VLA survey averaged by the histogram-fitting method with 125\arcsec$\times$125\arcsec\ boxes.  The rectangular box shows the extent of the left figure on this figure and the irregularly-shaped region schematically shows the outermost contour from the left figure.  The color scale for this figure is similar to that of Fig. \ref{padicomp}, with blue hues showing $\Delta\theta<0$\sdeg\ and yellow through red showing $\Delta\theta>0$\sdeg.  Assuming the Faraday rotation observed in the present survey has a $\lambda^2$ dependence, 1\sdeg\ is equivalent to an \rmeff\ of --220 rad m$^{-2}$.  Assuming this conversion factor for \dt, there is good agreement between the \rmeff\ measured by the present work and the more rigorous measurements of \citet{t86}. \label{rmcomp}}
\end{figure}

Comparing the present survey with that of \citet{t86} also provides a valuable check on the estimate of $RM$, since that work observed four frequencies and was able to derive reliable $RM$ measurements for the polarized emission associated with the Radio Arc.  Figure \ref{rmcomp} shows a comparison of the $RM$ measured by \citet{t86} and the \dt observed in the present survey.  Assuming that the Faraday rotation in the present survey has a $\lambda^2$ dependence, 1\sdeg\ in \dt is equal to a \rmeff\ of --220 rad m$^{-2}$.  After applying this conversion, there is remarkable similarity between the \rmeff\ of the present survey and the more rigorous measurements of \citet{t86}.  In particular, near (0\ddeg17,0\ddeg22) and (0\ddeg1,0\ddeg35), \citet{t86} find the $RM$ has a peak around +1000 rad m$^{-2}$, while the present survey finds a peak of $\Delta\theta\approx3\ddeg5\pm0\ddeg5\approx770\pm110$ rad m$^{-2}$ (averaging over a 125\arcsec$\times$125\arcsec\ box, similar to beam FWHM of 162\arcsec in Tsuboi et al. 1986).  Another similarity between the present measurement of \rmeff\ and the values measured by \citet{t86} is near (0\ddeg15,0\ddeg4), where the $RM$ switches to positive values;  \citet{t86} measure $RM\approx-250$ rad m$^{-2}$ while the present survey finds $\Delta\theta\approx1$\sdeg$\pm0\ddeg6\approx-220\pm130$ rad m$^{-2}$.  There is even agreement at the very northern tip of the Radio Arc, where the $RM$ switches back to negative values.  A detailed discussion of the structure in the \dt image is given in \S\ \ref{poln_results}.

\citet{r05} present an multifrequency study of the $RM$ of extragalactic sources seen in the central degrees of the Galaxy.  Only one source from that catalog is in the present survey and has significant polarization.  Measuring \dt and its error at that position finds $\Delta\theta=4.6\pm4.5\approx-1012\pm990$ rad m$^{-2}$.  Note that the error is large because this measurement is taken toward a source that occupies a single beam and is not an average over a large area.  \citet{r05} find a value of --568 rad m$^{-2}$, which is within the 1$\sigma$ errors of our measurement.

\citet{y97} presented detailed study of the polarization properties of the nonthermal filament, G359.54+0.18, at 6 and 3.6 cm.  Figure 3 of that work shows a map of the $RM$ across the filament, with three distinct, bright clumps with relatively uniform values each.  The morphology seen in the present survey is similar to that of \citet{y97}, although it had roughly three times higher resolution (4\arcsec compared to 12\arcsec in the present work).  The first clump, at RA,Dec (B1950) = (17:40:41,-29:12:30) has $RM\approx-2700$ rad m$^{-2}$, compared to $\Delta\theta\approx18\pm5$\sdeg\ $\approx-3960\pm1100$ rad m$^{-2}$ in the present survey.  The second clump, at (17:40:43,-29:12:40), has $RM\approx-2000$ rad m$^{-2}$, compared to $\Delta\theta\approx10\pm2$\sdeg\ $\approx-2200\pm440$ rad m$^{-2}$ in the present survey.  The third clump, at (17:40:44,-29:12:45), has $RM\approx-1500$ rad m$^{-2}$, compared to $\Delta\theta\approx7\pm3$\sdeg\ $\approx-1540\pm660$ rad m$^{-2}$ in the present survey.  Thus, in general, there is good agreement between the \rmeff\ of the present survey and the more rigorously measured values of \citet{y97}.  This comparison is discussed in more detail in \S\ \ref{filres}.

In summary, the polarized intensity and \rmeff\ of the present 6 cm survey shows good agreement other polarized intensity surveys and with more rigorously defined $RM$ measurements.  Systematic effects are not significantly contributing to the polarized intensity maps and \rmeff\ provides a reasonable estimate of the true $RM$ for polarized emission from compact, filamentary, and extended sources.

\section{Results}
\label{poln_results}
This survey of 6 cm polarized continuum emission probes the magnetic properties of the GC region and the intervening medium.  In this section, the distribution of the polarized emission and its Faraday rotation are described in detail.

\subsection{Extended Polarized Emission}
Figure \ref{poln_polc} shows a large-scale view of the polarized intensity of the 6 cm continuum at a 15\arcsec$\times$10\arcsec\ size scale.  The northern extension of the Radio Arc has polarized intensity as high as several mJy beam$^{-1}$ and spans the entire eastern edge of the survey, up to a latitude of $b\sim0\ddeg8$.  For latitudes up to $b=0\ddeg3$, this polarized emission has a total intensity counterpart, but north of this latitude there is no total intensity counterpart (see Ch. \ref{gcl_vla}), so the apparent polarization fraction often exceeds 100\%.  This polarized emission is clearly visible in single-dish observations of the region with polarization fractions of 20--40\% \citep[at $\sim3$\arcmin\ scales;][]{t86,h92}.

\begin{figure}[tbp]
\includegraphics[width=\textwidth]{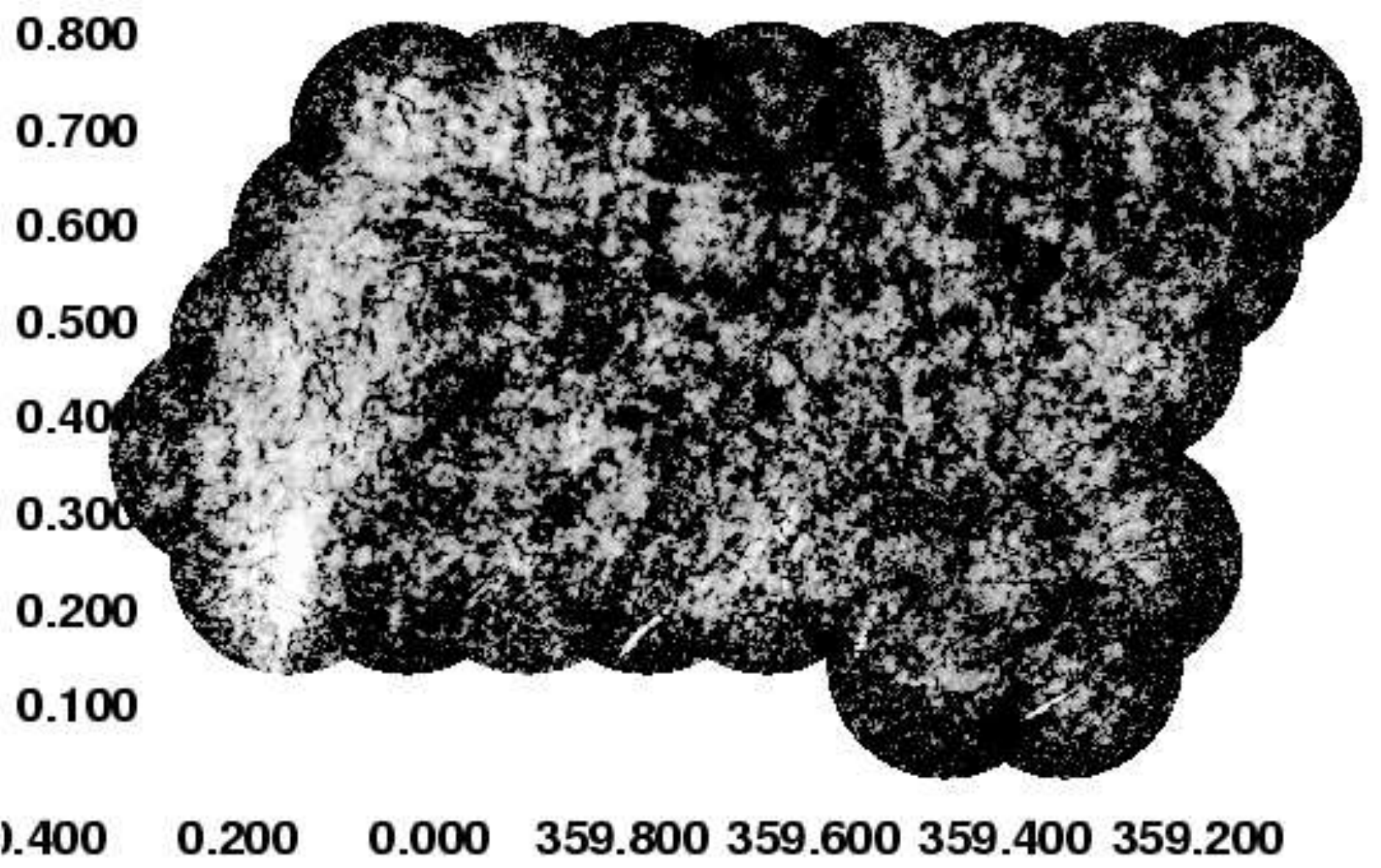}
\caption{Image of the signal to noise ratio for the 6 cm polarized continuum emission observed toward the GCL.  The image is similar to Figure \ref{poln_polc} except that the gray scale ranges from 1 to 10$\sigma$.  Much of the polarized emission that fills the survey is significant even on beam-size scales. \label{polcsnr}}
\end{figure}

Aside from the northern extension of the Radio Arc, much of the region is filled with significant polarized emission that has no counterpart in total intensity.  Figure \ref{polcsnr} shows the signal to noise ratio of the polarized flux.  The noise is measured for each field from the rms of pixel values outside the primary beam;  this is necessary because the primary beam is filled with emission.  The significance of the polarized emission per beam that is coincident with the northern extension of the Radio Arc is as high as 7$\sigma$.  Throughout the rest of the survey region, extended polarized emission per beam is up to 3--5$\sigma$.  The largest area without significant polarized emission is a 10\arcmin\ area near (359\ddeg6,0\ddeg7).  The polarized emission is visible to the present observations mostly because it has structure on angular scales smaller than $\sim5$\arcmin, the largest angular scale visible by the VLA at 6 cm.  As described below, Faraday rotation effects cause depolarization that breaks the polarized emission into small pieces \citep[e.g.,][]{g01}.  

\begin{figure}[tbp]
\includegraphics[scale=0.8]{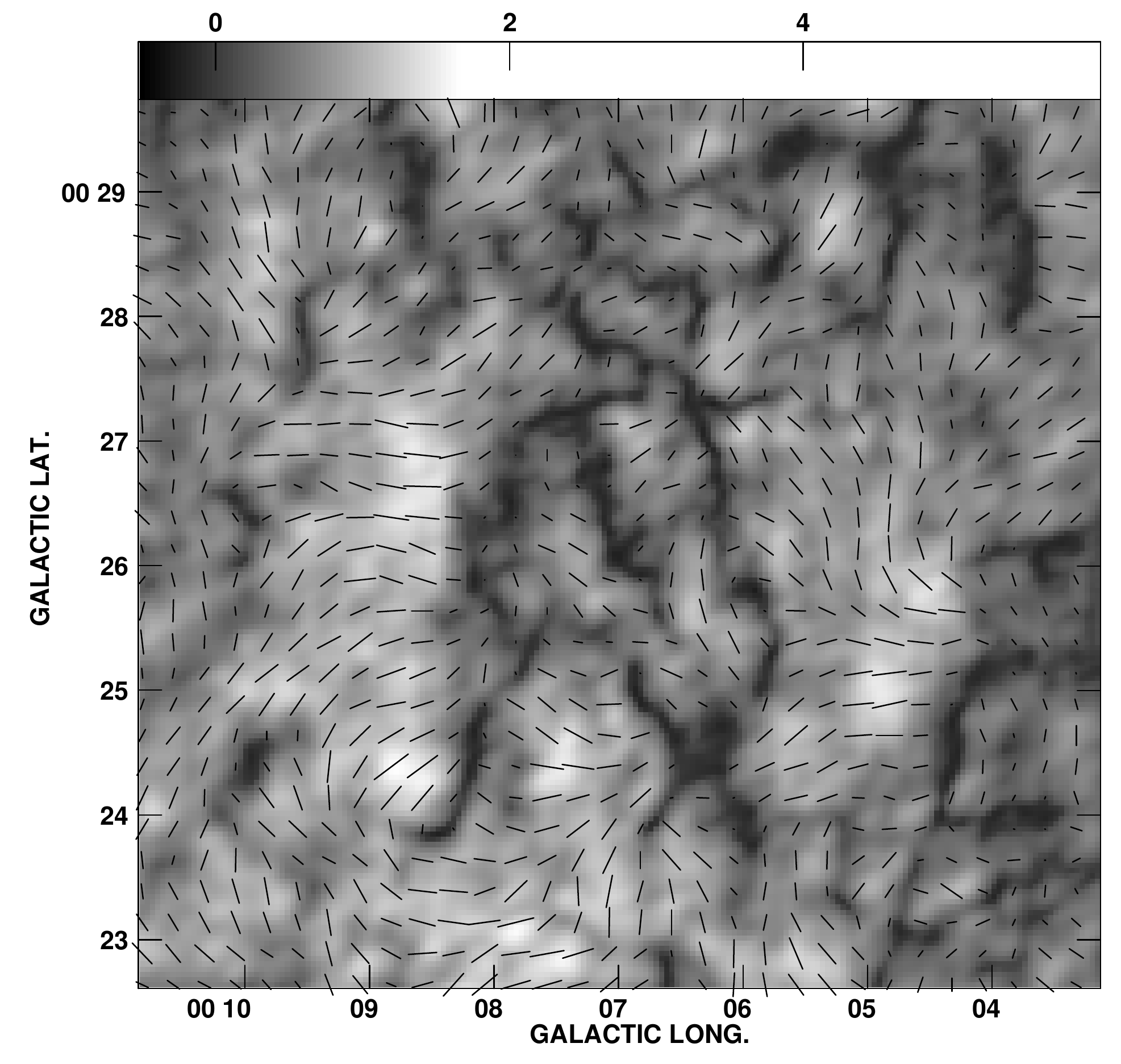}
\caption{Image of the 6 cm polarized intensity near the eastern edge of the survey, where depolarized ``canals'' are evident.  The lines show the polarization angle of the emission in IF 1.  The polarization angle tends to change by 90\sdeg\ across the canals, likely indicating beam depolarization by small-scale changes in the Faraday-rotating medium \citep{h04}. \label{canal}}
\end{figure}

The polarized intensity is visible on scales of a few arcminutes because it is laced with depolarized ``canals''.  These canals have been observed and studied extensively in other surveys of Galactic synchrotron emission \citep{w93,h00,g01,h04}.  Figure \ref{canal} shows an example of canals in the polarized emission near the eastern edge of the survey.  The canals seen there are like those seen in other work in that they are usually one beam wide ($\sim10$\arcsec) and the polarization angle changes across the canal by 90\sdeg\ \citep{h00,h04}.  Some canals are observed in the nonthermal filaments and are discussed in detail below.  Figures \ref{c3} through \ref{n11} show that the canals have no counterpart in total intensity emission;  the only signature of the depolarization is seen in polarized intensity images.

Single-dish observations have shown that the radio continuum emission throughout the GCL has a nonthermal spectral index, consistent with a synchrotron origin.  The range of spectral index values in the GCL suggests that the maximum, intrinsic polarization fraction should be near 75\% \citep{r79}.  The apparent polarization fraction is significantly less than the intrinsic value due to various depolarization effects.  Although there is no extended total intensity emission seen by the VLA, the polarization fraction can be estimated by comparing the brightest polarized intensity observed by the VLA to the GBT total intensity (Ch. \ref{gcsurvey_gbt}).  Convolving the VLA polarized intensity image to the 2\damin5-resolution of the GBT 6 cm data finds polarization fractions generally consistent with \citet{t86} and \citet{h92} with peak fractions reaching 25\% in the east and up to 10\% in the west (only up to $\sim$5\% coincident with the GCL-West).  
Since the depolarization is dominated by beam and depth depolarization effects, we can use the ratio of observed and intrinsic polarization fractions to constrain the degree to which these effects occur.  On 2\damin5 scales, the polarization fraction is reduced by factors of roughly 3 and 7 in the eastern and western parts of the survey, respectively from depth and beam depolarization effects.

\subsection{Polarized Filaments}
\label{filres}
This section describes the 6cm polarized emission properties of the NRFs detected in the present survey.

\subsubsection{G359.54+0.18 (NRF-C3)}
\begin{figure}[tbp]
\includegraphics[scale=0.37]{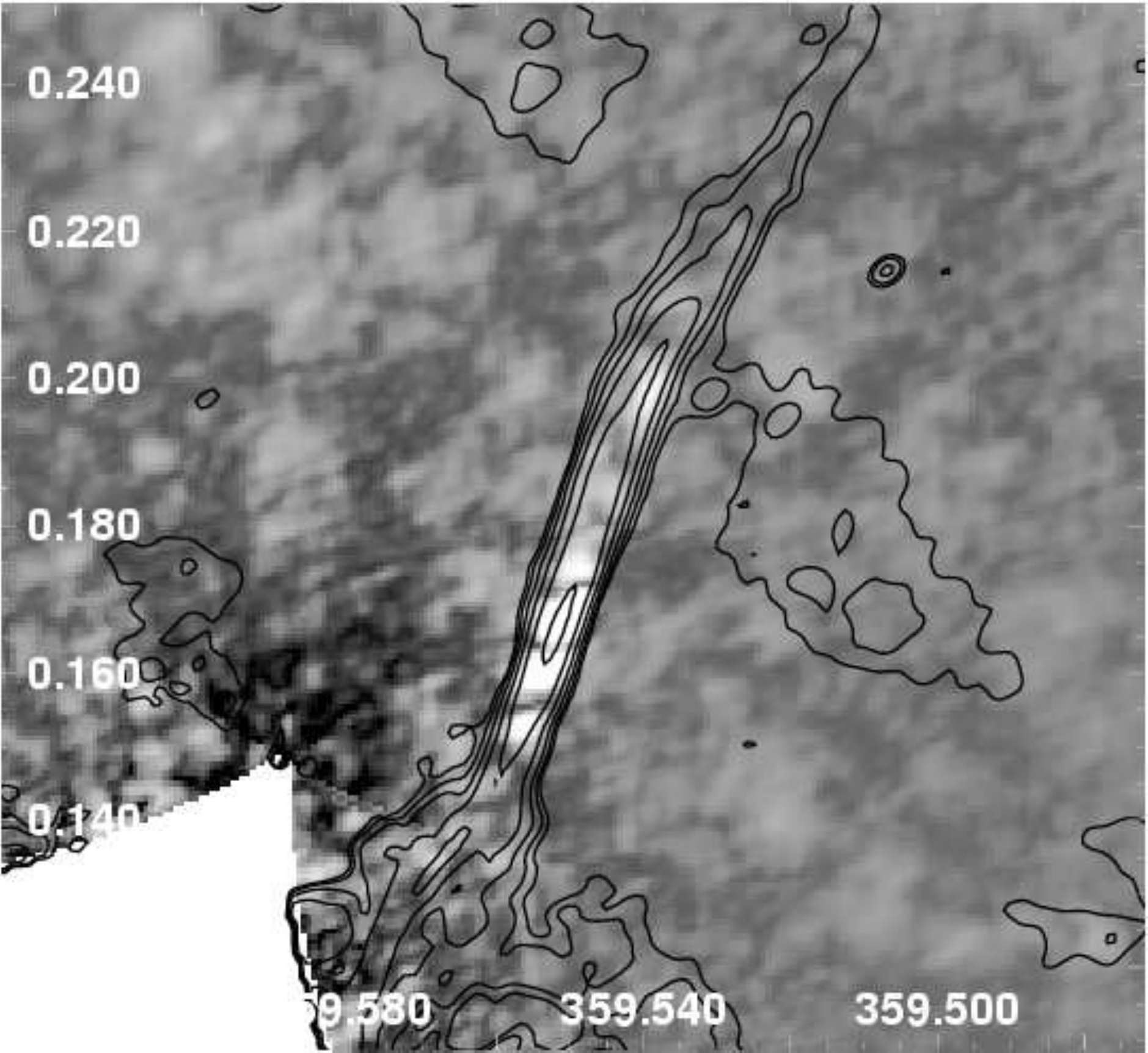}
\hfil
\includegraphics[scale=0.4,angle=270,bb=540 100 575 692]{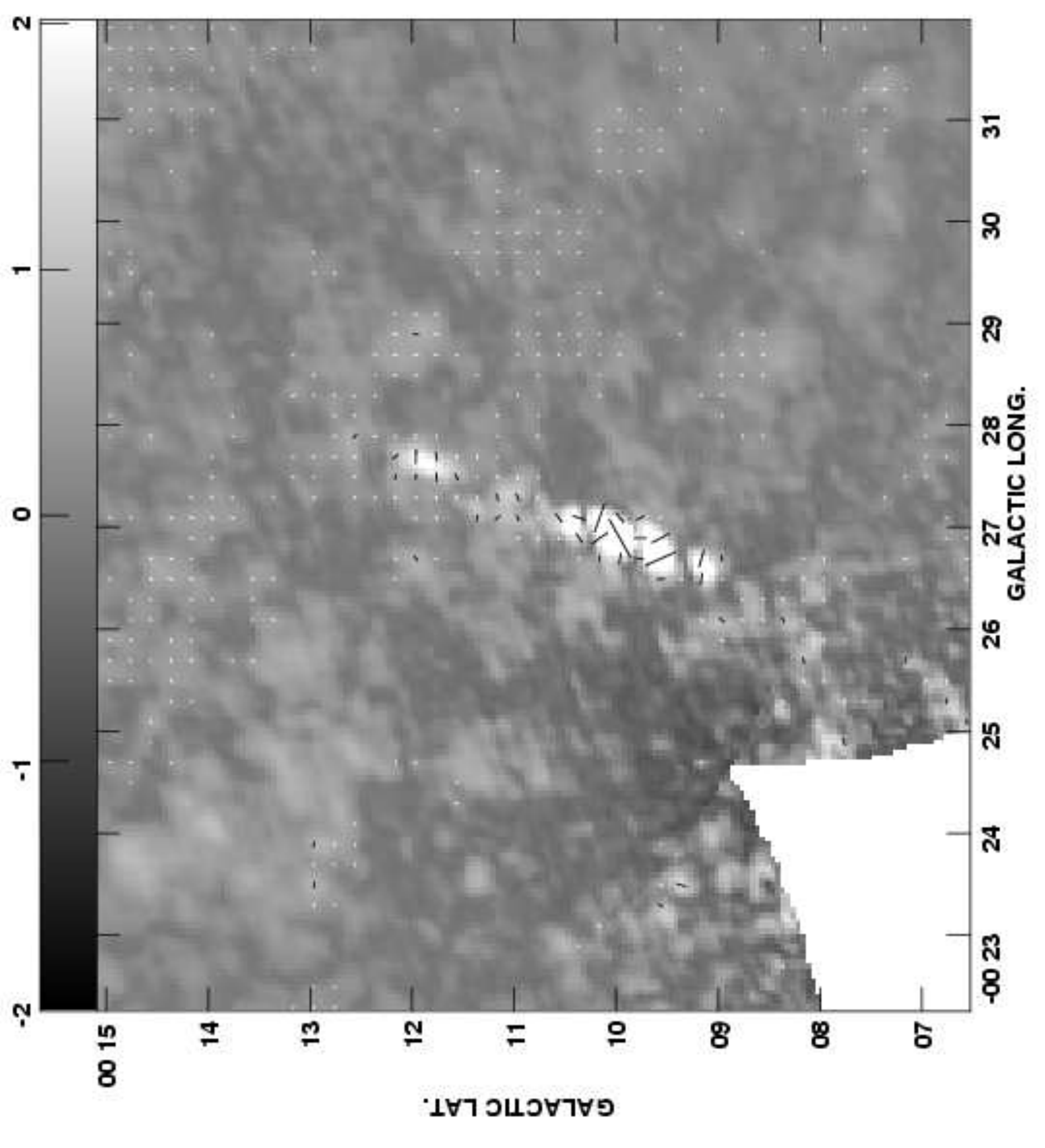}
\caption{\emph{Left}: Gray scale shows the 6cm polarized intensity from NRF-C3 (G359.54+0.18) for brightnesses ranging from -1 to 5 mJy beam$^{-1}$.  The contours show 6 cm total intensity at levels of $0.5*2^n$ mJy beam$^{-1}$, for $n=0-5$. \emph{Right}: Image of the same region with 6 cm polarization vectors shown.  Vectors are shown for regions with polarized intensity brighter than 0.2 mJy beam$^{-1}$. \label{c3}}
\end{figure}

Figure \ref{c3} shows the polarized emission from the nonthermal radio filament G359.54+0.18, also known as NRF-C3 \citep{y04}.  NRF-C3 is one of the brighter NRFs in total and polarized intensity, so it has been studied more than most NRFs \citep{b89,y97,r03}.  The 6 and 20 cm continuum emission observed as a part of this project are discussed in chapter \ref{gcl_vla}.  The peak 6 cm flux is about 18 mJy per 10\arcsec$\times$15\arcsec\ beam and the 6/20 cm spectral index is typically --0.8.

The polarized emission is broken into five rough ``islands'' of polarization about 30\arcsec\ across, separated from each other by depolarized canals.  Four of the islands are located along the brightest total intensity emission where the NRF is straight, while the fifth island is north of the others, near the northern bend in the NRF.  The middle three islands are the three seen most clearly in \citet{y97} and discussed in \S\ \ref{poln_comparison}.  The brightest polarized intensity from NRF-C3 is 6 mJy beam$^{-1}$ with a polarization fraction of about 35\%.  The other polarized islands have lower polarized intensities with polarization fractions ranging from 20 to 30\%.

Each of the polarized islands has a relatively constant \dt, so \rmeff\ is calculated for each one.  From south to north, the average \dt value for the islands are:  11$\pm$6\sdeg$=-2420\pm1320$ rad m$^{-2}$ (at 359\ddeg556,0\ddeg152), $7\pm3$\sdeg$=-1540\pm660$ rad m$^{-2}$ (at 359\ddeg554,0\ddeg163), $10\pm2$\sdeg$=-2200\pm440$ rad m$^{-2}$ (at 359\ddeg550,0\ddeg167), $18\pm5$\sdeg$=-3960\pm1100$ rad m$^{-2}$ (at 359\ddeg549,0\ddeg174), and $15\pm4$\sdeg$=-3300\pm880$ rad m$^{-2}$ (at 359\ddeg539,0\ddeg198).  As discussed below in \S\ \ref{padires}, the background polarized emission in the region has \dt$\approx1\ddeg5$, which is equivalent to an \rmeff\ of $-330$ rad m$^{-2}$, significantly less than that seen toward NRF-C3.

\subsubsection{G359.36+0.10 (NRF-C12)}
\begin{figure}[tbp]
\includegraphics[scale=0.37]{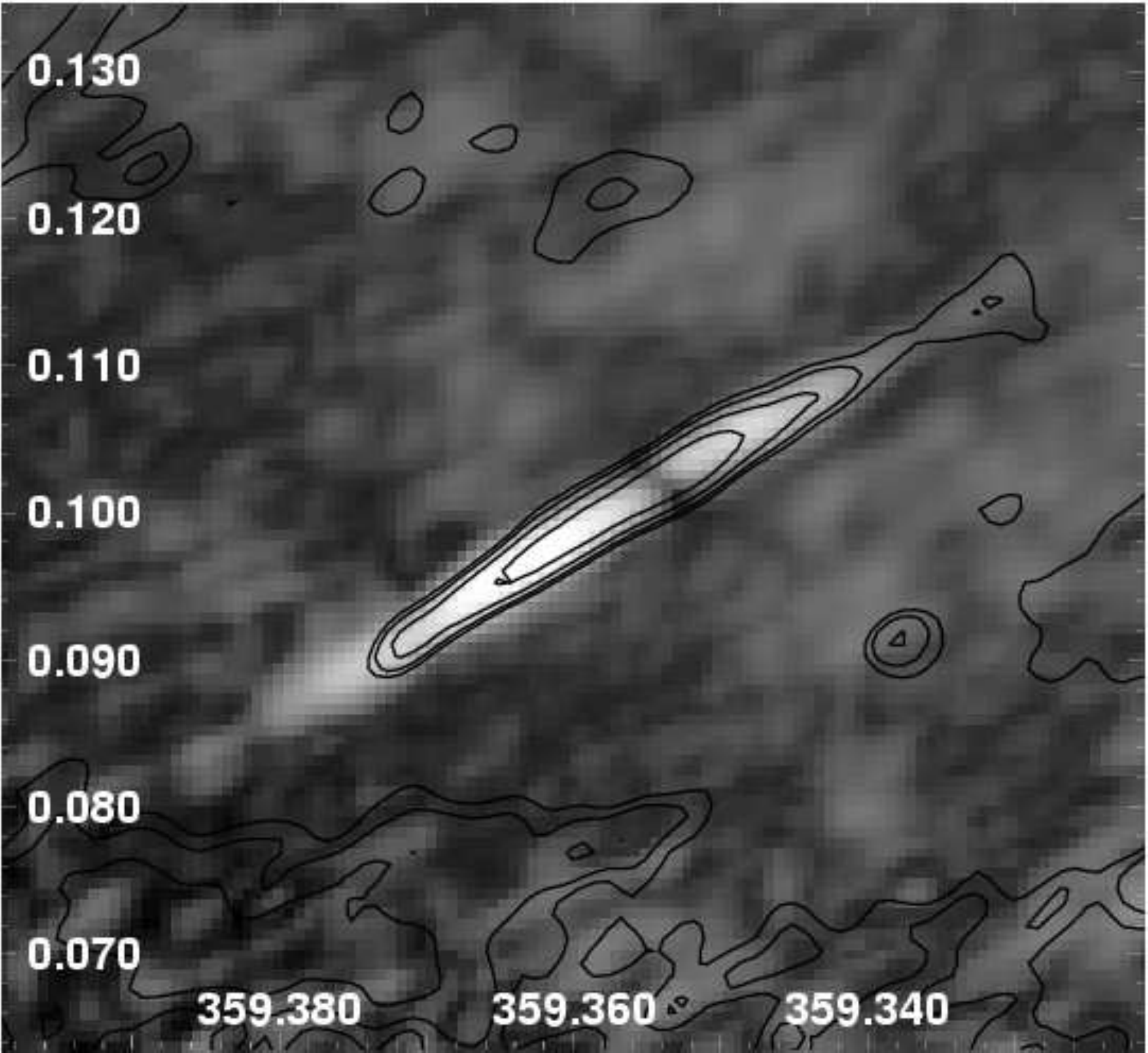}
\hfil
\includegraphics[scale=0.4,angle=270,bb=540 130 575 692]{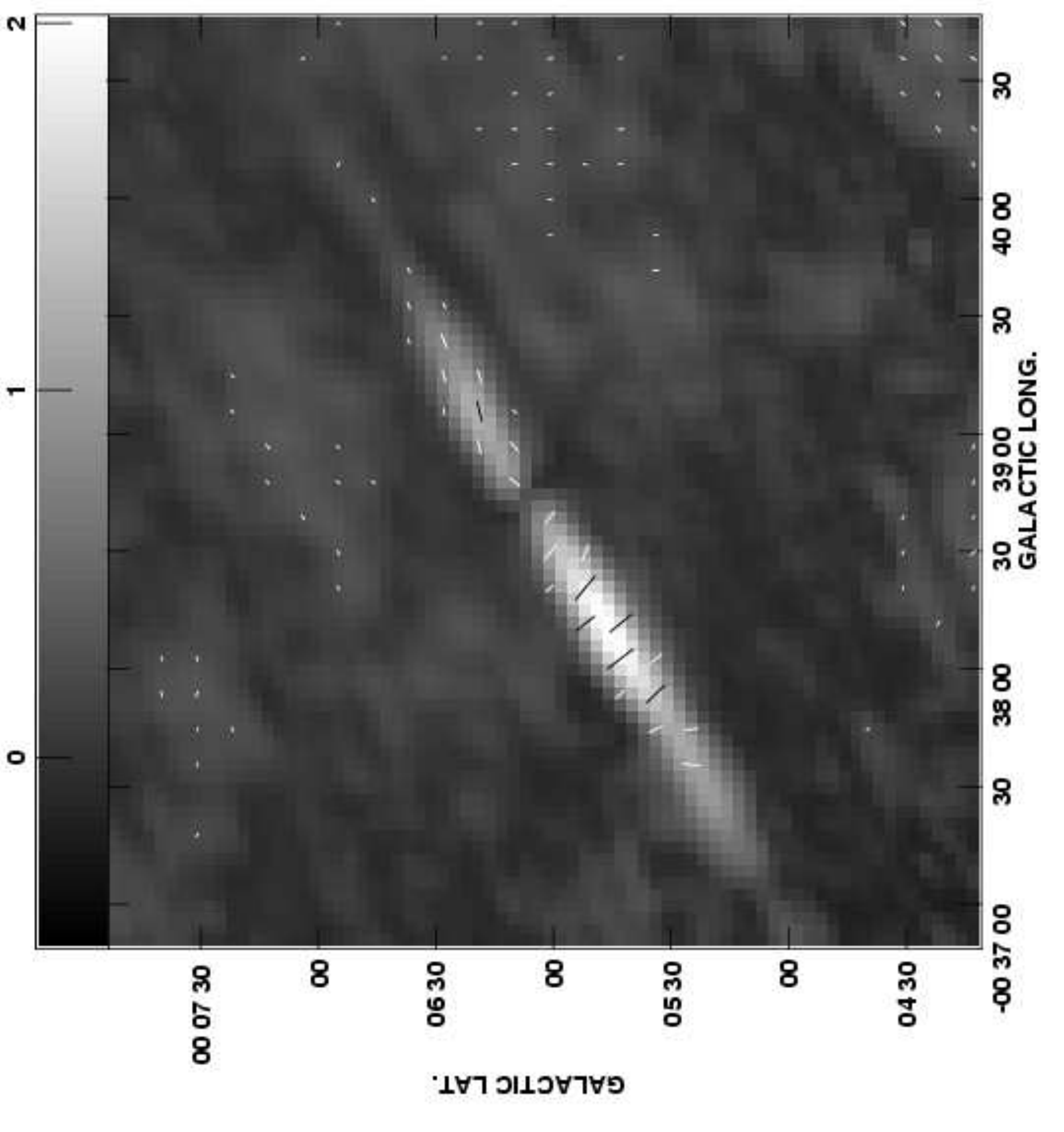}
\caption{\emph{Left}: Gray scale shows the 6cm polarized intensity from NRF-C12 (G359.36+0.10) for brightnesses ranging from -1 to 5 mJy beam$^{-1}$.  The contours show 6 cm total intensity at levels of $0.25*2^n$ mJy beam$^{-1}$, for $n=0-3$. \emph{Right}: Image of the same region with 6 cm polarization vectors shown.  Vectors are shown for regions with polarized intensity brighter than 0.2 mJy beam$^{-1}$. \label{c12}}
\end{figure}

The polarized intensity from nonthermal filament G359.36+0.10 is shown in Figure \ref{c12}.  This filament, also known as NRF-C12 \citep{y04}, is oriented at a 30\sdeg\ angle to the Galactic plane and is roughly perpendicular to filament RF-C11 that crosses it in projection.  The peak brightness at 6 cm is about 3.5 mJy per 10\arcsec$\times$15\arcsec\ beam and the 6/20 cm spectral index ranges from --0.5 to --1.8 (Ch. \ref{gcl_vla}).

Polarized emission is seen across the entire 4\arcmin\ length visible in 6 cm total intensity.  This is the first detection of linearly-polarized emission from G359.36+0.10 and confirms that it is a NRF.  There is one depolarization canal near the midpoint of the filament that divides it into eastern and western halves.  The peak, 6 cm polarized intensity of about 2 mJy beam$^{-1}$ is located along the eastern half, while the peak in 6 cm total intensity is located along the western half.  Thus, the polarization fraction changes along the length of NRF-C12 with a value of 30\% at the point of brightest total intensity and 70\% at the point of hightest polarized intensity.  The polarized emission is clearly detected at the faintest ends of the filament in total intensity such that the apparent polarization fraction in those regions is higher than 100\%.  

The \dt observed in the polarized emission of NRF-C12 gives an estimate of the $RM$ toward the filament.  The eastern and western halves have averages of \dt$\approx-4\pm2$\sdeg\ and $-2\pm2$\sdeg, respectively, giving \rmeff$=880\pm440$ rad m$^{-2}$ and $440\pm440$ rad m$^{-2}$.  The average value of \dt for the extended polarized emission in the region (discussed in \S\ \ref{padires}) is +4\sdeg, which gives \rmeff$=-880$ rad m$^{-2}$.

\subsubsection{G0.15+0.23 (NRF-N1)}
\begin{figure}[tbp]
\includegraphics[scale=0.37]{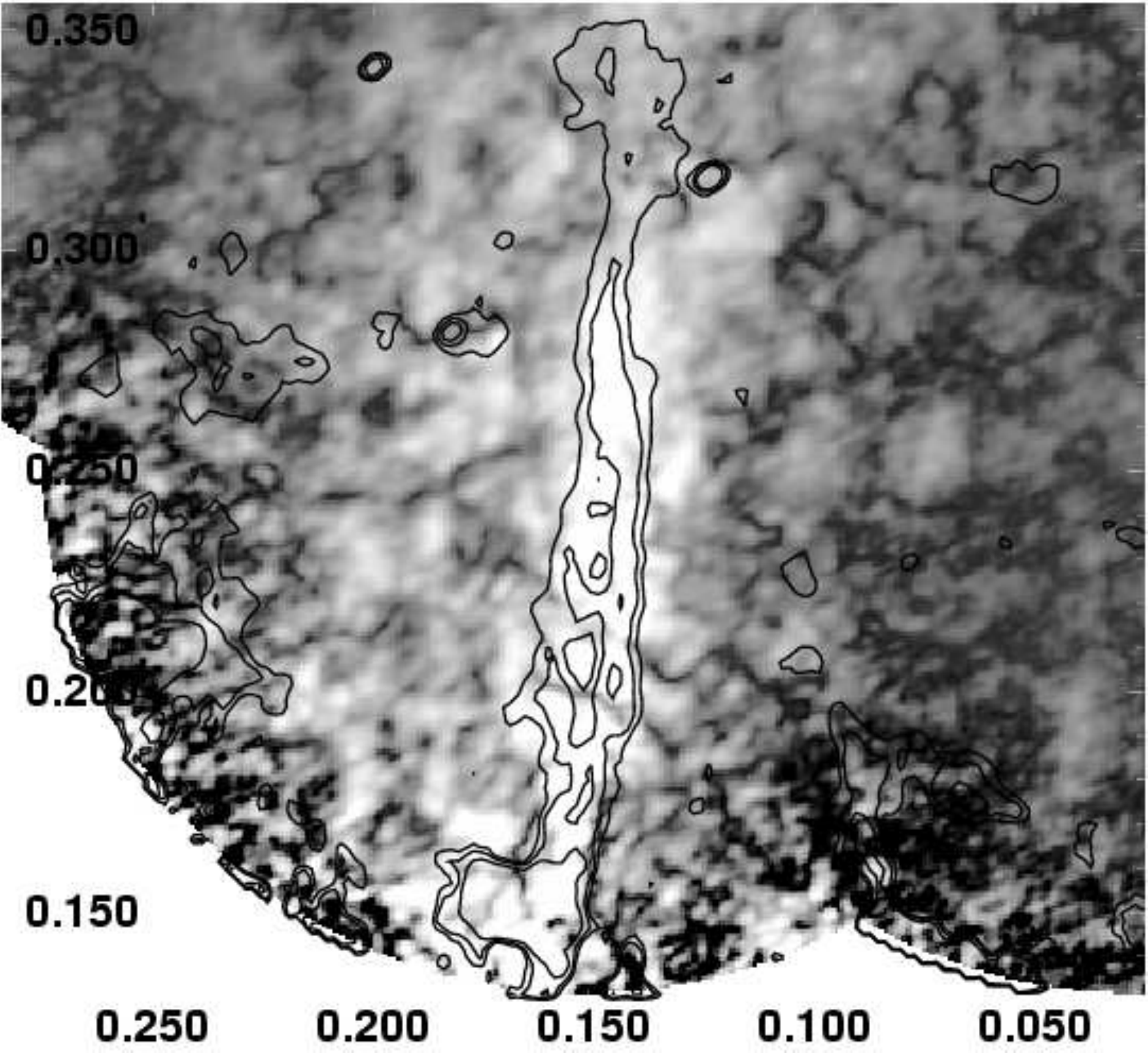}
\hfil
\includegraphics[scale=0.4,angle=270,bb=540 130 575 692]{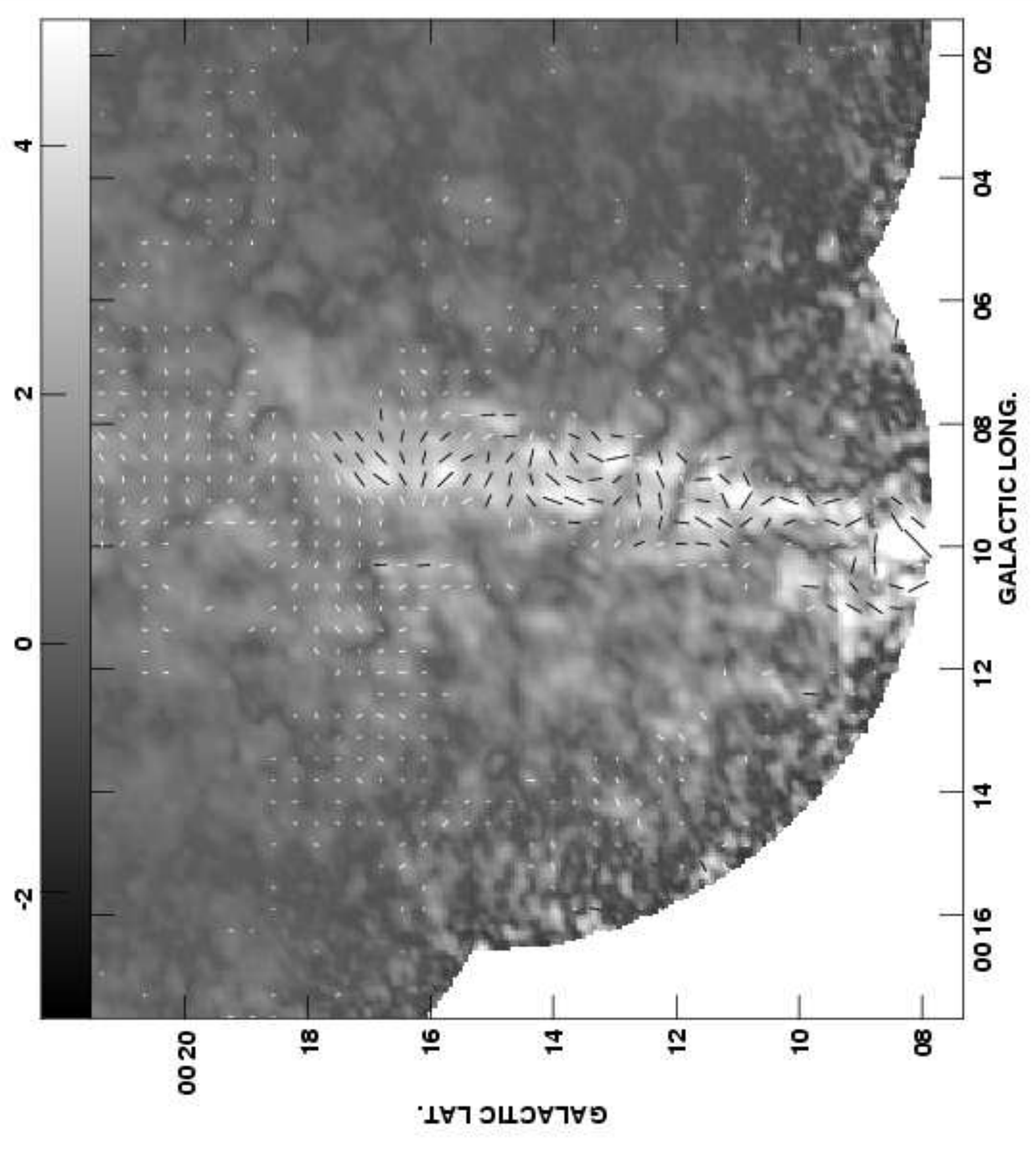}
\caption{\emph{Left}:  Gray scale shows the 6cm polarized intensity from NRF-N1 (G0.15+0.23) for brightnesses ranging from -1 to 5 mJy beam$^{-1}$.  The contours show 6 cm total intensity at levels of 1, 2, and 4 mJy beam$^{-1}$.  \emph{Right}:  Image of the same region with 6 cm polarization vectors shown.  Vectors are shown for regions with polarized intensity brighter than 1 mJy beam$^{-1}$. \label{n1}}
\end{figure}

Figure \ref{n1} shows the polarized emission from the nonthermal radio filament G0.15+0.23.  This filament, also known as NRF-N1 \citep{y04}, is oriented perpendicular to the Galactic plane and extends from the plane up to a latitude of 0\ddeg3.  Polarized emission fills the region near $l=0\ddeg15$ from the Radio Arc in the Galactic plane up into the GCL \citep{y97}.  Although NRF-N1 is in the midst of this extended polarized emission, no polarized emission has been specifically associated with the filament until the present 6 cm observations.  The region around NRF-N1 is also highly confused in total intensity images, making it difficult to isolate.  The 6 cm, total intensity brightness is as high as 5 mJy per 10\arcsec$\times$15\arcsec\ beam and the 6/20 cm spectral index ranges from +0.2 to --0.5 (Ch. \ref{gcl_vla}).

The polarized emission from NRF-N1 is not uniform, but has a clumpy structure formed by depolarization canals.  The brightest polarized intensity is about 5.5 mJy beam$^{-1}$.  In many locations, the polarized brightness is greater than the total intensity, so the polarized fraction exceeds 100\%.  One possibility is that the total intensity is more likely to be resolved out than the polarized intensity, since the structure of the polarized intensity is broken into many small pieces by depolarization canals.

The average \dt for NRF-N1 was measured for $b=0\ddeg16-0\ddeg30$ to be $-2\pm0\ddeg5$, equivalent to \rmeff$=440\pm110$ rad m$^{-2}$.  The average \dt and \rmeff\ is similar to the average of the surrounding extended polarized emission, as measured over adjacent regions with a similar latitude extent.  As described below, the average \dt over all latitudes of the present survey is slightly smaller, with \dt$\approx-1$\sdeg\ or \rmeff$\approx220$ rad m$^{-2}$.

\subsubsection{G0.08+0.15 (NRF-N2)}
\begin{figure}[tbp]
\includegraphics[scale=0.37]{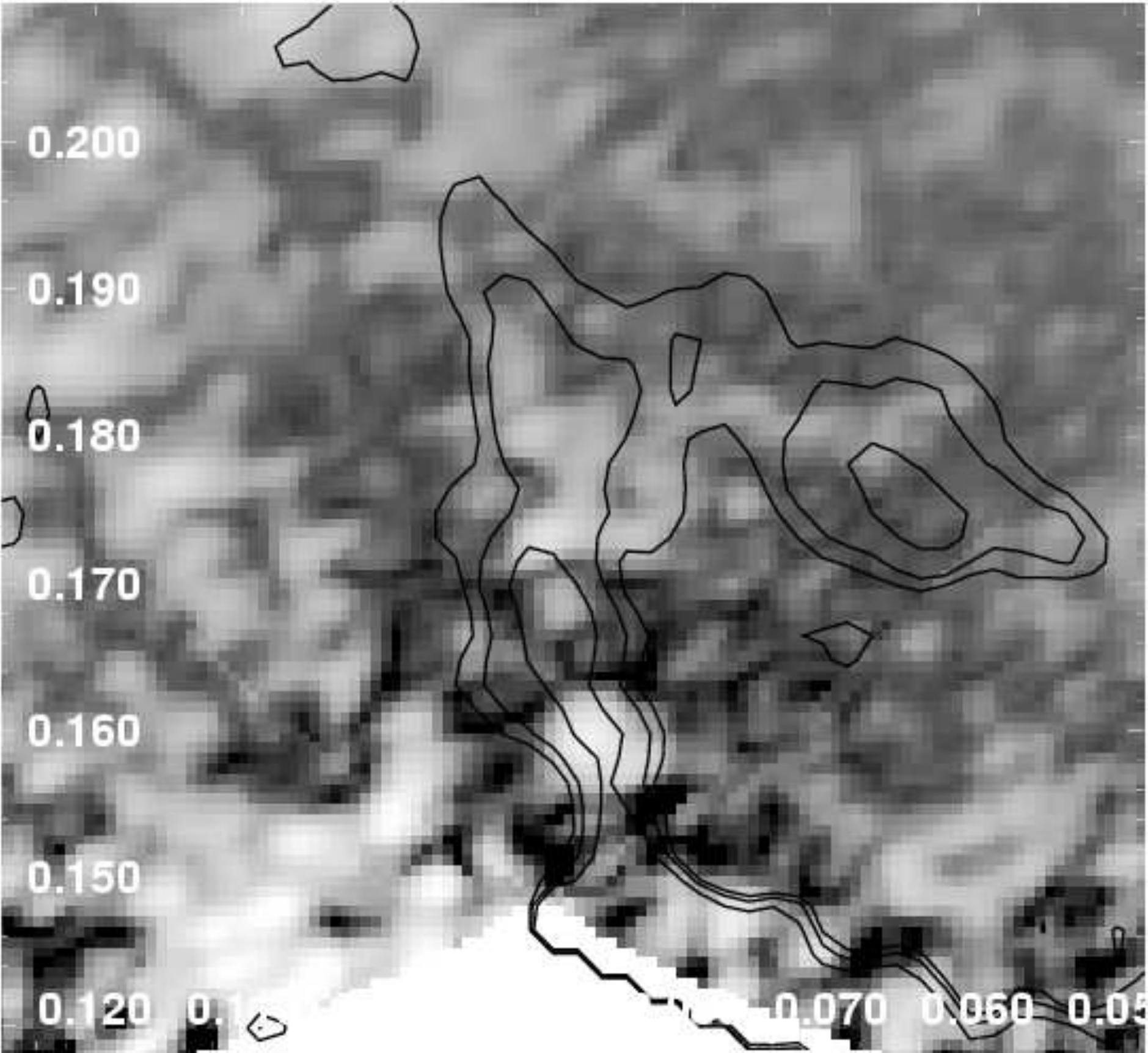}
\hfil
\includegraphics[scale=0.4,angle=270,bb=540 130 575 692]{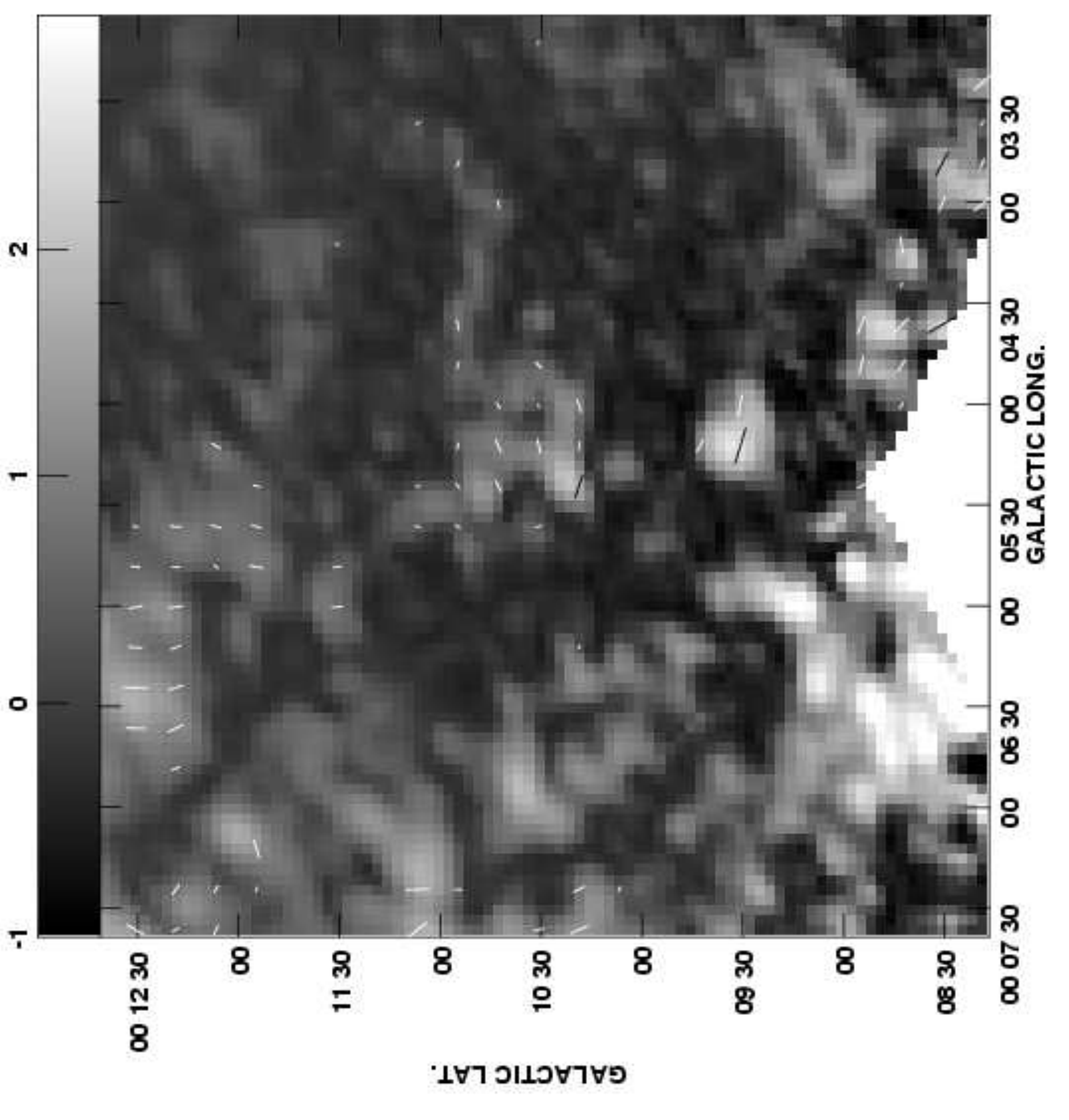}
\caption{\emph{Left}:  Gray scale shows the 6cm polarized intensity from NRF-N2 (G0.08+0.15) for brightnesses ranging from -1 to 5 mJy beam$^{-1}$.  The contours show 6 cm total intensity at levels of 1, 2, and 4 mJy beam$^{-1}$.  \emph{Right}:  Image of the same region with 6 cm polarization vectors shown.  Vectors are shown for regions with polarized intensity brighter than 0.2 mJy beam$^{-1}$. \label{n2}}
\end{figure}

Figure \ref{n2} shows the polarized emission from G0.08+0.15, also known as NRF-N2 \citep{y04} or the ``Northern thread'' \citep{l99}.  This filament extends several arcminutes roughly perpendicular to the Galactic plane, but only half of the filament lies in the coverage of the present survey.  In this region, the 6 cm brightness of NRF-N2 is about 6 mJy beam$^{-1}$ and a 6/20 cm spectral index ranging from --0.8 to --1.2 (Ch. \ref{gcl_vla}).

The 6 cm, polarized brightness image shows extended emission at the northern end of NRF-N2 with a brightness of 2 mJy beam$^{-1}$ and a region near the southern edge of the 6 cm survey with a brightness of 2.5 mJy beam$^{-1}$.  The southern polarized emission has the highest polarization fraction of $\sim50$\%.  The average \dt for the polarized emission from NRF-N2 is very poorly constrained, since it is near the edge of the survey, where the primary-beam corrections are large.

\subsubsection{G359.96+0.09 (NRF-N5)}
\begin{figure}[tbp]
\includegraphics[scale=0.37]{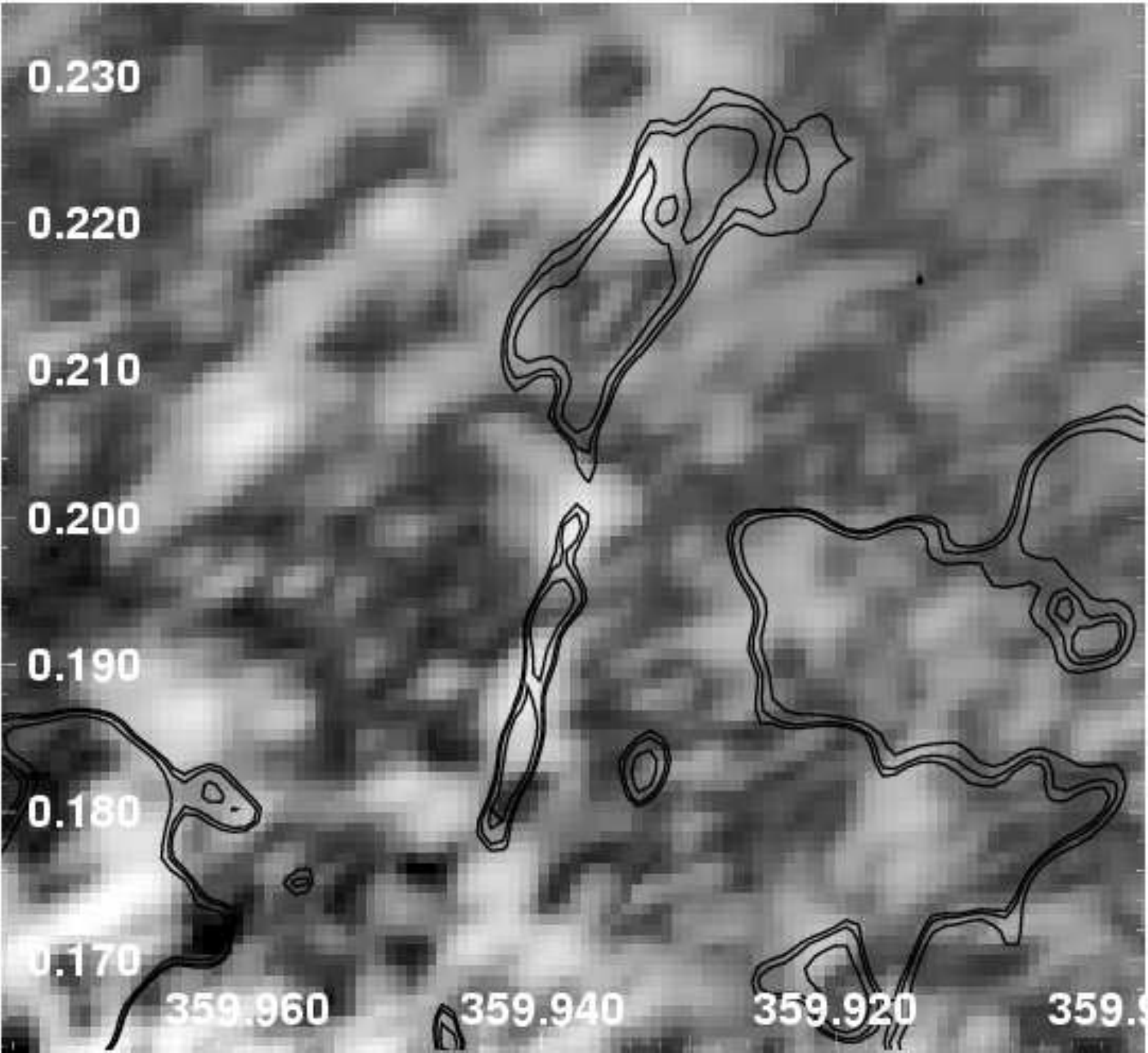}
\hfil
\includegraphics[scale=0.4,angle=270,bb=540 130 575 692]{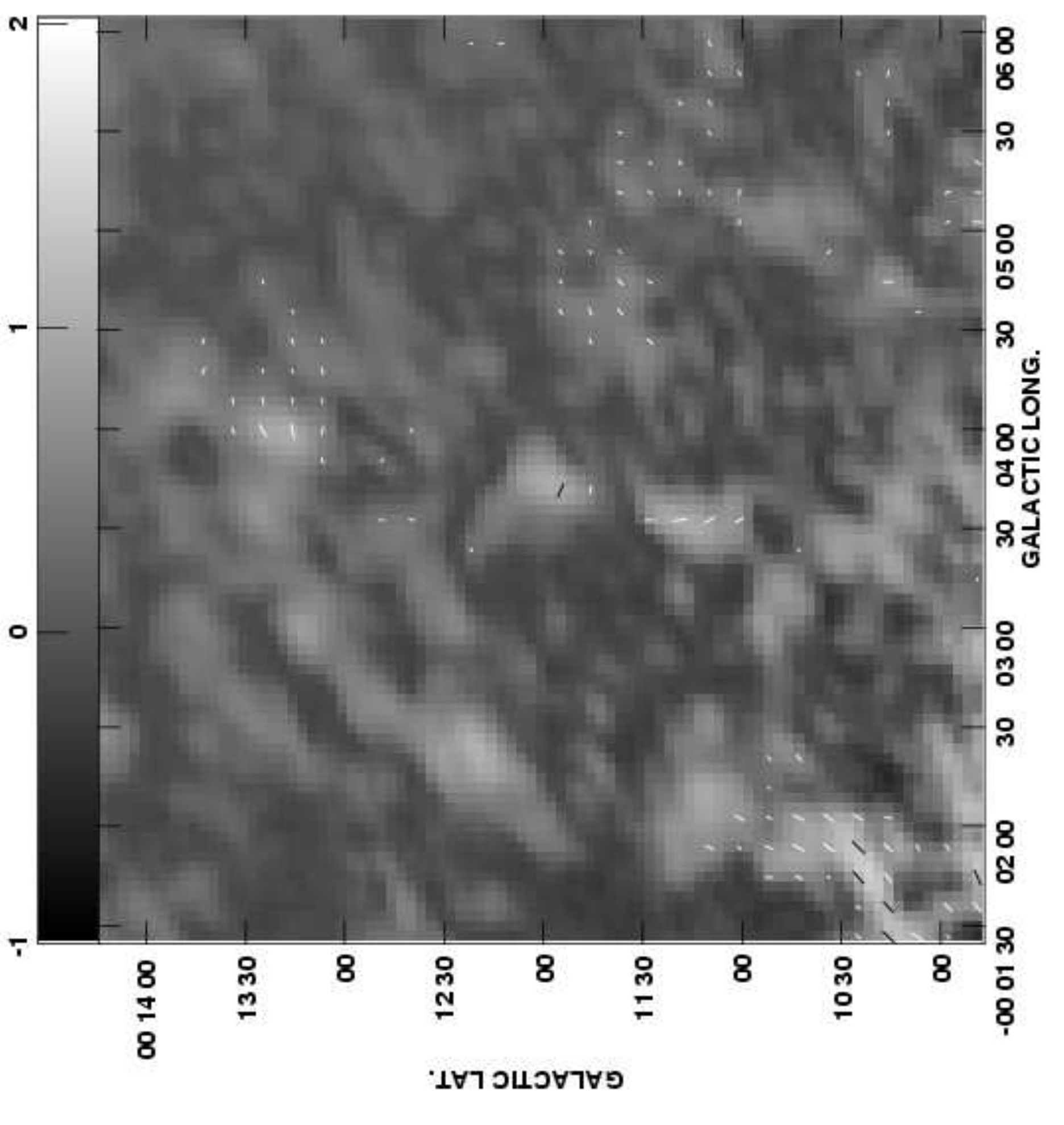}
\caption{\emph{Left}:  Gray scale shows the 6cm polarized intensity from NRF-N5 (G359.96+0.09) for brightnesses ranging from -1 to 5 mJy beam$^{-1}$.  The contours show 6 cm total intensity at levels of 0.1, 0.2, and 0.4 mJy beam$^{-1}$. \emph{Right}:  Image of the same region with 6 cm polarization vectors shown.  Vectors are shown for regions with polarized intensity brighter than 0.2 mJy beam$^{-1}$. \label{n5}}
\end{figure}

Figure \ref{n5} shows the 6 cm polarized continuum emission from the nonthermal radio filament G359.95+0.09.  The filament, also known as NRF-N5 \citep{y04} and ``the Southern Thread'' \citep{l99}, spans roughly 3\arcmin\ perpendicular to the Galactic plane.  The 6 cm peak brightness of NRF-N5 is not easy to constrain, since it is very faint and located in a noisy part of the survey, although it is constrained to be less than 1 mJy beam$^{-1}$ north of $b=0\ddeg16$.

The 6 cm polarized emission from NRF-N5 is found in three islands with peak brightnesses of $\sim$1 mJy beam$^{-1}$.  The Faraday rotation toward NRF-N5 gives \dt$=-5\pm4$\sdeg, which is equivalent to \rmeff$=1100\pm880$ rad m$^{-1}$.  At the longitude of NRF-N5, the extended polarized emission has \dt$\approx-2$\sdeg, or $RM\approx440$ rad m$^{-2}$.

\subsubsection{G359.79+0.17 (NRF-N8)}
\begin{figure}[tbp]
\includegraphics[scale=0.37]{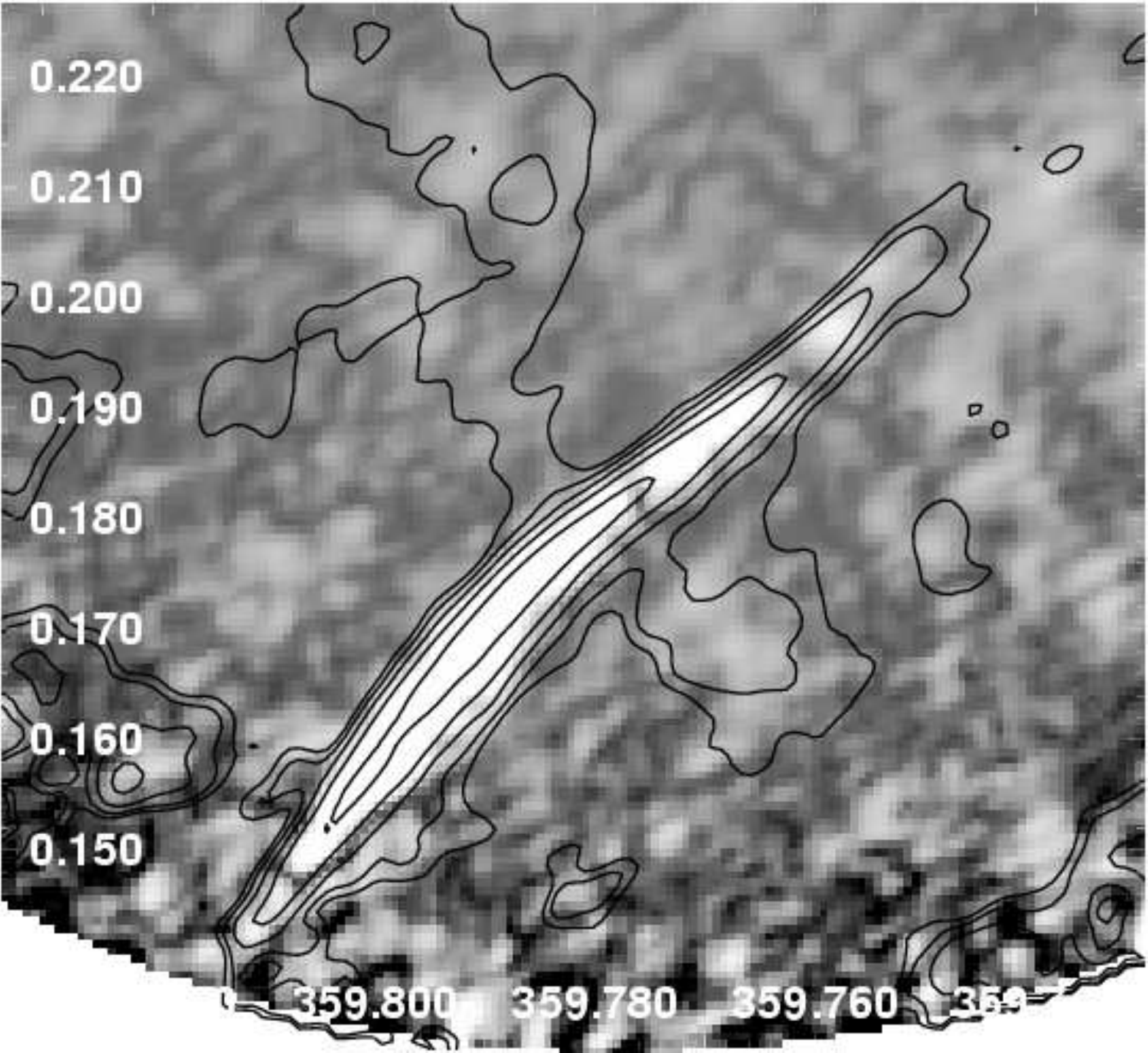}
\hfil
\includegraphics[scale=0.4,angle=270,bb=540 130 575 692]{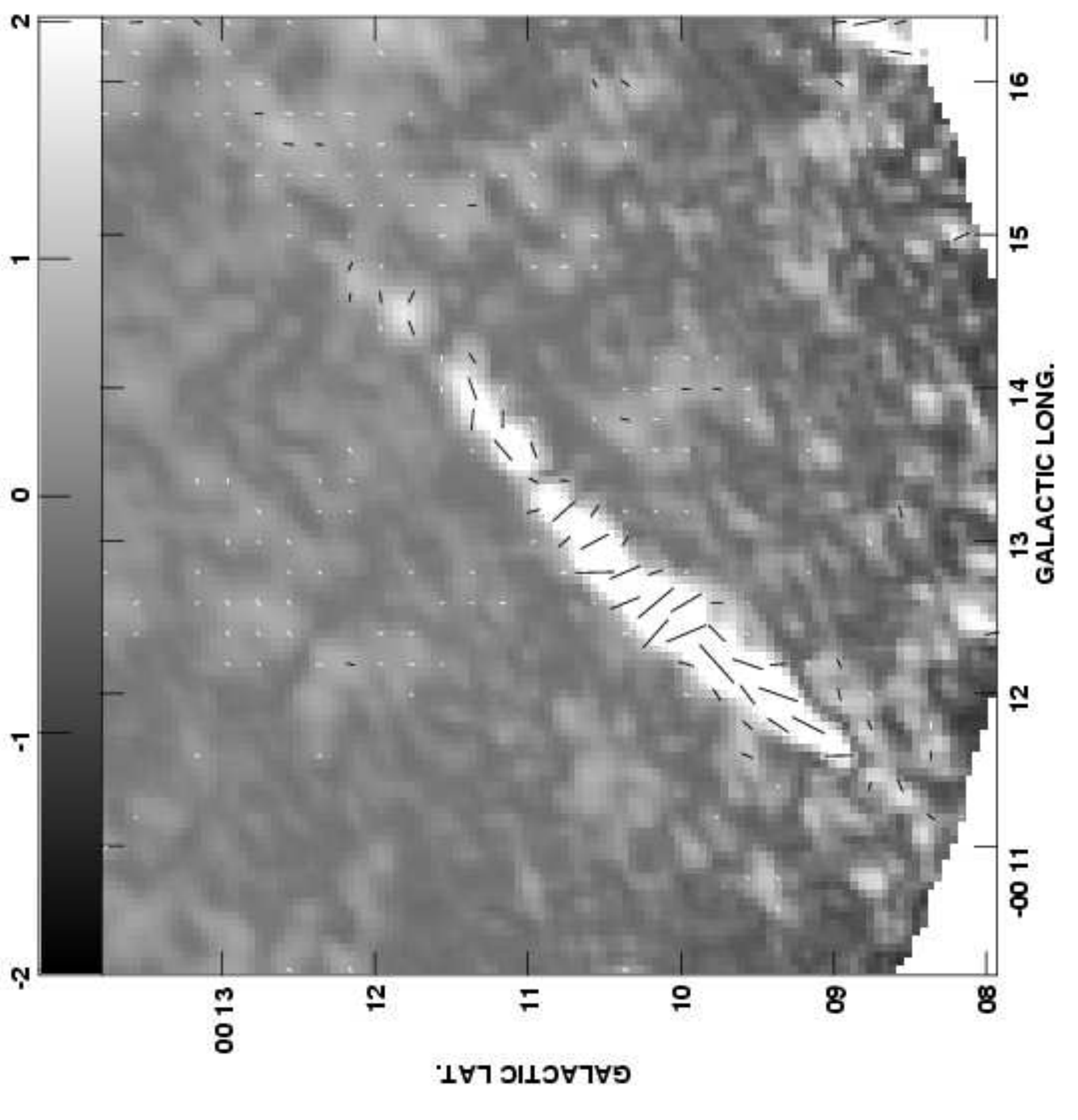}
\caption{\emph{Left}:  Gray scale shows the 6cm polarized intensity from NRF-N8 (G359.79+0.17) for brightnesses ranging from -1 to 5 mJy beam$^{-1}$.  The contours show 6 cm total intensity at levels of $0.5*2^n$ mJy beam$^{-1}$, for $n=0-4$. \emph{Right}:  Image of the same region with 6 cm polarization vectors shown.  Vectors are shown for regions with polarized intensity brighter than 0.2 mJy beam$^{-1}$.  \label{n8}}
\end{figure}

Figure \ref{n8} shows the polarized emission from G359.79+0.17, also known as NRF-N8 \citep{y04}.  NRF-N8 is among the brighter NRFs at 6 cm with peak brightnesses of 11 mJy beam$^{-1}$ and 6 mJy beam$^{-1}$ in total and polarized intensity, respectively.  The filament is oriented at a roughly 45\sdeg\ angle to the Galactic plane and is about 8\arcmin\ long.  The 6/20 cm spectral index ranges from --0.9 to --1.3, with a trend for a steeper index found toward its southern end (Ch. \ref{gcl_vla}).

The polarized intensity from NRF-N8 is broken in to several islands by depolarization canals.  Generally, the polarized brightness follows the total intensity brightness.  The maximum polarization fraction is 55\%, which is found at the peak brightness of two or three of the brightest polarized intensity islands.  Averaging over the polarized emission seen at 6 cm finds \dt$=-4\pm1$\sdeg, equivalent to \rmeff$=880\pm220$ rad m$^{-2}$.  At this longitude, the extended polarized emission typically has \dt$=-2$\sdeg, or \rmeff$=440$ rad m$^{-2}$.

\subsubsection{G359.85+0.39 (NRF-N10)}
\label{n10sec}
\begin{figure}[tbp]
\includegraphics[scale=0.37]{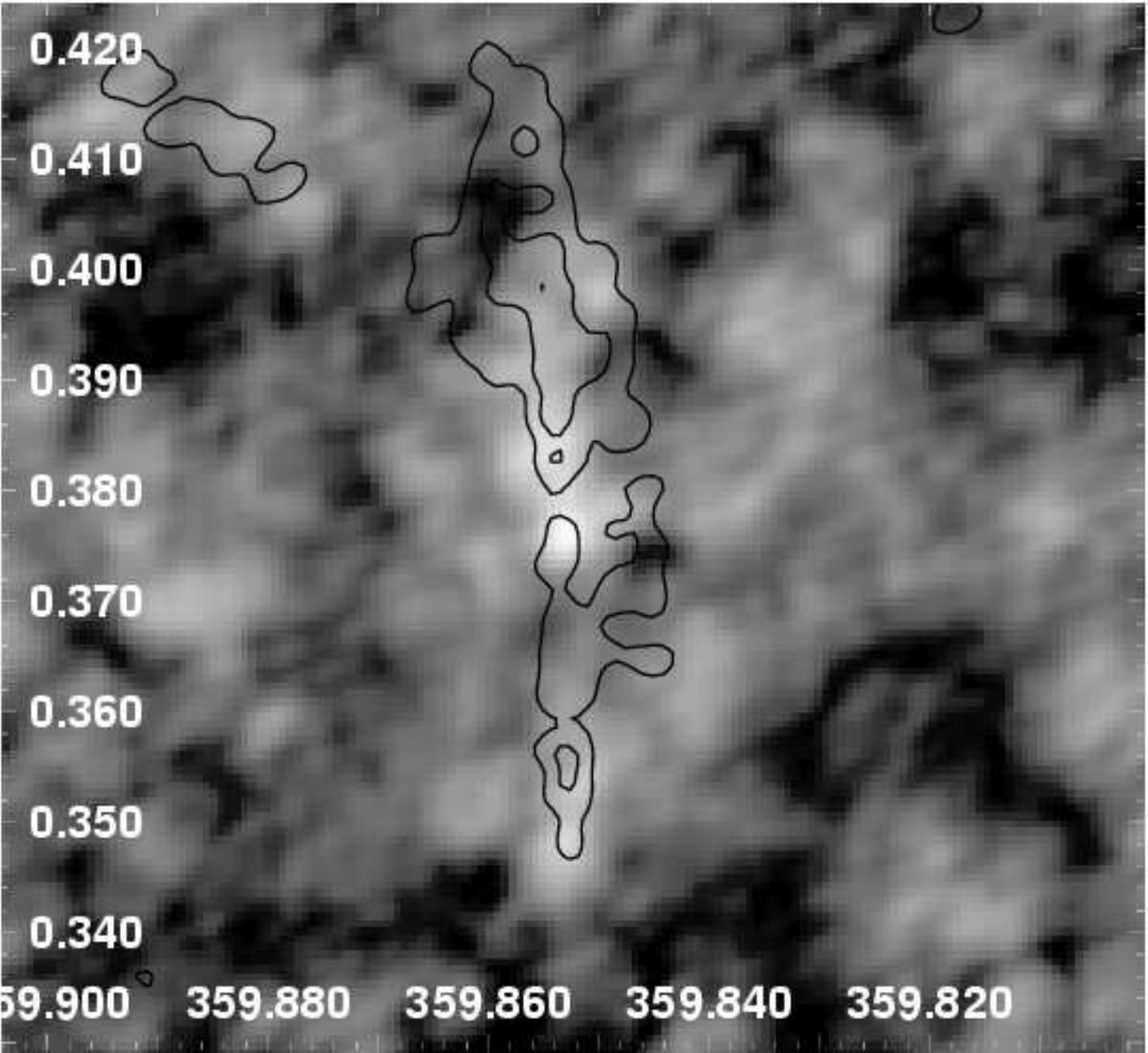}
\hfil
\includegraphics[scale=0.38,bb=340 30 575 692]{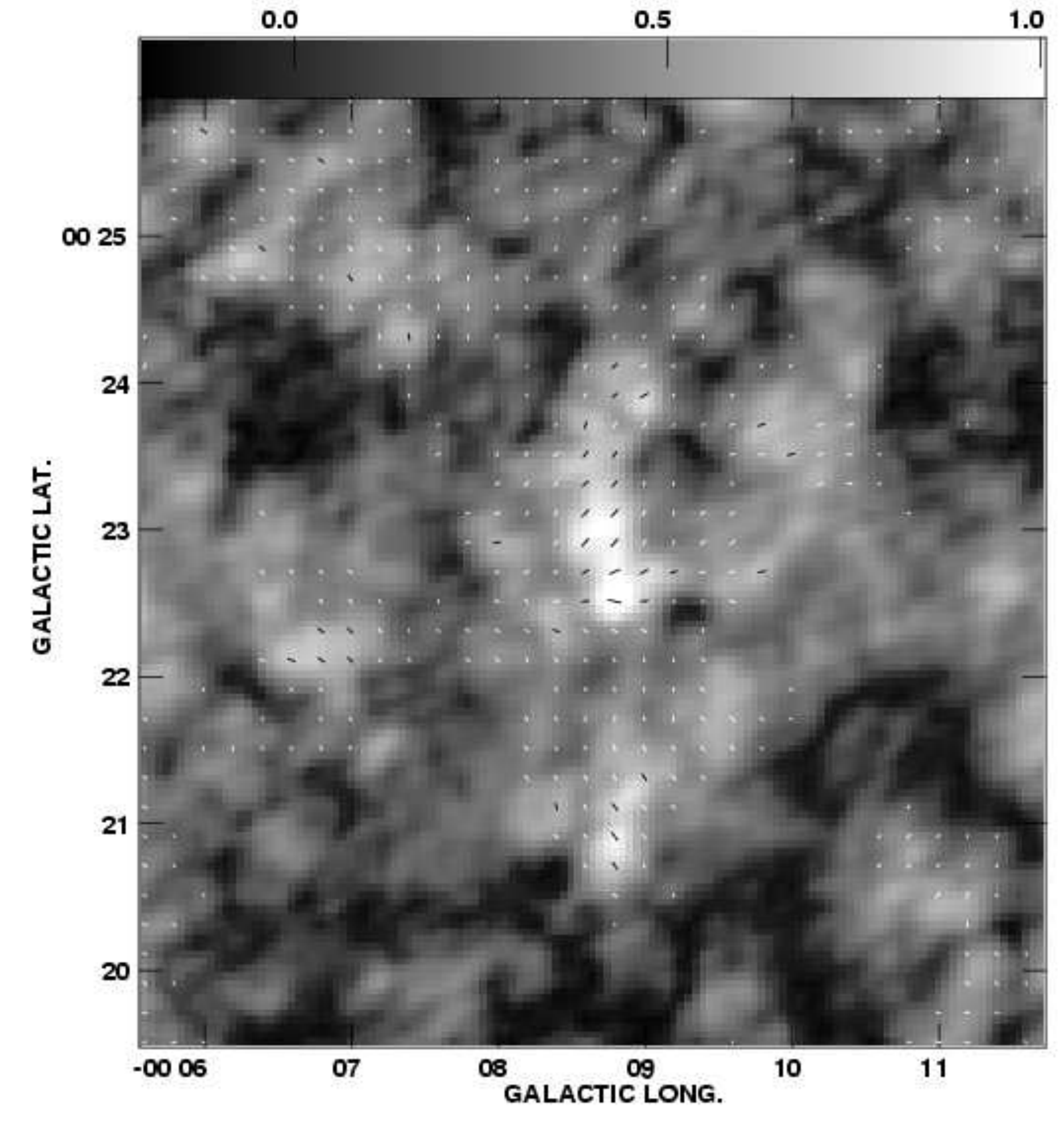}
\caption{\emph{Left}:  Gray scale shows the 6cm polarized intensity from NRF-N10 (G359.85+0.39) for brightnesses ranging from -1 to 5 mJy beam$^{-1}$.  The contours show 6 cm total intensity at levels of 0.5 and 0.7 mJy beam$^{-1}$.  \emph{Right}:  Image of the same region with 6 cm polarization vectors shown.  Vectors are shown for regions with polarized intensity brighter than 0.2 mJy beam$^{-1}$. \label{n10}}
\end{figure}

The nonthermal filament G359.85+0.39, also known as NRF-N10 \citep{y04} or the ``Cane'' \citep{l01}, is shown in Figure \ref{n10}.  NRF-N10 is composed of at least two parallel filaments oriented perpendicular to the Galactic plane with the longest subfilament (also the brightest) spanning about 5\arcmin.  NRF-N10 is at the northern edge of an apparent polarized loop of emission 10\arcmin\ across, which is located just north of Sgr A (see \S\ \ref{localized}).  The brightest, 6 cm total intensity is about 0.9 mJy beam$^{-1}$ and the 6/20 cm spectral index ranges from --0.6 to --1.5 (Ch. \ref{gcl_vla}).

The polarized brightness is not uniform or well correlated with total intensity.  The polarized intensity is found in a few isolated islands of emission along the brightest subfilament.  The brightest polarized emission is about 1.2 mJy beam and the polarization fraction is larger than 100\% at a few locations along the brightest subfilament.  The average \dt for the region with polarized emission is $-1\pm1$\sdeg, or \rmeff$=220\pm220$ rad m$^{-2}$.  The typical \dt value at that longitude is $-1\ddeg5$, which is equivalent to \rmeff$=330$ rad m$^{-2}$.

\subsubsection{G359.62+0.28 (NRF-N11)}
\begin{figure}[tbp]
\includegraphics[scale=0.37]{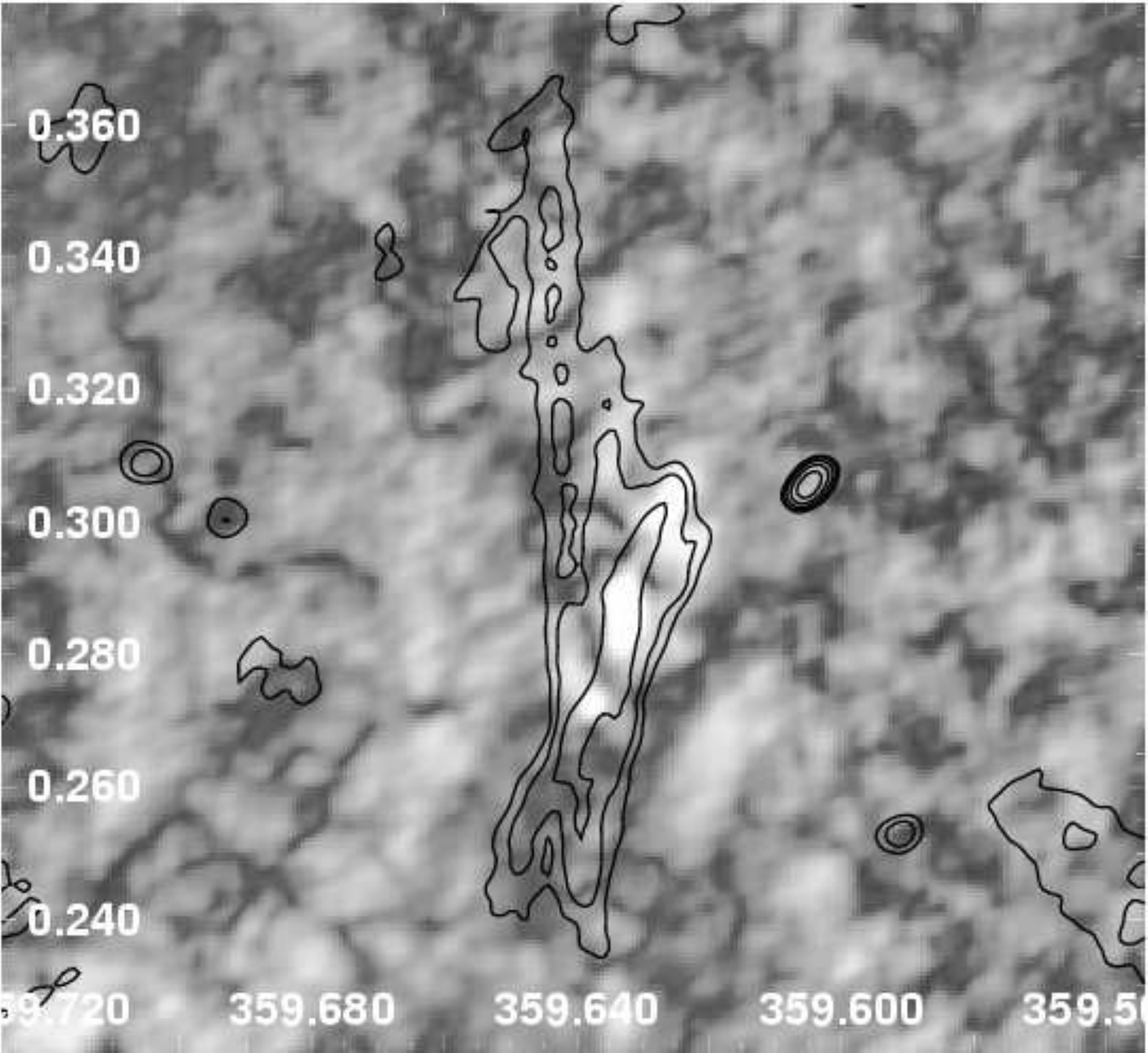}
\hfil
\includegraphics[scale=0.4,angle=270,bb=540 130 575 692]{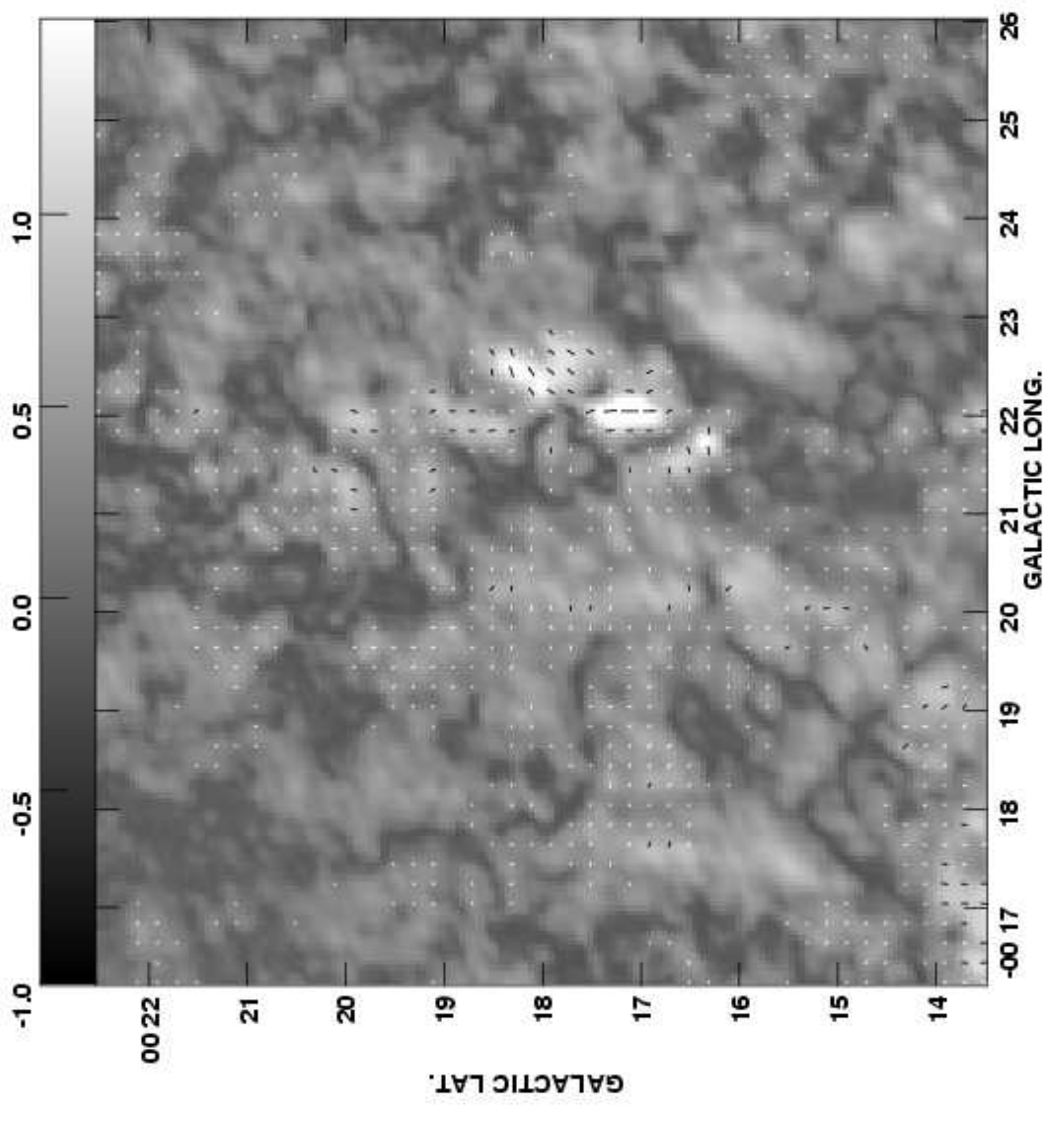}
\caption{\emph{Left}:  Gray scale shows the 6cm polarized intensity from NRF-N11 (G359.62+0.28) for brightnesses ranging from -1 to 5 mJy beam$^{-1}$.  The contours show 6 cm total intensity at levels of 0.5, 1, and 2 mJy beam$^{-1}$.  \emph{Right}:  Image of the same region with 6 cm polarization vectors shown.  Vectors are shown for regions with polarized intensity brighter than 0.2 mJy beam$^{-1}$. \label{n11}}
\end{figure}

Figure \ref{n11} shows the polarized emission from G359.62+0.28, also known as NRF-N11 \citep{y04}.  NRF-N11 seems to be composed of two sets of filaments with one, bending filament oriented about 70\sdeg\ relative to the Galactic plane (called N11a) and a group of at least two parallel filaments oriented nearly perpendicular to the plane (called N11b).  NRF-N11a is brighter than N11b with total intensities ranging from 2-3 mJy beam$^{-1}$;  the peak brightness of 3.6 mJy beam$^{-1}$ is found where the two sets of filaments intersect.  The vertical filaments have not been detected at 20 cm and thus have an unusually flat 6/20 cm spectral index ($>-0.1$), while the crooked filament has a 6/20 cm spectral index ranging from --0.1 to --0.3.

The brightest 6 cm polarized emission from the NRF-N11 complex is broken into three islands along the N11a subfilament.  This is the first detection of polarized emission from N11a, which confirms that it is a NRF.  The peak polarized brightness is 2 mJy beam$^{-1}$, where the polarization fraction is about 70\%.  As in NRFs C12 and N1, the polarized emission is brighter than the total intensity near the end of the filament, leading to apparent polarization fractions larger than 100\%.  There are two islands of polarized emission along the western subfilament of the N11b complex, but it is not clear that they are polarized counterparts to the filament seen in total intensity.

The average \dt for the polarized emission from NRF-N11a is $0\pm1$\sdeg, or \rmeff$=0\pm220$ rad m$^{-2}$.  There is weak evidence for changes in \dt across N11a, with \dt$=1\pm2$\sdeg\ (\rmeff$=-220\pm440$ rad m$^{-2}$) and $-1\pm2$\sdeg\ (\rmeff$=220\pm440$ rad m$^{-2}$) for the northern and central polarized islands, respectively.  The typical \dt at this longitude is --1\ddeg5, or \rmeff$=330$ rad m$^{-2}$.

\subsection{Structures in the Image of \dt}
\label{padires}
Images of the mean \dt provide a reliable way to study the changes in \rmeff\ across the GC region.  Figure \ref{padilg} shows two representations of the mean \dt found by averaging over 125-arcsec tiles with the histogram-fitting method.  This section describes the properties of the \dt images and correlations with objects seen in other observations, particularly the GCL.

\begin{rotate}
\begin{figure}[tbp]
\hspace{-2cm}
\includegraphics[width=\textheight]{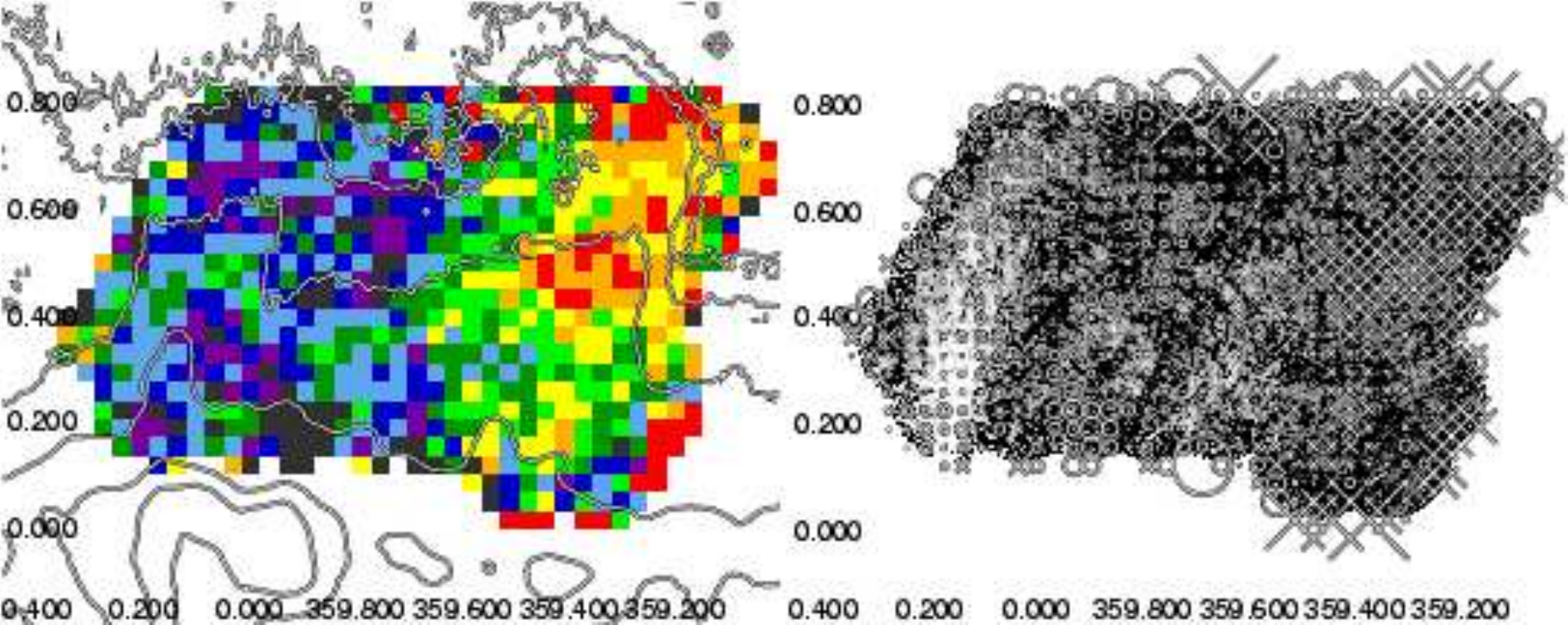}
\caption{\emph{Left}:  Image of the average \dt value over the entire 6 cm survey of the GCL, averaged over 125-arcsec tiles with the histogram-fitting method.  The color scale is similar to Fig. \ref{poln_padi}, but with the \dt values ranging from --10 to +10\sdeg. The contours show 6 cm intensity as observed by the GBT at levels of $3^n$*33 mJy per 150\arcsec\ beam, with $n=0-5$. \emph{Right}:  Gray scale shows the 6cm polarized intensity from 0 to 2 mJy, as in Fig. \ref{poln_polc}.  The symbols show the value of \dt as shown in the left panel, but with crosses showing positions with \dt$>0$ and circles showing positions with \dt$<0$. The size of the symbol is proportional to the value of \dt, with values ranging form -19 to 26\sdeg, or \rmeff\ values from 4180 to -5720 rad m$^{-2}$. \label{padilg}}
\end{figure}
\end{rotate}

\subsubsection{East-West Gradient}
\begin{figure}[tbp]
\begin{center}
\includegraphics[scale=0.7]{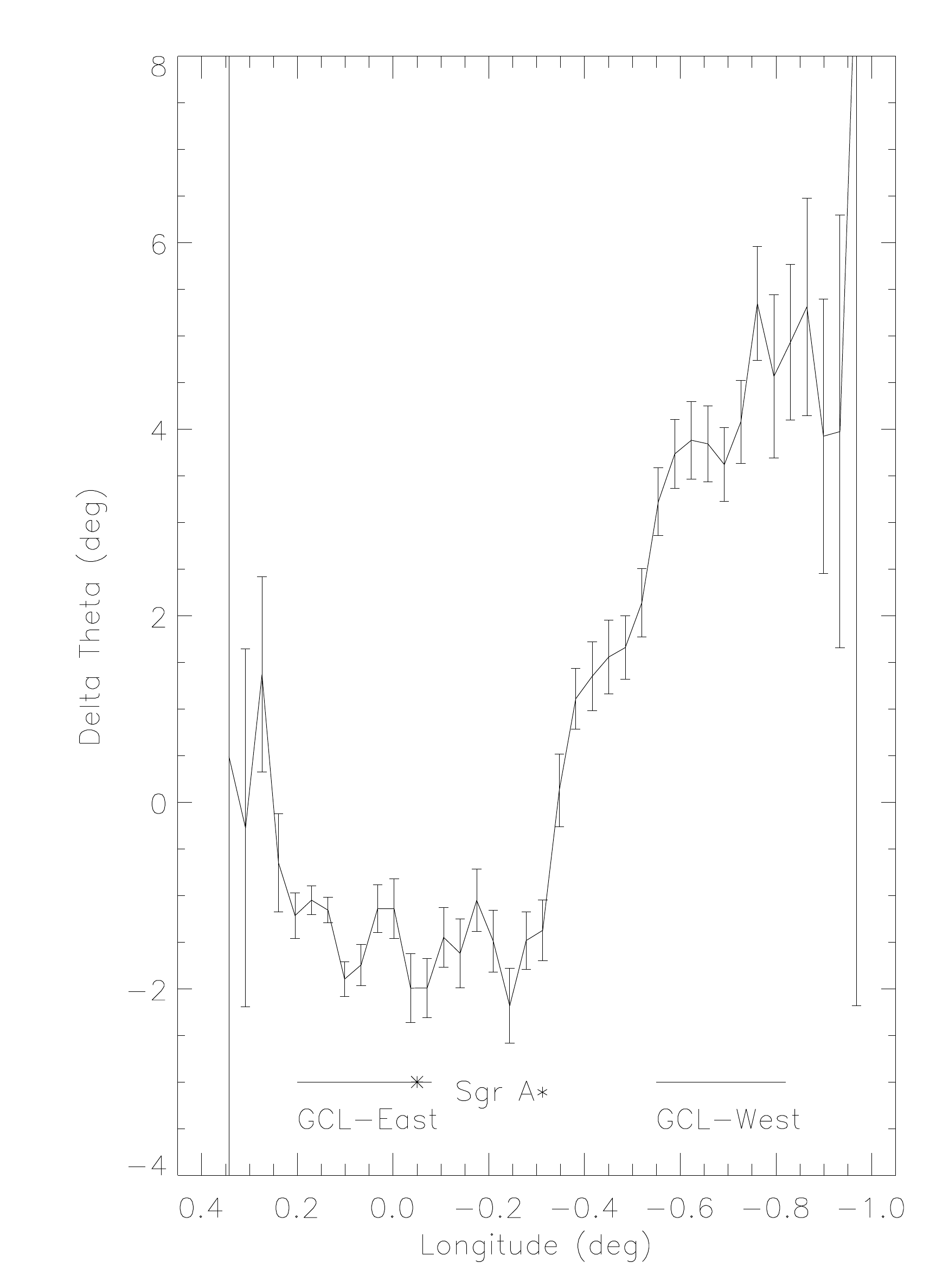}
\end{center}
\caption{Plot of the mean \dt for the polarization angle toward the GCL as a function of galactic longitude.  The mean value and error are found by the histogram-fitting method for the entire latitude range of the survey ($0\ddeg1<b<0\ddeg8$) in strips of width 125\arcsec.  For orientation, the longitude range of the GCL-East and GCL-West are shown with lines and the longitude of Sgr A* with a star symbol.  \label{padiplot}}
\end{figure}

The most obvious structure in the image of \dt is the large-scale gradient in the direction of Galactic longitude.  This gradient is visible in the full-resolution image (Fig. \ref{poln_padi}) and 125-arcsec mean image (Fig. \ref{padilg}).  The east side of the survey tends to have \dt less than zero and the west side, although noisier, has predominately positive \dt values.  This trend is seen more clearly in averages calculated over all latitudes, as done for Figure \ref{padiplot}, which shows only the longitude dependence of \dt.

\begin{figure}[tbp]
\begin{center}
\includegraphics[width=0.4\textwidth]{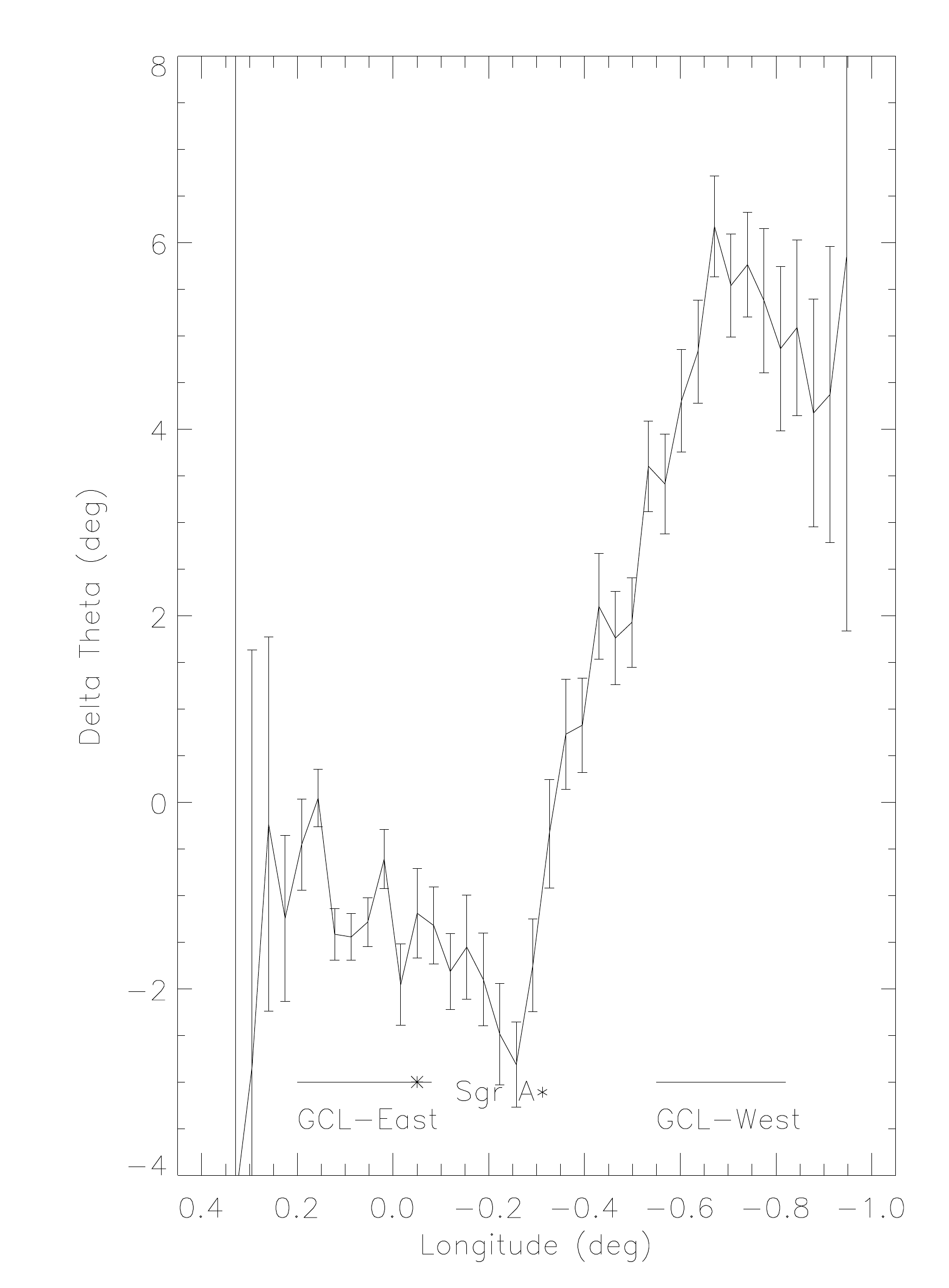}
\includegraphics[width=0.4\textwidth]{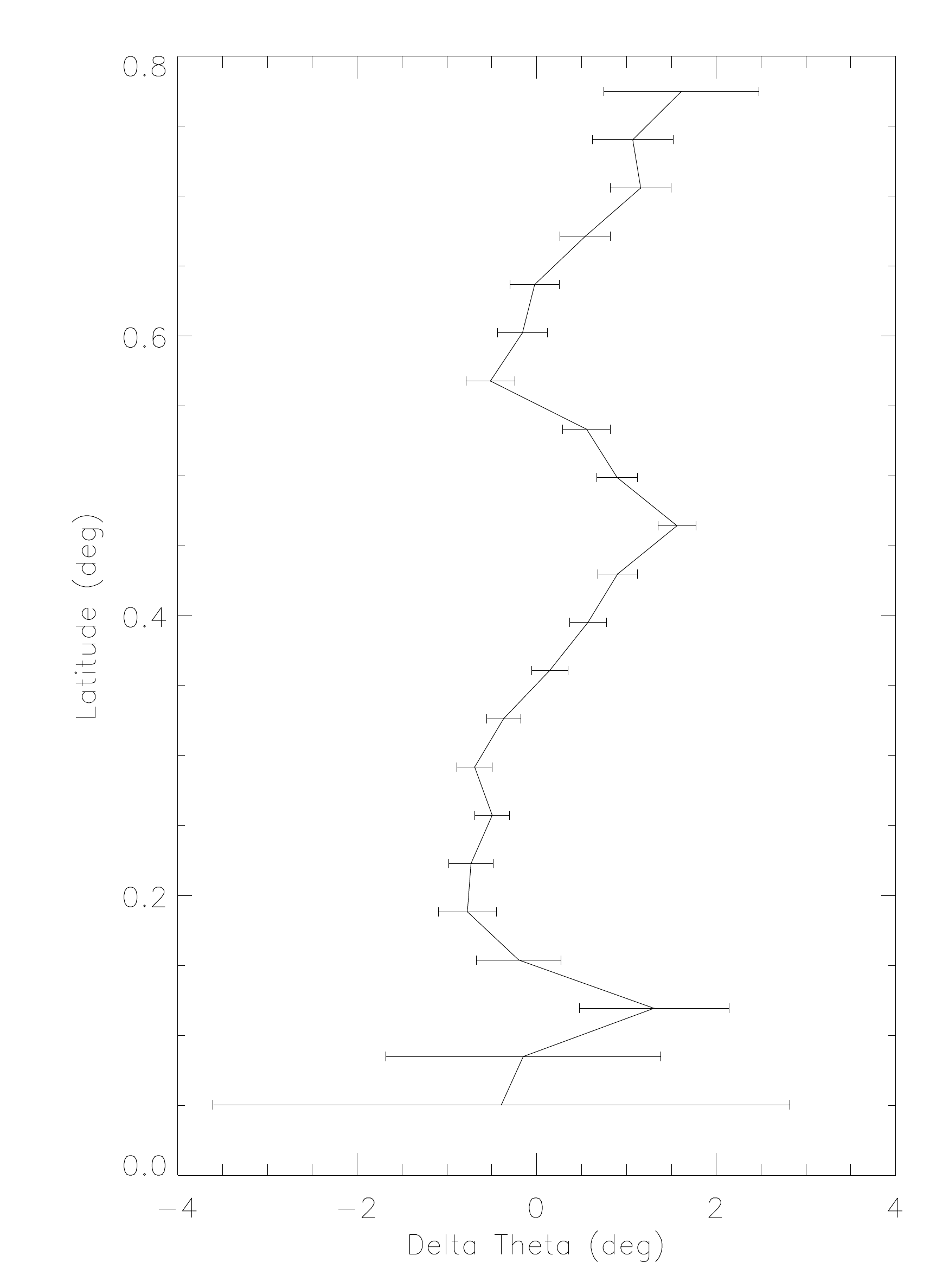}
\end{center}
\caption{\emph{Left}:  Plot of the mean \dt for the polarization angle, as shown in Fig. \ref{padiplot}, but for the northern half of the survey, with $0\ddeg4<b<0\ddeg8$.   \emph{Right}:  Plot showing the dependence of \dt with Galactic latitude.  The entire longitude range of the survey was incorporated to measure the average. \label{padiplot2}}
\end{figure}

The mean \dt changes sign at $l\approx-0\ddeg35$ with an uncertainty of $\sim0\ddeg02=1$\arcmin.  The longitude where \dt changes sign is independent of the range of Galactic latitude over which the average is taken.  The left panel of Figure \ref{padiplot2} shows the average \dt for the northern half of the 6 cm survey region.  No significant change in the longitude of the sign change in \rmeff\ was observed for various latitude ranges within the survey coverage.  The location of this central longitude of \dt is similar to the central longitude of the GCL (see Ch. \ref{gcl_all}).  This is noteworthy because both structures are offset toward negative longitudes from Sgr A*, the dynamical center of the Galaxy.

The shape of the \dt profile in longitude has three parts across the GCL: a relatively constant negative \dt value across the GCL-East, a sharp transition region near its central longitudes, and a relatively constant positive \dt value across the GCL-West.  At the GCL-East, the \dt has a value of $\sim-1.5$\sdeg\ for $0\ddeg2<l<-0\ddeg3$.  The switch of sign in \dt occurs for $-0\ddeg3<l<-0\ddeg55$, where \dt changes 5\ddeg5 over 0\ddeg15 of longitude.  At the GCL-West, the \dt values are $\sim4$\sdeg\ from $-0\ddeg55$ to the survey edge at $l\approx-0\ddeg95$.  The \rmeff\ values are +330 and --880 rad m$^{-2}$ for the east and west sides, respectively, with errors of roughly 50 rad m$^{-2}$.  The latitude dependence shown in the right panel of Figure \ref{padiplot2} shows significant variation, but no systematic gradient.  The large positive \dt values near $b=0\ddeg45-0\ddeg55$ are likely caused by the extended positive \dt feature near (--0\ddeg6,+0\ddeg5) (see \S\ \ref{localized}).  The \dt values range from --0.75 to 1.5, or \rmeff$=165$ to --330 rad m$^{-2}$.

As shown in the left panel of Figure \ref{padiplot2}, this shape is also seen when averaging over high galactic latitudes only, where there is less confusion from sources in the galactic disk.  Although, the rough shape is independent of galactic latitude, averaging over higher galactic latitudes shows a greater range of values for the \dt.  The value of \dt has a minimum of -3\sdeg\ near $l=-0\ddeg25$ on the east side, while the maximum is +6\sdeg\ near $l=-0\ddeg65$ on the west side.  The \rmeff\ values at these minimum and maximum \dt values are +660 and --1320 rad m$^{-2}$ for the east and west sides, respectively, with errors of roughly 75 rad m$^{-2}$.

\subsubsection{Localized \rmeff\ Features}
\label{localized}
\begin{figure}[tbp]
\includegraphics[width=\textwidth]{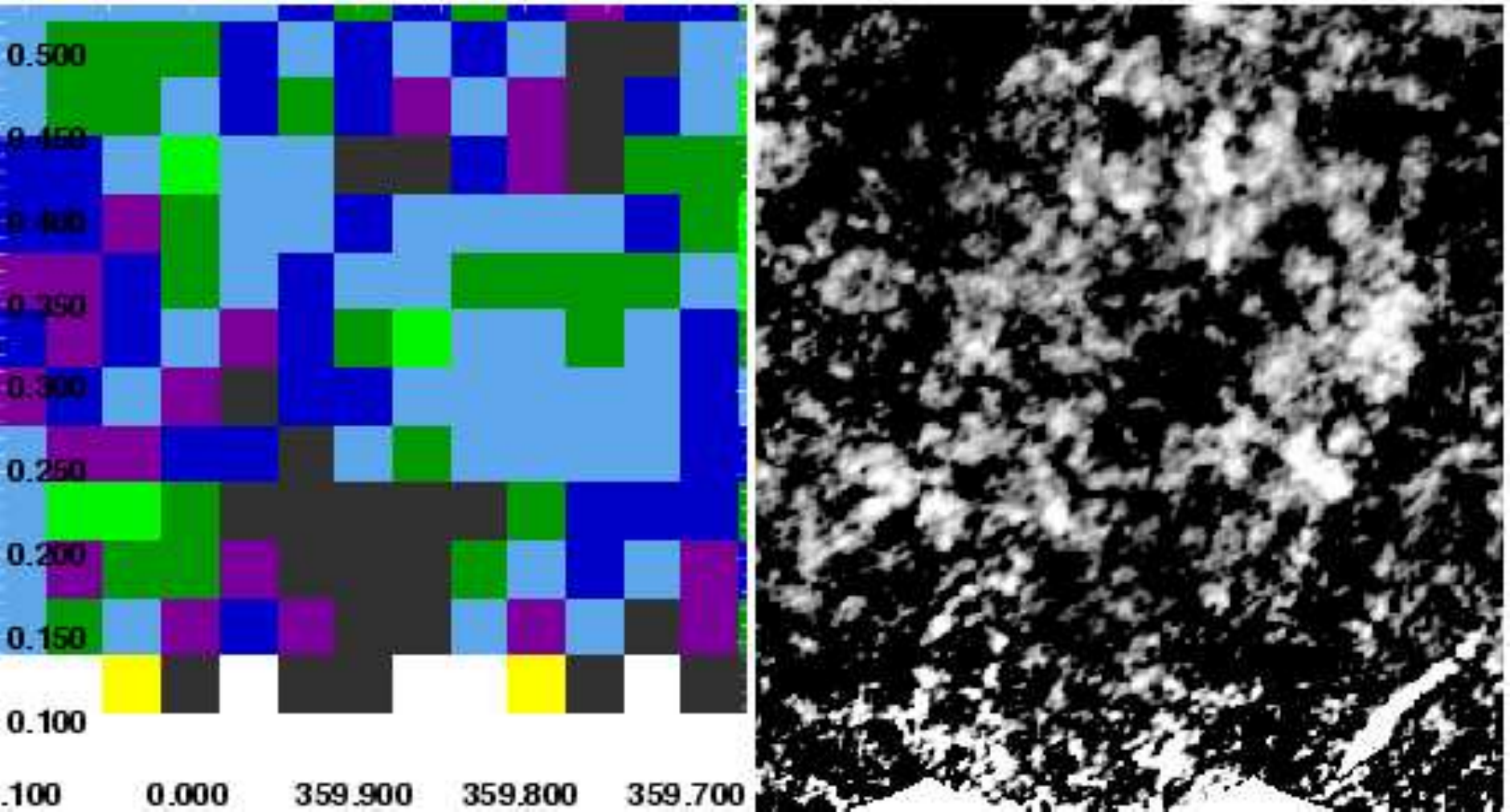}
\caption{\emph{Left}:  Image of the average \dt found by the histogram-fitting method for the region near the region of large negative \dt.  The color scale is identical to Fig. \ref{padilg}, which ranges from -10\sdeg\ to 10\sdeg.  \emph{Right}:  Image of the 6 cm polarized intensity for the same region as in the left panel.  The image is identical to that shown in Fig. \ref{poln_polc} with gray scale showing fluxes ranging from 0 to 1 mJy beam$^{-1}$.  The Cane NRF is located at the top of the loop at ($359\ddeg85$,$0\ddeg39$). \label{loop}}
\end{figure}

Aside from east-west gradient in the \rmeff\ values, there are three arcminute-scale features worth noting.  One of the regions with the largest negative \dt value is at the southern border of the survey, near (--0\ddeg1,0\ddeg2).  As shown in the \dt and polarized intensity images in Figure \ref{loop}, the large \dt covers a region about 8\arcmin\ across, just north of Sgr A.  The average \dt value for this region is $-5\ddeg4\pm0\ddeg9$, which gives an \rmeff$=1188\pm198$ rad m$^{-2}$.  The average \dt value for all latitudes near $l=-0\ddeg1$ is $\sim-1\ddeg5\pm0\ddeg3$.

This \dt feature does not have a direct counterpart in polarized intensity, but is located just south of a loop-like structure in polarized intensity that reaches up to $b\sim0\ddeg45$.  The polarized loop structure abruptly ends where the large \dt values begin, suggesting that the emission from the loop may be depolarized by the large $RM$ near $b=0\ddeg2$.  Coincidentally, a nonthermal radio filament is located at the peak of the loop structure;  the filament, called N10 or the ``Cane'' \citep{y04,l01}, is discussed in more detail in \S\ \ref{n10sec}.  In the present data, it is difficult to say whether the polarized loop traces an intrinsically loop-like structure or if depolarization makes it appear that way.

Figure \ref{padilg} shows that the region with the most negative value of \dt is on the west side of the survey, at (--0\ddeg6,+0\ddeg5).  Near this position there is an area about 8\arcmin\ across with a histogram-fitted mean \dt$=6\ddeg0\pm0\ddeg5$.  Figure \ref{padiplot} shows that the mean \dt value over all latitudes near $l\sim-0\ddeg6$ is $\sim4$\sdeg$\pm0\ddeg5$, so the value of \dt near (--0\ddeg6,+0\ddeg5) is unusually high.  The corresponding \rmeff\ in this region is $-1320\pm110$ rad m$^{-2}$.

Finally, Figure \ref{padilg} shows a ridge extending from (0\ddeg25,+0\ddeg4) to (0\ddeg0,+0\ddeg5) where the \dt is positive, which is the opposite of the average \dt in the region.  The value for the 125-arcsecond average along this ridge is $\sim$1\sdeg$\pm0\ddeg5$, as compared to the mean value of $\sim-1\ddeg5\pm0\ddeg3$ for all latitudes near $l=0$\sdeg.  This structure is seen in the $RM$ map of \citet{t86}, and a detailed comparison of that work to the present work is shown in Figure \ref{rmcomp}.  


\section{Discussion}
\label{poln_discussion}



\subsection{Localizing the Polarized Emission}
The primary goal of this observation is to study the polarized continuum emission from the GCL, so it must first be established that the polarized emission originates in the GC region.  However, the line of sight through the GC region also passes through roughly 8 kpc of the Galactic disk, so foreground emission and propagation effects will confuse and modify emission from the GC region.  Thus, an important first question to ask is:  Does any of the polarized emission seen in the present observations originate in the GC region?

The simplest way to find if the polarized emission originates in the GC region is to correlate the polarized sources with sources that are known to be in the GC region.  The polarized emission on the eastern edge of the survey has a compelling morphological connection to the Radio Arc, particularly in the south \citep{y88}.  The Radio Arc is believed to be within about 100 pc of the GC based on: (1) its uniqueness and projected position relative to the GC, (2) its interaction with molecular and ionized gas in the GC region \citep{y84,l99}, and (3) HI absorption measurements \citep{l89}.  The nonthermal filaments C3 (G359.54+0.18) and N8 (G359.75+0.17) are also detected in the present survey with polarized emission and are known to be within $\sim$200 and $\sim$2000 pc of the GC, respectively, based on HI absorption measurements \citep{r03}.  So there is strong evidence that the polarized, filamentary emission originates in the GC region.

Assuming that the NRFs are within a few hundred parsecs of the GC enables one to use them as probes of the $RM$ expected toward the region.  As shown in \S\ \ref{poln_comparison}, the $RM$ measured in the Radio Arc for $0\ddeg2<b<0\ddeg4$ ranges from 250--1000 rad m$^{-2}$.  Although the $RM$ toward the Radio Arc polarized emission north of $b=0\ddeg6$ has not been measured, the \rmeff\ measured with the present survey has a similar range of 200--600 rad m$^{-2}$.  The similarity between the $RM$ values in the region known to be in the GC and \rmeff\ north of $b=0\ddeg6$, in addition to the contiguous morphology in total intensity, strongly suggests that all the polarized emission found near the eastern edge of the survey is within a few hundred parsecs of the GC.

One concept in understanding the origin of Galactic synchrotron emission is that of the ``polarization horizon''.  The polarization horizon is the largest distance that polarized intensity can be seen without observational and Faraday rotation effects completely depolarizing it \citep{u03}.  On scales larger than a beam (e.g., for the extended emission observed here), the polarization horizon is largely determined by the amount of depth depolarization.  Depth depolarization occurs when the synchrotron emission and the Faraday rotating-medium are in the same volume in space, such that the polarization angle of radiation emitted from different depths in the source are perpendicular to each other and cancel out \citep{so98}.  For a medium with uniform, line-of-sight magnetic field, $B_{||}$, and electron density, $n_e$, the polarized intensity will cancel itself out after propagating a length, $L = \pi/0.81 \lambda^2 n_e B_{||}$ \citep{u03}.  Assuming a magnetic field equal to the uniform Galactic component of 2 $\mu$G \citep{b01} and a mean electron density of 0.07--0.08 cm$^{-3}$ \citep[predicted for line-of-sight to the GC, but excluding the central 100 pc;][]{t93,c04}, the polarization horizon distance for the present observation is 7 kpc.  However, the $n_e$ and $B$ is not uniform along the line of sight, but changes between the three spiral arm and inter-arm regions \citep{h99};  there is no contiguous 7 kpc region over which depth depolarization can occur.  Thus, we suggest that the polarization horizon distance is consistent with our detection of extended polarized emission from the GC region.  At 20 cm, the depolarization effects are roughly 11 times stronger than at 6 cm.  For $n_e=0.1$ cm$^{-3}$ \citep[predicted for this line of sight within a few kpc of the Sun;]{t93,c04}, the horizon distance is about 500 pc.  This is consistent with the lack of extended, 20 cm polarized emission correlated with objects in the GC region.


This study has made the first measurement of the $RM$ over such a large portion of the GC region.  Previous studies used single-dish observations and either did not cover such a large region or did not have multiple frequencies to estimate the $RM$ \citep{t86,h92}.  Thus, the degree-scale $RM$ structure and the arcsecond-scale depolarization canals have not been discussed before in this region.

\subsection{Filaments}
Eight nonthermal filaments are detected here in polarized emission, including three (C12, N1, and N11a) for the first time.  Generally, the brightest filaments in total intensity are detected in polarized emission, suggesting that number of confirmed NRFs is limited only by the sensitivity of the data.  The morphology of filaments (including their length, orientation, location) with polarized emission is very similar to that of the filaments with no detected polarized emission \citep{n04,y04}.  Most candidate NRFs are likely to be genuine NRFs.

The polarized emission from the NRFs generally correlates with the total intensity and peak polarizations for all NRFs are greater than 30\% at 6 cm.  The theoretical prediction for synchrotron polarization fraction is 75\% for a spectral index of --1 \citep[assuming $S_\nu\propto\nu^{\alpha}$;][]{r79}.  Thus, in some cases, the maximum depolarization toward an NRF is about $1-30/75 = 60$\% at 6 cm.  Since bandwidth depolarization isn't expected to be strong and beam depolarization only depolarizes on single-beam scales, depth depolarization is likely to dominate the depolarization toward the NRFs \citep{u03,h04}.

\begin{figure}[tbp]
\includegraphics[width=\textwidth]{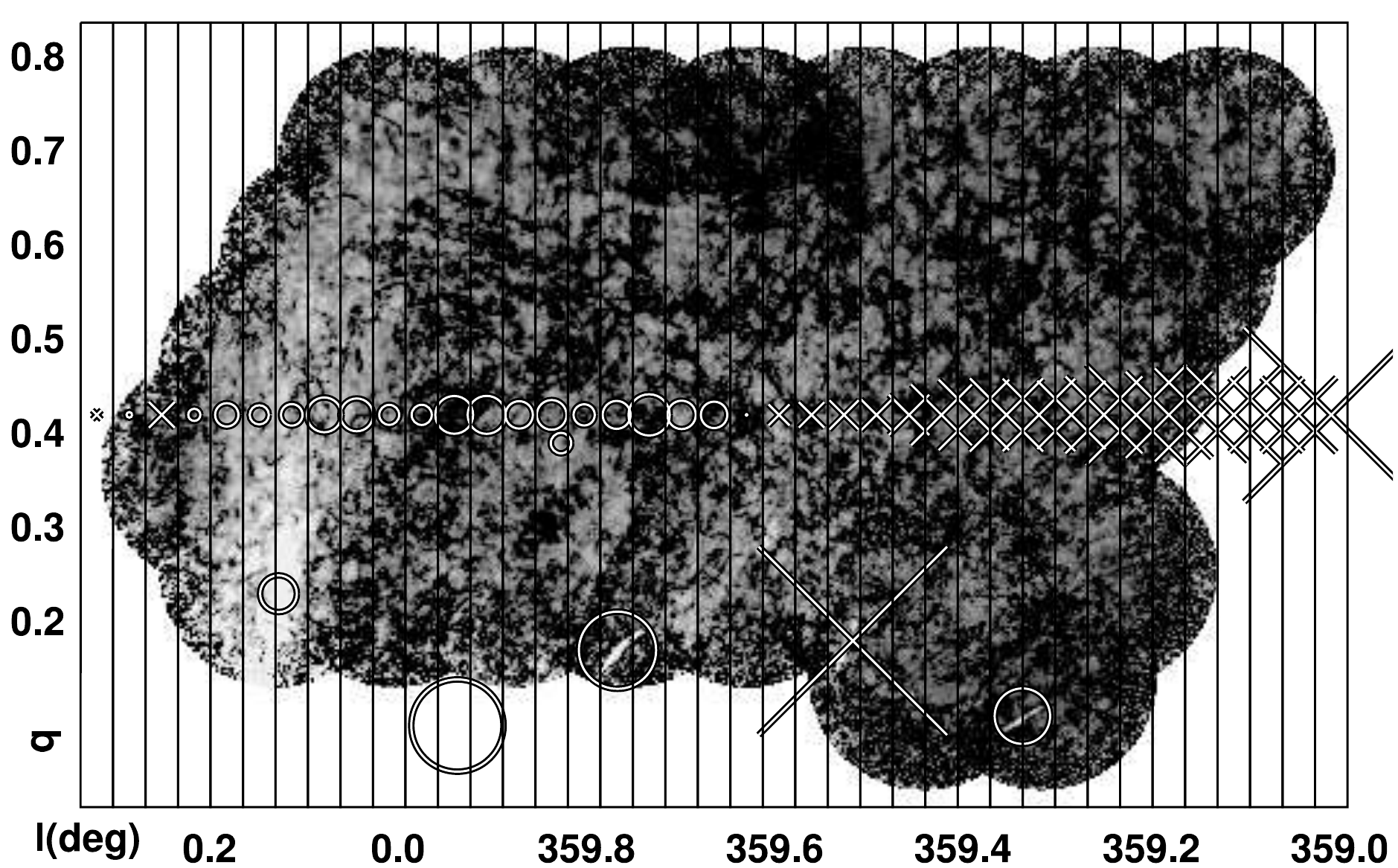}
\caption{Image of the 6 cm polarized intensity with symbols showing the mean \dt for NRFs in the survey region and the extended polarized emission.  The gray scale shows the 6cm polarized intensity from 0 to 2 mJy, as in Fig. \ref{poln_polc}.  The vertical rectangles show the extent of the region over which the \dt was averaged by the histogram-fitting method (also shown in Fig. \ref{padiplot}).  The cross symbols show positions with \dt$>0$ and circles showing positions with \dt$<0$. The size of the symbol is proportional to the value of \dt, with values ranging from -5 to 10\sdeg, or \rmeff\ values from 1100 to --2200 rad m$^{-2}$.  The symbols are located at the coordinates of the NRF or the center of the rectangular region over which the average was calculated.  Note that the mean \dt for NRF-N11 (G359.62+0.28) is zero, so no symbol is visible. \label{poln_rmschem}}
\end{figure}

As shown in Figure \ref{poln_rmschem}, in all but one case, the \rmeff\ of the filaments agree with that of their surrounding polarized emission.  East of $l\approx-0\ddeg35$, there are four filaments found to have good constraints on \rmeff:  N1, N5, N8, and N10.  All four of these have a positive \rmeff, just like that seen throughout the eastern half of the present survey.  West of $l\approx-0\ddeg35$, three are detected in linearly polarized flux:  N11, C3, and C12.   N11 has an \rmeff\ near zero and C3 has a negative \rmeff; both are consistent with the \rmeff\ near their respectively longitudes.  NRF-C12 has a very significant positive \rmeff, which makes it the only polarized filament detected in our survey with an \rmeff\ different from the \rmeff\ of its surrounding, extended polarized emission.

What might explain the fact that six of the seven NRFs seen here have similar \rmeff\ as their surrounding diffuse polarized emission?  There are two major contributions to the \rmeff\ observed toward these polarized features:  an $RM$ intrinsic to the object or an $RM$ due to a foreground screen.  The simplest possibility is that the NRFs and the diffuse polarized emission have little or no intrinsic $RM$ and that the observed \rmeff\ originates in a foreground screen.   In principle, this foreground screen could be near the GC or anywhere along the line of sight.  However, as discussed in \S\ \ref{rmdisc}, propagation through the Galactic disk outside the GC (i.e., through the Norma, Crux-Scutum, and Carina-Sagittarius arms) should induce a net $RM$ much smaller than the range observed here ($<100$ rad m$^{-2}$).  Thus, it is most likely that any $RM$ induced by a ``foreground'' is somewhere in the central few hundred parsecs, where the electron density is higher than in the disk \citep{t93,h06}.  It is also possible that the magnetic field is significantly stronger in the central few hundred parsecs, but that is a subject of active debate \citep{y84,m96,la05,b06}.

However, there is evidence that the NRFs' observed \rmeff\ may be intrinsic to them.  In particular, the \rmeff\ toward NRF-C12 has the opposite sign as that of its surrounding extended polarized emission.  This NRF shows that the $RM$ toward the region is not dominated by a simple screen, but that the filaments may have an intrinsic $RM$ created by the ISM physically near it or associated with it.  The idea that the $RM$ of NRFs is caused by local structure in the ISM of the GC region has also been suggested for NRF-C3 \citep{y97}.  Also, the nonthermal filament G358.85+0.47 (a.k.a. the ``Pelican'') shows changes in its $RM$ that coincide with morphological changes, possibly showing that the apparent $RM$ is directly related to the orientation of the NRF \citep{la99}.  Thus, there are indications that the $RM$ toward the GC region has a semi-regular structure, but that NRFs sometimes show deviations from that structure that may be related to their intrinsic orientation.


\subsection{Depolarized Canals}
The extended and filamentary polarized emission is broken up by depolarized canals, as has been seen in other polarized radio continuum surveys of the Galaxy \citep{y87,g01,h04}.  For some of the NRFs with strong polarized emission (e.g., NRF-C3), the canals are found to separate polarized islands with relatively-constant values of \rmeff.  In the case of NRF-C3, adjacent islands of polarized emission have \rmeff\ that differ by 2000 rad m$^{-2}$, a value comparable to the typical $RM$ of sources in the GC region.  A similar trend is observed in longer-wavelength observations of the Galactic plane, where $RM$ ranges up to 10 rad m$^{-2}$ and the change in $RM$ across a canal can be as large as 10 rad m$^{-2}$.

The polarization angle across the canals change by approximately 90\sdeg, as has been seen in other surveys \citep{h04,fl06}.  \citet{h00} found that depolarization canals were found in regions of rapid change in the polarization angle.  These two facts suggest that the depolarization is caused the averaging of changing polarization vectors within a single beam, or beam depolarization.  In particular, it has now been shown that when the polarization angle changes by 90\sdeg across a canal, the $RM$ must also change discontinuously on beam size scales \citep{fl06}.  This rapid change in the electron density or magnetic field may indicate the presence of shocks in the ISM.

Canals are present in radio continuum observations at 84 cm \citep{h00} and 6 cm (present work), which probe distances up to 500 pc and the GC region, respectively.  For both horizons, the change in $RM$ relative to the typical $RM$ ranges up to one.  This is somewhat surprising, since the two observations trace much different paths through the ISM.  

\subsection{Large-scale $RM$ Structure in the GC Region}
\label{rmdisc}
One of the most surprising results of this survey is the east-west gradient in \rmeff\ toward the extended polarized emission in the GCL.  Furthermore, similar \rmeff\ toward the extended polarized emission and the NRFs may have a similar origin or location.  In this section we discuss possible origins for the spatial structure in $RM$ measurements toward the GC region.

While we have shown that observed polarized emission may originate in the GC region, we need to more carefully consider the Galactic $RM$ structure and whether the \rmeff\ measured here originates in the GC region.  One simple estimate the $RM$ expected toward the GC region is found by observing extragalactic sources through the GC region.  \citet{r05} measured the $RM$ of six extragalactic sources within 1\sdeg\ of the GC range from roughly --1200 to 1500 rad m$^{-2}$.  Assuming that these are extragalactic sources with relatively little intrinsic $RM$, the $RM$ range toward the GC might be expected to be about half this range, or about --600 to 750 rad m$^{-2}$.  This range is similar to that inferred from our \dt images of the extended polarized emission, suggesting that it may be located near the GC.

What is the expected $RM$ contribution from the Galactic disk, outside of the GC region?  Pulsars are very useful probes of the Galactic electron and magnetic field distributions because their distances can be estimated and they are believed to have little intrinsic $RM$ \citep{r89}.  Pulsar observations have shown that the magnetic field in the disk alternates between counterclockwise and clockwise directions (as viewed from the North) in the arm and interarm regions, respectively \citep{b96,h99,h06}.  It is generally believed that there are three spiral arms that intersect the line of sight to the GC with a small pitch angle \citep[$\sim11$\sdeg;][]{ru03,h06}.  The frequent field reversals and near-perpendicular magnetic field alignment should tend to limit the range of $RM$ values observed toward the GC.  

Observationally, the ATNF Pulsar Catalogue \footnote{See \url{http://www.atnf.csiro.au/research/pulsar/psrcat}} \citep{ma05} shows 11 pulsars nearer to us than the GC and within 5\sdeg\ of it in projection that also have $RM$ measurements.  $RM$ values range from --240 to 111 rad m$^{-2}$, although the largest $RM$ values observed are toward pulsars with distances between 2 and 5 kpc.  \citet{h06} show that $RM$s measured toward pulsars in front of and within $\sim2$ kpc of the GC (i.e., well beyond the spiral arms) are much smaller.  Perhaps the best estimate of the $RM$ foreground of the GC is from J1739--3131, which is the nearest pulsar to the GC and in front of the GC ($D=7.7$ kpc) that is within 5\sdeg\ in projection \citep[$l=357.10$,$b=-0.22$;][]{ma05,h06}.  The $RM$ of J1739--3131 is 32 rad m$^{-2}$, which is significantly less than observed elsewhere in the disk and less than \rmeff\ observed toward most sources in the present survey.  Thus, the $RM$ induced by propagation from the GC region through the Galactic disk seems to be small in comparison to the values observed in the present survey.

The east-west gradient observed toward the GCL in \rmeff\ is not likely caused by a foreground to the GC region.  First, the large-scale pattern has a total range of about $330-(-880)=1210$ rad m$^{-2}$.  This is about five times more than the largest $RM$ observed for a pulsar along the line of sight to the GC region (see above).  Second, there is no signature of a foreground (e.g., an HII region) that could cause the large \rmeff\ toward either the eastern or western halves of the GCL.  In total intensity radio continuum and mid-IR images, the GCL appears relatively symmetric and none of the emission seen in the area of the GCL seems to be in the foreground (Ch. \ref{gcl_all}).  It is worth noting that any $RM$ observed toward polarized emission in the GC region is not likely to have much latitude dependence.  The scale height of Galactic free electrons is about 150 pc in the thin disk \citep{c04}, which, at the furthest spiral arm (Norma at $\sim5.5$ kpc), is taller than the GCL ($1\ddeg6$ versus 1\sdeg).  Also, the screen of electrons responsible for strong scattering of sources seen through the GC may be as tall as the GCL \citep[$\lesssim1$\sdeg][]{l98}.  In summary, the longitude gradient is too large to be caused by a Galactic foreground, although the lack of a latitude gradient in \rmeff\ does not constrain the origin of the \rmeff.

If the \rmeff\ signature originates in the GC region, then it may be physically related to the GCL and the NRFs.  In particular, there is a coincidence between the alignment of these three objects:  the central longitude of the GCL, the distribution of NRFs, and the longitude of the zero point of \rmeff.  First, as described elsewhere in this thesis, the GCL has been observed in radio continuum, radio recombination line, and mid-IR flux;  all components have a limb-brightened, shell-like morphology with a center ranging from $l=-0\ddeg25$ to $-0\ddeg45$.  The extensive, high-resolution 20 cm survey of \citet{y04} showed that there are dozens of candidate NRFs in the GC region with the highest density between $l=0\ddeg2$ to $-0\ddeg7$.  The central longitude of the distribution is between $l\sim-0\ddeg3$ and $-0\ddeg4$, similar to that of the GCL.  Finally, the \rmeff\ distribution has an east-west gradient that switches abruptly from positive to negative at $l\sim-0\ddeg35$.  The abruptness of the switch means that any foreground $RM$ under about 200 rad m$^{-2}$ would not significantly change the longitude of the zero point in \rmeff.   The alignment between the GCL, the NRFs, and the \rmeff\ pattern may be by chance, but it may also point to a connection between these three objects.  One obvious connection between them is magnetic fields.  The radio continuum in the GCL and the NRFs is from nonthermal synchrotron emission, while the \rmeff\ pattern is caused by a propagation through a magnetoionized medium.

The \rmeff\ pattern observed in the central degree is consistent with the idea that the Galactic disk ``drags'' magnetic flux relative to the halo \citep{n03,u85}.  This model assumes that large-scale pattern for the GC $RM$ is formed by the effect of galactic rotation on a primordial, frozen-in, vertical (poloidal) magnetic field.  As the disk rotates, the magnetic field in the disk is dragged away from us on the east side and toward us on the west side.  This pull creates a line-of-sight component of the magnetic field which has a checkerboard pattern in the sign of the $RM$;  the $RM$ will have opposite signs toward any two adjacent quadrants formed about the center of rotation (e.g, north-east, south-east, etc.).  Interestingly, this model can use the checkerboard pattern to measure the orientation of the magnetic field without the north-south degeneracy inherent to observations of polarization angle.  

\begin{figure}[tbp]
\includegraphics[width=\textwidth]{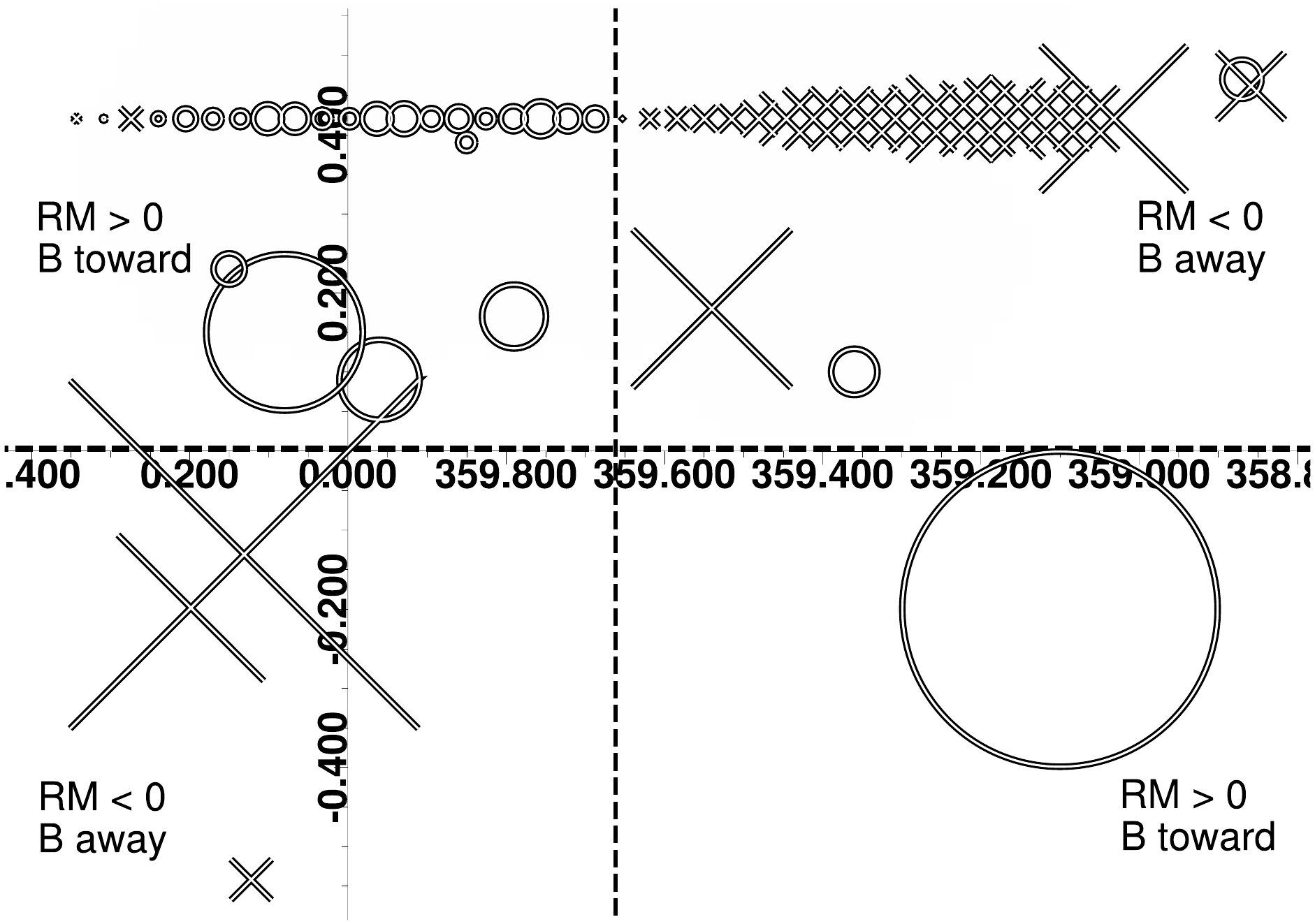}
\caption{A schematic diagram of all GC measurements of $RM$ and \rmeff\ for NRFs and the extended polarized emission.  The symbols and scaling are identical to that of Fig. \ref{poln_rmschem}, with circles showing $RM>0$ and crosses showing $RM<0$.  In addition to the measurements of \rmeff\ shown in Fig. \ref{poln_rmschem}, this figure shows measurements from the literature, specifically:  G0.08+0.15 \citep[a.k.a. ``the Northern thread''][]{l99}, the Radio Arc \citep[south of $b=0$\sdeg;][]{t86}, G0.87--0.87 \citep{re03}, G359.1--0.2 \citep[a.k.a. ``the Snake''][]{g95}, G358.85+0.47 \citep[a.k.a. ``the Pelican''][]{la99}.  The dashed lines split the region into quadrants that mostly have similar $RM$ values.  The checkerboard pattern is similar to what is expected when magnetic flux in the disk is dragged by gas motion \citep[see text;  also][]{n03}.  \label{poln_rmallschem}}
\end{figure}

Figure \ref{poln_rmallschem} shows a schematic of all $RM$ measurements toward sources believed to be in the GC region.  The \rmeff\ from Figure \ref{poln_rmschem} is shown along with observations of the extended polarized emission from the Radio Arc \citep{t86}, and other NRFs \citep{l99,g95,la99,re03}.  All of these objects are believed to be in the GC region.  North of the plane, the east-west switch is very well constrained by the present observations.  Toward the east, several observations show how the $RM$ switches sign across the plane, particularly in \citet{t86}.  The $RM$ in the southwest quadrant is not very well constrained, but the one measurement is consistent with a sign change relative to the north-west and south-east quadrants.  Thus, most measurements of the $RM$ are consistent with the idea that magnetic flux is dragged by gas in the disk.  The parity of the observed pattern is consistent with the magnetic field pointing from south to north.  

Although the $RM$ observations are suggestive, there are two complications to the flux-dragging scenario.  First, the $RM$ switches sign at $l\sim0\ddeg35$, which is offset from the center of rotation of the molecular gas in the disk by about 50 pc.  As described above, the longitude of the \rmeff\ zero point is robust to changes expected from the largest $RM$ observed in the Galactic disk ($300$ rad m$^{-2}$), so it seems to be intrinsic to the GC.  One way to explain the offset is that the $RM$ originates or is affected by the GCL, which is offset in a similar manner from Sgr A.  Chapter \ref{gcl_all} shows evidence supporting the idea that the GCL is an outflow from the central 100 pc.  It may be that the $RM$ signature in the GC region is affected by this putative outflow of gas.

A second complication to the flux-dragging scenario is the lack of a checkerboard $RM$ pattern toward extragalactic sources seen through the central several degrees.  \citet{r05} made accurate measurements of $RM$ toward dozens of extragalactic sources within several degrees of the GC in projection.  No checkerboard pattern is clearly seen in that data, although only six sources were studied in the central degree.  One possibility is that the number of measurements is inadequate to measure the $RM$ pattern in the central degree with statistical accuracy.  Gas motion is most rapid in the central degree, so the flux-dragging effect may be expected to be strongest there.  Alternatively, the $RM$ pattern may be directly related to the GCL and may not be seen outside of it.  Only three of the sources in \citet{r03} were seen through the GCL; two of the three have $RM$ values consistent with the pattern from the flux-dragging scenario.  At any rate, it is clear that the $RM$ pattern seen in the central degree does not extend greater than one or two degrees from Sgr A.


\subsection{Properties of the Polarized Emission and the Depolarizing Medium}
If the majority of the synchrotron-emitting and Faraday-rotating medium is in the GC, we can try to constrain its organization.  One possibility is that the polarized continuum emission and Faraday rotation occur in the same region, which would lead to internal depolarization, also called differential Faraday rotation \citep{b66,so98}.  Emission from a magnetized plasma will undergo a different amount of Faraday rotation in one region than in another such that the vector addition of different polarization vectors will partially cancel each other out.  In the simplest model, a uniform slab with electron density $n_e$ and regular magnetic field $B$ has an observed polarization fraction of:

\begin{equation}
P = P_i \left|\frac{\sin(2*RM*\lambda^2)}{2*RM*\lambda^2}\right|
\end{equation}

\noindent where $P_i$ is the intrinsic polarization fraction and $RM=0.81 \langle n_e\rangle> B_{||} L$ rad m$^{-2}$ with $n_e$ in cm$^{-3}$, $B_{||}$ in $\mu$G, and $L$ in pc.


There are a few interesting coincidences between the predictions of the uniform slab model and the present observations.  First, the apparent polarization fraction in the eastern half of the survey (particularly the northern extension of the Radio Arc) ranges from 20-30\%, which is about a factor of three smaller than the intrinsic polarization fraction expected from theory ($\sim75$\%).  This matches the prediction of the uniform slab model with $RM\approx330$ rad m$^{-2}$, as is observed in the region.  Second, the apparent polarization fraction toward the west half of the survey is relatively low ($\sim1-10$\%).  This is expected in the uniform slab model, which predicts an apparent polarization fraction of 1\% the intrinsic value for $\left|RM\right|\approx880$.  Although some predictions of the model break down when the polarization fraction is predicted near zero (i.e., no $RM$ can be measured if the polarization fraction is zero), the point remains that internal depolarization is expected to be much stronger for $RM$ values like that seen in the western half of the survey.  Interestingly, the apparent polarized intensity at the longitude with the lowest average \rmeff\ ($l\sim0\ddeg4$) is slightly enhanced, which is consistent with the idea that less internal depolarization is occurring there.  Internal depolarization has also been invoked to explain the change in the polarization fraction with frequency observed in the Radio Arc \citep{t86}.

Alternative explanations for the observed distribution of polarized intensity and \rmeff\ exist.  One possibility is that the eastern half of the survey is highly polarized because the Radio Arc injects energetic electrons into the region.  In this model, the GCL-West is different because it has no strong source of electrons.  Absorption of 74 MHz continuum toward the GCL-West  \citep{la05} provides another way to explain the lack of polarization observed there.  The 74 MHz absorption is caused by a thermal population of electrons, which could also cause stronger depolarization in the GCL-West as compared to the GCL-East.

In summary, the polarization fraction and \rmeff\ distributions through the survey have some consistency with the idea that the polarized emission and its depolarization come from the same volume of space.  It may be that the GCL-West has more depolarizing, ionized gas associated with it than in the GCL-East.  At any rate, the large \rmeff\ observed in the GCL-West provides a natural explanation for the lack of polarized emission in a region filled with synchrotron emission.  One possibility for the origin of the GCL-West's depolarizing medium is the shocked molecular cloud AFGL 5376 \citep{u94};  this idea is discussed in more detail in chapter \ref{gcl_all}.

\subsection{Other Extended Features in the \rmeff\ Image}
\label{poln_other}
The strongest positive \rmeff\ observed is in the portion of the survey nearest to Sgr A.  The \rmeff$\approx1200$ rad m$^{-2}$ is also unusual because the average \rmeff\ for the extended polarized emission at this longitude is $-330$ rad m$^{-2}$.  The large \rmeff\ region coincides with strong depolarization observed at the base of a loop-like structure that has a cap at $b=0\ddeg45$, where it coincides with G359.85+0.39 (a.k.a. NRF-N10 or the ``Cane'').  The \rmeff\ is large enough to cause significant depolarization, so it likely causes the observed depolarized emission;  the sign of the \rmeff\ indicates that the magnetic field points toward us.  The line of sight magnetic field is estimated by assuming the depth of the feature is equal to its width and that the electron density is around 10 cm$^{-3}$ \citep{l98}, giving $B_{||} = 7.4 * (10 \rm{cm}^{-3}/\langle n_e\rangle) (20 \rm{pc}/L)$ $\mu$G.  The coincidence of the large \rmeff\ with Sgr A may suggest an association with a magnetic structure near the massive black hole.  


The coincidence of the \rmeff\ feature near ($-0\ddeg6$,$0\ddeg5$) with the peak recombination line emission could mean that some of the ionized gas in the GCL is responsible for the observed \rmeff.  This is one of the few cases where the physical conditions in the recombination line-emitting and Faraday-rotating medium can be constrained.  Assuming that the ionized gas lies between us and the polarized emission, the line-of-sight magnetic field in $\mu$G is $B_{||} = 1.23\times10^{-3} n_e (RM/EM)$.  For the $EM=4570$\ pc cm$^{-6}$, $\left|RM\right|=1320$\ rad m$^{-2}$, and $n_e=950$\ cm$^{-6}$ observed at that location (see Ch. \ref{gcl_recomb}), the implied line of sight magnetic field is $B_{||}= 0.33 (T_e/3960 \rm{K})^{-3/2}$ $\mu$G.  This value is a lower limit, since the polarized medium may be inside the Faraday-rotating medium and $T_e$ is an upper limit.  This is also a lower limit to the total magnetic field strength in the area, since $B_{||} = B \cos\langle i\rangle$, where $i$ is the inclination angle of the magnetic field to the line of sight.

%% file: gcl_all_thesis2_astro-ph.tex
\chapter{A Study of the Galactic Center Lobe}
\label{gcl_all}

\section{Introduction}
\begin{figure}[tbp]
\includegraphics[width=\textwidth]{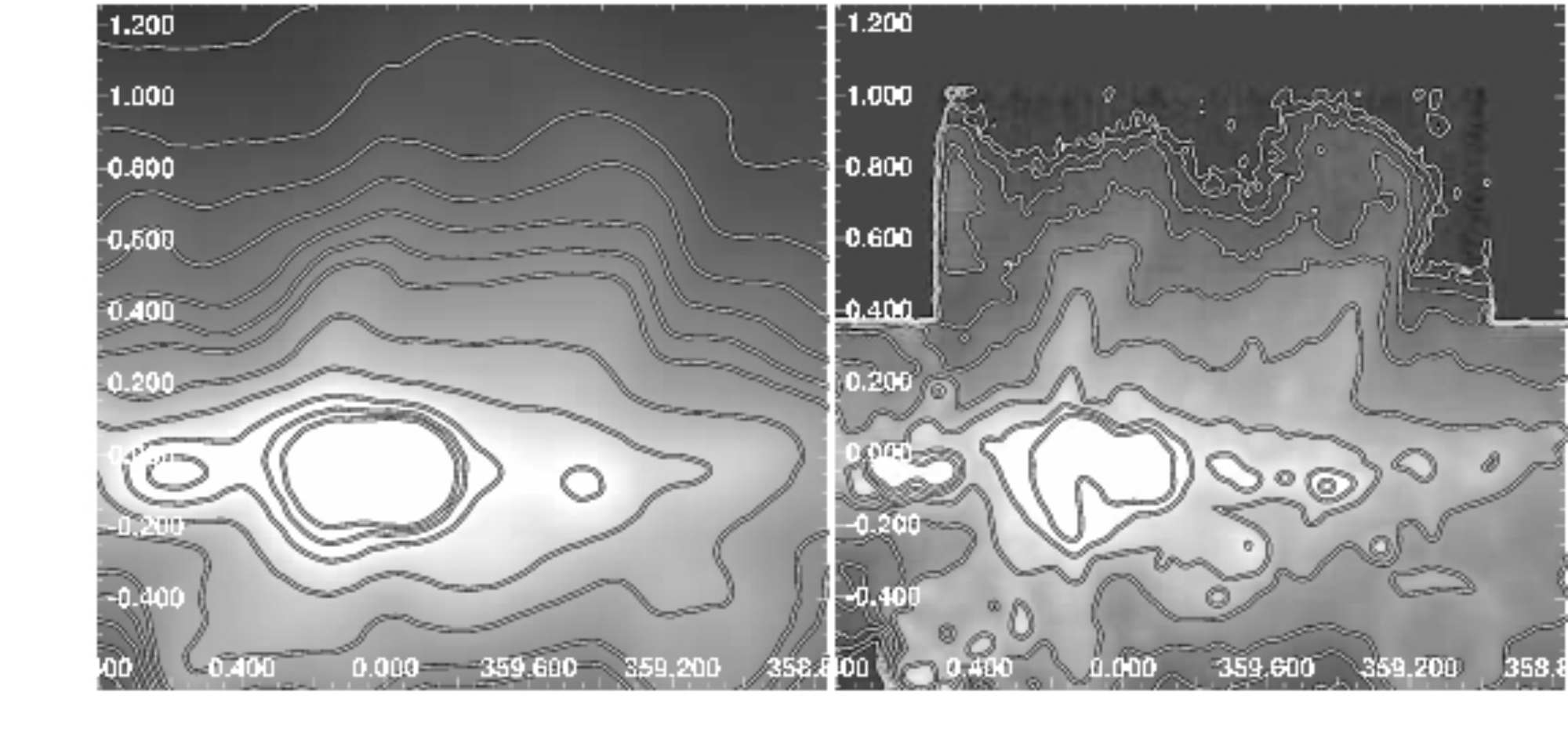}
\caption{\emph{Left}: 20 cm GBT image of the GCL.  Contours are at levels of 12, 14, \ldots, 24, 30, 40, 60, 80, 100 Jy beam$^{-1}$.  \emph{Right}: 6 cm GBT image of the GCL.  Contours are at levels of $0.02*2^n$ Jy beam$^{-1}$, with $n=0-8$.  \label{gcl620}}
\end{figure}

This chapter discusses in detail an object seen toward the Galactic center, called the Galactic center lobe (GCL).  Briefly, the GCL is a roughly 1\sdeg-tall, loop-like structure that spans the central degree of our Galaxy in radio continuum emission.  Figure \ref{gcl620} shows 6 and 20 cm images of the radio continuum emission from the GCL, as seen by the 100 m Green Bank Telescope (GBT).  The brightest parts of the emission are found at the eastern and western edges from the Galactic plane up to a latitude of about $0\ddeg5$.  The 20 cm images shows that the east and west sides are connected at the top, at a latitude of about $1\ddeg2$, which is equivalent to a distance of 165 pc, assuming it is in the GC region.

\section{Results from New Observations}
\label{all_new}
This section presents results from a multiwavelength observing campaign of the GCL.  New results from GBT radio continuum survey and \spitzer/IRAC observations;  results from previous chapters on radio recombination line (Ch. \ref{gcl_recomb}) and polarized radio continuum observations (\ref{gcl_vlapoln}) are also summarized.  Discussion of these results and how they comment on models of the GCL is presented in Sections \ref{points} and \ref{all_models}.

\subsection{Radio Continuum}
\label{all_gbtcont}
We surveyed the central two degrees of the galactic plane with the Green Bank Telescope (GBT) at 3.5, 6, 20, and 90 cm.  The calibration and analysis of these observations are described in detail in chapter \ref{gcsurvey_gbt}.  Here, we discuss results related to the GCL, including the radio continuum morphology, the spectral index distribution, and an estimate of the equipartition magnetic field.
 
The brightest emission seen in the GCL is consistent with previous observations, with a degree-scale, loop-like feature rising about a degree north of the Galactic plane.  The 6 and 20 cm surveys covered the brightest part of the GCL and are shown in Figure \ref{gcl620}.  The observed morphology is brightest at the eastern and western edges, but also shows clear evidence for a cap at the northern edge in the 20 cm image.  The 6 cm image in Figure \ref{gcl620} shows how the emission at the edges of the GCL are much broader north of $b=0\ddeg5$.  The contours highlight this broadening, showing how the GCL-West seems to split into two pieces that are separated by about $0\ddeg25\approx35$pc.  The GCL-East also may split at high latitudes, but at the very least it broadens a similar amount as seen in the GCL-West.  The distance from the east to the west side of the GCL is $0\ddeg8$, or approximately 110 pc.  The height of the GCL from the galactic plane is roughly $1\ddeg2$ or 165 pc.

\begin{figure}[tbp]
\includegraphics[width=\textwidth]{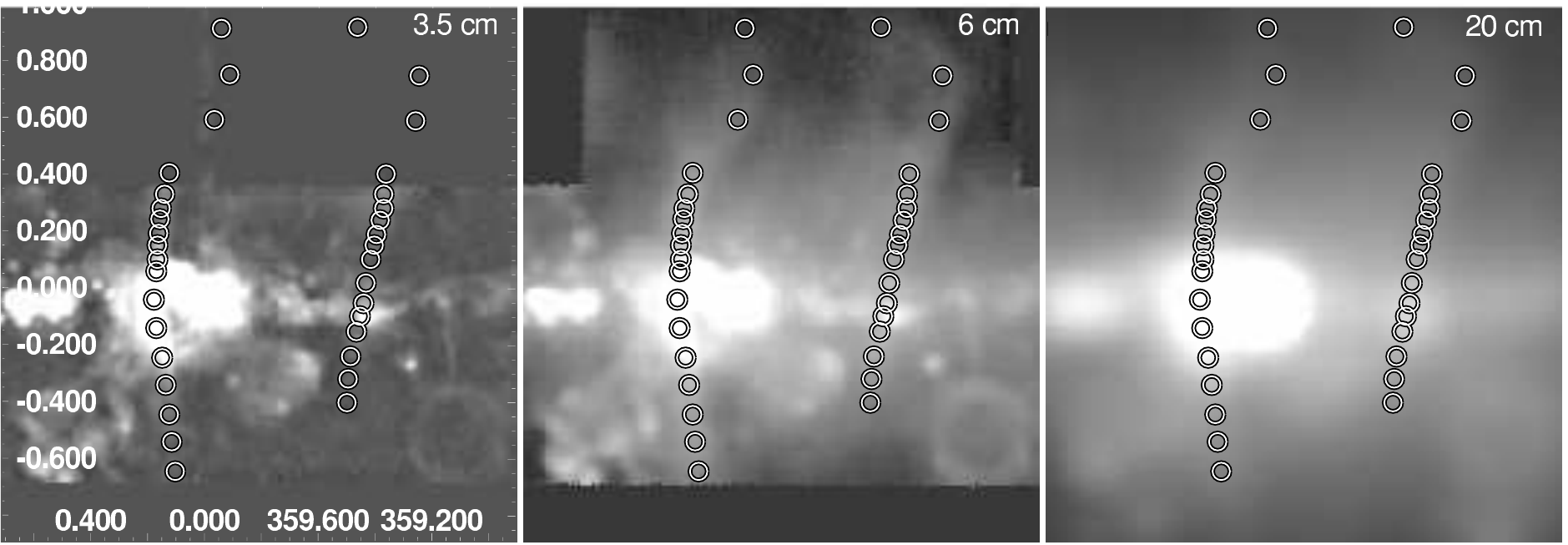}
\caption{GBT images showing the GCL at 3.5, 6, and 20 cm.  Circles in each image show the rough location at which spectral index measurements were made.  Note that at some locations, the GCL is confused with other well-known GC objects, such as Sgr C or the Arches;  these areas are labeled in Fig. \ref{gclspix} \label{gclspixpos}}
\end{figure}

One of the central goals of this study is to measure the radio spectral index of the GCL with high precision.  The spectral index was measured by taking slices across the 3.5, 6, and 20 cm images convolved to the same resolution (see Ch. \ref{gcsurvey_gbt}).  Figure \ref{gclspixpos} shows the locations where the spectral index was measured, which correspond to the peak flux of the eastern and western edges of the GCL.  At some locations, the GCL is confused with another object.  This confusion cannot be resolved without a detailed understanding of the nature of the GCL, which is discussed at the end of this chapter.  For now, we define the GCL by the roughly continuous ridge (e.g., Figs \ref{gcl620} and \ref{gclspixpos}) and specifically discuss where it is confused with other objects.

\begin{figure}[tbp]
\begin{center}
\includegraphics[width=\textwidth]{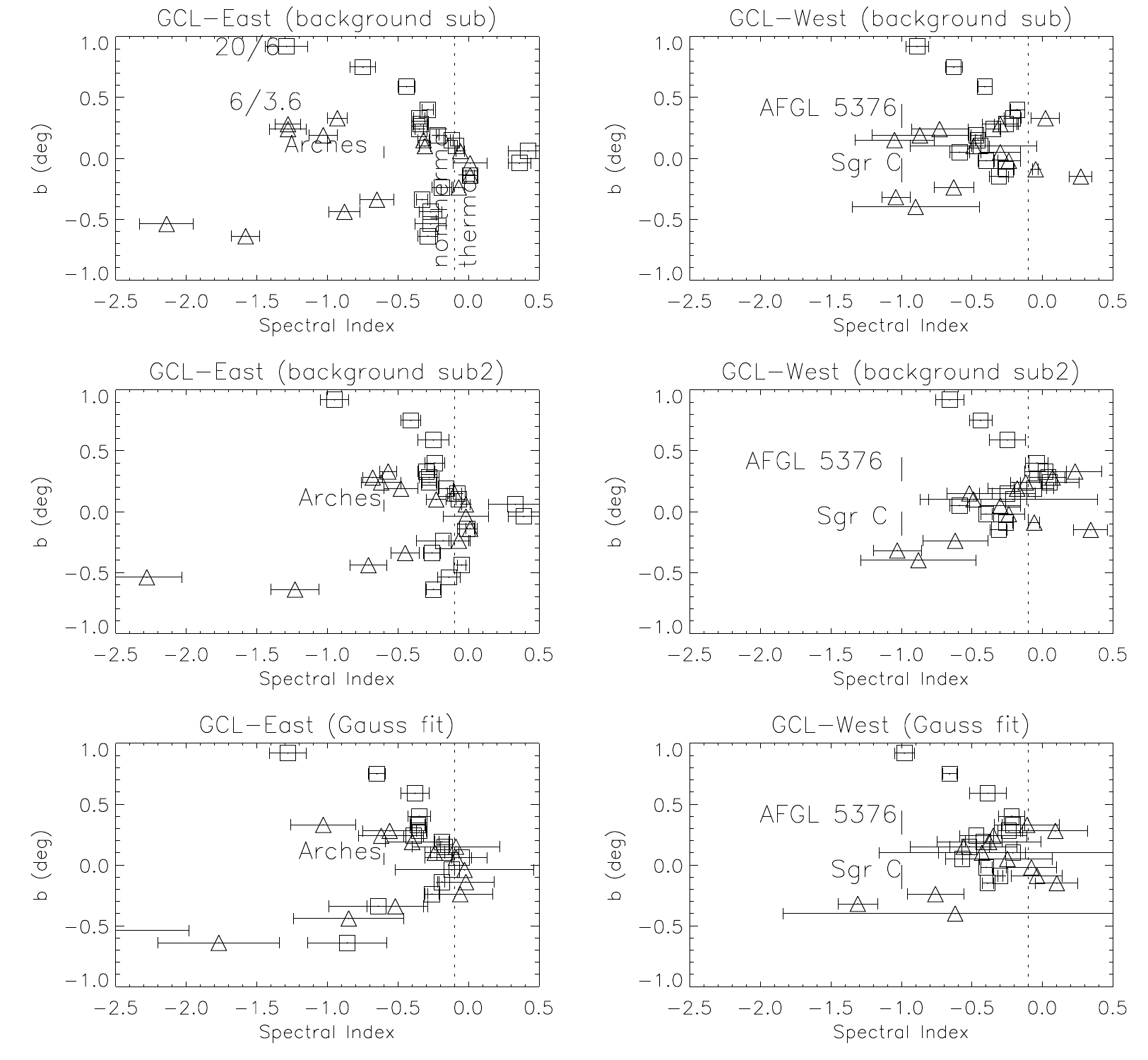}
\end{center}
\caption{\emph{Top panels}:  The 20/6 cm and 6/3.5 cm spectral indices (``$\alpha_{LC}$'' and ``$\alpha_{CX}$'', respectively) of the GCL as a function of Galactic latitude for the GCL-East and GCL-West.  The peak brightness is measured by the background-fitting technique estimating the background flux from the edges of the slice.  Square symbols show the 20/6 spectral index and triangles show the 6/3.5 spectral index.  The vertical dotted line shows a thermal-like spectral index of --0.1;  a nonthermal spectral index lies left of this line.  Ranges in galactic latitude where the GCL might be confused with other sources are labeled with vertical bars and the name of the confusing source.  \emph{Middle panels}:  Same as for the top panels, but with a background estimated close to the peak brightness of the GCL. \emph{Bottom panels}:  Same as for the top panels, but using the Gaussian-fit technique. \label{gclspix}}
\end{figure}

The arcminute resolution of the GBT makes confusion and background subtraction important and challenging problems.  To consider the biases introduced by this confusion, the slice analysis was conducted with two different techniques.  First, the background was estimated by fitting a linear background to parts of the slice without any emission (as described in Ch. \ref{gcsurvey_gbt}).  The results from this ``background-fitting'' technique depend on the definition of the background region, so the spectral index is calculated assuming a background very close to the peak emission and at the edges of the slice.  The second technique of measuring the spectral index of the GCL involves fitting a model to the slice profile.  The model consists of a linear background plus a Gaussian; all structure in the slice that does not match this model and does not appear to be noise are ignored.  The uncertainty in the brightness is estimated by calculating the rms deviation of the data relative to the best-fit model.  This technique is here referred to as the ``Gaussian-fit'' technique.  The spectral index measured at the peak continuum brightness of the GCL-East and GCL-West are shown in Figure \ref{gclspix} for both of these techniques and using two assumptions of the location of the background region.

Each of the techniques used to find the spectral index has its own biases.  The background-fitting technique calculates the spectral index across the entire slice, but picks a value at the location of the peak brightness at the shorter wavelength.  This tends to bias the spectral index upward, since the peak brightness in the shorter wavelength slice may include a noise peak.  In practice, this is not much of a problem because the shorter wavelength image is convolved to the resolution of the longer-wavelength image and tends to be very smooth.  The background-fitting technique can also be biased by the assumption of where the background lies.  If the background is chosen very close to the feature, it could subtract more of the extended background emission, which tends to be brightest at long wavelengths.  If long-wavelength emission is oversubtracted, the spectral index is biased toward positive values.  The Gaussian-fit technique is biased by the fact that the peak must be identified by eye and that all nonnoise, nonmodelable structures must be ignored.  This can reduce the usable part of the slice considerably and lead to an underestimate of the noise.  Also, the location of the best-fit Gaussian may shift (slightly) between the two wavelengths, so the peak flux may be an overestimate, which would also lead to an underestimate of the spectral index uncertainty.

Using a range of techniques to measure the spectral index in the GCL gives confidence in the trends that are common between them.  Inspecting Figure \ref{gclspix} shows several trends in the spectral index of the GCL.

\begin{enumerate}
 \item The values of $\alpha_{CX}$ and $\alpha_{LC}$ are significantly nonthermal, where not confused with other sources.  For the GCL-East, all spectral indices north of the Arches and south of $b\sim-0.3$ are significantly less than $-0.1$.  The indices are flat where the Arc is brightest (from $b\sim-0\ddeg3$ to $+0\ddeg1$).  For the GCL-West, the indices are consistent with a nonthermal origin, except at Sgr C and AFGL 5376.
 \item There is a clear steepening of $\alpha_{LC}$ at large positive latitudes in the both the GCL-East and GCL-West.  All three methods show that the indices in the northernmost three slices are nonthermal and progressively more negative toward higher latitudes.
 \item There is a steepening of $\alpha_{CX}$ at large negative latitudes in GCL-East and GCL-West.  In the GCL-East, the steepening is significant, if irregular.  In the GCL-West, the steepening is less clear, due to confusion with Sgr C, but it appears to get steeper in all three analysis techniques.
 \item $\alpha_{LC}$ tends to be flatter than $\alpha_{CX}$ far from plane in the GCL-East.  North of the Arches and south of $b\sim-0.3$, $\alpha_{CX}$ seems to steepen continually.  In contrast, $\alpha_{LC}$ seems to have a fairly regular value of $\sim-0.3$ (particularly noticeable in the top panels of Figure \ref{gclspix}).  Note that this cannot be excluded in the GCL-West, due to limited coverage at 3.5 cm and confusion with G359.8-0.8 at 20 cm.
 \item The values of $\alpha_{LC}$ and $\alpha_{CX}$ are flatter near AFGL 5376.  The $\alpha_{CX}$ index is consistent with thermal emission, but $\alpha_{LC}$ is significantly nonthermal, although less so than north or south of AFGL 5376.  Note that the middle panels show that $\alpha_{LC}$ is thermal, but this technique generally biases the index upward in confused regions.
\end{enumerate}

\begin{figure}[tbp]
\includegraphics[width=\textwidth]{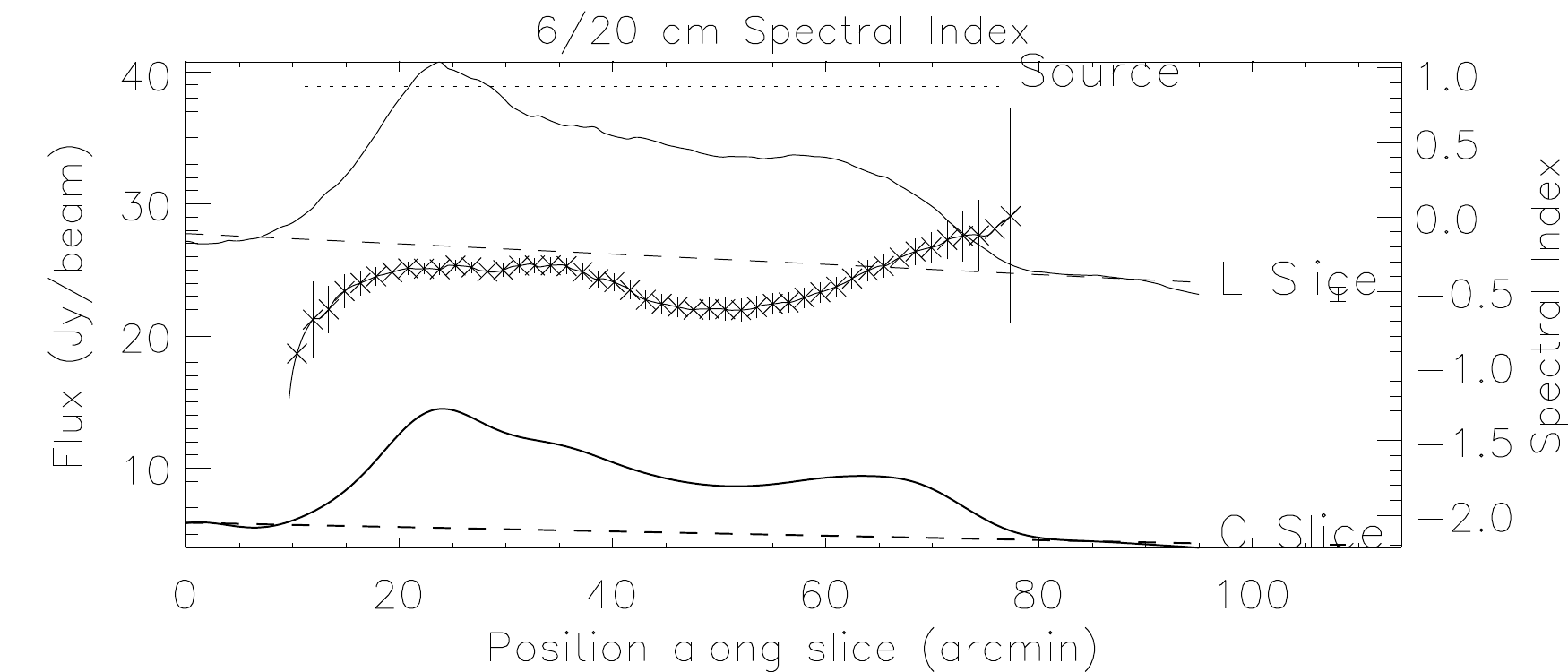}
\caption{Plot of the 6 and 20 cm brightness and their corresponding spectral index for a constant-longitude slice across the GCL at $b=0\ddeg24$ .  The dashed line shows the best-fit background line found from each slice by ignoring the range shown with the dotted line.  To reduce confusion, only points with spectral index error less than 1 are plotted.  The spectral index has maxima near the peak of the 6 and 20 cm brightness and a significant minimum toward the center of the GCL. \label{gclspixslice}}
\end{figure}

Figure \ref{gclspixslice} shows the flux and spectral index of a slice across the GCL at $b=0\ddeg24$ using the background-fit technique.  This slice shows a clear flattening of the spectral index near the peak brightness, as compared to the middle of the GCL.  Slices at all latitudes north of the Galactic plane show the same behavior.   In fact, the latitude dependence of the spectral index inside the GCL seems to change in a similar manner as at the peak flux of the GCL, with a nonthermal index throughout, but rapidly steepening for $b\gtrsim0\ddeg5$.  Aside from difference between the edge and center of the GCL, Figure \ref{gclspixslice} also shows how the spectral index tends to increase across the edge of the GCL.  Moving eastward across the GCL-East and westward across the GCL-West (toward the ``outside'' of the GCL) shows a slight increase in the spectral index in slices at several latitudes.  Thus, the peak of the 6 cm emission tends to lie outside of the 20 cm emission.  

From these trends, we conclude that the radio continuum emission associated with the GCL is dominated by nonthermal synchrotron emission.  From the brightness and spectral index of synchrotron emission, the equipartition magnetic field can be calculated \citep{m75,b05}.  For $b=0\ddeg6-0\ddeg9$, the spectral index between 20 and 6 cm at the edge of the GCL ranges from roughly --0.5 to --1.0 and the brightness varies from $\sim$0.25 to 0.1 Jy per $2\damin5$-beam.  Assuming a shell geometry (see \S\ \ref{shellsec}), the path length through the edge is about 50 pc.  From classical synchrotron theory, the equipartition magnetic field ranges from 44--55 $\mu$G, assuming the ratio of the proton-to-electron energy density ratio and the filling factor (i.e., $K/f$) is 100, and integrating over energies from 10 MHz to 10 GHz.  For the same assumptions, the revised derivation of the equipartition magnetic field by \citet{b05} gives a field strength of 38--53 $\mu$G.  The spectral index for $0\ddeg2<|b|<0\ddeg5$ at the edges of the GCL is about --0.3 and fluxes are about 1 Jy per $2\damin5$-beam.  Under the same assumptions, the equipartition field is 50 $\mu$G according to classical theory and the minimum field is 100 $\mu$G for the revised method appropriate for spectral indices flatter than --0.5 \citep{b05}.

The estimated minimum magnetic field in the edge of the GCL for $b>0\ddeg2$ ranges from 40 to 100 $\mu$G, for varying assumptions.  This field strength is similar or higher than that estimated for the diffuse nonthermal emission observed throughout the central few degrees at lower frequencies \citep[$\sim40$ $\mu$G, for similar energy density ratio;][]{la05}.  Previous work by \citet{r87} found a similar magnetic field strength of ``a few $10^{-5}$ G'' for the radio continuum emission from the entire GCL.

In the relatively unconfused emission of the GCL-East, there is a tendency for the 3.5/6 cm spectral index to be steeper than the 6/20 cm spectral index.  In fact, both spectral indexes steepen with increasing $|b|$, but the 3.5/6 cm index seems to steepen more rapidly.  The steepening of the spectra with $|b|$ is consistent with previous work, which modeled the changes as being caused mostly by adiabatic changes in a flux tube that expands at higher $|b|$ \citep{p92}.  The present observations also show that the higher frequency spectral index steepens more rapidly, which is better explained by synchrotron energy losses.  The time scale for synchrotron energy loss is $t_s=4.7\times10^5 B_{0.1}^{-3/2} \nu_5^{-1/2}$ yrs, where $B_{0.1}$ is in the field strength in units of 0.1 mG and $\nu_5$ is in units of 5 GHz \citep{l01}.  

The initial work of \citet{s85} found that the radio spectral index implied a thermal origin for the GCL.  One reason for the difference between that and the present work is that \citet{s85} measured the spectral index from images that had been unsharp masked \citep[``BGF technique'';][]{r87}.   The unsharp masking of the images may oversubtract long wavelength emission near the GCL, biasing the spectral index toward positive (thermal-like) values.  Also, the data used by \citet{s85} studied higher frequency emission (3 cm) than studied here, which could include more thermal emission.  Later work by \citet{r87} found that the radio continuum emission from the GCL is nonthermal.

\subsection{Radio Recombination Line}
\label{all_gbtrecomb}
The GCL was observed for radio recombination line emission with the Hat Creek Radio Observatory 26 m telescope (HCRO) and GBT.  The HCRO observations mapped the entire region, providing for a morphological study, while the GBT conducted deep, pointed observations, providing for a spectroscopic study of the GCL.  Here we describe the highlights of these observations, while the details of the analysis are described in chapter \ref{gcl_recomb}.

Morphologically, the radio recombination line emission has a similar, loop-like structure as seen in the radio continuum emission.  However, the recombination line emission is slightly smaller and tends to lie inside the continuum emission.  This is less clear in the western half of the GCL, where the components are coincident within their positional uncertainties.  Several pointings were made outside the radio continuum shell of the GCL and constrain the line brightness to be significantly less than observed inside the GCL.  

Spectroscopically, the radio recombination line is found to have unusually narrow widths and have relatively small velocities.  The line width of 13.5 \kms\ places a strict upper limit on the electron temperature of $3960\pm120$ K throughout the GCL;  this line can also be shown to have little or nor stimulated contribution.  Collisional line broadening of the lines is consistent with an electron density $n_e=950\pm270$ cm$^{-3}$, which combined with a mean electron density estimate of $n_e\approx 8.8$ cm$^{-3}$, gives a filling factor $f = (9\pm2)\times10^{-5} (T_e/3960\ \rm{K})^{1.3}$.  The mass of the ionized gas is $M=3\times10^5 (T_e/3960\ \rm{K})^{0.61}$ \msol\ and the ionizing flux is $N_{Ly} = 1.0\times10^{50} (T_e/3960\ \rm{K})^{1.22})$ s$^{-1}$.  The ionizing flux from known massive star clusters in the GC region can easily explain the ionization of the GCL.

The line velocities of the ionized gas in the GCL are generally within 20 \kms\ of the rest frequency, although its structure is more complex than previously observed in the GCL.  There is evidence for a gradient in the the line velocity from east to west across the GCL, consistent with Galactic rotation, but with a much smaller amplitude of 5 \kms\ over $0\ddeg7\approx100$ pc.  The radio recombination line is also clearly split by about 20 \kms\ in some parts of the GCL, which could be interpreted as expansion.  However, there is no regular trend in the split line, so we can only conclude that any radial expansion of the gas is less than $\pm10$\ \kms.

\subsection{High-Resolution Polarized Radio Continuum}
\label{all_vlapoln}
The 6 cm polarized continuum in the GCL was observed as a part of a larger VLA radio continuum survey.  The polarized continuum from $l=$359\ddeg2-0\ddeg2,$b=$0\ddeg2-0\ddeg7 was surveyed in two, adjacent 50 MHz bandpasses, which allowed an estimate of the rotation measure ($RM$) in the region.  Here we summarize the results of the observations, which are described in more detail in chapter \ref{gcl_vlapoln}.

Polarized emission filled with depolarized canals is found throughout the GCL, as has been seen in other polarized continuum surveys of the Galaxy \citep[e.g.,][]{h00}.  The polarized continuum brightness is also consistent with previous single-dish surveys, which showed a higher polarization fraction in the east and relatively little in the west.  Observations of background and foreground sources are consistent with the idea that the $RM$ observed in the diffuse emission toward the GCL is consistent with the GC region and is not biased by magnetic structure in the Galactic foreground \citep{r05,c04,ma05,h06}.  This suggests that the polarized intensity and $RM$ are likely to originate in or near the GC region.

\begin{figure}[tbp]
\includegraphics[width=\textwidth]{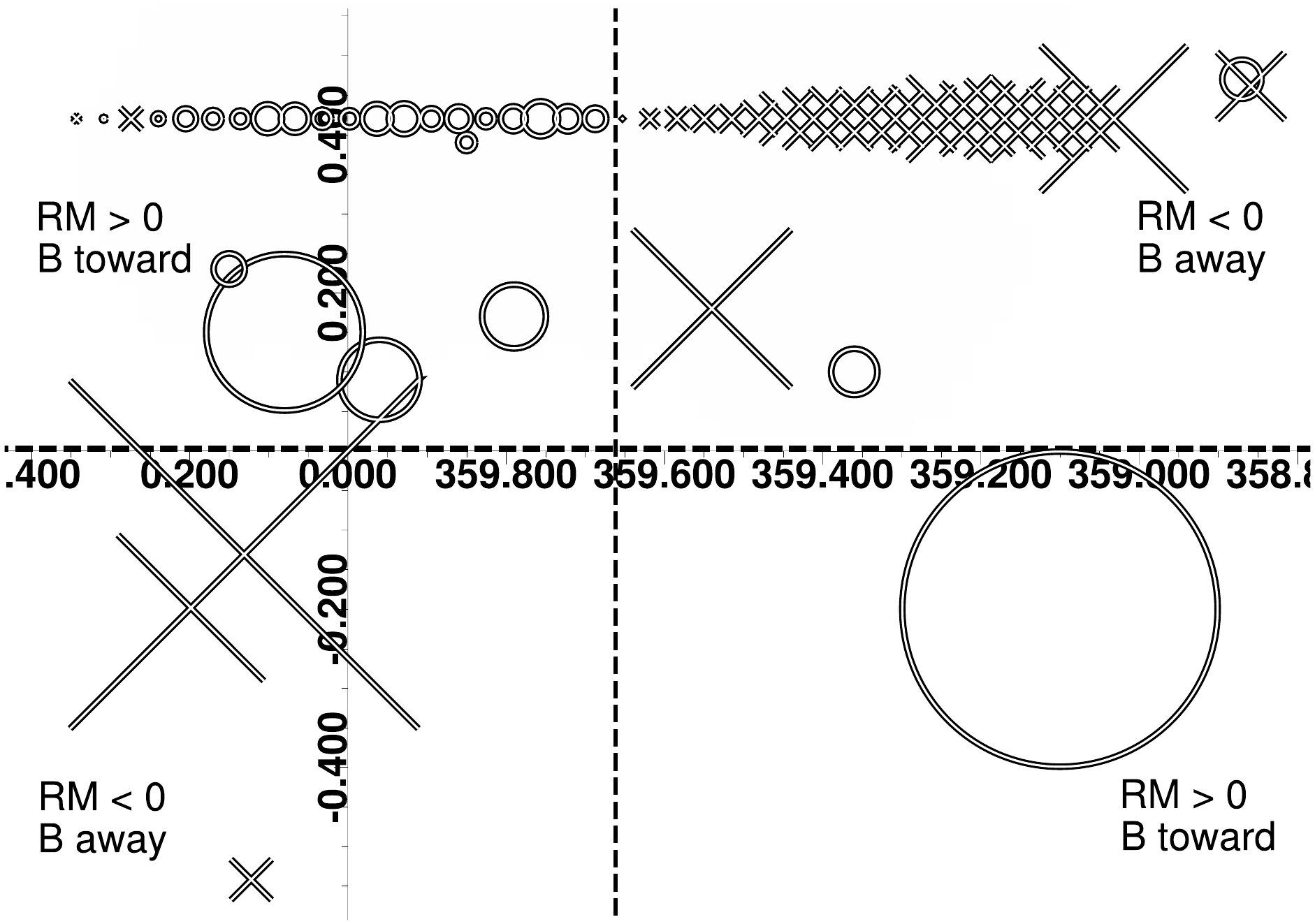}
\caption{A schematic diagram of all GC measurements of $RM$ and \rmeff for NRFs and the extended polarized emission.  The circles show $RM>0$ and crosses showing $RM<0$, with values ranging from --5\sdeg\ to 10\sdeg\ or \rmeff from 1100 to --2200 rad m$^{-2}$.  In addition to the measurements of \rmeff from our study, this figure shows measurements from the literature, specifically:  G0.08+0.15 \citep[a.k.a. ``the Northern thread''][]{l99}, the Radio Arc \citep[south of $b=0$\sdeg;][]{t86}, G0.87--0.87 \citep{re03}, G359.1--0.2 \citep[a.k.a. ``the Snake''][]{g95}, G358.85+0.47 \citep[a.k.a. ``the Pelican''][]{la99}.  The dashed lines split the region into quadrants that mostly have similar $RM$ values.  \label{all_rmallschem}}
\end{figure}

Studies of the $RM$ across the image shows a striking, large-scale gradient in Galactic longitude.  The \rmeff is about +330 rad m$^{-2}$ in the western half of the GCL and --880 rad m$^{-2}$ in the eastern half;  the mean \rmeff changes sign abruptly at $l=-0\ddeg35\pm0\ddeg02$.  Figure \ref{all_rmallschem} shows a schematic of all $RM$ measurements toward the diffuse polarized emission and NRFs.  In all but one case, the \rmeff of the NRFs agree with that of their surrounding polarized emission \citep[e.g.][]{l99,t86}.  One model for the formation of this structure suggests that the magnetic field in the GC region is dragged by gas in the disk, thus creating a line-of-sight component of the magnetic field with a checkerboard pattern about the center of rotation of the gas \citep{n03}.  Curiously, the center of this checkerboard pattern of $RM$ values is closer to the center of the GCL than to the center of Galactic rotation.

There are correlations between radio recombination line emission and the $RM$ observed in the GCL.  Chapter \ref{gcl_vlapoln} notes a clear morphological connection in the GCL-West, where a peak in the radio recombination line brightness near ($-0\ddeg6$,$0\ddeg5$) coincides with a peak in the estimated $RM$.  
On larger scales, the polarized radio continuum and radio recombination line emission show a rough anticorrelation, which again suggests a connection between the two.  If the radio continuum in the GCL is nonthermal synchrotron emission with a intrinsic polarization fraction of roughly 70\% \citep{r79}, the observed polarization fraction should be inversely related to the amount of depolarizing thermal gas in the GCL.  A model in which the Faraday-rotating and synchrotron-emitting gas are mixed together predicts the observed polarization fractions.

\subsection{Mid-IR Continuum}
\begin{figure}[tbp]
\includegraphics[width=\textwidth]{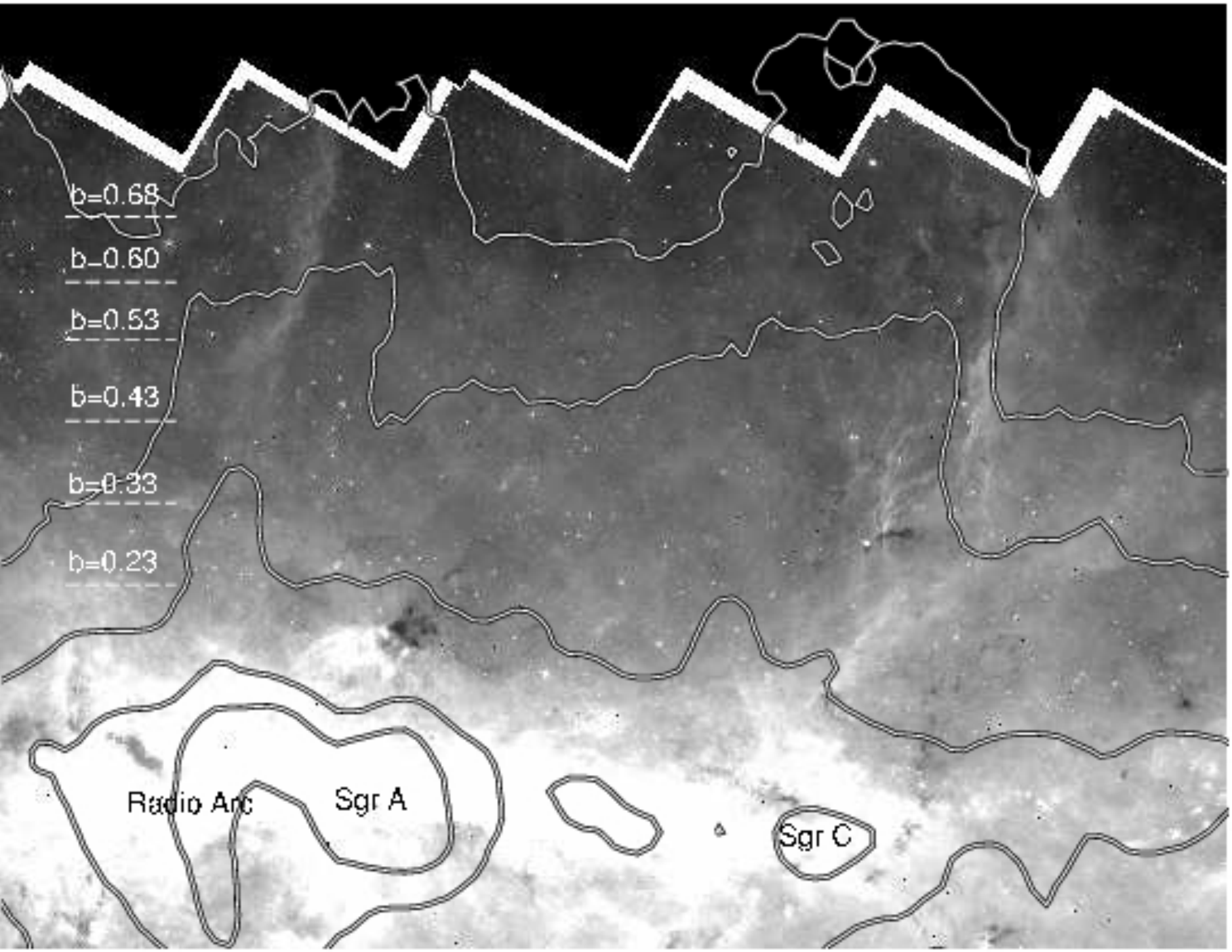}
\caption{Grayscale shows a portion of the \spitzer/IRAC 8 $\mu$m image and contours show GBT 6 cm continuum emission around the GCL.  Dashed lines show the latitudes at which slices were taken across the GCL for Fig. \ref{slices}; only a portion of the slices are shown here to avoid confusing the image.  \label{sp4GCC}}
\end{figure}

The central two degrees of our Galaxy was surveyed with \spitzer/IRAC in four bands from 2-8 $\mu$m.  \citet{s07} summarizes the calibration and major results of the survey.  Here we discuss the results of a morphological study of the GCL with \spitzer/IRAC, including a comparison to \msx\ and radio continuum observations.  Figure \ref{sp4GCC} shows \spitzer\ 8 $\mu$m image with contours of 6 cm emission as observed by the GBT.  The calibration and image merging algorithm is designed to smooth differences between different pieces of the map, and as a consequence the absolute zero-flux level of the map is not known.

\begin{figure}[tbp]
\includegraphics[width=0.5\textwidth]{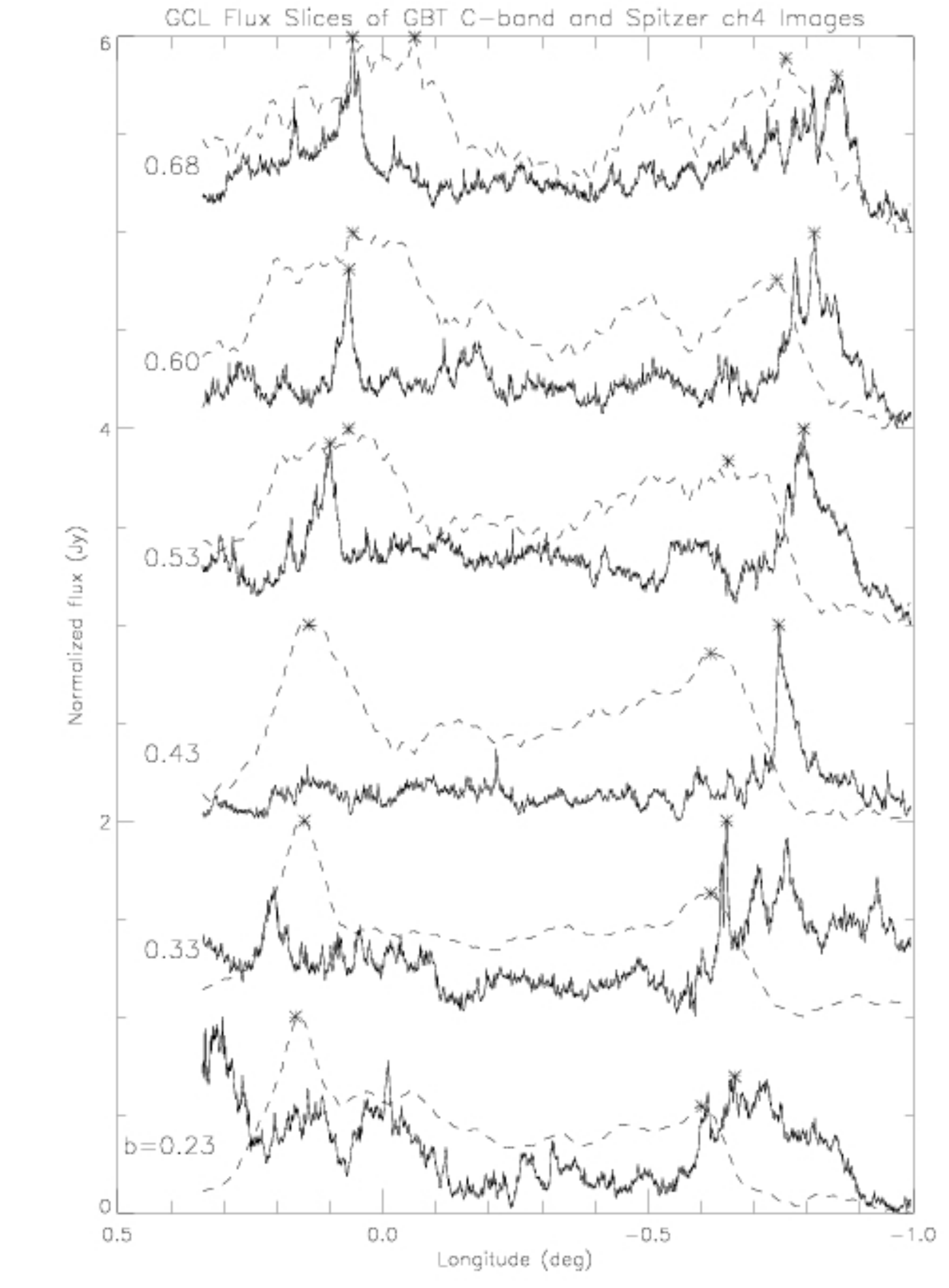}
\includegraphics[width=0.5\textwidth]{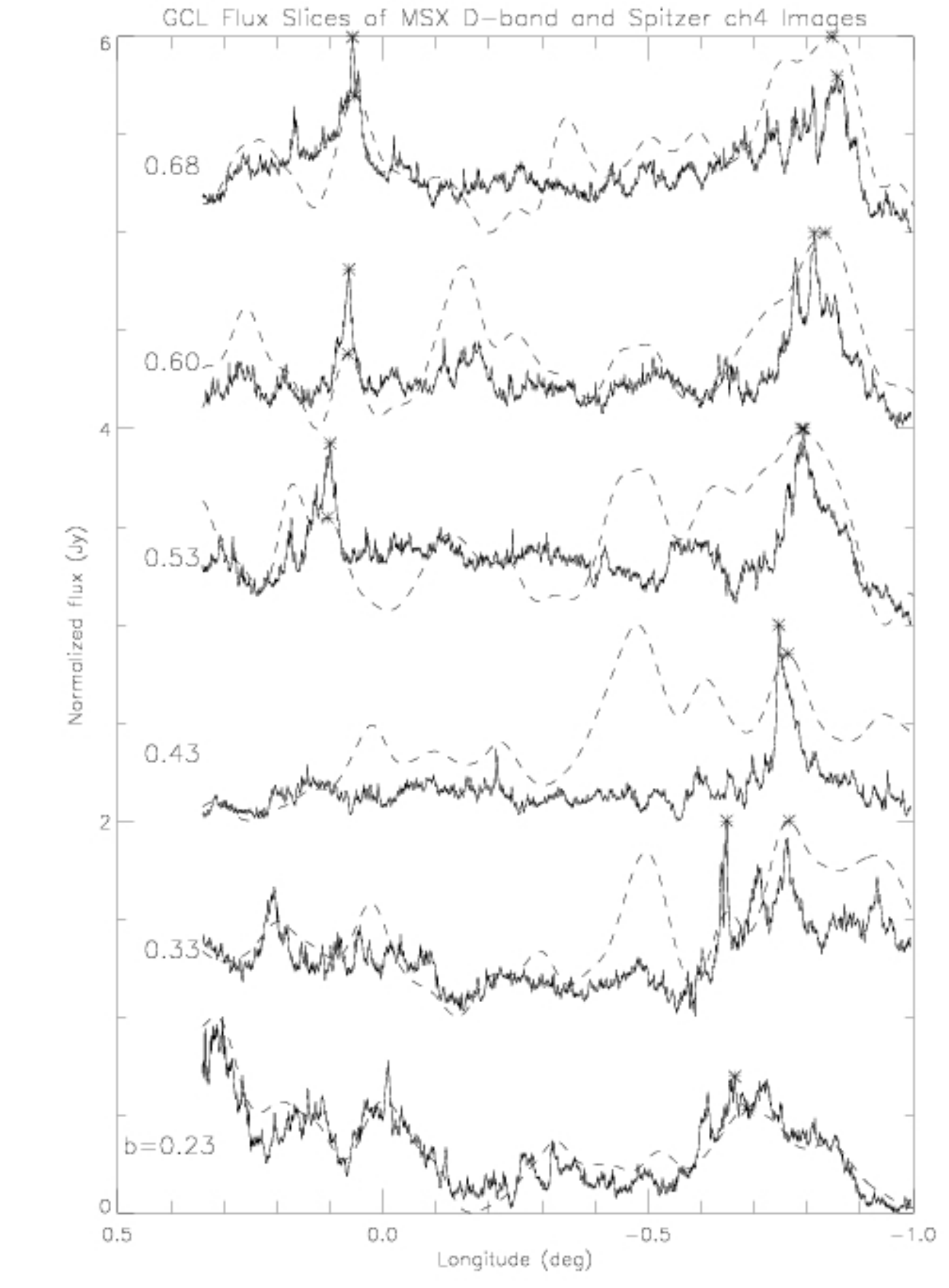}
\caption{\emph{Left}:  Normalized flux slices of GBT 6 cm continuum (dashed line) and \spitzer/IRAC 8 $\mu$m (solid line) for six latitudes across the GCL.  The bottom slices show the lowest latitude and subsequent slices are vertically offset by 1 for clarity.  The slices are over identical longitude ranges at a single galactic latitude, which is indicated for each pair.  The slices were calculated for a small range of latitudes and median filtered (by pixel) to remove compact sources.  The peak flux for the eastern and western halves of each slice are shown with stars, where they represent the peak flux of the GCL.  The lower latitude slices do not have a clear GCL counterpart in the mid-IR on the west side, so no stars are plotted there.  \emph{Right}: Same as the left side, but comparing smoothed \msx\ 15 $\mu$m (``D-band'') slices to the \spitzer/IRAC 8 $\mu$m slices.  \label{slices}}
\end{figure}

The 8 $\mu$m image shows that the GCL has a clear mid-IR counterpart with detailed correlation with the radio continuum structure.  The correlation of \msx\ 8 $\mu$m and 3 cm radio continuum has been discussed by \citet{b03}.  Using slices across the data, we give a more detailed comparison of the GBT radio continuum, \spitzer\ 8 $\mu$m, and \msx\ 15 $\mu$m morphology in Figure \ref{slices}.  The flux for slices across six latitudes are normalized and plotted together to show the relative location of the peak flux for each slice.  As shown elsewhere \citep{s07}, the \spitzer\ 8 $\mu$m band is dominated by PAH emission features, so it traces dust column density and the far-UV radiation field \citep{p04}.  The \msx\ 15 $\mu$m band does not have any strong PAH or line emission, so it is a good tracer of dust continuum emission \citep{a89,s07}.  Unfortunately, the GCL is not significantly detected in any two bands without PAH features, so no estimate of the dust temperature is given.

The slices confirm the visual impression that the radio continuum and 8 $\mu$m peaks are offset from each other.  The radio continuum emission is found closer to the GC than the 8 $\mu$m emission (on the ``inside'', assuming a shell geometry).  The positional offset is typically $0\ddeg05$ in the east and $0\ddeg1$ in the west for the range of latitudes where 8 $\mu$m emission is seen ($b=0\ddeg23$--$0\ddeg68$).  The peak-to-peak diameter of the 8 $\mu$m emission is $0\ddeg9\approx130$ pc.

The right side of Figure \ref{slices} compares the \msx\ 15 $\mu$m flux to the \spitzer\ 8 $\mu$m flux.  In general, the peak of the GCL structure in \msx\ 15 $\mu$m image is very near the peak of \spitzer\ 8 $\mu$m image.  Note that the GCL-East does not have a mid-IR counterpart in these two images for the southernmost three slices, and the peak in slices at $b=0\ddeg33,0\ddeg43,0\ddeg53$ around $l=-0.5$ are associated with AFGL 5376, which is inside the GCL-West.  Since the \msx\ image shows warm dust and \spitzer\ image shows PAHs from irradiated flux, we know that the GCL is host to warm dust and is being irradiated by far-UV photons.  Interestingly, the slice at $b=0\ddeg33$ shows a strong feature around $l=-0.6$ that has no \msx\ counterpart.  Figure \ref{sp4GCC} shows \spitzer\ 8 $\mu$m emission that has a striking filamentary structure oriented parallel to the GCL-West, but closer to the shocked molecular cloud, AFGL 5376 \citep{u94}.
The mid-IR emission associated with the GCL-East has an unusual twisted morphology as has been noted elsewhere \citep{m06}.

As discussed in \S\ \ref{all_gbtcont}, the shape of the GCL is slightly more complicated than a simple shell in the radio continuum.  The slices shown in Figure \ref{slices} show how the radio continuum from the edges of the GCL broadens or splits at higher latitudes.  Interestingly, the mid-IR emission is aligned west of the GCL-West, but in the middle of the broadened GCL-East radio continuum.  Regardless of this structure, the brightest radio continuum still tends to be closer to the GC than inside of the mid-IR emission.


\section{Conclusions on the Nature of the GCL from Multiwavelength Observations}
\label{points}
The multiwavelength observations described above lead to three major conclusions about the nature of the GCL.  

\subsection{The GCL is a Single Object}
\label{single}
The new observations presented here have found several characteristics of the GCL that show that it is a coherent object with a single origin.  The initial argument that the GCL had a single origin came from its shell-like structure, reminiscent of outflows and radio lobes seen in other galaxies.  This argument has been strengthened by the discovery of radio recombination line and mid-IR emission in the GCL with a similar morphology.  In particular, the radio recombination line emission is much brighter inside the radio continuum emission than outside of it.  The overall shape of the radio recombination line, radio continuum, and mid-IR emission around the GCL are similar in their tilted shape and offset from Sgr A*, strongly suggesting that they are associated with each other.

In addition to the morphological evidence, there are other characteristics of the GCL emission that indicates that its east and west sides have a common origin.  First, as described in \S\ \ref{all_gbtcont}, the 6/20 cm spectral index for the GCL-East and GCL-West have similar, nonthermal values.  This is noteworthy, since previous work had found that the emission was thermal, or that perhaps only the GCL-West was thermal \citep{s84,u94}.  Instead, we find that the majority of the radio continuum emission has a nonthermal synchrotron origin.  More importantly, the index for both edges of the GCL show a similar change with Galactic latitude, with significant steepening toward the highest latitudes.  Although the cause of this steepening is not known, the similarity between the GCL-East and GCL-West suggests that they have a similar physical origin.

Second, as discussed in \S\ \ref{all_gbtrecomb}, the radio recombination line measured several properties of the ionized gas associated with the GCL that suggest that it has a single origin.  The unusually narrow line width observed throughout the GCL implies very low electron temperature for the ionized gas.  It is unlikely that recombination line emission from a typical \hii region would show such narrow lines;  Galactic trends in electron temperature suggest that such low values are only expected near the GC \citep{a96}.  The intensity ratios of the lines detected in the east and west of the GCL are similar, implying that all the gas has a similar temperature and density.

Third, the change in polarized emission and 74 MHz absorption between the GCL-East and GCL-West \citep{b03b} may be explained by changes in the medium inside or in front of the GCL.  The radio spectral index in the GCL-East and GCL-West are consistent with nonthermal synchrotron emission, which should have intrinsic polarization fractions of about 70\% and emit strongly at 74 MHz.  The relatively low polarization fraction in the GCL-West may be caused by its having roughly twice the $|RM|$ as the GCL-East.  Furthermore, an internal depolarization model predicts the observed polarization fractions in the GCL-East and GCL-West, suggesting that the polarization-emitting and depolarizing media are in the same volume of space.  The coincidence of the radio recombination line emission with the radio continuum in the GCL-West might explain the larger $RM$ observed there.  Thus, the difference in polarization and 74 MHz absorption across the GCL can be explained by changes in the ionized gas associated with it and does not necessarily imply that the GCL-East and GCL-West are different objects.

\subsection{The GCL is Composed of Layered Shells}
\label{shellsec}
\begin{figure}[tbp]
\includegraphics[width=\textwidth]{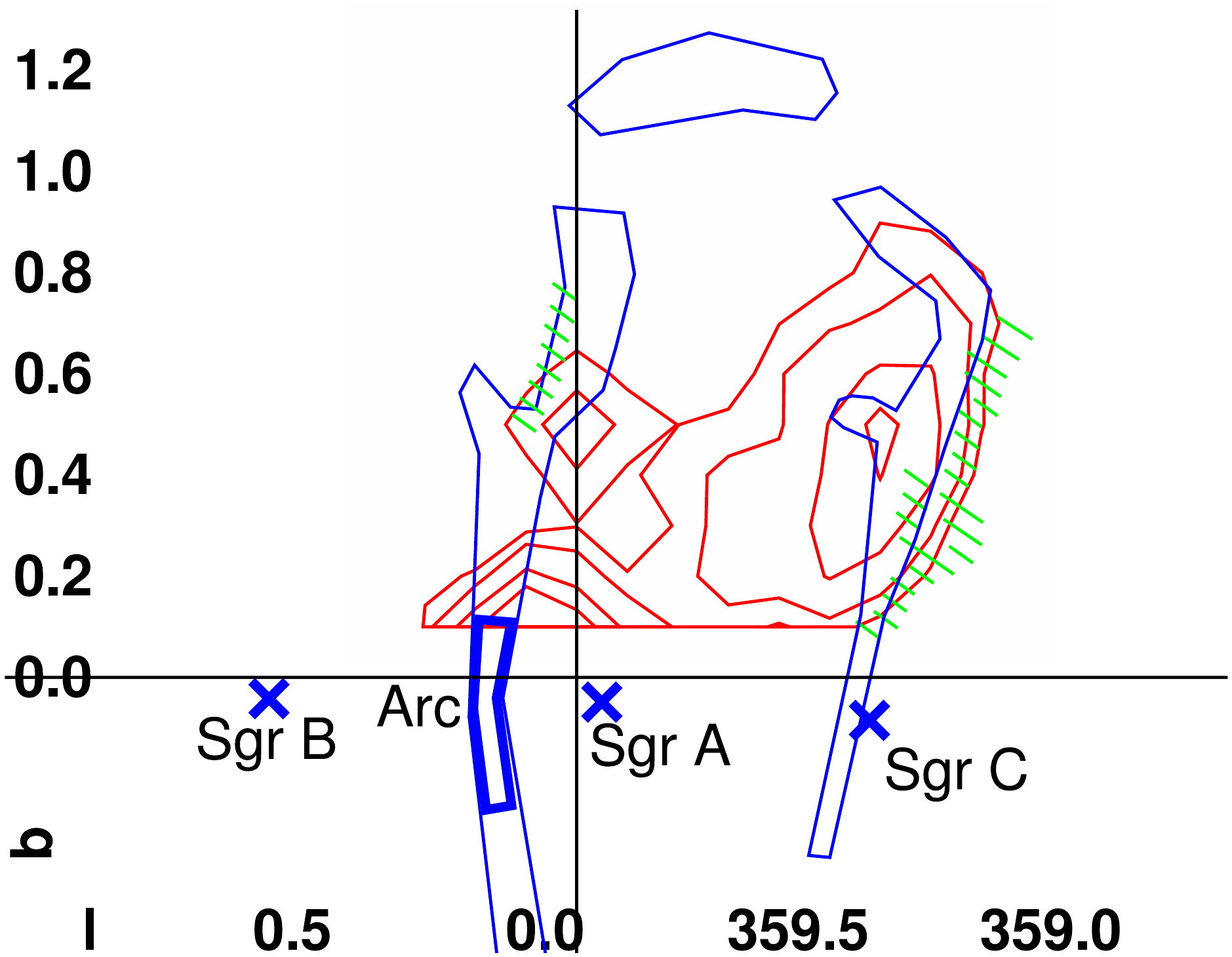}
\caption{Schematic of the multiwavelength structure of the GCL and GC region.  The red lines show the contours of radio continuum emission, the blue shows the alignment of radio continuum emission, and the green hatch marks show the mid-IR emission associated with the GCL.  The blue crosses show where the brightest \hii\ regions and the thick blue line shows where the brightest emission of the Radio Arc. \label{all_schem}}
\end{figure}

The emission from the three components of the GCL have a similar, limb brightened morphology, as would be expected from a shell.  Figure \ref{all_schem} summarizes the morphologies of the radio recombination line, radio continuum, and mid-IR components in the GCL.  All three components have a similar center longitude, while each component has a different edge-to-edge diameter.  The extent of the recombination line-emitting region is smallest, with a diameter of about $0\ddeg6\approx80$ pc.  The radio continuum and mid-IR component have diameters of about $0\ddeg8\approx110$ pc and $0\ddeg9\approx130$ pc, respectively.  This gives the appearance the radio line-emitting region being nested inside the radio continuum-emitting region, which is inside the mid-IR--emitting region.  One possible exception to this rule is in the relative alignment of the radio line- and radio continuum-emitting regions in the GCL-West, which are roughly coincident with each other.

For a quantitative study of the GCL, we need to apply our observations to a 3-d model for its structure.  An appropriate model for the structure is that of a ``telescope dome'' with a radius $r$, height $h$, and thickness $\delta$ \citep{b03}.  Table \ref{all_shell} shows the values of these parameters that best fit the multiwavelength structure of the GCL for all observations.  The height of GCL is only measured in the 20 cm GBT continuum observations to be about 165 pc.  Under this model, the volumes of the three components are roughly $1.0\times10^6$, $1.9\times10^6$, and $2.8\times10^6$ pc$^{-3}$, for the radio recombination line, radio continuum, and mid-IR shells (assuming equal heights).

To test the appropriateness of the shell model, the edge-to-center contrast is measured for each component and compared to that of an idealized model.  If the observing beam is much smaller than the shell thickness, the contrast can be calculated from geometry as $C=1+\frac{r}{2\delta}-\frac{\delta}{2r}$.  However, the effect of observing the GCL with a finite resolution reduces the contrast significantly.  To calculate the contrast from the idealized model, we created a model of a shell and convolved it with a Gaussian with a full width at half max of $B$.  From this simulation, the idealized contrast $C_{\rm{ideal}}$ is measured for comparison with the observed contrast.  Table \ref{all_shell} shows that the observations show reasonable agreement with the shell model, suggesting that the structure seen at those wavelengths originates in an intrinsically shell-like structure.  The change in contrast with resolution (e.g., between the different beam sizes of HCRO and GBT) is similar to the predicted change.  This simulation also helps to estimate the shell thickness, $\delta$, from the observations, which is generally about half the apparent width at half the peak brightness.

\begin{deluxetable}{lcccccl}
\tablecaption{Observed and Calculated Parameters for GCL, Assuming Shell Morphology \label{all_shell}}
\tablewidth{0pt}
\tablehead{
\colhead{Observation} & \colhead{$r$} & \colhead{$\delta$} & \colhead{$B$} & \colhead{$C_{\rm{ideal}}$} & \colhead{$C_{obs}$} & \colhead{Notes} \\
\colhead{} & \colhead{(pc)} & \colhead{(pc)} & \colhead{(pc)} & \colhead{} & \colhead{} & \colhead{} \\
}
\startdata
HCRO Recombination line & 40 & 15 & 23   & 1.5 & 1.5--3 & at $b=0\ddeg5$ \\
GBT Recombination line  & 40 & 15 & 5.8  & 2.1 & 3--7 & at $b=0\ddeg45$ \\
GBT 6 cm Continuum      & 55 & 15 & 5.8  & 2.4 & 2.2--2.6 & at $b=0\ddeg45$ \\
GBT 20 cm Continuum     & 55 & 15 & 21   & 1.8 & 1.2--1.4 & at $b=0\ddeg45$ \\
\emph{Spitzer} 8$\mu$m  & 65 &  5 & 0.08 & 5.2 & $>3$\tablenotemark{a} & at $b=0\ddeg56$ \\
\enddata
\tablenotetext{a}{The emission from the center of the GCL seems to be mostly foreground emission.  Thus, the observed contrast is a lower limit to the contrast associated with the shell itself.}
\end{deluxetable}

Some deviations from the idealized shell model are seen, however.  The contrast in the GBT 20 cm continuum is lower than the model predicts.  This can be explained if the GCL has some 20 cm emission inside of it.  Thus, the shell may not be hollow, but could contain some synchrotron-emitting gas.  Also, the GBT recombination line contrast is larger than the value predicted by the shell model.  As discussed in \S\ \ref{single}, there are several reasons to believe that the ionized gas is associated with the GCL and has a single origin, so it seems unlikely that the high contrast is caused by some unrelated foreground emission.  One possibility for the high apparent contrast is that the line-emitting shell is irregular, such that the front or back side (from our vantage) is not as thick as parts of the edges, such as at the GCL-West.

\subsection{The GCL is in the GC Region}
\label{ingc}
Since the discovery of the radio continuum emission of the GCL, its coincidence with the central parsecs of our Galaxy has suggested that it is located in the GC region.  While this coincidence is strengthened by the discovery of mid-IR and recombination line counterparts, more than morphological coincidence is needed for convincing evidence that the GCL is in the central few hundred parsecs of the Galaxy.  Here we summarize the evidence in favor of a GC location for the GCL.

Perhaps the strongest morphological connection between the GCL and the GC region is at the Radio Arc \citep{y84}.  The present radio continuum survey has confirmed previous work that found contiguous emission from the GCL-East to the Radio Arc \citep[shown schematically in Fig. \ref{all_schem};][]{y88}.  The Radio Arc (and other NRFs) are known to be within the central few hundred parsecs of the Galaxy from \hi\ absorption measurements \citep{l89,r03}.  This implies that the GCL is also in the central few hundred parsecs.

Our radio recombination line observations have also found an unusually low electron temperature that implies a high metal abundance for the ionized gas in the GCL.  The high metallicity for the GCL is consistent with that expected from the Galactic abundance gradient.  The Galactic gradient in metal abundance is caused by the higher star formation rate observed closer to the GC \citep{f04}.  We also estimate the $RM$ toward diffuse polarized emission in the GCL and find values consistent with the GC region.

Taken together, these observations show that the GCL is located in the central few hundred parsecs of our Galaxy.  We suggest that the appearance of the GCL spanning the central 100 pc in projection is likely to be true in actuality.

\section{Models for the Creation of the GCL}
\label{all_models}
Our new observations of the GCL strongly suggest that the GCL is a layered, shell-like structure that spans the central 100 parsecs of our Galaxy.  More importantly, the new observations provide a much more detailed view of its physical conditions.  In this section we discuss four models that have been proposed to explain the GCL.  The four models are:

\begin{enumerate}
 \item A small-scale starburst \citep{c92,b03}
 \item A jet from Sgr A* \citep{m01}
 \item Escape of the 8 keV, GC gas \citep{m04,be05}
 \item A magnetodynamic effect \citep{u85,su87}
\end{enumerate}

We begin by describing the physical properties of the GCL that these models must describe.  Then we proceed to show how well each model matches the GCL's observed properties.

\subsection{Energetics, Time Scales, and Physical Conditions of the GCL}
\label{physcond}
Assuming that the GCL is in the GC region, we can estimate the energy and time required to create or destroy it.  A minimum energy for the GCL is equal to its gravitational potential.  The approximate brightness-weighted mean height of the GCL (estimated from the brightness of mid-IR and radio recombination line emission) is about 50 pc north of the plane.  The gravitational potential difference between a galactocentric radius of 1--20 pc and 50 pc is equivalent to a velocity of about 40 \kms\ \citep[using the relations of][and $v_{\rm{esc}}=\sqrt{-2*\Phi}$]{b91}.  For the minimum mass estimated from radio recombination line observations, the potential energy difference between these locations is $E_{gr} = 5\times10^{51} * (T_e/3960\ \rm{K})^{0.61}$ ergs.  The total mass of the GCL is expected to be at least twice that in the ionized gas \citep[$M_{mol}>3\times10^5$ \msol;][]{s96,b03}, although this estimate is uncertain since there is no molecular line survey sensitive enough to find gas north of $l\sim0\ddeg3$ associated with the GCL.  The effect of the mass underestimate may be partially canceled by the assumption that all the mass originates between 1 and 20 pc.  If the swept-up gas originated further from the center of the expansion or if the expansion did not come from a single point, then the energy requirement would be reduced.  The degree to which these effects balance is not clear, but the true gravitational potential energy is likely to be within an order of magnitude of that calculated above.  The total energy required to create the GCL will be equal to the sum of its gravitational energy, kinetic energy, and energy lost due to other processes, such as resistance by magnetic fields \citep{v94}.

The gravitational potential energy of the GCL is larger than the kinetic energy associated with the line-of-sight expansion of the ionized gas in the GCL ($v_{\rm{los expansion}}\lesssim10$\ \kms).  Assuming a total mass twice the ionized mass and an expansion velocity of 10 \kms, its kinetic energy is $6\times10^{50}$ ergs s$^{-1}$.  Although the expansion of extragalactic outflows is fastest perpendicular to the plane, it is not clear how fast the GCL is expanding in that direction.  

The thermal pressure in the ionized gas of the GCL implied by the radio recombination line observations is $P/k=3.8\times10^6 (T_e/3960\ \rm{K})$ K cm$^{-3}$.  While this pressure is about 100 times larger than the total gas pressure near the Sun \citep{bl87}, it is not unusual for the GC region \citep{s92}.  Observations of molecular gas in the central degrees of the Galaxy show velocity dispersions that imply virial pressures around $P_{\rm{vir}}/k=nT_{\rm{vir}}=6\times10^6$ K cm$^{-3}$ \citep{ma04}.  The two components of the X-ray--emitting gas characterized by temperatures of $\approx10^7$ and $\approx10^8$ K have pressures in the range of $1-5\times10^6$ K cm$^{-3}$ \citep{k96,m04}.  This pressure is also similar to that of the equipartition magnetic field in the GCL of $\approx60 \mu$G \citep[$P_{\rm{B}}/k\approx1\times10^6$ K cm$^{-3}$;][]{m96}.  If the split line observed toward the center of the GCL is an indication of expansion of a shell, the ram pressure of the ionized gas with $n_e=1000$ cm$^{-3}$ and $v=10$ \kms\ (and $\mu=1.4$) is $8\times10^6$ K cm$^{-3}$.  Thus, it seems that the pressures in typical GC molecular clouds, X-ray gas, equipartition magnetic field, and the ionized gas in the GCL are roughly in equilibrium.

The energy associated with the GCL can also be estimated from the pressure observed inside it.  As described above, the pressure measured by radio recombination line, X-ray, molecular line, and other methods are consistent with a value of about $1-5\times10^6$ K cm$^{-3}$.  If the shell is not expanding and is in equilibrium with its surroundings, the energy of the shell is $E=PV$, where $V$ is the volume of the shell.  If the formation of such a shell was adiabatic, the total energy required to create it is $E=\frac{\gamma}{\gamma-1}PV$, where $\gamma$ equals $5/3$ if the gas is nonrelativistic and $4/3$ if relativistic \citep{d04}.  Assuming a pressure equal to the thermal gas pressure filling the volume enclosed by the mid-IR shell of the GCL, the formation of the GCL requires $E = 5/2 P V \approx 4\times10^{52} (T_e/3960\ \rm{K})$ ergs.  The energy required to form the shell is slightly larger ($E = 4 P V$) if it is assumed to be filled with relativistic plasma.  The formation energy for the GCL is roughly an order of magnitude larger than the minimum (gravitational) formation energy described earlier.  The consistency of these two independent constraints gives credence to the idea that the high gas pressure may be responsible for forming the GCL.  Note that this energy estimate does not account for the kinetic energy associated with the expansion of the GCL, if any.  The kinetic energy estimate given in \citet{b03} was about an order of magnitude larger than that given above, for an assumed expansion velocity of 100 \kms. \footnote{Note that this previous work calculated a total mass for the GCL an order of magnitude larger than the ionized mass found in the present work.  This relies on the analysis of molecular line data that covers Galactic latitudes less than $0\ddeg3$, in which emission from the plane is highly confused with emission that may be from the GCL \citep{s95}.}

The typical dynamical time ($t_{dyn}=R/V_{out}$) for a canonical outflow ranges from 0.1--10 Myr \citep{v05}.  The largest expansion velocity of the GCL allowed by recombination line observations ($\sim10$ \kms) gives a dynamical time to be greater than about 10 Myr, assuming constant expansion velocity.  However, if the expansion of the gas has decelerated, then this dynamical time will overestimate the actual formation time.  If the gas has decelerated from an initial expansion velocity, $v_{init}$, to 10 \kms, then its formation time is $t_{form} = 1.6 (\Delta r/50$\ pc$)(40$\ km s$^{-1}/v_{init})$ Myr.  As described above, the initial velocity and acceleration history of the GCL is not known, but for a range of scenarios the formation time is within an order of magnitude of 10 Myr.

If the GCL was formed by the expansion of hot gas, its cooling time scale is $t_{cool}=3 k T \sqrt{f} / \Lambda n_e $ yrs, where $\Lambda$ is the cooling coefficient and $f (\gtrsim0.1)$ is the filling factor for the hot gas \citep{m88,h90}.  For the hot gas observed in the central parsecs, with density and temperature similar to the gas that fuels extragalactic outflows \citep[$T\approx10^7$ K, $n_e\approx0.1$ cm$^{-3}$;][]{m04,k96}, we find $t_{cool} \approx 7\times10^7 \sqrt{f}$ \citep{m04}.  Since the cooling time scale is longer than its formation time scale under most models, the gas should expand adiabatically.  Gas hotter than $10^7$ K cools less efficiently, so its $t_{cool}$ is even larger.  The adiabaticity of the hot gas in the GC region is similar to that observed in the hot gas the powers extragalactic outflows \citep{v94}.  

It would also be useful to have an estimate of the survival time of the GCL, although existing observations do not constrain this well.  In general outflow models, the outflow begins in the ``snowplow'' phase, which lasts until the outflow has a height of few times the scale height of the ambient medium.  Eventually the outflow enters the ``blowout'' phase, where the pressure differential between the tenuous interior gas and dense exterior gas leads to a Rayleigh-Taylor instability that disrupts the shock front \citep{h90,su94}.  The contiguous shape of the GCL and its modest height of $\approx165$ pc \citep[only roughly twice the \hi\ scale height in the nuclear disk;][]{r82} suggest that it is still in the snowplow phase.  Unfortunately, without a measure of its expansion (if any) perpendicular to the line of sight, no constraint can be made on the time until it blows out of the disk.  For a sense of scale, an expansion velocity of 10 \kms\ will double the height of the GCL after 10 Myr.

The limits on the expansion velocity from radio recombination line observations is far smaller than the escape velocity for the GC region.  For a galactic potential including bulge, disk, and halo components, the escape velocity is about 900 \kms\ for galactocentric radii less than 500 pc \citep{b91}.  Thus, at its present velocity, the GCL will not break out of the Galaxy.  However, this assumes that the energy input to create the GCL was an impulsive event; it is also possible that the energy input may have had an extended duration and may still operate at present.  If the energy source that created the GCL continues to operate, then it may expand further or even accelerate, depending on the history of the energy input \citep{v05}.  If the GCL is formed by the expansion of hot gas, the ``escape temperature'' is $T_{\rm{es}}=1.1\times10^5 (v_{\rm{es}}/100$ \kms$) \rm{K} \approx 9\times10^6\ \rm{K}$  \citep{w95}, which is approximately 1 keV.  However, no hot, X-ray gas has yet been directly observed in the GCL, which makes the prediction of its ultimate fate difficult.  

Finally, the possible rotation of the ionized gas in the GCL may help constrain its origin.  The strip of GBT pointings across the GCL at $b=0\ddeg45$ shows a velocity gradient with a similar sign as the Galactic rotation (positive on the east, negative on the west).  If there is a trend for the ionized gas in the GCL to rotate like the disk gas, it would be consistent with the idea that the gas comes from the disk and was entrained in an outflow, as is observed in other galaxies \citep{s98,wa02}.  This entrainment could also explain the gradient in \rmeff observed in the polarized radio continuum under the flux-dragging scenario (see \ref{all_vlapoln}).  From east to west across the GCL, the mean gas velocity changes from +2.5 to --2.5 \kms; this is about a factor of 40 less than the linear velocity range of gas in the central 100 pc of the Galaxy \citep{b87,s04}.  Under the outflow model for the GCL, this gas likely originated in the plane and was carried away by the expansion of hot gas.  In extragalactic outflows, conservation of angular momentum reduces the velocity gradient from galactic rotation during expansion \citep{s98,s01}.  If angular momentum is conserved in the GCL, then the current angular momentum and rotation curve implies that the gas originated at a radius, $r_o \approx 2.5\ \rm{km\ s}^{-1} * 40\ \rm{pc}/100\ \rm{km\ s}^{-1} = 1\ \rm{pc}$.  However, there are two effects that could bias this estimate of $r_o$.  First, the range of gas velocities on the inner ``$x2$'' orbits could be much smaller than $\pm100$\ \kms \citep{bi91}.  Second, the angular velocity could be reduced by turbulence or resistance by magnetic fields, which would artificially lower the apparent line-of-sight gas velocity.  More sensitive observations at a range of latitudes are needed to see if the trend seen by the GBT extends over the the whole GCL.

\subsection{Starburst}
We now proceed to compare the observed characteristics of the GCL with those of specific models for nuclear outflows.  The first comparison is to the starburst outflow, which is one of the most widely observed causes for outflows in the local universe \citep{v05}.  Because of this, there are many specific observations of extragalactic outflows to which the GCL can be quantitatively compared.

Our estimate of the energy required to form the GCL ranges from a lower limit of $5\times10^{51}$ to a best estimate of $4\times10^{52}$ ergs, consistent with the energetics observed in small extragalactic starburst outflows.  Dwarf starburst galaxies have outflows with energies of $10^{50}-10^{54}$ ergs and ionized masses of $M=10^5-10^6$ \msol \citep{v05}.  The nearest starburst galaxy, the dwarf irregular IC 10, is host to a superbubble that may be considered an analogue of the GCL.  It is composed of a nonthermal radio continuum lobe with a diameter of about 200 pc that is filled with ionized gas \citep{y93,t05}.  The IC 10 superbubble is also similar to the GCL in that it has no clear sign of expansion; it has a kinetic (turbulent) energy of $5\times10^{52}$ ergs.  The energy expected to be deposited by stellar winds over its roughly $7\times10^6$-yr age is adequate to explain its formation, although supernovae may also have contributed \citep{t05}.  Note that canonical (i.e., massive) outflows have energies that are orders of magnitude higher, reaching levels of $10^{59}$ ergs.

The range of estimated energies required to create the GCL is roughly consistent with the energy expected from supernovae and stellar winds in the GC region.  The gravitational and thermodynamic energy estimates range from an equivalent of $(5-40)/\xi$ of the canonical, $10^{51}$-erg, type-II supernovae, where $\xi\approx0.2$ is the thermalization efficiency \citep{st03,v05}.  The supernova rate in the GC region has been estimated to be roughly $10^{-5}$ yr$^{-1}$ by (1) the need for an energy source for the diffuse, 0.8-keV gas in the GC region and (2) by scaling the Galactic supernova rate to the central 20 pc by mass \citep{m04,la02}.  For a formation time of about $10^7$ years, of order 100 supernovae are expected to occur in the central 20 parsecs.  Thus, the energy required to form the GCL is consistent with the energy expected from the expected GC supernova rate integrated over its nominal formation time.  Stellar winds may also make a significant contribution to the formation of the GCL.  \citet{c92} noted that the hot stars in the central parsec of the Galaxy have a wind power of roughly $5\times10^{37}$ ergs s$^{-1}$;  this power can account for the observed thermal energy of the GCL if it operates for $3\times10^7$ yr (assuming 50\% thermalization efficiency).  The estimate of stellar wind power is likely a lower limit, since it only considers the effect of about a dozen stars.  The mass loss by windy stars observed in the central parsec is now known to be a few times larger and other massive clusters of stars are observed within the span of the GCL \citep{g05,f99}.  Thus, the observed stellar wind power and nominal, GC supernova rate are adequate to power the GCL and it may be that no dramatic increase in star formation rate is needed to explain its formation.

The observed pressure in the GCL is similar to the analytic predictions of the pressure from the expected supernova rate.  Using the relations of \citet{c85}, \citet{v05} show that in the long term (or roughly 40 Myr after a burst of star formation), a given star formation rate will produce supernovae that inject momentum and create a pressure of $P/k = 3\times10^5 (SFR/$\msol yr$^{-1})(R_*/\rm{kpc})^{-2}$ K cm$^{-3}$.  For the observed star formation rate of 0.02 \msol yr$^{-1}$ in the central 100 pc of the Galaxy \citep{f04}, the predicted pressure is $P/k = 2.4\times10^6$ K cm$^{-3}$ as compared to the observed pressures of a few $10^6$ K cm$^{-3}$.  This suggests that the supernovae can explain the high pressure observed in the central 100 pc of the Galaxy.

The pressure profile observed from the central parsec out to the GCL is also consistent with a model powered by stellar winds \citep{h90,c92}.  The roughly constant gas pressure found within the GCL is consistent with earlier work that noted a constant pressure outside of $r\approx4$ pc rising by roughly two orders of magnitude in the central parsec \citep{c92}.  The pressure profile is consistent with that expected from energy injected by stellar winds in the central parsec.  In this model, the winds expand freely until they pass through a termination shock, after which the pressure drops to a relatively constant level.  The wind then continues to expand out to a contact discontinuity where the wind first meets the ISM.  In this outermost region, the outflow could look much like the GCL, with entrained mass and dust collected into a shell.  This model also predicts that the magnetic pressure will begin to dominate the total pressure outside the termination shock, which could explain the synchrotron-emitting shell of the GCL.  While this model found that stellar winds alone could be responsible for the observed pressure, accounting for the energy input from supernovae does not qualitatively change the results and would ease the energy requirements of the stellar winds.

A key test of any outflow model of the GCL should be able to explain the small velocities of its ionized gas.  Are the small velocities (at most $\sim20$ \kms) in the GCL consistent with observations of extragalactic outflows?  Typical edge-on extragalactic outflows, such as M82 and NGC 253, have expanding ionized gas shells with a line split by more than 200 \kms\ \citep{h90}.  The simplest explanation for the difference between the velocities observed in the GCL and canonical extragalactic outflows is the difference in their star formation rates.  \citet{m05} observe starburst outflow velocities that scale roughly as $SFR^{0.35}$, which suggests that the ratio of M82's outflow velocity to the putative GC outflow would be $(5/0.02)^{0.35}\approx7$ \citep{h90,f04}.  However, even without considering the difference in star formation rates, the small velocities observed in the GCL may not be inconsistent with extragalactic outflows when considering observational biases.  The inclination angles of the nearest extragalactic outflows makes the observation of line velocities difficult for galactocentric distances on the scale of the GCL ($\sim100$ pc).  Since these outflows typically accelerate with increasing galactocentric distance \citep{c85,su94}, the magnitude of the line split for distances comparable to the size of the GCL may be smaller than are presently considered ``normal'' outflow velocities.  In fact, \citet{s98} observe a line split in the M82 outflow of $\sim50$ \kms\ within 200 pc of the nucleus.  That work suggests that the flow in the inner 350 pc of M82 is highly collimated by ambient and entrained material, such that only the more distant parts of the outflow show large line splits.  Thus, the difference in star formation rate and the collimation of outflows at small size scales may help explain the small observed gas velocities in the GCL.

In the standard outflow theory presented in \citet{v05}, an energy-conserving shell has a velocity:
\begin{equation}
v_{shell} = 670 (\xi \dot{E}_{44}/n_o)^{1/3} r_{shell,kpc}^{-2/3}\ \rm{km\ s}^{-1}
\end{equation}
\noindent where $\xi$ is the thermalization efficiency of the mechanical energy and $n_o$ is the ambient electron density in cm$^{-3}$.  The observed star formation rate and pressures in the central 100 pc suggest an supernova energy deposition rate $\dot{E}_* = 1.4\times10^{40}$ ergs s$^{-1}$.  For a shell radius of 50 pc, $n_o=1-10$ cm$^{-3}$, and $\xi=0.2$ \citep{st03,v05}, the expansion velocity should be $v_{shell}=150-75$ \kms.  While these velocities are larger than observed in the GCL, this model does not account for the collimating effects of a stratified ISM.   While collimation likely reduces the line-of-sight velocity of the outflow, this effect is not well constrained observationally.  

\subsection{Jet}
Another commonly observed source powering nuclear mass outflows are active galactic nuclei \citep[AGNs;][]{v05}.  An AGN is a massive ($>10^6$ \msol) black hole that accretes mass from a dense disk of gas.  The accreted gas is highly luminous and magnetic processes in the accreting disk leads to the formation of powerful bipolar jets.  The black hole at the center of the Milky Way, Sgr A*, has a mass comparable to that seen in typical AGN \citep[e.g.,][]{fe01}.  This has raised the possibility that Sgr A* can behave like an AGN, emitting intense radiation and launching powerful jets.  Here we discuss some of the characteristics of jet-fuelled mass outflows and whether this model can explain the GCL.

Observations from radio to X-ray wavelengths have shown that, although Sgr A* does not currently behave like an AGN, it has the potential to do so.  Sgr A* is known for being remarkably underluminous, typically having a luminosity $\sim10^{-9}$ of its Eddington luminosity.  Not only does this imply that it is relatively inactive, but that the manner by which it emits radiation is much less efficient than seen in other massive black holes \citep{m01,mar05}.  Recent high resolution radio continuum observations have detected the intrinsic size of Sgr A*, but found no sign of a jet larger than about $10^{-6}$ parsecs \citep{bo04}.  However, the flaring behavior and spectral energy distribution of Sgr A* are best explained by a jet with a size roughly consistent with that upper limit at centimeter wavelengths \citep{fa00,fa04}.  More recently, several multiwavelength observing campaigns of Sgr A* have established that it has regular flares, which are likely tied to changes in its accretion rate or the way it accretes \citep{e06,mg05,y06}.  More controversially, there have been suggestions that Sgr A* has exhibited very powerful flaring, up to $10^6$ times its current X-ray luminosity, in the past few hundred years \citep{su98,r04}.  With good evidence for a jet in Sgr A*, there is strong motivation for the idea that a period of strong accretion activity could power a jet comparable to those seen in AGNs and could create the GCL outflow.

The extremely low luminosity of Sgr A* rules out any chance that it currently powers a jet that creates the GCL.  If a jet from Sgr A* formed the GCL, it must have been at some time long enough in the past such that the jet is no longer seen, but its effect on the ISM remains in the form of the GCL.  The minimum time since this jet existed is equal to the propagation time for the jet from Sgr A* to the GCL.  For a height of about 100 pc and a speed approximately equal to $c$, the minimum time is about 300 years.  The maximum time since any such jet was active is the disruption time for the GCL, which is not constrained.  Relativistic jets hit ambient galactic gas and drive mass outflows with velocities ranging from 100--1000 \kms \citep{t92,s97}.  Seyfert galaxies drive outflows with masses of ranging from $10^5-10^7$ \msol and energies from $10^{53}-10^{56}$ ergs.  Thus, AGN jets certainly have the capacity to power an outflow like the GCL.

There are a few characteristics of outflows that can help determine if they were formed as a result of an AGN jet.  In cases where the jet is directly observed to create an outflow, the radio emission from the jet correlates with optical line emission on scales of tens of parsecs.  This suggests that the relativistic jet shock-ionizes the ambient ISM where they interact \citep[e.g.,][]{sch03}.  Interestingly, the jet may have a random orientation relative to the galactic disk and in some cases points into the disk, with destructive results \citep{ce90}.  Active AGN are often identified by their intense optical/UV radiation, which create a bipolar ``ionization cone'' where ambient gas is photoionized \citep{ki00}.  Morphologically, a jet-fuelled outflow has a shape that can distinguish it from starburst outflows.  For example, radio galaxies like Cygnus A have a distinctive shape to their radio emission, with the brightest emission at the end of the radio lobe, where the jet interacts with the ISM \citep{p84}.  Jet-fuelled outflows also tend be have distinctive shapes because they are powered by very compact sources.  As such, the outflows tend to be much more narrow in the disk than beyond the disk, where collisions may reduce the outflow momentum and the ambient gas pressure is lower \citep{v05}.  

While many of the observed characteristics of the GCL match those seen in jet- and starburst-fuelled outflows, the morphology disfavors the jet model.  The GCL is brighter at its eastern and western edges than at its apex, which is inconsistent with the idea that a vertical jet from Sgr A* created it.  Also, the GCL is about 100 pc wide at its base, which is comparable to its height;  observations of jet-powered outflows tend to have more elongated structures and narrow bases.  These morphological points, combined with lack strong current activity in Sgr A*, suggests that a jet is unlikely to have created the GCL.

\subsection{8 keV Gas}
Another possibility for the formation of the GCL is the expansion of the hot gas observed in the central degree of the Galaxy.  X-ray observations of the central tens of parsecs have found two components of thermal gas with temperatures of about 0.8 keV and 8 keV \citep{k96,m04}.  As discussed above, the 0.8 keV gas is believed to be formed by the thermalization of stellar winds and supernovae and is gravitationally bound to the Galaxy.  However, the origin of the 8 keV gas is not known and it is too energetic to remain bound to the Galaxy \citep{m04,be05}.  It is possible that as this gas escapes from the Galaxy, it creates the structure we see as the GCL.  This mechanism operates similarly to the starburst mechanism, but may be distinct from it, since the hot gas may not be produced by stars.  The observed radiated power for the 8 keV emission integrated over the GC region is approximately $4\times10^{37}$ ergs s$^{-1}$ \citep{be06}.  Assuming that its radiated power is equal to its thermal power input, the mechanism that creates this hot gas could fuel the GCL if it operated for roughly $3\times10^7$ yr.

Determining if the escape of 8 keV gas can create a GCL-like outflow is difficult, since the origin of the gas is not well known.  One possible source for heating the gas is through viscous heating by molecular clouds \citep{be05,be06}.  In this model, infalling molecular clouds can convert gravitational energy into thermal energy by exciting Alfv\'en waves in the GC magnetosphere.  Using a reasonable estimate of the structure and strength of the GC magnetic field and molecular cloud properties, this model can generate enough energy to explain the 8 keV gas.  Unfortunately, the efficiency of the turbulent heating mechanism depends sensitively on the structure and strength of the GC magnetosphere, which is currently a topic of strong debate \citep{m96,b06}.  Furthermore, the model does not predict any unique observational test that can prove it is the source of the gas heating.  If these issues can be sorted out, then the power input by this turbulent heating mechanism can be more narrowly constrained and it will be possible to determine if it is energetic enough and operates long enough to power the GCL.

Approaching this question from another perspective, it may be useful to ask: is there any presently observable property of the GCL that could uniquely identify that it is caused by the escape of this 8 keV gas?  In most respects, the escape of 8 keV gas would have similar signatures as the starburst model.  In both cases, there is a overpressurized region that expands as it buoyantly escapes.  The principle difference between the two mechanisms is whether molecular clouds or stellar winds and supernovae produce the overpressurized gas.  Both mechanisms are correlated with star formation, so they are both likely to function over a similar physical extent (within the central molecular zone, $r\approx200$ pc) and under similar circumstances \citep[i.e., strongest during mass infall events;][]{st04}.  One possible way to distinguish between the two models may be to search for 8 keV gas correlated with the GCL.  If 8 keV gas is found preferentially inside the GCL (particularly away from the confusion of the Galactic plane), it would suggest that the escaping gas is powering the outflow.  \emph{Chandra} observations of a portion of the GCL will likely be conducted in the middle of 2007 and may answer this question.

\subsection{Magnetic Twist}
An early suggestion for creating the GCL suggested that it is formed from an outflow of gas powered by a specific magnetic field geometry \citep{u85,s87}.  The magnetic structure is called a ``magnetic twist'' and is formed by the interaction of orbiting gas clouds in a predominately poloidal (vertical) magnetic field.  Assuming that orbiting molecular clouds are partially ionized, the magnetic flux will be ``frozen'' into and dragged with the cloud.  In this case, the motion of the gas, as dictated by gravity and gas dynamics, can strongly modify the structure of the GC magnetosphere, with the field parallel to the disk in the plane and perpendicular outside of the plane.  Simulations have shown that the rotating, infalling gas can also twist and pinch the field toward the GC, such that a net vertical force is applied by the magnetic field on the rotating gas \citep[from the ${\bf J}\times {\bf B}$ force][]{u85,s87}.  It has been suggested that this vertical force is strong enough to power and outflow and create the GCL.

There are observations that seem to confirm some aspects of the magnetic twist model.  Observations of submm polarimetry of dust in the GC region have found evidence supporting the idea that molecular clouds modify the structure of the GC magnetosphere \citep{c03,n03}.  \citet{c03} observe that the orientation of the magnetic field in molecular clouds is more likely to be parallel to the plane of the Galaxy than outside of clouds.  They suggest that the magnetic field is intrinsically poloidal, but is sheared and dragged by the motion of molecular gas.  Furthermore, our observations of an east-west gradient in \rmeff towards the radio continuum of the the GCL is consistent with the line-of-sight magnetic field expected when gas in the disk drags a vertical magnetic field.

However, there are significant challenges to the magnetic twist model in general and in its application to the GCL.  First, the mechanism has only been tested in simulations of the GC region; no analytic work has shown that the effect occurs in a general case.  Second, the primary focus of simulations of the magnetic twist effect is to duplicate the morphology of the GCL, but it does not quantify how much power it can generate, so it is difficult to compare to observations.  Third, it isn't clear how the model would produce the layered structure of the GCL, and particularly the presence of ionized gas.  Finally, the model essentially requires that the GCL be symmetric about the center of rotation at Sgr A*, which is not the case.  Thus, while there is some reason to believe that molecular clouds can affect the structure of the GC magnetosphere, the magnetic twist model is not yet a viable model for a description of the GCL.

\subsection{Remaining Questions}
\subsubsection{AFGL 5376}
AFGL 5376 is an extended infrared source discovered by IRAS near the GC that is believed to be a shocked molecular cloud \citep{u90,u94}.  This hypothesis is based on the observation of shocks in molecular gas associated with AFGL 5376 and their coincidence with the infrared and radio continuum emission in the region.  The morphological connection between AFGL 5376 and the western half of the GCL is compelling and suggests a connection between these features \citep{u94}.  Comparing our radio continuum images to \spitzer\ mid-IR images strengthens this connection, since there are wispy, elongated structures in the mid-IR and radio continuum images coincident with shocks in the molecular gas.  While the morphology suggests that AFGL 5376 and the GCL are interacting, the direction of causality is unclear.  \citet{u94} note that the velocity jump across the shock in AFGL 5376 could ionize the gas and generate the radio continuum of the GCL.  While the GCL spectral index near AFGL 5376 is consistent with thermal emission, the GCL is a contiguous structure that is predominately nonthermal;  it would be an surprising if there were thermal emission confused with the nonthermal GCL that did not change its appearance.  More compellingly, we have shown that the GCL is filled with ionized gas, as seen in radio recombination line emission.  There is no clear way for AFGL 5376 to generate both the GCL and the ionized gas that fills it, so it seems that either they were both formed as a part of the same process or the GCL must have existed before AFGL 5376.


So can the connection between AFGL 5376 and the GCL be consistent with the concept of the GCL as a coherent shell?  One possibility is that AFGL 5376 is the result of a collision between a molecular cloud with the mass and magnetic field of the GCL.  The molecular gas in AFGL 5376 is peculiar in that it is massive ($10^6$ \msol) and moves at more than 122 km s$^{-1}$ counter to galactic rotation \citep{u94};  this makes it very likely that it will have an energetic collision with ambient, co-rotating gas.  From the mass estimate for the ``CO tongue'' associated with AFGL 5376, we estimate the average H$_2$ mass density of about $7\times10^{-22}$ g cm$^{-3}$ (assuming a depth similar to its width \footnote{This mass density gives a H$_2$ number density of about 200 cm$^{-3}$, which is less than the critical density to excite $^{13}$CO, and consistent with its nondetection by \citet{b87}.}).  Thus, the ram pressure of this feature is roughly $\frac{1}{2} \rho v^2 \sim 4\times10^{8} (v/122$ km s$^{-1})$ K cm$^{-3}$.  This pressure is two orders of magnitude larger than that observed in the GCL, so it seems unlikely that the collision of this counter-rotating gas with the shell of the GCL would cause the observed shock.

Alternatively, the molecular gas in AFGL 5376 is consistent with it being a part of the ``Expanding Molecular Ring'' (EMR), which may help understand its behavior.  As suggested by its name, the EMR is a system of gas in the central few hundred parsecs that seems to be radially expanding \citep[e.g.,][]{s95}.  However, it has been shown that the velocity structure is more simply explained by the gas dynamics of a barred gravitational potential \citep{m96,s04}.  In models of gas motion in a barred potential, AFGL 5376's position and velocity are consistent with it being on a self-intersecting orbit, such that the gas will eventually collide with itself.  Thus, it is possible that the shock in AFGL 5376 is caused simply by its motion through the gravitational potential in the GC region.  If so, it is not clear how this relates to the GCL, since there is no similar shocked molecular cloud on the eastern half of the GCL.

\subsubsection{Double Helix Nebula}
Recent mid-IR observations with \spitzer\ have found a pair of twisted filaments about 25 pc in length called the ``double helix nebula'' \citep[DHN;][]{m06}.  The twisting of the DHN strongly suggest that its structure is dominated by magnetic forces.  However, the DHN is clearly associated with the GCL because it makes up the brightest mid-IR emission in the GCL-East.  

The association of the DHN with the GCL is a challenge for the outflow model for the GCL for multiple reasons.  First, the model proposed for the DHN is that of a Alfv\'en wave propagating from the molecular gas in the central parsec of the Galaxy \citep{m06}.  This model suggests that the structure is unique and unrelated to the emission in the eastern half of the GCL.  Second, if the DHN is a part of the outflowing shell of the GCL, then the shell wall is clearly more than a simple, one-dimensional shock front.  Third, no similar mid-IR structure is seen in the eastern half of the GCL, which suggests that the two sides are not related to each other.

While the nature of the DHN is still unclear, there are ways in which it can fit in our proposed model for the GCL.  First, the model of the DHN as an Alfv\'en wave requires propagation from the central parsec, but there is little morphological evidence for such a propagation.  \citet{m06} identify two features in the 8 $\mu$m \msx\ map of the GC region that could indicate a path, but at least one of the features is also seen in optical H$\alpha$ emission and is thus likely to be in the foreground \citep{s07}.  Second, a wavy, filamentary structure like the DHN has been observed in optical emission lines at the shock front of the SNRs \citep[e.g., the Cygnus Loop][]{bl99}.  These structures, which arise naturally as the shock front propagates into an inhomogeneous medium, may also shape the putative shock front of the GCL outflow.  Finally, we note that the wavy structure of the DHN is mirrored by 6 cm continuum emission from G0.03+0.66, located adjacent to the DHN to the east (Ch. \ref{gcl_vla}).  G0.03+0.66 has no 20 cm counterpart, which suggests that it has a flat, thermal spectral index.  The nested, thermal radio emission with mid-IR emission is reminiscent of the general structure of the GCL described here, which suggests that the DHN may simply be fine-scale structure in the GCL.  However, more detailed analysis is needed before any clear conclusions can be made about the relationship between the DHN and the GCL.

\subsubsection{Asymmetries}
\label{asymmetries}
As described throughout this work, there are several asymmetries associated with the GCL that need to be explained by the models presented.  One asymmetry is the offset of the GCL toward the west of Sgr A*, the origin of which is not clear.  The center of the GCL is offset to the west of Sgr A* by $0\ddeg3$--$0\ddeg4\approx40$--$55$ pc or nearly half its width.  The starburst outflow model may best explain why such an offset can exist, since the center of the outflow is only constrained to be where the star formation is most vigorous \citep{v05}.  Thus, the central longitude of the GCL may trace the center of the star formation that powered the outflow.  This concept may also explain the possible lack of a southern counterpart to the GCL;  it may be that the star formation was offset north of the Galactic plane or the outflow was constrained by ambient molecular gas, as has been observed to some degree in other galaxies \citep{s98}.

The eastern offset of the GCL from Sgr A* may also help explain its internal asymmetries.  Assuming that the center of the GCL is the center of the outflow, one can imagine that the eastern and western sides of the outflow have encountered drastically different conditions during their expansion.  The large polarization fraction and connection with the unusual Radio Arc in the east of the GCL may be the result of the expansion of the outflow through the relatively dense and possibly stronger magnetic field in the central few parsecs.  Meanwhile, the western side of the GCL would expand relatively unimpeded through larger galactocentric radii.  Thus, in this sense, the offset of the GCL from Sgr A* may explain the peculiarities of its structure.

While the asymmetry of the GCL may be explained by an offset origin, it is not clear how this model can explain the possible offset in the GC magnetosphere.  Earlier we noted that the distribution of \rmeff observed toward radio continuum emission in the GC region has an east-west gradient that seems to be centered at the center of the GCL.  Seemingly independently, the spatial distribution of NRFs is found to be offset from Sgr A* by a similar amount \citep{y04}.  Although the structure of the GC magnetosphere is a topic of debate, most mechanisms are generally thought to be symmetric about Sgr A*.  The coincidence of the GCL with these two, independent indicators of the GC magnetosphere suggests that the magnetic field is connected to the GCL.  This would be consistent with the rough equipartition between magnetic pressure and the other pressures observed in the vicinity of the GCL.

\section{Conclusions}
We have conducted a multiwavelength observing program to study the GC lobe and determine if it is a signature of an outflow from the GC region.  The new observations show that the GCL has a layered structure with concentric mid-IR, radio continuum, and radio recombination line shells.  The morphological, spectral, and polarization properties of the GCL give compelling evidence in favor of it having a single origin and that it is in the GC region.  Based on the idea that the GCL is a contiguous shell located in the GC region, we calculate the energy and time required to create it and compare its characteristics to models for powering nuclear outflows.  The canonical starburst model is found to be an excellent fit to the properties of the GCL and that the energy output by current star formation rates are adequate to create it.  A jet from Sgr A* or a ``magnetic twist'' effect could also create the GCL, but some aspects of the GCL are not well fit by these models.  We also find that the escape of the observed 8 keV gas in the central tens of parsecs may create the GCL, although a detailed prediction is difficult since the origin of this hot gas is not known.

If the GCL is indeed formed by the outflow of gas from stellar winds and supernovae, it would be a confirmation that this exciting phenomenon is found not only in distant galaxies, but in our own.  While the scale of the putative GCL outflow is quite modest by extragalactic standards, it would be the closest example of the outflow phenomenon.  As such, it opens the possibility of studying an outflow at extremely high physical resolution and expanding our understanding of how they work.  In particular, the relatively small size of the GCL explores a range of outflow parameter space that is difficult to explore in other galaxies.  The fact that the GCL would very be difficult to detect in even the nearest spiral galaxies suggests that this type of outflow could be common in such systems.  
The recent identification of a method for the periodic infall of mass to the central parsecs suggests that the GC region goes through cyclical enhancements to the star formation rate \citep{st04}.  It may be that the GCL is just the most recent example of a long series of minor outflows from episodes of stellar activity in our GC region.

%% file: summary_astro-ph.tex
\chapter{Summary}
This thesis has used an observational approach to study a wide range of phenomena in the GC region.  The results apply to star formation regions and nonthermal filaments, with a particular focus on understanding the origin of the GCL.

Chapter \ref{gcsurvey_gbt} describes our multiwavelength, GBT survey of the central few degrees of the GC region.  The 3.5 and 6 cm observations have the best resolution, sensitivity, and coverage of any single-dish survey of the region.  This survey has been used to construct catalogs of compact and extended sources in the central degrees.  A detailed spectral index study has been performed for all extended sources in the survey region, allowing the separation and quantification of the contributions of thermal and nonthermal emission in the region.

In chapter \ref{gcl_recomb}, we study the radio recombination line emission from the GCL.  Observations from the HCRO show the that the emission from H109$\alpha$ is found from throughout the interior of the radio continuum shell of the GCL.  The velocity of the gas in the GCL ranges from --5 to +5 \kms\ and shows little spatial structure.  Pointed GBT observations of a range of H and He recombination lines provide a more detailed view of the gas conditions in the GCL.  These observations show that stimulated emission cannot account for the line ratios and that the gas is photoionized.  The gas has an electron density of about 1000 cm$^{-3}$ and the temperature is at most 3960 K;  this temperature is unusually low, but consistent with the high metallicity expected for the GC region.  A very low filling fraction of about $10^{-4}$ is implied by the data, which is consistent with observations of extragalactic outflows, where the gas is ionized at the surface of molecular material entrained in the outflow.

The VLA continuum observations of the GCL at 6 and 20 cm are presented in chapter \ref{gcl_vla}.  These observations give new, sensitive catalogs of the spectral index of compact and filamentary structures in the region.  We find evidence for spatial gradients on the spectral index of the filaments, which may indicate synchrotron aging in their electrons or changes in the magnetic field across the filaments.  Three candidate filaments are confirmed and two other unusual filament candidates are discovered.

In chapter \ref{gcl_vlapoln} we describe the linearly-polarized continuum seen in the 6 cm VLA survey of the GCL.  This work develops a new technique for estimating the rotation measure from standard, two-IF VLA data.  The polarized emission from the nonthermal radio filaments is described in detail.  Linearly-polarized emission is found throughout the field of view toward the GCL, which we suggest originates in the GC region.  The rotation measure of this polarized emission shows a striking gradient from Galactic east to west at all latitudes in the GCL.  This gradient is consistent with the rotation measure toward the filaments and other observations of the rotation measure in the GC region.  The rotation measure in the central two degrees forms a checkerboard pattern that has been interpreted as evidence for some large-scale structure in the GC magnetic field.  In particular, the pattern is consistent with a magnetic field that has some poloidal structure (pointing from south to north) that is dragged by the rotational motion of ionized gas in the Galactic disk.  However, the strength of the GC magnetic field is poorly constrained.

Finally, the observations of the GCL are summarized and compared to outflow models in chapter \ref{gcl_all}.  The data are consistent with the idea that the GCL is a three-dimensional structure located in the GC region.  The observations show that the GCL is composed of three nested shells of gas, with ionized, magnetized, and mid-IR--emitting components.  Several models for the formation of such a structure are presented.  The strength of star formation in the GC region, gas pressure, and ionized gas conditions in the GCL are consistent with the canonical model of a starburst outflow.  If this model for the GCL is valid, the outflow is smaller than typically observed (or can be observed) in other galaxies, which is consistent with the relatively moderate level of star formation in the GC region.  While alternate models for the formation of the GCL, including an AGN jet or magneto-dynamic processes, are not ruled out, the observed star formation rate in the GC region is sufficient to explain the formation of the GCL.

%% file: cv.tex
\chapter{Vita}

\begin{flushright}
Northwestern University\\
Department of Physics and Astronomy\\
Dearborn Observatory\\
2131 Tech Drive\\
Evanston, IL 60208
\end{flushright}
 
\begin{flushright}
Work Phone:  (847) 491-4661\\
Email:  claw@northwestern.edu\\
Web page: http://www.astro.northwestern.edu/$\sim$claw
\end{flushright}
 
\section*{Education}
\indent
 
Northwestern University, Ph.D., Physics and Astronomy (2007)
 
Boston University, M.A., Astronomy (2000)
 
University of Hawai`i, B.S. with distinction, Physics (1998)
 
\section*{Employment}
\indent
 
\begin{bf}Summer 2006 -- present:  \end{bf}Postdoctoral fellowship at the Universiteit van Amsterdam in the Netherlands working under Ralph Wijers.  My work centers on commissioning of the new radio telescope, LOFAR, in the area of radio transient searches.

\begin{bf}Summer 2002 -- Summer 2006:  \end{bf}Performed research with Prof. Farhad Yusef-Zadeh at Northwestern University.  My research has made use of X-ray and radio observations to study many aspects of the Galactic center environment.  In particular, we have studied on the X-ray emission from dense stellar clusters and molecular clouds, performed single-dish and interferometric radio observations, and a \emph{Spitzer}/IRAC infrared survey of the GC region.
 
\begin{bf}Fall 2003 -- present:  \end{bf}Provided tours of the historic Dearborn Observatory at Northwestern University approximately once per month.  Tours are open to the public and include viewings of objects in the night sky.

\begin{bf}Spring 2004:  \end{bf}Teaching Assistant at Northwestern University.
 
\begin{bf}Fall 2000 -- Spring 2002:  \end{bf}Astrophysicist at the Harvard-Smithsonian Center for Astrophysics.  I worked on the development, testing, and documentation of the CIAO software package for the \emph{Chandra X-Ray Observatory}.  My responsibilities also included supporting \emph{Chandra} users with their data reduction needs and problems.
 
\begin{bf}Fall 1999 -- Summer 2000:  \end{bf}Performed research with Prof. James Jackson of Boston University.  As a member of the BU-FCRAO Galactic Ring Survey, I helped with observing, processing, and analyzing data from the 14 m FCRAO telescope.
 
\begin{bf}Fall 1999 -- Spring 2000:  \end{bf}Taught lab sections under the Presidential University Teaching Fellowship.
 
\begin{bf}Fall 1998 -- Fall 1999:  \end{bf}Performed research as a Presidential University Graduate Fellow with Prof. Kenneth Janes of Boston University.  My research involved developing new systems of analyzing photometry of stellar clusters.
 
\begin{bf}Summer 1999:  \end{bf}Interned at Textron Systems Kaua`i, developing plans for a new electro-optics site at the Pacific Missile Range Facility on Kaua`i.
 
\begin{bf}Spring 1999:  \end{bf}Performed research as a Presidential University Graduate Fellow with Prof. Teresa Brainerd of Boston University.  My research with Prof. Brainerd made use of BU's supercomputer to study galaxy clustering in Keck images of QSO fields.
 
\begin{bf}Summer 1998:  \end{bf}Worked with Saroj Sahu in the High Energy Physics Group at the University of Hawai`i on the BELLE project.  BELLE's purpose was to detect CP violation at the KEK-B particle accelerator in Japan.  The work consisted of designing and testing data acquisition hardware and software.
 
\begin{bf}Fall 1997 -- Spring 1998:  \end{bf}Performed research as a Hawai`i Space Grant Fellow with Prof. James Heasley of the Institute for Astronomy in Honolulu.  My Space Grant research project, titled ``Examining the Relative Ages of Thick Disk Globular Clusters'', consisted of analyzing data taken at the UH 2.2 m telescope and the Canada-France-Hawaii Telescope on the globular cluster M71.
 
\section*{Observing Experience}
\indent
 
\begin{bf}Very Large Array:  \end{bf}Proposed and conducted continuum surveys of the Galactic center lobe at 6 and 20 cm.
 
\begin{bf}Green Bank Telescope:  \end{bf}Proposed and conducted polarized (synthesized) continuum observations of the Galactic center at 3.6 and 6 cm.  Assisted in observing 3.6, 6, 20, 90 cm continuum surveys of the entire GC region.  Proposed and conducted 6 cm radio recombination line observations.
 
\begin{bf}Chandra X-Ray Observatory:  \end{bf}Planned observing proposals with the ACIS instrument and reduced data.
 
\begin{bf}Spitzer Space Telescope:  \end{bf}Participated in proposal to survey the Galactic center.
 
\begin{bf}FCRAO 14 m:  \end{bf}Conducted remote observations for a survey of molecular transitions of $^{13}$CO and CS.
 
\begin{bf}Boston University-Lowell Perkins 72 inch:  \end{bf}Observed independently and cooperatively on optical and near-IR projects.
 
\section*{Fellowships and Honors}
\indent
                                                                                
\begin{bf}2005:  \end{bf}Awarded NRAO-GBT Student support grant toward the analysis of GBT polarized continuum surveys of the Galactic center.
 
\begin{bf}2004:  \end{bf}Awarded NRAO-GBT Student support grant toward the analysis of multiwavelength GBT surveys of the Galactic center.
 
\begin{bf}2004:  \end{bf}Awarded \emph{Chandra} archival research grant to develop a spatio-spectral analysis technique for \emph{Chandra} observations.
 
\begin{bf}2002 -- 2003:  \end{bf}Huang Fellowship at Northwestern University
 
\begin{bf}2002:  \end{bf}College Summer Research Program Participant (Illinois Space Grant Consortium and NASA)
 
\begin{bf}1999:  \end{bf}Boston University Presidential University Teaching Fellowship
 
\begin{bf}1998:  \end{bf}Boston University Presidential University Graduate Fellowship
 
\begin{bf}1995 -- 1998:  \end{bf}Four merit-based tuition waivers from the Department of Physics at the University of Hawai`i
 
\begin{bf}1997:  \end{bf}Hawai`i Space Grant Fellowship
 
\begin{bf}1997:  \end{bf}$\Sigma\Pi\Sigma$ (the honor society of the Society of Physics Students) member